Maunakea Spectroscopic Explorer

# The Detailed Science Case for the Maunakea Spectroscopic Explorer:

# The Composition and Dynamics of the Faint Universe

01.01.00.003.DSN
Version: A

Status: *Exposure draft*

2016-05-27

| **Prepared By:** | | |
|---|---|---|
| **Name(s) and Signature(s)** | **Organization** | **Date** |
| Alan McConnachie<br>MSE Project Scientist | MSE Project Office | 2016-05-27 |
| **Approved By:** | | |
| **Name and Signature** | **Organization** | **Date** |
| Rick Murowinski<br>MSE Project Manager | MSE Project Office | 2016-05-27 |
| | | |
| | | |
| | | |

## Change Record

| Version | Date | Affected Section(s) | Reason/Initiation/Remarks |
|---|---|---|---|
| A | 2016 May 27 | All | Exposure draft (public release) |
| | | | |
| | | | |
| | | | |
| | | | |
| | | | |

**Table of Contents**

# Preface

## P.1 Purpose and Scope

This is the MSE Detailed Science Case (DSC). It is the highest level document in the MSE document hierarchy, and forms the foundation for the Level 0 Science Requirements Document (SRD).

The MSE DSC provides the science narrative describing the principal envisioned science goals of MSE and describes its impact on a broad range of science topics. Science Reference Observations (SROs) describe in detail specific transformational observing programs for MSE that span the range of science described in the DSC. Science Requirements are defined as the science capabilities required to conduct the SROs. The SROs are included as Appendices to the DSC.

## P.2 A note on the structure of the DSC

The DSC is an extensive document covering a range of science topics. All chapters and all appendices can be read as stand-alone documents. All chapters begin with a 1 – 2 page synopsis that summarizes the content of the chapter. The first chapter provides a summary of the entire document.

A 10 page summary of the entire project aimed at the international astronomy community is presented in an accompanying document, “A concise overview of the Maunakea Spectroscopic Explorer”.

## P.3 Credit and Acknowledgements

This document represents contributions from over 150 scientists across the international astronomical community from the past 5 years. It has been compiled and edited by the MSE Project Scientist in close collaboration with members of the MSE Science Team. It is based on a large number of documents developed by members of the MSE Science Team and its precursor project, the Next Generation Canada-France-Hawaii Telescope (ngCFHT).

The MSE Project Office was established in 2014 after precursor studies led to the ngCFHT Feasibility Studies. The MSE Science Team then undertook a multi-stage process to better define the science capabilities of MSE and develop the associated science case. White Papers were drafted highlighting important science areas not necessarily previously considered by the ngCFHT studies. Phase 1 studies, under the coordination of three leads, were then conducted. The purpose of these studies was to identify the most compelling science areas for further development in three broad areas (Stars and the Milky Way, The Low Redshift Universe, The High Redshift Universe) and to present the science cases for each. Cooperation between these groups was encouraged for areas of overlap. This led to the first set of proposed Science Reference Observations. From this initial set, a shortlist of 12 SROs was selected for further development. These SROs form the basis for the derivation of the Science Requirements.

The main chapters of the Detailed Science Case are based on the compilation of material that has been developed for MSE during this multi-stage process, including the ngCFHT Feasibility Study. The final set of SROs is presented in as Appendices to the DSC.

All original science documents on which the DSC is based are available at http://mse.cfht.hawaii.edu/docs/sciencedocs.php

### P.3.1 Science team leads and coordinators

Carine Babusiaux (Stars and the Milky Way)
Michael Balogh (The low redshift Universe)
Simon Driver (The high redshift Universe)

Pat Côté (Lead author, the ngCFHT Feasibility Study 2012)

### P.3.2 Science Reference Observations leads and contributors

(Full author lists for each SRO are given in the Appendices)

*Leads*

Carine Babusiaux (SRO-3, Milky Way archaeology and the in situ chemical tagging of the outer Galaxy)
Michael Balogh (SRO-6, Nearby galaxies and their environments)
Helene Courtois (SRO-12, Dynamics of the dark and luminous cosmic web during the last three billion years)
Luke Davies (SRO-11, Connecting high redshift galaxies to their local environment: 3D tomographic mapping of the structure and composition of the IGM, and galaxies embedded within it)
Simon Driver (SRO-9, The chemical evolution of galaxies and AGN over the past 10 billion years (z<2)
Laura Ferrarese (SRO-7, Baryonic structures and the dark matter distribution in Virgo and Coma)
Sarah Gallagher (SRO-10, Mapping the inner parsec of quasars with MSE)
Rodrigo Ibata (SRO-4, Stream kinematics as probes of the dark matter mass function around the Milky Way)
Nicolas Martin (SRO-5, Dynamics and chemistry of Local Group galaxies)
Aaron Robotham (SRO-8, Evolution of galaxies, halos and structure over 12Gyrs)
Kim Venn (SRO-2, Rare stellar types and the multi-object time domain)
Eva Villaver (SRO-1 Exoplanets)

*Contributors*

Jo Bovy, Alessandro Boselli, Matthew Colless, Johan Comparat, Pat Côté, Kelly Denny, Pierre-Alain Duc, Sara Ellison, Richard de Grijs, Mirian Fernandez-Lorenzo, Laura Ferrarese, Ken Freeman, Raja Guhathakurta, Patrick Hall, Andrew Hopkins, Mike Hudson, Andrew Johnson, Nick Kaiser, Jun Koda, Iraklis Konstantopoulos, George Koshy, Andrew Hopkins, Khee-Gan Lee, Adi

Nusser, Anna Pancoast, Eric Peng, Celine Peroux, Patrick Petitjean, Christophe Pichon, Bianca Poggianti, Carlo Schmid, Prajval Shastri, Yue Shen, Chris Willot

### P.3.3 DSC Editors and principal contributors to new material in the DSC

Michael Balogh (Low mass galaxies)
Jo Bovy (CDM halo substructures)
Pat Côté (Low mass stellar systems)
Scott Croom (IFU studies of galaxy evolution)
Sara Ellison (Galaxy mergers)
Laura Ferrarese (The Virgo Cluster)
Sarah Gallagher (Reverberation mapping)
Rosine Lallement (The Interstellar Medium)
Nicolas Martin (The Local Group)
Patrick Petitjean (The Intercluster Medium)
Carlo Schimd (The Dark Universe)
Dan Smith (SKA synergies)
Matthew Walker (Nearby dwarf galaxies and dark matter)
Jon Willis (Galaxy clusters)

### P.3.4 Phase 1 study report co-authors

Carine Babusiaux, Michael Balogh, Alessandro Bosselli Matthew Colless, Johan Comparat, Helene Courtois, Luke Davies, Richard de Grijs, Simon Driver, Sara Ellison, Laura Ferrarese, Rodrigo Ibata, Sarah Gallagher, Aruna Goswami, Mike Hudson, Andrew Johnson, Matt Jarvis, Eric Jullo, Nick Kaiser, Jean-Paul Kneib, Jun Koda, Iraklis Konstantopoloulous, Rosine Lallement, Khee-Gan Lee, Nicolas Martin, Jeff Newman, Eric Peng, Celine Peroux, Patrick Petitjean, Christophe Pichon, Bianca Poggianti, Johan Richard, Aaron Robotham, Carlo Schmid, Yue Shen, Firoza Sutaria, Edwar Taylor, Kim Venn, Ludovic van Waerbeke, Chris Willot

### P.3.5 Science White Papers

Giuseppina Battaglia (The accretion history of the Milky Way halo through chemical tagging)
Pat Cote (Upgrades to the MSE Point Design Specifications)
Aruna Goswami (Study of old metal-poor stars)
Richard de Grijs (Spatially resolved stellar populations and star formation histories of the largest galaxies)
Pierre-Alain Duc (Probing the large scale structures of massive galaxies with dwarfs/GC/UCDs in a variety of environments)
Pat Hall, Charling Tao, Yue Shen, Sarah Gallagher (Opportunistic Transient Targeting)
Misha Haywood, Paola di Matteo (Bulge science with MSE)
Misha Haywood, Paola di Matteo (Galactic archeology: The outskirts of the Milky Way disk)
Matt Jarvis (SKA Synergies)
Iraklis Konstantopoulos (The Redshift Reject Rubric)
Iraklis Konstantopoulos (Dense, Low-Mass Galaxy Groups: A Phase-Space Conundrum)
Iraklis Konstantopoulos (Star Clusters as Chronometers for Galaxy Evolution)

Rosine Lallement (Galactic InterStellar Medium: the 3D era with MSE)
Patrick Petitjean (Three Dimensional Reconstruction of the InterGalactic Medium)
Charli Sakari, Kim Venn et al. (Integrated Light Spectroscopy at High Resolution with MSE)
Carlo Schmid et al.; updated October 2014 (MSE: a velocity machine for cosmology)
Arnaud Seibert (Milky Way structure and evolution, ISM)
Yue Shen (Multi-Object AGN Reverberation Mapping with MSE)
Sivarani Thirupathi (Tracers of Pre-galactic and extra galactic Lithium abundances in Milky Way: towards solving the lithium problem)
Yuting Wang (Optimal tests on gravity or cosmological quantities on large scales)
Yiping Wang (SFH and scale-length evolution of massive disk galaxies towards high-z)

### P.3.6 Other Contributions and Feedback

(not previously mentioned)

Ferdinand Babas, Steve Bauman, Elisabetta Caffau, Mary Beth Laychak, David Crampton, Daniel Devost, Nicolas Flagey, Zhanwen Han, Clare Higgs, Vanessa Hill, Kevin Ho, Sidik Isani, Shan Mignot, Rick Murowinski, Gajendra Pandey, Derrick Salmon, Arnaud Siebert, Doug Simons, Else Starkenburg, Kei Szeto, Brent Tully, Tom Vermeulen, Kanoa Withington

### P.3.7 MSE Science Team membership

Nobuo Arimoto
Martin Asplund
Herve Aussel
Carine Babusiaux
Michael Balogh
Michele Bannister
Giuseppina Battaglia
Harish Bhatt
SS Bhargavi
John Blakeslee
Joss Bland-Hawthorn
Alessandro Boselli
Jo Bovy
James Bullock
Denis Burgarella
Elisabetta Caffau
Tzu-Ching Chang
Andrew Cole
Johan Comparat
Jeff Cooke
Andrew Cooper
Pat Cote
Helene Courtois
Scott Croom
Richard de Grijs
Paola Di Matteo
Simon Driver
Pierre-Alain Duc
Sara Ellison
Ginevra Favole
Laura Ferrarese
Hector Flores
Ken Freeman
Bryan Gaensler
Sarah Gallagher
Peter Garnavich
Karoline Gilbert
Rosa Gonzalez-Delgado
Aruna Goswami
Puragra Guhathakurta
Pat Hall
Guenther Hasinger
Misha Haywood
Falk Herwig
Vanessa Hill
Andrew Hopkins
Mike Hudson
Narae Hwang
Rodrigo Ibata
Pascale Jablonka
Matthew Jarvis
Umanath Kamath
Lisa Kewley
Iraklis Konstantopoulos
George Koshy
Rosine Lallement
Damien Le Borgne
Khee-Gan Lee
Geraint Lewis
Robert Lupton
Sarah Martell
Nicolas Martin
Mario Mateo
Olga Mena
David Nataf
Jeffrey Newman
Gajendra Pandey
Eric Peng
Enrique Pérez
Celine Peroux
Patrick Petitjean
Bianca Poggianti

Francisco Prada
Mathieu Puech
Alejandra Recio-Blanco
Annie Robin
Aaron Robotham
Will Saunders
Carlo Schimd
Arnaud Seibert
Prajval Shastri
Yue Shen
Daniel Smith
C.S. Stalin
Else Starkenburg
Firoza Sutaria
Charling Tao
Karun Thanjuvur
Sivarani Thirupathi
Laurence Tresse
Brent Tully
Ludo van Waerbeke
Kim Venn
Eva Villaver
Matthew Walker
Jian-Min Wang
Yiping Wang
Yuting Wang
Jon Willis
David Yong
Gongbo Zhao

### P.3.8 ngCFHT Science Feasibility Study membership

(not previously mentioned)

Patrick Boisse, James Bolton, Piercarlo Bonifacio, Francois Bouchy, Len Cowie, David Crampton, Katia Cunha, Magali Deleuil, Ernst de Mooij, Patrick Dufour, Sebastien Foucaud, Karl Glazebrook, John Hutchings, Jean-Paul Kneib, Chiaki Kobayashi, Rolf-Peter Kudritzki, Damien Le Borgne, Yang-Shyang Li, Lihwai Lin, Yen-Ting Lin, Martin Makler, Norio Narita, Changbom Park, Ryan Ransom, Swara Ravindranath, Bacham Eswar Reddy, Marcin Sawicki, Luc Simard, Raghunathan Srianand, Thaisa Storchi-Bergmann, Keiichi Umetsu, Ting-Gui Wang, Jong-Hak Woo, Xue-Bing Wu

# Chapter 1 Wide field spectroscopy for the 2020s and beyond: An overview of the Maunakea Spectroscopic Explorer

## 1.1 Chapter Synopsis

- *A fully dedicated, 11.25m aperture wide field highly multiplexed spectroscopic facility*
- *The definitive deconstruction of all components and sub-components of the Galaxy through in situ chemo-dynamical analyses of millions of stars over the entire luminosity range of Gaia targets*
- *Linking galaxies to their surrounding large scale structure through emission and absorption line studies*
- *Measuring the dynamics of dark matter across all spatial scales*
- *An essential feeder facility for the Very Large Optical Telescopes (VLOTs)*
- *The "go-to" follow-up facility for the new generation of multi-wavelength imaging surveys*

The Maunakea Spectroscopic Explorer (MSE) has a unique and critical role in the emerging network of international astronomical facilities. As the largest ground based optical and near-infrared telescope aside from the TMT, the E-ELT and GMT (we refer to these facilities collectively as the Very Large Optical Telescopes, VLOTs), its wide field will be dedicated to highly multiplexed spectroscopy at a broad range of spectral resolutions. MSE is designed to enable transformational science in areas as diverse as tomographic mapping of the interstellar and intergalactic media; the in-situ chemical tagging of thick disk and halo stars; connecting galaxies to their large scale structure; measuring the mass functions of cold dark matter sub-halos in galaxy and cluster-scale hosts; reverberation mapping of supermassive black holes in quasars; next generation cosmological surveys using redshift space distortions and peculiar velocities. MSE is an essential follow-up facility to current and next generations of multi-wavelength imaging surveys, including LSST, Gaia, Euclid, WFIRST, PLATO, and the SKA, and is designed to complement and go beyond the science goals of other planned spectroscopic capabilities like VISTA/4MOST, WHT/WEAVE, AAT/HERMES and Subaru/PFS. It is an essential feeder facility for the VLOTs, namely E-ELT, TMT and GMT, and provides the missing link between wide field imaging and small field precision astronomy.

MSE is an 11.25m aperture observatory with a 1.5 square degree field of view that will be fully dedicated to multi-object spectroscopy. 3200 or more fibers will feed spectrographs operating at low (R ~ 2000 – 3500) and moderate (R ~ 6000) spectral resolution, and approximately 1000 fibers will feed spectrographs operating at high (R ~ 40000) resolution. At low resolution, the entire optical window from 360nm – 950nm and the near infrared J and H bands will be accessible, and at moderate and high resolutions windows with the optical range will be accessible. The entire system is optimized for high throughput, high signal-to-noise observations of the faintest sources in the Universe with high quality calibration and stability being ensured through the dedicated operational mode of the observatory. The discovery efficiency of MSE is an order of magnitude higher than any other spectroscopic capability currently realized or in development. Figure 1 is a rendering of a cut-away of MSE showing the system architecture and major subsystems.

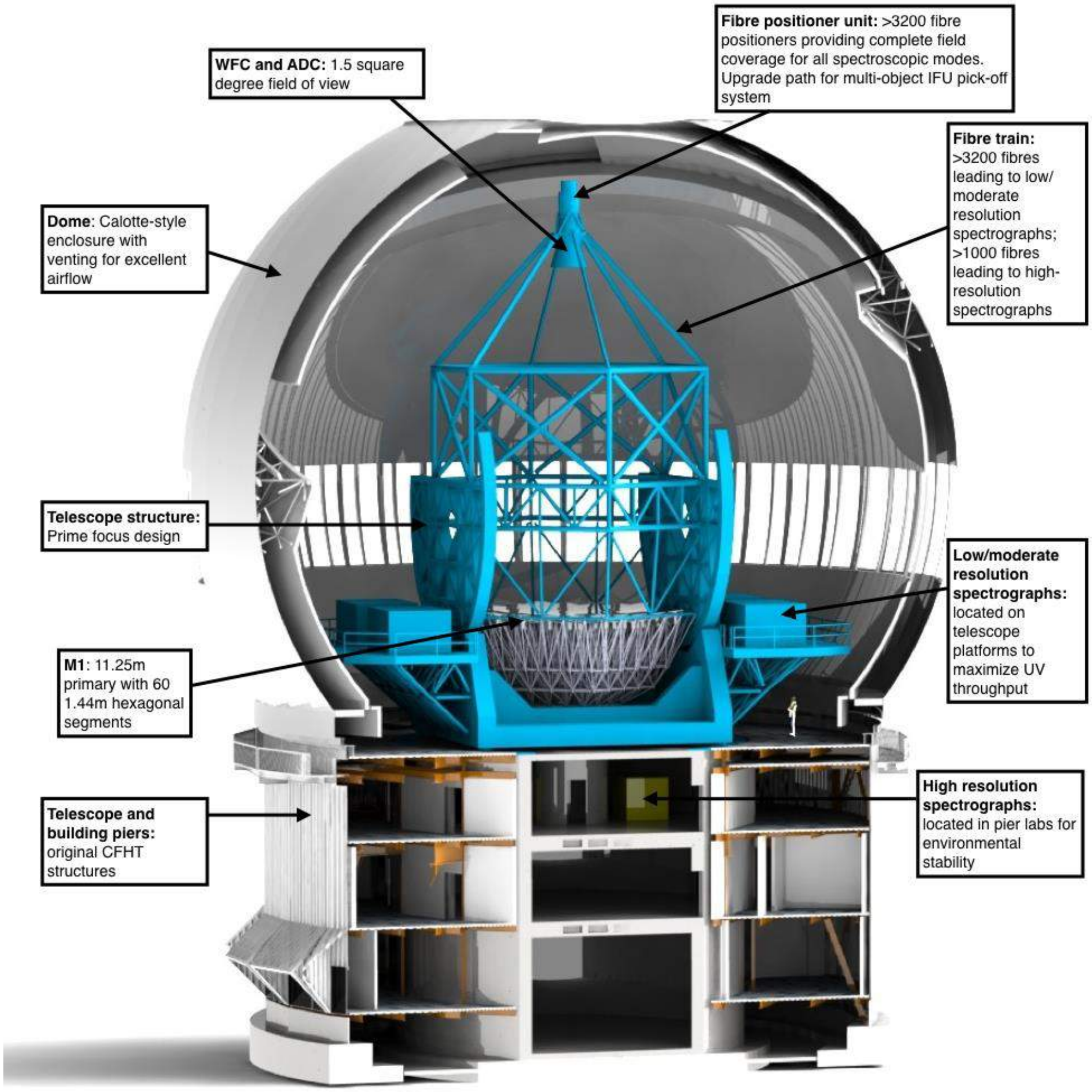

**Figure 1: Cut-away of MSE revealing the system architecture and major subsystems.**

MSE is a rebirth of the 3.6m Canada-France-Hawaii Telescope as a dedicated, 11.25m, spectroscopic facility within an expanded partnership extending beyond the original membership. The *Mauna Kea Comprehensive Management Plan* anticipates CFHT redevelopment. MSE will reuse the current building and pier (minimizing work at the summit) and the ground footprint will remain the same. MSE will benefit from CFHT's 40 years of experience on Maunakea and a support staff deeply rooted in the Hawaii Island community. CFHT/MSE is actively engaging the Hawaii community throughout the planning process and is committed to balancing cultural and environmental interests in the design of the observatory and the organization of the new partnership. With their support, MSE will begin construction once it has passed the Construction Proposal Review and the MSE partnership approves funding, and once all necessary permitting and legal requirements have been fulfilled. This includes the approval of a new master lease for the Mauna Kea Science Reserve.

MSE is a response to the internationally recognized need for highly multiplexed, deep, optical and near-infrared spectroscopy. The diverse and exciting scientific opportunities enabled by a wide-field, large-aperture, spectroscopic survey telescope have long been recognized by the international astronomical community (the initial calls for such a facility appeared more than 15 years ago). The suite of powerful photometric imaging and astrometric telescopes that have now started operations, or are due to begin imminently, have sharpened the need for wide-field spectroscopic follow up and have resulted in such a facility becoming the most glaring and important “missing capability” in the portfolio of international astronomical projects. Various national astronomical strategic plans now place the development of a facility that meets these needs at high priority.

*In summary, MSE is the realization of the long-held ambition for wide field multi-object spectroscopy at optical and near-infrared wavelengths. It is the result of diverse needs across diverse fields of astronomy all converging on a single capability. MSE occupies a critical nexus in the international network of astronomical facilities in the 2020s, and will enable stand-alone transformative science from our understanding and characterization of exoplanet hosts, to chemical tagging of the Milky Way, to the physics of supermassive black holes and the drivers of galaxy formation at cosmic noon. This document presents the context and the science that has driven the development and detailed design of MSE, and illustrates the very broad range of astrophysics that will be impacted by this unique and essential Observatory.*

## 1.2 The Composition and Dynamics of the Faint Universe

MSE will unveil the composition and dynamics of the faint Universe through chemical abundance studies of stars in the outer reaches of the Galaxy, through stellar population studies of distant and low mass galaxies, and through dynamical studies across all spatial scales from dwarf galaxies to the largest scale structures in the Universe. MSE is designed to excel at precision studies of faint astrophysical phenomena that are beyond the reach of 4-m class spectroscopic instruments. A broad range of spectral resolutions – including R = 2000, R = 6000 and R = 40000 – ensure that the specialized technical capabilities of MSE enable a diverse range of transformational astrophysics from stellar through to cosmic scales. Specifically, these include:

- (Chapter 2) The Origin of Stars, Stellar Systems and the Stellar Populations of the Galaxy
- (Chapter 3) Linking Galaxies to the Large Scale Structure of the Universe
- (Chapter 4) Illuminating the Dark Universe

Each chapter of the Detailed Science Case can be read as a stand-alone section, and each chapter starts with a 1 – 2 page synopsis of the content of that chapter. Reference is made throughout to a suite a “Science Reference Observations” (SROs), that are included as Appendices, and which can also be read as stand-alone sections. The SROs span the full range of science that MSE is expected to undertake, and highlight a range of possible operational modes. Importantly, they are compelling, high impact, transformational science programs that are uniquely possible with MSE, and it is from these that the science requirements – i.e., the fundamental design of the facility – are derived. They include the first large scale reverberation mapping experiment to probe the inner parsec of thousands of quasars; a measurement of the cold dark matter sub-halo mass function of the Galaxy through comprehensive and precision radial velocity mapping of tidal streams; derivation of the halo occupation function through the construction of eight photo-z selected survey cubes that probe a cosmologically relevant volume to low stellar mass and which cover the peak of the star formation history of the Universe; tomographic reconstruction of the interstellar medium in the Galaxy and the intergalactic medium to high redshift; chemical tagging of stars in the halo and thick disk of the Milky Way through in-situ analyses of stars that cover the entire magnitude range of Gaia targets.

The reader is referred to the relevant chapters or appendices for detailed discussion of the diverse science case for MSE. Here, we briefly highlight some of the key science themes in each area.

### 1.2.1 Theme 1: The Origin of Stars, Stellar Systems and the Stellar Populations of the Galaxy

*Science Reference Observations for The Origins of Stars, Stellar systems and the Stellar Populations of the Galaxy*

- *SRO-01: Exoplanets and stellar variability*
- *SRO-02: The physics of rare stellar types*
- *SRO-03: The formation and chemical evolution of the Galaxy*
- *SRO-04: Unveiling cold dark matter substructure with precision stellar kinematics*
- *SRO-05: The chemodynamical deconstruction of Local Group galaxies*

MSE will produce vast spectroscopic datasets for in situ stars across all components of the Galaxy at all galactocentric radii, across the full metallicity distribution of the Galaxy, and across the full luminosity range of Gaia targets. Exploration of our Galaxy provides us with the most detailed perspective possible on the formation of planetary systems, their host stars, and their parent stellar populations. At the largest scale, the properties of the stellar populations – including their spatial distributions, energy, angular momenta, and chemical properties – provide us with the fossil information necessary to decompose the Milky Way galaxy into its primordial building blocks, and in so doing reveal the history of its formation and the sites of chemical nucleosynthesis.

The key measurables are well-defined spectral line features for accurate stellar parameters (temperature, gravity, metallicity, rotational velocities, magnetic fields strengths, etc.), chemical abundances, and radial velocities. Such a detailed perspective necessitates the highest signal-to-noise data observed at high spectral resolution: at the highest resolution of MSE (R = 40 000), synthetic spectra show that lines are deblended at ~0.1Å, rotational and radial velocities can be measured to better than ~0.1 $kms^{-1}$ through cross-correlation of many lines (>100), and chemical abundances can be measured with Δ[X/Fe] <0.1 dex, when the signal-to-noise ratio (SNR) > 30. Here, the capabilities of MSE are unmatched amongst any existing or planned facilities. Further, an increasing fraction of astronomical resources are being devoted to characterizing the transient, moving and variable universe. MSE, thanks to its large aperture, wide field and high multiplexing, is well positioned to pioneer multi-object time domain optical/IR spectroscopic studies of faint stellar sources.

Specifically, MSE will provide complete characterization at high resolution and high SNR of the faint end of the PLATO target distribution (g ∼ 16) visible from the northern hemisphere to allow for statistical analysis of the properties of planet-hosting stars as a function of stellar and chemical parameters. With high velocity accuracy and stability, time domain spectroscopic programs will allow for highly complete, statistical studies of the prevalence of stellar multiplicity into the regime of hot Jupiters for this and other samples and also directly measure binary fractions away from the environment of the Solar Neighbourhood. Rare stellar types – including a menagerie of pulsating stars, but also halo white dwarfs and solar twins – will be easily identified within the dataset of millions of stars observed per year by MSE, and will allow for population studies that incorporate accurate completeness corrections and which examine their distribution within the Galaxy.

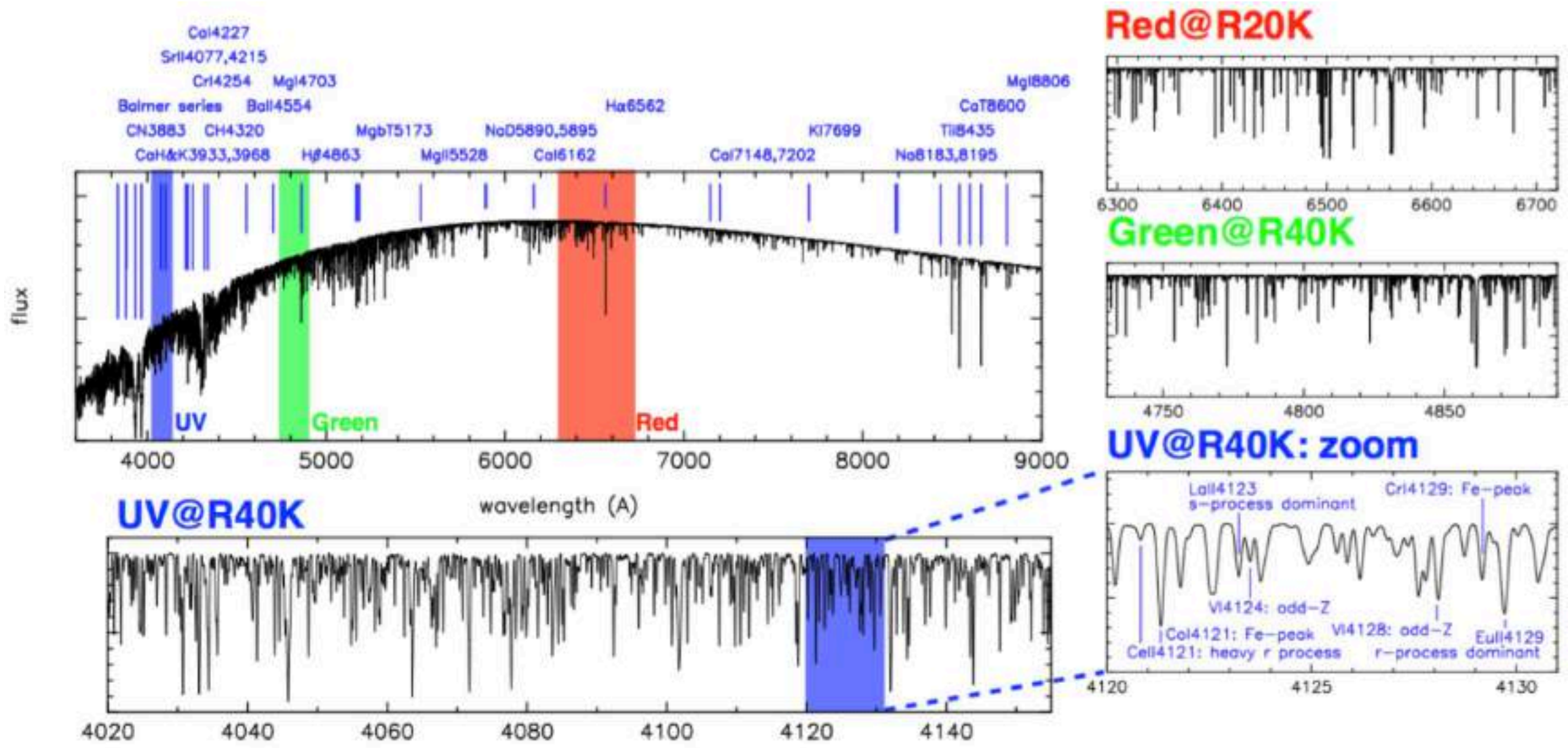

**Figure 2: Main panel shows the relative flux of a synthetic spectrum of a metal poor red giant star at the intermediate MSE spectral resolution of R∼6000, along with some of the strong line stellar diagnostics accessible at this resolution. Highlighted regions show the normalized flux in three windows observable with the high resolution mode of MSE. A magnified region of the UV window shows examples of the species that will be identified at high resolution. MSE chemical tagging surveys will identify species sampling a large and diverse set of nucleosynthetic pathways and processes. See Chapter 2 for more details.**

MSE will have unmatched capability for chemical tagging experiments. Recent work in this field has started to reveal the dimensionality of chemical space and has shown the potential for chemistry to be used in addition to, or instead of, phase space, to reveal the stellar associations that represent the remnants of the building blocks of the Galaxy. In particular, MSE will decompose the outer regions of the Galaxy – where dynamical times are long and whose chemistry is inaccessible from 4-m class facilities – into its constituent star formation events by accessing a range of chemical tracers that sample a large number of nucleosynthetic pathways. In so doing, MSE will provide a high resolution, homogenous sampling of the metallicity distribution function of the Galaxy. It will trace variation in the metallicity distribution across all components and as a function of radius. At the metal-weak end, [Fe/H] < -3, such an analysis reveals the distribution of the oldest stars in the Universe, and is a vital component to tracing the build-up of the Galaxy at the earliest times.

MSE will extend analysis of the resolved stellar populations of the Galaxy to the Local Group and beyond. Variable stars such as Cepheids can be spectroscopically studied throughout the Local Group utilizing the time-domain strengths of MSE, and massive stars will be used to trace the metallicity gradients of disks in galaxies in the next nearest galaxy groups. Such studies will probe the inside-out formation of galaxy disks through individual stellar metallicities observed at moderate spectral resolution. M31 and M33 are particularly prominent extragalactic targets for resolved stellar population analysis, and MSE will provide the definitive chemodynamical deconstruction of our nearest $L_*$ galaxy companion by a focused program targeting all giant branch candidates observed in M31 out to approximately half the virial radius of the galaxy. Such a spectroscopic panorama is unprecedented and a natural complement to the core Milky Way science that forms the drivers for MSE in this field.

MSE is uniquely suited to precision studies of the stellar populations of the Milky Way, and will provide a transformative impact in our understanding of the links between stellar and planetary formation, the sites of chemical nucleosynthesis, stellar variability, and the chemical structure of the Galaxy through in situ analysis of stars in all Galactic components. It will push these studies out to the nearest galaxies to provide panoramic complements to the driving Milky Way science. MSE will leverage the dataset created by PLATO for exoplanetary and stellar astrophysics, and will provide the essential complement to the Gaia space satellite by providing high resolution chemical abundances across the entire luminosity range of Gaia targets.

### 1.2.2 Theme 2: Linking Galaxies to the Large Scale Structure of the Universe

*Science Reference Observations for Linking Galaxies to the Large Scale Structure of the Universe*

- *SRO-6: Galaxies and their environments in the nearby Universe*
- *SRO-7: The baryonic content and dark matter distribution of the nearest massive clusters*
- *SRO-8: Multi-scale clustering and the halo occupation function*
- *SRO-9: The chemical evolution of galaxies and AGN*

MSE will provide a breakthrough in extragalactic astronomy by linking the formation and evolution of galaxies to the surrounding large scale structure, across the full range of relevant spatial scales, from kiloparsecs to megaparsecs. Within the ΛCDM paradigm, it is fundamental to understand how galaxies evolve and grow relative to the dark matter structure in which they are embedded. This necessitates mapping the distribution of stellar populations and supermassive black holes to the dark matter haloes and filamentary structure that dominate the mass density of the Universe, and to do so over all mass and spatial scales.

MSE is optimized for studying the evolution of galaxies, AGN and the environmental factors likely to influence their evolution. MSE will be the premier facility for unscrambling the non-linear regime (small scale clustering, mergers, groups, tendrils and filaments), the baryon regime (i.e., metallicity/chemical evolution), and the evolution and interplay of galaxies, AGN, and the IGM, incorporating environmental factors and the key energy production pathways. MSE will build upon the impressive legacy of current spectroscopic surveys in the near and far fields by going several magnitudes deeper, covering an order of magnitude larger sky area, and conducting both stellar population studies of the nearest galaxies and precision dynamical analyses across the luminosity function and across cosmic time.

In the local Universe, MSE will measure the dynamics and stellar populations of every baryonic structure in the Virgo and Coma clusters down to extremely low stellar mass. This will allow the physical association of dwarf galaxies, nucleated dwarfs, globular clusters and ultra-compact dwarfs to be probed through chemodynamical analysis of unprecedented datasets (both in size and completeness). Quenching processes operating in the cluster environment can be examined for all systems at all radii, and in particular analysis of the stellar populations of thousands of low mass galaxies all in the same cluster will give direct access to the relative roles of environmentally-driven and internally-driven processes across the luminosity function and to its faintest limits.

MSE will be able to conduct the definitive spectroscopic survey of the low redshift Universe for decades to come by obtaining spectra for every $z < 0.2$ galaxy in a cosmologically-relevant volume spanning 3200 $deg^2$, allowing every halo with mass $M_{halo} > 10^{12} M_{\odot}$ and four or more galaxies to be detected. Follow up in select regions will result in a sample covering 100 $deg^2$ spanning four decades in halo mass, accounting for the majority of stellar mass in the low-redshift Universe. This will allow the measurement of the shape of the stellar mass function to the scale of the Local Group dwarf galaxies, over a large volume and a range of environments.

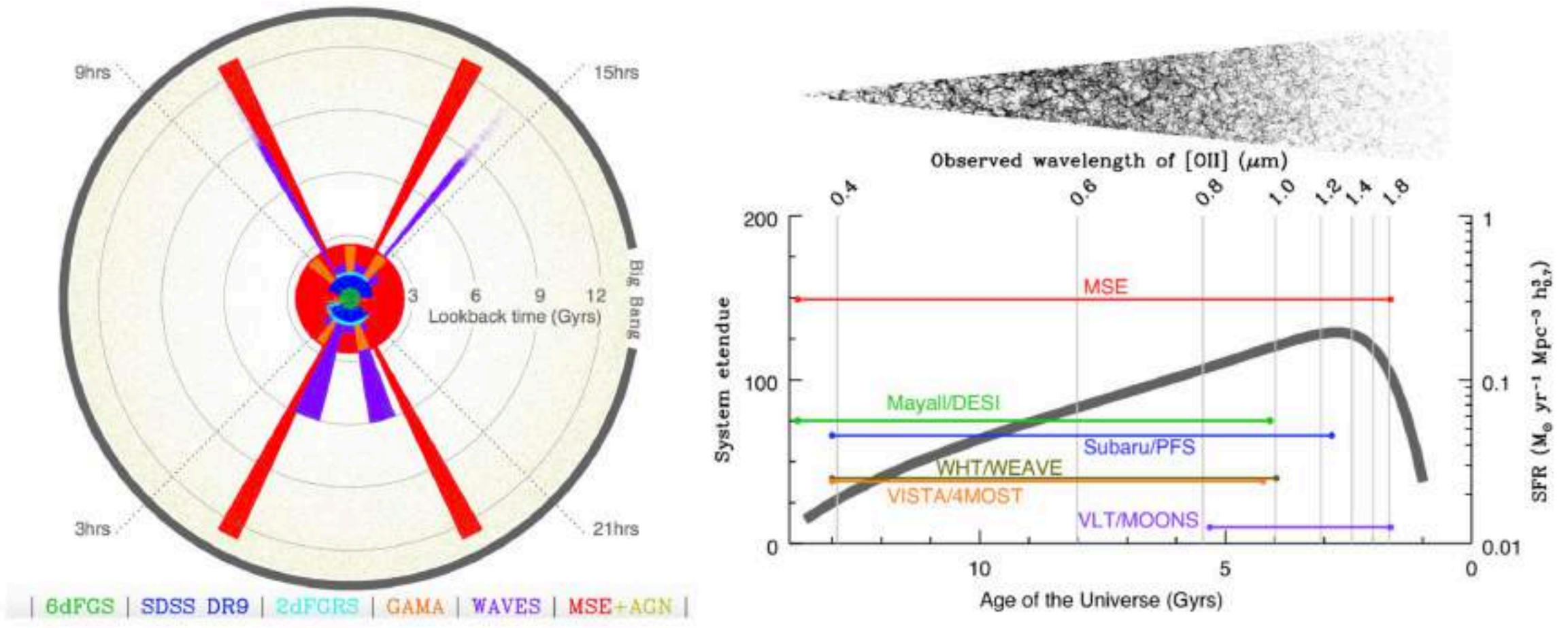

**Figure 3: Left panel: Cone plot showing an example MSE survey compared to other notable benchmark galaxy surveys. Cones are truncated at the redshift at which L* galaxies are no longer visible. Right panels: Lookback time versus cosmic star formation rate (right axis) using the parameterization of Hopkins & Beacom (2006). Grey lines indicate the wavelength of OII 3727Å. Also indicated on this wavelength scale is the wavelength range of all highly multiplexed spectroscopic facilities in development. These are offset vertically according to their system etendue (left axis). The light cone demonstrates the reach of MSE for extragalactic surveys using a homogeneous set of tracers at all redshifts. See Chapter 3 for more details.**

Uniquely, the broad wavelength range and extreme sensitivity of MSE allows extrapolation of this “low redshift” survey concept out to beyond cosmic noon. An ambitious set of photo-z selected survey cubes will allow MSE to measure the build up of large scale structure, stellar mass, halo occupation and star formation out to redshifts much greater than 2. By targeting (300 Mpc/h)$^3$ boxes, each cube will measure "Universal" values for an array of potential experiments. These survey volumes will trace the transition from merger-dominated spheroid formation to the growth of disks, covering the peak in star-formation and merger activity. This combination of depth, area and photo-z selection is not possible without a combination of future imaging surveys and MSE. As such, MSE will be able to produce the definitive survey of structure, halos and galaxy evolution over 12 billion years.

MSE will be unparalleled in its ability to connect the properties of galaxies to their surrounding large scale structure. It will link the properties and diversity of galaxies observed in the local universe to the evolutionary processes that have occurred across cosmic time. In so doing, MSE will fulfill three distinct roles in the 2025+ era for extragalactic astrophysics: dedicated science programs, coordinated survey programs and feeder programs for the VLOTs. All three roles are critical, transformational, and will lead to major advancements in the area of structure formation, galaxy evolution, AGN physics, our understanding of the IGM, and the underlying dark matter distribution.

### 1.2.3 Theme 3: Illuminating the Dark Universe

*Science Reference Observations for Illuminating the Dark Universe*

- *(SRO-03: The formation and chemical evolution of the Galaxy)*
- *(SRO-04: Unveiling cold dark matter substructure with precision stellar kinematics)*
- *SRO-06: The baryonic content and dark matter distribution of the nearest massive clusters*
- *SRO-10: Mapping the inner parsec of quasars through reverberation mapping*
- *SRO-11: Linking galaxy evolution with the IGM through tomographic mapping*
- *SRO-12: A peculiar velocity survey out to 1Gpc and the nature of the CMB dipole*

MSE will cast light on some of the darkest components of the Universe. The backbone of the large scale structures in which galaxies are embedded is dark matter, that is invisible in electromagnetic radiation. Embedded alongside galaxies in these large scale structures is the intergalactic medium, a major reservoir of baryons in the Universe but one which is best observed in the optical wavelength range through its absorption effects on background galaxies. Within galaxies, the interstellar medium is similarly probed through absorption studies in the optical, and the central supermassive black holes are generally only probed through indirect techniques. MSE will provide unprecedented perspectives on each of these components of the dark Universe, and provide a suite of capabilities ideally suited to providing critical data for next generation cosmological surveys.

Specifically, MSE will provide tomographic reconstruction of the three dimensional structures of the Galactic interstellar medium and the intergalactic medium through high signal-to-noise absorption studies along millions of sight lines. These programs will be orders of magnitudes larger than the currently state-of-the-art. Whether focused on stellar or galactic science, the relations between stars and galaxies and their surrounding media is a key issue in understanding the cycling of baryons. In the extragalactic regime, MSE can target the IGM at $z \sim 2-2.5$, to reconstruct the dark matter distribution and associated galaxies at high-z, allow detailed modeling of galactic scale outflows via emission lines, investigate the complex interplay between metals in galaxies and the IGM, and provide the first comprehensive, moderate resolution analysis of a large sample of galaxies at this epoch.

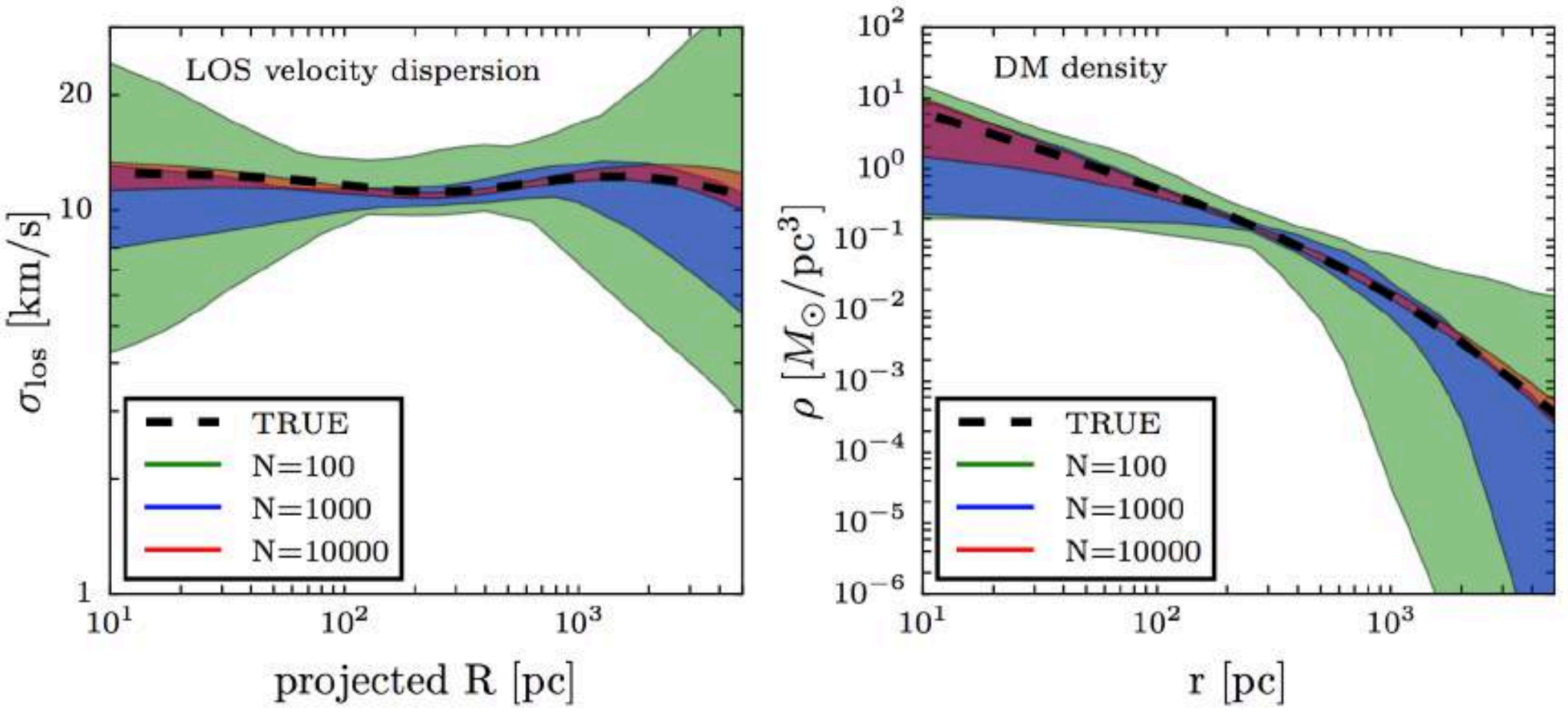

**Figure 4: Recovery of line-of-sight velocity dispersion (left) and dark matter density (right) profiles as a function of stellar spectroscopic sample size for dwarf galaxies, demonstrating the need for extensive datasets with MSE. Shaded regions represent 95% credible intervals from a standard analysis of mock data sets consisting of radial velocities for N = $10^2$, $10^3$ and $10^4$ stars (median velocity errors of 2 kms$^{-1}$), generated from an equilibrium dynamical model for which true profiles are known (thick black lines). See Chapter 4 for details.**

MSE will allow for an unprecedented, extragalactic time domain program to measure directly the accretion rates and masses of a large sample of supermassive black holes through reverberation mapping. This information is essential for understanding accretion physics and tracing black hole growth over cosmic time. Reverberation mapping is the only distance-independent method of measuring black hole masses applicable at cosmological distances, and currently only ~60 local, relatively low-luminosity AGN have measurements of their black hole masses based on this technique. A ground-breaking MSE campaign of ~100 observations of ~5000 quasars over a period of several years (totaling ~600 hours on-sky) to map the inner parsec of these quasars from the innermost broad-line region to the dust-sublimation radius will map the structure and kinematics of the inner parsec around a large sample of supermassive black holes actively accreting during the peak quasar era and provide a quantum leap in comparison to current capabilities. Further, the resulting well-calibrated reverberation relation for quasars offers promise for constructing a high-z Hubble diagram to constrain the expansion history of the Universe.

MSE is the ultimate facility for probing the dynamics of dark matter over every and all spatial scales from the smallest dwarf galaxies to the largest clusters. For dwarf galaxies in the vicinity of the Milky Way, MSE will obtain complete samples of tens of thousands of member stars to very large radius and with multiple epochs to remove binary stars. Such analyses will allow the internal dark matter profile to be derived with high accuracy and will probe the outskirts of the dark matter halos accounting for external tidal perturbations as the dwarf halos orbit the Galaxy. A new regime of precision will be reached through the combination of large scale radial velocity surveys using MSE with precision astrometric studies of the central regions using the VLOTs. In the Galactic halo, large scale radial velocity mapping of every known stellar stream accessible from Maunakea with high velocity precision will reveal the extent to which these cold structures have been heated through interactions with dark sub-halos and for the first time place strong limits on the mass function of all subhalos (luminous and non-luminous) in an $L_*$ galaxy. On cluster scales, MSE will use galaxies and globular clusters as dynamical tracers in exactly the same way as stars in dwarf galaxies, to provide a fully consistent portrait of dark matter halos across the halo mass function.

Finally, MSE provides a platform for enabling next generation cosmological surveys, particularly through large scale surveys of the internal dynamics of galaxies as part of future multi-object IFU developments. Such surveys allow for dynamical tests probing the scale of homogeneity, can test the backreaction conjecture in General Relativity, and provides a novel means of measuring gravitational lensing via the velocity field that complements standard measurements by being sensitive to very different systematics.

MSE will transform the study of some of the darkest components of the Universe. Absorption line studies, reverberation mapping programs, and precision velocity studies of stars and galaxies as tracer particles utilize aspects of the performance of MSE – sensitivity, stability and specialization – that are unique to the facility and which consequently allow unique insights into key aspects of the baryon cycle, black hole physics, and cosmology. MSE represents a suite of capabilities that is an essential platform for the development of cosmological programs that go beyond the current generation and which are relevant on the 2025+ timescale.

### 1.2.4 From science to requirements: mapping the science flowdown for MSE

Chapters 2 – 4 discuss in detail the anticipated science objectives of MSE, with the important caveat that it is impossible to predict how astrophysics will evolve over the next decade in response to discoveries and realizations that have not yet been made. Nevertheless, it is clear that diverse fields of research have converged in requiring dedicated large aperture multi-object spectroscopy, and it is clear that the capabilities of MSE will be a critical component of future lines of astronomical enquiry.

Reference is made throughout this document to a suite of Science Reference Observations presented in the Appendices. These have been defined by the international science team as specific, detailed, science programs for MSE that are transformative in their fields and which are uniquely possible with MSE. This last point is particularly important given the large number of spectroscopic resources that are becoming available and which are discussed in more detail in Section 1.5. The SROs have been selected to span the range of anticipated fields in which MSE is expected to contribute, and they have been developed in considerable detail. The Science Requirements for MSE – i.e., the highest level design requirements for the facility – are then defined as the suite of capabilities necessary for MSE to carry out these observations. The MSE architecture and technical requirements flow directly from, and deliver the capability described within, the Science Requirements. Figure 5 shows the science development procedure for MSE, and its relation to the key documents that define the design of MSE.

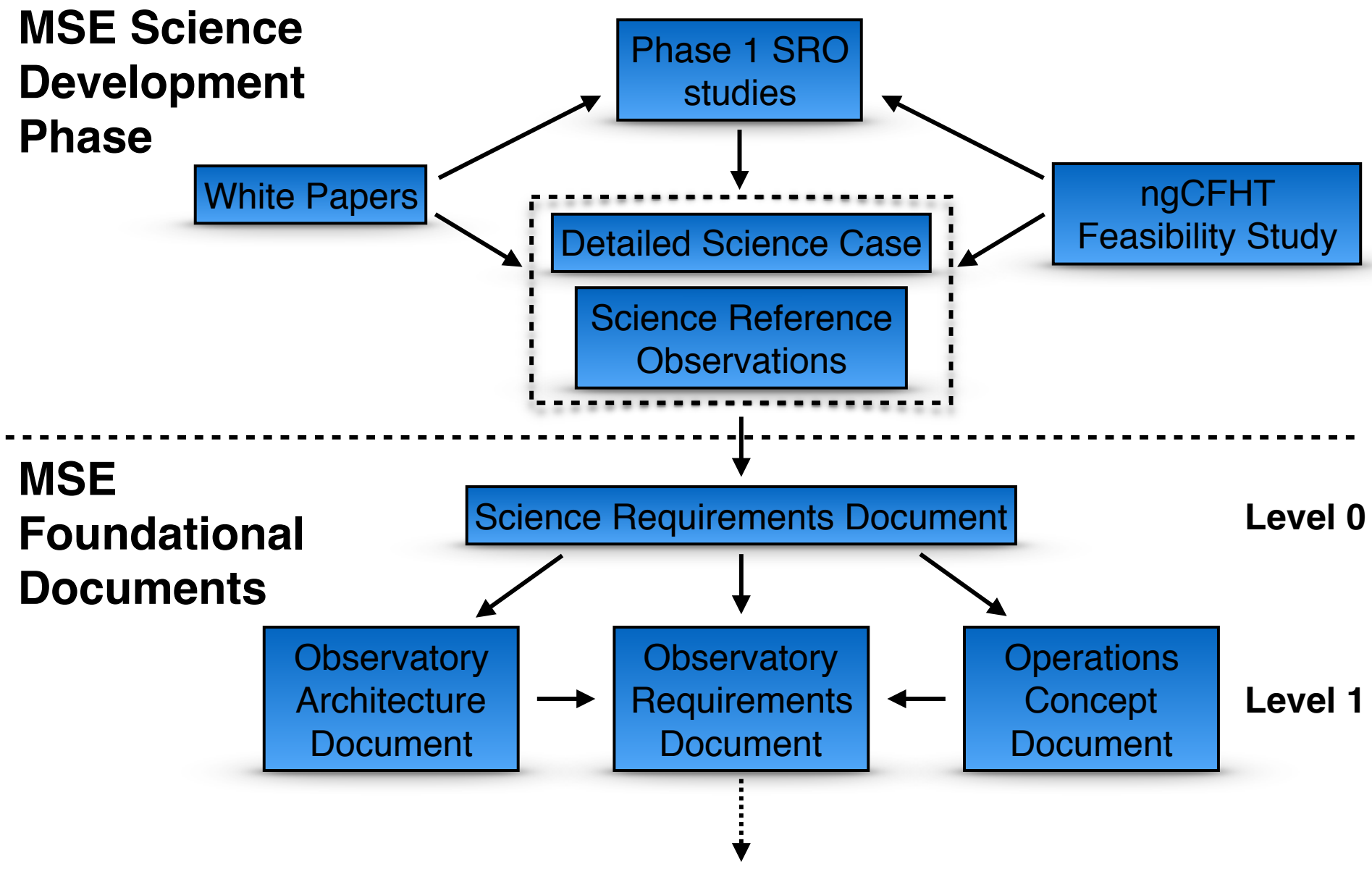

**Figure 5: Overview of the science development process to date for MSE, including the relationships to the main documents that define the design of MSE.**

We expect that many of the SROs discussed in this document will be precursors to observing programs that MSE will carry out. However, it is not necessarily the case that this is so, since come first light of the facility, a new set of forefront science topics might have emerged that the users of MSE will want to address. Nevertheless, the suite of capabilities that MSE will have, in response to the SROs as described in the Appendices, will ensure that MSE is at the cutting edge of astronomical capabilities in 2025 and beyond.

Table 1 cross references each SRO to each of the high level MSE science requirements. The science requirements are described and discussed in detail in the MSE Science Requirements Document.

**Table 1: Science Reference Observations cross referenced to Science Requirements. Dark green boxes indicate import impacts of the SRO on the requirement; light green indicates moderate impact of the SRO on the requirement; grey indicates minimal or no impact of the SRO on the requirement.**

| | | | Resolved stellar sources | | | | | Extragalactic sources | | | | | | |
|---|---|---|---|---|---|---|---|---|---|---|---|---|---|---|
| | | | DSC-SRO-01 *Exoplanet hosts* | DSC-SRO-02 *Time domain stellar astrophysics* | DSC-SRO-03 *Chemical tagging in the outer Galaxy* | DSC-SRO-04 *CDM subhalos and stellar streams* | DSC-SRO-05 *Local Group galaxies* | DSC-SRO-06 *Nearby galaxies* | DSC-SRO-07 *Virgo and Coma* | DSC-SRO-08 *Halo occupation* | DSC-SRO-09 *Galaxies and AGN* | DSC-SRO-10 *The InterGalactic Medium* | DSC-SRO-11 *Reverberation mapping* | DSC-SRO-12 *Peculiar velocities* |
| Spectral resolution | REQ-SRD-011 | *Low spectral resolution* | | R~2000 (white dwarfs) | | | | R~3000 | R~3000 | R~2000-3000 | R~3000 | | R~3000 | R~1000-2000 |
| | REQ-SRD-012 | *Intermediate spectral resolution* | | Any repeat observatio-ns | Essential, R~6500 | Essential, R~6500 | Essential, R~6500 | Velocities of low mass galaxies | Velocities of low mass galaxies | | | R~5000 | | |
| | REQ-SRD-013 | *High spectral resolution* | R>=40000 | R>=20000 | Essential, R~20-40K | Essential | Young stars | | Bright globular clusters | | | | | |
| Focal plane input | REQ-SRD-021 | *Etendue* | ~2000 sq. deg @g=16 | all-sky | 1000s sq. deg; g~20.5@ | 1000s sq.deg, g~22 with high v accuracy | 100s sq. deg, g~24 | 3200 (100) sq.deg, i<23 (24.5) | ~100sq.deg , i<24.5 | ~1000sq.de g, i<24.5 | ~300 sq.deg, i<25 | 40 sq.deg, r<24.0 | 7 sq. deg, 5000 targets to i=23.25 | all-sky |
| | REQ-SRD-022 | *Multiplexing at lower resolution* | | | | | | ~5000 galaxies/sq. deg | 100s targets (galaxies/GCs) / sq.deg | >5000 galaxies/sq. deg | 770 galaxies/sq. deg | | 600AGN/de g | 1000s galaxies/sq. deg |
| | REQ-SRD-023 | *Multiplexing at moderate resolution* | | | 1000s stars/sq.de g to g~23 | 1000s stars/sq.de g to g~23 | few - thousands stars/sq. deg | | | | | 500 galaxies/sq. deg | | |
| | REQ-SRD-024 | *Multiplexing at high resolution* | ~100 stars/sq.de g @ g = 16 | ~1000 stars/sq.de g to g=20.5 | ~1000 stars/sq.de g to g=20.5 | 1000s stars/sq.de g to g~23 | | | | | | | | |
| | REQ-SRD-025 | *Spatially resolved spectra* | | | | | | Goal | Goal | | | | | Yes |
| Sensitivity | REQ-SRD-031 | *Spectral coverage at low resolution* | | | | | | 0.37 - 1.5um | 0.37 - 1.5um | 0.36 - 1.8um | 0.36 - 1.8um | 0.36 - 1.8um (0.36 - | 0.36 - 1.8um | Optical emission lines |
| | REQ-SRD-032 | *Spectral coverage at moderate resolution* | | | Strong line diagnostics in optical | CaT essential | CaT essential | Goal: complete | Goal: complete | | | Goal: complete | | |
| | REQ-SRD-033 | *Spectral coverage at high resolution* | Strong lines for velocities; tagging | Strong lines for velocities | Chemical tagging | Strong lines for velocities | | | | | | | | |
| | REQ-SRD-034 | *Sensitivity at low resolution* | | | | | | i=24.5 | i=24.5 | i=25.3 | i=25 / H=24 | | i=23.25 | i=24.5 |
| | REQ-SRD-035 | *Sensitivity at moderate resolution* | | | g>20.5 | g~23 | i=24 | i=24.5 | i=24.5 | | | r~24 | | |
| | REQ-SRD-036 | *Sensitivity at high resolution* | g=16 @high SNR | g=20.5 | g=20.5 | g~22 | | | | | | | | |
| Calibration | REQ-SRD-041 | *Velocities at low resolution* | | | | | | v~20km/s | v~20km/s | v~100km/s | v~20km/s | | v~20km/s | v~20km/s |
| | REQ-SRD-042 | *Velocities at moderate resolution* | | | v~1km/s | v~1km/s | v~5km/s | v~9km/s | v~9km/s | | | v~20km/s (10km/s goal) | | |
| | REQ-SRD-043 | *Velocities at high resolution* | v<100m/s | v~100m/s | v~100m/s | v<1km/s | | | | | | | | |
| | REQ-SRD-044 | *Relative spectrophotometry* | | | | | | ~4% | | | | | Critical:3% | |
| | REQ-SRD-045 | *Sky subtraction, continuum* | few % | few % | few % | few % | <1% | <1% | <1% | <0.5% | <0.5% | <1% | <1% | <1% |
| | REQ-SRD-046 | *Sky subtraction, emission lines* | | | | important (CaT region) | important (CaT region) | critical | critical | critical | critical | critical | critical | |
| Operations | REQ-SRD-051 | *Accessible sky* | Plato footprint (ecliptic) | Gaia footprint (all sky) | Gaia footprint (all sky) | Gaia, PS1, HSC footprint | Northern hemisphere (M31, M33, dec~+40) | LSST overlap useful (10000 sq.deg) | NGVS footprint; dec +12 | LSST overlap useful (10000 sq.deg); Euclid (all sky) | LSST overlap useful (10000 sq.deg); Euclid (all sky) | LSST overlap useful (10000 sq.deg); Euclid (all sky) | all sky target distribution | all sky target distribution |
| | REQ-SRD-052 | *Observing efficiency* | maximise | maximise | maximise | maximise | maximise | maximise | maximise | maximise | maximise | maximise | maximise | maximise |
| | REQ-SRD-053 | *Observatory lifetime* | Monitoring >=years | Monitoring >=years | Survey >=5 years | Survey >=5 years | Survey ~100s nights | Survey ~ few years | Survey ~100s nights | Survey ~7 years | Survey ~100s nights | Survey ~100s nights | Monitoring ~5 years | Survey ~years |

## 1.3 MSE as a unique and diverse astronomical facility

### 1.3.1 The scientific capabilities of MSE

**Table 2: Summary of the scientific capabilities of MSE**

| | | | | | | | |
|---|---|---|---|---|---|---|---|
| **Accessible sky** | 30000 square degrees (airmass<1.55) | | | | | | |
| **Aperture (M1 in m)** | 11.25m | | | | | | |
| **Field of view (square degrees)** | 1.5 | | | | | | |
| **Etendue = FoV x π (M1 / 2)²** | 149 | | | | | | |
| **Modes** | **Low** | | **Moderate** | **High** | | | **IFU** |
| **Wavelength range** | 0.36 - 1.8 µm | | 0.36 - 0.95 µm | 0.36 - 0.95 µm ♯ | | | IFU capable; anticipated second generation capability |
| | 0.36 - 0.95 µm | J, H bands | | 0.36 - 0.45 µm | 0.45 - 0.60 µm | 0.60 - 0.95 µm | |
| **Spectral resolutions** | 2500 *(3000)* | 3000 *(5000)* | 6000 | 40000 | 40000 | 20000 | |
| **Multiplexing** | >3200 | | >3200 | >1000 | | | |
| **Spectral windows** | Full | | ≈Half | $\lambda_c/30$ | $\lambda_c/30$ | $\lambda_c/15$ | |
| **Sensitivity** | m=24 * | | m=23.5 * | m=20.0 ♮ | | | |
| **Velocity precision** | 20 km/s ♪ | | 9 km/s ♪ | < 100 m/s ★ | | | |
| **Spectrophotometic accuracy** | < 3 % relative | | < 3 % relative | N/A | | | |

♯ Dichoric positions are approximate

* SNR/resolution element = 2 ♪ SNR/resolution element = 5

♮ SNR/resolution element = 10 ★ SNR/resolution element = 30

The diverse science enabled by MSE and discussed in this document spans all astronomy, from planet formation, the microphysics of stars and the interstellar medium through to the dynamics of dark matter and the physics of black holes. What are the key capabilities of MSE that facilitate this impact and diversity, and how does it compare to other spectroscopic resources? Table 2 shows the relevant scientific capabilities of MSE. The telescope will have an 11.25m segmented primary (effective aperture of >10m) and a 1.5 square degree field of view. Three different spectral resolution settings are possible. At the lowest resolution, over 3000 spectra can be obtained in a single pointing that spans the entire optical spectrum into the near infrared (goal: H band). At moderate resolution, the same number of spectra can be obtained for approximately half the optical waveband. At the highest resolution, approximately 1000 spectra can be obtained per pointing for windows in the optical region of the spectrum. At all resolutions, MSE can push to the faintest targets, and will be well calibrated and stable over its lifetime. Located at the equatorial site of Maunakea ($\ell = 19.9°$), MSE will access the entire northern hemisphere and half of the southern sky, making it an ideal follow-up and feeder facility for a large number of existing and planned ground- and space-based facilities.

It is notoriously difficult to capture the science performance of an instrument/facility in a single metric. Nevertheless, for illustrative purposes we attempt to do this by defining a parameter, η, that loosely represents the ``discovery efficiency" of a MOS capability. Here,

$$\eta = \frac{D_{M1}^2\, \Omega\, N_{MOS}\, f}{IQ^2},$$

where

- $D_{M1}$ is the diameter of the primary mirror,
- $\Omega$ is the instantaneous field of view,
- $N_{MOS}$ is the number of simultaneous spectra,
- $f$ is the fractional time available for scheduling, and
- $IQ$ is the typical image quality (FWHM) delivered by the site and facility.

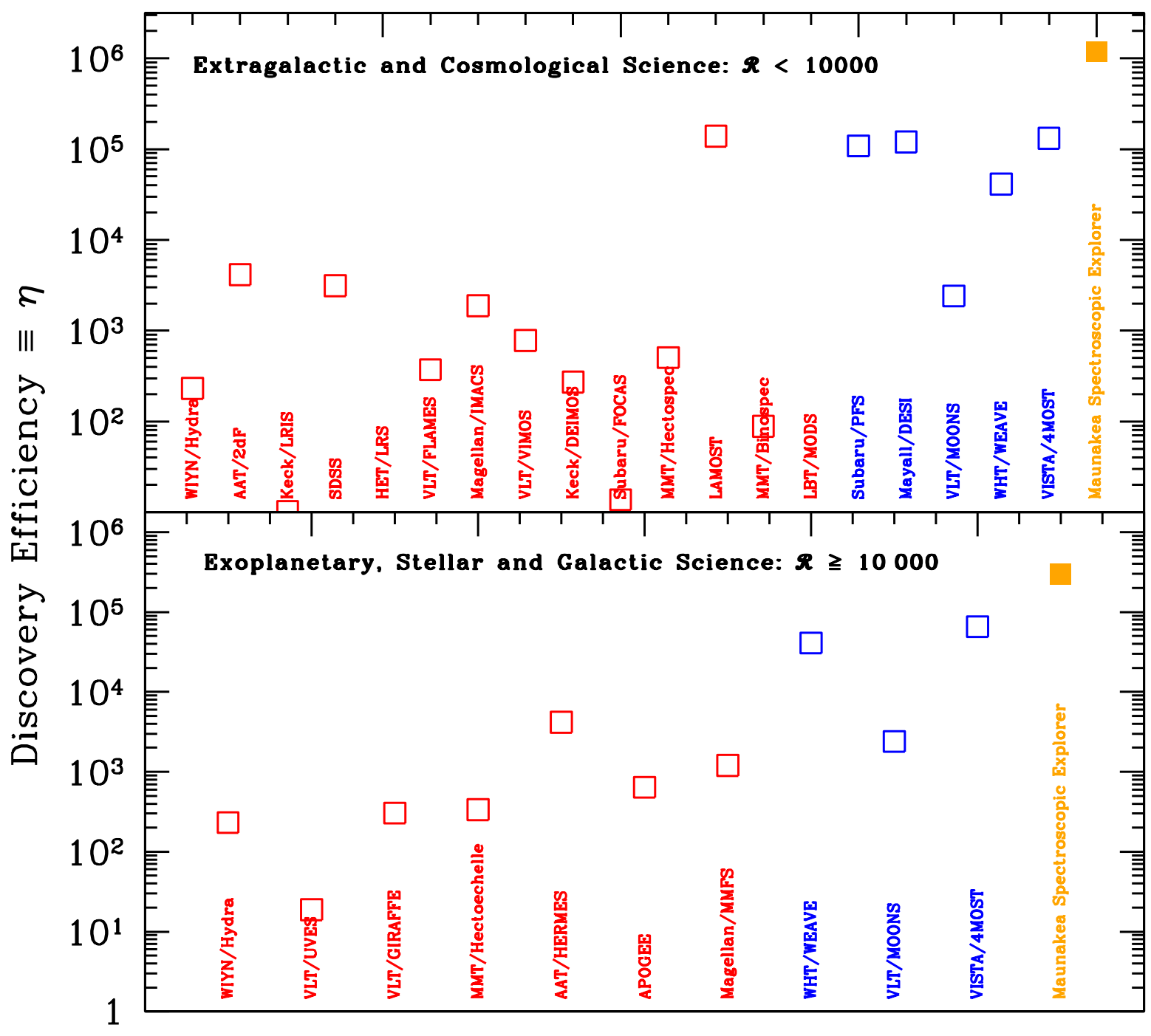

**Figure 6: The discovery efficiency of MSE relative to other existing and planned spectroscopic instruments, as defined in the text. The top panel shows instruments that operate at low and intermediate resolutions (here defined as R<10000) and the bottom panel shows instruments that operate at high resolution (R>10000). Judged by this metric, MSE is highly competitive across all of its modes of operation.**

Figure 6 compares this parameter for a number of facilities. Here, the facilities have been divided into ``low/intermediate" (1000 < R < 10000) and ``high" (R > 20000) resolution categories, although the exact resolution at which this split is made is arbitrary. Judged by this metric, MSE outperforms all other facilities by a wide margin for both high and low resolution science. However, given the uncertainties inherent in attempting to compare a large number of disparate facilities through a single metric, it is illustrative to consider the performance of the instruments listed in Figure 6 as it relates to specific science goals:

- At low resolution, consider the SRO described in Appendix J that will map the accretion processes occurring within the inner parsec of quasars through a large-scale reverberation mapping campaign. This program requires good SNR observations of a large number of relatively faint quasars at low/moderate resolution over an extended period of time. 10m class telescopes are therefore essential. At low redshift, many of the diagnostics to be monitored are in the optical region prior to being redshifted into the NIR, and so a large wavelength range is necessary. Only MSE and Subaru/PFS are therefore capable of this science, although the narrower wavelength range of Subaru/PFS means such studies are more limited in redshift space. Critical to this science, however, is accurate spectrophotometry and calibration, to be able to compare observations separated by up to 5 years in duration. Such precise observations are a strength of a dedicated, specialized facility where the entire system is optimized and stable, and is much more difficult to achieve for a facility instrument such as Subaru/PFS that moves on and off the telescope at regular intervals;

- At high resolution, consider the SRO described in Appendix D that will measure the cold dark matter sub-halo mass function of the Galaxy through precision velocity measurements of stars in tidal streams in the halo. Such observations require high SNR observations of very faint stars (g > 20), in order to be able to have the statistics necessary to probe the kinematic substructure of very diffuse and distant structures. Moderate or high resolution observations on 10m class facilities are required, and these additionally provide important metallicity diagnostics in order to isolate the populations of interest. In addition, however, data spanning hundreds or thousands of square degrees are required due to the very extended nature of the streams being probed, and in order to probe multiple streams at different locations in the Galactic halo. Field of view and multiplexing capabilities are essential. MSE is therefore the only instrument that can conduct this science program.

## 1.4 Key capabilities for MSE science

MSE is uniquely powerful for spectroscopic survey science due to a range of factors, including its large aperture, range of spectral resolution, excellent site quality, long lifetime and dedicated operations. Given the intense focus on wide field spectroscopy and the large number of instruments shown in Figure 6, it is worth discussing the key science-enabling capabilities for MSE in more detail.

### 1.4.1 Large aperture and high throughput

MSE is the fourth largest optical and near-infrared telescope to be built, after the E-ELT, TMT and GMT (we refer to these facilities collectively as the Very Large Optical Telescopes, VLOTs). The large aperture of MSE is fundamental to its importance within the network of international astronomical facilities. Firstly, smaller aperture telescopes cannot provide essential *follow-up* of faint sources identified by imaging surveys on large aperture telescopes like Subaru or LSST. Secondly, as a *feeder* facility for the VLOTs, MSE must be able to provide reasonable signal-to-noise measurements of faint sources that will be followed up with higher spatial resolution (e.g., IFUs), higher spectral resolution, and/or higher signal-to-noise on these giant facilities.

MSE will utilize the full light-gathering power of its large aperture in pursuit of its driving science goals. There are numerous excellent projects underway or under development that provide MOS capabilities for 4 m class apertures. MSE science, however, is focused towards an understanding of the *faint* Universe - such as intrinsically faint stars, the distant Galaxy, low mass galaxies, and the high redshift Universe - that is not accessible with smaller apertures. This opens up extensive new areas of research, such as *in-situ* chemodynamical studies of the distant Milky Way stellar halo, enabling all Galactic components to be studied in an unbiased manner. Indeed, MSE is the only facility that will be able to conduct detailed chemical abundance studies across the full luminosity range of targets identified by the ESA/Gaia mission. Further, for extragalactic science, the dynamics and stellar populations of dwarf galaxies that fall below the detection threshold of smaller-aperture facilities at low and moderate redshifts will be easily accessible, and MSE will allow the analysis of sub-$L_*$ galaxies out to high redshift. Again, the light gathering power of the large aperture is essential in allowing unbiased analyses of these galaxy populations that could not be achieved with smaller apertures.

### 1.4.2 Operation at a range of spectral resolutions

A core requirement of MSE is that it operates at a range of spectral resolutions and this is essential to its successful operation. Firstly, as already described, it enables a diverse range of scientific investigations, and MSE is expected to contribute equally to "Local Universe" science – where the brighter targets are generally more well-suited to high and intermediate spectral resolutions – and "Distant Universe" science – where the more distant faint targets generally require lower spectral resolutions. This diversity is essential to ensure success as a facility, in contrast to instruments (where a more limited range of capabilities is often acceptable due to a focus on a narrower range of science cases). Good examples of this latter approach are Mayall/DESI (that focuses upon the determination of spectroscopic redshifts of three different sub-classes of galaxies for precise determination of cosmological parameters) and AAT/HERMES (that is focussed upon the derivation of the chemical abundances of stars in specific components of the Galaxy).

As well as promoting scientific diversity, a range of spectral resolutions is essential to the operational efficiency of MSE. In particular, MSE will observe throughout the lunar cycle, with higher resolution observations of bright targets expected to dominate during bright time, and fainter extragalactic targets expected to dominate the observing during dark time. Operating with high efficiency is important for all facilities, but it is especially true for MSE where the accumulation of large datasets is a driving requirement for nearly all observing programs.

### 1.4.3 Dedicated operations over a long lifetime

MSE provides a specialized technical capability, namely large-aperture wide field extreme MOS. MSE will conduct spectroscopy of a vast number of astronomical sources to produce homogeneously-calibrated, well-characterized, datasets. For instruments that move on and off telescopes at regular intervals, data issues such as calibration, stability and reproducibility can be problematic since the instrument is not being left in a stable configuration. This is in contrast to the situation for MSE, where the basic operational philosophy of specialization enables high quality and stable data (e.g., fibre coupling will be left in place all the time, no prime focus changes, WFC or ADC removals, etc.). Indeed, this allows for science cases that would otherwise be very difficult to address on other instruments (for example, time resolved high-resolution spectroscopy and quasar reverberation mapping).

The impressive power offered by a large, homogeneous and well-characterised dataset such as that offered by MSE has been most successfully demonstrated by the Sloan Digital Sky Survey (SDSS). It should perhaps come as no surprise that the science that ultimately emerged from SDSS – and which continues to emerge – was far more diverse than anticipated at its outset. The importance of the spectroscopic component of SDSS can scarcely be overestimated: it is universally recognized that the science impact of SDSS would have been greatly diminished without its spectroscopic element, for the simple reason that broadband photometry alone provides only zeroth-order information on the physical properties of astrophysical sources. SDSS had a profound impact on topics as far ranging as small bodies in the solar system to the reionization of the Universe – a broad appeal that underlies its repeated ranking as the highest-impact telescope of the last decade (e.g. Madrid & Macchetto 2006, 2009; Chen et al. 2009)[1].

The overwhelming success of SDSS is despite the fact that the telescope is a relatively small-aperture facility by modern standards, and it is located at a site that, while good, does not compete in terms of median image quality with Maunakea. A large part of this success can therefore be traced to the extremely well-calibrated and well-characterized nature of the data. As the fourth largest OIR telescope in operation in the 2020s, and the only large aperture facility fully dedicated to wide field multi-object spectroscopy at equatorial latitudes, MSE can be viewed as an evolution of the SDSS concept. The major differences are that MSE is realized on a facility that has ~25 times larger aperture than SDSS and that is located at arguably the best telescope site on the planet. The science potential for MSE is therefore significant.

An additional analogy between SDSS and MSE lies in the fact that, in contrast to many of the instruments or facilities listed in Table 2, MSE is not built to address a single science case or to conduct a single survey. Rather, MSE is built to provide a single, missing, important, set of capabilities (large aperture dedicated wide field MOS at a range of spectral resolutions). MSE is designed and optimized to excel at providing these capabilities, and in so doing enables a vast range of science and strategic surveys. Similarly, SDSS has demonstrated flexibility and responsiveness to science over its lifetime though successive generations of the telescope addressing new and strategic science goals: the Legacy Survey; SEGUE I and II, Marvels, BOSS, eBOSS, APOGEE I and II. Over its long lifetime, MSE is expected to address a very large range of science cases, and instrument/telescope upgrades will occur. But like the SDSS over the last few decades, MSE will remain the world's premier resource for the spectroscopic exploration of the Universe.

### 1.5 Competition and synergies with future MOS

Figure 6 shows a large number of MOS instruments and facilities, many of which will be operating on timescales that overlap with MSE. Here, we provide a brief overview of the major new projects in this area.

[1] A search on the Astrophysical Data System for publications that mention "SDSS" in their abstract – presumably those papers that either rely on SDSS data for analysis, interpretation or comparison – returns over 11000 papers with over a quarter of a million citations.

### 1.5.1 4-m class MOS

**Table 3: Summary of new optical and infrared multi-object spectroscopic instruments and facilities.**

| *Class* | *Facility / Instrument* | *First light (anticipated)* | *Aperture (M1 in m)* | *Field of View (sq. deg)* | *Etendue* | *Multiplexing* | *Wavelength coverage (um)* | *Spectral resolution (approx)* | *IFU* | Dedicated facility |
|---|---|---|---|---|---|---|---|---|---|---|
| Comparison | **SDSS I - IV** | Existing | 2.5 | 1.54 | 7.6 | 640 | 0.38 - 0.92 | 1800 | Yes | Yes |
| 4-m | **Guo Shoujing / LAMOST** | Existing | 4 | 19.6 | 246 | 4000 | 0.37 - 0.90 | 1000 - 10000 | No | Yes |
| | **AAT / HERMES** | 2015 | 3.9 | 3.14 | 37.5 | 392 | windows | 28000, 50000 | No | No |
| | **WHT / WEAVE** | 2017 | 4 | 3.14 | 39.5 | 1000 | 0.37 - 1.00 | 5000 | Yes | Yes |
| | | | | | | | windows | 20000 | | |
| | **VISTA / 4MOST** | 2017 | 4 | 2.5 | 31.4 | 2400 | 0.39 - 0.95 | 5000 | No | Yes |
| | | | | | | | windows | 18000 | | |
| | **Mayall / DESI** | 2018 | 4 | 7.1 | 89.2 | 5000 | 0.36 - 0.98 | 4000 | No | Yes |
| 8-m | **VLT / MOONS** | 2018 | 8.2 | 0.14 | 7.4 | 1000 | 0.8 - 1.8 | 4000 | No | No |
| | | | | | | | windows | 20000 | | |
| | **Subaru / PFS** | 2019 | 8.2 | 1.25 | 66 | 2400 | 0.38 - 1.26 | 2000 | No | No |
| | | | | | | | 0.71 - 0.89 | 5000 | | |
| 10-m | **MSE** | 2024 | 11.25 | 1.5 | 149 | 3468 | 0.36 - 1.8 | 3000 | Second generation | Yes |
| | | | | | | | 0.36 - 0.95 50% coverage | 6500 | | |
| | | | | | | | windows | 40000 | | |

**LAMOST (2011):** The Large Sky Area Multi-Object Fiber Spectroscopic Telescope (LAMOST) – also known as the Guo Shoujing Telescope – is a 4m reflecting Schmidt telescope with a 5 degree diameter field of view with a dedicated suite of spectrographs capable of carrying out R ~ 1500 spectroscopy for ~ 4000 objects per field. It is undertaking two main surveys, the LAMOST ExtraGAlactic Survey (LEGAS) and the LAMOST Experiment for Galactic Understanding and Exploration (LEGUE).

**AAT/HERMES (2015):** The High Efficiency and Resolution Multi-Element Spectrograph (HERMES) is a four channel fiber fed spectrograph on the 3.9m Anglo-Australian Telescope (AAT) that uses the 392 fiber 2 Degree Field (2dF) fiber positioner system. HERMES operates at R = 28 000, although a higher resolution of R ~ 50 000 is possible using a slit mask at the cost of considerable light loss. The main program being undertaken by HERMES is the ``Galactic Archaeology with HERMES" (GALAH) survey, that targets a total of 1 million stars brighter than V ~14 mag, measuring a radial velocity for each as well as chemical abundances for ~15 different elements.

**WHT/WEAVE (2017):** The WHT Enhanced Area Velocity Explorer (WEAVE) is based on the successful 2dF positioner system of the AAT, modified to allow ~1000 fibers to be positioned over a 2 degree field for the 4.2m William Herschel Telescope. The spectrographs provide R ~ 5000 over the full optical wavelength range (370 – 1000 nm), and R ~ 20000 over a more restricted range. WEAVE also possesses notable IFU capabilities, specifically 20 deployable IFUs each with thirty seven fibers with a diameter of 1.3” each, as well as a large format IFU with 540 fibers with diameters of 2.6” each.

**VISTA/4MOST (2017):** The 4-meter Multi-Object Spectroscopic Telescope (4MOST) is a 4 square degree field of view facility. 2400 targets can be observed per field, feeding both high resolution (R~20000) and intermediate resolution (R~5000) spectrographs. 4MOST will operate as a dedicated facility enabling a wide range of spectroscopic surveys, and in many operational respects is a 4m analogy to MSE.

**Mayall/DESI (2018):** The Dark Energy Spectroscopic Instrument (DESI) will go on the 4-m Mayall telescope. It is focused on performing a Stage IV Dark Energy measurement of baryonic acoustic oscillations by targeting a total of 25 million luminous red galaxies, emission line galaxies and quasars over a period of 5 years. 5000 objects can be obtained per 8 square degree field of view, at a spectral resolution of R = 4000 by feeding 10 identical three-arm spectrographs. DESI will also operate additional surveys during bright time, including targeting some 10 million stars brighter than V ~ 18.

### 1.5.2 8-m class MOS

**Table 4: MSE in comparison to other planned MOS instruments on 8-m class telescopes.**

| | **8 - 12 m class facilities** | | | | | | |
|---|---|---|---|---|---|---|---|
| | **VLT / MOONS** | | **Subaru / PFS** | | **MSE** | | |
| *Dedicated facility* | No | | No | | Yes | | |
| *Aperture (M1 in m)* | 8.2 | | 8.2 | | 11.25 | | |
| *Field of View (sq. deg)* | 0.14 | | 1.25 | | 1.5 | | |
| *Etendue* | 7.4 | | 66 | | 149 | | |
| *Multiplexing* | 1000 | | 2400 | | 3468 | | |
| *Etendue x Multiplexing* | 7400 *( = 0.014)* | | 158400 *( = 0.31)* | | 517100 *(=1.00)* | | |
| *Observing fraction* | < 1 ? | | 0.2 (first 5 years)<br>0.2 - 0.5 afterwards ? | | 1 | | |
| *Spectral resolution (approx)* | 4000 | 20000 | 2000 | 5000 | 3000 | 6500 | 40000 |
| *Wavelength coverage (um)* | 0.8 - 1.8 | windows | 0.38 - 1.26 | 0.71 - 0.89 | 0.36 - 1.8 | 0.36 - 0.95<br>50% coverage | windows |
| *IFU* | No | | No | | Second generation | | |

**VLT/MOONS (2018):** The Multi-Object Optical and Near Infrared Spectrograph (MOONS) is a 500 square arcminute field of view spectrograph for the 8.2m VLT operating at R ~ 5000 and R ~ 20000 and capable of taking 1000 spectra simultaneously (nominally, 500 object and 500 sky spectra). In addition to its field of view, it differs from most other MOS spectrographs by targeting the red/NIR wavelength range, 0.8 – 1.8μm, making it ideally suited for studies of the bulge and other Galactic fields with high reddening, as well as the high redshift Universe.

**Subaru/PFS (2019):** The Prime Focus Spectrograph (PFS) for the 8.2m Subaru telescope (Vives et al. 2012; Sugai et al. 2012) is a 2400 object spectrograph with a 1.25 square degree field of view that covers the optical/NIR wavelength range (0.38 – 1.26μm). It primarily operates at R ~ 2000, although there are plans for an additional R ~ 5000 mode in the region of the Calcium Triplet. A "Strategic Survey Program" will use this instrument for ~300 nights distributed over a 5-year period to conduct a program to explore the nature of dark energy through observations of emission line galaxies, and a Galactic Archaeology program for the derivation of velocities and stellar parameters for ~1 million stars (Takada et al. 2014).

## 1.6 The future-scape of MSE

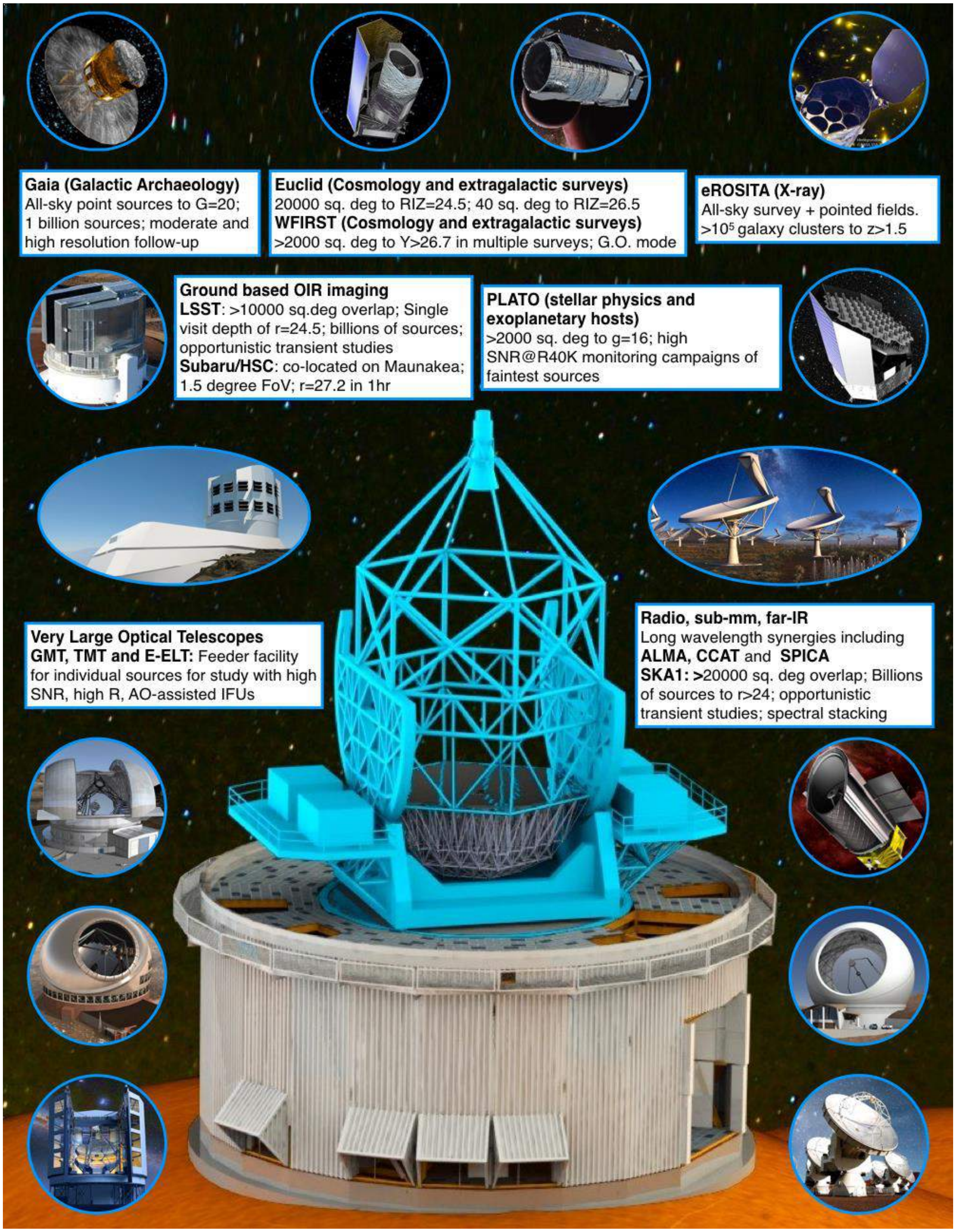

**Figure 7: Some of the most notable next generation facilities that, together with MSE, will help define the international network of astronomical facilities operational beyond the 2020s.**

Astronomy is entering a new, multi-wavelength realm of big facilities. Figure 7 shows some of the most notable of these new observatories. Collectively, they represent many billions of dollars of investment and many decades of development by large teams from the international community. They probe questions as diverse and fundamental as the origin of life and the composition of our Universe, and their scientific exploitation has already begun. At the time of writing, ALMA is now in routine operation, and the first data release from Gaia is imminent. These are no longer “next generation” ambitions, but instead are the cutting edge of what is possible now. Perhaps most excitingly, our view of the Universe is no longer limited only to the electromagnetic spectrum, such that the term “multiwavelength” is no longer sufficient to completely describe the windows into the Universe that we have at our disposal. Indeed, the detection of the first gravity waves in February 2016 by LIGO promises to provide a fundamentally new way of exploring the Universe.

Each of the facilities shown in Figure 7 has a well-defined science case and can operate as a stand-alone facility. However, the recent history of astronomy demonstrates that it is through the combination of data from these new facilities and extant telescopes that many of the major advances will be made. The science that will ultimately emerge from this collaboration is far more diverse than we can currently anticipate.

The development of MSE explicitly recognizes the emerging existence of an international network of astronomical facilities that will define the future of frontline astrophysics. MSE is positioned to be a critical hub in this network, with scientific capabilities that naturally complement and extend the capabilities of the facilities shown in Figure 7. Many of the synergies are obvious, many cannot yet be anticipated, but all are important to the future health of astronomy. Here, we discuss MSE in this global context of astronomical capabilities.

### 1.6.1 Wide field imaging of the Universe

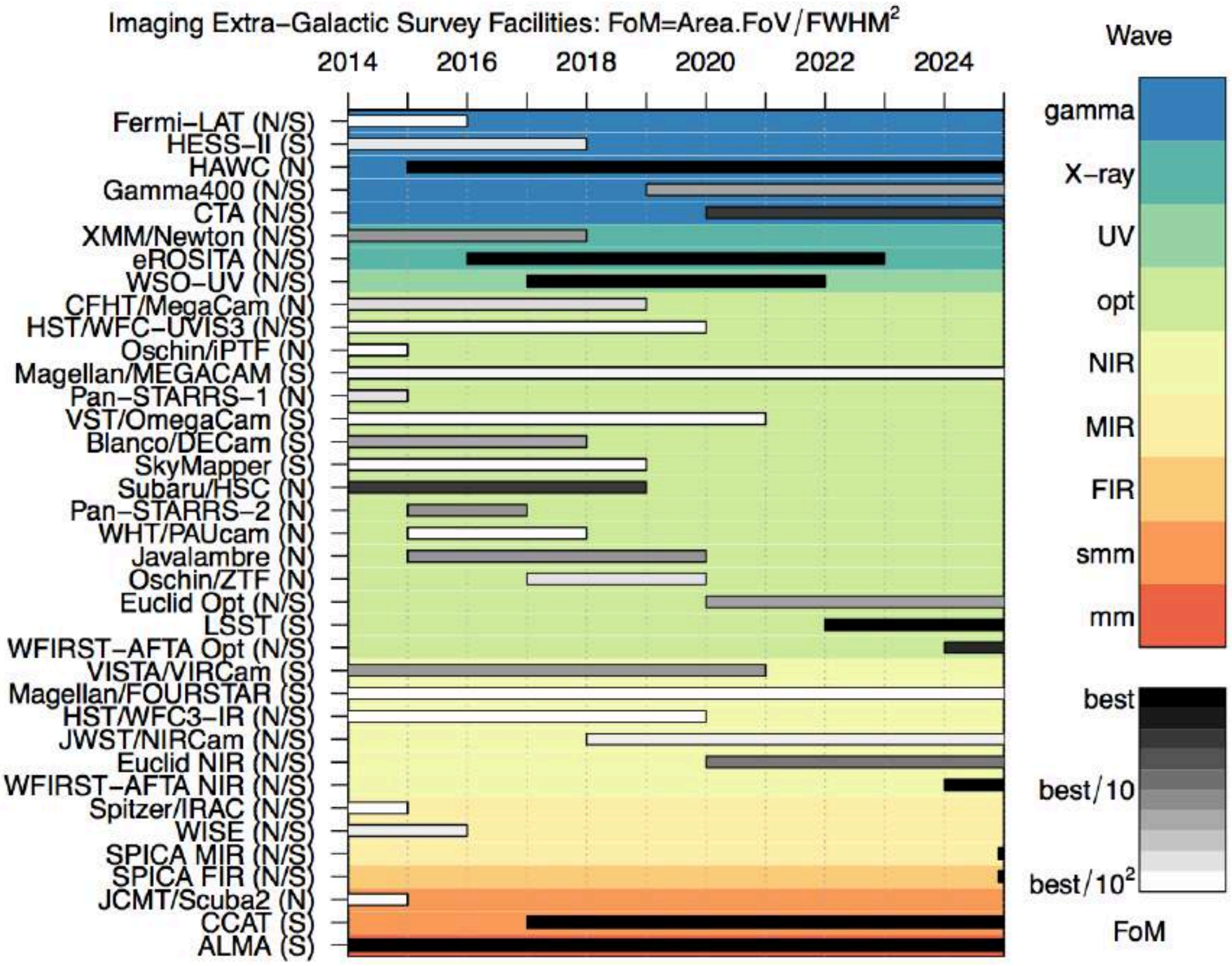

**Figure 8: Calendar comparing various imaging survey capabilities across the Gamma ray to sub-mm spectrum (as indicated by color). The Figure of Merit for these surveys is calculated using Area x FoV / (IQ)$^2$ and is indicated by gray scale.** ***Figure from Meyer et al. (2015).***

Figure 8 shows the timeline for a large number of optical and NIR imaging surveys (green) as well as at shorter (blue/purple) and longer wavelengths (orange/red). Complete, deep, optical imaging of the sky will shortly be available through a combination of survey programs. In the northern hemisphere, the entire 3π steradians of sky visible from Hawaii has been mapped using the PanSTARRS (PS1) telescope in the grizy filters to 5-σ AB limiting point-source magnitudes of (21.9, 21.8, 21.5, 20.7, 19.7) mags, respectively. In the south, SkyMapper is undertaking its Southern Sky Survey, an all-hemisphere survey (in uvgriz) that will be approximately 0.5 – 1 mag deeper than the SDSS. Hyper-Suprime-Cam (HSC) on the Subaru telescope on Maunakea is the largest mosaic camera ever commissioned on an 8m-class telescope, and is currently conducting a 5 year strategic survey to map ~1500 square degrees along the celestial equator to unprecedented depth. From the south, the Dark Energy Survey is using the Dark Energy Camera (DEC) on the 4m Blanco telescope to map an area of 5000 square degrees. Many other extensive wide field imaging surveys on 4 – 8m telescopes already exist, both at optical wavelengths (e.g., the CFHT Legacy Survey and other CFHT programs; the Subaru SuprimeCam imaging archive) and at NIR wavelengths (e.g. the UKIDSS northern sky survey using UKIRT; various surveys on VISTA).

Figure 7 demonstrates that the most powerful ground-based imaging capability on the horizon is the Large Synoptic Survey Telescope (LSST). This facility is the top-ranked, ground-based facility in the 2010 decadal survey for US astronomy. To be situated at Cerrro Pachon, LSST is a 6.2m (effective aperture) telescope with a field of view of 3.5 square degrees that will see first light in 2021. Its primary mission is to undertake a 10-year program to monitor ~25 000 square degrees, building up a deep image of the sky through the co-addition of ~1000 exposures at each position. Each individual exposure lasts ~ 15s and has a single-pass depth of r ~ 24.5. The final, co-added, LSST survey depth is r ~ 27.5.

The discovery space of LSST is enormous. The design of the surveys has been guided by four main science themes – "Taking an inventory of the Solar System", "Mapping the Milky Way", "Exploring the Transient Optical Sky", and "Probing Dark Energy and Dark Matter". In practice, however, LSST can be expected to impact every area of astronomy. Over the 10 year baseline, LSST will take ~ 5.5 million images and will have measured ~ 7 trillion single epoch sources, for a total of ~ 37 billion distinct objects. It is a requirement that photometric zeropoints are stable to ~ 10 millimags; with nearly 1000 visits per patch of sky, LSST will probe the variable and transient optical sky with unprecedented accuracy and cadence.

The scientific synergies between MSE and deep imaging surveys such as LSST, Subaru/HSC, PS1 and others, that together map the entire sky, are extensive. It is worth emphasizing that the majority of the billions of sources that will be photometrically identified will be faint, far outside the capabilities of 4-m class spectroscopy. As an 11.25m aperture facility, MSE obtain spectroscopic data for sources across the full magnitude range of PS1, and will be able to obtain spectra for sources identified in a single pass of LSST. Smaller aperture spectroscopic facilities are unable to exploit the discovery potential of these surveys to the same degree as MSE, and emphasizes the strong need for large aperture, dedicated MOS.

While there is a long heritage of the obvious synergies between optical and infrared imaging and spectroscopic surveys, new modes of follow-up arise from the dedicated spectroscopic survey capabilities of MSE that highlight the diversity of science that it enables. Due to their cadence, both PS1 and LSST are powerful transient discovery machines; LSST expects to issue $10^5$ – $10^6$ transient alerts per night. Similarly, SKA and other multi-wavelength survey facilities will also identify very large number of transients, and follow-up of photometrically identified optical/near-IR counterparts of radio, X-ray and gamma-ray transients will be essential.

MSE is uniquely placed to undertake a systematic, long term opportunistic science program in support of LSST, SKA and other transient surveys (see Chapter 4). Given its equatorial location, MSE provides good access to both hemispheres, and there will be ~ 1500 square degrees per night of LSST imaging accessible to MSE, in which we estimate there will be ~ 300 new SNe and >75000 variable stars (Ridgway et al. 2014, ApJ, 796, 53), as well as many more exotic transient phenomena. For any given pointing of MSE, one or two fibers on average can be allocated to sources recently identified as transients by LSST (or other surveys). Given MSE is always obtaining spectra for many thousands of targets per hour, then this quickly amortizes to ensure that MSE will dominate a new regime of faint transients (probing very early or very late stages, as well as large distances and low luminosities), Indeed, MSE will likely dedicate more of its time to transient spectroscopy than any other 10m class telescope.

Inspection of Figure 8 reveals that, while LSST dominates wide field imaging from the ground, imaging from space is dominated by two missions: Euclid and WFIRST. Euclid, scheduled for launch in 2020, is an ESA-led mission to understand the nature of dark energy and dark matter through weak lensing and galaxy clustering. It is a 1.2m NIR/optical telescope that will image a minimum area of 15 000 square degrees at |b| > 30 degrees during its anticipated 5-year lifetime. Limiting AB magnitudes are expected to be 24 in YJH, and 24.5 in a single, broad optical filter (RIZ). Overall, more than a billion galaxies will be observed. WFIRST is a complementary, NASA-led mission that has a considerably larger aperture than Euclid (at 2.4m) and that is equipped with a wide field imager, an integral field unit, and a coronographic imager for exoplanet imaging. In terms of imaging, one of its primary missions will be to map $\sim$ 2000 square degrees of sky to depths of J $\sim$ 26.7, in addition to a set of ultra-deep degree-scale fields. Precision photometry and astrometry will be obtained for all sources. Importantly, up to 25% of WFIRST time is likely going to be P.I.-time, ensuring that deep pointed surveys of various fields will occur. Euclid and WFIRST will together provide an unprecedented wide-field NIR dataset that, while focused primarily on cosmological science questions, will have a lasting legacy for all of astronomy.

Examples of the sort of scientific opportunities produced by Euclid and WFIRST for MSE are numerous and discussed throughout this document. A driving theme of extragalactic science with MSE is linking galaxies to their surrounding large scale structure (Chapter 3). As such, the emergence of structure on kiloparsec scales is of great interest. Driver et al (2013) proposed that galaxies form via two stages, firstly bulge formation via some dynamical hot process (i.e., collapse, rapid merging, disc instabilities and/or clump migration), and secondly disc formation via a more quiescent dynamically cool process (i.e., gas in-fall and minor-merger accretion events). The pivotal redshift for these processes is at z $\sim$ 1.5. At later times, during periods of low interaction rates (see Robotham et al 2014), bars emerge along with disc-buckling and pseudo-bulge formation. In the present day Universe, systems adhere to the Hubble sequence, at earlier times galaxies appear more lie train-wrecks, as unveiled by the Hubble Space Telescope. While the Hubble Space Telescope (and its successor the James Webb Space Telescope) provide (and will provide) high quality imaging, the fields of view are very small and only the COSMOS survey extends to more than a square degree. While this provides insight into the morphologies of galaxies at very high redshifts, the epochs and process by which each structure (bulge, bar, disc) emerge is empirically unconstrained.

Euclid and WFIRST will provide extremely deep high-spatial resolution imaging ($\sim$ 0.2”) and will discern structure to sub-kpc scales out to z > 2.5 at rest-optical wavebands over extremely large areas. This will allow a direct measure of the epochs at which the various structures emerge (e.g., bulge, disc formation) and how they evolve (i.e, growth of spheroids, bulges, and discs). However, to fill in the void between the very nearby surveys such as SDSS and GAMA and the very distant surveys with HST and JWST requires the *combination* of Euclid/WFIRST imaging with photometric-redshifts (LSST) and MSE spectra. Through this collaboration, we can extract star-formation rates, metallicities, and can confirm pair membership to establish merger rates.

The combination of data from Euclid/WFIRST/LSST (and SKA; see Section 1.6.3) with MSE spectral analysis will provide a complete blueprint of galaxy evolution from the present epoch to the peak of the cosmic star-formation era (i.e., z = 0 to z ~ 2.5). Samples sizes need not be large; the crucial element will be the ability to obtain reasonable SNR spectra (~ 30) for high-z (z ~ 2.5), faint (i ~ 25 mag) systems with a high-level of completeness. MSE is the only spectroscopic facility that provides these capabilities.

### 1.6.2 The era of Gaia

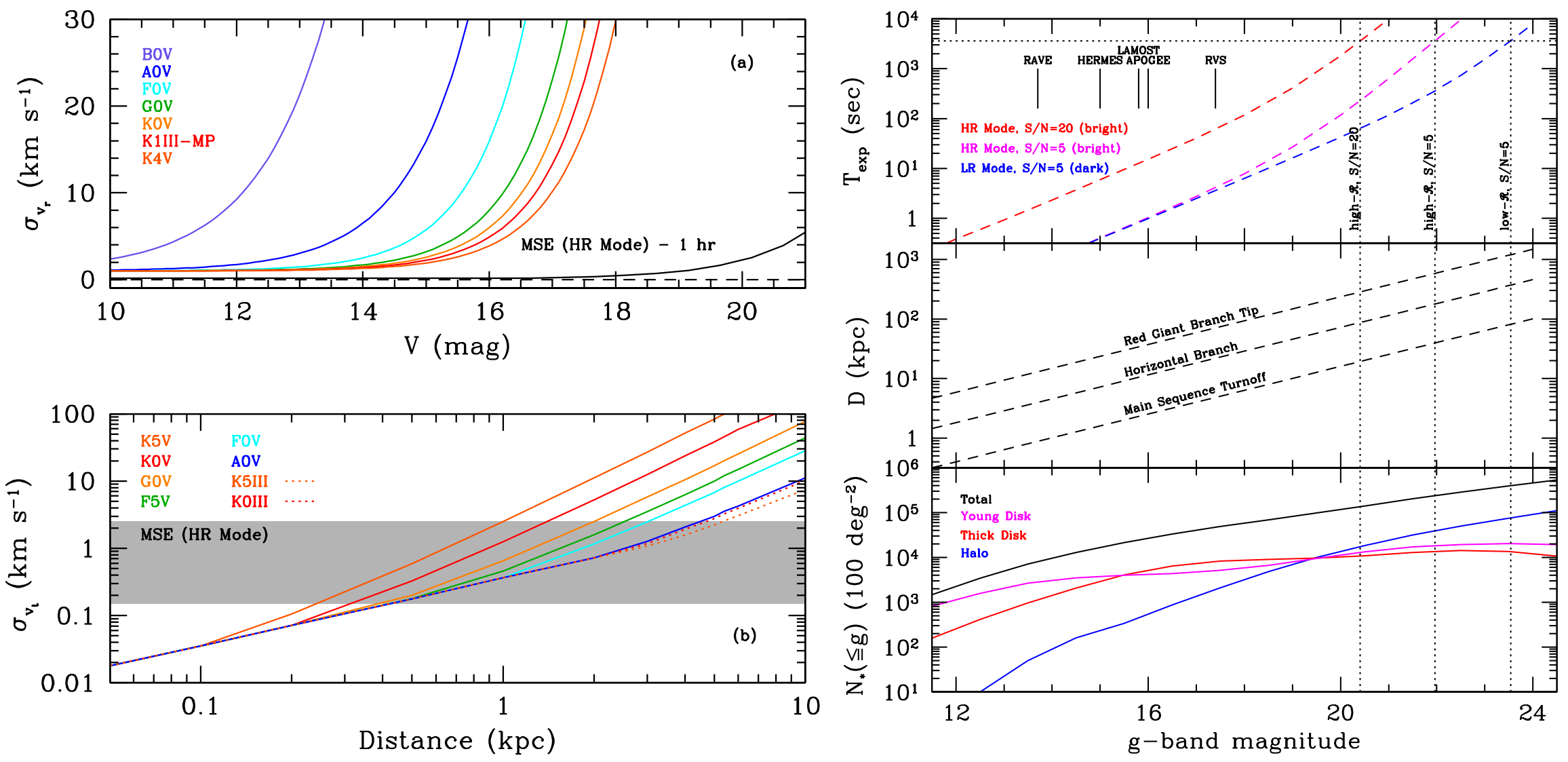

**Figure 9: Top left: Expected radial velocity accuracy for the RVS instrument aboard Gaia for stars of different spectral types. The lower black curve shows the radial velocity accuracy expected for a G2V star observed for a total of 1 hour with MSE in high-resolution mode. Bottom left: Typical errors in transverse velocity as a function of distance for stars of different spectral type observed with Gaia. The shaded grey region shows the approximate errors calculated for G2V stars with g = 20.5 with MSE. Top right: Approximate exposure times required for MSE to reach target signal-to-noise ratios, plotted as a function of g-band magnitude. The horizontal dotted line indicates the nominal survey exposure time of 1 hour. Middle right: Depth reached for old, metal-poor halo populations as a function of g-band magnitude for three different tracer populations. Bottom right: Cumulative number counts for Milky Way stars in a 100 deg$^2$ region at the north Galactic cap (black curve). Star counts for three separate Galactic components (young disk, think disk and halo) are shown in colour.**

Gaia is a landmark astrometric space mission that has as its primary focus a detailed understanding of the structure and composition of our Galaxy. Launched in 2014, it has begun its 5 year mission to produce a three-dimensional stellar map of the Milky Way with unprecedented precision. Gaia is conducting an all-sky survey to measure the positions of roughly 1 billion stars, approximately 1% of the entire stellar content of the Milky Way. In fact, all astronomical point sources brighter than V ~ 20 mag will be catalogued, and objects brighter than V ~ 15 mag will have their position measured to better than 20 microarcseconds (the approximate width of a human hair viewed at a distance of 1000km). Beyond astrometry, Gaia will obtain multi-band photometry for all sources, and it is additionally equipped with a Radial Velocity Spectrometer (RVS) that will measure velocities for objects brighter than V ~ 17 mag (roughly 150 million stars) to an accuracy of 1 – 10kms$^{-1}$. Basic astrophysical information – including interstellar reddening and atmospheric parameters – will be acquired for the brightest

∼ 5 million stars. Chemical abundance information will also be provided for a few elements (i.e., Mg, Si, Ca, Ti and Fe for stars of spectral type F-G-K) for stars brighter than 12th magnitude.

The Gaia dataset is a unique and comprehensive resource for astronomy, given the unprecedented accuracy with which it will measure the 3-D positions of Galactic stars. There is a near perfect synergy between MSE and Gaia on multiple fronts, and these synergies feature heavily in Chapter 2 in particular.

The importance of ground-based spectroscopy to supplement the Gaia data is well recognized in the international community, particularly as it relates to the chemical tagging of individual stars (Chapter 2). For chemical evolution studies, Gaia will identify hundreds of millions of stars fainter than G ∼ 17 for which the RVS cannot provide any spectral information. In response to the immediate need for deep spectroscopic observations of faint stars for chemical abundance analysis, some 300 co-investigators are involved in the ESO-Gaia survey (Gilmore et al. 2012) since 2011. This is a dedicated effort to use the VLT's high-resolution, multiplexed spectrographs to observe $\sim 10^5$ stars with FLAMES/GIRAFFE ($R \sim 20000$, $V < 20$) and $\sim 10^4$ stars with FLAMES/UVES ($V < 15$) over a period of 5 years. Results from this survey are already informing critical aspects of next generation spectroscopic surveys, including MSE, specifically informing trade-offs between resolving power, wavelength range and multiplexing. As discussed in Chapter 2 these trades are particularly important for chemical tagging studies since we aim to measure the abundances of enough chemical elements, with sufficiently diverse nucleosynthetic origins, with a sufficiently high precision.

The general goal of chemically tagging stars in the Galaxy is common between MSE and other high resolution spectroscopic surveys, such as VISTA/4MOST and WHT/WEAVE. These other surveys will provide essential insight into the nearby Galaxy and the main disk in particular. However, only MSE has the resolution and throughput to conduct these experiments throughout the Galaxy, via in-situ chemical analysis of every component and sub-component, and across the entire luminosity range of Gaia targets.

In terms of data on the dynamics of stars, the astrometric accuracy of Gaia is only matched by similarly accurate radial velocities for the very brightest subset of its targets. For relatively faint stars, MSE is required in order to provide precise radial velocities to complement the precise transverse velocities measured by Gaia. Figure 9 illustrates this with a comparison of the anticipated velocity accuracy of MSE compared with the RVS on Gaia, including showing the distance out to which certain stellar tracers can be observed and what this corresponds to in terms of contributions from different Galactic components. We return to discussion of Gaia synergies in Chapter 2.

### 1.6.3 The Square Kilometer Array

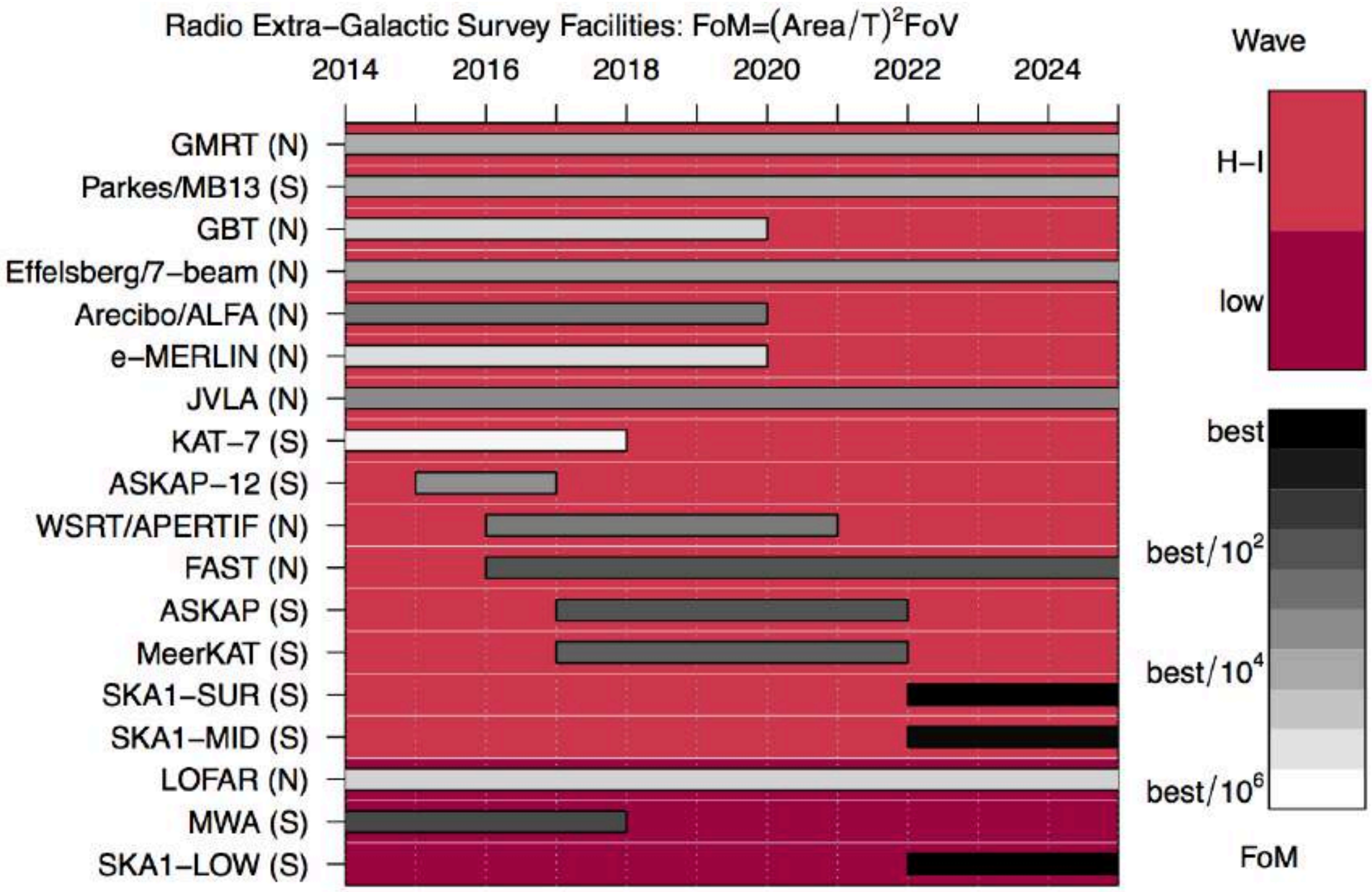

**Figure 10: Calender comparing various radio survey facilities in operation now and the near future. The Figure of Merit is calculated using $(Area/T)^2$ x FoV / $IQ^2$ and is shown as grayscale for each survey. Note that "H-I" corresponds to facilities that are able to observe HI in the local Universe i.e., that can observe frequencies as high as 1.4 GHz.** ***Figure from Meyer et al. (2015).***

Figure 10 shows the major radio surveys and facilities currently operational or planned to be operational prior to 2025. Inspection of this figure shows that the SKA will have a profound impact. Among the many science goals of this transformational telescope, the SKA has the capability of detecting Milky Way-type galaxies via synchrotron radiation into the epoch of reionization, finding AGN of all types and luminosity, and tracing the neutral hydrogen content of galaxies to z ∼ 2.

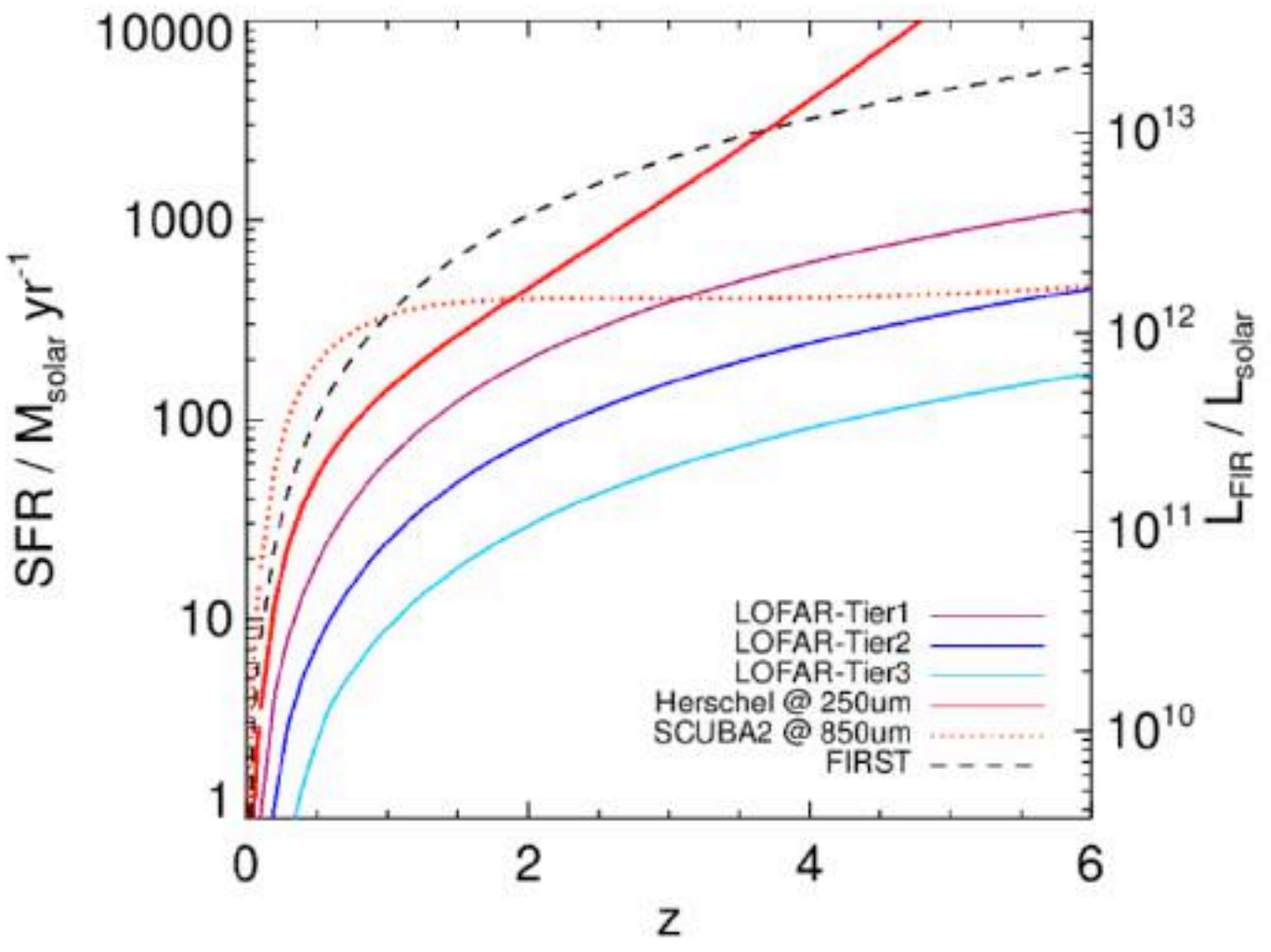

**Figure 11: Star formation rate sensitivity of selected far-infrared observatories and radio continuum surveys as a function of redshift. The red lines show the confusion-limited SFR sensitivity of the *Herschel Space Observatory* at 250 μm and the SCUBA-2 instrument on the JCMT at 850 μm (solid and dashed, respectively). The black dashed curve shows the SFR sensitivity of radio surveys from the J-VLA over Stripe 82 (Heywood et al. *in prep;* solid line), and the colored lines show the three tiers of the LOFAR Surveys KSP; Tier 1 (dotted line), Tier 2 (dot-dashed line), and Tier 3 (dot-dot-dot-dashed). Spectroscopic redshifts are necessary to provide the x-axis for real data in this plot, and will be provided for LOFAR by WHT/WEAVE. MSE and SKA can expect similar synergies in this field. *Figure from Smith (2015).***

Optical spectroscopy is crucial to maximise the scientific output of the SKA, and to gain the biggest leap in our understanding of galaxy formation and cosmology. For example, the extreme sensitivity of the SKA means that it will be able to detect star-forming galaxies in radio continuum emission to high (z > 1) redshifts. However, the lack of spectral features in the radio continuum means that additional optical spectra are required. While photometric redshifts provide reasonable precision, they are insufficient for most other science, for example quantification of the galaxy environment, or the measurement of ages and metallicity etc. Discrimination of AGN from star formation activity likewise requires optical emission line diagnostics.

A good example of the type of synergy we can expect between SKA and MSE in this arena can be seen with the proposed WEAVE – LOFAR survey (P.I. Daniel Smith), to be conducted using the 4-m class WHT/WEAVE spectroscopic facility. Figure 11 shows the star formation rate sensitivity of the three teirs of the LOFAR radio continuum survey compared to other surveys. The sensitivity of Tier 3 of the LOFAR survey is extreme and will allow the routine detection of sub-millimeter-like galaxies at z > 5. However, the critical redshift information on the x-axis of Figure 11 is lacking without optical spectroscopy and here WEAVE will provide essential data. Full details of the WEAVE – LOFAR survey are available elsewhere (Smith 2015). MSE and SKA can expect equally strong synergies as they relate to the SKA radio continuum surveys.

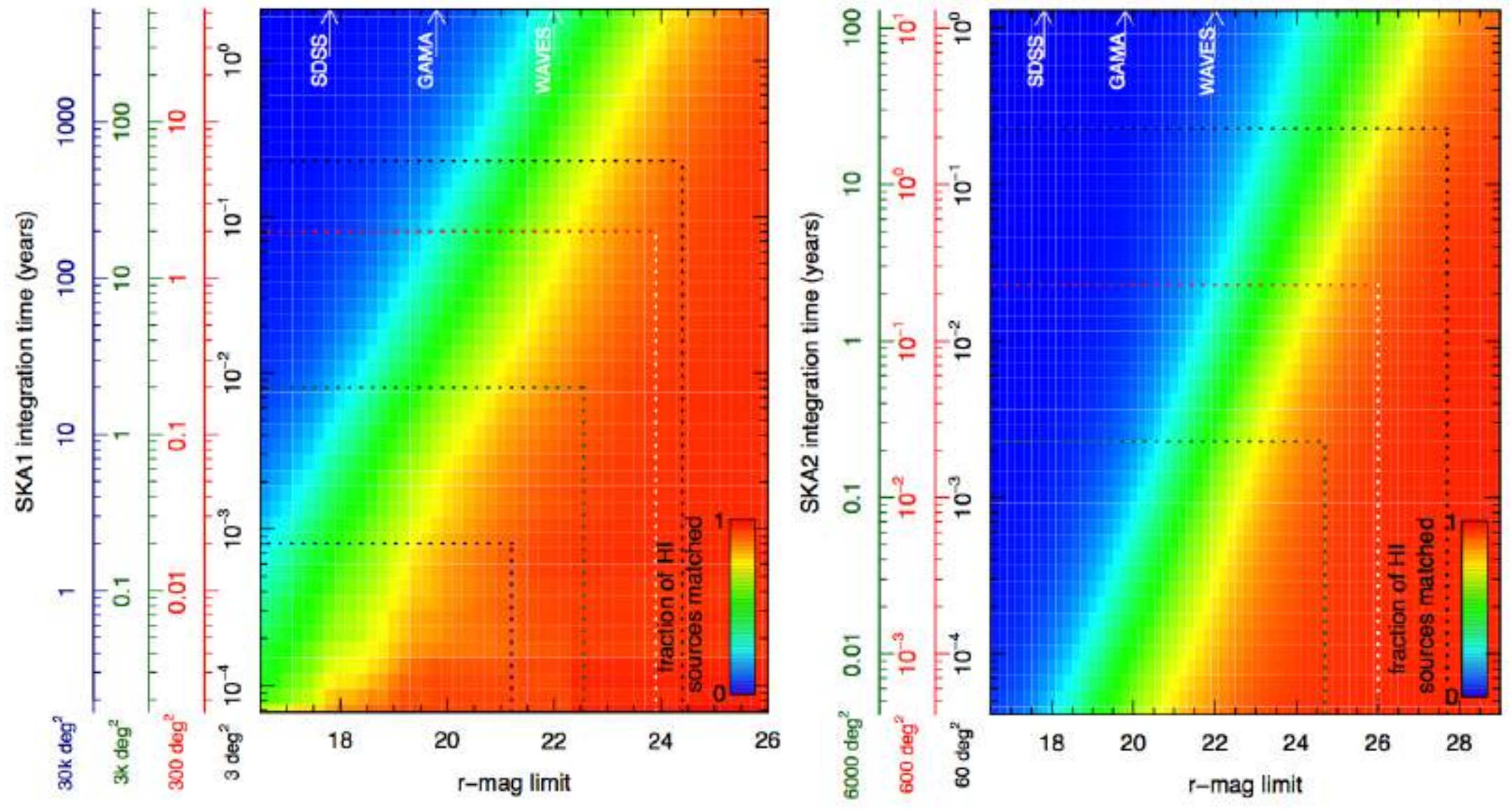

**Figure 12: Fraction of HI SKA survey sources (left panel, SKA1; right panel, SKA2) that are detected in an r-band apparent magnitude-limited sample. For SKA1, four surveys are considered (including the now defunct SKA-SUR mode): 3 deg$^2$ using SKA1-MID (black), 300 deg$^2$ using SKA1-SUR (red), 3000 deg$^2$ using SKA1-SUR (green), and 30000 deg$^2$ using SKA1-SUR (blue). For SKA2, three surveys are considered: 60 deg$^2$ (black), 600 deg$^2$ (red), and 6000 deg$^2$ (green). Dotted lines indicate the r-magnitude limit needed to achieve matches for 90% of sources in each SKA survey area (given 2000 hrs of integration for SKA1-MID & SKA2 surveys, and 2 years of telescope time for SKA1-SUR surveys). Dashed lines indicate 90% completion thresholds. *Figure from Meyer et al. (2015).***

The challenge posed by the sensitivity of SKA for optical astronomy is summarized in Figure 12. This shows the results of some simulations presented in Meyer et al. (2015), in which they calculate the fraction of HI sources that will be identified in SKA that will also be identified in the optical, as a function of r-band limiting magnitude. Several different configurations for SKA 1 (left panel) and SKA 2 (right panel) are considered (including the now defunct SKA-SUR mode). Clearly, even to match to SKA 1 detection thresholds requires optical data extending to r ~ 24. While easy to achieve with imaging, this is at the limit of 4-m spectroscopic capabilities; for example, the detection limits for one of the most ambitious extragalactic surveys to be conducted on VISTA/4MOST, called WAVES, is marked on Figure 12, and even this is below the 90% completeness threshold for several of the possible configurations. In addition, the construction of large statistical datasets to capitalize on the SKA 1 data requires the survey speed of the spectroscopic facilities to be high, and this scales with the aperture of the facility. While the sensitivity limits of SKA 2 are uncertain at this stage due to the range of possible options open to SKA after SKA 1 is complete, it is obvious that 4-m class spectroscopic facilities will be unable to provide the spectroscopy necessary to best match these data. Very large aperture facilities such as MSE are therefore essential.

Numerous other synergies exist between MSE and SKA. For example, even the SKA will not be sensitive enough to directly detect HI in all but the gas rich galaxies out to z ~ 2. Therefore, a key focus is on stacking galaxies with known redshifts. Here, spectroscopic redshifts are far more valuable than photometric redshifts. MSE could provide the ideal spectroscopic sample to carry out HI stacking analyses of galaxies, subdivided by various galaxy properties such as age, morphology, redshift etc.

MSE and SKA naturally provide measurements of distinct tracers to trace the underlying density distribution of the Universe via galaxy redshifts catalogues (e.g., precise redshifts for luminous red galaxies from MSE and precise redshifts for lower mass, gas- rich galaxies detected in HI with the SKA). These galaxies trace the underlying dark-matter distribution with vastly different bias, thus allowing the so-called “multi-tracer” technique (Seljak 2009) to be used in order to overcome cosmic variance effects (e.g., Ferramacho et al. 2014). Such an approach is necessary since it is increasingly apparent that cosmology is becoming limited by systematics not statistics (see Chapter 4).

A further area of direct synergy between MSE and SKA is to obtain spectroscopic redshifts of distant galaxies that are used for weak lensing. In much the same way as for optical surveys, redshift information adds significant power to weak lensing analyses, allowing the growth of structure to be traced. Radio weak lensing is in itself extremely complementary to optical weak lensing surveys, as the systematic uncertainties are different, e.g., the wavelength-dependent PSF is known analytically in radio interferometry but the source density is generally lower (until SKA2). Thus the same MSE spectroscopic surveys that will be used to supplement optical weak lensing surveys will play the same role in radio weak lensing with the SKA.

### 1.6.4 Feeding the giants

The Thirty Meter Telescope (TMT), the European Extremely Large Telescope (E-ELT) and the Giant Magellan Telescope (GMT) will become the premier astronomical OIR facilities for detailed, high spatial resolution views of the faintest astronomical targets when they see first light at the start of the 2020s (TMT & E-ELT – 2024; GMT – 2025, although science operations are planned earlier with a reduced number of segments). Together, each of these ≥USD1B facilities can access the entire sky with unprecedented collecting areas and with fields of view of order a few arcminutes.

Essential to the efficient scientific exploitation of these forefront observatories is target identification. Ideally, this will use a coordinated suite of supporting facilities to ensure these forefront facilities maximize their science impact by targeting the most scientifically compelling phenomena.

MSE will occupy an important role in the era of the VLOTs through its ability to provide statistically significant samples of OIR spectra of relatively faint sources identified in wide field imaging surveys and wide field surveys at other wavelengths. Targets can be selected from this sample based on criteria specific to the individual science case, using spectral information derived over the same wavelength to which the VLOTs are sensitive. Observations with these giant facilities can then focus on higher SNR, higher spectral resolution and or higher spatial resolution (in the case of spatially resolved sources). Given the plethora of sources identified by current and future wide field surveys at faint magnitudes, this type of filtering is essential for nearly all science programs.

The wide field perspective of MSE and the high precision small field perspectives of the VLOTs will naturally encourage the development of combined science programs between the facilities. For example, Chapter 4 describes precision measurements of the dark matter mass profiles of Milky Way satellites through spatially complete radial velocity surveys extending out to very large radius. Here, it is essential to sample stars that cover the full range of orbital parameters. The VLOTs, with fields of views that are relatively small in comparison to many Milky Way dwarf galaxies, can nevertheless provide critical astrometric data in the central regions of the dark matter halo. This allows the derivation of tangential velocities for a subset of stars that provides significant leverage in distinguishing cored profiles from cusped profiles. On larger astrophysical scales, mass measurements of galaxy clusters using complete kinematic data for cluster members can be obtained using MSE, and compared to precision lensing mass measurements for the same clusters using the VLOTs (Chapter 4).

## 1.7 MSE and its capabilities: a priority in international strategic planning

The plethora of deep imaging and astrometric surveys at optical wavelengths, and of survey missions at other wavelengths, has resulted in significant focus turning to how to obtain complementary OIR spectral data for the (literally) billions of sources that these surveys will identify. Wide field spectroscopy is now seen as a critical "missing link" in the international portfolio of astronomical facilities, especially at large apertures where MSE is the only facility in active development. For example:

- [Australia] The Australian Astronomy Decadal Plan 2016 – 2025 (Australian Academy of Science 2015) explicitly recognizes the importance of large aperture wide field MOS: "Building on Australia's world-leading expertise in optical multi-object spectroscopy, development of an 8-metre class optical/IR wide-field spectroscopic survey telescope towards the end of the decade would complement Australian optical astronomers' leading priority of access to a multi-purpose 8-metre class optical observatory. Such a facility, most likely constructed by an international partnership, would address several of the fundamental astronomical questions of the coming decade, including the nature of dark matter and dark energy, the formation and evolution of the Milky Way and how stars and galaxies process chemical elements. It would also provide follow-up spectra of objects identified by the SKA and imaging telescopes like the US-led Large Synoptic Survey Telescope (LSST)."

- [Canada] The Long Range Plan 2010 (Pritchet et al. 2010) notes that the 10m class telescope equipped with an extremely multiplexed spectrograph would "…have a transformative impact in a wide range of fields, including stellar structure and evolution, large-scale structure, galaxy formation and evolution, dark matter, AGN physics, and the epoch of reionization. Furthermore, … [it] … would be a unique resource for follow-up spectroscopy, both for the European Gaia satellite mission, and also for LSST and Euclid/WFIRST". The report goes on to describe the science case for such a facility as "unassailable".

  Subsequently, the Mid-Term Review of LRP2010 (Thacker et al. 2016) comments directly on MSE and notes that "The scientific and collaborative opportunities available to MSE partners are a direct indicator of the strategic relevance of MSE to the future astronomy landscape." They make the following recommendation "The MTRP strongly recommends that Canada develop the MSE project, and supports the efforts of the project office to seek financial commitments from Canadian and partner institute sources."

- [ESA, ESO] The Report by the ESA-ESO Working Group on Galactic Populations, Chemistry and Dynamics (Turon et al. 2008) made some prescient recommendations on the need to improve wide-field spectroscopic capabilities worldwide, with an eye towards maximizing the scientific legacy of the Gaia dataset. These recommendations include the development of dedicated, highly multiplexed, 4m- and 8m-class spectroscopic telescopes, noting that ``Our terms of reference were to propose a set of recommendations to ESA and ESO for optimizing the exploitation of their current and planned missions. However, the Galaxy is an all-sky object; in fact, from the ground, the outer parts of the Galaxy are best observed from the Northern hemisphere, as the extinction is on-average lower there. In parallel with [*earlier recommendations*] there is a real need for dedicated highly multiplexed spectrographs in the northern hemisphere".

  More recently, the prioritisation of the ESO science programme for the 2020s has been completed. Highly multiplexed spectroscopy has received strong community backing due to its broad importance for a vast range of science[2]. A Working Group has been established to investigate the science case and synergistic opportunities and practical requirements for a dedicated ground-based wide field spectroscopic survey telescope in the 2020s, that is expected to report in mid-2016.

- [USA] The "Astro 2010 Decadal Review" (Blandford et al 2010) noted that ``The properties of dark energy would be inferred from the measurement of both its effects on the expansion rate and its effects on the growth of structure (the pattern of galaxies and galaxy clusters in the universe). In doing so it should be possible to measure deviations from a cosmological constant larger than about a percent. Massively multiplexed spectrographs in intermediate-class and large- aperture ground-based telescopes would also play an important role."

  Subsequently, the US National Research Council commissioned a study entitled "Optimizing the US Ground Based Optical and Infrared Astronomy System" (Elmegreen et al. 2015) in which an official recommendation includes "The National Science Foundation should support the development of a wide-field, highly multiplexed spectroscopic capability on a medium- or large-aperture telescope in the Southern Hemisphere to enable a wide variety of science, including follow-up spectroscopy of Large Synoptic Survey Telescope targets. Examples of enabled science are studies of cosmology, galaxy evolution, quasars, and the Milky Way"

[2] http://www.eso.org/public/about-eso/committees/stc/stc-85th/public/STC-551_Science_Priorities_at_ESO_85th_STC_Mtg_Public.pdf

MSE is located at arguably the premier site on the planet for ground-based optical and infrared astronomy, with access to the entire northern sky and more than half of the southern sky at reasonable airmass. The science presented herein reflects several years of effort by the MSE Science Team in the development of transformational science cases that demonstrate the unique, high impact and exceptionally diverse science enabled by large aperture, wide field MOS. MSE is designed to provide a natural next step beyond what is envisioned for the current and imminent generation of MOS instruments. The "stand-alone" science potential of MSE is awesome, but moreover the strategic importance of MSE within the international network of astronomical facilities cannot be overstated. This is reflected by the strong backing that large aperture, wide field MOS has on the international scene, and for which MSE is the realization of that ambition.

# Chapter 2 The origin of stars, stellar systems and the stellar populations of the Galaxy

## 2.1 Chapter synopsis

- *Exoplanetary host characterization in the era of PLATO*
- *Wide field stellar spectroscopy in the time domain*
- *Population studies of rare stellar types across all Galactic components*
- *The detailed chemical evolution of stars across the entire luminosity range of Gaia targets*
- *In-situ chemical tagging of the distant stellar halo and the interfaces between components*
- *The spatial distribution of the first stars through analysis of the metal weak Galaxy*
- *The chemodynamical deconstruction of the nearest L* galaxy*

*Science Reference Observations for The Origins of Stars, Stellar systems and the Stellar Populations of the Galaxy*

- *SRO-01: Exoplanets and stellar variability*
- *SRO-02: Rare stellar types and the multi-object time-domain*
- *SRO-03: Milky Way archaeology and in situ chemical tagging of the outer Galaxy*
- *SRO-04: Unveiling cold dark matter substructure with precision stellar kinematics*
- *SRO-05: The chemodynamical deconstruction of Local Group galaxies*

MSE will produce vast spectroscopic datasets for in situ stars across all components of the Galaxy at all Galactocentric radii, across the full metallicity distribution of the Galaxy, and across the full range of luminosity of Gaia targets. Exploration of our Galaxy provides us with the most detailed perspective possible on the formation of planetary systems, their host stars, and their parent stellar populations. At the largest scale, the properties of the stellar populations – including their spatial distributions, energy, angular momenta, and chemical properties – provide us with the fossil information necessary to decompose the Milky Way galaxy into its primordial building blocks, and in so doing reveal the history of its formation and the sites of chemical nucleosynthesis.

The key measurables are well-defined spectral line features for accurate stellar parameters (temperature, gravity, metallicity, rotational velocities, magnetic fields strengths, etc.) chemical abundances, and radial velocities. Such a detailed perspective necessitates the highest signal-to-noise data observed at generally high spectral resolution: at R = 40 000, synthetic spectra show that lines are deblended at ~0.1 Å), rotational and radial velocities to better than ~0.1 km/s through cross-correlation of many lines (>100), and chemical abundances with Δ[X/Fe] <0.15 dex (0.1 dex), when SNR>30. Here, the capabilities of MSE make it unmatched amongst any existing or planned facilities. Further, an increasing fraction of astronomical resources are being devoted to characterizing the variable universe. MSE, thanks to its large aperture, wide field and high multiplexing, is well positioned to pioneer the field of wide-field, large aperture, time domain optical/IR spectroscopy.

Specifically, MSE will provide complete characterization at high resolution and high SNR of the faint end of the PLATO target distribution (g ~ 16) visible from the northern hemisphere to allow for statistical analysis of the properties of planet-hosting stars as a function of stellar and chemical parameters. With high velocity accuracy and stability, time domain spectroscopic programs will allow for highly complete, statistical studies of the prevalence of stellar multiplicity into the regime of hot Jupiters for this and other samples and also directly measure binary fractions away from the environment of the Solar Neighbourhood. Rare stellar types – including a menagerie of pulsating stars, but also halo white dwarfs and solar twins – will be easily identified within the dataset of millions of stars observed per year by MSE, and will allow for population studies that incorporate accurate completeness corrections and which examine their distribution within the Galaxy.

MSE will be unmatched as a chemical tagging experiment. Recent work in this field has started to reveal the dimensionality of chemical space and has shown the potential for chemistry to be used in addition to, or instead of, phase space, to reveal the stellar associations that represent the remnants of the building blocks of the Galaxy. MSE will push these techniques forward to help realize the "New Galaxy", as originally envisioned by Freeman & Bland-Hawthorn (2002). In particular, MSE will decompose the outer regions of the Galaxy – where dynamical times are long and whose chemistry is unaccessible from 4-m class facilities – into its constituent star formation events by accessing a range of chemical elements that sample a large number of different nucleosynthetic pathways. In so doing, MSE will provide a high resolution, homogenous sampling of the metallicity distribution function of the Galaxy. It will trace variation in the metallicity distribution across all components and as a function of radius. At the metal-weak end, [Fe/H] < -3, such an analysis reveals the distribution of the oldest stars in the Universe, and is a vital component to tracing the build-up of the Galaxy at the earliest times.

MSE will extend analysis of the resolved stellar populations of the Galaxy to the Local Group and beyond. Variable stars such as Cepheids can be spectroscopically studies throughout the Local Group utilizing the time-domain strengths of MSE, and massive stars will be used to trace the metallicity gradients of disks in galaxies in the next nearest galaxy groups. Such studies will probe the inside-out formation of galaxy disks through individual stellar metallicities observed at moderate spectral resolution. M31 and M33 are particularly prominent extragalactic targets for resolved stellar population analysis, and MSE will provide the definitive chemodynamical deconstruction of our nearest $L_*$ galaxy companion by a focused program targeting all giant branch candidates observed in M31 out to approximately half the virial radius of the galaxy. Such a spectroscopic panorama is unprecedented and a natural complement to the core Milky Way science that forms the drivers for MSE in this field.

*MSE is uniquely suited to precision studies of the stellar populations of the Milky Way, and will provide a transformative impact in our understanding of the links between stellar and planetary formation, the sites of chemical nucleosynthesis, stellar variability, and the chemical structure of the Galaxy through in situ analysis of stars in all Galactic components. It will push these studies out to the nearest galaxies to provide panoramic complements to the driving Milky Way science. MSE will leverage the dataset created by PLATO for exoplanetary and stellar astrophysics, and will provide the essential complement to the Gaia space satellite by providing high resolution chemical abundances across the entire luminosity range of Gaia targets.*

## 2.2 The physics of stars: from exoplanetary systems to rare stellar types

MSE is a critical component for stellar astrophysics in the 2020s that leverages the datasets that will be obtained by the multitude of facilities described in Chapter 1. Amortised over its lifetimes, MSE will cover large fractions of the Galactic volume and survey many millions of stars *per year*. The sheer increase in sample size will create countless opportunities for – and will motivate original science questions in – stellar structure, stellar processes, and local environment. The discovery of significant samples of rare objects through blind surveys, e.g., Li-rich stars, C-rich stars, solar twins and solar analogues, peculiar AGB stars, metal-line white dwarfs, or even potentially first star remnants (e.g., Rameriz et al. 2010, Liu et al. 2014, Carollo et al. 2014, Datson et al. 2014, Venn et al. 2014, Koester et al. 2014, Aoki et al. 2014) will have a transformative impact in modeling their phenomena. This extends to increasing the sample sizes of objects in nearby dwarf galaxies, where differences in their environments and histories can provide important constraints for stellar nucleosynthesis, e.g., progenitor mass and metallicity dependent yields from Asymptotic Giant Branch (AGB) stars and supernova (SNe) events (e.g., Shetrone et al. 2003, Venn et al. 2004, Herwig 2005, Heger & Woosley 2010, Nomoto et al. 2013, Frebel et al. 2014, Schneider et al. 2014).

In many ways, stellar astrophysics forms the backbone of modern astronomy. Stellar evolutionary theory, guided by measurements for fundamental parameters from high-resolution spectroscopy, underpins our models of galaxy evolution and cosmology. Furthermore, it is now recognized that the formation of planetary systems is closely connected to that of the stellar hosts, as evidenced by the apparent high efficiency of planet formation among metal-rich stars. In the coming decade, a number of the key questions in stellar astrophysics will need to be addressed using very large, systematic spectroscopic surveys conducted at high spectral resolution: i.e., roughly an order of magnitude higher than that of the SDSS ($R \sim 1800$). MSE, as an ambitious, wide-area stellar survey, will uncover rare and important stellar types, such as solar twins and faint, metal-poor white dwarfs associated with the Milky Way thick disk or halo. The detection and characterization of such objects using $R \geq 20000$ optical spectroscopy is required to deepen out understanding of, e.g., planetary formation processes and their dependence on environment, the chronology of disk and halo formation, and the end stages of stellar evolution for low-mass stars. Targeted programs will explore the evolution and chemistry of massive stars in the local volume, and MSE can pioneer the field of large aperture, wide field time-domain stellar spectroscopy. The latter has the potential to improve dramatically our understanding of stellar multiplicity (including the interconnections between companions of all sorts, from low-mass stars, to brown dwarfs and exoplanets) and pulsating, eclipsing or eruptive stars.

In the field of stellar astrophysics, the transformational role of MSE is based on three distinct atributes: first, in the collection of spectra for stellar samples of unprecedented size, depth and resolution; second, in the spectral monitoring of time variable sources; third, in accessing stellar populations belonging to Galactic components that are difficult to access with current spectroscopic instruments due to limited aperture and multiplexing.

The stellar populations that are most difficult to access using current spectroscopic surveys include those in the outer halo, outer disk, and faint stars in the bulge. The MSE high resolution mode (R ≥ 20,000) can provide high precision in stellar parameters, radial velocities, rotational velocities, chemical abundances, and isotopic ratios, which are often needed for detailed stellar modeling and observational testing. For example:

- *In the outer halo,* high resolution spectra of unique stars (e.g., chemically peculiar stars, remnants of first stars, RR Lyrae stars, and halo dwarfs) can provide new information for stellar nucleosynthetic yields, stellar pulsation theory, and ages for halo white dwarfs (e.g., Ivans et al. 2003, Aoki et al. 2013, 2014, Yong et al. 2013, Cohen et al. 2013, Drake et al. 2013, Kilic et al. 2005, Kalirai 2012).
- *In the outer disk,* young metal-poor stars found in star forming regions can be compared to the typical outer disk field stars, providing valuable information on the outer disk star formation environment (e.g., Friel et al. 2010, Frinchaboy et al. 2013, Carrero et al. 2015).
- *In the Galactic bulge,* stars tend to be more metal-rich than in the solar neighbourhood and likely to have formed early, thus high resolution spectroscopy of the faint stars (dwarfs, and those in high extinction regions) provide unique constraints for stellar nucleosynthesis, stellar interiors, and stellar atmospheres at higher metallicities, in higher density environments, and with unique chemical compositions (e.g., Fulbright et al. 2006, 2007, Bensby et al. 2010, 2011, Garcia Perez et al. 2013, Johnson et al. 2014; see also the discovery of ancient metal-poor stars in the bulge by Howes et al. 2015).

The nature and structure of the Galactic components referred to above is a primary focus of the complementary Galactic Archaeological surveys discussed in Section 2.3, but the underlying questions are unique to stellar astrophysics:

1. *Is the Sun unusual because it hosts a planetary system?* Detailed spectroscopic analyses for a handful of "solar twins" have shown that their chemistry differs from that of the Sun, i.e., they contain higher fractions of refractory elements. In the case of the Sun, such elements may have been consumed by its surrounding planetary system (e.g., Rameriz et al. 2010). However, the interpretation of this result is difficult because the effects are subtle and there are alternative explanations for the observed abundance patterns (e.g., chromospheric activity, stellar seismology, or even first ionization potential effects). Because fewer than a few dozen solar twins are presently known, a larger sample is essential if we are to understand the origin of this difference;
2. *Time domain stellar spectroscopy.* A large number of imaging surveys are exploring and characterizing the variable sky down to very faint magnitudes Eminent among these endeavors is PLATO, that will survey nearly half the sky and monitor sources down to g~16 (with the core sample being brighter than this limit). Spectral monitoring of many of these variable sources, and many of the PLATO fields, will be required to understand their nature and can be expected to reveal a rich diversity of interesting astrophysical phenomena. Multi-epoch spectral monitoring over wide areas will in turn require a large-aperture, highly-multiplexed spectrograph. Obvious applications in the stellar regime include spectroscopic follow-up of transit-selected planetary host candidates, binary stars, pulsating and/or eclipsing variables (see Chapter 4 for discussion of extragalactic transient and variable phenomena);
3. *How old is the Galactic halo?* White dwarfs can be used to age-date their host stellar

population using cooling curves. Unfortunately, they are so faint that most known white dwarfs are restricted to the solar neighbourhood. A handful (2 – 4) of white dwarfs has been found with high proper motions, suggesting they are members of the Galactic halo, yet only ~ 30 pc away. The ages of 11 – 12 Gyr measured for these stars (Kilic et al. 2012, Bergeron et al. 2005) are in good agreement with white dwarf cooling curve ages in the globular clusters NGC 6397 and M4 (Hansen et al. 2004, 2007). Nevertheless, a much larger sample of more distant halo white dwarfs (with medium-resolution spectroscopy) could be combined with detailed white dwarf model atmospheres calibrated using nearby objects (having high-resolution spectroscopy) to provide an independent measurement of the age distribution for the Galactic halo;

4. *Where are the minority populations in the Galactic Bulge?* Stars in the Galactic bulge tend to be metal-rich, as would be expected when star formation occurs rapidly (e.g., Lecureur et al. 2007, Fulbright et al. 2007, McWilliam et al. 2008). However, microlensing studies toward the bulge (Bensby et al. 2010, 2011) and some recent spectroscopic surveys (Gonzalez et al. 2011) find some stars with lower metallicities. Very recently, a sample of extremely metal poor stars have been identified in the bulge (Howes et al. 2015). Clearly, very large sample sizes are needed to identify and characterize the most metal-poor stellar populations in the bulge;
5. *How does environment impact the chemical evolution of disk galaxies?* Precise stellar parameters for massive stars can be determined using low-resolution spectroscopy and narrowband photometric indices (e.g., Kudritzki et al. 2012; Firnstein & Przybilla 2012). Because these stars are bright, they can be studied individually in galaxies out to ~ 4 Mpc: i.e., at the distance of the Sculptor and M81 groups. In principle, metallicity and reddening maps can be determined for the disks of star-forming galaxies in order to test the effects of environment on disk evolution. While individual galaxies can be examined with present-day instruments, a comprehensive study will require homogeneous observations for many stars belonging to a large, statistically robust sample of galaxies.

In what follows, we describe a number of MSE legacy science programs in the field of stellar astrophysics. Some of these programs will require datasets that cover a large fraction of the Galactic volume and that sample all major Galactic components (including the halo, the thin and thick disk, and the bulge). Others require observations with specific cadence, and others are smaller, targeted programs that could complement larger scale surveys.

### 2.2.1 Exoplanetary host characterization and their environments in the 2020s

**Science Reference Observation 1 (Appendix A)**

**The characterization and environments of exoplanet hosts**

*MSE will provide spectroscopic characterization at high resolution (R ~ 40000) and high SNR (~100) of the faint end of the PLATO target distribution ($m_V$ ~ 16), for statistical analysis of the properties of planet-hosting stars as a function of stellar and chemical parameters. MSE is able to cover the full ~three-quarters of the PLATO sky (~15000 square degrees) accessible from Maunakea in less than 800 hours. Such a survey will provide homogeneous spectral typing, metallicity estimates and chemical abundances for all PLATO targets at the faint end. Such information will otherwise remain unavailable for most of this important sample. With high*

*velocity accuracy (~100ms$^{-1}$) and stability, MSE will conduct a monitoring campaign of ~100 square degrees in the PLATO footprint, potentially in the northern long-duration field. This time domain spectroscopic program will directly measure binary fractions away from the Solar Neighbourhood and will allow for a well characterized statistical studies of the prevalence of stellar multiplicity into the regime of hot Jupiters. Such information is important to understand the physical mechanisms governing stellar multiplicity, planet migration and their dependence on host star properties.*

At the time of writing, 3411 planets have been confirmed in 2550 planetary systems[3], using a variety of methods (1) radial velocity searches; (2) photometric transits; (3) astrometric searches; (4) direct imaging; (5) transit timing variations; (6) pulsar timing; (6) microlensing surveys. Transit surveys are now responsible for more than three quarters of detections, thanks in large part to the overwhelming success of the Kepler space mission (Borucki et al. 2010). Different search methods probe different planetary parameters, and each approach has its own advantages and biases. However, by combining results from the various methods, it is possible to build up a surprisingly complete picture of planetary systems. For instance, radial velocities provide important constraints on planetary masses, while planetary radii can be measured using photometric transits. Thus, it has been possible in recent years to lay important groundwork in the burgeoning field of comparative planetology.

Even the relatively well-characterized population of giant planets shows a surprising diversity in properties, with orbital configurations of all kinds observed: i.e., very short orbital periods, large eccentricities, large inclinations, etc. Thanks to Kepler, we now able to detect planets with masses and radii not much larger than those of the Earth (and even identify some intriguing multiple-planet systems containing rocky planets). In general, a planet's fundamental parameters (e.g., radius, mass, density and orbit) provide the basic data needed for a characterization of extrasolar planets. These allow an assessment of the overall nature of a given planet (i.e., whether it is a gas giant, a Neptune-like system, or a terrestrial planet) and provide important insights into its interior structure. A measured orbit also allows us to investigate the dynamical evolution of the planet, and to set constraints on the current theories of planetary formation.

[3] Data from exoplanets.eu, May 25, 2016

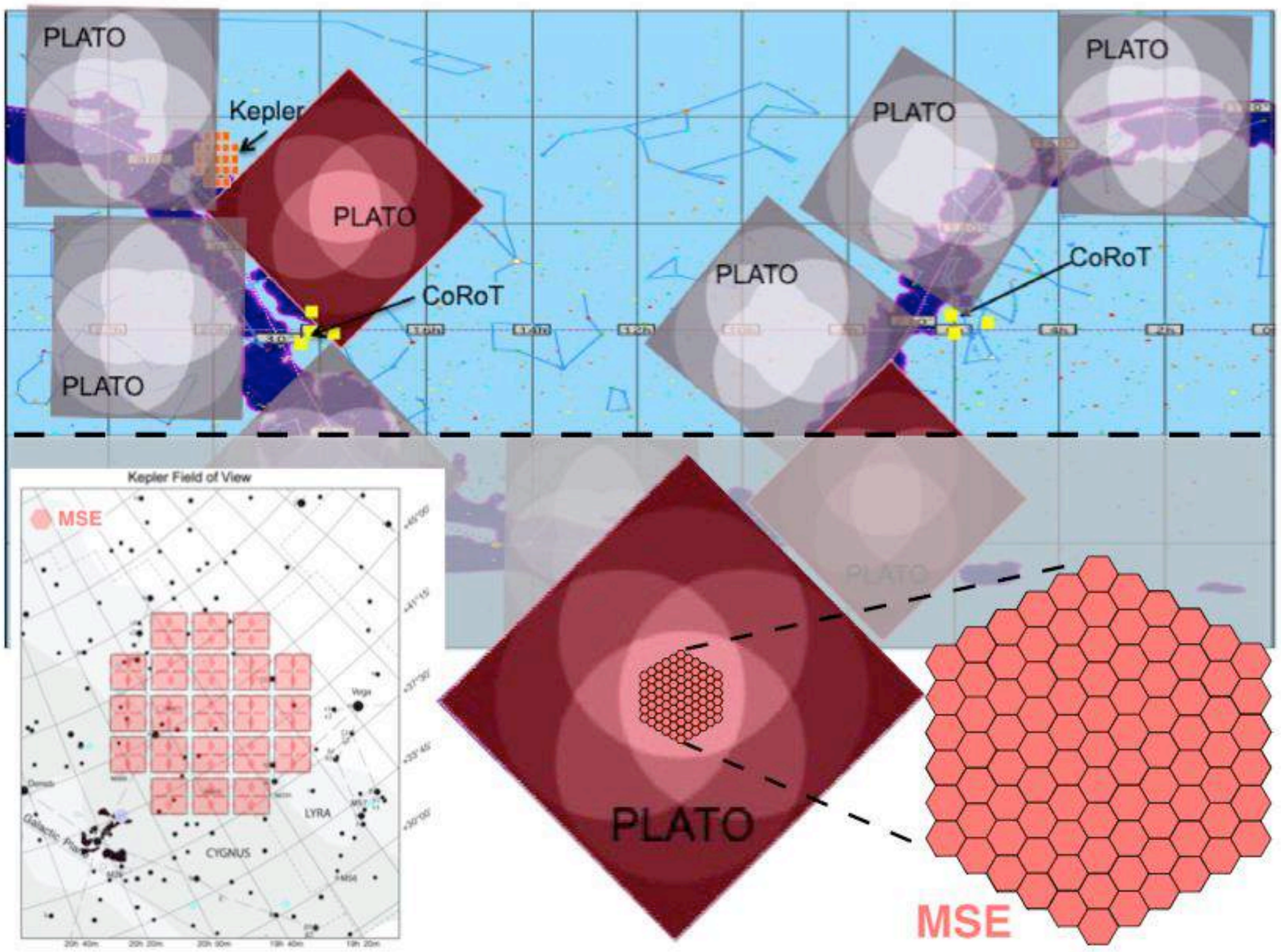

**Figure 13: Top panel: schematic diagram showing the possible layout of the PLATO observing fields, relative to CoRoT and Kepler, in celestial coordinates, from Rauer et al. (2014). The PLATO field of view is 48.5 x 48.5 square degrees, and it will observe roughly half the sky to $m_V \sim 16$. The area above the dashed line is easily accessible to MSE (roughly ¾ of the PLATO fields, or ~15000 square degrees). MSE can observe the faint end of the PLATO target distribution at high spectral resolution (R~ 40000) and very high SNR (~ 100) in short exposures (<300s), with a velocity precision of order ~100m/s. Possible observing strategies for MSE in the field of exoplanetary hosts include (a) covering all of the PLATO targets with single exposures; (b) targeting the Kepler survey fields, which will still have limited spectral coverage at the faint end in the 2020s. In the lower left panel, we show that the total ~115 square degree survey area can be covered, with high spatial completeness, in 84 MSE pointings; (c) a monitoring campaign of a subset of PLATO targets. In the lower right panels, we show a possible layout of 91 MSE fields covering the central 136 square degrees of the northern PLATO long duration field (that would take MSE about 8 hours to observe).** ***(Figure adapted from Rauer et al. 2014).***

Major thrusts of exoplanet research given recent discoveries include (but are not limited to):

1. Surveys sensitive to planets in the lower part of the mass – radius diagram, covering an extended range in orbital period. A major goal continues to be the discovery of an Earth-like planet in the habitable zone of a G-type star. For Doppler searches, this requires a radial velocity precision of about 10 cm $s^{-1}$ over a period of several years. For a photometric detection, it requires a measurement of the transit signal of a few x 0.001%.
2. The detection of giant planets beyond the snow line (3 – 5 AU), including the discovery of Jovian analogs.
3. A thorough assessment of how the bulk properties of planets depend on the properties of their stellar hosts. This requires detection of planets (and other companions) around a large sample of diverse stars, from F- to M-type dwarfs, as well as moderately evolved stars.

4. A determination of the uniqueness of our Solar System, and a comparison of its properties to those of planetary systems in different environments and at different evolutionary stages.

No single facility is capable of addressing all of the issues. For instance, a number of ongoing or planned surveys are focusing on improved detection limits via extended time coverage and/or increased radial velocity precision. At present, there are about fifteen instruments worldwide that can deliver a radial velocity accuracy of ~10 m $s^{-1}$ or better. Some have been monitoring bright stellar samples for more than a decade, and some can achieve a precision of ~1 m $s^{-1}$: i.e., HARPS (ESO 3.6m), HIRES (Keck 10m), SOPHIE (OHP 1.9m) and HARPS-N (TNG 3.6m). Thus, the domain of massive planets will be relatively well explored in a decade's time. For lower-mass planets (or those having long orbital periods), ESPRESSO is in active development. This is an ultra-high-precision radial velocity spectrograph for the VLTs that aims to achieve a precision of a few x 10 $cms^{-1}$ (Pepe et al. 2010).

For MSE, the radial velocity precision attainable in high-resolution mode is of order ~100 m $s^{-1}$. While this will certainly not compete with dedicated Doppler-search instruments, MSE's large aperture, wide field and extreme multiplexing will allow it to make a unique contribution to the question of how exoplanet properties depend on environment, a recurring theme in the above list, and to examine the statistics of massive planetary companions.

For illustration, we consider the role MSE might play in the era of PLATO. PLATO (*PLanetary Transits and Oscillation in stars*) is a medium-class ESA mission with a targeted launch date of 2024. PLATO has an enormous field of view (2232 square degrees), a large dynamic range (observing stars from 4 – 16 magnitude) and it will observe roughly half the sky (see a schematic diagram of its possible sky coverage in Figure 13, adapted from Rauer et al. 2014). For the brightest subsample, (4 – 11 magnitude, estimated to be around 85000 stars), it will provide accurate planetary parameters, such as density, mass, radius and age, down to Earth-size objects. Two long pointings (highlighted in Figure 13) are included in the observing strategy to detect and characterize planets in the habitable zones of solar-like stars. Roughly 1 million stars will be observed with PLATO to 16 magnitude, and massive planets will still be detected at this faint limit, providing statistically robust samples for investigation.

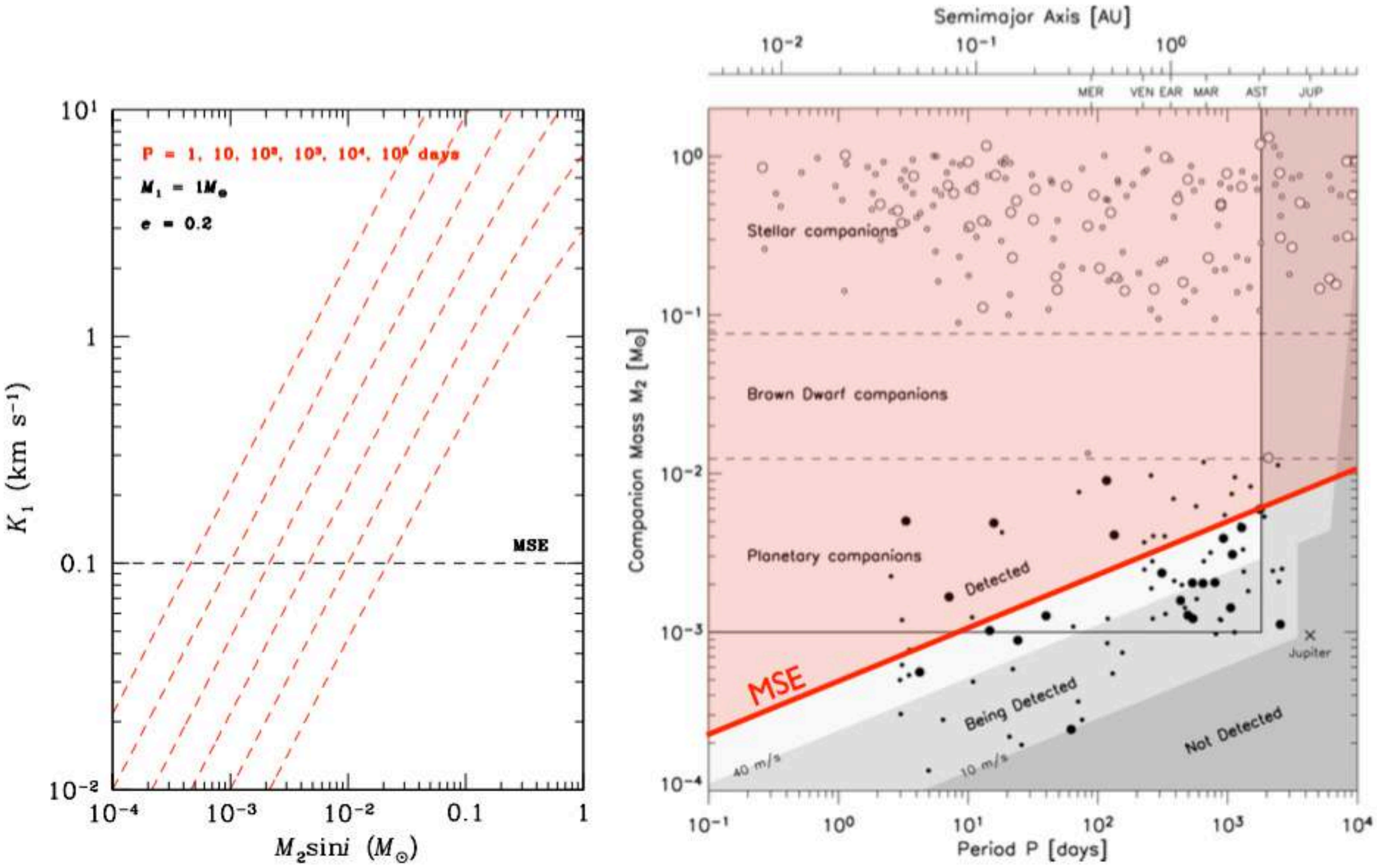

**Figure 14: Illustration of the possible role of MSE in characterizing the low-mass companions of solar-type stars. The panel on the left shows velocity induced by a companion plotted against (minimum) companion mass. The dashed red curves show the reflex velocities expected for companions with orbital periods between 1 and $10^5$ days, as indicated in the figure. All calculations assume a 1 M¤ primary and an orbital eccentricity of e = 0.2. The dashed horizontal line shows the expected typical precision of MSE radial velocity measurements for stars brighter than g ~16.5. The panel on the right shows the sensitivity of MSE in terms of companion mass and orbital period.** ***Figure adapted from Grether & Lineweaver (2006).***

There are numerous, unique, observational strategies for MSE to complement PLATO, and detailed analysis will be required to determine the optimal survey parameters. Figure 13 shows a schematic diagram of the PLATO sky coverage, of which roughly 15000 square degrees is accessible to MSE (the non-shaded area in Figure 13). While this is a large area, these targets are among the brightest that MSE will study, and only relatively short exposures (<300seconds) are required in order to obtain high SNR (~100) at high resolution (R~40000). Indeed, such a program could even be reserved for poor conditions (e.g., poor seeing, bright time), such is the relative ease with which it can be conducted given the capabilities of MSE. The entire PLATO survey area could be covered by MSE in less than 800 hours, and this would provide detailed stellar parameters and chemical abundance information for all PLATO targets at the faint end where such information will otherwise remain unavailable for most of the sample.

Alternatively, or in addition, monitoring campaigns of select regions of sky will provide essential time domain spectroscopic information to complement the extensive photometric monitoring of these stars. In this respect, it is worth highlighting that, even by the mid-2020s, the spectroscopic data available for the faint end of the Kepler sample will still be limited due to the relatively faint limit of this survey (~17 magnitude). Excellent synergies currently exist between Kepler and MOS programs on smaller facilities, with perhaps the best example being the APOGEE – Kepler campaigns (see Fleming et al. 2015, Rodrigues et al. 2014, and references therein) and the same can be expected with MSE for fainter targets. The lower left panel of Figure 13 shows how MSE could survey the entire Kepler region in ~84 pointings. Similarly, regions of the PLATO sky can be identified for monitoring by MSE: the lower right panels of Figure 13 shows a zoom-in of the northern long duration PLATO field, and a possible MSE survey configuration in its central parts is highlighted. The 91 MSE fields shown cover approximately 132 square degrees, and can be observed by MSE in less than 8 hours. Repeat visits to these fields, perhaps monthly or with a variable cadence, could be used to develop an extensive characterization of the time variable spectral characteristics of the faint PLATO stellar populations.

A survey of this sort would have a two-fold purpose, and the deep, time domain MOS data provided by MSE would be a unique contribution to ground-based support of these exoplanetary space missions. First, it would provide for a large sub-sample of PLATO (or Kepler) targets, homogeneous spectral typing and metallicity estimates for all stars using a common set of spectral features. There is emerging evidence that the correlation between planetary frequency and host metallicity (see Section 2.2.3) exhibited by massive planets may not hold for the lowest-mass planets, so homogeneous metallicity estimates for the host stars of planetary systems of all sorts, based on high-SNR, homogeneous, optical/IR spectroscopy, would be an important dataset to complement the extensive transit observations. Secondly, at the faint end PLATO will be sensitive to massive planets, especially “hot Jupiters”, whose formation is thought to involve scattering and/or migration processes. The physical interpretation of these systems would benefit from a thorough characterization of stellar companions of all sorts — from the regime of low-mass stars down to brown dwarfs. Although MSE’s velocity precision would not be sufficient to detect exoplanets themselves, it would be more than adequate to perform a complete census of low-mass stellar companions and brown dwarfs. MSE should take advantage of its stability and high quality calibrations to conduct long-term monitoring programs that other MOS instruments to measure stellar multiplicity and augment the PLATO science goals. Figure 14 illustrates graphically the detection efficiency of MSE for these studies as a function of companion mass and period.

Finally, these examples have considered the specific case of the PLATO mission. However, there are several other opportunities for MSE to provide valuable contributions to exoplanetary research by using its unique set of capabilities. For example, a major component of exoplanetary research will continue to be exoplanet detection via direct imaging studies. In such surveys, panoramic spectroscopy for millions of Galactic stars would be invaluable for optimizing the target selection by providing accurate stellar parameters, such as effective temperature, surface gravity, and also [Fe/H] or some elementary abundances of key elements; continued radial velocity monitoring would also allow a part of the companion populations in the target stars to be identified and characterized, as described above. A particularly exciting approach would be to select imaging targets using a combination of multi-band imaging and wide-field spectroscopy — to identify stars of various ages using Ca II H and K strength, Li abundances, rotational velocities, and/or UVW space motions. Combined with direct imaging data and radial velocity monitoring, this would allow a more complete characterization of the time evolution of planetary systems.

### 2.2.2 Stellar astrophysics with time domain multi-object spectroscopy

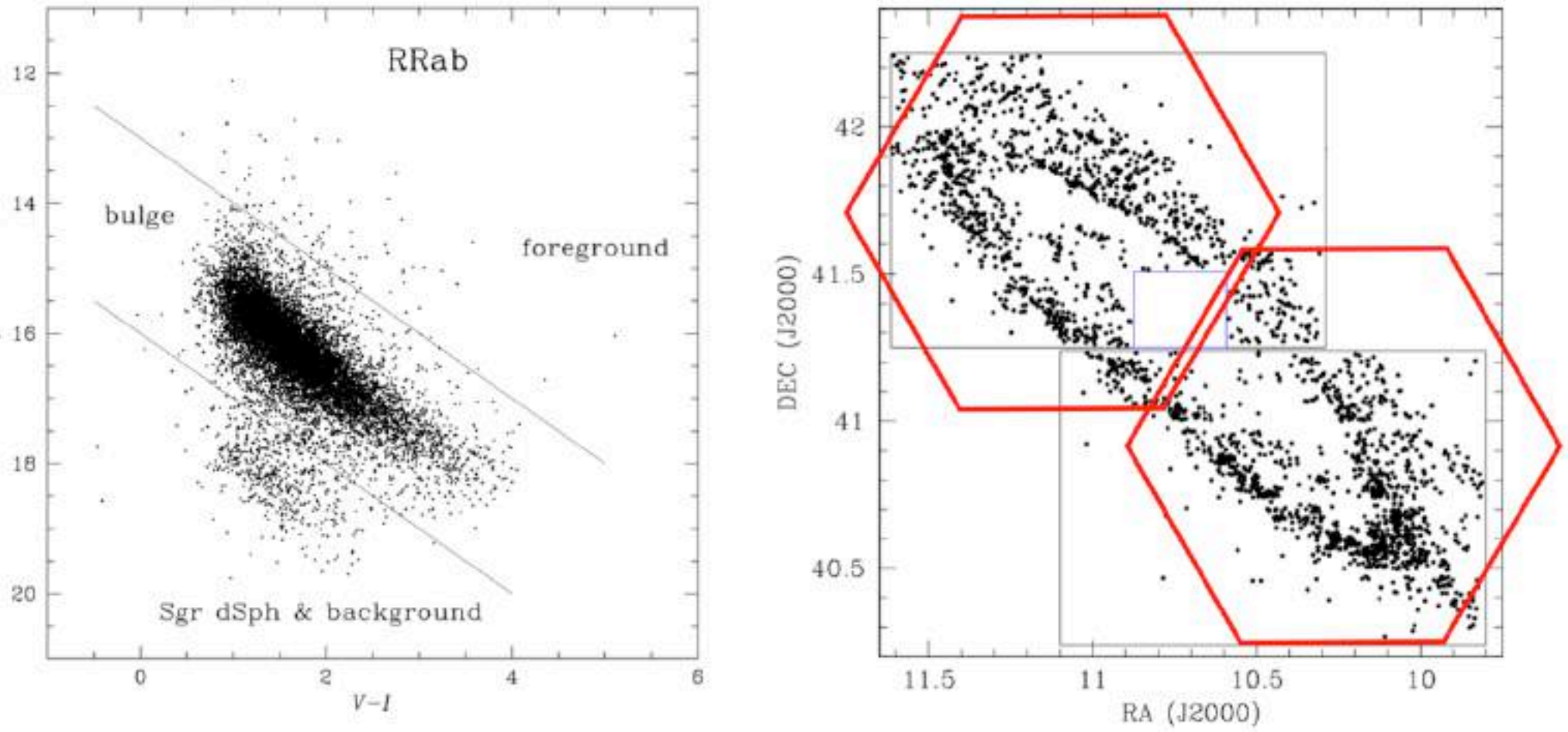

**Figure 15: Left panel: Composite colour-magnitude diagram for 11 124 RR Lyr variables of type RRab from the OGLE-III survey, which carried out an 8 year monitoring campaign of a ~100 square degree region in the direction of the Galactic bulge. A roughly five-fold increase in the number of bulge RR Lyr variables is expected from the VVV survey (Minniti et al. 2010, Hempel et al. 2014). *Figure from Pietrukowicz et al. (2012)*. Right panel: The distribution of Cepheids in M31 from the POMME survey (Pixel Observations of M31 with MEgacam). The red hexagons show possible MSE fields for spectroscopic monitoring campaigns. Note that the Cepheid distribution is similar to that of the brightest stars in M31, although there are small differences due to reddening variations, gas content and changes in underlying star formation history. *Figure adapted from Fliri & Vals-Gabaud (2012).***

For the study of variable stars — including eclipsing and pulsating variables — the yields from the next generation of imaging telescopes (Pan-STARRS, Skymapper, ATLAS, LSST, etc) promise to be immense. As an indication of what the future may hold, the left panel of Figure 15 shows results from a search for RR Lyrae variables based on the OGLE-III microlensing survey of the Galactic bulge. This survey, covering a total area of ~100 deg$^2$, discovered more than 16 000 RR Lyrae variables of all types, producing phased light and colour curves for each object (Pietrukowicz et al. 2012). The VISTA Variables in the Via Lactea survey (VVV; Minniti et al. 2010, Hempel et al. 2014) is expected to increase the bulge RR Lyrae sample to 40 000 – 70 000. The gains expected for other types of variables will be equally dramatic; for example, the right panel of Figure 15 show results from the POMME survey (Fliri & Vals-Gabaud 2012) which used CFHT to monitor the disk of M31 over a period of about 2.5 months. This figure shows the distribution of more than 2500 Cepheids with periods in the range 2 to 80 days, highlighting the familiar ring-like distribution of massive, bright stars in M31. In short, the coming decade should see a revolution in the study of variable stars.

Cepheids, RR Lyraes and other types of pulsating stars are not only cornerstones of the cosmic distance ladder, but also important testbeds for stellar evolutionary models. While there is much to be learned from photometry alone, the addition of multi-epoch spectroscopy would make possible a number of investigations that have heretofore been impossible. For instance, RR Lyrae variables in the bulge region have magnitudes in the range $15 \le I \le 18$ and velocity semi-amplitudes of $A \sim 20$ km s$^{-1}$ or more, while Cepheids in M31 with periods longer than 3 – 4 days have $19 \leq r \leq 23$ and $A \geq 15$ km $s^{-1}$. These magnitudes and pulsation velocities demonstrate the importance of spectroscopic monitoring campaigns with MSE: i.e., repeated radial velocity measurements, at a resolution of $R \sim 6500$ (i.e., $\varepsilon_v \sim 3$ km $s^{-1}$), for hundreds or thousands of variables in selected fields. Such data will be combined with contemporaneous, multi-colour, photometric data to carry out Baade-Wesselink studies for stellar samples of unprecedented size. Because the Baade-Wesselink method represents a direct route to the measurement of fundamental parameters including stellar radius and distance, such programs would have obvious implications for our understanding of the mechanisms that drive stellar pulsation and the evolution of stars through the instability strip. In the case of M31, a number of other attractive targets for spectroscopic monitoring would also be available including bright, eclipsing binaries (which would provide direct mass measurements for stars at the upper end of the main sequence) and novae, where pioneering efforts in spectral classification based on a handful of objects in Local Group galaxies has revealed an unexpected diversity in spectral properties (Shafter et al. 2012).

For the role MSE might play in the study of explosive transients, including high-z supernovae, see Chapter 4. For more discussion of MSE and Local Group galaxies, see Section 2.4.

**Science Reference Observation 2 (Appendix B)**

**Rare stellar types and the multi-object time-domain**

*The role of MSE for stellar astrophysics is based on three distinct attributes: first, in the collection of spectra for stellar samples of unprecedented size, depth and resolution; second, in the spectral monitoring of time variable sources; third, in accessing stellar populations belonging to Galactic components that are difficult to access with current spectroscopic instruments due to limited aperture and multiplexing. The sheer increase in sample size will create countless opportunities for – and will motivate original science questions in – stellar structure, stellar processes, and local environment. The discovery of significant samples of rare objects through blind surveys, e.g., Li-rich stars, C-rich stars, solar twins and solar analogues, peculiar AGB stars, metal-line white dwarfs, or even potentially first star remnants will have a transformative impact in modeling their phenomena, specifically by allowing for population studies that incorporate accurate completeness corrections and which examine their distribution within the Galaxy. This also extends to increasing the sample sizes of objects in nearby dwarf galaxies, where differences in their environments and histories can provide important constraints for stellar nucleosynthesis, e.g. progenitor mass and metallicity dependent yields from AGB and SNe events. Here, we outline a baseline set of programs and analyses that will provide a transformative legacy for MSE in the broad arena of stellar astrophysics.*

### 2.2.3 Solar Twins: The Sun and Solar System in Context

Solar-type stars are obvious targets in the search for exoplanetary systems. In recent years, possible connections between the "architecture" of planetary systems and the fundamental parameters of their host stars, including their chemical makeup, have emerged as key constraints on proposed mechanisms of planet formation.

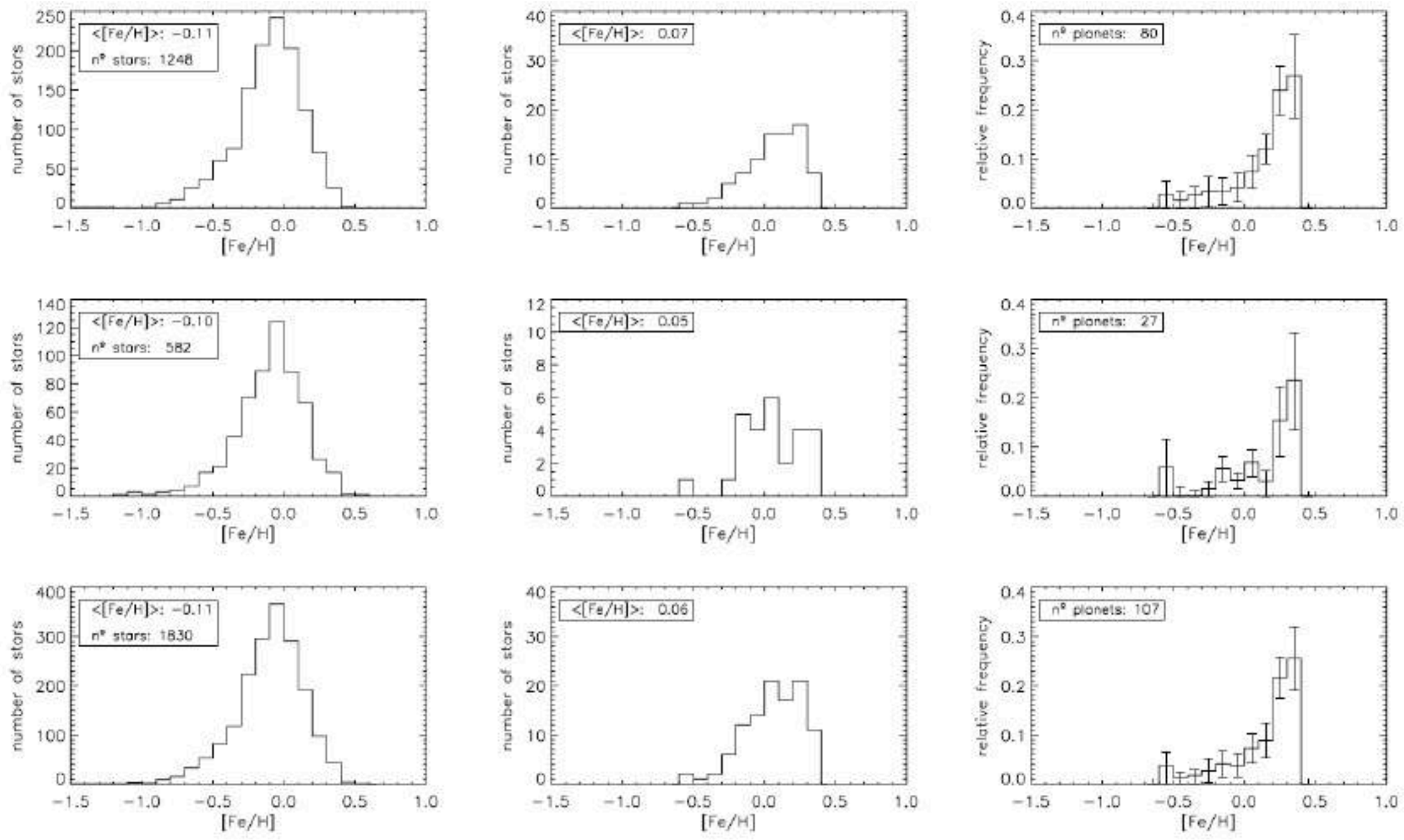

**Figure 16: Metallicity distributions for the CORALIE sample (top panels), the HARPS sample (middle panels), and the union of the two samples (bottom panels), as presented in Sousa et al. (2011). The left panels shows the full samples distribution of [Fe/H], while the center plots for each sample shows the planet host [Fe/H] distribution. The right plots shows the frequency of planets for each [Fe/H] bin.** ***Figure from Sousa et al. (2011).***

It has been recognized for some time that the stars found to host planets in radial velocity surveys – usually high mass, giant planets – tend to have rather high metallicities (e.g., Gonzalez 1997). With the ever-increasing number of Doppler-selected exoplanetary systems, it has been shown many times that the shift in metallicity — usually characterized by [Fe/H] — between planet-hosting FGK dwarf stars, and those known not to host closely orbiting giant planets, is of high statistical significance. Most studies agree (e.g., Fischer & Valenti 2005) that, in the majority of cases, the metal-rich nature of the giant-planet-hosting stars is intrinsic to the star itself. This suggests that giant planets are preferentially formed in metal-rich environments. Indeed, data on 582 low-activity FGK stars in the HARPS volume-limited sample confirm that planet-hosting stars tend to be enhanced in metals (Figure 16). Intriguingly, this trend is strongest for stars with giant planets; stars which host Jovian-mass planets tend to be more metal-rich than those stars which have only Neptunian-mass planets or smaller (e.g., Ghezzi et al. 2010; Buchhave et al. 2012).

Recent results measuring the metallicities of a large number of stars hosting small planets suggest that the planets can be divided into three regimes, namely terrestrial planets, gas-dwarf planets (small planets with lower mean densities), and gas-giants (Buchhave et al. 2014) where the planets in the terrestrial planet regime have a metallicity consistent with solar. This has been confirmed by an analysis by Buchhave & Latham (2015), who analyse a sample of stars hosting transiting terrestrial planets and compare them to an equivalent sample without transiting planets. They find no statistical difference between the average metallicity of the two samples. These results suggests that metallicity plays a role not just in the formation of giant planets, but also in the distribution of planetary masses within exosolar systems.

Additional, intriguing, chemical abundance trends have been noted in stars that may be related to the presence of planets. Using a technique in which spectral lines in "solar twins" (i.e., stars having stellar parameters similar to those of the Sun) are compared in a careful differential abundance analysis, Melendez et al. (2009) found that the chemical composition of the Sun is anomalous with respect to most (85%) of the solar twins. Compared with its twins, the Sun exhibits a deficiency of refractory elements (i.e., those with high condensation temperatures, $T_c$) relative to volatile elements (those with low $T_c$). This finding may be a sign that planet formation occurred more efficiently around the sun compared with the majority of solar twins. Furthermore, within the context of this scenario, it seems likely that the observed abundance patterns are specifically related to the formation of terrestrial planets: i.e., while refractory material is missing in the sun, those dust-forming elements are overabundant in meteorites and terrestrial planets.

If this supposition is correct, the consequences are exciting: it points the way to using chemical abundance distributions as diagnostics of inner, terrestrial-type planetary architectures. If the chemical signature found in the Sun is indeed due to the formation of terrestrial planets, then the study of metal-rich solar analogs and F-type dwarfs can provide further clues to planetary formation mechanisms (Ramirez et al. 2010). In short, exploring the possibility that planet mass may be correlated with the chemical composition of the stellar host, using a large and homogeneous stellar sample, is of utmost importance.

These studies have demonstrated that there are subtle, but fascinating, patterns in the chemical abundance distributions of stars hosting exoplanets. Such patterns have potential significance beyond a simple empirical shift towards higher metallicities. Probing the trends in chemical abundance distributions requires an ability to derive stellar parameters and chemical abundances for many different elements in a very large sample of solar-type stars. With a sample of ~100000 or more FGK main-sequence stars, the intrinsic dispersions of particular chemical families could be measured. This would make it possible to establish whether or not the peculiar signatures seen in the small numbers of planet-hosting and non-planet-hosting stars to-date are found in a general census of solar-type main-sequence stars. In particular, the availability of targets from PLATO (as well as the Kepler fields) is a resource to be exploited by MSE. At a resolution of R ≥ 20 000, a well-calibrated, uniform spectral dataset could be used to extract accurate chemical abundance distributions of a wide variety of elements, covering all nucleosynthetic origins and chemical behaviours.

### 2.2.4 White Dwarfs as Probes of Galactic Formation and Evolution

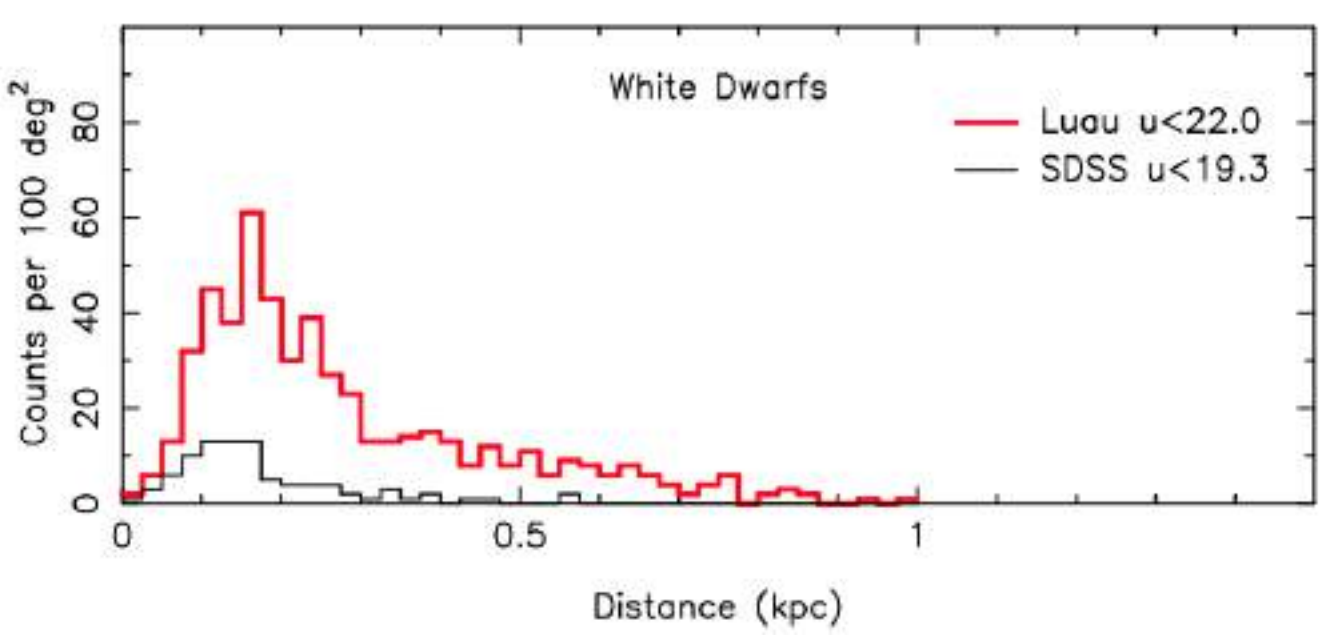

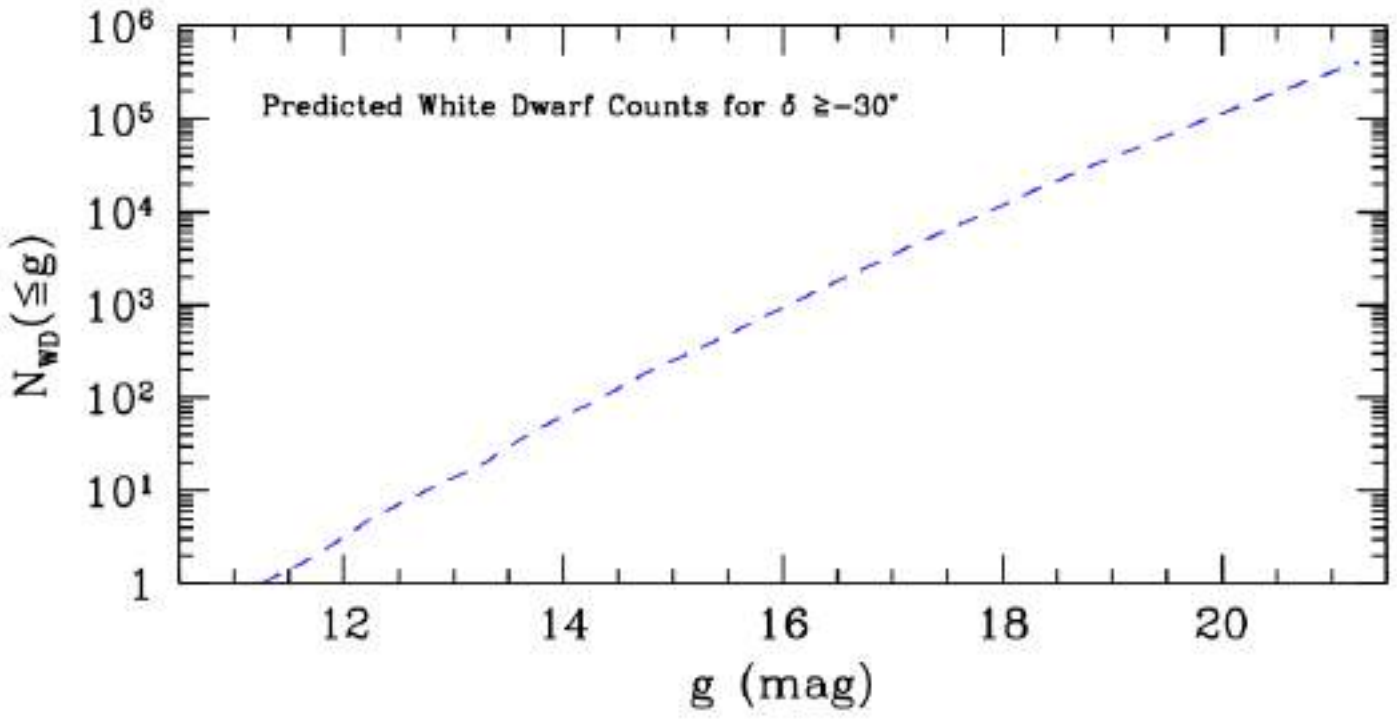

**Figure 17: Top panel: Bescanon Galactic model simulation of the number of white dwarfs detectable per 100 square degrees in the CFHT Legacy for the u-band all-sky universe (Luau) survey, compared to SDSS. In the near future, we can expect the photometric sample of white dwarf candidates to increase significantly, particularly for the more distant Galaxy. Bottom panel: Bescanon Galactic model of the cumulative luminosity function of Galactic white dwarfs. There are approximately 170 000 white dwarfs brighter than g ~ 20.4 and at declinations greater than -30 degrees accessible by MSE.**

Prior to the SDSS, only ~2500 white dwarfs were known. The SDSS DR4 catalogue of white dwarfs (Eisenstein et al. 2006) more than doubled this number, all with spectroscopic confirmations. The growth of the SDSS spectroscopic white dwarf sample has continued (DR7 – Kleinman et al. 2013; DR10 – Kepler et al. 2015a), and including the most recent sample based on SDSS DR12 (Kepler et al. 2015b) now has more than 30 000 members. The SDSS catalogues have been used for statistical studies of white dwarf stellar physics and parameters (e.g., Kepler et al. 2007; Tremblay et al. 2011) as well as for studies of the ages of the Galactic thin and thick disks (e.g., Harris et al. 2006). The catalogue includes rare sub-classes of magnetic and pulsating white dwarfs, and white dwarfs in binary systems (which have also been identified by the UKIDSS survey, Steele et al. 2011). These latter systems can be further studied using time-resolved spectroscopy in order to identify compact, post common-envelope binaries: these objects — which are the precursors of cataclysmic variables, X-ray binaries and possibly SNe Ia — allow the various formation mechanisms proposed for these systems to be carefully examined.

At present, SDSS catalogues contain few white dwarfs belonging to the Galactic halo (i.e., the SDSS “20 pc” sample is estimated to be only 80% complete, Holberg et al. 2008). However, analyses of four, inner-halo white dwarfs moving through the solar neighbourhood (Kilic et al. 2012; Kalirai 2012) yield ages in good agreement with those of old stellar clusters having measured white dwarf ages (e.g., NGC 6397 and M4; Hanson et al. 2007, 2004). This is an area where MSE can be expected to have a profound impact on white dwarf studies: MSE can identify new white dwarfs in the Galactic halo, which are critical for improved age estimates, as well as the discovery of additional rare objects, e.g., white dwarfs polluted by accretion of terrestrial objects, magnetic stars, etc.

The measurement of precise stellar parameters for white dwarfs is only possible by combining photometry, from surveys like SDSS, Gaia, Pan-STARRS or LSST, with follow-up spectroscopy. Recently, a new survey initiated at CFHT, called the “Legacy for the u-band all-sky Universe” (Luau) is surveying a large fraction of the northern hemisphere to a u band depth considerably deeper than SDSS (u ~ 24.5), that will be an ideal dataset for identifying white dwarf candidates out to much larger radius than SDSS, including a higher fraction of Galactic halo dwarfs (see the top panel of Figure 17).

MSE is the *de facto* provider of spectroscopy for this and any other white dwarf sample, given its much deeper limiting magnitude than SDSS, Gaia, or 2 – 4m class telescopes in general. It will play an invaluable role by providing good SNR spectra (15 – 50 was shown to be ideal in the analysis of SDSS spectra by Tremblay et al. 2011) for a volume-limited sample: e.g., within a distance of ~100 pc. Indeed, the bottom panel of Figure 17 illustrates the possibility for transformational gains with MSE relative to SDSS for white dwarf studies, potentially through a program such as that described in SRO-02. The dashed curve shows the predicted, cumulative number of white dwarfs down to $g \sim 21$ for declinations $\delta \geq -30^\circ$. Down to this limiting magnitude, we expect roughly ~8 white dwarfs per MSE field, averaged over this $\sim 3 \times 10^4$ $deg^2$ region. For a subset of the newly discovered white dwarfs, proper motions and distances from Gaia would further improve the precision of the measured stellar parameters. In most cases, intermediate-resolution spectroscopy, R = 6500, is sufficient. Indeed, spectra at R = 2000 would be adequate for most white dwarfs (i.e., DA and DB) where $\log g \sim 8$ and the H and He lines used for stellar parameter determinations are very broad. Higher resolution would be particularly interesting for other subtypes. For example, those with metal lines, which may have been polluted by the accretion of planets, could be used to determine the chemical composition of the former rocky planets. Needless to say, it would also be possible to analyse those systems with magnetic fields to a level of precision that is impossible with SDSS.

### 2.2.5 Fundamental parameters for high mass stars: The chemical evolution of disk galaxies

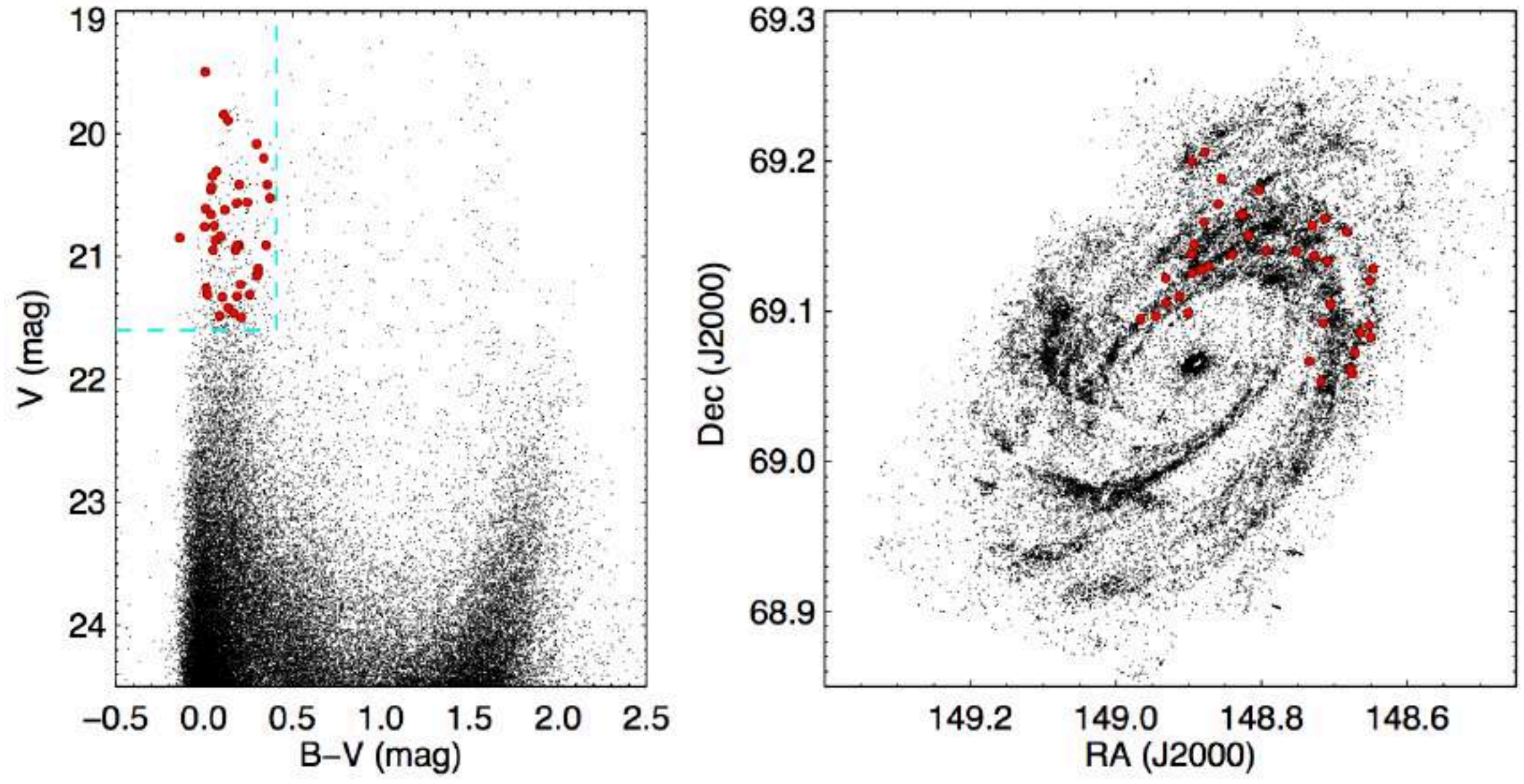

**Figure 18: Left panel: HST/ACS colour-magnitude diagram used to select blue supergiant star candidates in M81. Right panel: Location of the blue supergiant star targets in M81. *Figure from Kudritzki et al. (2012).***

Both high-and low-resolution spectroscopic surveys can play an important role in the derivation of fundamental parameters for massive stars. Detailed stellar parameters and chemical abundances for massive, hot stars can be determined with exquisite precision from high-resolution spectroscopy (e.g., Firnstein & Przybilla 2012), while medium-resolution spectroscopy can be used to measure basic parameters and metallicities. For example, Kudritzki et al. (2012) have shown that Keck/LRIS spectroscopy (R ~ 2000) can be used to measure reliable temperatures and luminosities for hot, massive stars by fitting the Balmer jump and Balmer line profiles. Because these stars are intrinsically bright (i.e., $M_V > -6$), medium-resolution spectra can be used to examine stars in distant galaxies (e.g., star forming galaxies in the Sculptor and M81 groups). Spectral templates can then be used to estimate metallicities in individual stars, and thus examine directly the metallicity gradients of disk galaxies.

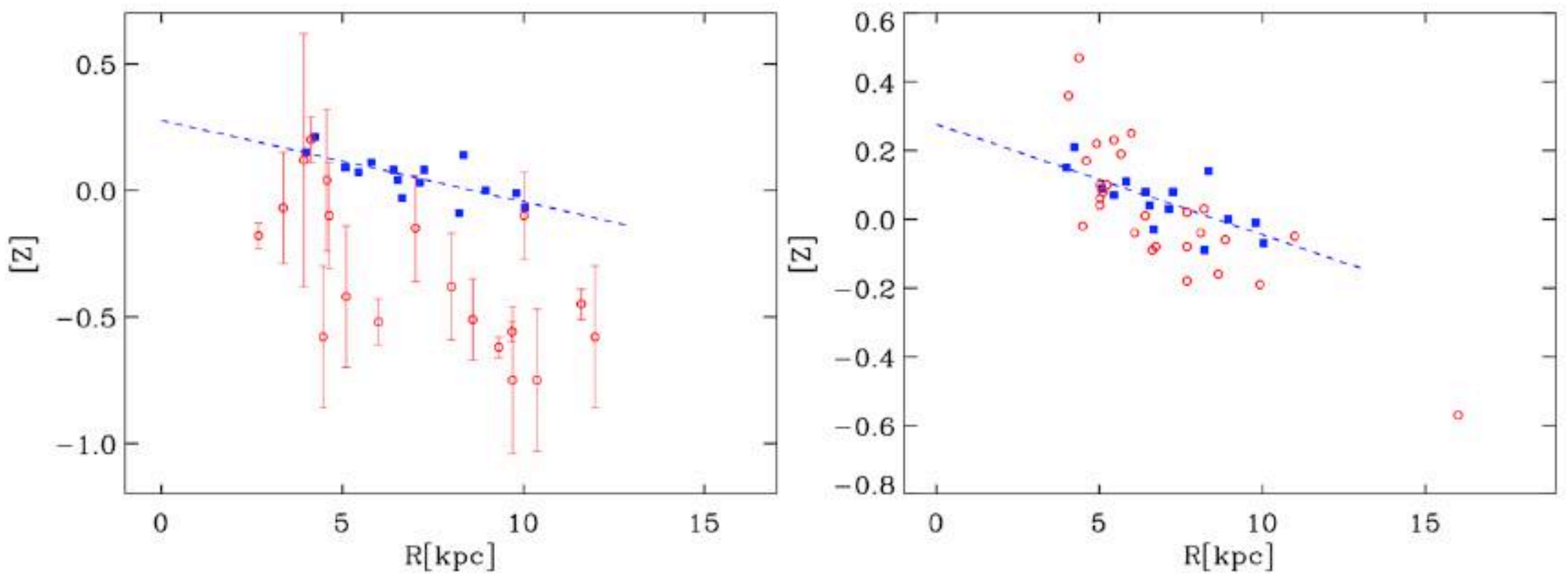

**Figure 19: Left panel: Comparison of metallicities in M81 deduced from blue supergiant stars with planetary nebulae oxygen abundances (blue and red, respectively). Right panel: Comparison of metallicities deduced from blue supergiant stars with oxygen abundances derived from HII regions (blue and red, respectively). The random uncertainties in these measurements are ~0.1 – 0.2 dex.** ***Figure from Kudritzki et al. (2012).***

In the simplest picture of disk formation, in which a disk forms through "inside out" gravitational collapse, one expects the metallicity to decrease as a function of Galactocentric radius (see Cheng et al. 2012). On the other hand, the existence of radial flows (e.g., Spitoni & Matteucci 2011) or the presence of a turbulent disk at early times (Brooks et al. 2009; Dekel et al. 2009) could alter this picture significantly. Strong radial migration mechanisms might also wash out an existing gradient in the stars (Roskar et al. 2008; Schoenrich & Binney 2009), while radial breaks may indicate either dynamical interactions within disks, or merger events in outer disks (e.g., Pohlen & Trujillo 2006; Younger et al. 2007; Bigiel et al. 2010). The measurement of disk metallicity gradients from individual stars means not only that coverage is more homogeneous than would be possible with H II regions, but it ensures that metallicity and reddening are sampled together, rather than relying on just oxygen abundances. Needless to say, metallicity measurements in a large and diverse sample of galaxies beyond the Milky Way will provide crucial environmental information that can be used to disentangle competing disk formation scenarios.

To illustrate the transformational role that MSE could play in this field, we consider the case of M81, the nearby Sab galaxy at a distance of ~ 3.5 Mpc. In this galaxy, stars as faint as V ~ 21.5 have been observed with Keck/LRIS at SNR ~ 50, and the metallicity gradients were found to be consistent with the oxygen abundance gradients from HII regions, with comparable uncertainties (see Figure 18). It is interesting that the gradient in M81 seems to have evolved over the last ~5 Gyr, as the oxygen gradient derived from PNe is steeper (see Figure 19). With MSE, a dedicated survey of nearby disk galaxies would be possible, including galaxies spanning a range in group and field environments. Previously, this kind of work was restricted to UV and high-resolution spectroscopic analyses; new techniques, combined with a wealth of high-quality spectra from MSE, would allow an ambitious, comprehensive, survey to be mounted from the ground.

With MSE, high-resolution spectroscopy could be used to provide a large statistical sample of chemistries and rotation rates of massive stars. These data would be useful in addressing issues related to rotational mixing (e.g., Maeder & Meynet 2010; Ekstrom et al. 2012). Since the inclination angle for any individual star is rarely known, a large statistical sample is one of the best ways to look for correlations between stellar line strengths/chemical abundances and stellar rotation rate (e.g., Hunter et al. 2009; Brott et al. 2011). Rotation is a fundamental parameter in the analysis of massive stars since it is thought to affect core masses, and therefore, ages, of these stars. In addition, rotation drives the mixing of nuclear processed material from a stellar interior to its surface, and it is thus one of the key observational constraints in stellar astrophysics (see, e.g., Venn et al. 2002). For very massive stars, rotation rates are thought to affect Eddington luminosities, and therefore mass loss rates as well, and ultimately to lead to the development of Wolf-Rayet stars and luminous blue variables. There is also some speculation that rotation is affected by stellar metallicity (Penny & Gies 2009), and that these parameters combine to influence stellar evolution predictions, including evolutionary masses. A large statistical sample of measured chemical abundances and rotation rates for massive, bright stars in M31, M33, and other Local Group, northern hemisphere star-forming galaxies would be the obvious way in which to test this prediction. These stars typically have $V \leq 18$ and are thus easily within reach of MSE in its high resolution mode.

Of course, rotation is also important for the (main sequence) evolution of lower mass stars, as it affects directly convective mixing efficiencies and, ultimately, the interpretation of surface chemistry measurements (including the lithium isotopic abundances that are compared to Standard Big Bang predictions; Asplund et al. 2006; Suda et al. 2011). With a large sample of accurate, homogeneous rotation rates measured for many thousands of bright stars with MSE in high-resolution mode, it would be possible to study these effects statistically.

## 2.3 The chemodynamical evolution of the Milky Way

**Science Reference Observation 3 (Appendix C)**

**Milky Way archaeology and the in situ chemical tagging of the outer Galaxy**

*MSE will carry out the ultimate spectroscopic follow up of the Gaia mission. No other planned or proposed survey will be able to chemical tag stars at the outskirts of the Milky Way and down to the faintest Gaia stars. Our view of the Milky Way will be completely revisited imminently with the first release of the Gaia data. MSE is designed to have all the key elements to be the essential spectroscopic tool for Galactic Archaeology in the post-Gaia area, able to probe the Galaxy through chemical and dynamical studies of individual stars across all Galactic components at all radii in the Galaxy, and through absorption line studies of the intervening interstellar medium.*

*We highlight here the main questions that we currently know MSE will uniquely be able to answer as well as the essential elements for its implementation. This includes extensive discussion of trades between spectral resolution and wavelength coverage to maximize MSE's utility to efficiently probe chemical space.*

### 2.3.1 Gaia and the near field of galaxy formation

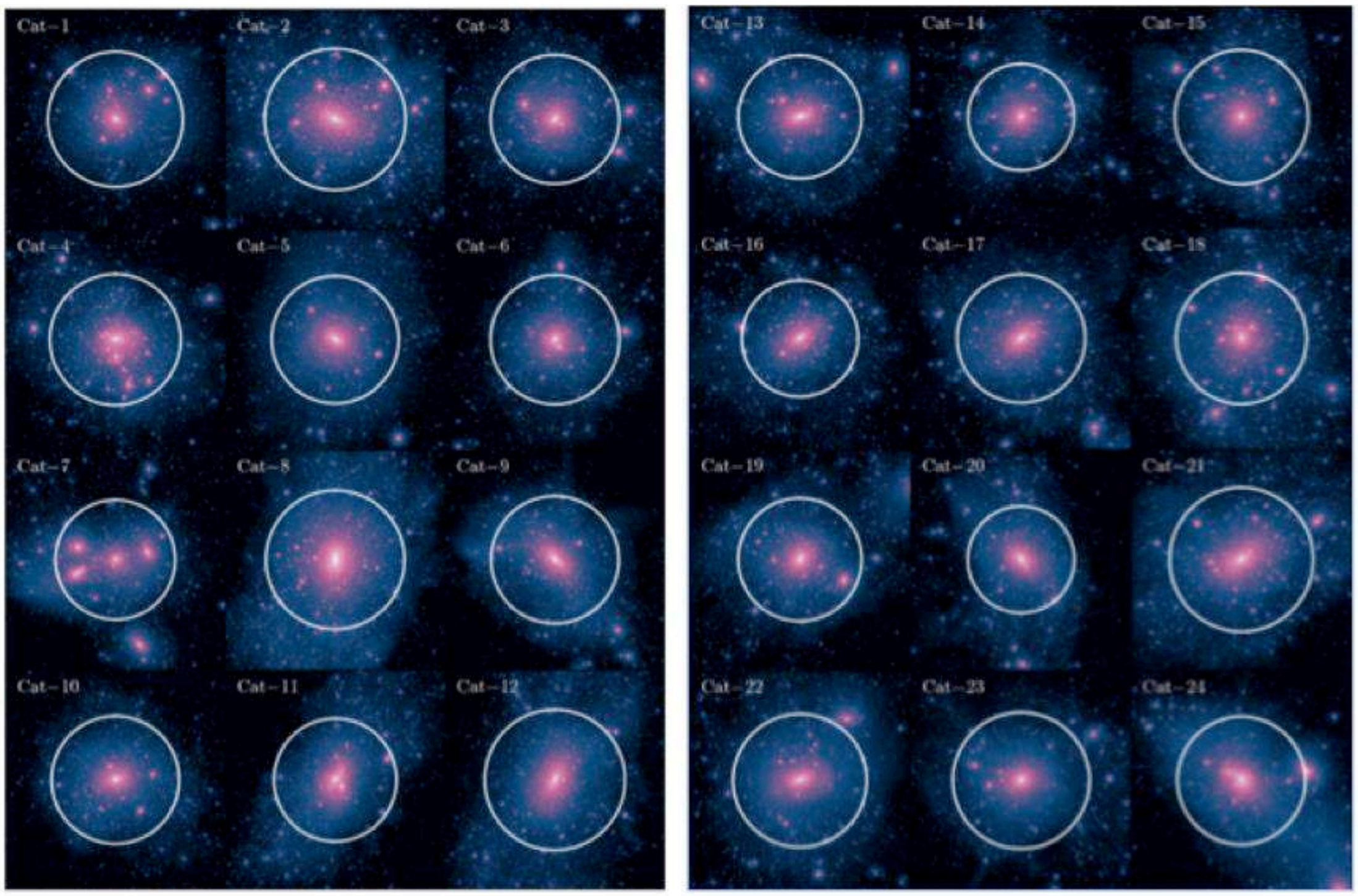

**Figure 20: Projected dark matter distribution of a suite of high resolution simulations of Milky Way mass dark matter halos at z=0 from the Caterpillar Project. The underlying cosmology is based on results from Planck (The Planck Collaboration et al. 2014), and candidate halos were originally selected for resimulation based on a simulation of a much larger volume. Box width in each panel is 1Mpc, and the white circles mark the virial radius.** ***Figure from Griffen et al. (2016).***

The measurement of the fluctuations in the cosmic microwave background (CMB) allow us to determine key cosmological parameters, including the total density, matter density and baryonic density of the universe. The picture that emerges is of a Universe dominated by a cosmological constant, with a mass budget largely dominated by non-baryonic matter (Komatsu et al. 2011). The overall success of this Lambda Cold Dark Matter (ΛCDM) cosmological paradigm on large scales is impressive. By following cosmological simulations of the growth of the primordial fluctuations that gave rise to the overdensities where structures form, ΛCDM is now able to make predictions on smaller (i.e., galactic) scales. Figure 20 shows a suite of high resolution simulations of the structure of dark matter on the scale of galaxies the mass of the Milky Way, where the underlying cosmology is based on results from The Planck Collaboration et al. (2014). It is on these scales that observations and theory show important discrepancies. The details of galaxy formation and evolution within this framework are not well understood due to the complexities of baryonic physics, and so it is on these small scales that the prevailing cosmological paradigm faces its toughest challenges.

Advancing our understanding of the galaxy formation process in general requires detailed observations of individual systems, and, naturally, our Galaxy provides the most important laboratory for exploring chemodynamical evolution on these scales. Indeed, the Milky Way is now recognised as being a complex ensemble of stellar populations, each having characteristic age and metallicity distributions, and unique dynamical properties.

Eggen et al. (1962) were the first to show that stellar abundances and kinematics could be used to understand the formation and evolution of our Galaxy, and thus guide ideas of galaxy formation in general. Indeed, their analysis of 221 very nearby stars remains, arguably, the single most influential observational paper on galaxy formation. In it, they proposed that metal-poor stars in the halo of the galaxy were formed during the rapid collapse of the protocloud that eventually became the Milky Way. An alternative proposal was offered by Seale & Zinn (1978), whose analysis of the stellar populations of a number of Galactic globular cluster led them to infer that they were formed in independent "protogalactic fragments" that later assembled the outer parts of the Galaxy. The current paradigm — whereby gas collapses in dark matter structures that form disks and eventually merge together or are accreted by larger systems — contains key elements of both of these original scenarios.

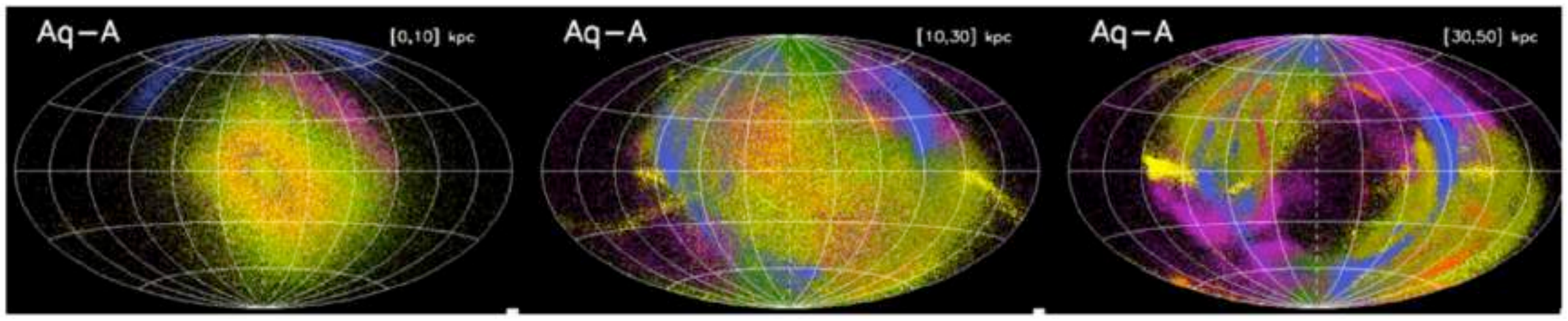

**Figure 21: Simulation of the distribution of field red giant branch stars on the sky at various distances from the Sun for the stellar halo of the simulation Aq-A. The different colours correspond to stars originating in different progenitors. Different progenitors cannot be distinguished between 10 and 30 kpc in photometric surveys alone, and velocities and chemical abundances are essential to reveal the individual events.** ***Figure from Helmi et al. (2011).***

These seminal papers demonstrated that information relating to the chemical and dynamical characteristics of the early protogalaxy, its constituent building blocks, and the subsequent evolutionary processes that have produced the Milky Way (and its subcomponents) can be explored through chemodynamical studies of modern-day stellar populations. Contemporary studies of the dynamics of stars in the Galaxy show this is approach is powerful: Helmi (2001) and others showed that stars that were accreted into the Milky Way from the same merger event can still be identified through their energies and angular momenta, even if they no longer form an obvious stellar stream across the sky (in contrast to coherent, present-day features such as the Sagittarius and Orphan streams, and other prominent substructures). Figure 21 shows a realization of the expected stellar distribution of a "Milky Way" galaxy halo from the Aquarius consortium. Depending on the distance regime being examined, it is clear that unraveling the individual events that contributed to the formation of this component requires information beyond the three spatial coordinates. Extending these concepts into chemical space, Freeman & Bland-Hawthorne (2002) proposed identifying stars that were born from the same gas cloud by the measurement of abundances for many chemical species, and "tagging" those with similar abundance patterns i.e., identification of Galactic "building blocks" by chemistry alone.

The chemical signatures, energies and angular momenta of the individual stars in the Galaxy thus encode its formation and evolutionary history, and can provide us with the most precise understanding of the processes through which the Milky Way was assembled. In contrast to the early study of Eggen et al. (1962), MSE will provide these parameters for *millions* of stars throughout every component and subcomponent of the Galaxy, and contribute to the creation of the most detailed dataset ever assembled for a single galaxy.

The analysis of Eggen et al. (1962) was based on the Solar neighbourhood; the halo stars that were observed are that very small subset that happen to be at a point in their orbit where they are passing close to the Sun. More than 50 years later, the high resolution large scale spectroscopic surveys that exist or are imminent – in particular AAT/HERMES and VISTA/4MOST – are similarly probing the very nearby Universe where the thin disk component is dominant. A critical component of Galactic science with MSE in comparison to *every* other existing or proposed spectroscopic facility is the ability to access the detailed chemodynamical signatures of every Galactic component and sub-component through *in-situ analysis* of individual stars.

The capabilities of MSE in this field are perfectly matched to the capabilities of Gaia, a landmark astrometric space mission that has as its primary focus a detailed understanding of the structure and composition of our Galaxy. Launched in 2014, it has begun its 5 year mission to produce a three-dimensional stellar map of the Milky Way with unprecedented precision. Gaia is conducting an all-sky survey to measure the positions of roughly 1 billion stars, approximately 1% of the entire stellar content of the Milky Way. In fact, all astronomical point sources brighter than $V \sim 20$ mag will be catalogued, and objects brighter than $V \sim 15$ mag will have their position measured to better than 20 microarcseconds. Beyond astrometry, Gaia will obtain multi-band photometry for all sources, and it is additionally equipped with a Radial Velocity Spectrometer (RVS) that will measure velocities for objects brighter than $V \sim 17$ mag (roughly 150 million stars) to an accuracy of $1 - 10 kms^{-1}$. Basic astrophysical information – including interstellar reddening and atmospheric parameters – will be acquired for the brightest $\sim 5$ million stars. Chemical abundance information will also be provided for a few elements (i.e., Mg, Si, Ca, Ti and Fe for stars of spectral type FGK) for stars brighter than G = 12.

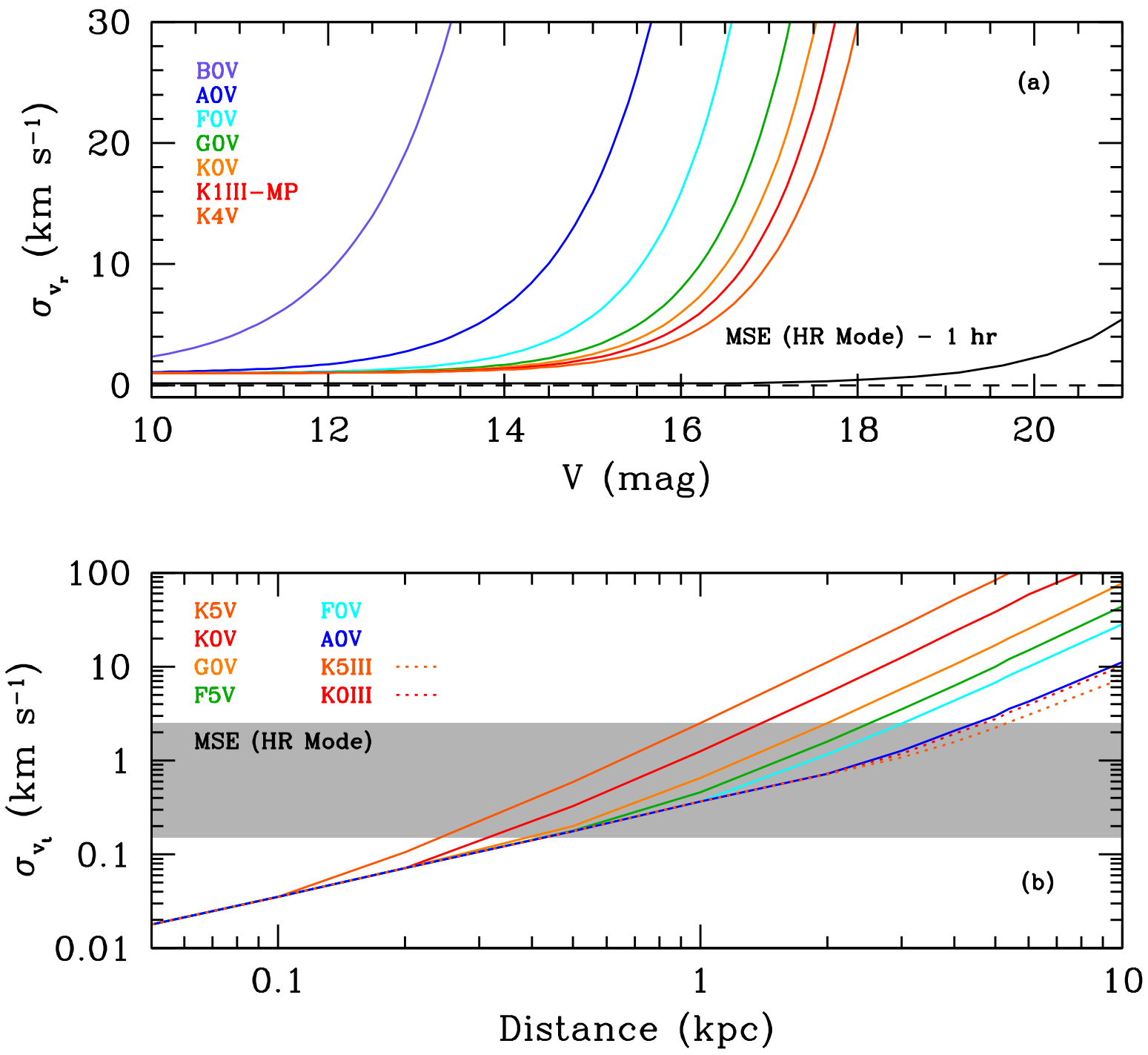

**Figure 22: (a) Expected radial velocity accuracy for the RVS instrument aboard Gaia for stars of different spectral types. The lower black curve shows the radial velocity accuracy expected for a G2V star observed for a total of 1 hour with MSE in high-resolution mode. (b) Typical errors in transverse velocity as a function of distance for stars of different spectral type observed with Gaia. The shaded grey region shows the approximate errors calculated for G2V stars with g = 20.5 observed with MSE.**

The importance of ground-based spectroscopy to supplement the Gaia data is well recognized in the international community, particularly as it relates to the chemical tagging of individual stars (Section 2.3.2). For chemical evolution studies, Gaia will identify hundreds of millions of stars fainter than G ~ 17 for which the RVS cannot provide any spectral information. Indeed, the astrometric accuracy of Gaia is only matched by similarly accurate radial velocities for the very brightest subset of its targets. For relatively faint stars, MSE is required in order to provide precise radial velocities to complement the precise transverse velocities measured by Gaia. Figure 22 illustrates this with a comparison of the anticipated velocity accuracy of MSE compared with the RVS on Gaia. Clearly, the kinematic data provided by MSE is an essential complement to the astrometric data from Gaia across the entire magnitude range of Gaia sources.

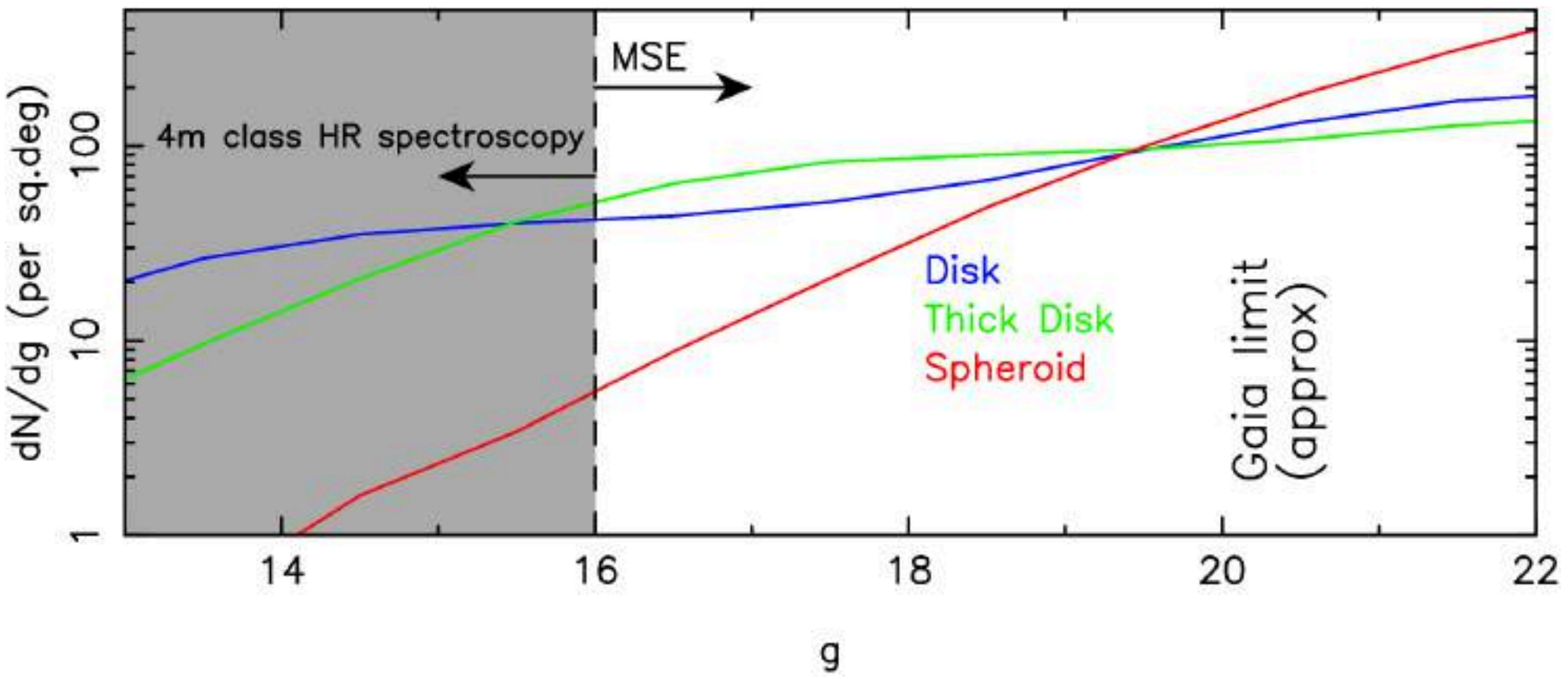

**Figure 23: Differential star counts as a function of magnitude for the three main Galactic components, based on the Bescancon model of the Galaxy (Robin et al. 2003, A&A, 409, 523) for a 100 square degree region in the vicinity of the north Galactic cap. The shaded region indicates the magnitude range accessible at high resolution to 4m class spectrographs (typically operating at R ~ 20 000 – 40 000). MSE is the only facility able to access the thick disk and spheroid at high resolution in the regions of the Galaxy in which they are the dominant components. MSE targets at high resolution span the full luminosity range of targets that will be identified with Gaia.**

The grasp of MSE for exploration of the Galaxy is awesome. Figure 23 shows the differential stellar number density of each of the main components of the Galaxy, as calculated from the Besancon model (Robin et al. 2003, A&A, 409, 523) for a 100 square degree region around the north Galactic cap. In the magnitude regime that is best accessible at high resolution with 4m class spectrographs (shaded area of Figure 23), the thin disk is the dominant component: even at the Galactic pole, thin disk stars out-number halo stars in the range 14 ≤ g ≤ 16 by a factor of 15:1, and this ratio increases dramatically with decreasing |b|. For example, AAT/HERMES (that has a brighter magnitude limit of V = 14) estimate that only 0.2% of their targets will be halo stars[4]. Fainter than g ≥ 16, the thick disk becomes the dominant component at higher latitudes and is easily accessible with MSE even with minimal pre-selection of targets. As an 11m aperture facility, MSE will obtain good SNR at high resolution in reasonable exposure times even in the magnitude range where the stellar halo is dominant (i.e., g ≥ 19.5 at higher latitudes).

[4] http://www.mso.anu.edu.au/galah/survey_design.html#selection_criteria

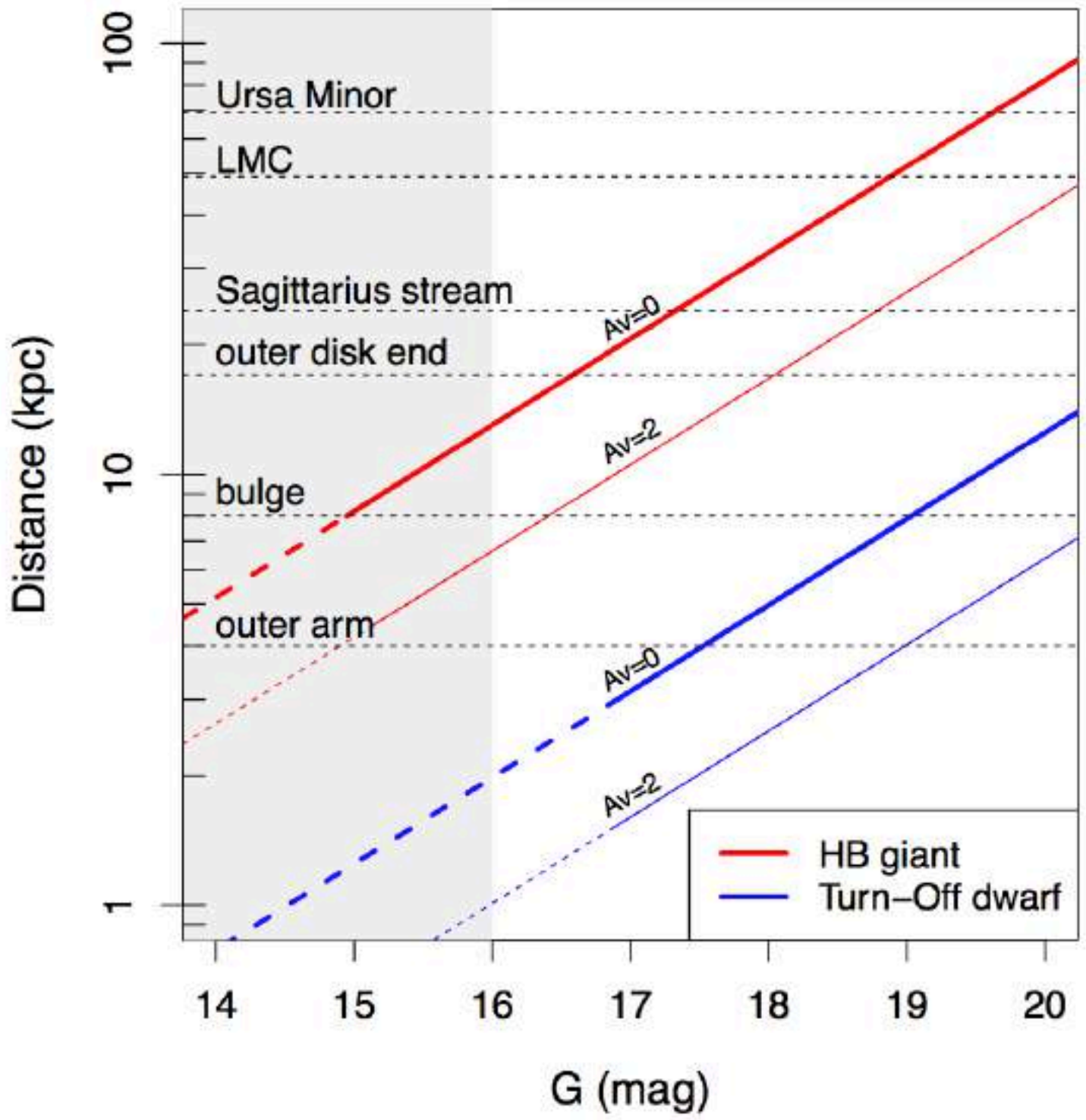

**Figure 24: Distance range probed as a function of magnitude by using two different stellar tracers: horizontal Branch stars (in red) probe to greater distances at a fixed magnitude than turn-off stars (in blue). Note that tip of the red giant branch stars (not shown) probe beyond the Milky Way virial radius. The shaded grey area corresponds to the magnitude range covered by the 4m-class spectroscopic surveys, and is therefore not a priority for MSE. Horizontal dashed lines indicate the distance threshold required to probe some prominent Galactic structures and satellites. Extinction shifts the red and blue lines to the right as illustrated with the $A_V$ = 2 mag presented here. For each colored line, the dotted section represents the distance – magnitude range where Gaia parallax accuracies will be better than 20%. The solid line therefore represents the distance – magnitude range where distances derived through spectroscopy will be essential to complement Gaia.**

Figure 24 shows the effective distance range probed by MSE as it extends down the magnitude range of Gaia sources. Here, distance as a function of (Gaia) magnitude is shown for two possible stellar tracer populations, a main sequence turn-off star (blue lines) and a horizontal branch giant star (red lines). The two different lines per tracer indicate the effect of different assumptions regarding Galactic extinction. Horizontal dashed lines correspond to approximate distance thresholds for probing various stellar components and/or satellites. The dotted segments of the colored lines indicate the regimes in which Gaia parallaxes will have better than 20% accuracy. Beyond this, spectroscopy will provide important constraints on the distances of the targets.

The potential role for MSE in furthering our understanding of the Galaxy in the post-Gaia era is extensive and covers all components and sub-components of the Galaxy, and it is not the intent of this document to detail all possible science programs. Representative subjects include:

- The Interstellar Medium and carriers of Diffuse Interstellar Bands

MSE will be the only instrument able to observe at high resolution millions of objects with known distance, allowing an outstanding step forward in the knowledge of the multi-phase galactic ISM.

- The Galactic disk
    - The outer disk
      Extragalactic observations have revealed all the complexity of the outer regions of galaxy disks: truncated/anti-truncated surface brightness profiles, breaks in metallicity profiles, U-shaped age profiles, complex star formation histories, etc. For the Milky Way, its outer regions are still largely unknown. MSE will be able to reach stars beyond the edge of the outer disk. It will quantify its boundary and substructures, provide a detailed chemical description of the outer disk, and trace its formation history and its link with the inner disk
    - Disc dynamics
      MSE is crucial to map the 3 dimensional velocity field beyond the extended solar neighbourhood. A detailed 3 dimensional map of the velocity field contains not only information the mass distribution of the Galaxy, it also holds the signatures of the ongoing perturbations of the disk, such as the spiral arms or the central bar and their back reaction. It allows measurement of the various pattern speeds at play in the Galaxy and the location of their resonances.

- The Galactic bulge
    - Bulge dwarfs
      The inner regions of the Milky Way keep trace of the early phases of formation of the Galaxy, and of its subsequent evolution. No high resolution survey is yet able to reach the bulge dwarfs, but the discovery potential is enormous given that the bulge is likely home to some of the oldest stars in the Galaxy. The main competitor of MSE for the bulge science case will be VLT/MOONS (near-infrared over a relatively small field, and therefore particularly efficient for the bulge in the most crowded and extincted regions).
    - The bulge star formation history
      The bulge is too far to use Gaia astrometry to derive ages in the bulge. For this, one relies entirely on spectroscopy. MSE is the only high-resolution spectrograph that can reach the bulge turn-off. This is at $V \sim 20$ in Baade's Window and $H \sim 17.5$, therefore too faint for VLT/MOONS.
    - Linking the inner galactic substructures to the outer Galaxy
      The detailed abundance distribution of the bulge and its outskirts, and its comparison with those obtained for the halo and the disk studies, will reveal the continuity/ discontinuity between these populations. The bulge is densely populated and different populations dominate at different longitude/latitudes. Reconstructing the distribution of its chemistry, kinematics and age profiles therefore requires good statistics over a large area, up to $|b| > 10-20$ degrees).
    - Searching for primordial populations
      It is possible that the bulge contains the remains of the most ancient mergers that shaped the core of our Galaxy (e.g., Howes et al. 2015). Stars in the inner Milky Way are kinematically well mixed, and so chemical abundances are

required to reveal their origins. The structure and kinematics of the bulge is that of a barred system presenting a boxy/peanut shape. However, there are also indications of the presence of a primordial structure within the inner galactic regions, but its relative mass and connection with the local old structures is unknown. We are dependent upon detailed chemical studies for many elements in order to identify the presence of a separate primordial bulge component and distinguish it, or link it, with the structures we identify at our solar neighbourhood.

- The Stellar Halo
    - Structure and substructure
      Observationally, there is evidence that the Milky Ways stellar halo contains plenty of substructure. There are also evidences for different formation mechanisms to be at play in the stellar halo of the Milky Way, with an inner and outer component having significantly different structure, metallicity and kinematic properties (e.g. Carollo et al. 2007). Kinematics and chemistry of the Galactic halo up to its virial radius will provide us with strong constraints on the formation and evolution of our Galaxy.
    - In-situ versus accreted stars in the inner halo
      MSE will definitively quantify the portion of halo in-situ stars versus accreted ones, measure potential gradients in the in-situ population and record the chemical imprint of the accreted progenitor formation history.
    - The first stars
      The unprecedented size and quality of the stellar spectroscopic dataset of MSE will enable a detailed analysis of the metal-weak tail of the halo metallicity distribution function (MDF).

Examination of Figure 23 and Figure 24 emphasizes the crucial role of MSE for understanding the otherwise least accessible parts of our Galaxy i.e., the faint and distant regimes where the outer disk, thick disk and especially stellar halo are dominant. This is not to say that MSE will not provide critical data for examination of the thin disk or bulge; it is clear from the list above that MSE will have significant impact on the study of these components. However, it does recognize the need to optimally deploy the spectroscopic resources available in the 2020s to best advance our understanding of the evolution of the Milky Way, and to optimize MSE so it is of best utility to those science cases for which it can be uniquely, and powerfully, suited . Further, only MSE can complement Gaia in the faint regime by *providing the detailed chemistry and dynamics of target stars over the entire magnitude range of sources observed by this satellite*; a prediction of the expected source density of Gaia targets accessible to MSE is shown as an Aitoff projection in Figure 2.

As listed above, a major source of Gaia – MSE synergy is in the understanding of the interstellar medium through absorption line studies of intervening gas clouds along sight-lines to bright stars. MSE is literally unmatched in this field, and extensive discussion of this important aspect of the structure of our Galaxy can be found in Chapter 4. In the remainder of this section, we focus discussion on the resolved stellar populations of the Galaxy.

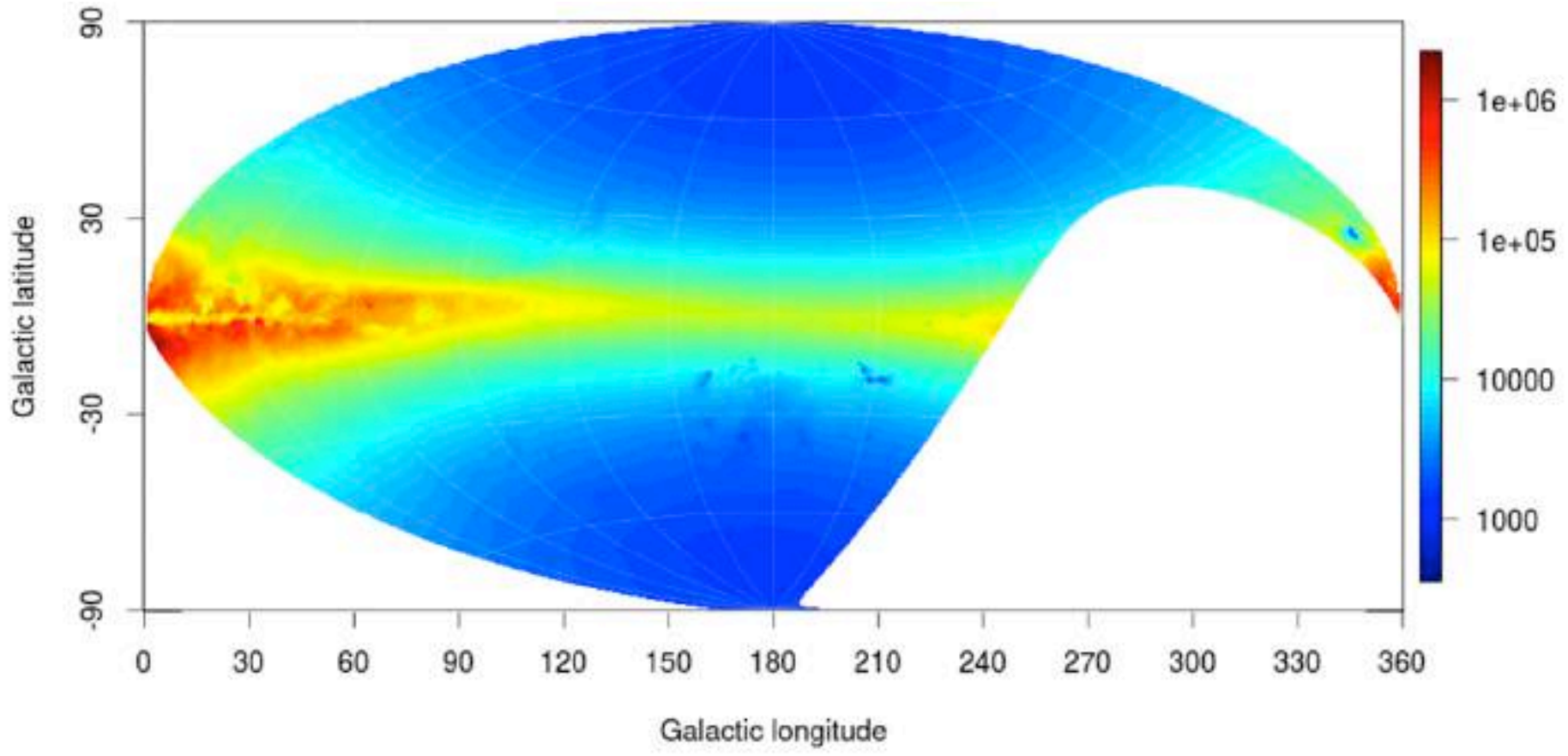

**Figure 25: Aitoff projection of the predicted sky density of targets (per deg$^2$) to G=20 observed by Gaia that will be accessible to MSE (δ > −30◦).** ***Input data credit: Gaia DPAC CU2.***

**Science Reference Observation 4 (Appendix D)**

**Unveiling cold dark matter substructures with precision stellar kinematics**

*As part of an extensive Galactic archaeology mission, MSE will measure the accurate velocities of millions of stars in stellar streams distributed throughout the Galactic halo. Stellar streams are dynamically cold features that have been stripped from globular clusters and dwarf galaxies. However, according to ΛCDM, the Milky Way halo should host many thousands of dark sub-halos. As these orbit in the Galactic potential, they will produce a heating effect on cold stellar structures such as stellar streams. By mapping the kinematics of a large number of streams, extending over large areas of sky and at a range of radii from the center of the Galaxy, MSE will be able to measure the heating caused by dark sub-halos and constrain their mass distribution. The discovery of even a single extended cold stellar stream places strong limits on the allowed mass distribution of these otherwise invisible structures.*

### 2.3.2 Chemical tagging of the outer Galaxy

The halo contains some of the oldest stars in the Galaxy — including those that formed only a few Myr after the Big Bang (Frebel & Norris 2015). Both the kinematics and chemical compositions of these old stars provide us with strong constraints on the formation and evolution of our Galaxy. Many of the "proto-galactic fragments" whose existence in the outer halo was implied in the seminal work of Searle & Zinn (1978) have since been identified photometrically in the SDSS, most notably in the "field of streams" (Belokurov et al. 2006; see Figure 26 for a recent visualization of the data from Bonaca et al. 2012). Such streams are generally only observable if the time that has passed since they merged with the Galaxy is short compared to the dynamical timescales at their location. Streams in the outer halo tend to be longer lived and thus easier to recognize as coherent structures (Johnston et al. 1999).

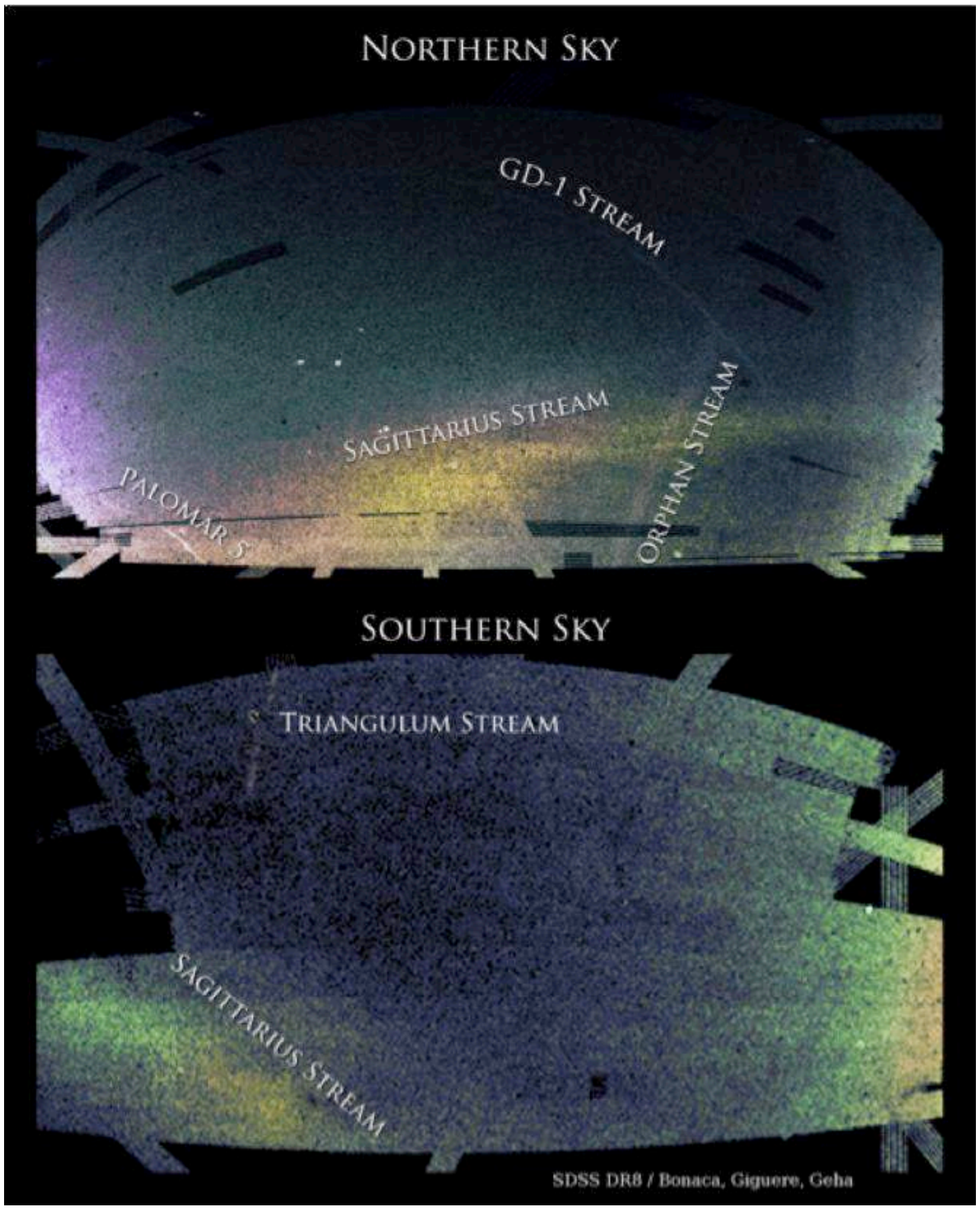

**Figure 26: Stellar density maps of the Milky Way halo in Celestial coordinates (right ascension increases to the left, declination increases to the top) in the footprint of the SDSS imaging survey. A matched filter analysis reveals a large number of stellar streams and other substructures, the most prominent of which are labeled. *Figure from Bonaca, Giguere & Geha (2012).***

A critical weapon in the arsenal available to MSE lies not just in dynamical analyses, but in the derivation of detailed chemical abundances. Recognising the suggestion by Freeman & Bland-Hawthorne (2002) that “the major goal of near-field cosmology is to tag individual stars with elements of the protocloud”, MSE will provide abundance ratios for elements formed through multiple nucleosynthetic channels for many millions of targets. Of course, this applies not just to the stellar halo but also to other Milky Way components, including the disk, thick disk and bulge. As discussed in the preceding section, the study of the chemical properties of the halo has relied, with a few exceptions, on local samples of halo stars that pass near enough to the Sun to be observable at high spectral resolution, whereas MSE will move shift the paradigm towards in situ analyses.

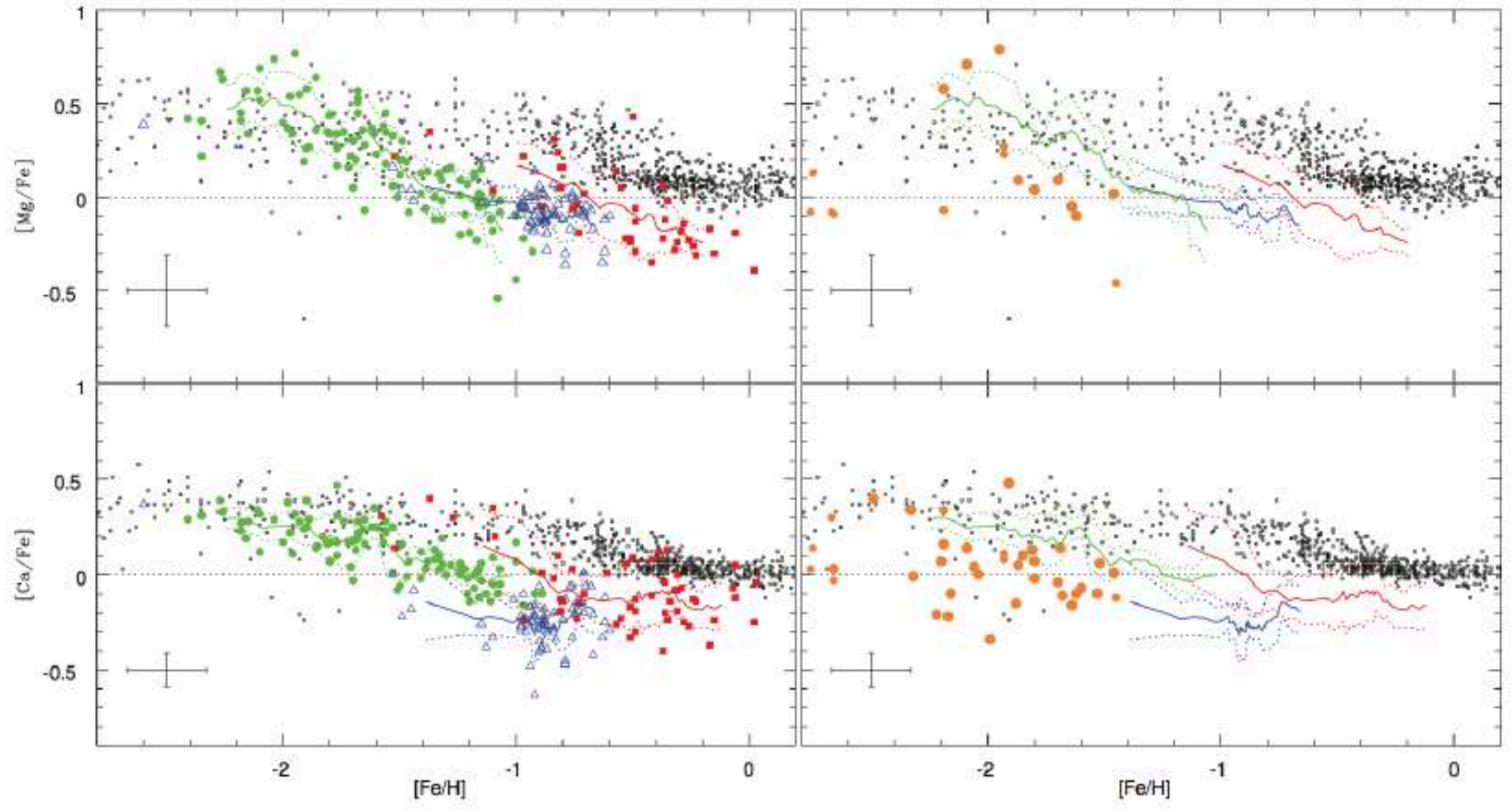

**Figure 27: Alpha elements as a function of [Fe/H] in a variety of nearby dwarf galaxies. Left panel: Sagittarius (red/gray filled squares: Sbordone et al. 2007; McWilliam & Smecker-Hane 2005), Fornax (blue/dark-gray open triangles: Letarte et al. 2010) and Sculptor (green/light-gray filled circles: Hill et al. 2012). Right panel: Sextans (orange/light-gray filled circles: Kirby et al. 2010; Aoki et al. 2009; Shetrone et al. 2003). In all panels, solid lines show the average trend and dotted lines show the 1-σ dispersion for each dwarf galaxy. The underlying points show stars in the Milky Way (Cayrel et al. 2004; Venn et al. 2004).** ***(Figure from Hill et al. 2012).***

Figure 27 provides a taster of the power of large chemical abundance datasets in helping to unravel the formation of Milky Way and its satellites. Here, Hill et al. (2012) show the alpha element distribution as a function of [Fe/H] for Milky Way stars (black points; Cayrel et al. 2004, Venn et al. 2004) as well as several prominent Milky Way dwarf spheroidal satellites (Sagittarius – red/gray filled squares – Sbordone et al. 2007, McWilliam & Smecker-Hane 2005, Fornax – blue/dark-gray – open triangles – Letarte et al. 2010, Sculptor – green /light-gray filled circles – Hill et al. 2012, Sextans – orange/light-gray filled circles – Kirby et al. 2010, Aoki et al. 2009, Shetrone et al. 2003). It is clear that the dwarf spheroidals show a different distribution in this parameter space from the Milky Way (and potentially from each other). Specifically, the “knee” in the distribution of Milky Way stars, at which point the alpha element abundance is broadly constant with decreasing [Fe/H], informs us on the earliest star formation epochs that made the stars that we see in this Galactic component. Type II supernovae are alpha-rich, and explode only a few-to-tens of Myrs after a star formation event. Type Ia supernovae are iron rich in comparison, and take at least a few hundred Myrs prior to contributing to the chemical abundances. The position of the knee in the distribution is therefore a measure of the intensity of early star formation, since a knee at higher metallicity indicates more intense star formation (i.e., more Type II supernovae contributed to both the alpha and iron abundance prior to Type Ia supernovae contributing to the iron abundance only). The dwarf spheroidals, in comparison to the Milky Way halo, appear to have evolved more passively since their knee (if indeed one is even present!) is at considerably lower metallicity.

While the insights available through chemical abundance studies are powerful, large statistical datasets of high resolution chemical abundances are only now emerging, prominent among which is SDSS/APOGEE and AAT/HERMES. As such, there has been a wealth of recent analyses discussing the practical application of chemical tagging to these datasets. Key questions that have been addressed, and which are essential for MSE development to acknowledge, include:

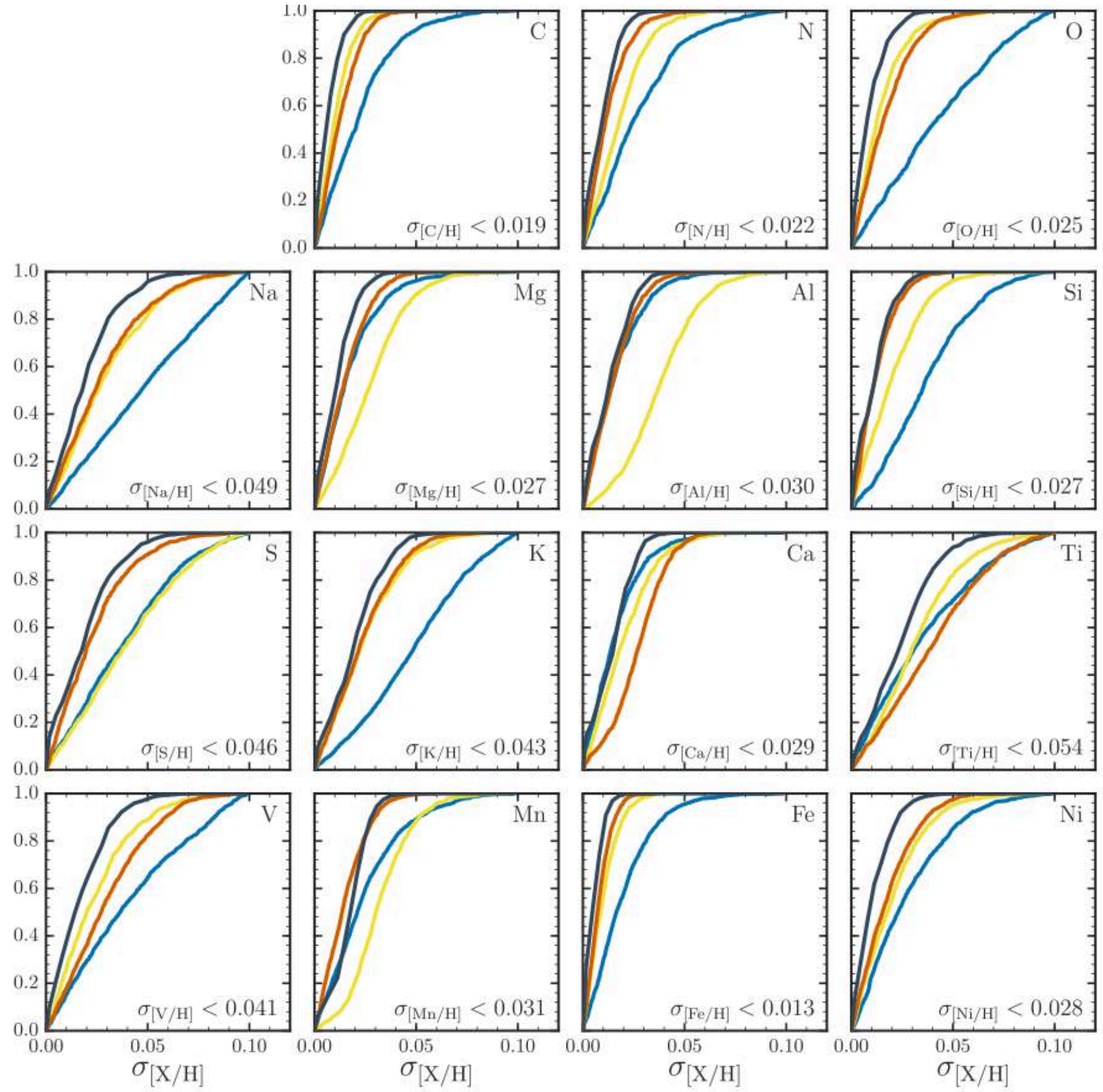

**Figure 28: Cumulative posterior distribution functions for the intrinsic abundance scatter in 15 elements obtained for three clusters (M67 in red, NGC 6819 in yellow, NGC 2420 in blue) based on analysis of SDSS/APOGEE spectroscopy. The cumulative distribution function for all three clusters is shown in black, and the combined 95% upper limit is given in each panel. The intrinsic scatter of each element in these clusters (star formation sites) is extremely small, as demanded for chemical tagging to be successful. (*Figure from Bovy 2016).***

*i. The chemical homogeneity of star formation events*

A fundamental premise of chemical tagging is that stars born in an individual cluster (which henceforth we adopt as shorthand for "a star formation event", regardless of whether the event forms a recognizable stellar cluster) share a single chemical "fingerprint", such that there is a chemical signature that can be used to relate the stars even if they do not cluster strongly in phase space. Until recently, however, this had not been tested to high accuracy (although see work by De Silva et al. 2007, 2009; Ting et al. 2012; Friel et al. 2014; Önehag et al. 2014). Traditional techniques to determine stellar abundances have an accuracy that, empirically, appears to be limited to $\sim$0.1 dex for whatever abundance is being measured. Generally, this accuracy reflects systematic uncertainties in the underlying stellar models used to interpret the data and in understanding the instrument-dependent signatures present in the data.

Bovy (2016) undertakes a novel analysis of three different star clusters observed in the SDSS/APOGEE dataset that attempts to measure the chemical homogeneity, or otherwise, of each cluster, with minimal recourse to models. Rather, the working hypothesis of this analysis is that a homogeneous stellar cluster is one in which all the stars form a one dimensional sequence, with any variations between members due solely to stellar mass (as well as observational uncertainties).

Bovy (2016) test their hypothesis at the level of the stellar spectra, and results are shown in Figure 28. Here, the solid black line is the overall posterior probability of the spread in abundance for each of the 15 elements indicated, and colored lines correspond to the individual clusters (M67 in red, NGC 6819 in yellow, NGC 2420 in blue). This figure shows that, for each element in each cluster, the intrinsic spread between the stars in the cluster is at the level of a few hundredths of a dex, at most. This precision is nearly an order of magnitude better than can be achieved using standard techniques, and implies that the fundamental assumption of chemical tagging is sound.

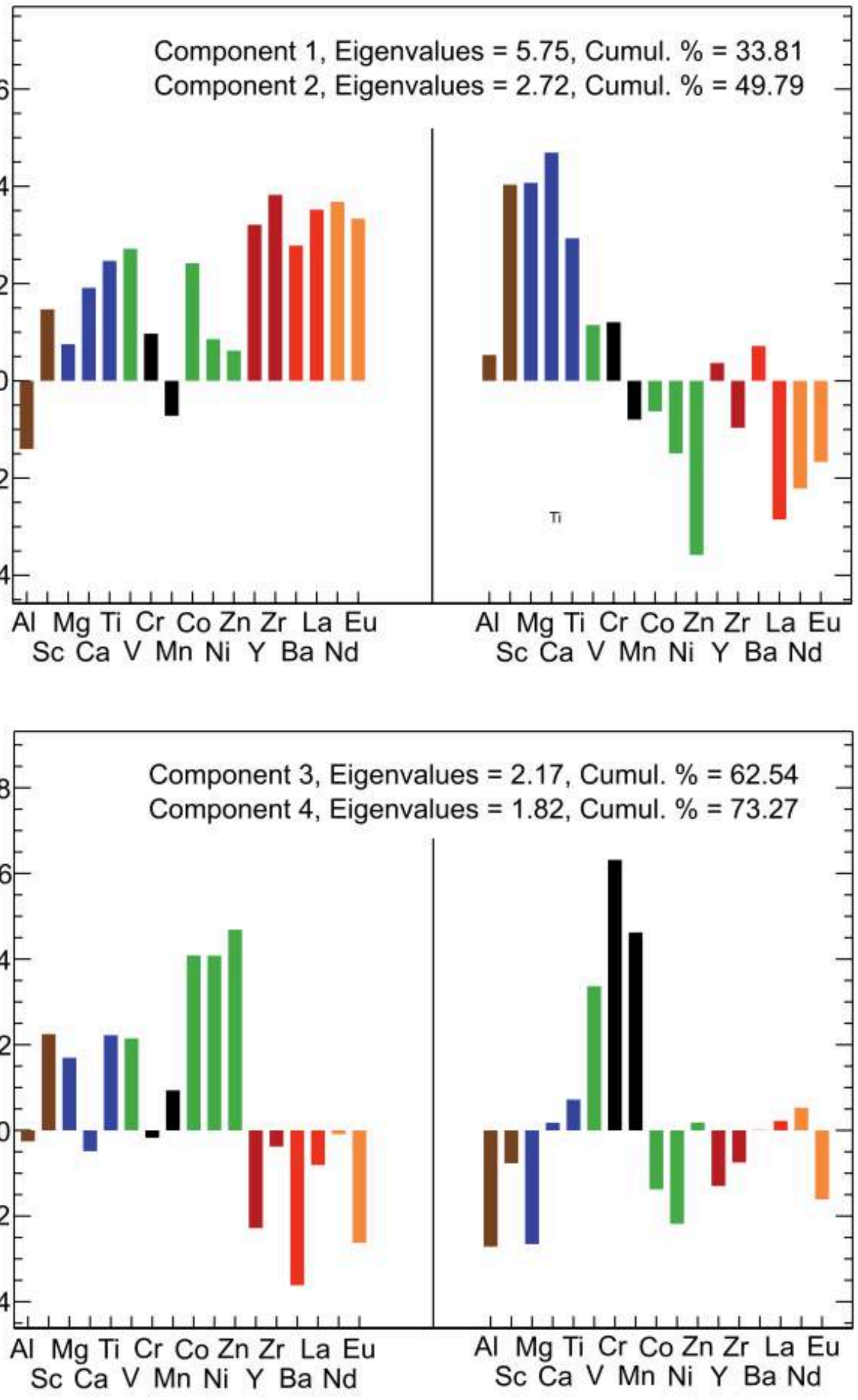

**Figure 29: The four principal components of a 17 element abundance space for a metal poor halo sample from Barklem et al (2005).** ***Figure from Ting et al. (2012).***

### *ii. The dimensionality of chemical abundance space*

The application of chemical tagging requires the measurement of a large number of chemical species per star. However, exactly how many measurements are required, and what species are most valuable, is a complex question. For example, two different species that are found to vary in lock-step with each other as a function of any other variable will provide considerably less discriminating power than two species whose behavior over a large sample of stars is less correlated. Clearly, the dimensionality of chemical abundance space is linked to the number of unique pathways by which chemical enrichment can take place.

Ting et al. (2012) present a landmark study of the structure of chemical abundance space as a pathfinder study for the AAT/HERMES spectrograph. These authors conduct a principal component analysis of the abundance space of stars observed in several different environments, including the solar neighbourhood, halo metal poor stars, etc. For the metal poor stars (-3.5 < [Fe/H] < -1.5 dex), they use a dataset of nearly 300 stars from Barklem et al. (2005) and the First Stars Survey (Cayrel et al. 2004, Francois et al. 2007, Bonifacio et al. 2009) in which there are measurements for 17 different chemical elements.

Figure 29 shows the main 4 principal components of the metal-poor stars analysed by Ting et al. (2012). Here, neutron capture elements are different shades of red (dark red represent lighter s-process elements, red represents heavier s-process elements and orange are mostly r-process elements), light odd-Z elements are dark brown, alpha elements are blue, iron-peak elements are green, and Cr and Mn are black. In full, they find that there are typically 7 – 9 dimensions in chemical abundance space for the solar neighbourhood and metal poor stars, and that dwarf galaxies may have extra dimensions, perhaps associated with their longer star formation timescales.

As well as leading to a better understanding of the structure of chemical space, the principal components shown in Figure 29 are a fascinating way to reveal the nature of the sites of nucleosynthesis. For example:

- *The first principal component* in Figure 29 reveals the presence of a site that produces both r process elements and alpha process elements (perhaps r-process core collapse SNe?)
- *The second principal component* reveals the presence of a process that produces an anti-correlation between alpha elements with iron peak and neutron capture elements (perhaps normal core collapse SNe?)
- *The third principal component* reveals a process producing an anti-correlation between alpha elements and iron peak elements with neutron capture elements (perhaps hypernovae?)
- *The fourth principal component* shows a strong contribution to Cr and Mn (both of which are synthesized in the incomplete Si-burning region).

In addition to providing insight into the chemical space that MSE will explore, the analysis of Ting et al. (2012) also shows the power of principal component analysis to chemical abundance studies. Ultimately, the dimensionality of chemical abundance space will be able to be explored more fully by MSE using data for millions of stars across all components and sub-components, in contrast to the ~300 stars in Figure 29.

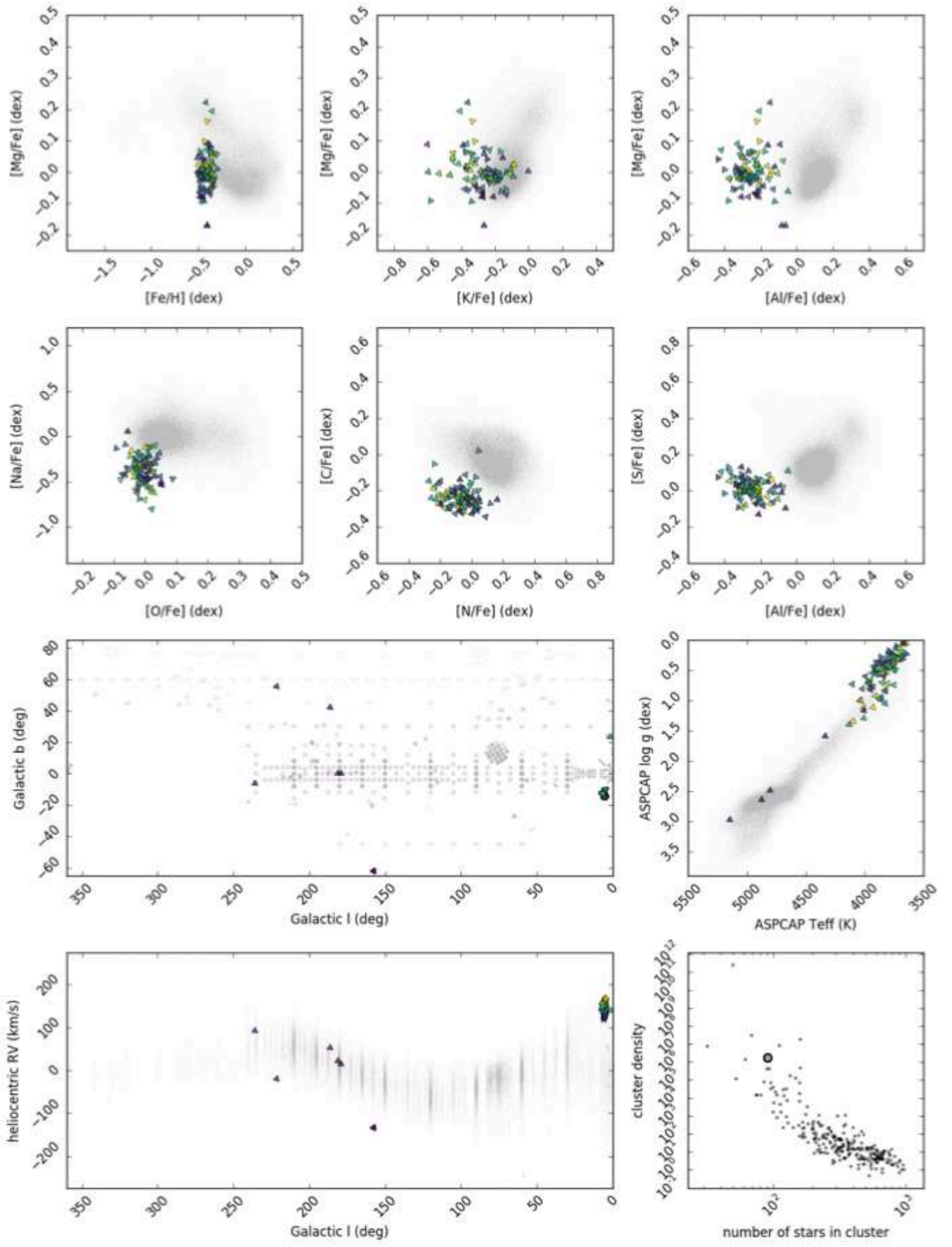

**Figure 30: Greyscale points show projections of approximately $10^5$ individual stars from the SDSS/APOGEE survey. The top set of panels show projections in abundance space only. Clustering analysis of this (15-dimensional) space reveals substructures. One of these most prominent substructures is shown as the large points. The lower panels show the structure of this feature in phase space coordinates that were not used for its identification. In fact, this feature is due to the Sagittarius dwarf spheroidal galaxy.** ***Figure from Hogg et al. (2016).***

### *iii. Sampling rate and practical cluster mass limits for chemical tagging*

Ting et al. (2015) examine some of the survey considerations for successful chemical tagging. Key parameters they consider include:

- Sampling rate, that is the number of targets in the survey compared to the possible number of targets ($N_{target}/N_{sample}$). Higher sampling is required in order to be sensitive to lower mass clusters;

- The number of "distinct" regions of chemical space in which the clusters are distributed, $N_{cells}$. The relevant region of chemical space to be explored will have a finite volume, in which each of the clusters will occupy a unique point. However, once measurement errors are considered, it is clear that some clusters may become indistinguishable from other clusters in chemical space, particularly if the number of clusters is high, the volume of chemical space is small, and/or measurement errors are large, i.e., the dimensionality of chemical space, and typical measurement uncertainities, are both critical parameters in defining $N_{cells}$;

- The number and mass distribution of individual star clusters ("star formation events") contributing to the signal. Clearly, measurement of these quantities is a fundamental goal of chemical tagging experiments.

The analysis of Ting et al. (2015) is focused towards studies of the stellar disk. Here, dynamical times are relatively short and additional dynamical processes can complicate the analysis. For example, radial migration can act to move stars out of the survey volume in which they were born (or introduce new stars that did not form in this region into the survey volume), thereby acting to make identification of the relevant, individual clusters, more difficult. Shorter dynamical times also means that phase space coordinates become increasingly difficult to apply as additional constraints.

Despite these additional difficulties, Ting et al. (2015) conclude that, through strategic observational programs, it is possible to statistically reconstruct the slope, high mass cut-off and evolution of the cluster mass function for studies that focus on the stellar disk of the Milky Way for surveys that sample of order 1 million stars (presumably, the most challenging environment as described above).

A recent analysis by Hogg et al. (2016) is a nice demonstration of chemical tagging in action: the top panels of Figure 30 shows various projections of $10^5$ stars observed with SDSS/APOGEE. The darkest points show a prominent sub-grouping in this space, identified via an algorithm that identifies clustering in N-dimensional space. The lower panels show these same stars in various projections of phase space. Even although the phase space coordinates were not used to identify these stars, it is clear that the stars do form a coherent structure. In fact, these stars belong to the Sagittarius dwarf galaxy.

These recent analyses of different aspects of chemical tagging emphasise the importance of carefully planned observing programs to (i) best sample the structures being probed (ii) to best probe the numerous dimensions of chemical abundance space (iii) to best use phase-space information in addition to chemical abundance information, and (iv) to best enable precision determination of individual abundances. SRO-03 describes in detail these considerations as they apply to MSE, with specific emphasis on important trades between multiplexing, spectral resolution and wavelength coverage. Baseline Galactic Archaeology programs are described in SRO-03 and SRO-4, that will produce high quality, precision abundances for a large number of chemical elements in millions of stars that span all the main components of the Galaxy, focusing on the interface of the thick disk and halo, and that reaches to the edge of the Milky Way.

### 2.3.3 The metal weak tail of the Galaxy: first stars and the progenitors of the Milky Way

The Galactic Archaeology programs described in SRO-03 and SRO-04 will enable a vast range of science on the formation of the Galaxy and stellar nucleosynthesis, and will provide a holistic perspective on the formation of the Galaxy through multi-parameter analyses. We have already emphasised that a multitude of science cases spanning all aspects of the Galaxy are possible with such a program. In keeping with the focus of this document on science cases that are uniquely possible with MSE, we now discuss in more detail a particularly compelling component of SRO-03 that is focused on the metal-poor Galactic halo.

The unprecedented size of the stellar spectroscopic dataset for MSE will enable the definitive analysis of the metal-weak tail of the halo metallicity distribution function (MDF). This key observable has a direct bearing on models for the formation of the first stars, and on the dark baryonic content of galaxies. The first stars to be formed after the Big Bang were formed with the "primordial" chemical composition: i.e., hydrogen and helium, plus traces of lithium. A protogalactic cloud consisting of such a gas may have had difficulty in providing cooling mechanisms efficient enough to allow the formation of low-mass stars. Several theories on star formation postulate the existence of a "critical metallicity" below which only extremely massive stars can form (Bromm & Loeb 2003; Schneider et al. 2011 and references therein). Other theories invoke fragmentation to produce low-mass stars at any metallicity (Nakamura & Umemura 2001; Clark et al. 2011; Greif et al. 2011).

The implication of these considerations for the baryonic content of galaxies is obvious: if the first generation of massive stars that reionized the universe formed along with low-mass stars, a large fraction of these would now be present as old, cool white dwarfs. On the other hand, a small fraction of these (essentially those of mass less than 0.8 $M_{\odot}$) would still be shining today and can in principal be observed. Further, if there is a critical metallicity, then the metal-weak tail of the MDF ought to show a sharp drop at this value. MSE can examine both the chemical abundance distributions of old stars to search for nucleosynthetic signatures of these first stars, but also construct a precise MDF to search for evidence for a critical metallicity.

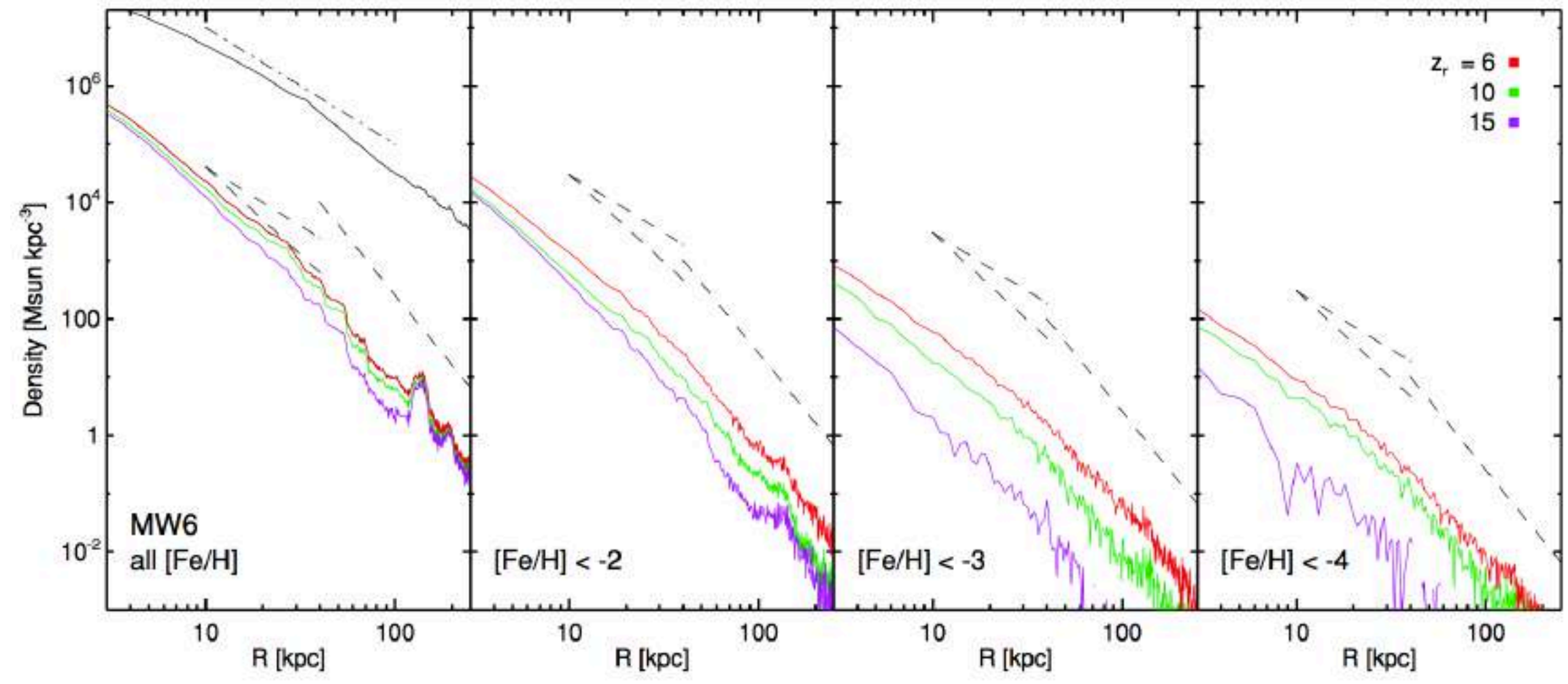

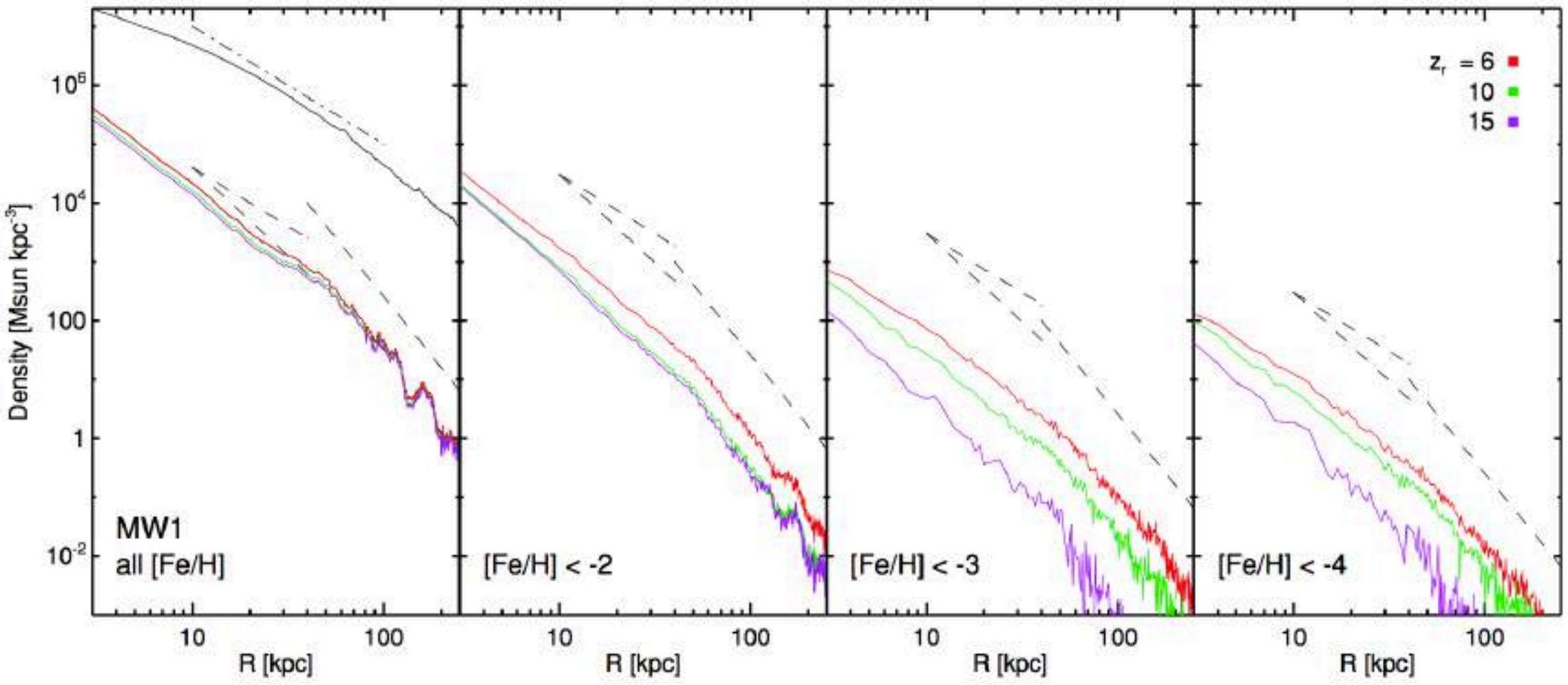

**Figure 31: Stellar mass density profiles for two different realizations of a cosmologically-consistent Milky Way mass halo (top and bottom panels), for all stars (left panel) and three metallicity cuts (from left to right), [Fe/H] < −2, −3, and −4, and three redshifts of re-ionization in each panel (as described by the inset text). The dashed lines mark power laws with slopes −2, −3, and −4.** ***Figure from Tumlinson (2010)***

Detailed modeling of the formation of a Milky Way mass galaxy in which the distribution of these first stars are monitored reveal fascinating insights into the expected growth of the Milky Way with time. Tumlinson (2010) simulate a suite of six Milky Way analogues and incorporate baryonic processes on top of the dark matter "scaffolding" to allow analysis of the evolution of the implied stellar populations. They find that the fraction of stars dating from the oldest epochs is a strong function of radius within the Galaxy and with metallicity. The former trend reflects the inside out growth of dark matter halos, and implies the oldest stars are some of the most tightly bound to the Galaxy. The latter trend reflects the trend of increasing metallicity with time.

Figure 31 shows the radial distribution of stars in a couple of the different halo realisations from Tumlinson (2010) as a function of metallicity. Stars that formed at redshifts z = 6 – 10 likely have metallicities more metal poor than [Fe/H] = -3 dex. To test the validity of these models, and to probe the formation of the Galaxy and its stars at the earliest times, it is therefore important not just to find metal poor stars, but to develop a detailed understanding of the shape of the full MDF and its spatial variation in the Galaxy, including out to very large radius. Such a program requires in situ analysis of large numbers of metal poor stars, and is a core science goal of MSE.

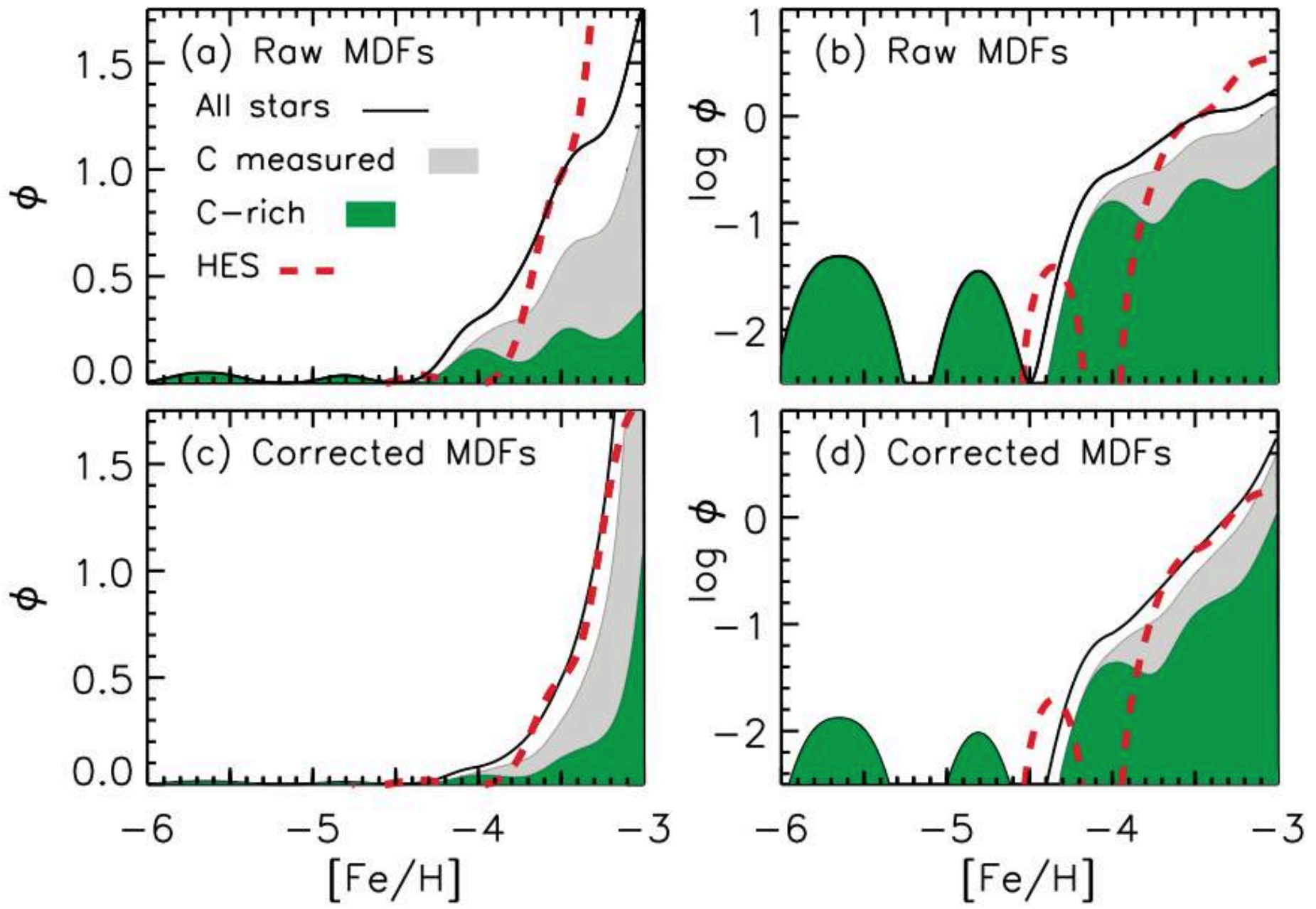

**Figure 32: The metallicity distribution function, [Fe/H], based on the high-resolution abundance analysis of Yong et al. (2013a) at the metal-weak end, [Fe/H]<-3 dex. Left panels show a linear scaling, and right panels show a logarithmic scaling. The upper and lower panels refer to the raw data and those following completeness corrections on the range −4.0 < [Fe/H] < −3.0, as described by Yong et al. (2013b). Green and grey color-coding is used to present the contribution of C-rich and C-normal stars for which measurement was possible, respectively. The dashed line shows the Hamburg-ESO MDF based on the data of Schorck et al. (2009).** ***Figure from Frebel & Norris (2015).***

To put the potential of MSE studies of the metal-weak tail of the Galaxy in context, Figure 32 shows the state-of-the-art MDF for stars more metal-poor than [Fe/H= -3 dex. For homogeneity, this is based on a high resolution analysis by Yong et al. (2013a). Also shown as a dashed line is the MDF derived from the Hamburg-ESO Survey (Schorck et al. 2009). The Hamburg-ESO survey is recognized as a landmark study of the metallicity of the Galaxy halo, and is based on a total of 1638 stars. The Yong et al. (2013a) sample has 86 stars with [Fe/H] < -3 dex, of which 32 have [Fe/H] < -3.5 dex. Current surveys will increase the known sample of metal-poor stars significantly over the next few years, and will produce important lists of candidate metal poor stars for spectroscopic follow-up (e.g., the SkyMapper project – which has already found a star with [Fe/H] ~ -7, Keller et al. (2014) – and the Pristine project on CFHT(P.I. E. Starkenburg). However, full characterisation of the MDF through in-situ spectroscopic analyes requires the large aperture, high resolution, highly multiplexed capabilities of MSE. Through a program such as that described in SRO-03, MSE will provide the definitive study of the metal-weak structure of the Galaxy, by providing metallicities for a sample of several million stars at resolution R > 20 000, allowing a complete, homogeneous, characterization of the halo MDF down to [Fe/H] ~ –7.

## 2.4 The Local Group as a time machine for galaxy evolution

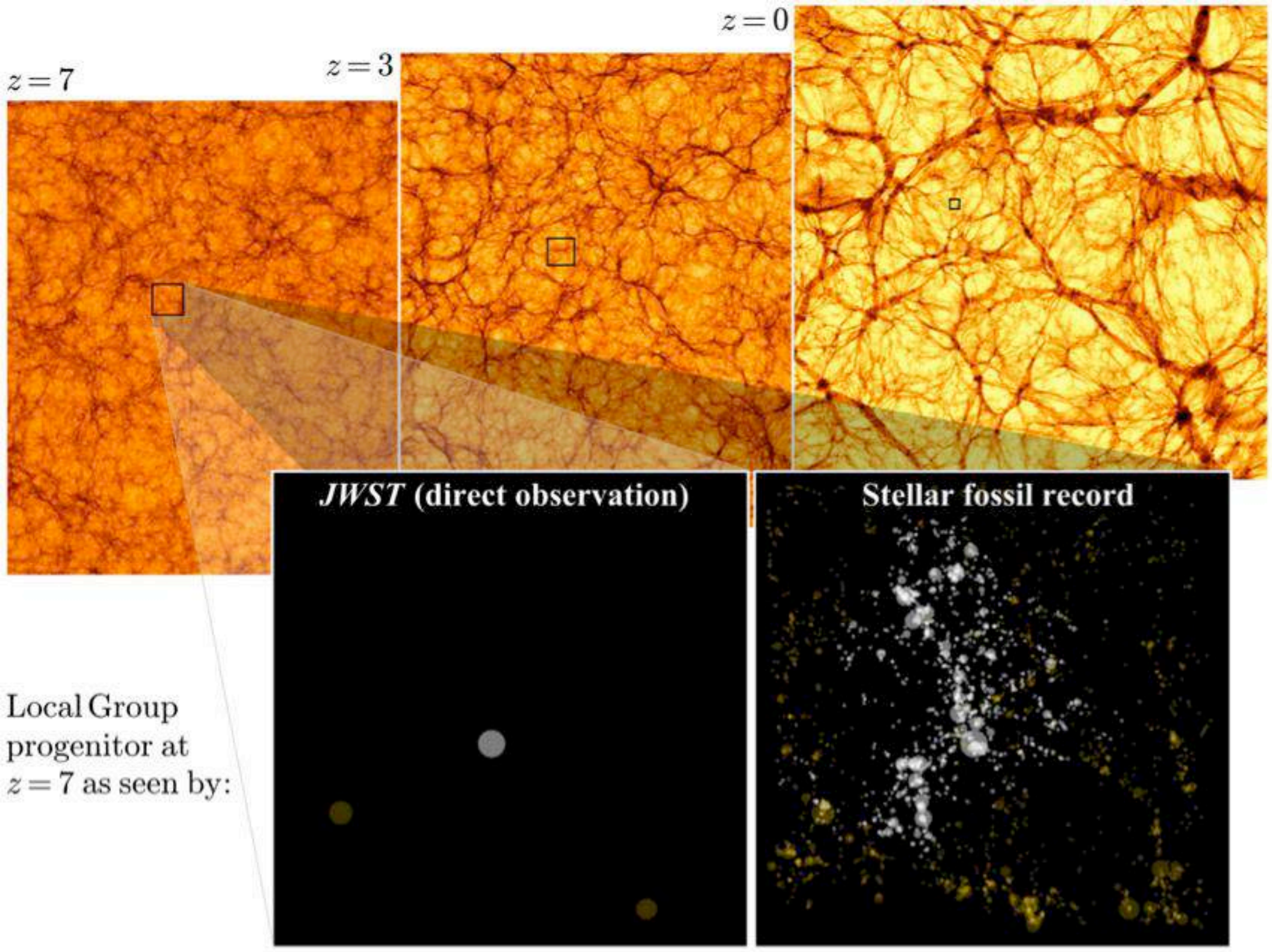

**Figure 33: Density slices of the Illustris simulation at different redshifts, each with a side length of 106.5 co-moving Mpc. The square in each panel indicates the co-moving size of a Local Group-type structure at each redshift. These are comparable in size to the HST Ultra-Deep Field. The zoom-in panels shows an expected image of the galaxies visible at z = 7 ± 0.02 using JWST (left panel), and through the stellar fossil record in the Local Group (right panel; white symbols indicate structures still present in the Local Group at z = 0, whereas gold symbols represent objects that are no longer present in the Local Group at z=0).** ***(Figure from Boylan-Kolchin et al. 2015).***

The collection of galaxies in the immediate vicinity of the Milky Way — the Magellanic Clouds, the Andromeda Galaxy, Triangulum, and the ~100 currently known dwarf galaxies that constitute the "Local Group" — are the nearest examples of galaxies spanning a wide range of morphological types. Generally speaking, they are the only galaxies in the Universe — aside from the Milky Way itself — for which large numbers of individual stars can be spectroscopically observed from the ground. As such, these nearby systems offer unique insights into galaxy formation and evolution

The size of the Local Group is of order a couple of Mpc in diameter, as defined by the zero-velocity surface where the gravitational attraction of the galaxies of the Local Group exactly balances the Hubble flow (McConnachie 2012). A recent analysis of the formation of Local Group-like structures in cosmological simulations shows that, at z = 7, the co-moving linear size of the Local group is 7Mpc (i.e., a volume of ~350Mpc). At early epochs, the Local Group therefore probes a cosmologically representative volume of the Universe. At z ≤ 3, the co-moving size of the Local Group exceeds that of the Hubble Ultra-Deep Field, and is similar to the volumes that will be probed by JWST (see Figure 33).

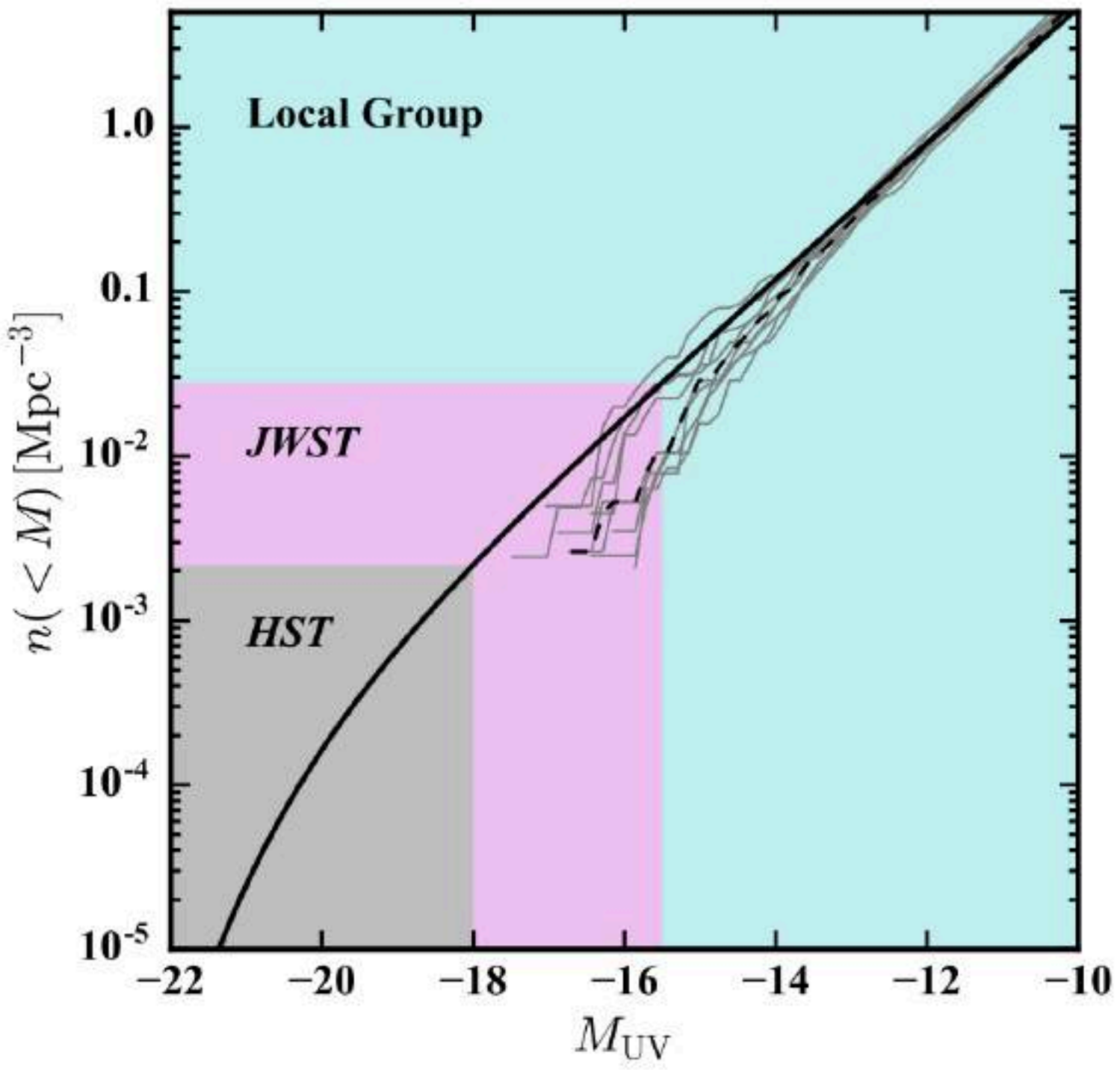

**Figure 34: The solid black line represents the cumulative luminosity function of galaxies at z = 7 (Finkelstein et al. 2015), extrapolated beyond $M_{UV}$= -18 to faint magnitudes. The think grey lines represent the cumulative luminosity functions in simulations of Local Group environments (the ELVIS simulations, see Garrison-Kimmel et al. 2014), and the dashed black line represents the mean. The approximate detection limits of HST, JWST and the Local Group are shown, highlighting the complementarity of the near and far-field regimes for a full understanding of galaxy evolution across the luminosity function.** ***(Figure from Boylan-Kolchin et al. 2016).***

The complementarity of studies of the Local Group population of galaxies to high redshift studies of galaxy evolution with JWST, WFIRST and other future facilities, is clear. At early times, the Local Group and JWST explore similar volumes. The Local Group contains representative numbers of faint galaxies that will never be detectable at high redshift with JWST, and whose formation can be explored through their stellar fossil record. Meanwhile, JWST is sensitive to galaxies that are intrinsically brighter and, therefore, rarer. This is shown explicitly in Figure 34.

Local Group galaxies subtend large angles on the sky and consist of a multitude of stars that are, individually, quite faint. A complete exploration of the deep field of the Local Group therefore requires wide fields, extreme sensitivities, and high multiplexing. MSE is therefore essential for this field, be it for the lowest mass or highest mass galactic systems that are present, as we now discuss.

**Science Reference Observation 5 (Appendix E)**

**The dynamics and chemistry of the Local Group galaxies**

*With MSE, we will be able to conduct a systematic survey of the kinematics and chemistry of*

*Local Group galaxies within 1 Mpc with a level of detail that has never been achieved before and cannot be achieved anywhere else. Gaia is gathering unprecedented astrometric measurements that provide the ability to map the six-dimensional phase-space structure of a galaxy will likely usher in a "Golden Age" of Milky Way research. However, without essential complementary information from the nearest galaxies, conclusions we may draw about the process of galaxy formation from a single galaxy may be premature. It is in this context that we propose to conduct the ultimate chemodynamical decomposition of Local Group galaxies across the galactic luminosity function, including a large population of faint dwarfs as well as the sub-$L_*$ galaxy M33, and the $L_*$ galaxy M31. These galaxies represent the obvious stepping stones between our highly detailed description of the Milky Way, and the low-resolution studies of more distant galaxies, in the realm of a few Mpc and beyond, where we can obtain significant samples of galaxies (as a function of galaxy type, environment, mass, etc).*

### 2.4.1 Dark matter and baryons at the limits of galaxy formation

The Local Group's dwarf galaxies include the nearest, smallest, oldest, darkest and least chemically-enriched galaxies known. Their extreme properties have made these objects the focus of investigations ranging from galaxy formation to cosmology and even particle physics. Several generations of imaging surveys, combined with targeted spectroscopic follow-up, have dramatically improved knowledge of these galaxies' luminosity functions and spatial distributions, internal structure and kinematics, chemical abundance patterns and star-formation histories (Mateo 1998, Tolstoy et al. 2009, McConnachie 2012, and references therein).

Over the last decade, large-scale imaging surveys (e.g., SDSS, Pan-Starrs, Dark Energy Survey, PAndAS, etc.) have nearly quadrupled the number of known dwarf galaxies in the Local Group, revealing tens of M31 satellites as well as a vast population of "ultra-faint" Galactic satellites that reaches astonishingly low luminosities of $L_V \sim 10^{2-3} L_\odot$. (e.g., Willman et al. 2005, Zucker et al. 2006, Belokurov et al. 2007, McConnachie et al. 2009, Koposov et al. 2015, Laevens et al. 2015, Bechtol et al. 2015, Drlica-Wagner et al. 2015). The ability of large imaging surveys to deliver data sets that are simultaneously deep, wide and homogeneously processed has also enabled discoveries of low surface brightness features such as M31's metal-poor stellar halo (Chapman et al. 2006, Kalirai et al. 2006), numerous stellar streams in the Milky Way (e.g., Belokurov et al. 2006), distorted morphologies in the outermost regions of several dwarf galaxies (e.g., Muñoz et al. 2010), and distant tracers of the Galactic potential (e.g., blue horizontal branch stars; Brown et al. 2010, Deason et al. 2012).

On the spectroscopic front, over the same timeframe multi-object spectrographs at 6 − 10m class telescopes have revealed intriguing details about the chemical and dynamical properties of these stellar populations. Velocity dispersion profiles measured from hundreds to thousands of stars per galaxy imply domination by dark matter at all radii (Walker et al. 2007). In many cases, even the oldest stellar populations in these systems can be divided into multiple components that provide clues about formation and evolution: typically a relatively metal-rich, spatially compact and kinematically cold sub-population is embedded within – and traces the same dark matter potential as – a relatively metal-poor, kinematically hotter sub-population (Tolstoy et al. 2004, Battaglia et al. 2006, Battaglia et al. 2011). The faintest Galactic satellites not only extend the galactic luminosity – metallicity relation by at least three orders of magnitude in luminosity (Kirby et al. 2013), they also host significant numbers of extremely metal-poor stars ([Fe/H] < -3) that encode information about early supernovae events (Kirby et al. 2008, Frebel et al. 2010, Simon et al. 2015). Furthermore, chemical abundance patterns are uncovering the extent to which the Galactic halo grew by devouring the siblings of its surviving satellites (Tolstoy et al. 2009, and references therein), and most recently are even helping to settle decades-long debate regarding the astrophysical mechanisms that lead to "rapid" neutron-capture production of heavy elements (Ji et al. 2015, Roederer et al. 2016).

Yet due to the lack of a dedicated large-aperture, wide-field, multiplexing spectroscopic facility, imaging surveys have dramatically outpaced their spectroscopic counterparts thus far. While recent and/or future large-scale spectroscopic surveys target primarily extragalactic sources (e.g., SDSS, BOSS, DESI) and/or Galactic stars (e.g., SEGUE, APOGEE, Gaia, Gaia-ESO), none has a significant component devoted to building large, uniform and spatially complete data sets for the resolved stellar populations comprising the Local Group's dwarf galaxies.

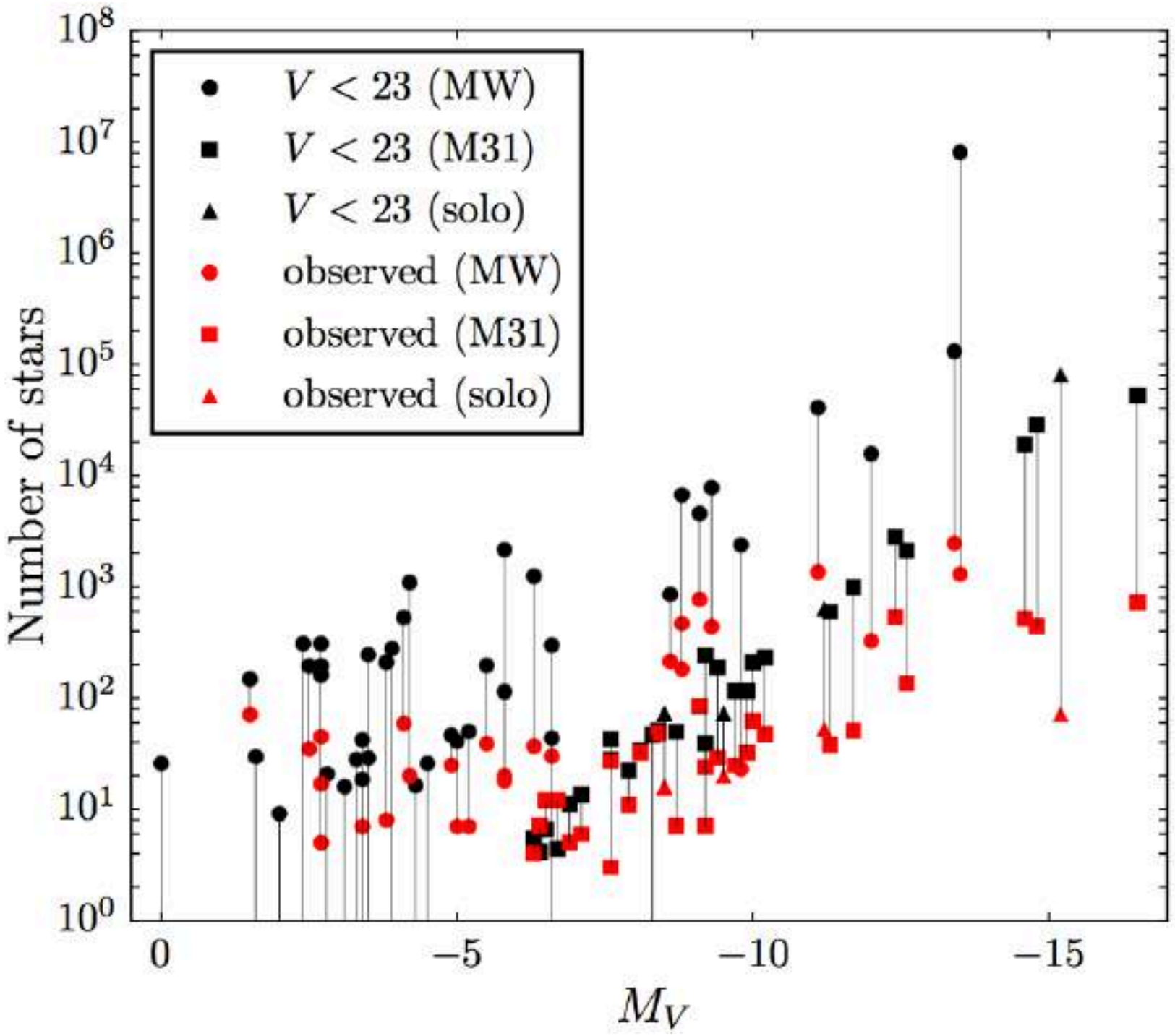

**Figure 35: Possible stellar-spectroscopic sample sizes for known Local Group dwarf galaxies with $M_V \geq -16.5$. As a function of galactic luminosity, black points indicate number of stars brighter than a fiducial magnitude limit of V < 23.0. Connected red points indicate sizes of samples available in the current literature. Marker types specify whether the dwarf galaxy is isolated (triangles; see the *Solo* survey of Higgs et al. 2016) or a satellite of the Milky Way (circles) or M31 (squares). Data from McConnachie (2012); Muñoz et al. (2005); Geha et al. (2006); Koch et al. (2007); Lewis et al. (2007); Martin et al. (2007); Simon & Geha (2007); Mateo et al. (2008); Walker et al. (2009a); Fraternali et al. (2009); Carlin et al. (2009); Koch et al. (2009); Belokurov et al. (2009); Geha et al. (2010); Willman et al. (2011); Koposov et al. (2011); Ho et al. (2012); Tollerud et al. (2012); Frinchaboy et al. (2012); Kirby et al. (2013); Collins et al. (2013); Walker et al. (2015a,b); Kirby et al. (2015); Walker et al. (2016); Torrealba et al. (2016).**

This technological shortcoming of Local Group astronomy is a unique opportunity for MSE, as its unprecedented combination of depth, wide field of view, spectral resolution and multiplexing will be transformative in this field. In order to help quantify the improvements one can expect, we consider the spectroscopic sample sizes that would be achievable in an MSE survey of Local Group galaxies. For all known systems less luminous than M32 ($M_V \geq 16.5$), we estimate the number of stars brighter than a fiducial magnitude limit of $V \leq 23$. Figure 35 compares these numbers to the largest spectroscopic sample sizes that are currently available in the literature for each observed galaxy. In most cases an MSE survey would grow the available spectroscopic sample by more than an order of magnitude. For the "ultrafaints" ($M_V > -5$), this means samples reaching hundreds to one thousand member stars. For the Milky Way's "classical" dwarf spheroidals ($-7 \leq M_V \leq -13$) and more luminous but farther objects like NGC 185, NGC 205 and NGC 6822, it means samples reaching into the tens of thousands of member stars.

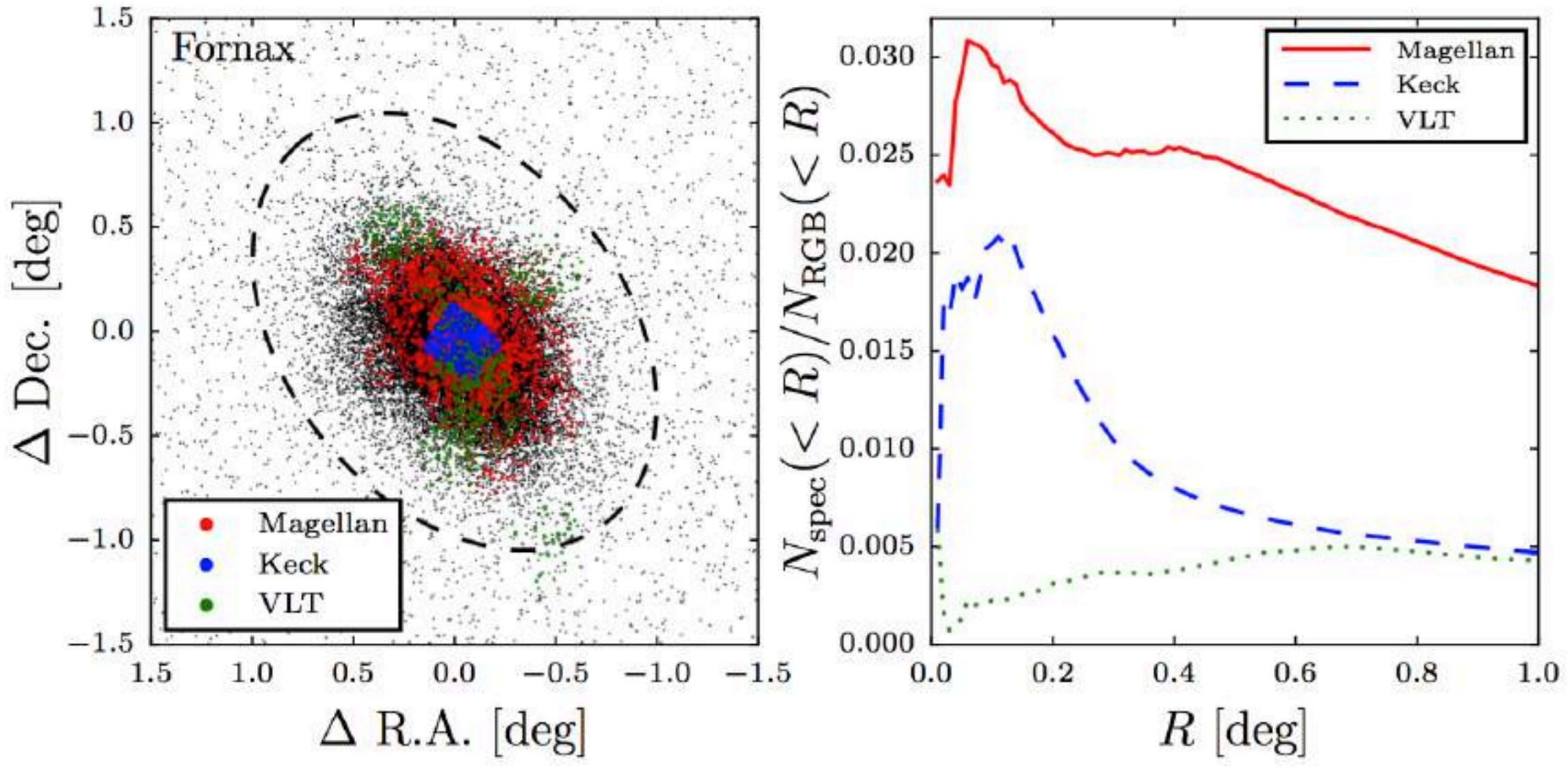

**Figure 36: Examples of observational selection bias in currently available spectroscopic samples for individual dwarf galaxies. *Left:* Spatial distribution of candidate red giant branch (RGB) stars with V ≤ 23 mag (black points) in the Fornax dwarf spheroidal (Bate et al. 2015), compared with the distribution of stars with published spectroscopic measurements from surveys with VLT/FLAMES (Battaglia et al. 2006), Magellan/MMFS (Walker et al. 2009b) and Keck/Deimos (Kirby et al. 2010). The dashed ellipse represents the "tidal" radius from the best-fitting King model to RGB candidates (Bate et al. 2015). *Right:* Ratio of number of spectroscopically observed stars to RGB candidates within an angular separation *R* from the center of Fornax.**

It is not just the size of the datasets that MSE will impact. Given the modest fields of view (≤ 0.5 degree) that are available for conducting multi-object spectroscopy at the largest-aperture telescopes available today, nearly all published spectroscopic data sets for Galactic satellites suffer from spatial incompleteness; more specifically, selection biases that leave outer regions either undersampled or neglected altogether. Even though outer regions are of keen interest for tracing stellar population gradients (Harbeck et al. 2001, McConnachie et al. 2007), investigating the outer structure of dark matter halos (Walker et al. 2007), identifying kinematic signatures of tidal disruption (Muñoz et al. 2006) as well as evolution of internal halo structure (El-Badry et al. 2015) and even measuring systemic proper motions (Kaplinghat & Strigari 2008; Walker et al. 2008), the relatively low fractions of bona fide members at large radius has made thorough observations of these regions prohibitively expensive.

Figure 36 demonstrates the degree to which this problem afflicts even Fornax, for which independent surveys with VLT/FLAMES, Magellan/MMFS and Keck/Deimos have delivered the richest available spectroscopic data sets for any Local Group dwarf galaxy (N ~ 3000 members; Battaglia et al. 2006; Walker et al. 2009b; Kirby et al. 2010). The left-hand panel displays the spatial distribution of red giant branch (RGB) candidates with V ≤ 23 (Bate et al. 2015), compared to distributions of spectroscopically observed stars from each survey. The right-hand panel shows how the number of spectroscopically observed stars within a given separation from Fornax's center compares to the number of RGB candidates. Figure 36 makes clear that even the "richest" existing data sets are woefully deficient: not only is the fraction of spectroscopically-observed RGB candidates less than 3% within Fornax's estimated tidal radius, but < 10% of the spectroscopically observed stars lie beyond one half-light radius (~ 0.25 degrees) from the center of Fornax. This selection bias propagates into biased estimates of chemodynamical quantities like velocity dispersion and mean metallicity and chemical abundance patterns, all of which exhibit radial gradients out to the limits of current samples. Furthermore, the centrally-biased spatial sampling can entirely miss important features that become apparent only at large radius (e.g., diffuse tidal debris that can be detected out to ~ 10 half-light radii from the center; Muñoz et al. 2006).

In contrast to the current situation, MSE's wide field will enable the first spatially complete surveys of the Milky Way's relatively large ($0.1 \leq R_{half} \leq 1$ degree) satellites. MSE's unprecedented capabilities will be particularly well-suited to observing even the most enigmatic Galactic satellites like Crater 2 which, albeit relatively luminous at $M_V \sim -8$, remained hidden until this year due to its large size ($R_{half} \sim 0.5$ degree) and correspondingly low surface brightness (31 mag/arcsec$^2$; Torrealba et al. 2016).

How does this improvement in the size and completeness of datasets help studies of the nearest dwarf galaxies? One source of keen interest in these spectroscopic samples is the information they convey about the dark matter content of dwarf galaxies (see also Chapter 4). Such information is crucial for investigations ranging from tests of cosmological models (e.g., those built on "cold" vs "warm" vs "self-interacting" dark matter) as well as interpreting (non-) detections of dark matter decay and/or annihilation. In order to gauge the effects of sample size on inferences about dark matter, we compare results from dynamical analyses of three mock data sets consisting of $N = 10^2$, $N = 10^3$ and $N = 10^4$ stars. In each case the artificial sample is drawn from a phase-space distribution function describing a stellar population that follows a Plummer (1911) surface brightness profile and traces a gravitational potential dominated by a Navarro, Frenk & White (1997) dark matter halo. The analysis uses standard Bayesian procedures to fit simultaneously for the velocity anisotropy, surface brightness and dark matter density profiles (a by-product is specification of the line-of-sight velocity dispersion profile), similar to the procedures described by Geringer-Sameth et al. (2015) and Bonnivard et al. (2015).

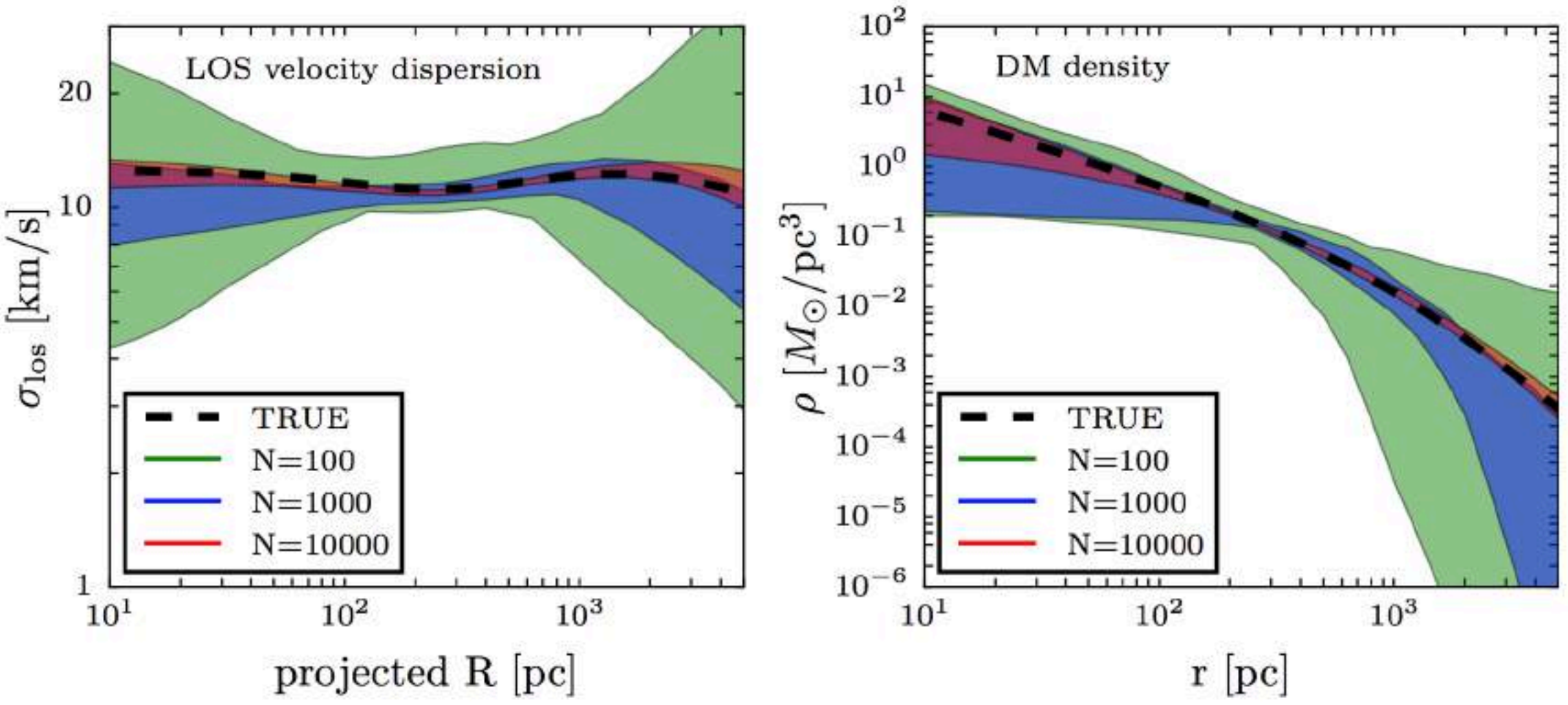

**Figure 37: Recovery of intrinsic line-of-sight velocity dispersion (left) and dark matter density (right) profiles as a function of spectroscopic sample size. Shaded regions represent 95% credible intervals from a standard analysis (based on the Jeans equation) of mock data sets consisting of line of sight (LOS) velocities for *N* = $10^2$, $10^3$ and $10^4$ stars (median velocity error 2 kms$^{-1}$), generated from an equilibrium dynamical model for which true profiles are known (thick black lines).**

Figure 37 compares the resulting inferences for velocity dispersion and dark matter density profiles, displaying bands that enclose 95% credible intervals in each case. Given the samples that MSE can provide, inferences about kinematics and dark matter content of dwarf galaxies will become dramatically more precise. We will infer the dark matter densities of ultrafaint satellites with precision similar to what is achieved today only for the most luminous classical dwarfs, for which we will infer density profiles with unprecedented precision from pc to kpc scales. This improvement will render *dynamical analyses limited by systematics (e.g., triaxiality, non- equilibrium kinematics) instead of statistics*.

As a practical illustration of the sheer power of MSE for chemodynamical studies of nearby dwarf galaxies, we consider NGC 6822, one of the nearest dwarf irregular galaxies at a distance of 500 kpc. It is one of the more intriguing targets for detailed study because of ongoing disturbances in its HI velocity field and very active star formation. There is some evidence for young stellar populations associated with infalling HI clouds, and for deviations from circular disk rotation. However, the large angular scale of the system (~ 1 degree across) and the likelihood that the substructures are represented by only a small fraction of the stars means that the system remains poorly understood, even with a small (but steadily growing) sample of stellar spectra from VLT/FLAMES and Keck/DEIMOS.

Figure 38 shows the spatial distribution of red giant branch candidates in NGC6822 based on CFHT/MegaCam imaging, compared to the MSE field of view. The number of potential targets in NGC6822 available for spectroscopy with MSE dwarfs the capabilities of existing multi-object spectrographs on 8-10m-class telescopes. For instance:

- the number of potential “red star” targets (i.e., stars with ages > 500 Myr, most of which are likely older than 1 Gyr) is:

- ~2500 stars *above* the TRGB, including ~$10^3$ carbon stars (18 < I < 19.5).
- ~7500 stars within 0.5 mag of the TRGB (19.5 < I < 20)
- ~30000 stars between 0.5 and 1.5 mag of the TRGB (20 < I < 21).

- the number of potential "blue star" targets is:
  - ~1000 stars with ages < 50 Myr (15.5 < g < 18.5)
  - ~1000 stars with ages 50 – 200 Myr (18.5 < g < 20.5)
  - ~2000 stars with ages 200 – 400 Myr (20.5 < g < 21.5)

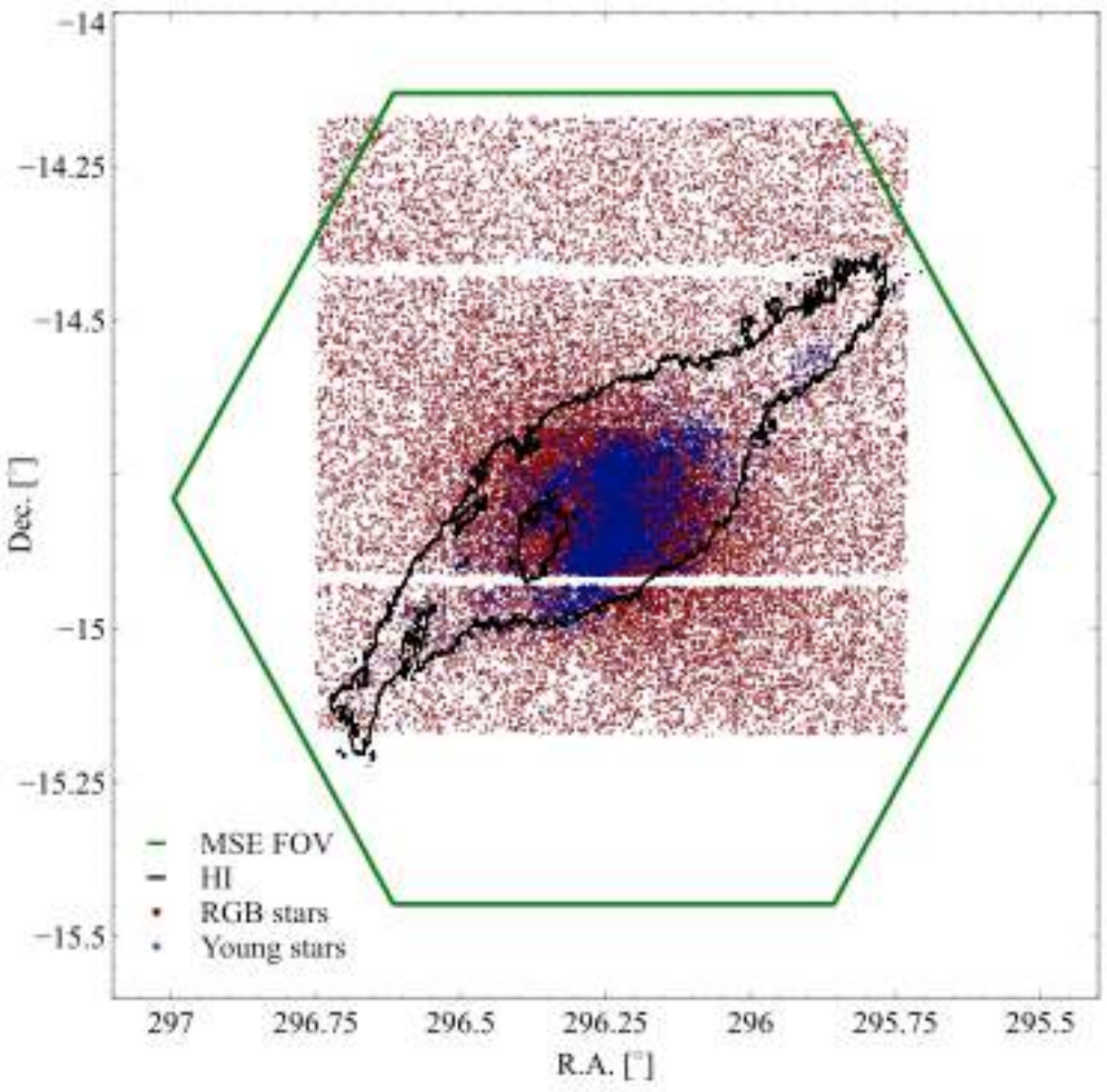

**Figure 38: Spatial distribution of red giant branch stars (red dots) in the barred irregular galaxy NGC6822 from CFHT/MegaCam observations (see Higgs et al. 2016). The contours show the boundary of the HI disk from de Blok & Walter (2006). The MSE field of view is shown for comparison; clearly, MSE is ideal for wide field spectroscopic studies of Local Group dwarf galaxies.**

With these stellar densities and spatial distributions, it would be possible, for example, to measure radial velocities accurate to better than 5 $kms^{-1}$ for every AGB and red supergiant star, and nearly all RGB stars within 0.5 mag of the TRGB, in just one night on MSE, at medium resolution and with SNR ~ 10 – 20. A single pointing, with three different configurations of the 3200 fibres, each observed for roughly a 1 hour, would provide a complete census of these bright red stars. With samples of > $10^4$ stars, sub-populations representative of just 1% of the total number of stars would become detectable with high significance. To obtain spectra further down the RGB would take more time, but not unimaginably so. For instance, at I = 21, SNR ~ 10 per hour would be achieved in the red/near- IR region, meaning that homogeneous, high-quality kinematic and metallicity results for ~ 30 000 stars could be achieved in only 10 nights: i.e., 6 – 8 hours per setup, with 10 configurations needed to reach every target. If the shallower survey were repeated multiple times over a period of several years, the sensitivity and multiplexing would allow unprecedented studies of variability and evolutionary changes for stars in the late phases of evolution. Similarly, intermediate-resolution radial velocities could be obtained for every main-sequence star in the galaxy younger than ~ 400 Myr in just ~ 3 nights at SNR ~ 10 – 15 for g < 21.5.

### 2.4.2 The chemodynamical deconstruction of the nearest L* galaxy

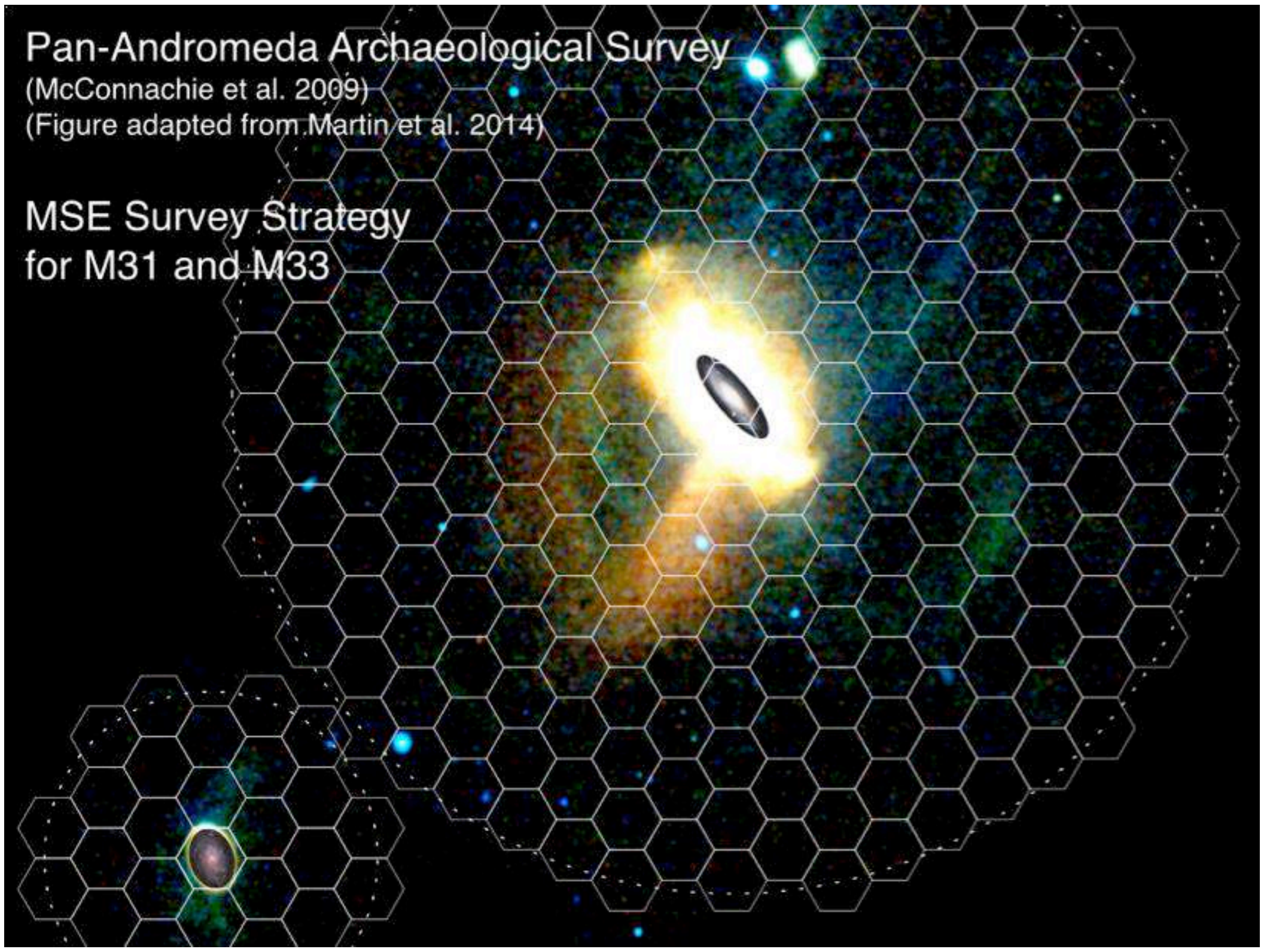

**Figure 39: Spatial distribution of candidate red giant branch stars in the environs of M31 and M33, as identified from colour-magnitude cuts from the Pan-Andromeda Archaeological Survey. Dashed circles highlight maximum projected radii of 150 kpc and 50kpc from M31 and M33 respectively. The tiling strategy for an MSE survey of this region is overlaid.**

With MSE, we will be able to conduct a systematic survey of the kinematics and chemistry of Local Group galaxies within 1 Mpc with a level of detail that has never been achieved before and cannot be achieved anywhere else. On the scale of (large) spiral galaxies, Gaia is gathering unprecedented astrometric measurements with the accuracies needed to produce a stereoscopic and kinematic census of about one billion stars in our Galaxy. This dramatic advance in our ability to map the six-dimensional phase-space structure of a galaxy will likely usher in a "Golden Age" of Milky Way research. The consequences of this endeavor for MSE have been discussed in preceeding sections.

Gaia will provide a very detailed view of the structure and dynamics of our Galaxy and this information will have to be put into context by comparing the results with observations of other galaxies. Without this essential complementary information, any conclusions that we may draw about the process of galaxy formation from observations of the Milky Way would be premature. It is in this context that the capability of MSE to provide the ultimate chemodynamical decomposition of the two closest Local Group spirals (M31 and M33) is most important. These galaxies represent the obvious stepping stones between our highly detailed description of the Milky Way, and the low-resolution studies of more distant galaxies, in the realm of a few Mpc and beyond, where we can obtain significant samples of galaxies (as a function of galaxy type, environment, mass, etc.). By ~2020, M31 and M33 will have been studied in detail by many existing and imminent instruments/observatories. Nevertheless, a gaping chasm in our understanding of these structures will remain: i.e., their kinematics and chemistry from resolved stellar populations. We note that Subaru/PFS intends to observe approximately 66 square degrees of the halo of M31 (50 individual exposures; Takada et al. 2014). However, it is clear from inspection of Figure 39, discussed below, that a comprehensive understanding of the chemodynamics of M31 and M33 require full spatial coverage and high completeness, especially in the target-rich inner regions of these galaxies.

MSE will make a pivotal contribution to the subject by undertaking a large spectroscopic survey of these two Local Group galaxies. The goal would be to obtain a statistically significant sample of stars belonging to their constituent structural components. The scientific aim of this "Galactic archaeology" theme is to understand the galactic formation process, which is clearly an open-ended endeavour. Our power to address the key issues (e.g., thick disk formation, halo formation, streams, chemical enrichment history), depend on the spectral resolution that the survey affords.

Figure 39 shows the spatial distribution of candidate RGB stars in the environs of M31 and M33, as identified by colour-magnitude selection from 400 square degrees of contiguous *gi* imaging with CFHT/MegaCam as part of PAndAS. Colour-coding corresponds to the colour of the RGB stars, such that redder RGB stars (likely higher metallicity) appear red, and bluer RGB stars (likely metal-poor) appear blue. Dashed circles correspond to maximum projected radii of 50, 100 and 150 kpc from M31, and 50 kpc from M33. PAndAS resolves point sources at the distance of M31 (D ~ 780 kpc; McConnachie et al. 2005) to g $\simeq$ 25.5 and i $\simeq$ 24.5 at SNR ~ 10. More than $10^7$ stellar sources are shown in Figure 39. The typical colour of tip of the RGB stars in M31 is (g − i) ~ 1.3 so $g_{TRGB}$ ~ 22.5. PAndAS therefore reaches to (nearly) the horizontal branch level, providing photometry of sufficient depth that potential spectroscopic targets for a 10m facility could be selected.

In Figure 39, the effective surface brightness of the faintest visible features is of order 33 mag arcsec$^{-2}$. This corresponds literally to a few RGB stars per square degree. Note that the disk of the Milky Way is located to the North so there is increasing contamination in the colour-magnitude of the RGB locus by foreground dwarfs; the reddest RGB stars are particularly affected by this source of contamination. Young, blue stellar populations — and even intermediate-age populations such as asymptotic giant branch (AGB) stars — are not present in the outer regions of M31 in any significant numbers. Thus, any spectroscopic study of the outer regions of the M31 halo will necessarily concentrate on the older, evolved, RGB population. For this reason, we consider separately surveys of the *outer halo* (characterized by a low surface density of evolved, giant star candidates) and the *inner galaxy* (with a high surface density of targets from a mixture of stellar populations):

- **An outer halo survey of M31/M33:** This program aims at obtaining a *complete, magnitude-limited, spectroscopic census* of every star in the outer regions (40–150 kpc) of an $L_*$ galaxy halo to provide complete kinematics for every star, supplemented by metallicity estimates for most stars and alpha-abundances for the brightest ones.

Ultimately, such a survey will allow us to deconstruct a nearby galactic halo into its accreted "building blocks". Such a survey will provide the ultimate testbed of the hierarchical formation of L* galaxies and further yield the optimal data set to constrain the dark matter content of M31 and M33;

- **Faint stellar populations in the inner regions of M31/M33:** This program has as its focus the study of the disk/thick-disk/halo transition region to measure the extent of the disks, characterize the relationship between these components, and to determine the role of mergers in the disk and inner halo evolution.

The richness of this dataset can be anticipated by considering what has already been achieved for M31 and M33 using Keck/DEIMOS (hundreds of pointings in the vicinity of M31 targeting RGB star candidates, for a total of > 10 000 stars, primarily by the SPLASH and PAndAS collaborations; e.g., Dorman et al. 2012, Collins et al. 2011, and references therein). Keck/DEIMOS has a 5 x 16 arcmin field of view and is able to observe of order 100 targets per pointing, making it the most well suited large aperture spectrograph currently available for studies of M31.

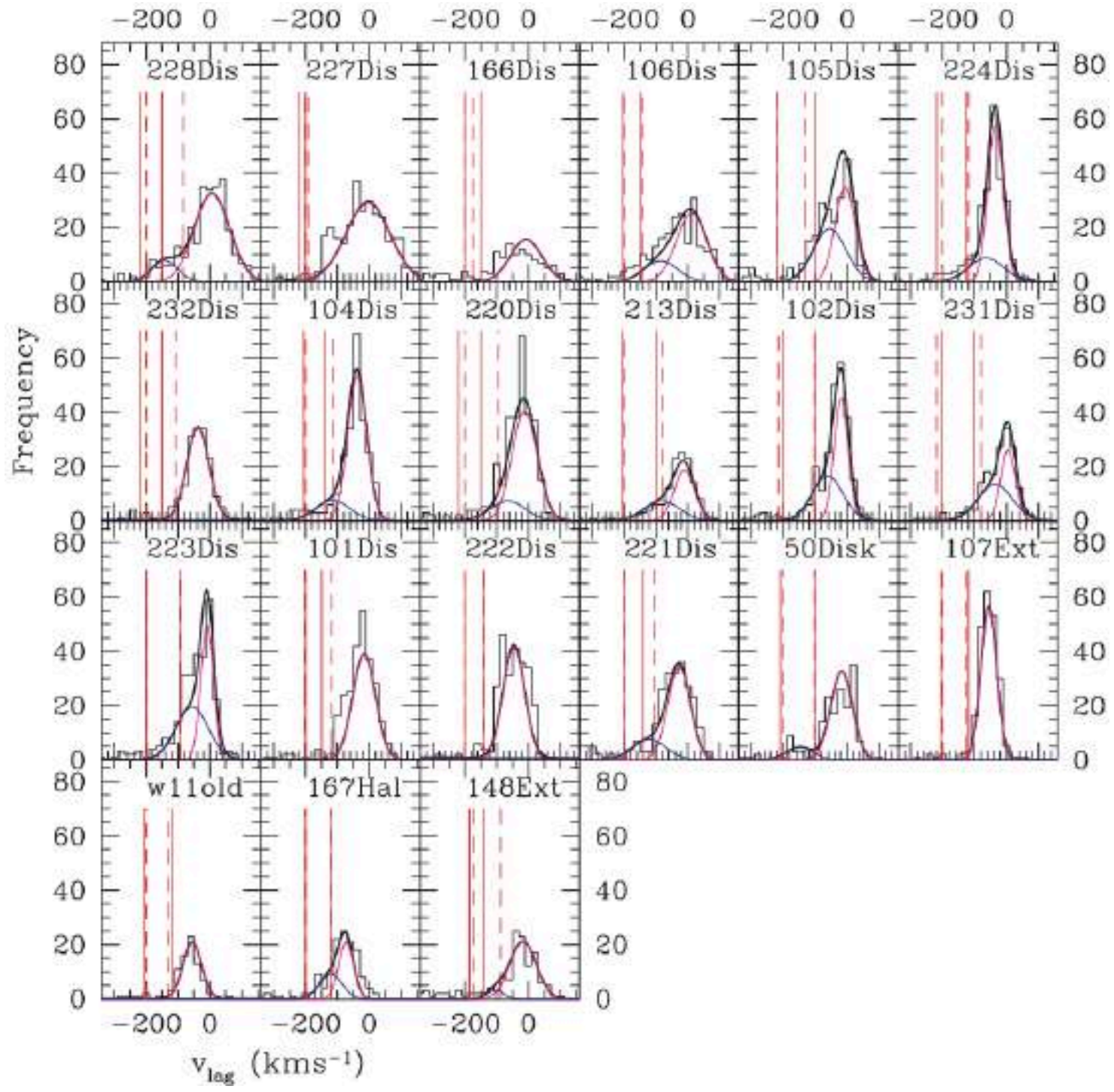

**Figure 40: A selection of fields from Keck/DEIMOS spectroscopic studies of the M31 disk region along the South West major-axis of M31, shown in order of increasing (projected) radius. Gaussian fits indicating the thin and (where applicable) thick discs are shown as magenta and blue curves, respectively. Dashed and solid vertical lines indicate selection criteria for the kinematic identification of thick disk stars. *(Figure from Collins et al. 2011).***

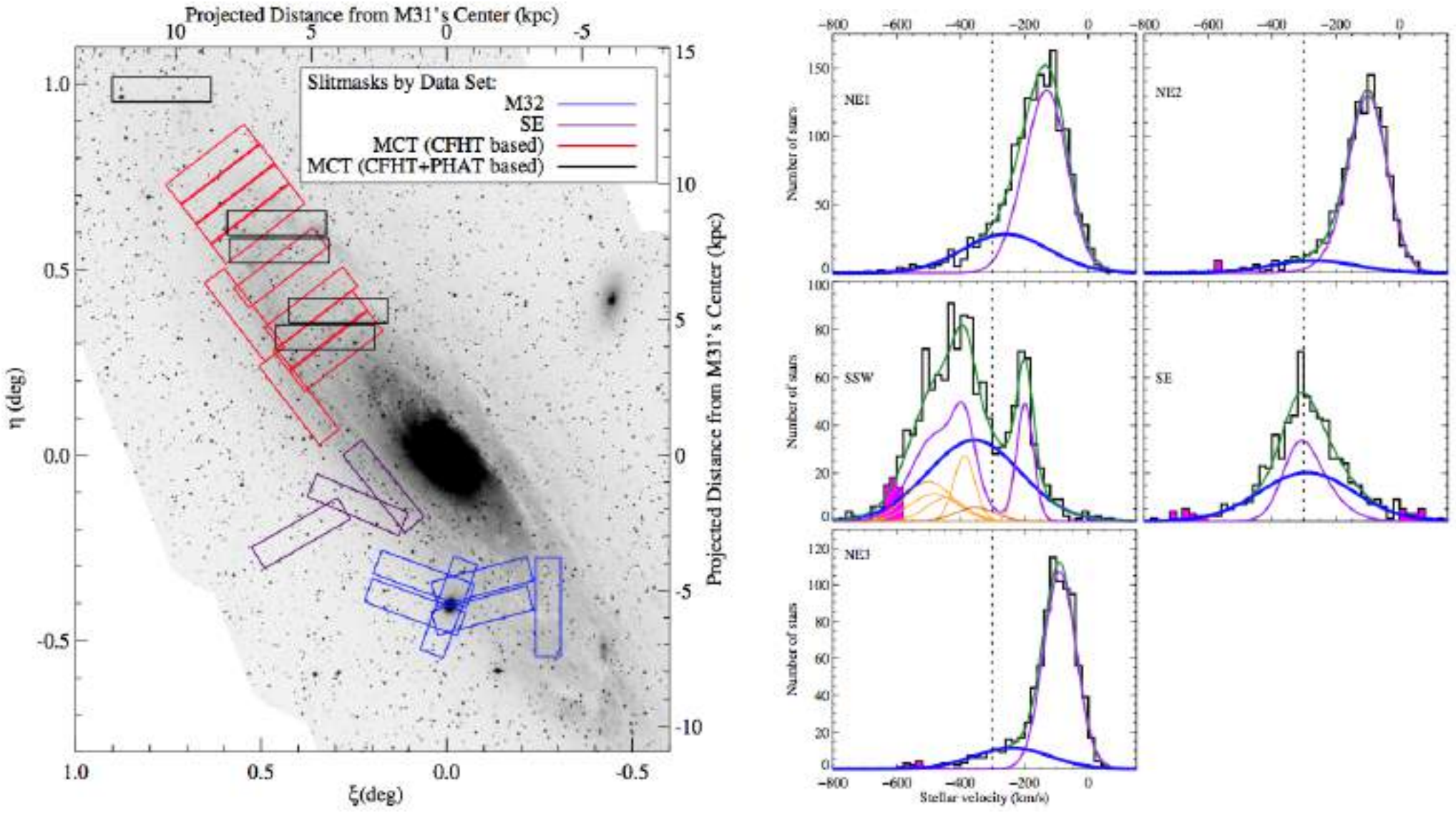

**Figure 41: Left panel: The positions of twenty-four Keck/DEIMOS multiobject slitmasks from the SPLASH survey of M31 overlaid on the Choi et al. (2002) KPNO Burrell Schmidt B-band mosaic image of M31. Note that this image covers approximately 1.5 degrees on a side, to be compared with the ~25 x 25 degree image shown in Figure 39. Right panels: Maximum likelihood fits of a kinematically hot spheroid to each of five sub-regions of the survey, after excluding the velocity range encompassing the giant stellar stream (Ibata et al. 2001) and its associated tidal debris (shaded pink). Violet lines show the cumulative region cold component; blue show the best-fit spheroid Gaussian; green show the sum of these two components. The dashed lines show the systemic velocity of M31 relative to the MW. Individual subregion cold components are shown in orange in the SSW region panel, but left out of the other panels for clarity. The complexity of the disk region of M31 is clear. *(Figures from Dorman et al. 2012).***

Figure 40 and Figure 41 show velocity histograms for RGB stars in the vicinity of the disk of M31, from spectroscopic studies of this region by Collins et al. (2011) and Dorman et al. (2012). Figure 40 shows fits of two Gaussians to a large number of fields, that Collins et al. (2011) suggest is evidence of a thick disk, with a velocity dispersion that is nearly 50% higher than for the thin disk, and with a scale height that is approximately 3 times larger than the thin disk. A detailed dynamical analysis of the relation between the thin and thick disks based on individual stars has previously only been possible for the Milky Way, but here the perspective is very different and, as a result, very complementary.

Dorman et al. (2012) examine the disk of M31 and considerably smaller radius than Collins et al. (2011), and show that there are significant contributions from the inner spheroid of M31. The locations of their fields and the resulting velocity histograms are shown in Figure 41; it is clear that this region is complex to analyse and maximum likelihood analyses suggest that multiple components are required per field. They find evidence that the spheroid rotates, and more closely resembles that of an elliptical galaxy than a typical spiral galaxy bulge.

In much the same way as discussed previously for the dwarf galaxies, all of these analyses are based on kinematic subsamples that are very sparse and highly incomplete. More than 10 million RGB candidates are present in the PAndAS map of M31 shown in Figure 39, and yet more than a decade of research using Keck/DEIMOS has barely sampled 1% of the possible candidates. It is interesting to note that, even with the order of magnitudes difference in the stellar density of M31 in the inner regions compared to the diffuse and distant regions shown in Figure 39, MSE is absolutely necessary: at small radius, the multiplexing of M31 members is extreme, and in the outskirts, the contamination from foreground stars is extreme, even with reasonable preselection.

Dorman et al. (2012) note that "the literature is full of vocabulary such as 'bulge,' 'spheroid,' 'inner spheroid,' 'outer spheroid,' 'disk,' 'thin disk,' 'thick disk,' 'extended disk,' and so on. There is not yet a consensus on the best combination of these nouns to represent M31". Given the extreme dynamical complexity of this galaxy, resulting at least in part from a vigorous history of galaxy mergers, the next major paradigm shift will result from a coherent, holistic view of its kinematics and chemistry across its entire spatial extent, and which MSE is uniquely able to provide.

# Chapter 3 Linking galaxies to the large scale structure of the Universe

## 3.1 Chapter synopsis

- *Connecting the properties of galaxies to the surrounding large scale structure;*
- *Connecting galaxy diversity in the local Universe to their evolution across cosmic time.*

*Science Reference Observations Linking Galaxies to the Large Scale Structure of the Universe*

- *SRO-6: Galaxies and their environments in the nearby Universe*
- *SRO-7: The baryonic content and dark matter distribution of the nearest massive clusters*
- *SRO-8: Multi-scale clustering and the halo occupation function*
- *SRO-9: The chemical evolution of galaxies and AGN*

MSE will provide a breakthrough in extragalactic astronomy by linking the formation and evolution of galaxies to the surrounding large scale structure, across the full range of relevant spatial scales, from kiloparsecs through to megaparsecs. Within the ΛCDM paradigm, it is fundamental to understand how galaxies evolve and grow relative to the dark matter structure in which they are embedded. This necessitates mapping the distribution of stellar populations and supermassive black holes to the dark matter haloes and filamentary structure that dominate the mass density of the Universe, and to do so over all mass and spatial scales.

MSE is optimized for studying the evolution of galaxies, AGN and the environmental factors likely to influence their evolution. MSE will be the premier facility for unscrambling the non-linear regime (small scale clustering, mergers, groups, tendrils and filaments), the baryon regime (i.e., metallicity/chemical evolution), and the evolution and interplay of galaxies, AGN, and the IGM, incorporating environmental factors and the key energy production pathways. MSE will build upon the impressive legacy of current spectroscopic surveys in the near and far fields by going several magnitudes deeper, covering an order of magnitude larger sky area, and conducting both stellar population studies of the nearest galaxies and precision dynamical analyses across the luminosity function and across cosmic time.

In the local Universe, MSE will measure the dynamics and stellar populations of every baryonic structure in the Virgo and Coma clusters down to extremely low stellar mass. This will allow the physical association or otherwise of dwarf galaxies, nucleated dwarfs, globular clusters and ultra-compact dwarfs to be probed through chemodynamical analysis of unprecedented datasets (both in size and completeness). Quenching processes operating in the cluster environment can be examined for all systems at all radii, and in particular analysis of the stellar populations of thousands of low mass galaxies all in the same cluster will give direct access to the relative roles of environmentally driven processes and internally driven processes across the luminosity function and to its faintest limits.

MSE will be able to conduct the definitive spectroscopic survey of the low redshift Universe for decades to come by obtaining spectra for every $z < 0.2$ galaxy in a cosmologically-relevant volume spanning 3200 deg$^2$, allowing every halo with mass $M_{halo}>10^{12}M_{\odot}$ to be detected with four or more galaxies. Follow up in select regions will result in a sample covering 100 deg$^2$ spanning four decades in halo mass, accounting for the majority of stellar mass in the low-redshift Universe. This will allow the measurement of the shape of the stellar mass function to the scale of the Local Group dwarf galaxies, over a large volume and a range of environments.

Uniquely, the broad wavelength range and extreme sensitivity of MSE allows extrapolation of this "low redshift" survey concept out to beyond cosmic noon. An ambitious set of photo-z selected survey cubes will allow MSE to measure the build up of large scale structure, stellar mass, halo occupation and star formation out to redshifts much greater than 2. By targeting $(300 \text{ Mpc/h})^3$ boxes, each cube will measure "Universal" values for an array of potential experiments. These survey volumes will trace the transition from merger-dominated spheroid formation to the growth of disks, covering the peak in star-formation and merger activity. This combination of depth, area and photo-z selection is not possible without a combination of future imaging surveys and MSE. As such, MSE will be able to produce the definitive survey of structure, halos and galaxy evolution over 12 billion years.

*MSE will be unparalleled in its ability to connect the properties of galaxies to their surrounding large scale structure. It will link the properties and diversity of galaxies observed in the local universe to the evolutionary processes that have occurred across cosmic time. In so doing, MSE will fulfill three distinct roles in the 2025+ era for extragalactic astrophysics: dedicated science programs, coordinated survey programs and feeder programs for the VLOTs. All three roles are critical, transformational, and will lead to major advancements in the area of structure formation, galaxy evolution, AGN physics, our understanding of the IGM, and the underlying dark matter distribution.*

## 3.2 Extragalactic surveys with MSE

### 3.2.1 Galaxy evolution and the link to large scale structure

The basic model of galaxy formation has been in place for more than twenty years (White & Frenk 1991). Dark matter dominates the evolution of structure in the Universe, and it is well described by the gravitational growth of fluctuations observed in the Cosmic Microwave Background. This gives rise to the cosmic web of filaments and clusters we see in the galaxy distribution. The clustering, spatial structure, and redshift evolution of these galaxies is sensitive to the underlying cosmological model, the very nature of dark matter, and the highly complex mechanism of galaxy formation.

Whereas radiative cooling of baryons leads to efficient star formation, various mechanisms must work to reheat, or expel, this gas in order to match the low efficiency of star formation observed today. Large spectroscopic surveys, covering cosmologically relevant volumes with homogeneous selection and good calibration, have demonstrated that the distribution and growth rate of galaxies is largely decoupled from that of dark matter haloes (Behroozi, Wechsler & Conroy 2013; see Figure 42). This is likely due to the highly nonlinear nature of cooling, star formation and heating processes such as photoionization, supernovae feedback, stellar winds and energy output from supermassive black holes. These phenomena and others operate with different strengths, over different spatial scales, and their relative effectiveness evolves with time.

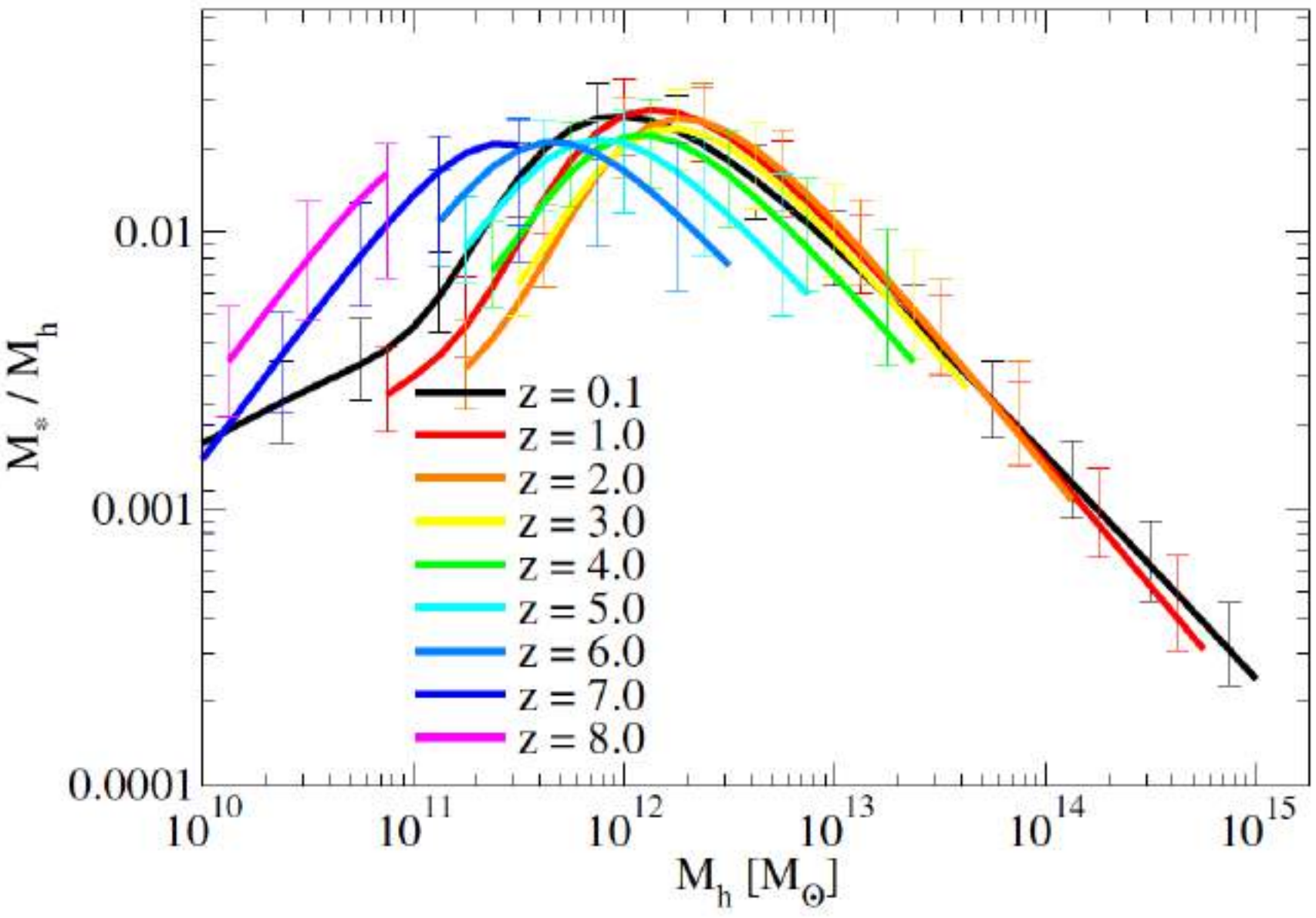

**Figure 42: The evolution of stellar mass fractions (stellar-to-halo mass ratio) as a function of halo mass. The complex, non-linear nature of this plot shows the growth of stellar mass is largely decoupled from that of the dark matter *(Figure from Behroozi et al. 2013)*.**

While the broad picture of galaxy formation and evolution is coming into focus, it is clear that we are still far from understanding how galaxies end up with the masses, structures, dynamics and stellar populations that they have today. Linking the luminous mass to dark matter halos is critical, since many of the outstanding problems can be grouped broadly in two related themes, namely, *the abundance and structure of dark matter halos, particularly at low mass scales*, and *star formation, quenching, and the roles of environment and feedback.*

A general class of models, collectively known as the "halo model", has been devised to attack galaxy formation from a statistical viewpoint. Among these models, the conditional luminosity function (CLF) and stellar mass function (SMF), both as a function of halo mass, and the halo occupation distribution (HOD), are the most popular. Briefly, these approaches employ a few simple physical quantities to make predictions for the clustering properties, luminosity (or stellar mass) functions of galaxies, as well as galaxy-galaxy lensing results. Models that simultaneously reproduce these observables are then representative of the underlying association between galaxies and dark matter halos. Thus, the halo model provides a powerful, statistical description of the relationship between the halo mass and its baryonic content.

Applied to SDSS data at z ~ 0, techniques like HOD modeling (Zhang et al. 2007) and abundance matching (Yang et al. 2008), have demonstrates that galaxy formation efficiency peaks in halos of $M_h \sim 10^{12} M_{\odot}$. At higher redshift, the constraints are poorer, as they either rely on either photometric redshifts (e.g., Wake et al. 2011; Leauthaud et al. 2012; Coupon et al. 2011), or spectroscopic surveys over small areas (e.g., Abbas et al. 2010; Knobel et al. 2009). Nevertheless, these can be used to trace the evolution of star formation efficiency since about z ~1 over a wide range of masses. The cosmic star formation rate is observed to be rising with time between z ~ 8 and z ~ 3, where it remains high and roughly constant until z ~ 1 (Hopkins &

Beacom 2006; Seymour et al. 2008; Kistler et al. 2009; Magnelli et al. 2011; see Figure 43). Since ΛCDM models predict a declining infall rate at all epochs, this observation has proved difficult to explain theoretically (e.g., Weinmann et al. 2011; Krumholz & Dekel 2011). Better constraints on halo occupation at these redshifts are needed, and these require the deep, spectroscopic surveys that MSE will provide.

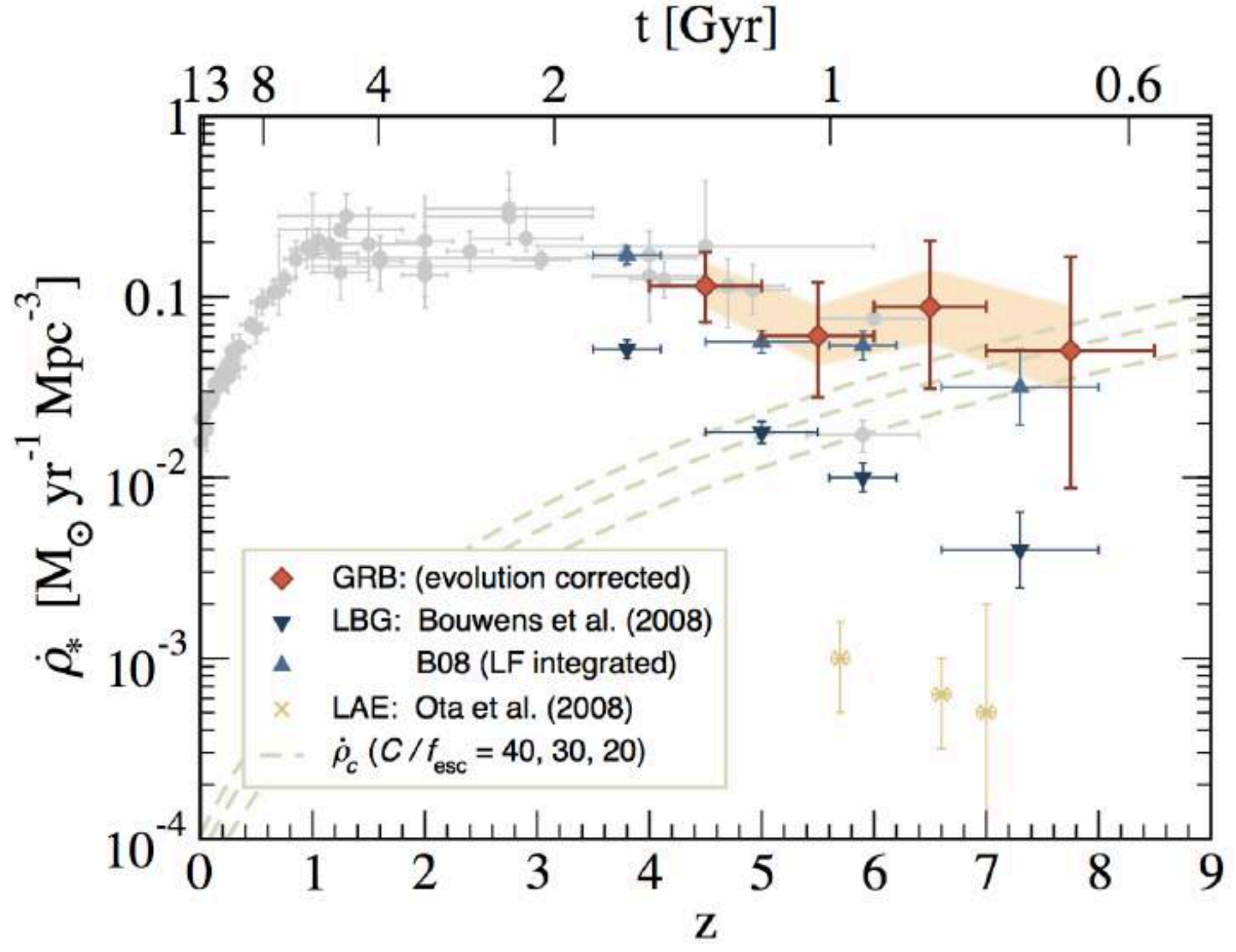

**Figure 43: The cosmic star formation history. Shown are the data compiled in Hopkins & Beacom (2006) (light circles) and contributions from Lyα Emitters (LAE) (Ota et al. 2008). Recent luminous blue galaxy data is shown for two UV luminosity function integrations: down to 0.2L*$_{z=3}$ (down triangles; as given in Bouwens et al. 2008) and complete (up triangles). *Swift* gamma-ray burst inferred rates are diamonds, with the shaded band showing the range of values resulting from varying the evolutionary parameter be- tween α = 0.6 − 1.8. Also shown is the critical star formation density from Madau et al. (1999; dashed lines, top to bottom) *(Figure from Kistler et al. 2009).***

A key component of HOD models, and many *ab initio* galaxy formation models (Bower et al. 2006), is that galaxies that are central in their dark matter halo evolve differently from those that are satellites. For central galaxies that are dominant in their dark matter halo, their evolutionary history is strongly correlated with their total stellar mass at the epoch of observation (e.g., Brinchmann et al. 2004, Salim et al. 2007, Noeske et al. 2007, Daddi et al. 2007, Rodighiero et al. 2011, Whitaker et al. 2012, Sobral et al. 2014, Speagle et al. 2014; see Figure 44). However, satellite galaxies of a given stellar mass appear to have ceased forming stars at an earlier epoch, and this leads to the well-established correlation between galaxy properties (age, star formation rate, and perhaps morphology) and environment (e.g., Peng et al. 2010) that is observed locally (e.g., Peng et al. 2010; Weinmann et al. 2010) and at higher redshifts (e.g., Poggianti et al. 2006; Cooper et al. 2006; McGee et al. 2011; Cooper et al. 2010; Balogh et al. 2011).

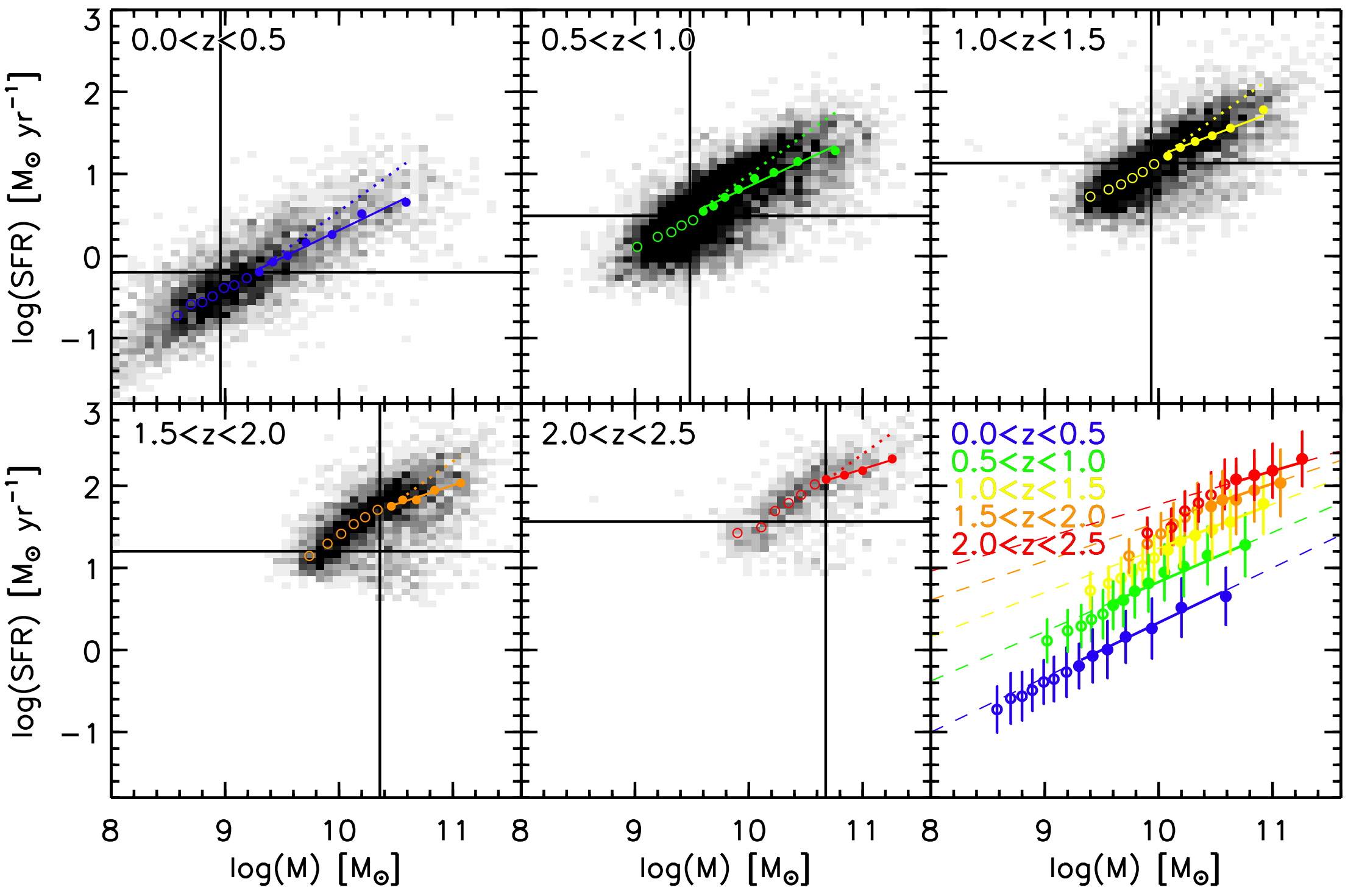

**Figure 44: SFR mass sequence for star-forming galaxies from z = 0 to z = 2.5. Dotted lines show linear fits, solid lines show non-linear slopes. The running medians and scatter are color-coded by redshift, with a power-law fit above the mass and SFR completeness limits (solid lines in bottom, right panel)** ***(Figure from Whitaker et al. 2012).***

Several different processes are thought to contribute to the central – satellite distinction. For example, the amount of cold gas in galaxy disks is insufficient to sustain star formation at observed rates for a Hubble time, and it is necessary that galaxies be "fueled" by gas on larger scales. For central galaxies that are dominant in their halo, this could come either through cooling of a hydrostatic hot halo, or through "cold flows" i.e., cold, clumpy gas accreted for example along filaments. Both modes are likely important, at different epochs and for different mass galaxies (Birnboim & Dekel 2003; Keres et al. 2005; Dekel et al. 2009). However, satellites are not expected to have access to as much of this material, leading to a rapid depletion of fuel for star formation (Larson et al. 1980; Cattaneo et al. 2006).

Large-scale environment likely influences both the mass accretion rate (e.g., Wechsler et al. 2002; Maulbetsch et al. 2006; Gao & White 2007) and the way that gas is accreted (Cattaneo et al. 2008; Dekel et al. 2009). The evolution of most galaxy properties (e.g., luminosity, colour, morphology) is sensitive to the gravitational and hydrodynamic interactions between galaxies, and the details of these interactions are determined at least in part by the environment of the galaxies. These effects are generally complex, nonlinear and often subdominant to more local effects like feedback.

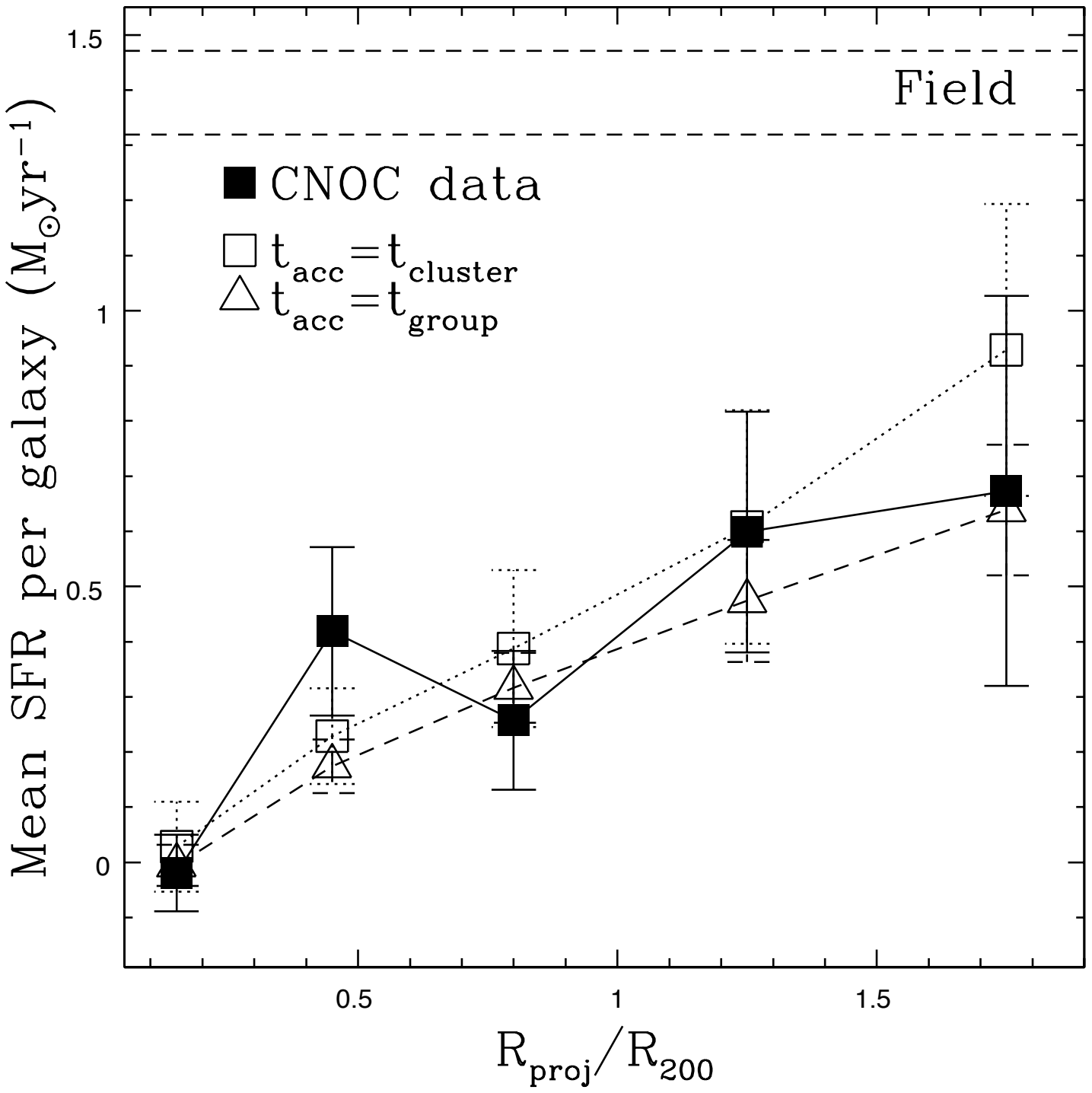

**Figure 45: The mean star formation rate (SFR) per galaxy as a function of projected radius for galaxies in the CNOC1 cluster sample (solid squares) compared with model predictions. The horizontal dashed lines correspond to the field SFR, bracketed by its 1-σ dispersion.** ***(Figure from Balogh et al. 2000).***

Distinguishing which of the multitude of process that affect galaxy evolution are dominant, and where, remains a challenge, particularly since most of these effects are likely a function of galaxy mass. Thus, *a complete understanding of galaxy evolution requires studying central and satellite galaxies across a range of scale in environments, from massive clusters (e.g., Park & Hwang 2009) to small-mass, galactic-scale halos (Ann et al. 2008)*. A correlation with large-scale environment might be expected because of the halo assembly bias seen in dark matter simulations (e.g., Maulbetsch et al. 2007; Gao & White 2007), although this effect has not been observed in the galaxy population at low redshift (Tinker et al. 2011). A correlation with location within a halo is expected, and observed, as a result of a correlation between accretion time and present-day location (e.g., Ellingson et al. 2001; Balogh et al. 2000; see Figure 45). Finally, independent correlations with local density could be related to tidal interactions and mergers between close pairs of galaxies (e.g., Lin et al. 2007; Ellison et al. 2008; Park & Choi 2009; Patton et al. 2011).

Spectroscopic observations of *millions* of galaxies are needed to disentangle the effects of environment from the other parameters driving galaxy evolution. Indeed, after many years of speculation based on small samples with poorly understood selection criteria, this field was transformed by the large, homogeneous survey data provided by the SDSS. The sample size and wavelength coverage allowed exploration of correlations with both environment and the stellar mass of the galaxy. This immediately demonstrated that galaxy properties (such as morphology, age, metallicity, etc.) depend strongly on their stellar mass. Many of the trends with environment reported previously turned out to be at least partly due to the fact that the stellar mass function depends on environment (Balogh et al. 2001, Baldry et al. 2006). After accounting for this, however, a strong environmental dependence remains: the fraction of "red sequence" galaxies (at fixed stellar mass) is larger in dense environments than in the general field (e.g., Baldry et al. 2006). In fact, the fraction of red galaxies is well described by a relatively simple function of just stellar mass and local density, as shown in Figure 46.

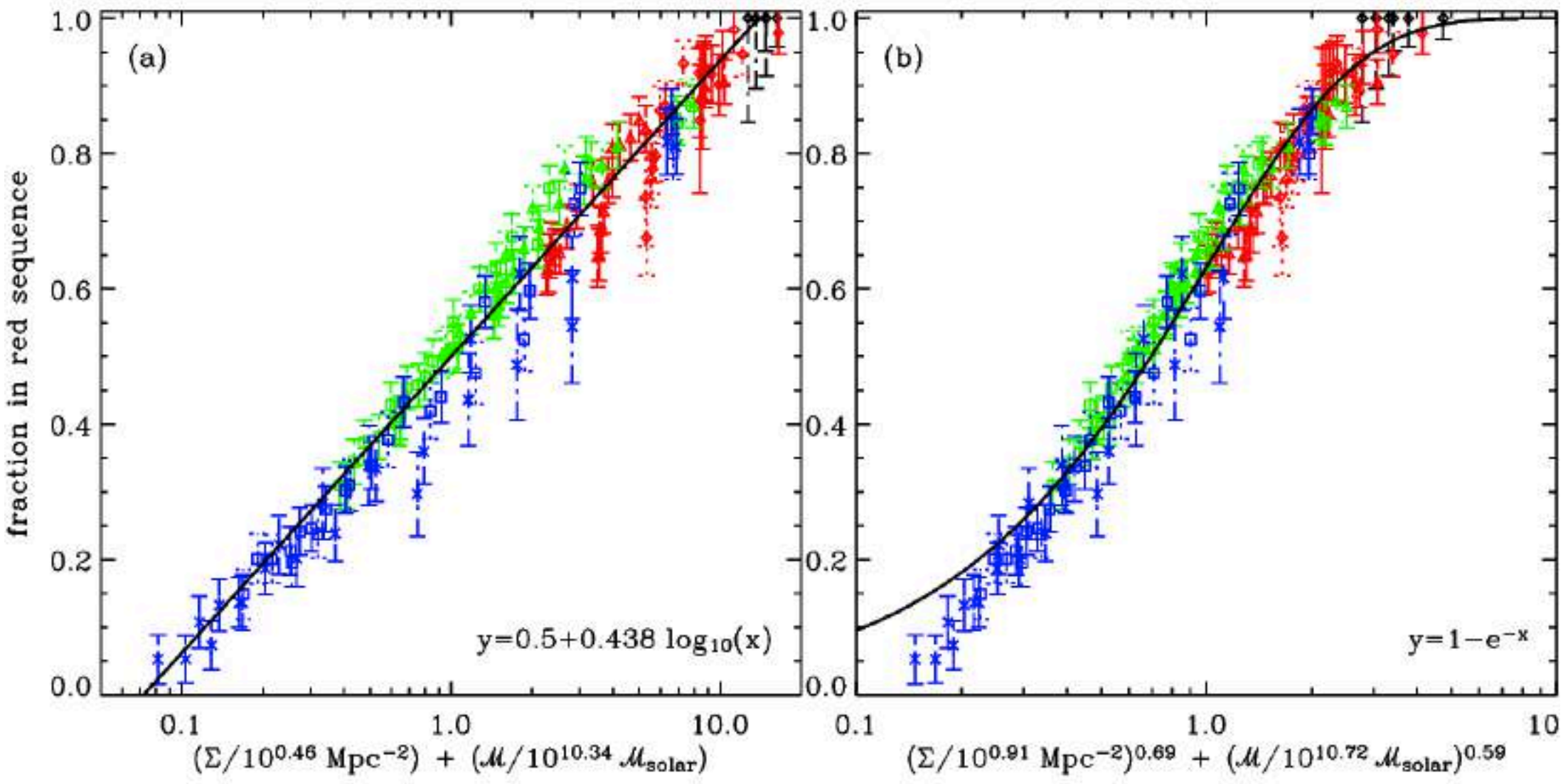

**Figure 46: The fraction of red (passive) galaxies is shown as a function of two different combinations of stellar mass, M, and local density, Σ. The symbols represent different mass ranges, from $10^9 < M/M_* < 10^{10}$ in blue to $M/M_* > 10^{10.8}$ in red. The red fraction depends on both environment and stellar mass, in a surprisingly simple way. *(Figure from Baldry et al. 2006).***

Since all galaxies must have been blue, star-forming systems at some point, the red fraction must be linked somehow to the decline, or quenching, of star formation in blue cloud galaxies. To understand the nature of this transformation, one must either identify galaxies in the process of moving between the two populations (e.g., in the "green valley"), and/or trace the redshift evolution of these correlations. Locally, the relatively low gas fractions and star formation rates, combined with low infall rates into groups and clusters, mean that the fraction of galaxies in the process of transformation today is probably quite small. This could be why observations have shown the properties of star-forming galaxies at $z < 0.2$ to be at most weakly dependent on environment (Balogh et al. 2004; Wolf et al. 2009; Vulcani et al. 2010; McGee et al. 2011; Wijesinghe et al. 2012). Thus, although these trends are now measured quite precisely, we still have very little insight into the physical mechanisms that might be driving any such transformation. A simple assumption is that satellite galaxies are deprived of the gaseous halos and cold flows that normally perpetuate star formation, an assumption that is at least partly supported by numerical simulations (e.g., Kawata & Mulchaey 2008; McCarthy et al. 2008; Cen 2011; Bahe et al. 2012). However, a simple implementation of this effect in *ab initio* galaxy formation models leads to a predicted environmental signature that is much too strong when compared with observations (e.g., Weinmann et al. 2006; 2012).

Peng et al. (2010, 2012) have taken the observation of Baldry et al. (2006) a few steps further and presented an especially attractive, simple model in which galaxy evolution is driven by two "quenching rates" that are independent of one another; one related to stellar mass and the other to local density. From this they are able to predict how the stellar mass functions of star–forming and passive galaxies will evolve to z ∼ 3. This is shown in Figure 47, where the points show the fraction of passive galaxies as a function of stellar mass for low and high density regions, and the curves show the predicted behavior of this quantity with redshift.

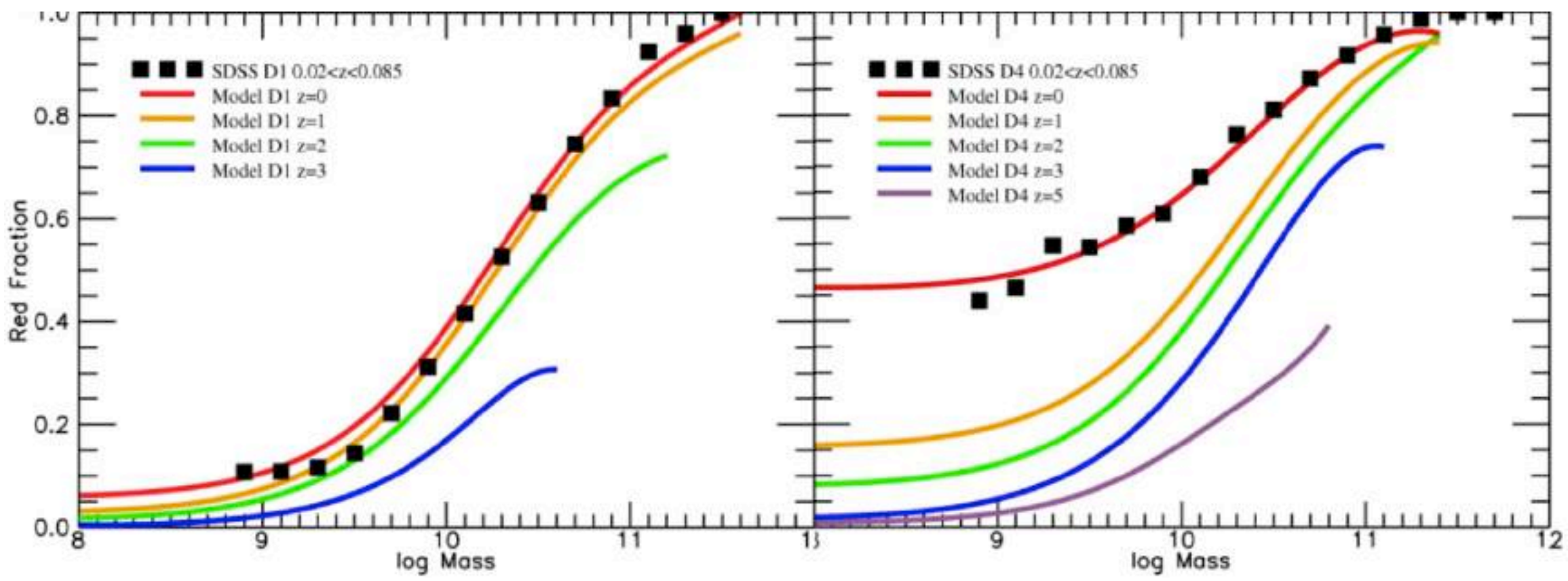

**Figure 47: The fraction of red (passive) galaxies as a function of stellar mass for the low and high density quartiles of the density field (left and right panels, respectively). The model parameters are fixed to reproduce the SDSS data (black squares) at z∼0 (red curve), while the other curves show predictions for higher redshifts.** ***(Figure from Peng et al. 2010).***

It is important to measure these correlations with comparable, or better, precision at higher redshifts in order to establish the timescales of transformation and solve this problem. Essentially, including a third axis for cosmic time in Figure 46 allows a direct measurement of the rate at which galaxies evolve in different environments (McGee et al. 2009). Even more directly, the higher star formation rates and infall rates into clusters at z > 0.5 should lead to a substantial, and measureable, population of galaxies caught in the midst of their transformation. In the model of Peng et al. (2010), for example, normalization of the "transition galaxy" mass function, relative to that of the star–forming population, is equal to the transition timescale divided by the inverse of the average specific star formation rate at $M_*$.

Of course, such measurements have been attempted in the past, beginning with Butcher & Oemler (1978), and continuing with large redshift surveys today (e.g., Poggianti et al. 2006; Cooper et al. 2006; Park & Hwang 2009; Cooper et al. 2010; McGee et al. 2011; Balogh et al. 2011; Raichoor & Andreon 2012). The best constraints to date have come from the zCOSMOS and DEEP2 surveys due to their depth and relatively high sampling completeness. However, even combined these surveys consist of 70 000 redshifts covering only ∼ 2 deg$^2$ and spanning ∼ 9 Gyr in cosmic time. The sample size alone (per Gyr) is ∼ 50 times smaller than the SDSS, and the small areas severely limit the range of environments probed.

MSE will provide a breakthrough in the field of galaxy formation by linking galaxy formation and evolution to the surrounding large scale structure, across the full range of relevant spatial scales (from kpc through to Mpc scales). The paradigm shift it will enable will likely exceed in importance that which resulted from SDSS, due in part to its ability to produce datasets *of the* *statistical power of SDSS per Gyr of cosmic time, with similar completeness and limiting stellar mass.* High completeness levels in MSE extragalactic surveys will be essential for studies of environment to robustly identify structures over a large dynamic range in spatial scale. Given the importance of correctly identifying neighboring galaxies when studying environmental effects (see Park et al. 2008), sampling rates will need to be high: the study of strong, small-scale environments is highly compromised if the completeness is ~50 per cent or less (Park & Hwang 2009).

### 3.2.2 Spectroscopic redshift surveys

Understanding the processes discussed in the previous section necessitates observing programs that span redshift space, to reconstruct directly the statistical assembly of galaxy populations. The overarching goal of galaxy evolution is to link the high redshift regime – where we observe galaxies at an earlier stage in their evolution – to the diversity of galaxy properties observed in the local Universe. Current surveys that span the low ($z < 0.3$, e.g., GAMA/SDSS), moderate ($z \sim 1$, e.g., zCOSMOS, Lilly et al 2007) and high redshift ($z > 3$, e.g., VVDS, Le Fevre et al. 2013) Universe come from hugely different telescopes. At low redshift, the extra-galactic field is dominated by the SDSS (2.5m, FoV ~ 3 degrees) and the Anglo-Australian Telescope (4m, FoV ~ 2 degrees). At higher redshifts, a mixture of large facilities dominate (≥8m, FoV << 1 deg). Low and high redshift surveys also tend to target photometric data from different facilities. Historically, co-moving volumes for high redshift surveys are tiny (comfortably dominating the error term for any "Universal measurement") and tend to be highly incomplete. In particular, surveys that use Lyα to obtain redshifts have very poor velocity accuracy (due to the complex nebula component) and probe a minority of available galaxies within a survey volume (~20%, VVDS, Le Fevre et al. 2013). As a result of this work, however, we now have a good picture of how *massive* galaxies evolve relative to the dark matter over the redshift range $0 < z < 1$.

#### 3.2.2.1 Capabilities at low redshift

In any redshift survey the local Universe plays a special role, as it will always be here that the lowest-mass galaxies can be observed at the highest physical spatial resolution. High-quality observations of nearby galaxies serve as the anchor for all higher redshift surveys that aim to study galaxy evolution, and in many cases they provide the only way to connect statistical population parameters at higher redshift with the underlying physical mechanisms. In addition, the nearest baryonic systems are generally spatially extended, and contain within them the fossil records of hierarchical structure formation in the local universe. Thus, they provide a complementary approach to galaxy evolution for decoding the history of mergers, interactions, star formation and chemical enrichment that have operated in different environments, and across a vast range in galaxy mass. Here, the primary role of MSE will be to obtain more detailed information not accessible at higher redshift.

An important aspect of a large fraction of extragalactic studies at very low redshift with MSE is the analysis of stellar populations. Good relative flux calibration between fibers is therefore crucial. This is in recognition of the fact that one of the biggest advantages the nearby Universe holds for understanding the physics of galaxy formation and evolution is the large angular size of stellar systems. At first light, close-packing some fibers enables MSE to efficiently study the spatially resolved properties of low mass and/or distant (z~0.1) galaxies whose spatial extent is limited to a few arcseconds. Full IFU capabilities – an envisioned second generation capability for MSE – are required to fully exploit the large angular size of stellar systems in the nearby universe. IFU capabilities will allow MSE to take full advantage of its aperture and significantly outperform other planned IFU facilities in probing the gradients and outskirts of nearby galaxies (see discussion in Section 3.3.3).

As discussed throughout this chapter, it is clear that a range of spectral resolutions is paramount for effective analysis of the properties of the closest galaxies: strong science cases exist for R~2000 (specifically, redshifts of faint galaxies and measurements of nebular emission line strengths); R~5000 (dynamics of galaxies); and R>>10000 (absorption line abundance analysis and dynamics of globular clusters). Actual science surveys conducted in this area will interweave multiple science programs, and so it will be desirable to have flexibility in the choice of resolution, with the maximum survey efficiency achieved if multiple resolutions can be observed simultaneously. In addition, broad wavelength sensitivity is required over the optical window: it is crucial to maintain sensitivity over 3600 – 7500Å, to enable measurement of the [OII]3727Å emission line for the measurement of metal abundance, and the Hα emission line out to z~0.1. Extending the wavelength coverage (for example to cover Paβ 1.282μm for a more accurate determination of the dust attenuation, and thus star formation rate, in the gaseous component of galaxies) is also desirable.

#### 3.2.2.2 Capabilities at high redshift

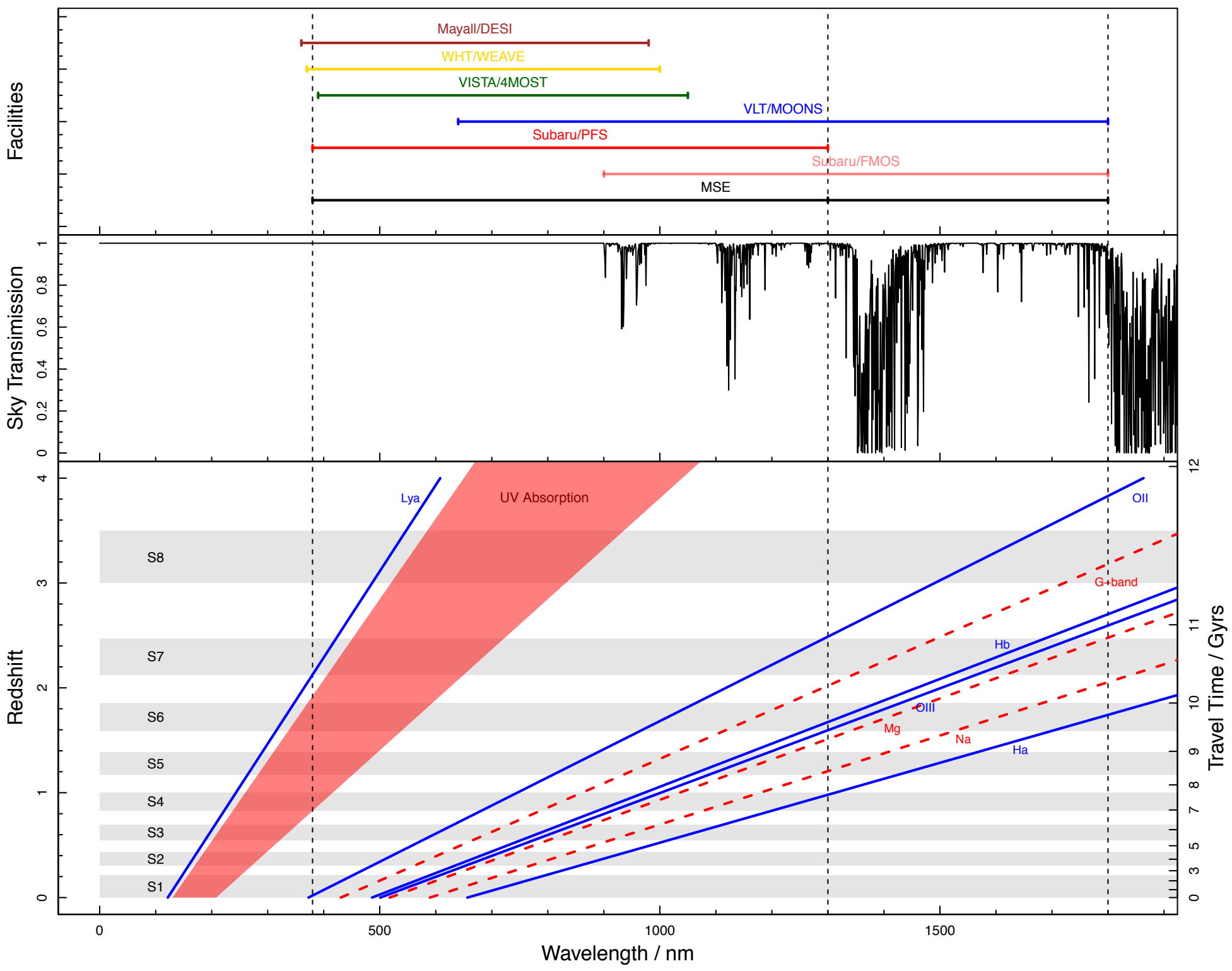

**Figure 48: The bottom panel shows the shift of key spectral features versus lookback-time (left-side axis) or redshift (right-side axis) with the baseline wavelength range for MSE indicated (grey rectangles indicate survey cubes relating to DSC-SRO-08). The top panel shows the range covered by other notable facilities and the middle panel shows the night sky spectrum.**

What are the needs of MSE as they relate to studying high redshift? i.e., at distances greater than a few hundred megaparsecs and hence epochs prior to the present, z > 0.1, look-back time > 1Gyr). This is the extra-galactic spatially unresolved regime spanning a broad range of masses (typically with stellar masses $M_* > 10^9 M_\odot$), environments (pairs, groups, clusters, tendrils, filaments and voids), and epochs (0 to ~12.5Gyrs, i.e., z < 5.5).

The relevant astrophysical measurements MSE provides in this regime are low and intermediate resolution spectroscopy spanning the optical-near-IR range. Processes which generate energy at these wavelengths are predominantly star-formation and accretion, potentially triggered via environmental factors such as dynamical interactions, mergers and/or interactions with the inter-galactic (IGM) or intra-cluster medium (ICM). Complimentary external information will naturally arise from panchromatic imaging facilities from the X-ray to the radio, providing potentially high-spatial resolution imaging and potentially spatially resolved spectroscopy for well-defined sub-samples. The most competitive facilities compared to MSE for studying galaxy evolution in this regime are Subaru/PFS and VLT/MOONS. At the 4m aperture level, ESO/4MOST and AAT/AAOmega will make important contributions. Within this context, MSE – as a high throughput, large aperture facility – is best placed to conduct unique science beyond Subaru and VLT, best compliment LSST and Euclid and best bridge from the 8m to 30+m gap as a feeder facility. A key aspect that we continually emphasize is this document for all science programs is to have close communications with the key imaging facilities, and in particular, LSST, Euclid, WFIRST, and the SKA.

The z > 0.1 case does not drive the capabilities of MSE towards high spectral resolution; R ~ 3000 is sufficient for almost all astrophysical objectives because of the inherent motion within the structures being studied (the exception to this is resolving the Lyα forest, although this requires extremely high resolution, R > 150000). Similarly, IFUs are not as critical as for the nearby Universe, since the resolution of the majority of higher redshift targets are comparable to the native seeing of the site. However, extensive wavelength range, spanning the entire optical to NIR range is an important science enabling capability, and the science cases described in this section do drive the red end of the wavelength range into the near infrared. Figure 48 shows how key astrophysical tracers for extragalactic astronomy drift as a function of wavelength. From a science perspective, there is no obvious science-driven NIR cutoff insofar as the NIR cutoff effectively sets a redshift limit beyond which analyzing galaxies using a homogeneous set of tracers is not possible. In addition to wavelength coverage, the high fiber density of MSE and the ability to close pack fibers is important, since survey speed scales with fiber density for many of the envisioned science programs (such as SRO-08, SRO-09), and it is important to overcome fibre-collision bias, a serious concern for the driving clustering studies.

### 3.2.3 Bringing it together: the role of MSE

Environmental considerations on a range of spatial scales demonstrate that, within the ΛCDM paradigm, *it is fundamental to understand how galaxies evolve and grow relative to the dark matter structure in which they are embedded*. At the heart of all of MSE's investigations into galaxy evolution is a detailed, fully sampled, mapping of the distribution of galaxies, their stellar populations and supermassive black holes, to the dark matter haloes and filamentary structure that dominate the mass density of the Universe, and to do so over all mass and spatial scales. These must be anchored to precision measurements of the closest galaxies, since it is only in the nearby Universe that we can access the lowest mass galaxies, conduct high signal-to-noise analyses of galactic stellar populations, and obtain spatially-resolved insights into their dynamical and chemical evolution.

The capabilities of MSE – its sensitivity, etendue, look-back time, wavelength range and dedicated operations – are such that a fundamental goal for the facility must be to provide a holistic view of galaxy evolution connecting the historically distinct regimes of the nearby Universe with the high redshift Universe. MSE is ideally suited to the design of observing campaigns with equal co-moving volumes from low redshift (z ~ 0) to high redshift (z ~ 3). At all redshift slices, MSE campaigns will reach to low stellar mass, will cover a representative volume of the Universe, will be spatially complete, and will have statistical uncertainties sub-dominant even when galaxies are binned by their properties. Such datasets definitively answer a host of science questions over 12 Gyrs in look-back time:

- Detailed kinematical and stellar population analysis of the central regions of galaxies to explore supermassive black hole scaling relations across the full mass and environment regimes and exploring correlations between the properties of low mass stellar systems such as globular clusters and the large scale properties of the host galaxies;

- Spatially-resolved chemodynamic studies probing merger remnants at large galactocentric radius combined with detailed halo dynamics for each target galaxy;
- Spatial clustering and dynamics over scales ranging from kpc to hundreds of Mpc, to link galaxies to the underlying dark matter distribution. Specifically, to identify galaxy locations and dynamics within bound haloes, and filamentary structures, and the distribution of those haloes relative to one another;
- Measurement of merger rates as a function of redshift, galaxy types, environment and over a wide range in galaxy mass.
- Properties of stellar populations, including ages, star formation rates, chemical abundance and initial mass function parameters, particularly at low masses where the galaxy mass – halo mass relation is most uncertain;
- The occurrence and strength of AGN activity, expected to play a significant role in galaxy formation by providing an important source of energy that inhibits star formation;

In this chapter, we discuss in detail some of the specific scientific opportunities that MSE will enable for understanding galaxy formation and evolution. We begin by discussing detailed studies of the nearest galaxies across all mass scales (Section 3.3) before discussing the evolution of galaxies across cosmic time (Section 3.4). The science described provides a comprehensive and transformative contribution to our understanding of the relation of galaxies to their surrounding large scale structure, and connects the galaxy diversity observed in the nearby Universe to their evolution observed across cosmic time. Some topics specific to the Intergalactic Medium and the detailed physics of black hole accretion are discussed in detail elsewhere, in Chapter 4.

## 3.3 Galaxies across the hierarchy

**Science Reference Observation 6 (Appendix F)**

**Galaxies and their environments in the nearby Universe**

*To understand the physical drivers of galaxy evolution it is necessary to map the distribution of stellar populations and supermassive black holes to the dark matter haloes and filamentary structure that dominate the mass density of the Universe. This requires deep, spatially complete spectroscopic surveys over wide areas, ideally suited to the capabilities of MSE. We propose a wide-area survey S1-W, that obtains redshift-quality spectra for every z < 0.2 galaxy over a 3200 deg$^2$ area. This is efficiently achieved if target selection is based on deep imaging catalogues spanning the NUV to NIR, so that precise photometric redshifts are available. This survey will, among other things, allow every halo in the area with mass $M_{halo}>10^{12}M_\odot$ to be identified with a group of four or more galaxies. We propose to follow up select regions within this survey with longer exposures, resulting in a sample covering 100 deg$^2$ to a depth of at least i<24.5, and spanning four decades in halo mass. This will allow us to measure the shape of the stellar mass function to the scale of the Local Group dwarf galaxies, over a large volume and a range of environments. A subset of fibers will be used to obtain long integrations on bright galaxies, yielding a final sample of >50000 high signal-to-noise ratio spectra, for which detailed stellar*

*population synthesis analysis is possible. This will be the definitive spectroscopic survey of the local Universe for decades to come.*

### 3.3.1 The lowest mass stellar systems in the Local Universe

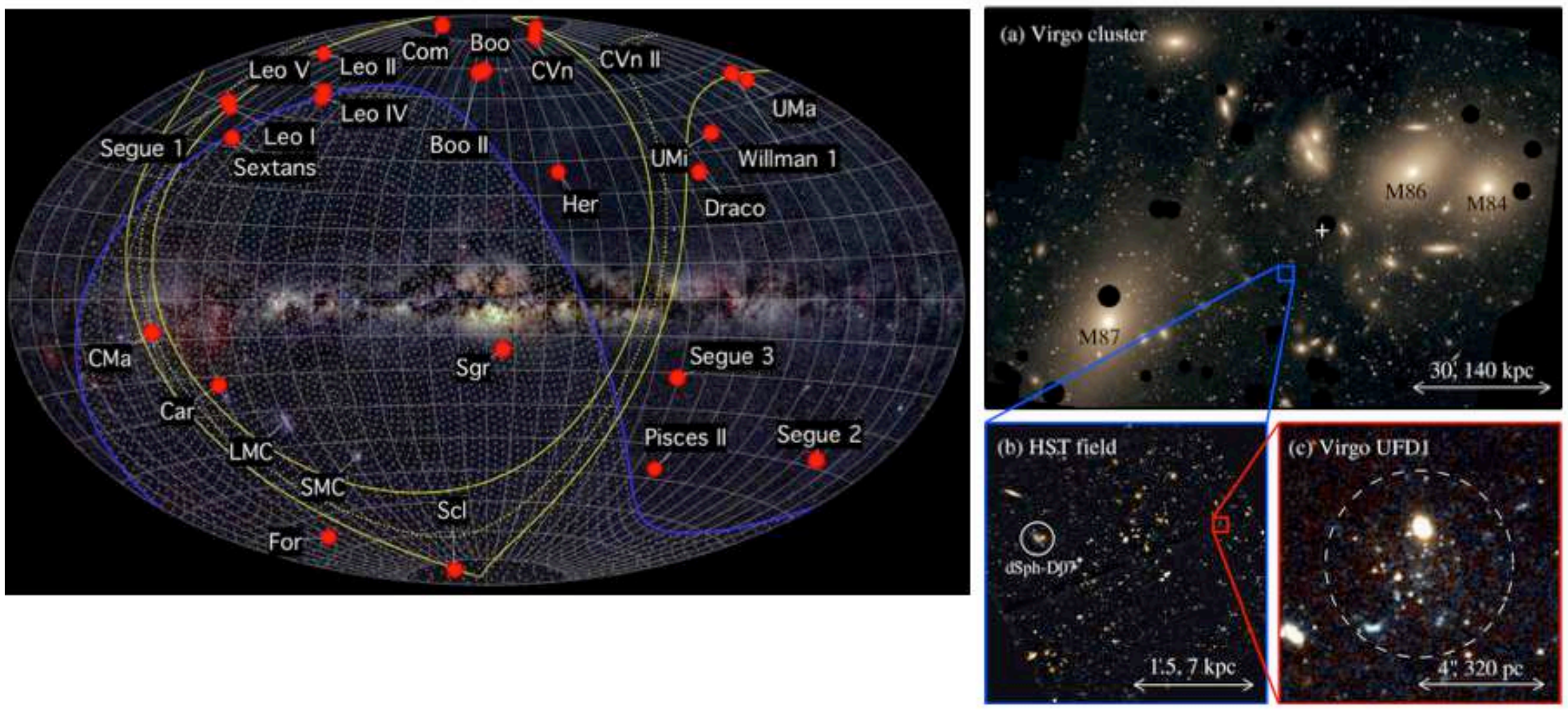

**Figure 49: Left panel: Distribution of low-mass galaxies associated with the Milky Way. The sample of such systems has nearly doubled in recent years thanks to the Dark Energy Survey (Image credit, A. Frebel, MIT). Right panel: An ultra-faint dwarf galaxy – comparable in luminosity to the low-mass Galactic satellites shown in the previous panel – in the Virgo Cluster. Many thousands of such systems are expected to be discovered by future imaging facilities like LSST, Euclid and WFIRST *(Figure from Jang & Lee 2014).***

The menagerie of known stellar systems has grown significantly in recent years, with reports of several new classes of compact, low-mass stellar systems. These discoveries include "extended" or "diffuse star clusters" (e.g., Brodie & Larsen 2002; Huxor et al. 2005; Peng et al. 2006), "ultra-faint dSph galaxies" (e.g., Belokurov et al. 2006; Figure 49) and "ultra-compact dwarfs" (UCDs; Drinkwater et al. 2003). Other types of low-mass, compact stellar systems include the more familiar globular clusters and nuclear star clusters (see, e.g., the recent compilation of Mihos et al. 2015 shown in the left panel of Figure 50).

At this time, a reliable census of such stellar systems in the local universe — a prerequisite for understanding their properties and origin — is lacking, but it is abundantly clear that these faint and compact objects will play a key role in the coming decade in answering some important open questions in astrophysics. Unfortunately, their faintness and compact nature means that they are difficult to identify and study. Although clever and efficient techniques have been developed to efficiently identify candidates systems from high-resolution UV-visible-IR imaging (e.g., Munoz et al. 2014; Liu et al. 2015), confirmation and characterization of these systems will always require optical spectroscopy.

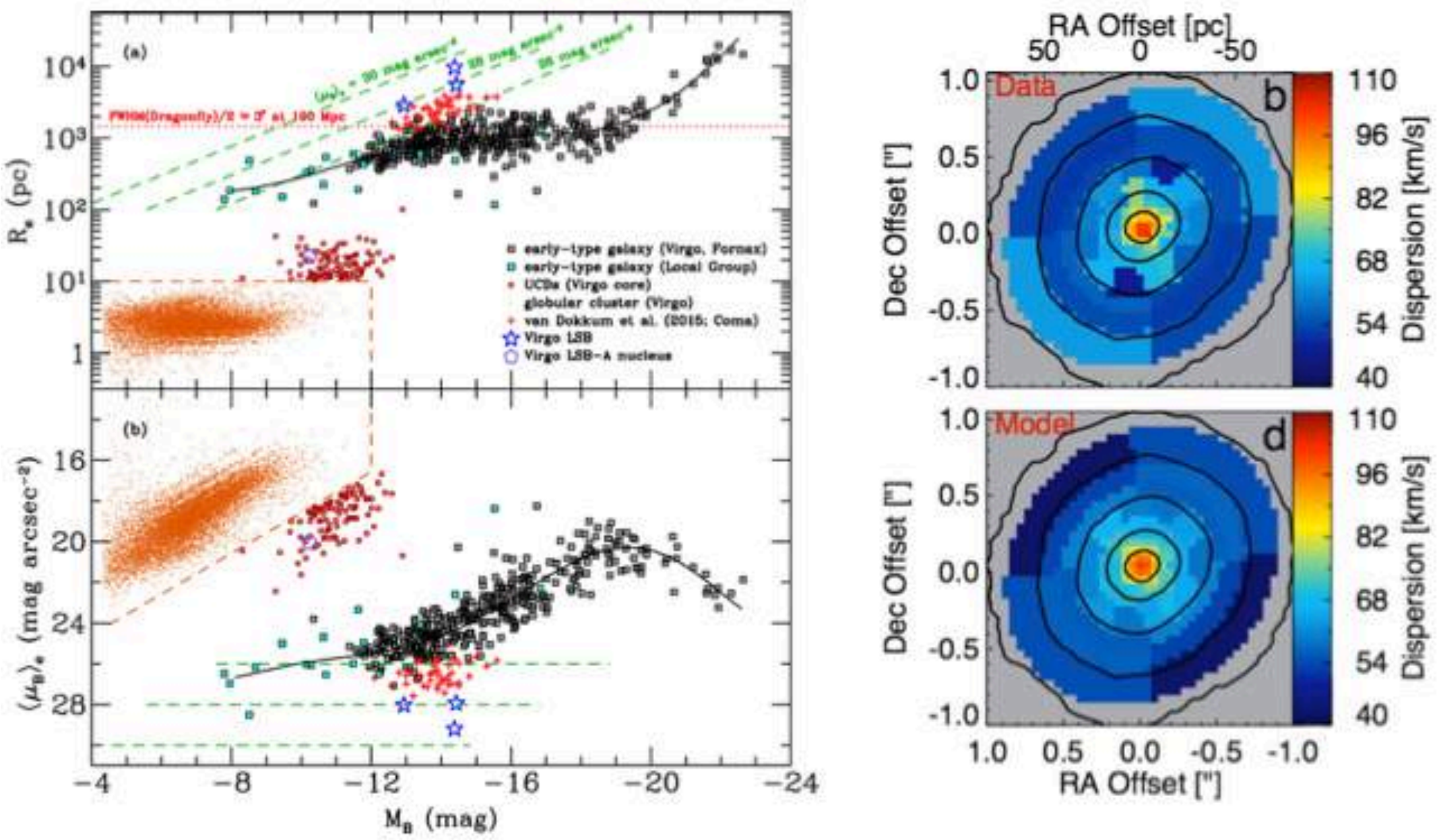

**Figure 50: Left panel: Magnitude–size–surface brightness relations for a collection of dynamically hot stellar systems, mostly in the Virgo and Fornax clusters (*Figure from Mihos et al. 2015*). Right panel: Comparison of the observed and simulated two-dimensional velocity dispersion map for M60-UCD1, one of the most massive UCDs currently known. The central velocity dispersion "spike" is the signature of a 21 million solar mass black hole embedded in this low-mass ($1.2\times10^8 M_\odot$) and compact ($R_e$ = 24 pc) stellar system *(Figure from Seth et al. 2015).***

The next generation of imaging facilities (LSST, Euclid, WFIRST) — which collectively offer complete SED coverage over the optical and IR regions, good sensitivity and high angular resolution — can be used to select large and well-characterized samples of nearby candidates, but, as yet, no planned facility will be capable of detailed and efficient spectroscopic follow up. With its large field of view, high sensitivity, high multiplexing, and range of spectral resolutions, MSE would spectacularly fill this role, and would allow astronomers in the 2020s to address a number of important open issues. These include:

i. *The Abundance of Compact Stellar Systems*.

A subset of compact systems (the ultra-faint dSphs) are thought to correspond to the low-mass, dark matter halos predicted by cosmological simulations. At present, the observed numbers of such systems fall well short of the predictions of "standard" cosmological models, which has potentially profound implications for the nature of dark matter (i.e., whether it is cold, warm, mixed, decaying, etc.) and for gas cooling, star formation and feedback in the low-mass halos. The space densities of other types of systems, such as UCDs, bears directly on the efficiency of tidal stripping in low-mass galaxies. Unfortunately, spectroscopic confirmation of carefully selected targets is essential if we are to build a complete census of such systems over a range of environment. Even in the richest nearby clusters, such as Virgo, this is a monumental task that requires low-resolution spectra (with SNR $\sim$ 10 at R $\sim$ 2,000 and g $\sim$ 24 – 25) for *hundreds of thousands* of candidates distributed over $\sim$100 deg$^2$ fields.

ii. *Galaxy Disruption and Transformation*

There is mounting evidence that many, and perhaps most, UCDs have formed via tidal stripping of low-mass galaxies that initially contained a central nuclear star cluster. Independent lines of evidence additionally suggest that at least some nuclear star clusters were, in turn, assembled through multiple mergers of globular clusters in low-mass halos. Finally, it is possible that tidal effects acting on initially compact star clusters might inflate their velocity dispersions and effective radii so that they could masquerade as ultra-faint dSphs. A common element in these different scenarios is the transformation and possible disruption of low-mass stellar systems by tidal effects. Testing these suggestions across a range of mass and environment will hinge on our ability to acquire spectroscopy for large and representative samples of compact stellar systems. In this case, desirable data products would include radial velocities (for membership and dynamical studies of sub-populations), ages and metallicities (for tests of formation scenarios) and internal velocity dispersions (for dynamical mass estimates and tests of dark matter vs. tidal heating).

iii. *Chemical Enrichment and Star Formation in Extreme Environments*

Some types of stellar systems are noteworthy in being rare and extreme environments. For instance, UCDs and nuclear star clusters are among the densest environments known in the local universe and can provide insights into how star formation and chemical enrichment proceeds under such circumstances. They are also extreme environments for exploring the form of the initial mass function, which mounting evidence suggests may vary with mass and/or environment.

iv. *Dark Matter on Large and Small Scales*

Orbiting within the gravitational potentials of their host galaxies, compact stellar systems are dynamical test particles that can be used to probe the distribution of dark matter within low- and high-mass galaxies. The technical requirements for spectroscopy (most notably, spectral resolution) will depend strongly on the choice of host galaxy, with higher resolution needed for the lowest-mass systems. In a related issue, spectra for any compact sources associated with fine structures seen in some galaxy halos (such as tidal streams, plumes and shells), can help shed light on the accretion histories of these galaxies.

v. *Supermassive Black Holes and Their Environments*

The exciting discovery of a supermassive black hole (BH) in M60-UCD1, one the most massive UCDs in the Virgo Cluster (Seth et al. 2015**)**, has provided compelling evidence that tidal stripping of low-mass galaxies is a viable formation channel for UCDs. Equally important, BHs like the one in M60-UCD1 provide a rare opportunity to explore the BH mass function and the occupation function in low- and intermediate-mass galaxies. While the secure detection of a BH in a UCD, globular cluster or nuclear star cluster will always require AO-assisted or space-based IFU spectroscopy (from, e.g., TMT or JWST), the pre-selection of candidate systems must be made on the basis of moderate-SNR, intermediate-resolution, integrated-light spectroscopy; a powerful niche of MSE.

### 3.3.2 The evolution of low mass galaxies

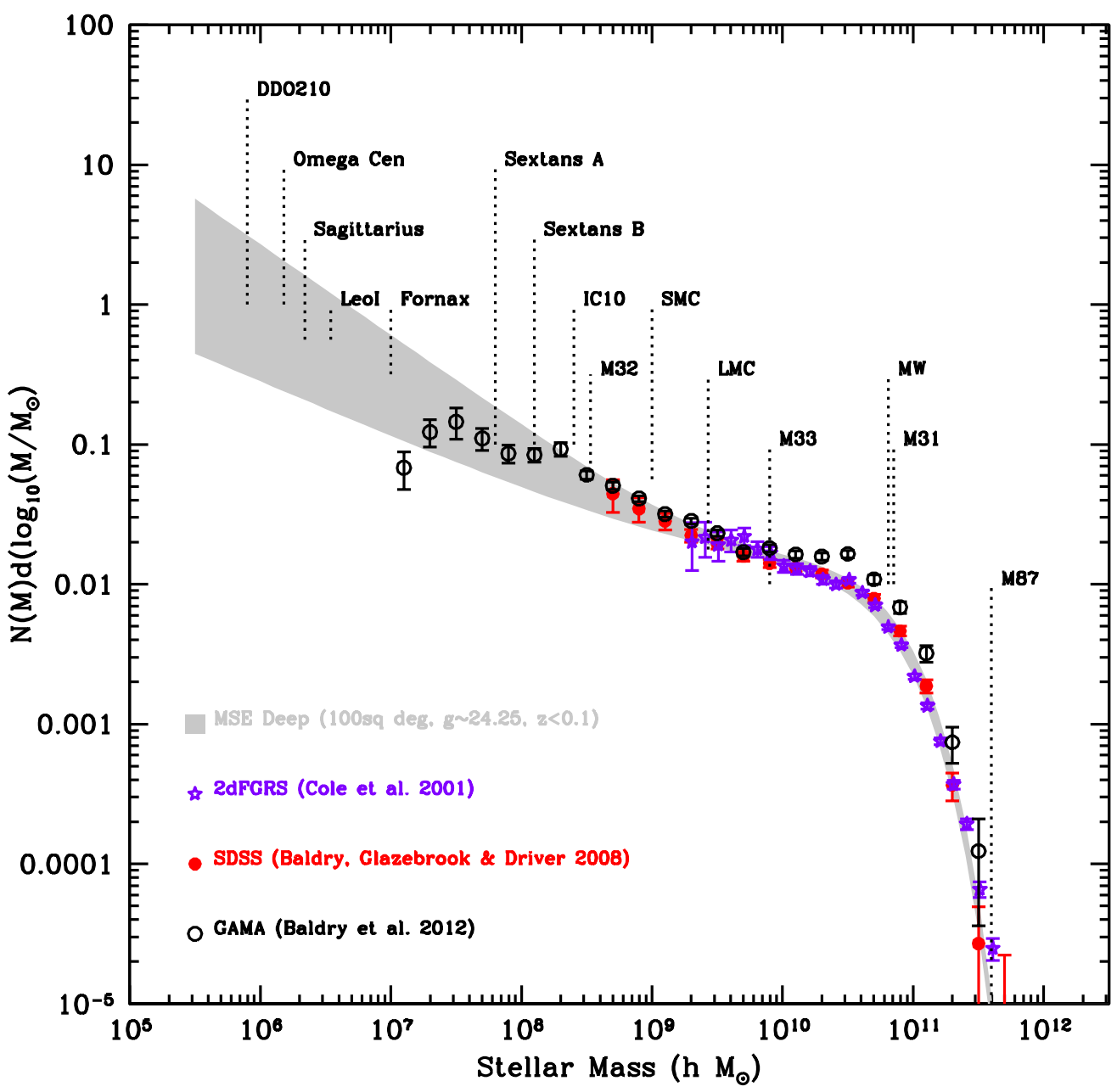

**Figure 51: The local stellar mass function as measured by SDSS, 2dFGRS and GAMA. GAMA is the deepest of these surveys, and probes robustly down to the scale of M32; lower mass measurements are actually lower-limits due to surface brightness limitations. The shaded region shows the precision and stellar mass coverage anticipated with MSE through SRO-06.**

In the past 15 years, imaging and spectroscopy of large, unbiased samples of nearby galaxies have provided a good, empirical description of massive galaxy formation and evolution. Specifically, these observations have allowed the identification and detailed characterization of fundamental scaling relations, such as those between stellar mass, SFR and metallicity. The large sample sizes allowed the measurement of the scatter in these relationships, and of their dependence on other parameters, such as morphology, AGN activity and environment (Kauffmann et al. 2003; Tremonti et al. 2004; Balogh et al. 2004). Measurements of galaxy clustering, weak lensing, and velocities have enabled the relationship between galaxies and their dark matter haloes to be mapped in some detail (e.g., Leauthaud et al. 2012; van Uitert et al. 2016). Through observations like these, which inform and constrain theoretical models, our understanding of the physics driving galaxy evolution has undergone tremendous growth.

However, the extension of these statistical studies into the dwarf regime is relatively uncharted territory, due to the flux-limited nature of most surveys. Our knowledge of the very lowest mass galaxies is restricted to those in the Local Group (see discussion in Boylan-Kolchin et al. 2012), and the local observational census is likely incomplete (Tollerud et al. 2008, Koposov et al. 2008, McConnachie 2012). Shallow, wide-area surveys such as the SDSS (York et al. 2000) and the 2dF galaxy redshift survey (2dFGRS; Colless et al. 2001) are only sensitive to massive galaxies and intermediate-mass dark matter halos (Eke et al. 2006; Yang et al. 2007). And while a handful of very low-mass systems have been found in both these surveys, neither the 2dFGRS nor SDSS were able to identify host halos below $M_h \sim 10^{12} M_{\odot}$, thereby barely sampling the dominant halo mass at their sensitivity limits.

The GAMA survey, with spectroscopy completed in 2014, improved upon the SDSS in two main respects. First, the increased depth of the survey extended the limits of the measured stellar mass function by more than an order of magnitude, from $M_* \sim 5 \times 10^8 M_\odot$ to less than $M_* \sim 10^7 M_\odot$ (see Figure 51). Secondly, GAMA obtained spectra for nearly 100% of the magnitude-selected target sample, independent of small-scale clustering. This sampling is critical to robustly link galaxies with their dark matter haloes, and to distinguish satellite galaxies from the dominant galaxy in each halo. Though covering only 300 square degrees, this survey demonstrated that low-mass galaxies evolve in a fundamentally different way from their more massive counterparts (Lara-Lopez et al. 2013a, 2013b, Bauer et al. 2013). The high spatial completeness, moreover, enabled an unprecedented study of satellite galaxy evolution (Prescott et al. 2011, Figure 52), and merger rates (Robotham et al. 2014).

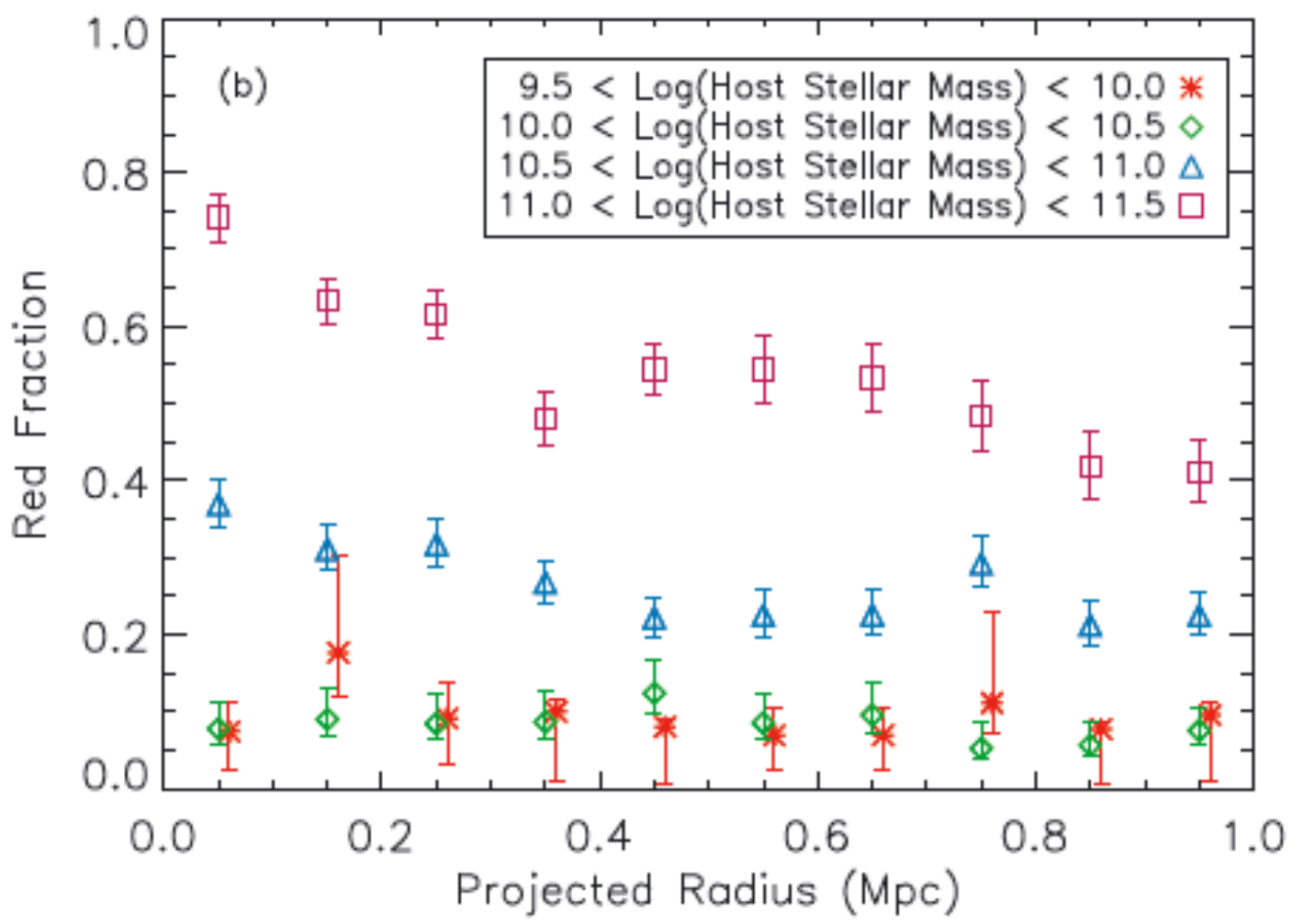

**Figure 52: The red fraction of satellite galaxies as a function of projected radius, for four different host stellar mass ranges in the GAMA survey (Prescott et al. 2011). These results showed that quenching of star formation is most efficient in massive systems, and acts within ~500kpc of the host. This points to relatively slow mechanisms of gas removal, like strangulation (Larson et al. 1980; Balogh et al. 2000) as the main driver of this differential evolution. (*Figure from Prescott et al. 2011*).**

Building on these new insights into the dwarf galaxy regime, the next generation of galaxy surveys in the nearby universe must target large, homogeneous samples of the lowest mass galaxies. The combination of high sensitivity, wide field of view, and survey capability provided by MSE is perfectly suited to play this role. To achieve an understanding of low-mass galaxies that rivals what we know for their more massive counterparts, MSE should conduct a comprehensive survey of several thousand galaxies in the mass range $10^7 < M_*/M_\odot < 10^9$, spanning a range of environment (from rich cluster cores to filaments and voids) and including all morphological classes.

SRO-06 describes a possible observing program with MSE that meets the above requirements and that allows for transformative advances in many areas. SRO-06 extends well into the regime of dwarf galaxies (i.e., to stellar masses of $M_* \sim 10^7 M_\odot$ at z < 0.1 and down to an ultimate, low-z mass limit of $M_* \sim 10^{5.5} M_\odot$), enabling studies of halo occupancy, halo dependent mass functions, halo substructure, and star formation efficiency, described in more detail below. Galaxies will be selected using well-defined criteria to minimize selection bias, with complete sampling of the mass function down to very low masses and very small separations. Through such a survey, MSE will impact a vast range of science areas pertaining to galaxy evolution, particularly at low masses. SRO-07, discussed later, describes a survey of galaxy clusters that augments many of the science goals of SRO-06 through extension to more extreme environments, including our nearest neighbours, Coma and Virgo.

#### 3.3.2.1 Halo occupation and the formation of the stellar mass function

*i. Stellar mass function*

The most fundamental measurement that MSE will make as part of SRO-06 is to extend the observed stellar mass function to masses below $10^8 M_\odot$, with a cosmologically representative, unbiased, spatially complete spectroscopic sample. Figure 51 shows the current constraints on the z < 0.1 galaxy SMF from GAMA (Baldry et al. 2012), the SDSS (Baldry et al. 2008), the 2dFGRS (Cole et al. 2001), and the predicted performance for MSE (SRO-6). The stellar mass function is of particular interest because numerical simulations have long argued that halos such as the Local Group should contain thousands of galaxies, rather than the ~100 that are known (Moore et al. 1999; Klypin et al. 1999). Indeed, observations find that the Local Group mass function is relatively flat (Pritchet & van den Bergh 1999, Driver & de Propris 2003, Ferrarese et al. 2016). There are two principal mechanisms that have been proposed to reconcile this apparent discrepancy (Benson et al. 2003), namely supernova feedback (whereby SNII explosions heat and expel a significant fraction of the baryons from lower mass halos), and early photo-ionization (in which the primordial baryons in low-mass halos are heated by external sources such as AGN and Pop III stars, and are thereafter unable to cool due to low gas metallicities). These two mechanisms give quite distinct, qualitative predictions for the faint-end slope of the galaxy SMF, as the former mechanism leaves a relic stellar population in every halo while the latter does not. Thus, by extending existing work to observations of galaxies more than an order of magnitude lower in mass, MSE will provide leverage on decoupling the effects of these heating mechanisms from one another.

*ii. Dark matter halo occupation*

In the ΛCDM paradigm, galaxy formation takes place within dark matter halos. Measurement of the overall abundance of such halos as a function of mass, dN/d(log M) is among the most important outstanding issues in modern cosmology. This quantity is sensitive to the form of the inflationary perturbations and to the nature of the dark matter particle itself: i.e., whether it is collisional or collisionless, decaying, self-interacting, warm (light), or cold (massive). Moreover establishing its functional form at low redshift with high precision is critical for measuring its evolution, which in turn is sensitive to the cosmological parameters.

Linking the luminous mass to dark matter halos is critical, both to understand the halo abundance and to understand galaxy formation. A key focus for SRO-06 is therefore a comprehensive sampling of both the halo mass function *and* the stellar mass function to the faintest possible limits. For a fully sampled, r ≤ 24 mag nominal survey, it should be possible to detect ~200 halos per 100 $deg^2$ with $\log(M_h)/M_\odot = 10.75 \pm 0.5$ dex. This is 50 times greater than the density found in the GAMA survey (Robotham et al. 2011) — the deepest precursor for halo measurements (although the small area limits its utility for the purpose of cosmological constraints) — and more than an order of magnitude beyond the mass range probed by the SDSS and 2dFGRS datasets (Rines et al. 2007; Tempel et al. 2014).

#### 3.3.2.2 Galaxy growth and transformation

The hierarchical assembly of galaxies through merging is a fundamental prediction of cold dark matter models, and plays a particularly important role in the evolution of galaxy sizes and densities (e.g., Trujillo et al. 2006). As discussed in Section 3.3.4, merger rates can be estimated from the frequency of close pairs, potentially including morphological information (Robotham et al. 2014; Casteels et al. 2014). The highly complete spectroscopic survey described in SRO-06 is ideal for these analyses, in order to identify close pairs in velocity as well as space. Such measurements provide not only a test of the underlying dark matter model prediction, but also, together with the in-situ star formation rate, a complete empirical description of galaxy growth rates at a given epoch.

In addition to merging, galaxies owe much of their present mass to in-situ star formation. How star formation is regulated, or even stopped, in a galaxy is an unanswered question of fundamental importance. For massive galaxies we have developed a good picture of how the star formation rate depends on stellar mass, halo mass and epoch (e.g., Behroozi et al. 2013). However, there is evidence that these relationships change at low masses, in a way that is not predicted by current state-of-the-art simulations (**Figure 53**). MSE will extend the analysis shown in **Figure 53** by at least an order of magnitude in stellar mass, over a much larger area. Furthermore it will be possible to measure any dependence on halo mass, which is a prediction of most satellite quenching models but remains controversial observationally (e.g., McGee et al. 2009; Wetzel et al. 2013; Vulcani et al. 2010; Rasmussen et al. 2012; Lu et al. 2012).

Two major phenomena – AGN and stellar feedback – are thought to bear primary responsibility for the late-time quenching of star formation, with additional physical processes playing a role in high-density environments. The feedback of AGN has been proposed as a possible origin of the quenching of the star formation activity in massive galaxies (Bluck et al. 2014), while that of supernovae and massive stars dominates in dwarf systems (but see discussion below). The kinetic energy injected in the ISM by the nuclear black hole in massive galaxies and by stellar winds in dwarfs is able to remove the gaseous component feeding the star formation process.

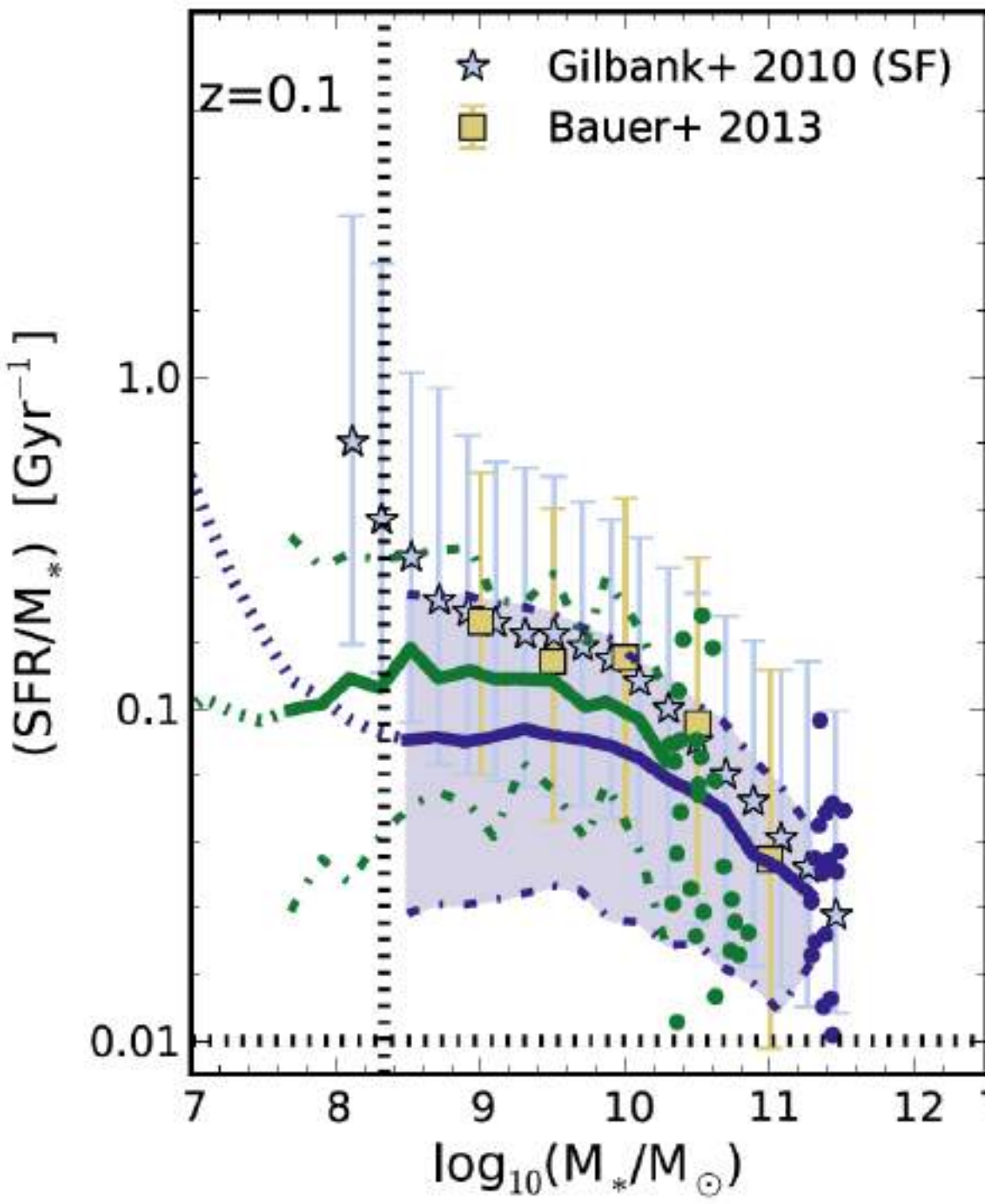

**Figure 53: Predictions of the sSFR-mass relationship from the EAGLE hydrodynamic simulations, compared with observations at low redshift. The purple and green solid lines are low and high-resolution simulation results, respectively; the dashed lines indicate where resolution effects may be important. There is indication of a discrepancy with the SDSS Stripe 82 analysis of Gilbank et al. (2010) at the lowest stellar masses probed.** ***(Figure from Furlong et al. 2015).***

*i. Do dwarf galaxies host AGN?*

The role of black holes in galaxy formation is probably one of the most debated topics in the field (see the Annual Review article by Kormendy & Ho 2013). Current observations are largely limited to high mass galaxies, where the majority of galaxies are considered to host a black hole. However, several hundred low mass and/or bulgeless galaxies are now also suspected of having central black holes (e.g., Greene & Ho 2004, Reines et al. 2013, Satyapal et al. 2014). As demonstrated by the work at higher masses, large samples are required to study the demographics of this population. Current work is focusing on fine-tuning selection techniques and obtaining follow-up observations to confirm successful identification of an AGN (e.g., Secrest et al. 2015). The coming years promise to pin down the optimal selection techniques, likely through a combination of optical spectroscopy and mid-IR imaging (e.g., from WISE). New, large spectroscopic samples of low mass galaxies, with good quality imaging, will be needed to exploit this groundwork.

From analysis of emission line ratios with MSE, it is possible to determine the occupation fraction of AGN in low mass galaxies, and determine whether supermassive black holes relate more closely to the spheroidal (bulge) component or to other properties of the host (e.g., total gravitational mass). The large sample of objects identified by MSE will offer the opportunity to study the occupation fraction of AGN in low mass galaxies, determine whether SBHs relate more closely to the spheroidal (bulge) component or other properties of the host (e.g., total gravitational mass), and discover how the AGN affects the structure (on both global and nuclear sclaes) of the host galaxy.

ii. *The role of environment*

In high-density environments, such as clusters and groups, gravitational interactions or interactions with the hot and dense ICM trapped within the potential well of the over-dense region (such as ram pressure stripping and thermal evaporation) can efficiently remove the gaseous component of galaxies, transforming star forming systems into quiescent objects. The study of these physical processes is crucial for understanding galaxy evolution. However, analysis of satellites in the Local Group has demonstrated that quenching of star formation has a complex dependence on stellar mass that is not understood. **Figure 54** (from Fillingham et al. 2015; see also Geha et al. 2012) shows that while the effectiveness with which star formation is quenched in satellite galaxies steadily declines with decreasing stellar mass in the mass regime probed by all large spectroscopic surveys, at lower masses the trend appears to reverse. While the effect is subtle, this differential measurement (comparing satellite galaxies with central galaxies) is a potentially powerful way to constrain feedback parameters (e.g., McGee, Bower & Balogh 2014; Balogh et al. 2016). A statistically significant sample of groups with Local Group halo masses is therefore a critical complement to these local studies, to determine the universality, or environmental dependence, of the remarkably low efficiency of star formation observed in the very local Universe.

Only MSE is able to make this measurement over a cosmologically relevant volume, and with a homogeneous selection of galaxies over the full mass range as described in SRO-06.

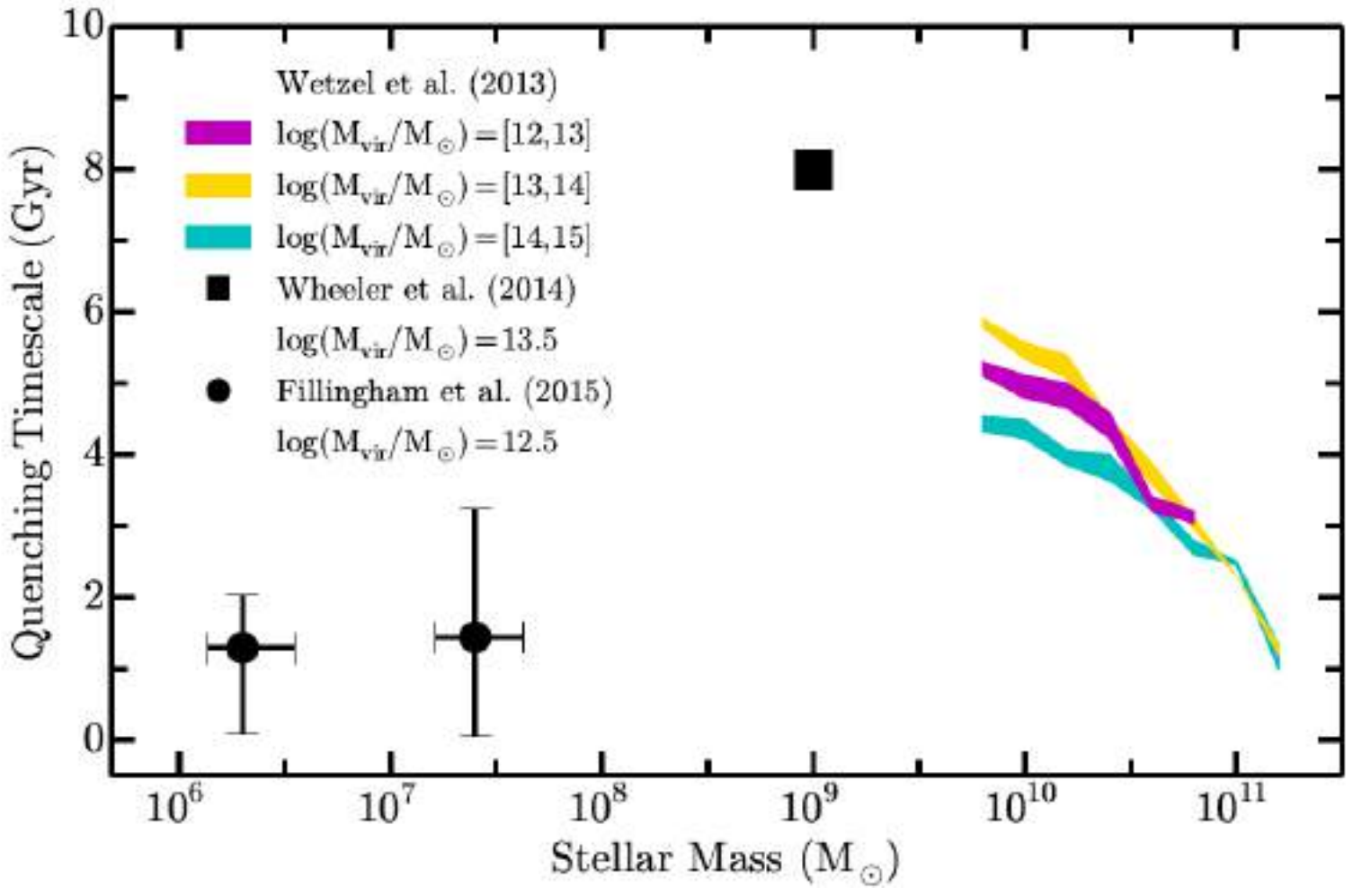

**Figure 54: Dependence of the inferred star formation quenching timescale of satellite galaxies as a function of their stellar mass. Short quenching timescales lead to high fractions of galaxies with little or no star formation at the epoch of observation. Large spectroscopic surveys like SDSS and GAMA have been instrumental in uncovering the trend in decreasing timescale with increasing stellar mass shown as the colored lines at the high-mass end. However, this and other analyses of the Local Group show that the lowest-mass satellites may respond very differently to environment *(Figure from Fillingham et al. 2015).***

#### 3.3.2.3 Stellar populations

While relatively low SNR spectra are useful for probing dynamics and large scale environment, a significant advance in our current understanding relating to the evolution of baryons in galaxies requires high quality data to measure the star formation and enrichment histories of the stellar populations. Colors are useful for distinguishing currently star-forming galaxies from passive ones, but to break the age-metallicity degeneracy and so obtain more precise ages of the dominant red population, one needs absorption line indices from high SNR (> 25) spectra for a significant sample of the galaxies. Thus, SRO-06 will transform our understanding of several related issues concerning the mass dependence of fundamental scaling relations of galaxies, and the role of feedback in regulating the evolution of baryons in galaxies. Highlights include:

i. *Chemical enrichment*

The mass-metallicity relation derived from SDSS galaxies extends to log $M_*/M_\odot = 8.5$ (Tremonti et al. 2004). Pushing to lower masses has previously only been possible with either a handful of nearby galaxies (e.g., Lee et al. 2006) or stacking analyses (Andrews & Martini 2012), and has extended this relation down by a further factor of ten in mass, to suggest that the metallicity continues to fall with mass. However, at low masses, the chemical enrichment is likely to become increasingly stochastic, necessitating large samples of individual measurements that are carefully modeled (and not just stacked) to account for, e.g., varying levels of nitrogen abundance, which can be elevated in dwarfs (Berg et al. 2011). Intriguingly, the lowest mass Lyman Break Galaxies currently detectable at $z = 2$ also have anomalously high N/O abundances, so that understanding the origin for these chemical deviations has a potentially broad impact.

At low masses, we expect chemical enrichment to become increasingly stochastic and sensitive to factors such as winds, infall and environment (Kirby et al. 2013), potentially with a metallicity "floor" where self-enrichment is driven by a few generations of stars (Sweet et al. 2014). Here, again, the environmental dependence of these scaling relations is mass-dependent (Ellison et al. 2009; Cole et al. 2014). Quantifying the role of these factors requires individual metallicity measurements for a cosmologically-relevant sample of dwarfs. Moreover, outliers in mass – metallicity space can help us quantify the fraction of dwarfs that have a tidal origin (e.g., Croxall et al. 2009), which may represent a significant (but currently poorly constrained) fraction of the dwarf population (e.g., Ploeckinger et al. 2014 and references therein).

*ii. Tracking the variation in the initial mass function*

The role of the initial mass function (IMF) for studies of galaxy formation and evolution can hardly be overstated, as the assumptions regarding its slope and mass range are the basis for the interpretation of most galaxy integrated observable properties at any redshift. Recent work (van Dokkum & Conroy 2010; Capellari et al. 2012; Smith et al. 2012a) has suggested the intriguing possibility that the slope of the stellar IMF varies as a function of galaxy mass and/or morphological type (see Figure 55). But this evidence remains controversial, and because of the small samples for which such measurements are available today, it is unknown to what extent the IMF depends on spatial scale, galaxy type, mass, redshift or environment. It is possible to put useful constraints on the high-mass end of the initial mass function using combinations of Hα emission lines and galaxy colour (Kennicutt 1983; Baldry & Glazebrook 2003; Gunawardhana et al. 2011). And if spectroscopy with SNR>100 is available, a measurement of the IMF shape at the low-mass end also becomes possible via the dwarf-sensitive NaI 8183Å, 8195Å and FeH 9916 Å (Wing-Ford) lines (van Dokkum & Conroy 2010; Cappellari et al. 2012; Smith et al. 2014).

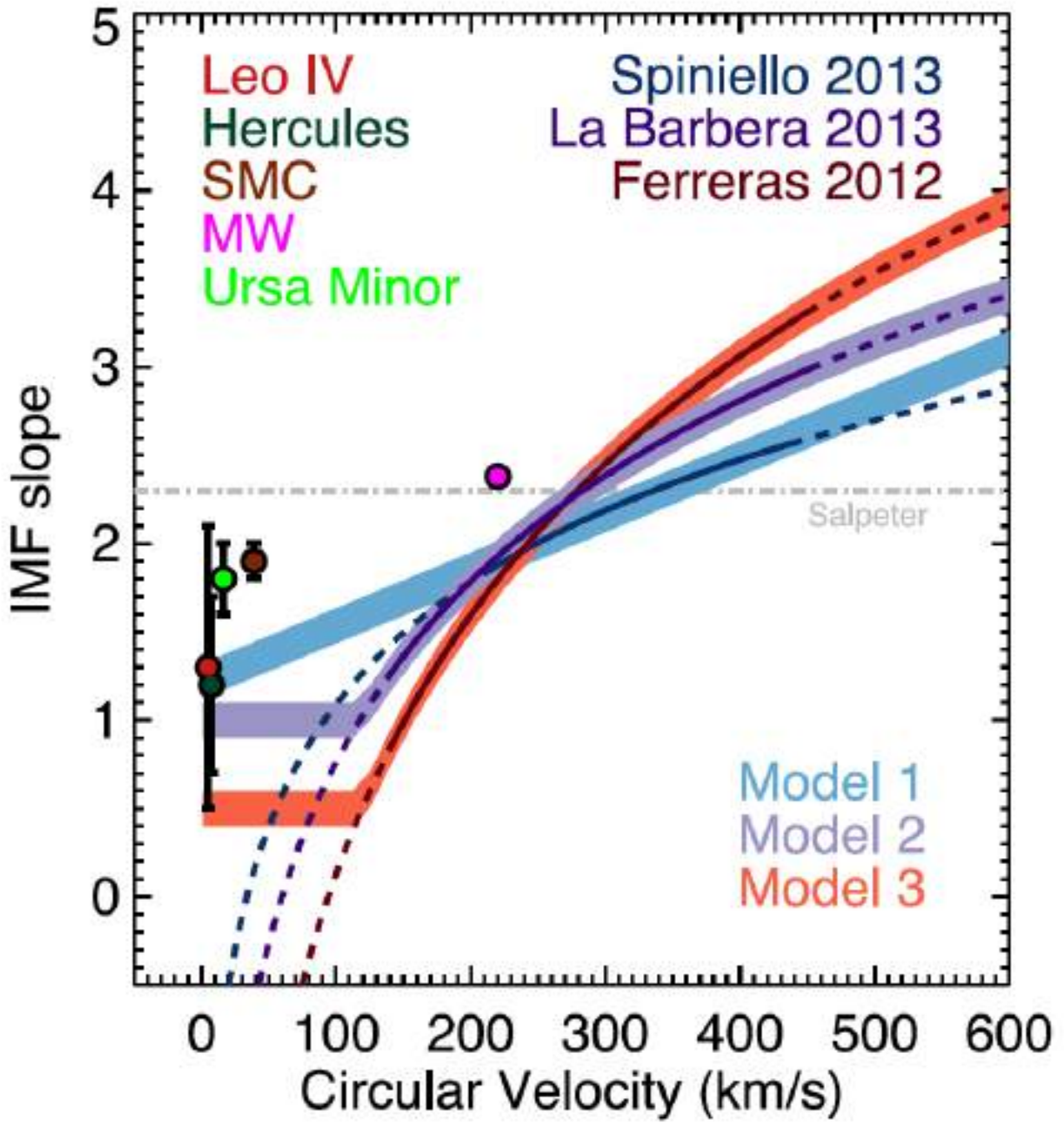

**Figure 55: Compilation of observed relations (black solid lines) for how the slope of the IMF varies with galaxy circular velocity, and their extrapolations (black dashed lines). Also shown are individual measurements for some Local Group galaxies, based on resolved star counts. This remains a tremendously important and controversial topic, with data limited to small samples of inhomogeneously selected galaxies. *(Figure from McGee, Goto & Balogh 2014)*.**

### 3.3.2.4 The evolution of low mass galaxies: summary

MSE will be the preeminent facility for studies relating to missing satellites, group finding, merger rates, the study of the galaxy populations with respect to halo mass, and the halo mass function and SMF. It allows the sampling of the galaxy stellar mass function right down to the Jeans limit of $M_* \sim 10^{5.5}$ $M_\odot$ within environments ranging from rich clusters to group, filament, and voids. Critical factors for success will be *complete* sampling over representative volumes at a spectral resolution sufficient to obtain ± 20 $kms^{-1}$ accuracy, in established regions with high-quality, multi-wavelength data. Suitable regions include the equatorial GAMA 9h, GAMA12h and GAMA 15h regions (Driver et al. 2011) that contain deep GALEX, VST, VISTA, WISE, HERSCHEL, GMRT, and ASKAP data. High completion is critical because many low-mass halos will typically contain only a few galaxies and all of these galaxies will need to be sampled to attain robust velocity dispersions (i.e., halo masses). Inevitably, this will require a multiple-pass survey to unpick the close pairs, triplets and compact groups. A reasonable resolution (i.e., R ~ 2000) is needed to measure the velocity dispersions of the lowest-mass groups (note that the Local

Group velocity dispersion is 60 kms$^{-1}$). For full characterization of the galaxy population within each halo, ancillary multi-wavelength data would be required such as that assembled by the GAMA team (for an overview, see Driver et al. 2016), and to which a large number of future facilities will contribute, as discussed in detail in Chapter 1.

### 3.3.3 Spatially resolved studies of galaxies

Integral field spectroscopy is arguably the most powerful way to capture the complexities of galaxy formation, to examine how the dark matter, stars, neutral gas, ionized gas and super-massive black holes all interact with each other. There are a large number of observables that are difficult or impossible to obtain from single fibre surveys and that allow direct tests of the latest galaxy formation theory. Observables from integral field spectroscopy include gas and stellar internal and bulk kinematics, the spatial distribution of star formation, stellar metallicity and abundance gradients, stellar age gradients, gas phase metallicity distributions, resolved ionization diagnostics and many others.

**Table 5: A summary of current and future MOS IFU instruments. For each instrument and telescope combination we list the actual or expected date of commissioning, wavelength range, telescope aperture, field-of-view, number of fibres, fibre size (diameter), the number of IFUs and the spectral resolution (R = λ/dλ in FWHM). KMOS is the only system using image slicers rather than fibres, so we give the number of spatial elements in the slicer and the slicer spatial sampling at *.**

| *Instrument* | *Telescope* | *Date* | *Wavelength range (nm)* | *Telescope diameter (m)* | *Field of view (sq.deg)* | *Number of fibres* | *Fibre size (")* | *Number of IFUs* | *Spectral resolution* |
|---|---|---|---|---|---|---|---|---|---|
| FLAMES | VLT | 2003 | 370 - 950 | 8 | 0.42 | 300 | 0.52 | 15 | 11000 - 39000 |
| SAMI | AAT | 2012 | 370 - 740 | 3.9 | 1 | 819 | 1.6 | 13 | 1700 - 4500 |
| KMOS* | VLT | 2013 | 778 - 2456 | 8 | 0.12 | 4704* | 0.2* | 24 | 2800 - 4800 |
| MaNGA | SDSS | 2014 | 360 - 1000 | 2.5 | 3 | 1423 | 2 | 16 | 2000 |
| HET | VIRUS | 2017 | 350 - 550 | 10 | 0.37 | 33600 | 1.3 | 150 | 800 |
| WEAVE | WHT | 2018 | 370 - 1000 | 4.2 | 2 | 1000 | 1.3 | 20 | 5000 |
| HECTOR | AAT | 2020 | 370 - 900 | 3.9 | 3 | 8500 | 1.6 | 100 | 4000 |
| **MSE** | | **~2025** | **360 - 1300** | **11.25** | **1.5** | **3400** | **≤1.2** | **50** | **6500** |

The power of integral field spectroscopy has been widely recognised within the astronomical community with most large telescopes now having integral field spectroscopy capabilities of some type. We are already at the stage of having completed at least two generations of major integral field surveys (e.g., SAURON/ATLAS3D, Cappellari et al. 2011a;CALIFA, Sanchez et al. 2012) and the first generation of major multi-object integral field surveys are now underway (SAMI, Croom et al. 2012; MANGA, Bundy et al. 2015; KMOS, Stott et al. 2016). Future projects are already at an advanced stage (WEAVE, Balcells et al. 2010; HECTOR, Bland-Hawthorn 2015), but the combination of large aperture, wide field-of-view, high spectral resolution and excellent image quality make MSE a unique prospect in the integral field spectroscopy domain.

A significant number of large-scale monolithic IFUs are now available on large-telescopes, and so a single monolithic IFU using all the fibres of MSE is unlikely to be competitive. For example, compare the ~3400 fibers of MSE to the ~90000 spectra per shot of the MUSE IFU on VLT (Bacon et al. 2015; however, we note that the limited field of view of MUSE and other instruments could mean that very large monoliths could still prove of significant interest). However, with a large number of deployable IFUs MSE would be an immensely powerful instrument. In Table 5 we list the main characteristics of current and planned multi-object integral field spectrographs. VIRUS on the HET, although not a deployable IFU system, has a huge multiplex that gives it some of the same advantages as a deployable system. FLAMES is the only comparable system with truly deployable IFUs on an 8 – 10m class telescope, and this has an order of magnitude less fibres and an order of magnitude smaller field-of-view. As a result a MOS IFU system on MSE sits in a truly unique part of parameter space. Such a system is planned as a second generation capability and here we discuss some aspects of the motivation and implementation.

#### 3.3.3.1 The IFU perspective of galaxies

Given current instruments like MUSE of VLT and Keck CWI with monolithic IFU capabilities, it is the MOS IFU capability on a large aperture that will set MSE apart. As such, the science cases described below focus on the science case for MOS IFU, rather than a large monolithic IFUs (that could in principle also be fitted to MSE).

The current and near-future generations of major IFU surveys on smaller telescopes will likely have targeted of order ~100 000 local, z < 0.1, galaxies by 2025. In this context the major advantages for MSE will be high surface brightness sensitivity coupled with the image quality of Maunakea that allows MOS IFU spectroscopy to push beyond the local Universe for more than just the highest mass galaxies. A second substantial focus will be on local low mass galaxies ($\log(M_*/M_\odot) < 10^8$), that are the building blocks of higher mass galaxies and are the most influenced by their environment.

i. *The evolution of galaxy morphology*

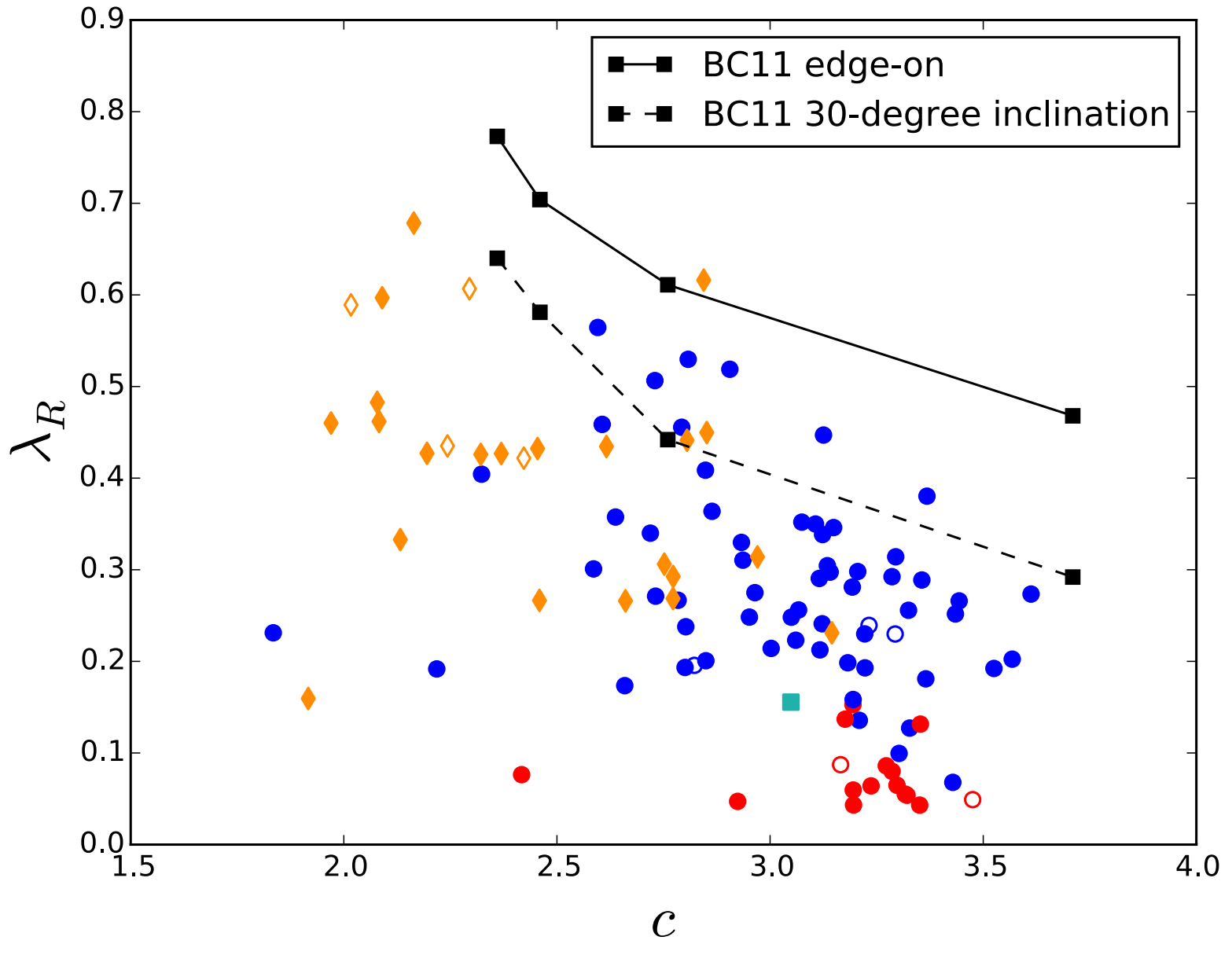

**Figure 56: $\lambda_R$ vs. concentration ($C = R_{90}/R_{50}$) for the SAMI galaxies (Fogarty et al. 2015) that are slow rotators (red), fast rotators (blue) and spiral galaxies (orange). The black tracks are from Bekki & Couch 2011 and show the**

**evolutionary path of spirals (top left) subject to multiple interactions and thereby transformed into an S0 (bottom right), 5Gyr later.** ***(Figure from Fogarty et al. 2015).***

The morphological structure of galaxies is one of their most fundamental properties. We know that environment modifies galaxy morphology, that mass and morphology are connected and that galaxy morphologies evolve over time. However, tackling the underlying physical processes which drive these trends is hard. Integral field spectroscopy, particular through stellar kinematics, has the power to uncover the physics. After all, the photometric morphology of a galaxy is defined by the distribution of orbits that its stars live on.

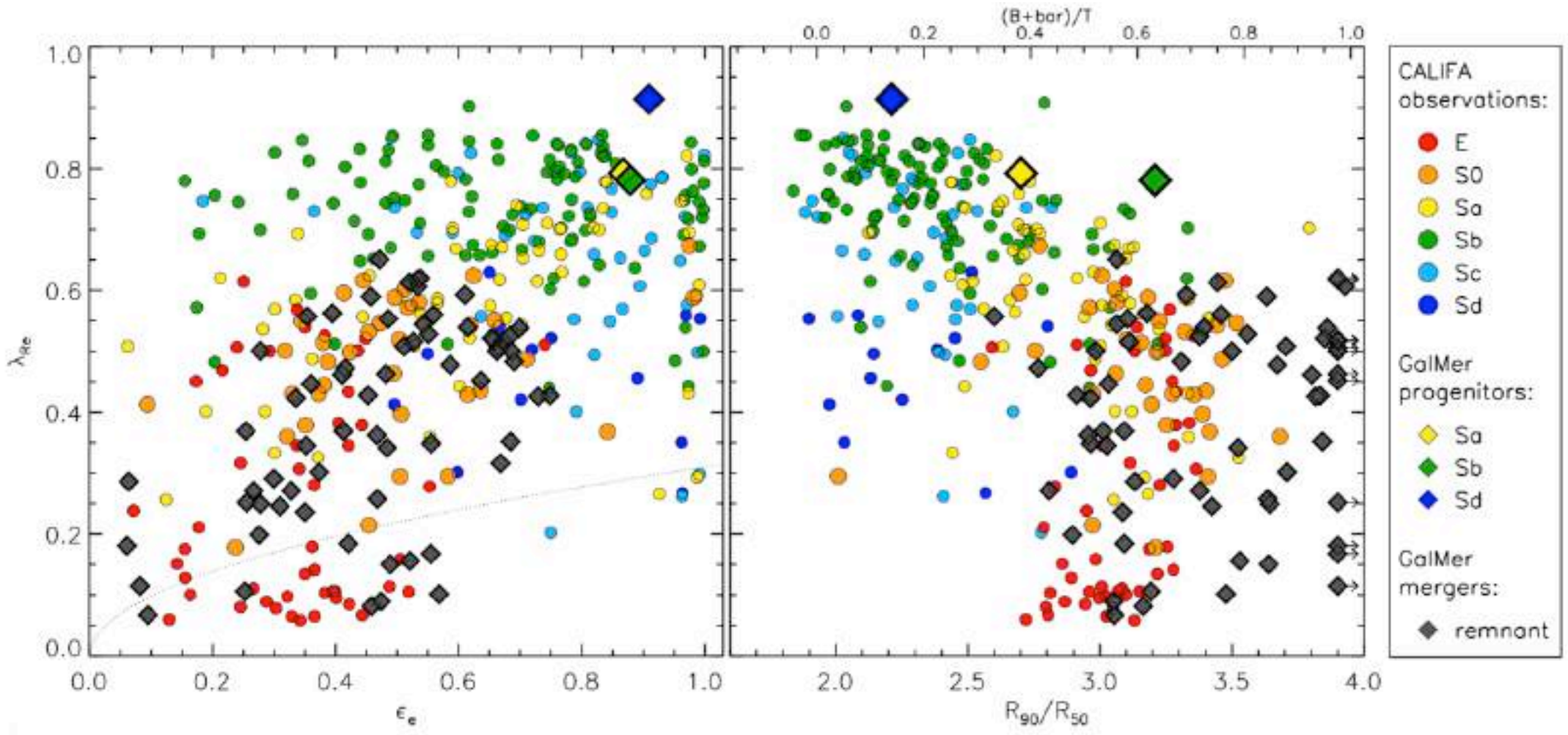

**Figure 57: $\lambda_R$ vs. ellipticity (left) and concentration (C = $R_{90}/R_{50}$) (right) for CALIFA galaxies, colour coded by morphology (circles). These are compared to remnant galaxies in merger simulations (diamonds*). (Figure from Querejeta et al. 2015).***

We know that high density environments contain more S0 galaxies and fewer spirals (e.g. Dressler, 1980) and the transformation of spirals into S0s must play an important role. The S0 fraction in clusters doubles from ~20% at z ~ 0.8 to ~40% by the present day (Dressler et al., 1997; Couch et al., 1998), with a corresponding decline in the spiral fraction. The trend is similar, but twice as strong, in groups (Just et al., 2010), suggesting that groups efficiently drive transformations. Revealing the physics behind such transformations requires stellar kinematics as only this can discriminate between hydrodynamical mechanisms (that should not produce an increase in random support) and gravitational interactions (that will build-up pressure supported systems). Simulations of group interactions or minor mergers already provide possible pathways from spirals to S0s (Bekki & Couch, 2011; Bournaud et al., 2005), and are qualitatively similar to observations (Figure 56; Fogarty et al. 2015). However, alternative interpretations are possible, such as major mergers (Querejeta et al. 2015; see Figure 57).

A fundamental challenge currently is that we are comparing current spiral galaxies to current S0s. To properly understand this transformation we need to look at earlier spiral galaxies, as their properties may change with cosmic time. Direct scaling from current surveys suggests that it will be relatively straight-forward to measure stellar kinematics in $L_*$ or greater galaxies out to z ~ 0.3 which already covers 3.5 Gyr of dynamical evolution. Higher mass galaxies can be targeted at even higher redshift. MSE will have the unique ability to measure stellar kinematics for large samples of galaxies beyond the local Universe and as a result directly examine the evolution of disk structure. As a result we will be able to directly map earlier spirals to current spirals and S0s.

Work from ATLAS3D has shows that morphology may be better defined in terms of dynamical properties (Emsellem et al., 2011; Cappellari et al., 2011b). Early type galaxies can now be classified as *fast rotators* (rotation dominated) and *slow rotators* (dispersion dominated). The global fraction of slow rotators is relatively constant at ~15% of early types, although they increase in number towards the centre of clusters (Cappellari et al., 2011b; Fogarty et al., 2014). Slow rotators may undergo more major mergers and build up a higher fraction of their mass from the accretion of satellites (Khochfar et al., 2011), but the cluster environment is not optimal for this to happen. Again, in order to understand the growth of slow rotators directly we need to examine how the population evolves with redshift. This will only be possible with large-scale integral field surveys beyond the local Universe. The frequency of slow rotators in different regimes can be used as a strong test of current cosmological hydrodynamic simulations that struggle to converge on the amount of angular momentum in simulated galaxies (e.g., Scannapieco et al., 2012).

Size evolution is another key feature of the galaxy population, particular in the most massive galaxies. Once again, kinematics allows us to understand the structure of such galaxies. Here the challenge is to obtain sufficient spatial resolution to be able to resolve rotation curves and minimize the transfer of rotational velocity to measured dispersion via beam smearing. This is possible for the most massive galaxies in the optical, as demonstrated by van der Wel & van der Marel (2008) using FORS on the VLT, however to make this a reality on MSE smaller fibres that better sampled the seeing would be required.

*ii. Modulating star formation*

Numerous physical processes are able to modulate star formation across cosmic time. This can broadly be separated into factors that are associated with galaxy environment and galaxy mass (Peng et al., 2012). However, we need to move beyond these broad categorizations towards a more physically motivated picture, where we understand the differing contributions from each physical process.

Environmental quenching is the suppression of star formation in high density regions (e.g. Lewis et al., 2002). While the existence of such quenching has been known for some time, the detailed physical mechanisms remain hard to discern. The different processes that can contribute to quenching include ram-pressure stripping (e.g., Gunn & Gott, 1972), strangulation (Larson et al., 1980, Bekki, 2009), dynamical interactions and mergers. Crucially, these processes have different observational signatures. Ram-pressure stripping leads to concentrated star formation, as gas is removed from the outer disk first, and may cause shocks and extra-planar gas. In contrast, strangulation leads to suppression of star formation across the whole disk. Interacting galaxies can have their gas funneled into the central regions, triggering nuclear star formation, and they may be dynamically disturbed. The processes are likely to depend on whether a galaxy is the central galaxy or a satellite in its parent halo, as well as the mass of the parent halo and local galaxy–galaxy interactions (e.g. Davies et al. 2015).

IFU spectroscopy from MSE can provides (1) spatially resolved Hα emission to map star formation on time–scales of $10^7$yr; (2) Balmer absorption gradients from stellar populations that traces star formation on $10^9$yr time–scales; (3) stellar and gas-phase kinematics to identify galaxies that are dynamically disturbed or misaligned; (4) shock detection via emission line diagnostic ratios such as [N II]/Hα, [O III]/Hβand [S II]/Hα (e.g., Kewley et al. 2006). This need to be done in tandem with detailed 3D environmental measurements so that large-scale and small-scale environment can be defined, as well as close pairs of galaxies that are interacting.

Mass-related quenching of star formation may be caused by AGN (Croton et al., 2006) or related to morphology with bulges providing support against the formation of central gas disks (Martig et al., 2009). IFU spectroscopy allows us to pick apart the separate components of a galaxy (Johnston et al., 2014) and measure the separate stellar population characteristics of bulge and disk independently. We can also identify the sphere of influence of an AGN and determine whether the stellar populations and current star formation are influenced by the presence of an AGN.

MSE will have the surface brightness sensitivity to detect Hα to trace star formation until the emission line hits the spectral range of the instrument (1.3 microns) at z ~ 1. As the star formation rate increases back to higher redshift (Hopkins & Beacom, 2006) we will be able to directly test where this star formation is happening. Is the increased in the normalisation of the star formation main sequence driven primarily by star formation in disks? When, and in which circumstances, does star formation become more clumpy (as seen in current infrared IFU surveys; Genzel et al. 2011)?

### *iii. The role of outflows and inflows*

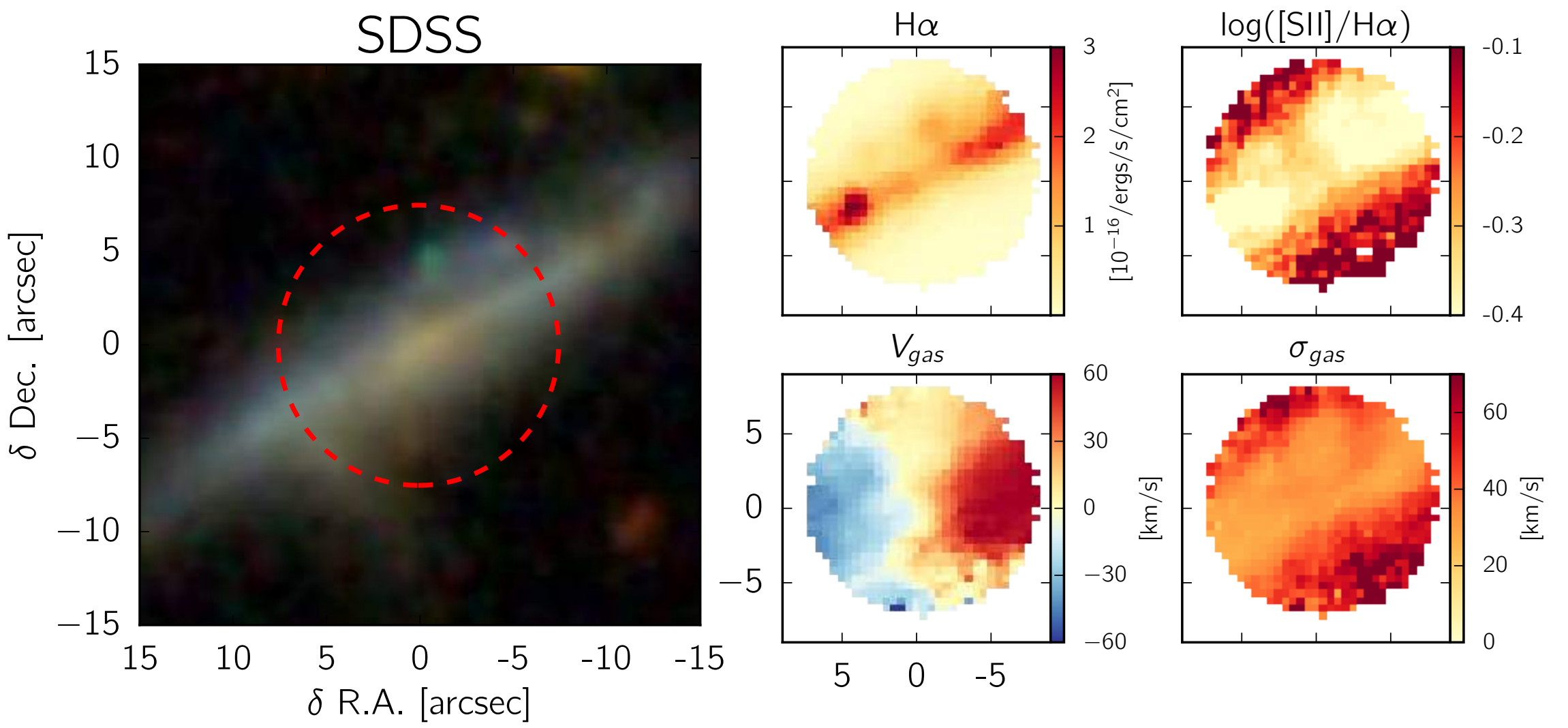

**Figure 58: SAMI maps of wind galaxy GAMAJ115927.23-010918.4. The dashed red circle in the SDSS 3-colour image (right panel) shows the SAMI field-of-view (15 arcsec diameter). The increased velocity dispersion, $\sigma_{gas}$, and line ratios, [S II]/Hα, off the disk plane, as well as the disturbed velocity field, are all indicative of galactic winds.** ***(Figure from Ho et al. 2016).***

Many star forming galaxies showing evidence for outflows (e.g. Heckman et al., 1990; Veilleux et al. 2005), but their full impact on the process of galaxy formation in still uncertain. New local IFU surveys have already started to address this issue (e.g., Ho et al. 2016, see Figure 58). A high fraction of outflows are being found using a combination of gas kinematics and gas ionization diagnostics. Some of these have current star formation rate densities below the canonical 0.1 $M_{\odot}$ $kpc^{-2}yr^{-1}$ expected to drive winds, and stellar ages and star formation time–scales appear consistent with the winds being driven by bursty star formation episodes. Central to such analyses is high spectral resolution to separate out star forming and wind components from complex emission line profiles. Shock models (e.g., Rich et al. 2011) can be used to estimate galactic wind properties such as shock luminosity, the shock velocity, the dynamical timescale of the wind, and the contribution of shocks to the global energy budget.

The role of gas accretion is fundamental to the build-up of galaxies. It is gas accretion that fuels the ongoing process of star formation. The misalignment between gas and stellar kinematics provides a potent means to identify and characterize gas accretion because internally produced gas (e.g., via stellar mass loss) will follow the kinematics of the stars, while externally accreted gas need not follow the motion of the stars. Misalignments have been analyzed in early–type galaxies (Davis et al. 2011), and this is now being extended to all galaxy types by ongoing local IFU surveys. The gas misalignment distribution and can be used to constrain the dynamical relaxation, depletion and accretion timescales of gas in galaxies (Davis & Bureau 2016). With MSE pushing such measurements beyond the local Universe we will be able to directly see whether the increased star formation to high redshift is driven by a increase in externally accreted gas, and in which situations this is most significant.

A complementary approach to studying gas accretion is via gas-phase chemical abundance gradients. Normal spiral galaxies generally show clear metallicity gradients, becoming enriched by a factor of 10 from $\sim$ 8 – 10 kpc to their central regions. Gas inflows, either from external accretion or triggered by interactions, can drive low metallicity gas towards the central regions of a galaxy. This results in flatter metallicity gradients (Kewley et al., 2010; Rich et al., 2012).

*iv. The role of dynamical interactions and mergers*

In the standard hierarchical picture of galaxy formation and evolution, the mass of galaxies is built up hierarchically through successive mergers and through the accretion of gas from the intergalactic medium. At high redshift galaxy mergers may be the primary assembly route for massive galaxies and drive rapid star formation. Theoretical predictions of the merger rate are highly uncertain (e.g., Jogee et al. 2009, Hopkins et al. 2010) due to the inability of models to precisely map galaxies onto dark matter halos and sub-haloes (see Lotz et al. 2011 for a discussion). Measurement of dynamical disturbance can be a route to identifying mergers (e.g., Shapiro et al. 2008). This is currently being done locally with nearby samples, and at high redshift with star-forming galaxies in the infrared (e.g., with KMOS), but MSE will have the ability to track dynamical disturbance uniformly from z = 0 – 1 in Hα, uncovering the importance of mergers and disk instability in the evolution of their star formation density (Hopkins & Beacom 2006). Kinematic data can also be used to calibrate and compared other approaches such as close pairs of galaxies (e.g., Robotham et al. 2014) and morphological disturbance (e.g., Conselice et al. 2003).

Cosmological hydrodynamical simulations suggest that it is possible to connect higher–order stellar kinematic moments $h_3$ (skewness) and $h_4$ (kurtosis) in galaxies to their assembly history (Naab et al., 2014). These simulations predict that fast rotating galaxies with gas–rich merger histories show a strong $h_3$ – (V/σ) anti-correlation, while fast rotators that experienced gas–poor mergers do not, due to the absence of a dissipative gas component. The higher-order stellar kinematics of from MSE can be used to diagnose the formation pathways for the galaxy population. As well as extending this work to higher redshift, MSE can push down in stellar mass for local galaxies. As a result we can start to understand whether dwarf galaxies have the same formation routes as more massive galaxies in terms of their merging, or discern where in the hierarchy physical processes change.

*v. Cosmology via scaling relations, peculiar velocities and lensing*

Scaling relations have long been used as a cosmological tool. There is a growing realisation that using scaling relations (such as the fundamental plane or Tully-Fisher relation) to map cosmic flows has great potential to probe the gravitational growth of structure and test general relativity. Such tests are largely orthogonal to other methods targeted at dark energy usually based on high redshift spectroscopic surveys (e.g., Subaru/PFS, Mayall/DESI, etc.). Integral field spectroscopy allows us to reduce the scatter in scaling relations and generate more physical quantities, such as moving from the fundamental-plane to the mass-plane (Cappellari et al., 2013, Scott et al., 2015, Cappellari, 2016). The ability to capture both bulge and disk components means that combined scaling relations such as $s_{0.5}$ (Kassin et al., 2007) can be used for all galaxies, as demonstrated by Cortese et al. (2014).

Integral field spectroscopy of galaxy velocity fields has great potential as a lensing tool (de Burgh-Day et al., 2015a,b). Measurement of shear on velocity fields can be made to much higher precision that normal weak lensing based on shape measurements. As a result individual galaxy-galaxy lensed systems can in principle provide direct measurements of dark matter halo profiles. A second combination of IFU spectroscopy and lensing is to use the rotation curves derived from IFU spectroscopy to reduce the shape noise in lensing measurements by comparison to the Tully-Fisher relation (Huff et al. 2013, Blain 2002). Integral field spectroscopy of SN host galaxies can help to shed light on the differences between SN populations (e.g. Sullivan et al., 2006), and whether this is dependent on the global properties of a galaxy or the local properties at the location of the SN. Understanding such differences is critical to reducing the systematics in SN cosmology measurements. Chapter 4 provides a more detailed discussion on next generation cosmology surveys with IFUs.

### 3.3.3.2 Instrument and survey design: practical considerations

If we assume a baseline of 3400 fibres, then this would map to ~ 56 IFUs if each one had 61 fibres (~11 arcsec diameter assuming 1.2" arcsec fibres and a 75% fill factor). In practice a variety of IFU sizes would be most appropriate (e.g. Bundy et al., 2015) to match the sizes of galaxies in any design reference survey. A resolution of R = 6500 over much of the optical would in general be much preferred, compared to the lower R = 2000. This is because the higher resolution allows a variety of science including the kinematics in low mass galaxies, measurement of dispersion in gas disks and separating different kinematic components in outflows.

*Supporting data:* The connection of IFU data from MSE together with a variety of ancillary data will greatly enhance the science outcomes. Ideally targets should be selected from regions that have well defined environmental measurements, cover a broad range in stellar mass, and have substantial multi-wavelength data available. The GAMA survey (Driver et al., 2011) currently provides the best regions in this regard, with deep (r < 19.8) and highly complete spectroscopy and carefully matched 21–band photometry (Driver et al., 2016) from UV (GALEX) to far-IR (Herschel). However, a large-scale MSE IFU survey will need to reach beyond the current GAMA survey. Spectroscopic programs like WAVES (wavesurvey.org on the VISTA telescope using 4MOST will provide suitable target catalogues, but of course MSE galaxy redshift surveys could also do this. Tying an MSE IFU survey to new HI surveys using instruments such as ASKAP will be immensely valuable, allowing us to access gas fraction and the large-scale angular-momentum of galaxies beyond the optical radius. At the same time new deeper, higher resolution image data will become available from LSST and Euclid.

*Sensitivity limitations*: The two largest constraints on the information content in IFU data is the surface brightness of the source and the spatial sampling. Higher spatial sampling (to the limit of the seeing) is possible by having smaller resolution elements (i.e. smaller fibres). However, with smaller fibres the SNR per element is reduced, and in some cases can reach the point at which data is read-noise rather than background limited. As one of the unique aspects of MSE will be its ability to push to higher redshift over a large field-of-view, this naturally pushes the design to the smallest fibres that are still background limited. Then the surface area that can be targeted depends on the number of fibres than can be taken by the spectrographs. Smaller fibres take less room on the spectrograph slit, but not sufficiently less to fully compensate for the reduced field- of-view with smaller fibres. As a result the sky area that can be sampled by the IFUs directly depends on the number of spectrographs in the system. The area where spatial resolution is most demanded is at the highest redshift. At the extreme end of the MSE redshift range for Hα the sensitivity to stellar continuum will be poor (due to surface brightness dimming). If only targeting emission lines at this redshift, then fibres could be small enough to optimally sample the seeing, as the surface brightness sensitivity limits will be less severe. This points to two sets of fibres, one that will be focussed on surface brightness sensitivity in the continuum (say ~1” diameter) and a second that optimally samples the seeing (say ~0.2 − 0.3”). Detailed modeling of the target galaxy populations should be done to optimise this fully. Reaching the seeing limit of the telescope also places constraints on the design of the telescope and corrector optics.

*Positioning IFUs in the focal plane:* With less than 100 IFUs and a field-of-view of 1.5 deg diameter, some consideration needs to be given to how best to position IFUs in the focal plane. With a typical target redshift of z ~ 0.2 − 0.3 there will be substantial clustering across the field of view, which points to a flexible positioning system where each IFU has a relatively large patrol region. Detailed simulations will be required to examine how different positioning technologies impact the efficiency of IFU surveys.

### 3.3.4 The growth of galaxies through mergers

In the hierarchical structure formation paradigm, galaxy merging is expected to play a crucial role in the build-up of galaxy mass. However, the role of galaxy mergers extends far beyond their nominal delivery of mass, thanks to the dynamical processes triggered during the interaction and coalescence phases of the merger. In brief, tidal torques lead to the formation of a bar, which presents a mechanism through which angular momentum can be drained from the gas, which is then funneled towards the galactic centre. In turn, this increased central gas density has long been predicted by simulations to trigger starbursts, alter the distribution of metals and induce AGN activity (e.g., Barnes & Hernquist 1991; Springel et al. 2005; Cox et al. 2008; Hopkins et al. 2006, 2010; Torrey et al. 2012). In the low redshift universe ($z < 0.2$), large spectroscopic surveys of galaxies have led to a detailed assessment of the role of mergers amongst relative massive galaxies. For example, there is now unequivocal evidence that mergers can lead to increased star formation rates (e.g., Lambas et al. 2003, Nikolic et al. 2004, Scudder et al. 2012, Patton et al. 2013, Scott & Kaviraj 2014) and can trigger AGN activity (Ellison et al. 2011, Khabiboulline et al. 2014, Satyapal et al. 2014).

Nonetheless, many questions remain unanswered by the current generation of galaxy surveys, including the role of low mass galaxies (either with one another, or as accreted satellites) to these processes, the mechanism that leads to galaxy quenching (and the timescale for this shut-down of star formation) and how these effects evolve out to higher redshifts. Addressing these questions requires extending large, homogeneous, highly complete spectroscopic surveys to higher redshifts, and a carefully planned synergy with future multi-wavelength facilities. MSE is perfectly positioned to contribute to this potentially high impact field of galaxy evolution for several reasons. First, its high spatial completeness will yield a highly effective identification of spectroscopic galaxy pairs, and avoid many of the biases present in, for example, the SDSS (e.g., Patton & Atfield 2008). Second, the sensitivity to lower stellar masses will permit a more complete view of the merger process at low redshift, by allowing us to quantify the relative role of “minor” mergers, and dwarf-dwarf mergers in a statistically significant way. Finally, a combination of deeper detection thresholds and coverage of the near-IR will permit a homogeneous assessment of properties such as star formation rates and metallicities, using a consistent set of diagnostics across a broad wavelength range.

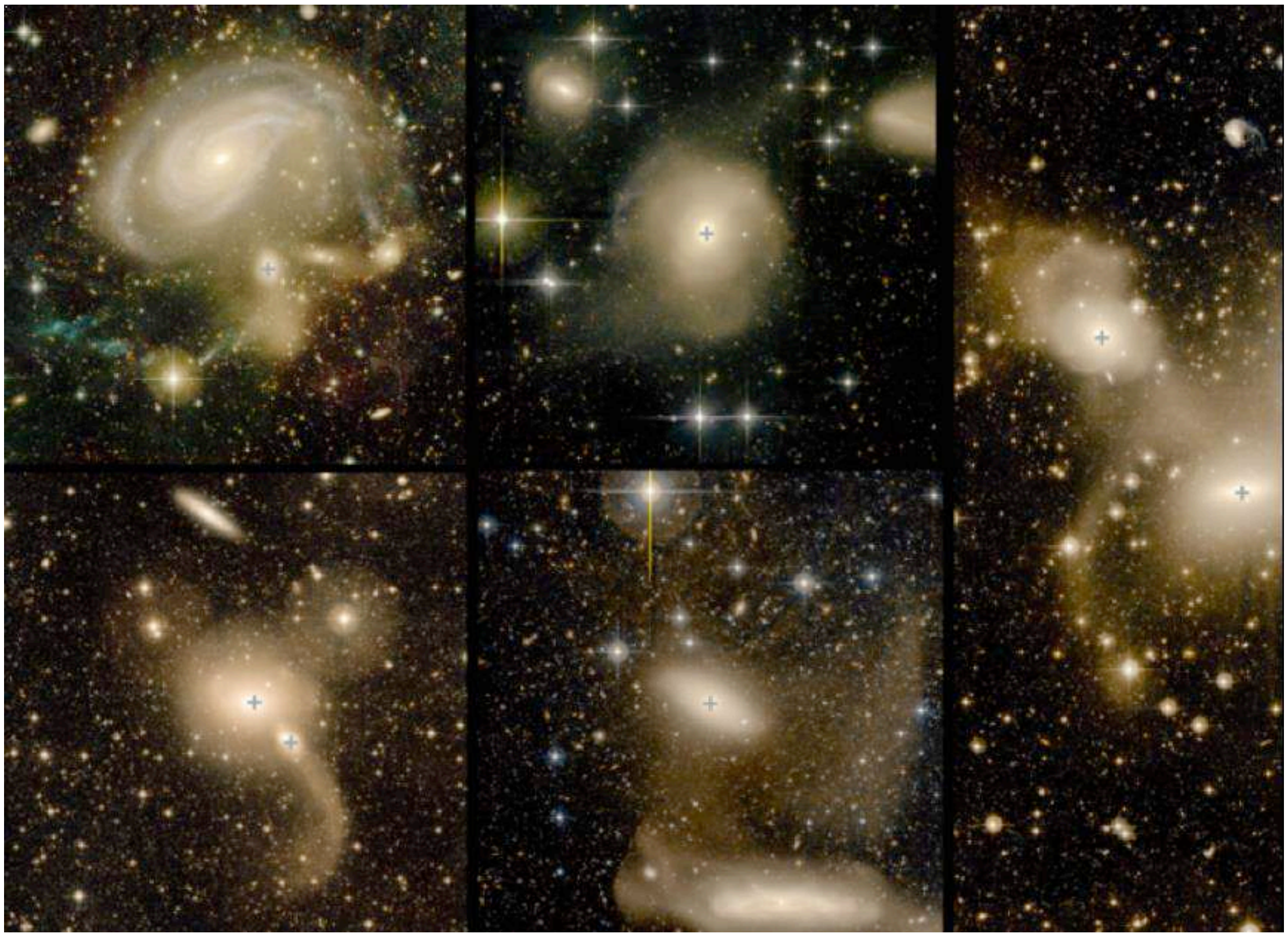

**Figure 59: Examples of early-type galaxies from the ATLAS3D survey that are currently involved in a tidal interaction with a nearby massive companion and exhibiting prominent tidal tails. Clockwise, composite true colour images of NGC 770, NGC 680, NGC 2698/99, NGC 5507 and NGC 5574/76 *(Figure from Duc et al. 2015).***

### 3.3.4.1 The identification of merger samples

In the most nearby galaxies, on-going, or recently completed interactions between galaxies are readily recognized through morphological disturbances, e.g., Figure 59. Thanks to very deep imaging, which is necessary in order to detect faint surface brightness features and stellar substructures, contemporary surveys are revealing that such structures are common in nearby galaxies (e.g., McConnachie et al. 2009, Tal et al. 2009, Martinez-Delgado et al. 2010, Duc et al. 2015). Detailed modeling of the morphologies of shells and tidal features permits a reconstruction of the progenitors and orbital details of the merger (Barnes 2016, Mortazavi et al. 2016, Privon et al. 2013, Barnes & Hibbard 2009). However, these detailed studies of individual, or small samples, of very nearby galaxies do not provide a global picture of the merger phenomenon, nor does deep imaging (currently) represent a feasible mechanism for the wholesale identification of large samples of mergers.

In order to more generally characterize both the frequency of galaxy interactions, and their impact on galactic properties, automated selection techniques that can be applied to the current generation of large imaging and spectroscopic surveys have been developed. These selection techniques generally fall into two categories. The first approach is to look for morphologically disturbed galaxies, through non-parametric assessment of galaxy morphologies, using metrics such as such as Gini-$M_{20}$ and asymmetry (Conselice et al. 2003, Lotz et al. 2004, Pawlik et al. 2016, Peth et al. 2016). Although traditionally applied to images obtained in the optical, these techniques can equally well be applied to galaxy images obtained at other wavelengths, such as the IR or even radio, where they are potentially even more effective at discerning persistent morphological disturbances (Lelli et al. 2014, Psychogyios et al. 2016). However, despite their widespread use, the automated morphological identification of mergers has several drawbacks. For example, the identification of a merger is sensitive to the both the depth of the image and the nature (e.g., mass ratio and orbital geometry) of the merger itself (Lotz et al. 2010a,b, Ji, Peirani & Yi 2014). Indeed, some mergers, such as those that are gas poor may have very weak asymmetries and minor mergers may barely perturb the more massive component. Moreover, there is no definitive separation in asymmetry parameter space between mergers and `normal' galaxies, such that contamination by clumpy or irregular galaxies is always problematic.

The alternative approach is to identify kinematic (or spectroscopic) pairs of galaxies from spectroscopic surveys. This technique selects likely mergers based on close projected pairs of galaxies with similar line-of-sight velocities (e.g., Patton et al. 2002, Carlberg et al. 2000, Woods et al. 2010, Patton et al. 2011, Wong et al. 2011, Lambas et al. 2012, Kampczyk et al. 2013, Davies et al. 2015). The availability of spectroscopic data is a further boon to the characterization of galaxies involved in the merger, permitting measurements of the star formation rate (SFR), gas and stellar metallicities and classification of AGN.

### 3.3.4.2 Lessons from the low redshift universe and future directions

The properties of low redshift galaxy pairs are now well characterized, thanks to a combination of large samples and supporting multi-wavelength data. **Figure 60** summarizes several of the key results that have emerged from studies of SFR (Nikolic et al. 2004; Alonso et al. 2007; Ellison et al. 2008; Scudder et al. 2012; Patton et al. 2011, 2013), gas phase metallicity (O/H, e.g., Ellison et al. 2008; Michel-Dansac et al. 2008; Scudder et al. 2012; Gronnow et al. 2015) and AGN fractions (e.g., Alonso et al. 2007; Woods & Geller 2007; Koss et al. 2010; Ellison et al. 2011; Sabater et al. 2013; Satyapal et al. 2014). In brief, these results have lent observational support to the predictions of simulations in which metal-poor gas from the disk outskirts flows to the centre simultaneously diluting the gas phase metallicity, triggering new star formation and providing material for AGN accretion.

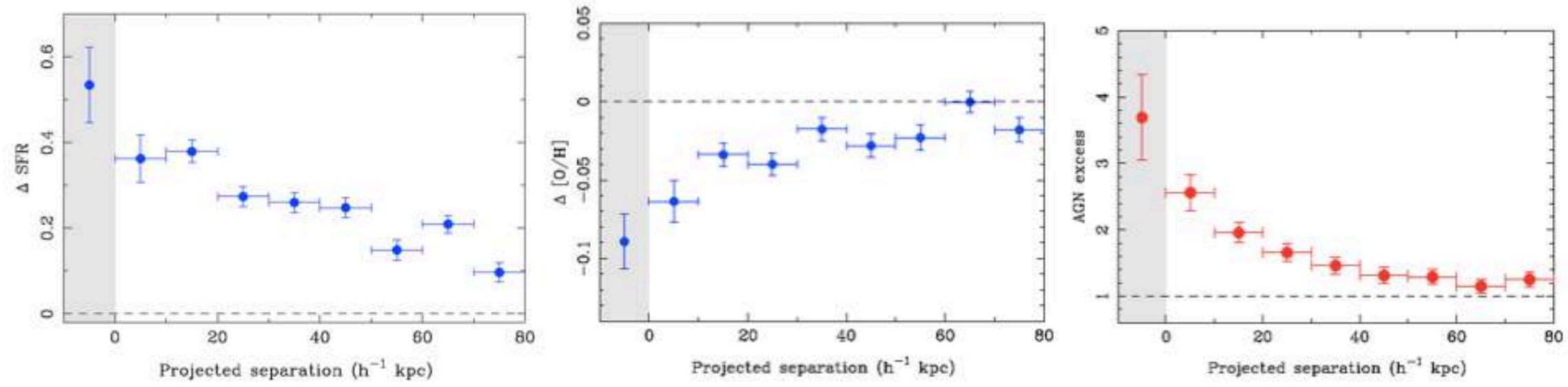

**Figure 60: Results summarizing the changes experienced by galaxies in mergers at z < 0.2, based on spectroscopic pairs selected from the SDSS, as a function of projected separation. Comparisons are made differentially with a**

**control sample of non-interacting galaxies matched in stellar mass, redshift and local environment. The grey shaded panel shows the results for fully coalesced "post-mergers". Left panel: The SFR enhancement in pairs relative to controls on a log scale. The typical enhancement of SFR in close pairs is a factor of a few, peaking in the post-mergers. Middle panel: The gas phase metallicity, relative to a control sample on a log scale. There is a gradual decrease in the metallicity towards smaller separations, with the most extreme dilutions occurring in the late stage post-mergers. Right panel: The frequency of optically selected AGN in pairs relative to controls. Again, the excess increases steadily with decreasing pair separation, peaking in the post mergers. *(Figure from Ellison et al. 2013).***

Despite the progress in our understanding of the impact of mergers on galaxy evolution, there are some aspects of this field that are poorly addressed with existing datasets. For example:

i. *What is the role of mergers amongst low mass galaxies?*

The limiting spectroscopic magnitude for the SDSS is approximately 18 in the r-band. Consequently, the majority of galaxies for which detailed properties such as SFR and O/H have been measured are limited to typically log $M_* < 9$. Moreover, the limited dynamic range of the SDSS means that minor mergers are relatively under-represented, with the pairs sample dominated by those whose masses are within a factor of a few of their companion (Ellison et al. 2008). These limitations mean that SDSS provides a good view of neither the majority of the minor merger (or satellite accretion) regime, nor the characterization of mergers between low mass galaxies. And yet, the dominance of low stellar masses by number indicates that majority of mergers should be dominated by such interactions at all redshifts (De Lucia et al. 2006; Fakhouri et al. 2010). Based on a small sample of ~100 low mass (log $M_* < 9.7$) galaxies in pairs, Stierwalt et al. (2015) have recently shown that "dwarf" interactions can also lead to enhanced SFRs, at a level similar to their more massive counterparts. Although the GAMA survey can improve upon these statistics, being sensitive to deeper magnitudes, a full census of the contribution to cosmic star formation ideally requires extending observations into the true "dwarf" regime. With its far deeper spectroscopic limits, MSE will provide the same level of statistical rigour as SDSS, but for the characterization of merger effects down to masses log $M_* \sim 6.5$. A survey of dwarf-dwarf mergers would also address whether interactions play a significant role in the creation of central supermassive black holes (SMBH), or can trigger accretion onto an existing SMBH.

ii. *What is the impact on the galactic gas reservoir during and after a merger?*

The combination of optical surveys with all sky missions at different wavelengths ranging from the UV (e.g., GALEX) to the mid-IR (e.g., WISE) has enabled a scientific yield well beyond the capabilities of a single wavelength regime. However, a significant piece of the puzzle has remained elusive, the measurement of atomic gas on industrial scales. Although the Arecibo ALFALFA Survey (Giovanelli et al. 2005) has provided a major boon in this direction, detections at 21cm continue to be counted in the thousands, rather than the hundreds of thousands, and the vast majority of low redshift galaxies in surveys such as the SDSS, 2dF and GAMA do not have HI masses available. This is a major short-coming, since the delivery, availability, consumption and possible expulsion of gas are likely to be key features of a complete picture of galaxy evolution (e.g., Bothwell et al. 2013). A further impediment for the study of HI gas in galaxy mergers is that the workhorse telescope for 21cm surveys, Arecibo, has a beam width of almost 4 arcmin, far too large to resolve close pairs. Targeted VLA observations of galaxy pairs is currently feasible, but extremely expensive and also does not provide a large control sample for comparison (e.g., Scudder et al. 2015).

The advent of the SKA (and its pathfinders) will revolutionize the field of galaxy formation by providing unprecedented quantities of HI masses for the galaxies surveyed by MSE. The impact for galaxy merger studies will be the ability to investigate the role of interactions on the galactic gas reservoir. Some simulations have predicted mergers may be responsible for replenishing, or even boosting (through halo gas condensation) the cool gas content of galaxies (e.g., Moster et al. 2011; Tonnesen & Cen 2012; Rafieferantsoa et al. 2015). On the other hand, the bursts of star formation known to take place in mergers would be expected to deplete the gas, and the triggered AGN could provide significant feedback. It is therefore far from clear whether mergers re-invigorate star formation, or may ultimately quench it. So far, no change in the HI gas fraction of mergers has been detected in the SDSS (Ellison et al. 2015). However, 21cm measurements for SDSS mergers number in the tens, and are affected by beam confusion. With MSE and SKA, it could be possible to determine *resolved* HI mass for thousands of galaxy mergers (and post-mergers) and allow a measurement of gas consumption to be made.

iii. *Mergers at high redshift*

After the success of surveys such as SDSS for characterizing mergers at $z < 0.2$, the next obvious step is the extension to higher redshifts. The best statistical assessment of merger induced SFRs beyond $z > 0.2$ comes from the GAMA survey (Davies et al. 2015), but the redshift range of GAMA extends only to $z \sim 0.5$. At higher redshifts, samples of spectroscopic pairs are frequently smaller (e.g., Xu et al. 2012; Schmidt et al. 2013;), or rely on either visual classification or less certain redshifts (e.g., Chou et al. 2011; Wong et al. 2011). Beyond $z \sim 0.5$, there is the additional complication that many of the key emission lines used for inferring SFRs, O/H and the presence of an AGN are shifted into the IR. Although alternative diagnostics exist, homogeneity is crucial for uncovering the often subtle changes observed in mergers; ideally we would like to be able to study the evolution of merger-induced star formation with a uniform set of data from $z \sim 0$ to $z \sim 1$ and beyond. MSE is potentially superbly positioned to complete our picture of galaxy mergers to such redshifts, where different gas fractions and morphologies could potentially dramatically change the efficiency of gas flows.

## 3.4 Stepping through the hierarchy: the nearest galaxy groups and clusters

### 3.4.1 Clusters as laboratories of galaxy evolution

**Science Reference Observation 7 (Appendix G)**

**Baryonic structures and the dark matter distribution in Virgo and Coma**

*Galaxy clusters act as (nearly) closed boxes, retaining the entire thermal and gravitational history of their assembly and growth. Here we propose an ambitious survey targeting all baryonic structures (mostly galaxies and globular clusters) brighter than $g \sim 24.5$ within the virial radii of the two nearest large clusters, Virgo and Coma. The brightest subset of sources ($g < 22$; some 50 000 sources in Virgo) will also be observed at high resolution. With this unprecedented complete dataset, we will map the dark matter distribution of halos on the*

*largest scales by examining the line of sight velocities of all baryonic tracer particles in these clusters. This will provide a* precise measurement of the dark matter mass distribution in the clusters, and reveal coherent structure in phase space that can be used to infer the accretion history within the cluster potential. *The MSE Virgo+Coma survey will be unrivalled for exploring all aspects of galaxy evolution, especially with respect to stellar populations and low mass galaxies. This is because the Virgo and Coma clusters afford us the best opportunity to study the galaxy population in its totality, from quiescent to star forming galaxies, from the high density cores to the low-surface brightness haloes, across a wide mass range, on all spatial scales, in a controlled environment in which all other baryonic substructures are also well characterized.*

Galaxy groups and clusters are extraordinarily important astrophysical laboratories. As the only places in the Universe where all relevant components — dark matter, hot gas, cold gas and stars — can be studied in a self-consistent manner, groups and clusters afford us the opportunity to observe the interactions between galaxies and the assembly of large scale structure, thus establishing an important link between predictive ΛCDM theory and the complex (but observable) hierarchical process of galaxy formation (Figure 61).

Galaxy groups – both within the infall regions of massive galaxy clusters, and more isolated systems – represent the earliest stage in this hierarchy. Within infalling groups, galaxies make the important transition from central galaxies in their own halo, to gas-starved satellite galaxies subject to a wide range of new, external forces. At the extreme of a continuum distribution of halo masses, galaxy clusters accrete individual galaxies and larger subclumps from their outskirts. With the most recent arrivals found predominantly near a cluster's virial radius, investigating the galaxy population as a function of clustercentric distance provides a way to reconstruct the history of structure assembly and its effect on galaxies (e.g. Balogh et al. 2000; McGee et al. 2009).

Clusters and groups have been extensively studied for over 30 years, and the field is now ready for the next technological/observational breakthrough. The situation is especially critical for galaxy groups: while the vast majority of stellar mass at the present day is contained in groups (with massive clusters representing just 2% of the stellar mass in the present-day universe, e.g. Eke et al. 2005), we know very little about the galaxy dynamics, stellar populations or the transformational physics that drive their evolution. In clusters, testing cosmological hydrodynamical simulations hinges on collecting detailed information — ages, metallicities, systemic velocities as well as internal kinematics — for baryonic structures from the cluster core out to several virial radii. As an example, the simulations predict a depletion of both hot and cold gas and a decline in the star-forming fraction in galaxies as far out as 5 virial radii (e.g. Bahé et al. 2013), but observations that extend to such large distances are limited to a handful of clusters or superclusters (e.g., Merluzzi et al. 2010, Mahajan et al. 2011, Smith et al. 2012b, Haines et al. 2011).

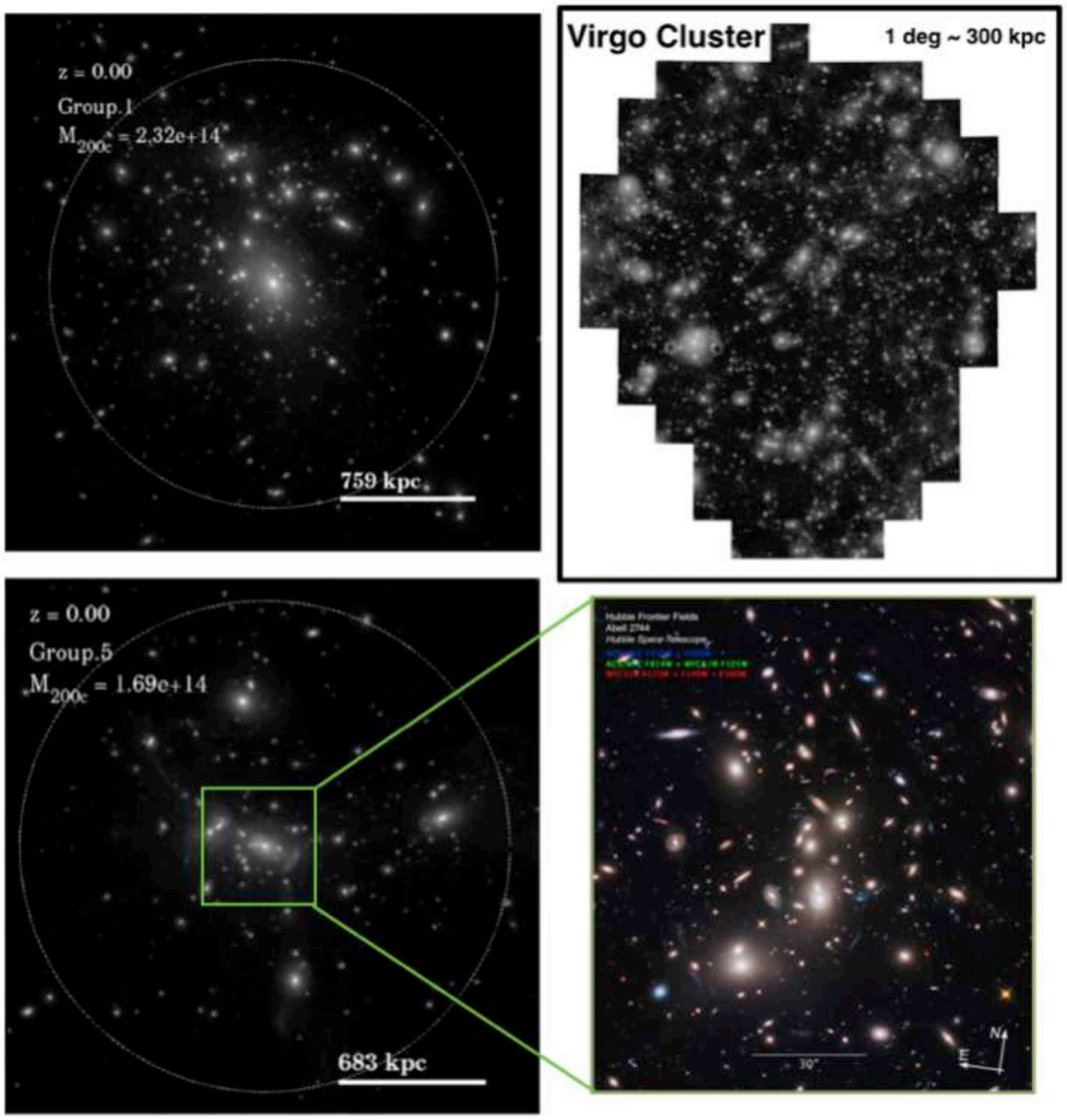

**Figure 61: Cosmological hydrodynamical simulations are now resolving galaxies well into the "dwarf" regime (M ~ $10^8$ $M_{\odot}$), matching the depth of the latest state-of-the art deep imaging surveys. The left two panels show two massive (virial mass in excess of $10^{14}$ $M_{\odot}$) clusters in the Illustris simulations (Mistani et al. 2016), with virial radius indicated by the dashed circle. The panels to the right show, at the top, a mosaic image of the Virgo cluster, extending out to one virial radius, obtained with CFHT/MegaCam; and, at the bottom, the HST image of the core of one of Hubble Frontier Field clusters, Abell 2744 (z ~ 0.308; for scale this region corresponds to the area within the green box in the simulated cluster to the left). By establishing cluster membership, studying the stellar populations, chemical enrichment histories and internal dynamics of thousands of cluster members, MSE will allow for seamless comparison between the simulations and the data.**

The much needed observational breakthrough is the availability of high quality, deep, homogeneous spectroscopy for massive datasets. The ability of MSE to reach faint objects, coupled with a vast multiplexing factor, makes MSE an ideal tool for understanding the growth of the hierarchy of galaxy environments. Cluster environments are a key component of the major science theme of MSE of linking galaxies to the large scale structure of the Universe; mongst the science investigations MSE will enable are:

- *A membership census in groups and clusters*

Spectroscopic redshifts are essential in building clean samples of baryonic structures

(galaxies and globular clusters) belonging to groups and clusters. In its low resolution mode (R ~ 2000), MSE can efficiently accomplish the task thanks to its massive multiplexing capability. Detailed cluster dynamical analysis is not only fundamental for understanding the interactions between galaxies, the intracluster gas and the dark matter halo. Galaxies can be used as tracers to measure the cluster mass and velocity anisotropy profiles, providing insight into the nature of dark matter, constraining alternative theories of gravity (Biviano et al. 2013), and exploring the halo formation process and history (e.g. Klypin et al. 2014). Finally, dynamical mass estimates of clusters play an important role in understanding the relationship between various observational mass tracers (e.g., X-ray, Sunyaev–Zel'dovich, optical richness) and the mass of the underlying dark halo, thus improving the accuracy with which the galaxy cluster abundance evolution can be measured and related to the predicted evolution of the dark matter halo mass function. These themes are explored in more detail in Chapter 4.

- *A study of integrated stellar populations, star formation rates (SFR) and enrichment histories, extending from the most massive systems to dwarf galaxies*.

  In high mass galaxies, known scaling relations between stellar mass, SFR and metallicity have long been used as important constraints to theoretical models. High SNR, medium resolution (R ~ 6500) MSE spectroscopy will extend these scaling relations to the dwarf galaxies regime ($M < 10^9\ M_\odot$), the building blocks from which more massive systems are formed. By targeting a statistically significant sample of thousands of dwarf galaxies, spanning a range of environments, it will be possible to explore the processes (e.g., dynamical interactions, black hole accretion, supernova feedback) driving chemical enrichment and regulating star formation. Internal galaxy dynamics — using IFUs or fiber bundles available on MSE as second generation capabilities — will elucidate the roles of tides, past merger events, and stripping processes in the evolution of cluster galaxies. Section 3.3.3 has a full discussion of the galaxy evolution science case for IFU capabilities on MSE.

- *A precision measurement of the Initial Mass Function*

  Knowledge of the IMF is a *sine qua non* condition for the interpretation of the integrated spectra of galaxies. A survey of rich clusters in combination with the survey described in SRO-06 will efficiently provide a definitive measurement of IMF-sensitive line indices for galaxies as a function of their mass, morphology, and across all possible environments.

- *High resolution (R ~ 20000) abundance analysis of compact stellar systems — compact galaxies, stellar nuclei, Ultra Compact Dwarfs (UCDs) and globular clusters (GCs)*. Several lines of evidence point to (some of) these systems as the end-result of extreme evolution — tidal stripping and threshing — within a cluster environment. MSE will allow us to test this hypothesis by providing precise estimates of ages, metal abundance, recent star formation and AGN activity for thousands of compact systems, investigating possible links amongst baryonic substructures that have traditionally been though of as distinct from each other.

In the next decade, MSE will be as transformational for cluster studies as SDSS was in the decade past. Most of our modern understanding about nearby galaxy clusters comes directly from the SDSS and 2dFGRS: the surveys homogeneity and control of systematics are what allowed us to begin to understand the interplay between galaxies and their environments. However, the limiting magnitude, moderate signal-to-noise ratio, moderate resolution and spatial incompleteness at small separations of SDSS are severe limitations in the utilization of SDSS data for studies of galaxy clusters. Although other surveys (e.g. GAMA) have probed deeper than SDSS, they only targeted relatively small areas of the sky or a selected few clusters. MSE will overcome these limitations. Most critically, its field of view and multiplex capability will allow MSE to directly couple detailed information on kpc scales or smaller (associated with star formation and galactic winds for example) to structure on scales of several Mpc and larger — a task that has historically posed severe observational challenges. Finally, MSE will provide *the* spectroscopic dataset to match the next generation of deep imaging surveys — Euclid, LSST and eROSITA — and complement current and next generation of ground based observatories, in particular ALMA, SKA, and JWST.

### 3.4.2 The Virgo and Coma Clusters

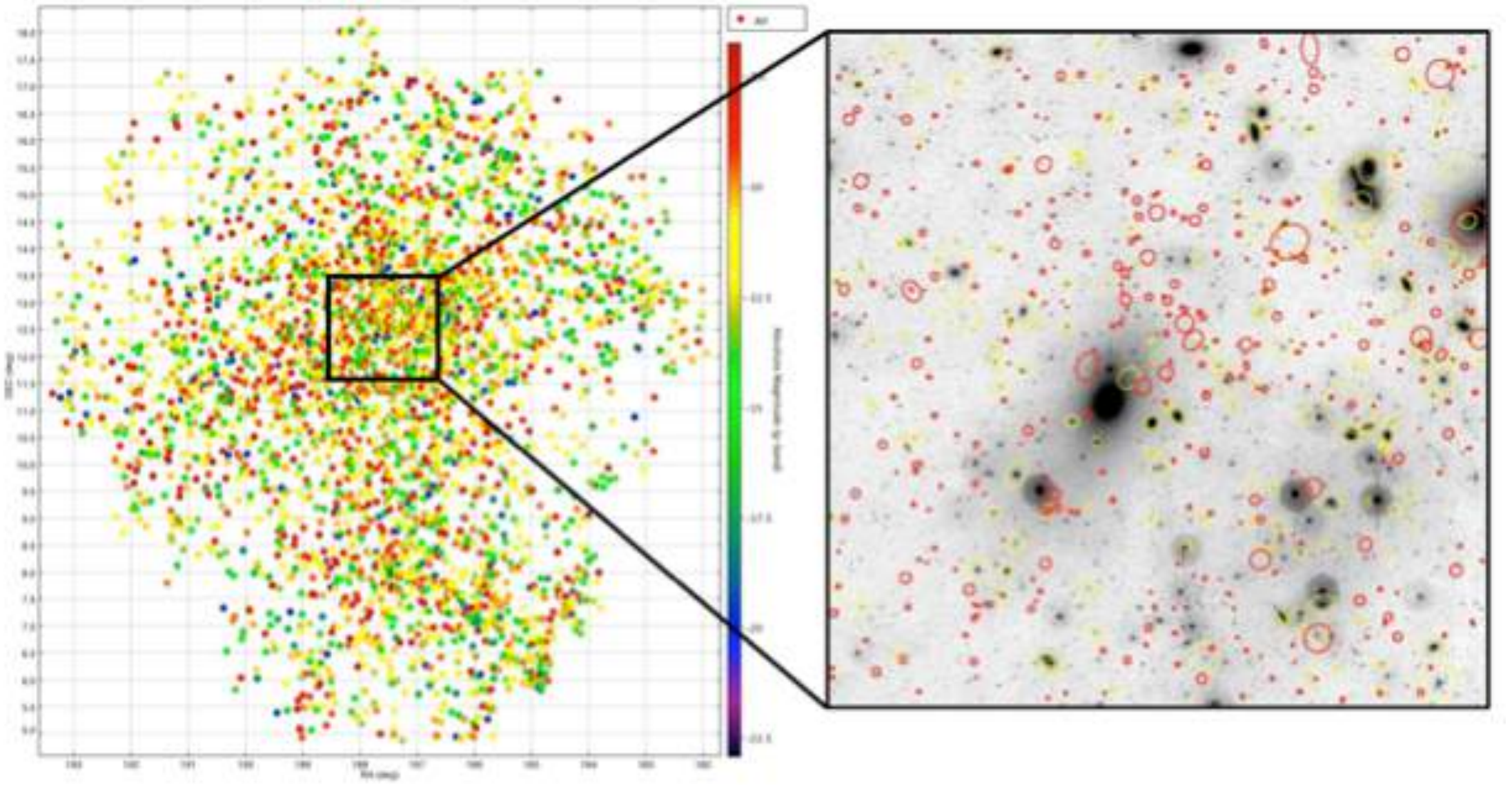

**Figure 62: The left panel shows the spatial distribution of galaxies that are cluster member located within one virial radius of the Virgo cluster, color coded according to their total magnitude as indicated by the bar at the right. The CFHT image of the core of the cluster (a 2 square degree area surrounding M87) is shown to the right, where galaxies are identified by yellow or red ellipses according to whether they were known or not prior to the CFHT NGVS survey.**

In addressing the problematics outlined in Section 3.4, local galaxy clusters — in particular the Virgo and Coma clusters — have special significance. According to simulations, the majority of galaxies that are already massive and evolved at $z \geq 2$ will end up in haloes with masses $M_h > 10^{14} M_\odot$ today (Poggianti et al. 2013); rich clusters at low redshifts are therefore the repository of the majority of the descendants of the massive galaxy population studied at high redshifts, and it is where we should look to trace their evolution. Additionally, while studies at high redshifts are largely confined to massive galaxies ($M_* > 10^{10.5} M_\odot$), it is only through detailed studies of the local volume that we can hope to understand the role of gas dynamics, cooling, star formation and feedback in the hierarchical assembly of baryonic substructures.

The Virgo and Coma clusters afford us the best opportunity to study the galaxy population in its totality, from quiescent to star forming galaxies, from the high density cores to the low-surface brightness haloes, across a wide mass range, on all spatial scales, in a controlled environment in which all other baryonic substructures are also well characterized. At a distance of 16.5 Mpc and with a gravitating mass of $M_{200}$ = 4.2 $\times 10^{14}$ $M_{\odot}$, the Virgo Cluster is the dominant mass concentration in the local universe, the centre of the Local Supercluster, and the largest concentration of galaxies within ~35 Mpc. Its proximity allows galaxies to be studied on scales of only 100pc (1 arcsec at the distance of Virgo). Coma has a gravitating mass of $M_{200}$= 2.6 $\times 10^{16}$ $M_{\odot}$ (Kubo et al. 2007) within its virial radius ($r_{200}$= 2.8 Mpc = 1.6 deg, compared to 1.6 Mpc = 5.5 deg for Virgo), but given its larger distance, a factor ~5 farther away than Virgo, subtends a smaller area on the sky. In Coma, MSE could observe most of the virialized region within a single pointing, and this greater efficiency nearly compensates for the longer integration times needed to reach the same physical depth as in Virgo. In Virgo, ~70 separate MSE pointings would be needed to cover the cluster out to one virial radius.

Both Virgo and Coma have been extensively studied at virtually all wavelengths. For Virgo, a new catalogue of likely cluster members (based on photometric and morphological criteria) reaching $g$ = −7 mag (stellar masses ~5 $\times 10^{5}$ $M_{\odot}$) is now available based on CFHT/MegaCam data (the Next Generation Virgo Cluster Survey, NGVS, Ferrarese et al. 2016, Figure 62). Detailed structural parameters, reaching surface brightnesses of 30 mag arcsec$^{-2}$ ($g$-band), exist for all cluster members. A 90% complete census of GCs, as well as many UCDs, is available within one virial radius (Figure 63). In the optical, Coma has been imaged by both ground (CFHT, Subaru) and space-based (HST) telescopes (Adami et al. 2006; Yamanoi et al. 2012; Carter et al. 2008), although not to the depth or spatial coverage that exists for Virgo. Extensive multiwavelength data is also available for both clusters.

Given the spatial extent of galaxies in Virgo and Coma, and the different density of various baryonic substructures within the cluster, a range of MSE observing modes and configurations will be needed to cover all science goals. These science goals are described in detail in SRO-07, and include:

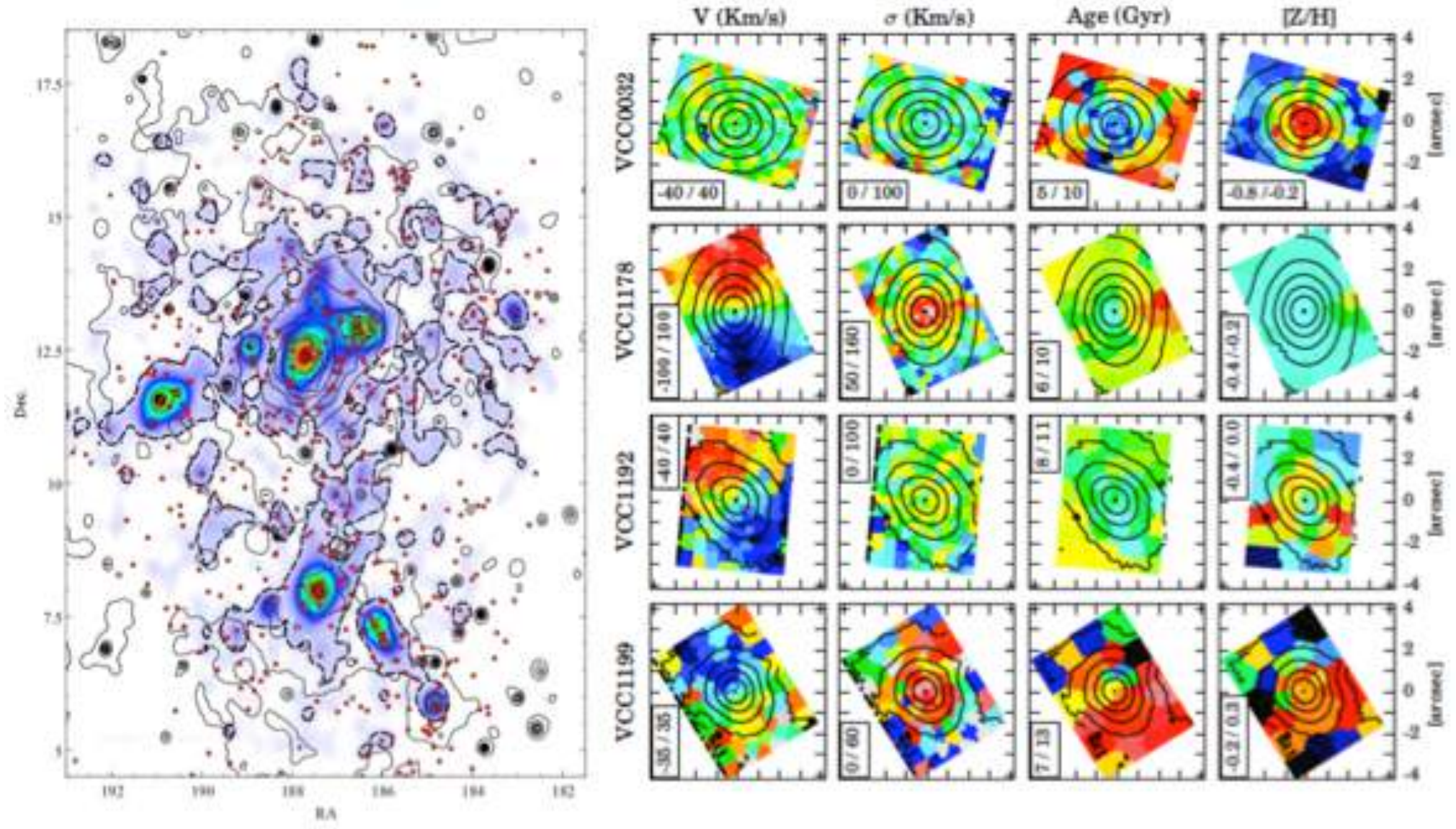

**Figure 63: Left panel: the surface density of GCs within one virial radius of the Virgo cluster (Durrell et al. 2014). Orbiting in the cluster graviational potential, GCs can be used as dynamical tracers to study the underlying mass distribution.** ***(Figure from Durrell et al. 2014)*****. Right panel: Gemini/GMOS IFU data for a sample of compact galaxies in the Virgo cluster (Guerou et al. 2015). Based on the ages, metallicities and internal dynamics inferred from the IFU data, the authors suggest a tidally stripped origin for these systems. These data are at present limited to a handful of (mostly) massive galaxies; extending them to a sample of several hundred galaxies, all the way to the dwarf galaxy regime, would allow us to identify evolutionary trends linking galaxies across a wide mass range and as a function of environment.** ***(Figure from Guerou et al. 2015).***

- *The star formation and enrichment history of dwarf galaxies (*$M_* \sim 10^7 - 10^9\ M_\odot$). As the building blocks from which more massive systems are assembled, dwarf galaxies are key to our understanding of hierarchical structure formation. However, the processes driving the chemical enrichment, the level and frequency of AGN activity, and the role of feedback and environmental effects on their evolution are still largely unknown. The effective radii of galaxies in the mass range of interests are 10 to 20 arcsec in Virgo, and 2 to 4 arcsec in Coma. Single fiber, moderate resolution (R ~ 6500) observations would yield gas phase metallicities (provided the necessary wavelength coverage can be achieved), star formation and enrichment histories, as well as provide diagnostic tests to assess the presence of AGNs. This would allow us to extend the known scaling relations between stellar mass, SFR and metallicity to the dwarf galaxy regime, and provide renewed constraints on cosmological models.

- *The nature of compact stellar systems.* In Virgo, a single MSE configuration per field would yield systemic velocities and high resolution spectra for all stellar nuclei, GCs and UCDs brighter than $g$ = 22 mag. This includes 400 stellar nuclei in galaxies with masses as low as $10^6\ M_\odot$ and spanning a range of environments, all known UCDs, and over 50,000 GCs within one virial radius of the cluster. At the highest MSE resolution (R ≥ 20000) the spectra would yield a detailed kinematical and stellar population analysis and allow us to seek evolutionary connections between the nuclei, globular clusters, UCDs, and the large scale properties of the host galaxies.

- *Mapping the dark matter distribution*. Faint objects orbiting in a galaxy gravitational

potential, including dwarf galaxies, GCs, planetary nebulae and UCDs, are powerful tracers of the underlying (total) mass distribution. At large distances from the galaxy cores, where the low surface brightness of stellar haloes makes spectroscopic observations prohibitively expensive, these tracers can be used to map the outer kinematics. A map showing the spatial distribution of over 50,000 globular clusters within the virial radius of the Virgo cluster is shown in the left panel of Figure 63 (Durrell et al. 2014). Sytemic velocities for these clusters (which would be a natural by-product of the program described in the previous bullet, but could also be achieved with a low resolution, single fiber condifuration) would lead to a precise measurement of the dark matter mass distribution, and also reveal coherent structure in phase space that can be used to infer the accretion history within the cluster potential.

- *A study of the internal kinematics of cluster galaxies.* Here and in the following examples, we will refer specifically to Virgo; the generization to Coma is trivial and indeed a Coma survey will generally benefit from the smaller projected extent of the cluster and higher density of targets. The Virgo cluster contains about 700 galaxies (DM subhalos) with stellar masses larger than $M_* \sim 10^9 M_\odot$ (see Figure 62). The internal velocity dispersions of these galaxies vary from $\sigma_v \sim 15 – 20$ km s$^{-1}$ at $M_* = 10^9 M_\odot$ to $\sigma_v \sim 300$ km s$^{-1}$ for the most massive galaxies. Spatially resolved, medium resolution spectroscopy (R ~ 2000 for galaxies that are more massive than $M_* \sim 6 \times 10^9 M_\odot$ and have $\sigma_v \sim 75$ km s$^{-1}$, and R ~ 6500 for the less massive systems), possibly using an IFU or fiber bundle, covering the galaxies from their core to the low-surface brightness regions out to several effective radii, would provide 2D maps of the age, metallicity and internal kinematics (see right panel of Figure 63). The information, available for several hundred galaxies spanning nearly three orders of magnitude in mass and a range of cluster environments, would be transformational. It would elucidate evolutionary trends along the Hubble sequence and allow us to emplore the nature of the perturbing mechanisms (tidal stripping, threshing, ram-pressure stripping, starvation, etc.) that shape galaxies within a cluster environment.

To appreciate the power of MSE for a spectroscopic survey of this sort, consider the time required to conduct an equivalent survey with existing or soon to be commissioned multi-object spectrographs. As detailed in SRO-07, a high resolution abundance analysis of all baryonic structures brighter than $g \sim 22.0$ within a virial radius of the Virgo cluster (100 deg$^2$) would require about 400 hrs of dark time and yield spectra for ~50,000 sources. It would therefore be possible to complete this survey in, at most, two observing seasons with MSE, assuming that Virgo is observable during dark time for an average of ~ 200 hrs per year. A comparable programme with the two VLT high resolution spectrographs, VLT-FLAMES/GIRAFFE (132 fibres, 0.126 deg$^2$ FOV) and VLT-MOONS (1000 fibres, 0.14 deg$^2$ FOV) would require 24 and 22 years, respectively. The same program using MMT-Hectochelle (240 slits, 0.013 deg$^2$ FOV) or Magellan-M2DF (128 slits, 0.02 deg$^2$ FOV) would be unfeasible due to the small FOV of the instruments.

## 3.5 Galaxies across redshift space

### 3.5.1 Redshift evolution of clustering and halo models

| **Science Reference Observation 8 (Appendix H)** |
|---|
| **Multi-scale clustering and the halo occupation function** |

*We propose an ambitious design of nine photo-z selected survey cubes that will allow MSE to measure the build up of large scale structure, stellar mass, halo occupation and star formation out to z = 5.5. By targeting (300 Mpc/h)$^3$ boxes, each volume will measure "Universal" values for an array of potential experiments. At lower redshifts we will directly observe halo abundances below $10^{12}M_\odot$, which means we can measure the occupation of halos and their abundance over a four decade range in halo mass, accounting for the majority of stellar mass in the low-redshift Universe. At higher redshifts our survey volumes will trace the transition from merger-dominated spheroid formation to the growth of disks, covering the peak in star-formation and merger activity. This combination of depth, area and photo-z selection is not possible without a combination of LSST and MSE. As such, MSE will be able to produce the definitive survey of structure, halos and galaxy evolution over 12 billion years.*

The dark matter halos that are believed to host massive galaxies make up more than 85% of the galaxy's total mass. An essential aspect of galaxy evolution is the underlying complex baryonic processes that take place within these dark matter halos. Processes such as gas infall and outflow, star formation, metal enrichment, negative feedback to star formation, black hole growth, and galaxy merging, should all work differently for different dark matter halo masses at different cosmic times. However, directly measuring dark halo masses either requires observations that are very challenging (especially at high redshift, e.g., Conselice et al. 2003; Swinbank et al. 2006; Mandelbaum et al. 2006), or rely on unusual and rare circumstances such as strong gravitational lensing events (e.g.,Treu 2010).

Thankfully, halo clustering is a strong function of halo mass, with more massive halos being more strongly clustered. Galaxies of different stellar masses and different star formation histories will correlate with dark matter halos of different masses and this relationship will evolve with time (e.g., Foucaud et al. 2010; Wake et al. 2011; Lin et al. 2012). Studies of galaxy clustering in large, local surveys have shown how clustering at $z \sim 0$ does depend significantly on several specific properties. These include luminosity (e.g., Norberg et al. 2002a; Zehavi et al. 2005), colour or spectral type (e.g., Norberg et al. 2002b; Zehavi et al. 2002), morphology (e.g., Guzzo et al. 1997), stellar mass (e.g., Li et al. 2006) and environment (e.g., Abbas & Sheth 2006). In recent years, deep spectroscopic surveys have made it possible to extend these investigations to $z \leq 2$, yielding results on how these dependencies evolve with time (e.g., Le Fevre et al. 2005; Coil et al. 2006; Meneux et al. 2008, 2009). More luminous and massive galaxies tend to be more clustered, and this segregation seems stronger at $z \sim 0$ than at $z \sim 1$.

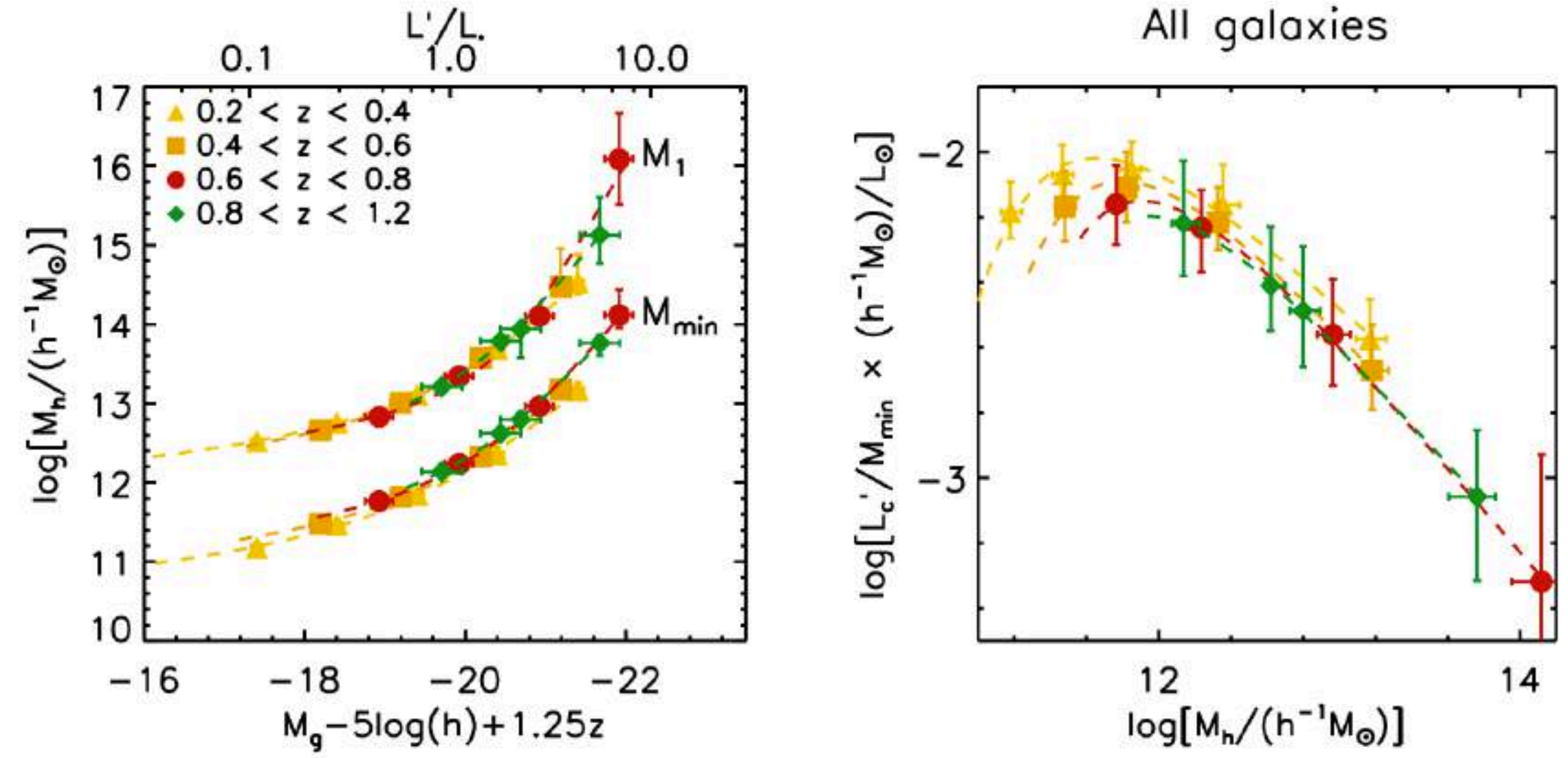

**Figure 64: Left panel: Halo mass estimates, $M_{min}$ and $M_1$, for all galaxy samples from the CFHTLS-wide survey, as function of luminosity threshold, corrected for passive redshift evolution to approximate stellar mass selected samples. The dashed lines correspond to the relation between central galaxy luminosity, $L_c'$, and the halo mass stated in Zehavi et al. (2011). Right Panel: Light-to-halo mass ratios, $L_c'$, /$M_{min}$, with identical parameters as those fitted with the same relation but as function of halo mass.** ***(Figure from Coupon et al. 2012).***

These results support a scenario in which the stellar mass of a galaxy is essentially proportional to the mass of the most recent dark matter halo in which it was the central object (Conroy et al. 2006; Wang et al. 2006). Investigating such results at higher redshift is essential, as the relationship between the galaxies and their halos is expected to be more straightforward. However, samples at z > 2 are limited and based mainly on multi-colour photometric selections or photometric redshifts and suffer from limitations in sample size, area covered and incompleteness. In order to compare the correlation function $\xi(r_p, \pi)$ and its projection $w_p(r_p)$ at high redshifts (z > 2) with similar measurements at z < 2, a deep, accurate, complete and unbiased sample of spectroscopic redshifts over a large area is required.

It is natural to use a halo-based prescription where galaxies form by the cooling of gas within bound and virialized clumps of dark matter. Galaxies occupy dark matter halos following a HOD model. In turn, the HOD fully describes the bias in the distribution of galaxies with respect to the underlying dark matter distribution, in terms of the probability distribution that a halo of virial mass $M_h$ contains N galaxies of a given type (Berlind & Weinberg 2002). HOD models can be used to directly investigate the changing relationship of dark matter and luminous matter, to separate contributions from satellite and central galaxies, and relate them to the masses of the dark matter halos. Until now, the majority of analyses conducted using HOD modeling to interpret galaxy clustering have been based either on large, low-redshift surveys such as the SDSS (e.g., Zehavi et al. 2011), or at higher redshifts, smaller (~ 1deg$^2$) deep fields such as COMBO-17, VVDS and COSMOS (e.g., Phelps et al. 2006; Abbas et al. 2010; Leauthaud et al. 2012). Photometric redshifts enable clustering measurements in the framework of the halo model over a larger redshift baseline and larger area (e.g., Coupon et al. 2012), but such studies suffer from a lack of accuracy at all scales required to constrain the models. In particular, the dark matter halos in the lower-mass regime suffer from the lack of precision on the line-of-sight dimension due to the reliance on photometric redshifts. Such imprecision tends to bias the results for faint galaxies toward being satellite galaxies in large halos, rather than central galaxies in low-mass halos.

The clustering approach to extract information on the dark matter halo masses has proven an efficient way to identify the physical processes driving galaxy evolution at the dark-matter scale. For instance, it is now established that the fraction of luminous-to-dark matter is a strong function of dark-matter halo mass (e.g., Foucaud et al. 2010; Wake et al. 2011; Leauthaud et al. 2012). This fraction decreases toward lower masses, because the dynamical time is long and supernovae feedback is strong, preventing efficient star formation. The fraction also decreases toward higher masses, where star formation is prevented by the cooling time and strong AGN feedback. Thus, the efficiency of star formation peaks at an intermediate mass range ($\sim 10^{12} M_\odot$ at $z \sim 0$), and the location of this peak evolves to higher masses at higher redshift. Figure 64, extracted from Coupon et al. (2012), demonstrates that HOD modeling can probe the light-to-halo-mass ratio. This figure also shows that the technique is limited in the lower mass regime, especially at high redshifts, due to the lack of reliable redshifts at the faint end.

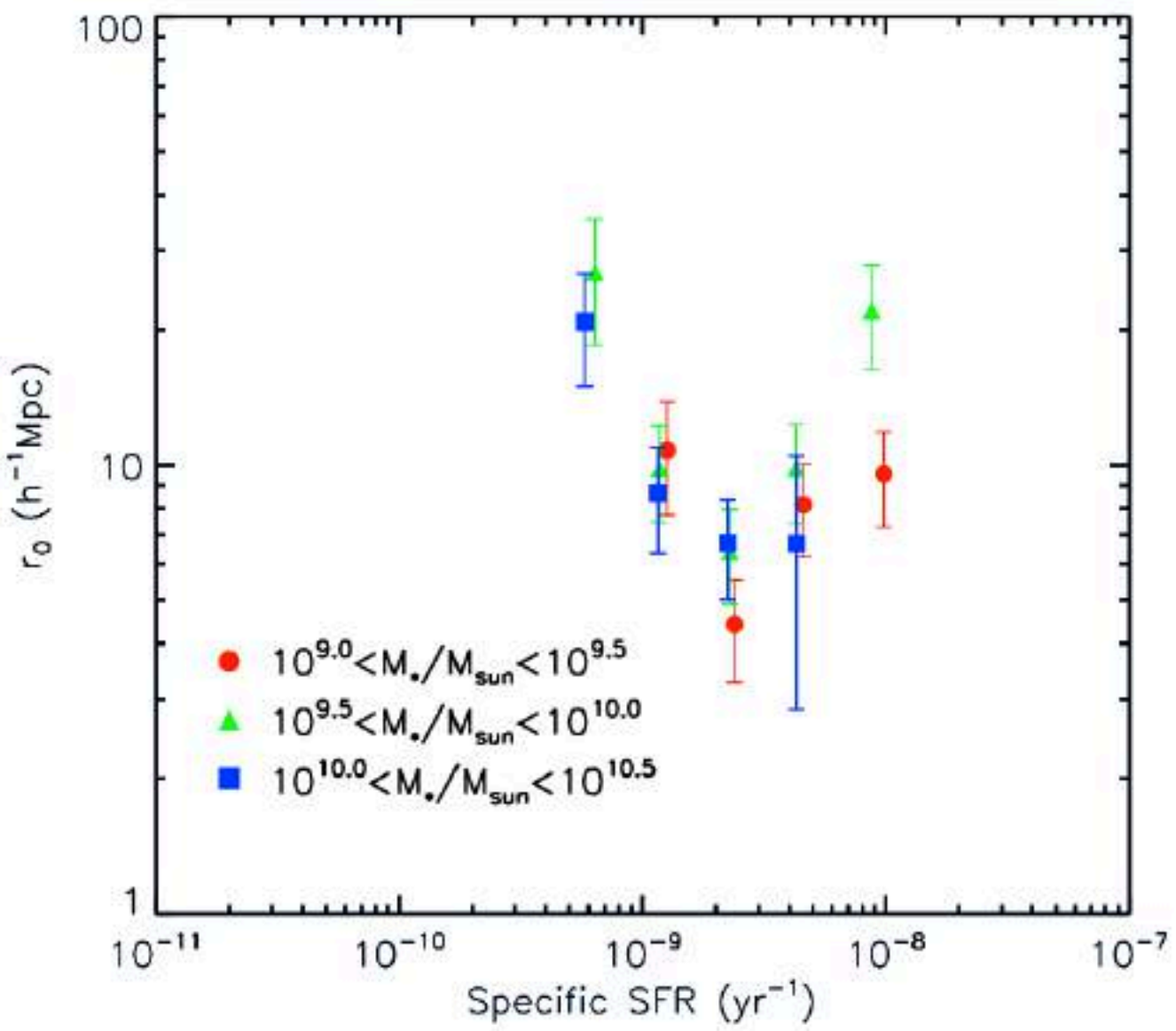

**Figure 65: The relationship between the correlation length, $r_0$, and specific star formation rate of sBzK galaxies in three different stellar mass bins: 9.0 < log($M_*/M_\odot$) < 9.5 (blue circles), 9.5 < log($M_*/M_\odot$) < 10.0 (green triangles), 10.0 < log($M_*/M_\odot$) < 10.5 (orange squares).** ***(Figure from Lin et al. 2012)*****.**

Interestingly, when looking at the clustering properties of galaxies selected according to their star-formation rates, even more complex phenomena can be explored. Figure 65 illustrates the evolution of the correlation length with different specific star formation rates (sSFR). Two distinct regimes are apparent, with galaxies having sSFR $\leq 2 \times 10^{-9}$ yr$^{-1}$ showing increasing correlation power with decreasing specific star-formation rates, and galaxies above this threshold show the opposite behaviour. The positive correlation observed at high sSFR reflects the "main sequence" evolution of galaxies, with galaxies in deeper dark matter halos showing more efficient star formation. On the other hand, the anti-correlation observed at low sSFRs can be understood as an environmental effect similar to that seen at lower redshifts, i.e., these are likely to be galaxies that are falling into the denser environments where their star formation activity is more effectively suppressed. Large samples of galaxies with accurate spectroscopic redshifts are required to make further progress in understanding these trends. For instance, in the context of HOD models, the previous correlation between sSFR and correlation length could be understood as a physical processes in action for satellite and central galaxies. It is also essential to probe higher redshifts to be able to understand when, and how, such relationships were established.

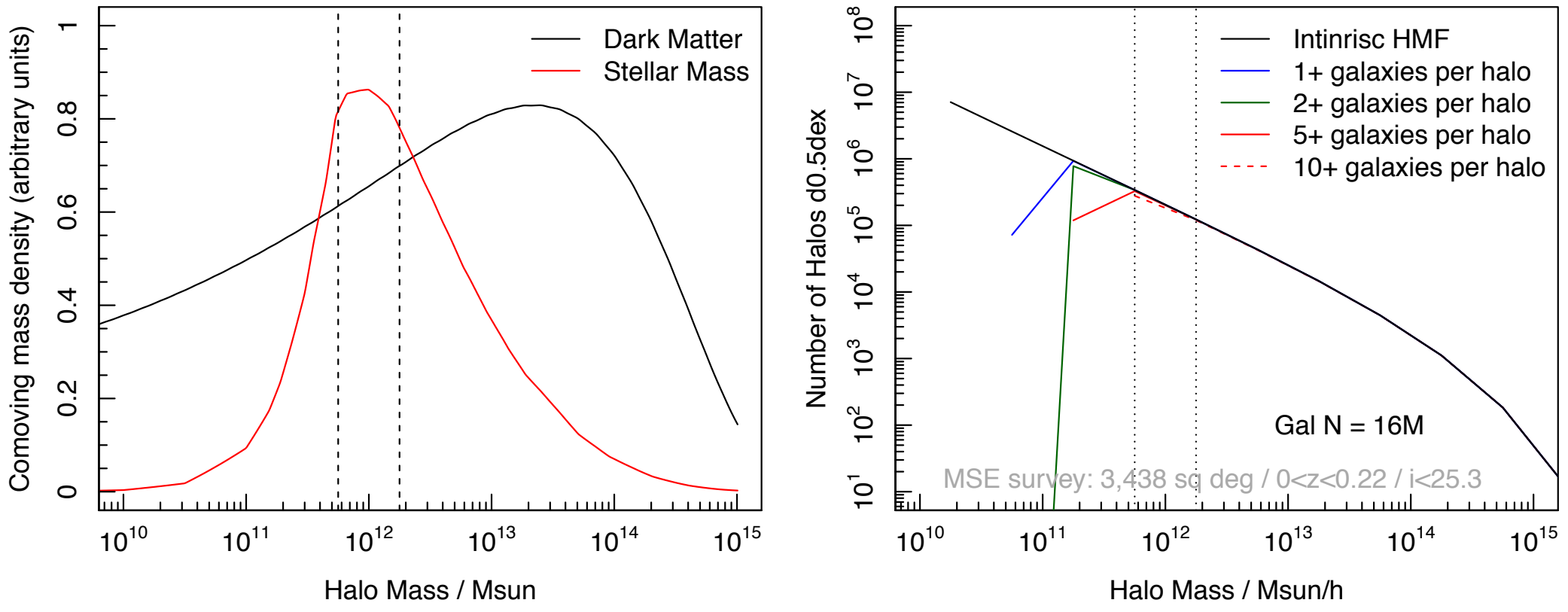

**Figure 66: Left panel presents the distribution of dark matter and stellar mass as a function of halo mass. It is clear that we expect it to be highly dominated by $10^{12}M_*$ halos. Right panel presents the expected galaxy occupation frequency as a function of halo mass. Our proposed survey will sample the $10^{12}$ $M_\odot$ halo mass regime with better than 5 galaxies per halo, allowing for per-halo dynamical mass measurements.**

The critical dark matter halo mass range in terms of stellar mass content is the decade around $10^{12}M_\odot$ (see left panel of Figure 66). This halo mass range contains our own Milky-Way halo and that of our nearest large spiral galaxy M31. To robustly detect groups with a low false-positive rate we require at least 5 galaxies to be observable within a group/halo, and to measure dynamical halo mass within a factor ~3 accuracy requires 10 or more galaxies to be observed within a group/halo (Robotham et al 2011). Also, it is important to reduce "cosmic variance" to a negligible level (Somerville et al. 2004) and so include a fair sample of rare, high- and low-density regions in the survey.

Combining the science cases mentioned above to coexist within the same survey volume (both sky coordinates and redshift window aligned) requires a large sky area and a robust photo-z selection in order to improve the efficiency of the volume overlap. Without a photo-z selection (i.e. only an apparent magnitude selection) we would naturally find more massive clusters at higher redshifts only, and we would have a comparatively small volume that contains the full range of $10^{12}M_\odot$ – $10^{15}M_\odot$ halos (see Figure 16 in Robotham et al 2011). Also, without a photo-z selection it becomes extremely inefficient to sample faint galaxies at low redshift, with fainter apparent magnitudes naturally pushing the peak in the n(z) distribution to higher redshifts.

SRO-08 describes a possible observing program with MSE that enables a vast range of science goals simultaneously in addition to tackling the key issues of HOD studies and galaxy environments, as a function of redshift to large look-back times, that employs a homogeneous and consistent survey strategy across all of redshift space. The basic goal is to observe 3200 square degrees of the Northern extra-galactic sky (overlapping fully with the proposed LSST survey region and substantially with the current SDSS footprint) between redshift 0 and 0.2 and down to a limiting magnitude of $i_{AB}$=23 selected from LSST standard depth multi-year survey (see SRO 7 survey S1-W). To efficiently probe down to low mass halos it is advantageous to observe as large an area as possible, and 3200 sq deg over the suggested redshift extent minimally contains a 300 x 300 x 300 Mpc/h co-moving cube. This means our volume is large enough that we reach Universal homogeneity in all three dimensions (Scrimgeour et al 2013, Driver & Robotham 2010). By observing to i = 23 within z = 0.2 we expect to be deep enough that essentially all $10^{12}M_\odot$ mass halos are both detected and have a reasonable mass estimate, with 10 or more galaxies identified.

By repeating the basic survey design (in particular the target volume) for this low redshift HOD and halo abundance focused science case, we can open up a new suite of science. Figure 67 shows a basic possible observing strategy, where a co-moving 300 Mpc/h cube is targeted in eight separate redshift windows using a mixture of LSST photo-z selection (S1 to S7) and Euclid photo-z selection (S8). Each survey region proposed would be targeted down to a different i/Y-band limit but with ~100% completeness within each volume (certainly better than 95%, i.e. ~SDSS). Such a suite of surveys allows detailed halo occupation modelling out to z = 5, spanning the rapid increase and slow decline in universal star-formation (top panel of Figure 67), the era of merger dominated mass build-up (z > 2), the transition into galaxy disk formation and in-situ star-formation dominating build-up (z < 2) and the epoch of rapid large-scale structure formation (z > 2). For obtaining robust "Universal values" for merger rates and star formation history the co-moving volume analysed and the stellar mass depth observed is key. By selecting common 300 Mpc/h cubes we will have sub per-cent sample variance independent of look-back time. By aiming to be complete to stellar masses at least 1 dex below $M_*$ for S1 to S4 we can definitely explore the interplay between mass build up through merger and star formation (see Robotham et al 2014 for a z=0.2 version of this experiment). Beyond this range we are limited by the high quality photo-z i < 25.3 sample provided by LSST. Despite this, we can still probe the dominant component of stellar mass ($M_*$) and its halo occupation distribution out to z = 2 with S5 and S6. This takes our galaxy evolution analysis out to 10 Gyrs in consistent co-moving volumes that have high statistical quality. Further, OII becomes visible out to z = 3.8 i.e. it becomes a common emission feature that S1 to S8 can all observe, offering a consistent star-formation tracer over nearly 12 Gyrs from a single facility. Access to both Ly-α and OII makes possible direct measurements of the feedback of gas into the inter-stellar medium. We will also be able to observe Hα, and construct a full BPT diagram out to z = 2 (i.e. S1 to S6).

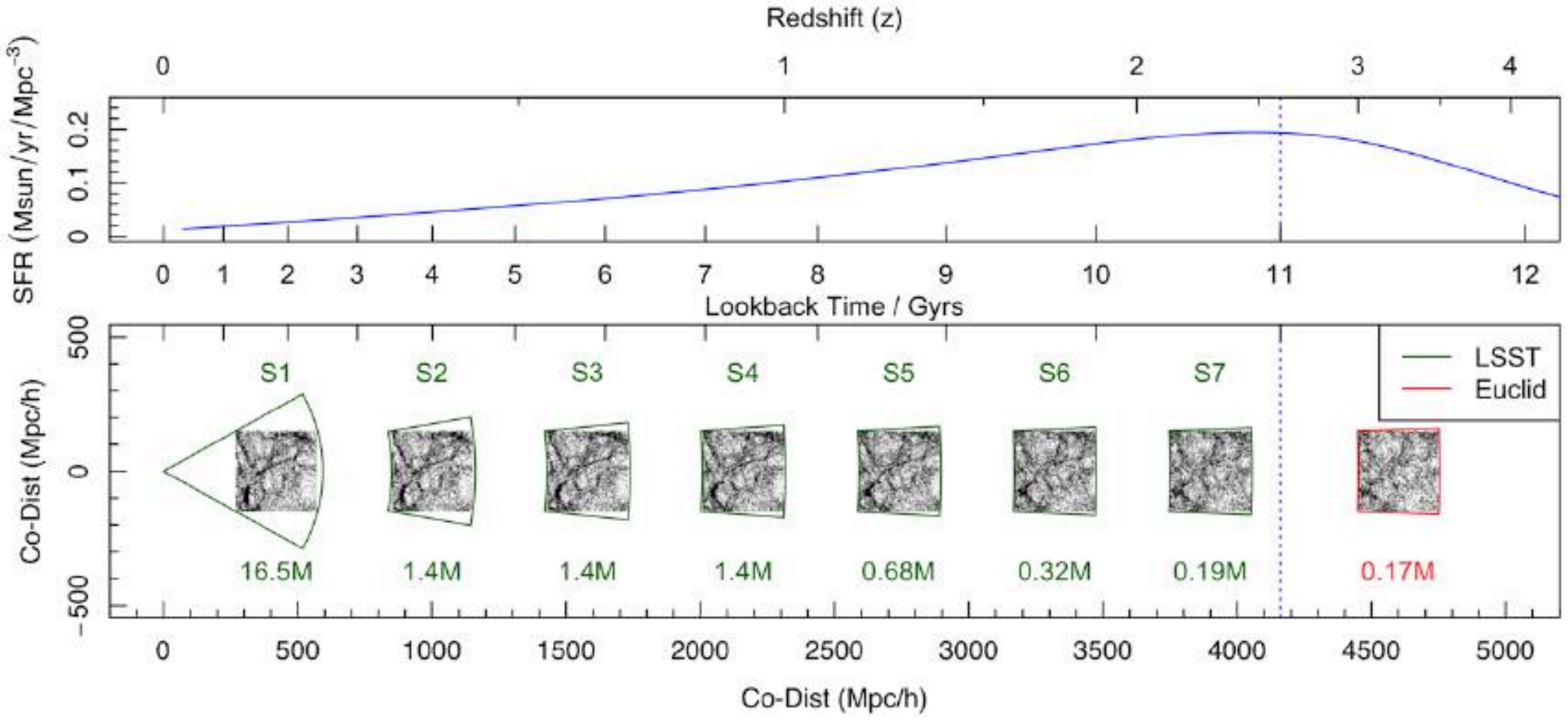

**Figure 67: A proposed range of equi-comoving MSE survey cubes (S1 to S8). The top panel shows the cosmic SFH from Hopkins & Beacom 2006. The lower panel shows the proposed surveys, showing the growth in large-scale structure in TAO simulated (50 Mpc/h)$^3$ survey cubes.**

It is clear that such a survey suite opens up a vast amount of science that is not accessible to any other facility. A non-exhaustive summary is given below:

i. Measure halo occupation (through group finding) below $10^{12}M_{\odot}$. This combined with an HOD analysis will definitively uncover the interplay between stellar mass and halos, and will put any detailed study of individual halos (the future Milky Way and M31 archeology studies described in Chapter 2 into a proper cosmological context.
ii. Directly observe the evolution of the massive end of the halo mass function out to z = 1 (half the age of the Universe). This is step beyond cluster count cosmology, and will be conducted with a homogenous selection with LSST (the selection function is one of the major limitations of current cluster cosmology work).
iii. Study the evolution of the large-scale structure and cosmic web out to z = 5 with a consistent galaxy tracer with homogenous bias.
iv. Study the merger rate and star-formation history for all galaxies down to $M_*/10$ (covering the converged majority of stellar mass) out to z = 1.
v. Study the close clustering and merger rate of massive galaxies ($10^{11}M_{\odot}$) out to z = 5.
vi. With photometry from LSST we will be able the majority of the stellar component of the cosmic spectral energy distribution (CSED) out to z = 2.
vii. Together with other next generation telescopes (most significantly LSST, Euclid, WFIRST, SKA, the VLOTs), these surveys will have a vital role in:
    a. Measuring morphological evolution to z = 1
    b. Studying the interplay between gas, dust, stars and environment
    c. Providing IFU targets for 30+m telescopes for well understood samples at multiple epochs
    d. Offering the sample superset for high S/N observations of individual galaxies in order to ascertain the evolution of metals through cosmic time.

The main competition to MSE in doing such a combined survey is Subaru/PFS (Takada et al 2013)

and VLT-MOONS (Cirasulo et al 2012). PFS will have a spectral range 380 – 1300 nm, and it will nominally be conducting z < 1, i < 21.5 and 1 < z < 2, i < 23.9 photo-z pre-selection surveys (based on shallower HSC data). In both regimes, the science programs MSE will undertake are at least a magnitude deeper (a factor ~3 in stellar mass) and over much larger areas (fixed 16 square degrees for PFS, 400 to 30 square degrees over the same range for MSE. The huge advantage of MSE compared to all of the proposed PFS surveys is that we will be extending to the scale of homogeneity in all three dimensions. For many of the science cases laid out here (and foreshadowed on a smaller scale by PFS) the definitive measurement will be made by MSE, with no further appreciable improvements to be made by moving to larger survey volumes (since many extragalactic measurements are dominated by sample variance not Poisson statistics of the sample itself).

VLT/MOONS will have a spectral range 680 – 1800nm and is focusing on surveys at z > 1; tying together low and high redshift surveys using equivalent tracers at all redshifts is therefore a unique strength that is possible only with MSE. Further, MSE has a potentially huge advantage in its ability to use the LSST and Euclid/WFIRST photometric source catalogues. These telescopes will only become available post 2020, and offer a paradigm shift in available optical image quality over large areas. In particular, using LSST allows us to conduct surveys S1 (3200 square degrees) to S7 (22 square degrees) with identical photometry and photo-z technique. Such homogeneity will improve the robustness of almost every type of analysis made. Given the fast survey speed of MSE, once LSST and Euclid data become available no other facility will be able to keep up with the observing speed of MSE, i.e., it will dominate the next generation of spectroscopic surveys from 0 < z < 4.

### 3.5.2 The chemical evolution of galaxies across Cosmic time

**Science Reference Observation 9 (Appendix I)**

**The chemical evolution of galaxies and AGN over the past 10 billion years**

We propose a high-SNR (SNR>30) spectroscopic study of carefully selected volume-limited samples from the present epoch to the peak of star-formation 10 billion years earlier. Each volume-limited sample should contain 10k galaxies uniformly spanning the broadest possible stellar-mass range, and in regular 1Gyr intervals (i.e., 10 volume-limited samples of 10k galaxies resulting in a total survey of 100k galaxies). The volume-limited samples could be selected either from photo-z catalogues identified by LSST, EUCLID or WFIRST, or from a redshift pre-survey of the selected regions (i.e, such as that proposed by Science Reference Observation 8). Key requirements for this survey are high throughout, near-IR spectral coverage (extending as red-ward as physical possible), but relatively low to modest resolving power (R~1000). Integration times are likely to extend up to ~50hrs for the faintest highest redshift systems. Hence, the full study will require 50 – 100 nights of observations (assuming a multiplex factor i.e., ~1000 targets per field-of-view), resulting in a legacy sample for multiple and diverse studies of the chemical evolution of galaxies and AGN.

The chemical evolution of galaxies and AGN over cosmic time, and in particular the gas- and stellar-phase metallicities, is a topic of significant current attention (e.g., Zahid et al. 2014, and references therein). Relatively strong evolution has been reported in the stellar mass – metallicity relation (Tremonti et al. 2004) from nearby samples to those close to the peak of cosmic star formation history. See, for example, samples assembled by Zahid et al. (2013) and Figure 68. However, the sample sizes are relatively modest and arguably biased towards the highest star-forming systems at each epoch. This is because the selection is typically made short-wards of the 400nm break for the very high-redshift intervals, and as such probes only the most UV-bright galaxies. Locally we know that the specific star-formation rate, stellar mass, and metallicity form a fairly complex surface (e.g., Lara-Lopez et al. 2013a). Moreover, gas-phase metallicity (usually the only measurement accessible in high-z surveys) is harder to interpret than the stellar-phase metallicity, as the former are very sensitive to recent infall and outflows. Ideally one would wish to study the sSFR – stellar mass – metallicity trifecta using both gas and stellar- phase metallicities, for representative populations at regular time steps. By assuming the earlier population evolves into the later population one can build up a complete picture of the global evolution of both the gas and stellar-phase metallicites over time, and the impact of both sSFR and stellar mass on metallicities (and vice-versa) at a range of epochs. The required infall and outflow models can then be determined through comparisons to simple analytic models, with potentially some additional constraints coming from the comparison of the emission line-widths (sensitive to dynamics and outflows) to the absorption line-widths (sensitive to just the dynamics) using multi-Guassian line-fitting (e.g., McElroy et al. 2015).

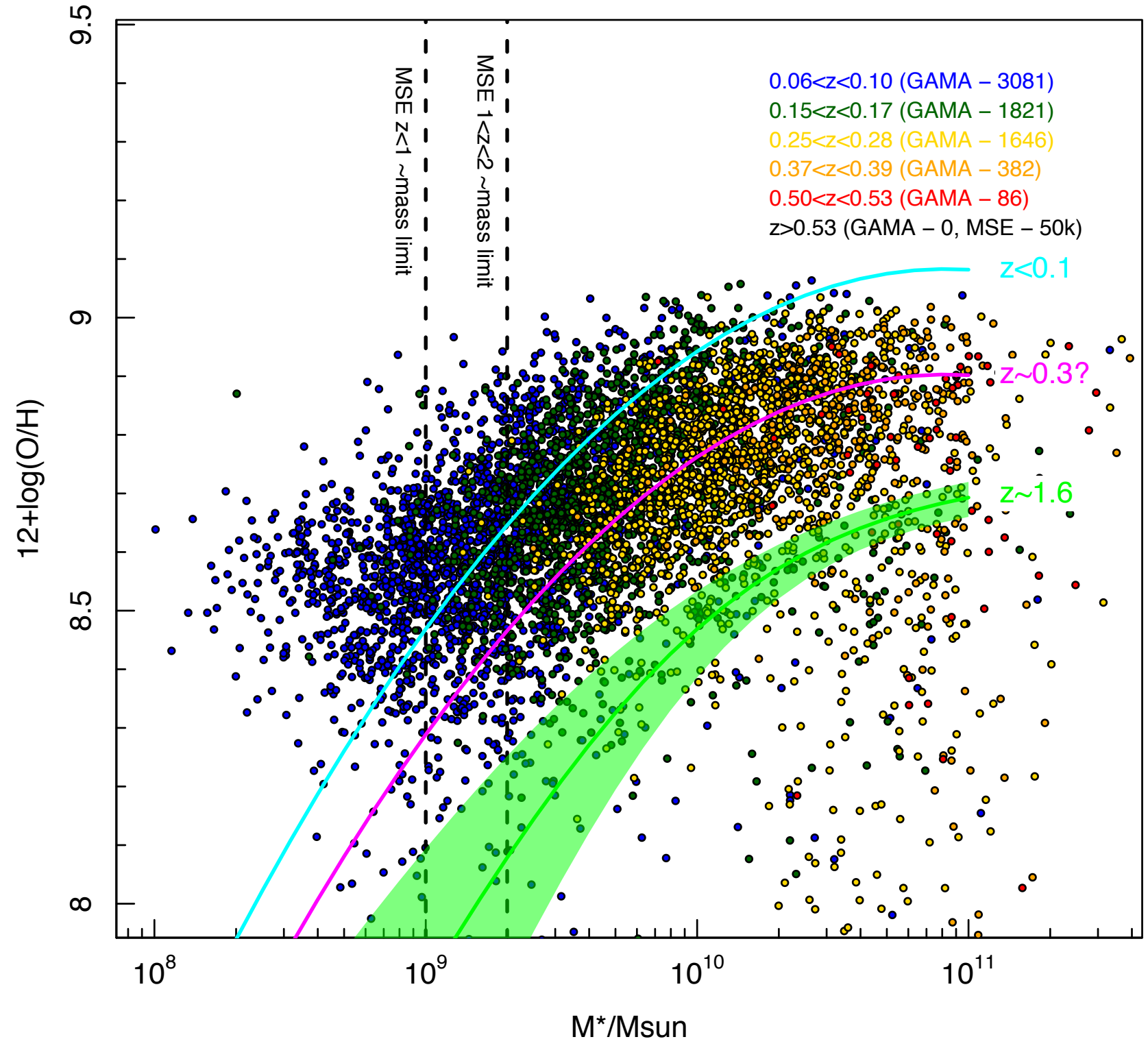

**Figure 68: The mass-metallicity relation seen by GAMA and with the redshift ranges discussed in SRO-09 shown ($0.1Z_{sol}$ lies at the lower limit of the plot). MSE will probe mass limits similar to GAMA at z<0.1 and z<0.2 but out to z < 1 and z < 2 respectively. The z < 0.1 mass – metallicity relation from SDSS+GAMA (Lara-Lopez, et al. 2013a) is displayed as the cyan line, the magenta line show the same relation normalized to the 0.25<z<0.39 sample and the green polygon shows the stacked galaxy z∼1.6 relation from Zahid et al. (2013). These observations clearly predict**

**a strong evolution in the M – Z relation as a function of redshift. MSE will probe this evolution using individual galaxies, in a comparatively unbiased sample out to z∼2.**

The measurement of stellar- phase metallicities, essential for the above analysis, require significantly higher-S/N spectra than are typically attained in surveys such as zCOSMOS, VVDS and Deep2, since they are reliant on absorption rather than emission line measurements. Modeling of line-broadening through dynamics and outflows also requires significant signal-to-noise levels in the spectral features and for the features to be clean (i.e., away from night sky lines). As one pushes towards higher-redshift this later aspect becomes harder because of the preponderance of telluric features, etc. This can potentially be overcome by using UV metal-lines (e.g., Sommariva et al., 2012; Figure 48 showed the drift of key spectral features with look-back time). MSE is able to provide essential data to further this field by constructing a high-SNR sample of a series of well selected and sufficiently extensive samples to address a number of compelling questions by:

i. Providing a fully empirical description of the sSFR-stellar mass metallicity relation for systems with masses greater than $M_* \sim 10^9 M_\odot$ to the peak of the cosmic star-formation history at z ∼ 2 (see Figure 43). This empirical relation would include measurements of both the gas and stellar-phase metallicities, removing any biases due to recent gas infall/outfall. Such an empirical result would provide an ideal benchmark for the further calibration and development of numerical (hydro-dynamical) and semi-analytic models, which as yet have very few high-z metallicity constraints.
ii. Quantifying the decline in the abundance and frequency of AGN activity within the normal galaxy population at all redshifts to z∼2 (the epoch of peak AGN activity) using high SNR emission line diagnostics, such as Baldwin, Phillips & Terlevich (BPT) diagrams and WHα versus [N II]/Hα (WHAN) line strength diagnostics. This would enable the exploration of the co-evolution of galaxies and AGNs by measuring both the star formation history of the galaxy population, the decline in the luminosity density of AGN and any coupling between these two.
iii. Measuring and comparing the gas-phase and stellar-phase metallicities to constrain the degree of pristine gas infall/outflow as a function of look-back time. Some models suggest galaxies initially evolve according to a closed-box model after which fresh pristine gas arrives significantly reducing the gas-phase metallicity. The clear signature of a closed-box model is a stellar-phase metallicity approximately half that of the gas-phase. Detailed comparisons of the stellar and gas phases will provide constraints on the degree of infall and outflow that can then be monitored over the range of time sampled.
iv. Performing detailed stellar population measurements to determine the progression of stellar-evolution. Earlier epochs of star-formation are typically hidden within later waves of star-formation, however detailed analysis of the complete spectrum on a pixel-by-pixel rather than line basis can reveal the full star-formation histories. These histories of later volume-limited samples need to mesh with what is seen in the earlier volume-limited samples enabling detailed test of our star-formation models. It is also worth nothing that such an analysis as outlined in iii and iv, combined with ASKAP/SKA HI measurements (at the lower-z end), will provide a comprehensive test of the star-formation history and baryonic composition of individual galaxies, placing strong constraints on likely galaxy evolution scenarios.
v. Using multiple star-formation tracers (e.g., Hα v UV v mid or far-IR) to study the onset of

star-formation with dynamical separation and/or environment.

To identify AGN, star-forming, and inert galaxies, measure their chemical composition, and to achieve the objectives listed above, requires not only a robust redshift measurement, but also sufficient signal-to-noise spanning a broad range of emission and absorption features. To robustly distinguish AGN from star-forming galaxies requires SNR > 5 in the rest-frame 370 – 680nm (OII through to Na), to measure gas-phase metallicity (0.1 $Z_{\odot}$), and to measure star formation rates to modest levels (1$M_{\odot}$/yr), requires SNR ~ 10 – 15 (Choi et al. 2014). To recover stellar-phase metallicites using, for example, Lick indices (400 – 650nm) or UV metals lines (130 – 200nm) to 0.1 $Z_{\odot}$, we require S/N ~ 20 – 30. Here, MSE can be a game changer.

Massive samples are not critical, although in order to fully map out a 3D structure in sSFR, stellar mass and metallicity does require volume-limited samples of ~10k systems. This assumes an average of 20 galaxies per bin and ~ 8 bins each in SF, stellar-mass and metallicity. SRO-09 describes a possible MSE observing program where $S/N_{continuum}$ ~30 spectra from 0.4 – 1.8 micron are obtained for 10 samples each containing 10k galaxies. These samples are spaced at regular Gyr time steps from the present epoch to a look-back time of 10 Gyrs. *This samples provies from the peak of the star-formation activity at z~2, to the present epoch*. In total, the proposed project would target 100 000 galaxies, which would be either spectroscopically pre-surveyed with MSE itself, or selected from the upcoming photo-z surveys of LSST, EUCLID and WFIRST.

The potential of MSE here is to provide significant statistics to have a transformation impact and fully map the evolution of the mass-metallicity relation to the peak of the star-formation activity at z~2. Existing facilities are capable of constructing modest samples but only with the high throughput, high fiber density and extensive spectral coverage that MSE can provide, do we have the potential to move beyond detections to robust statistics. These observations will span, not just the most luminous systems, but those systems which contain the bulk of the stellar mass at the current epoch (and hence the bulk of metals in the Universe).

# Chapter 4 Illuminating the Dark Universe

## 4.1 Chapter Synopsis

- *The Universe in absorption: linking stars and galaxies to their surrounding media*
- *The evolution of supermassive black holes through multi-object reverberation mapping*
- *Exploring the Universe through massive transient datasets*
- *A precision machine to measure the dynamics of dark matter halos on all spatial scales.*

*Science Reference Observations for Illuminating the Dark Universe*

- *(SRO-03: The formation and chemical evolution of the Galaxy)*
- *(SRO-04: Unveiling cold dark matter substructure with precision stellar kinematics)*
- *SRO-10: Mapping the inner parsec of quasars through reverberation mapping*
- *SRO-11: Linking galaxy evolution with the IGM through tomographic mapping*
- *SRO-12: A peculiar velocity survey out to 1Gpc and the nature of the CMB dipole*

MSE will cast light on some of the darkest components of the Universe. The backbone of the large scale structures in which galaxies are embedded is dark matter, that is invisible in electromagnetic radiation. Embedded alongside galaxies in these large scale structures is the intergalactic medium, a major reservoir of baryons in the Universe but one which is notoriously difficult to observe directly in the diagnostic-rich optical wavelength range, other than through its absorption effects on background galaxies. Within galaxies, the interstellar medium is similarly difficult to detect, and the central supermassive black holes are generally only probed through indirect techniques. MSE will provide unprecedented perspectives on each of these components of the dark Universe, and provide a suite of capabilities ideally suited to providing critical data for next generation cosmological surveys.

Specifically, MSE will provide tomographic reconstruction of the three dimensional structures of the Galactic interstellar medium (ISM) and the intergalactic medium (IGM) through high signal-to-noise absorption studies along millions of sight lines, that are orders of magnitudes larger than that which is currently possible. Whether focused on stellar or galactic science, the relations between stars and galaxies and their surrounding media is a key issue in understanding the cycling of baryons. In the extragalactic regime, MSE can target the IGM at $z \sim 2 - 2.5$, to reconstruct the dark matter distribution and associated galaxies at high-z, allow detailed modeling of galactic scale outflows via emission lines, investigate the complex interplay between metals in galaxies and the IGM, and provide the first comprehensive, moderate resolution analysis of a large sample of galaxies at this epoch.

MSE will allow for an unprecedented, extragalactic time domain program to measure directly the accretion rates and masses of a large sample of supermassive black holes through reverberation mapping. This information is essential for understanding accretion physics and tracing black hole growth over cosmic time. Reverberation mapping is the only distance-independent method of measuring black hole masses applicable at cosmological distances, and currently only ~60 local, relatively low-luminosity AGN have measurements of their black hole masses based on this technique. MSE will conduct a ground breaking campaign of ~100 observations of ~5000 quasars over a period of several years (totaling ~600 hours on-sky) to map the inner parsec of these quasars from the innermost broad-line region to the dust-sublimation radius. These observations will map the structure and kinematics of the inner parsec around a large sample of supermassive black holes actively accreting during the peak quasar era and provide a quantum leap in comparison to current capabilities. Other extragalactic time domain programs can be expected to produce some of the largest spectroscopic datasets on transient phenomena that exist, that will be essential resources as the international community devotes increasing observing time to this relatively unexplored domain.

MSE is the ultimate facility for probing the dynamics of dark matter over every-and-all spatial scales from the smallest dwarf galaxies to the largest clusters. For dwarf galaxies in the vicinity of the Galaxy, MSE will obtain complete samples of tens of thousands of member stars to very large radius and with multiple epochs to remove binary stars. Such analyses will allow the internal dark matter profile to be derived with high accuracy and will probe the outskirts of the dark matter halos accounting for external tidal perturbations as the dwarf halos orbit the Galaxy. A new regime of precision will be reached through the combination of large scale radial velocity surveys using MSE with precision astrometry for tangential velocities in the central regions using the VLOTs. In the Galactic halo, large scale radial velocity mapping of every known stellar stream accessible from Maunakea with high velocity precision will reveal the extent to which these cold structures have been heated through interactions with dark sub-halos and so for the first time place strong limits of the mass function of all subhalos (luminous and non-luminous) in an $L_*$ galaxy. On cluster scales, MSE will use galaxies and globular clusters as dynamical tracers in exactly the same way as stars in dwarf galaxies, to provide a fully consistent portrait of dark matter halos across the halo mass function.

Finally, MSE provides a platform for enabling next generation cosmological surveys, particularly through large scale surveys of the internal dynamics of galaxies as part of future multi-object IFU developments. Such surveys can provide dynamical tests probing the scale of homogeneity, can test the backreaction conjecture in General Relativity, and provides a novel means of measuring gravitational lensing via the velocity field, that complements standard measurements by being sensitive to very different systematics.

*MSE will transform the study of some of the darkest components of the Universe. Absorption line studies, reverberation mapping programs, massive transient datasets and precision velocity studies of stars and galaxies as tracer particles utilize aspects of the performance of MSE – sensitivity, stability and specialization – that are unique to the facility and which consequently allow unique insights into key aspects of the baryon cycle, black hole physics, the time domain, and cosmology. MSE represents a suite of capabilities that is an essential platform for the development of cosmological programs that go beyond the current generation and which are relevant on the 2025+ timescale.*

## 4.2 The Universe in absorption

### 4.2.1 The Interstellar Medium

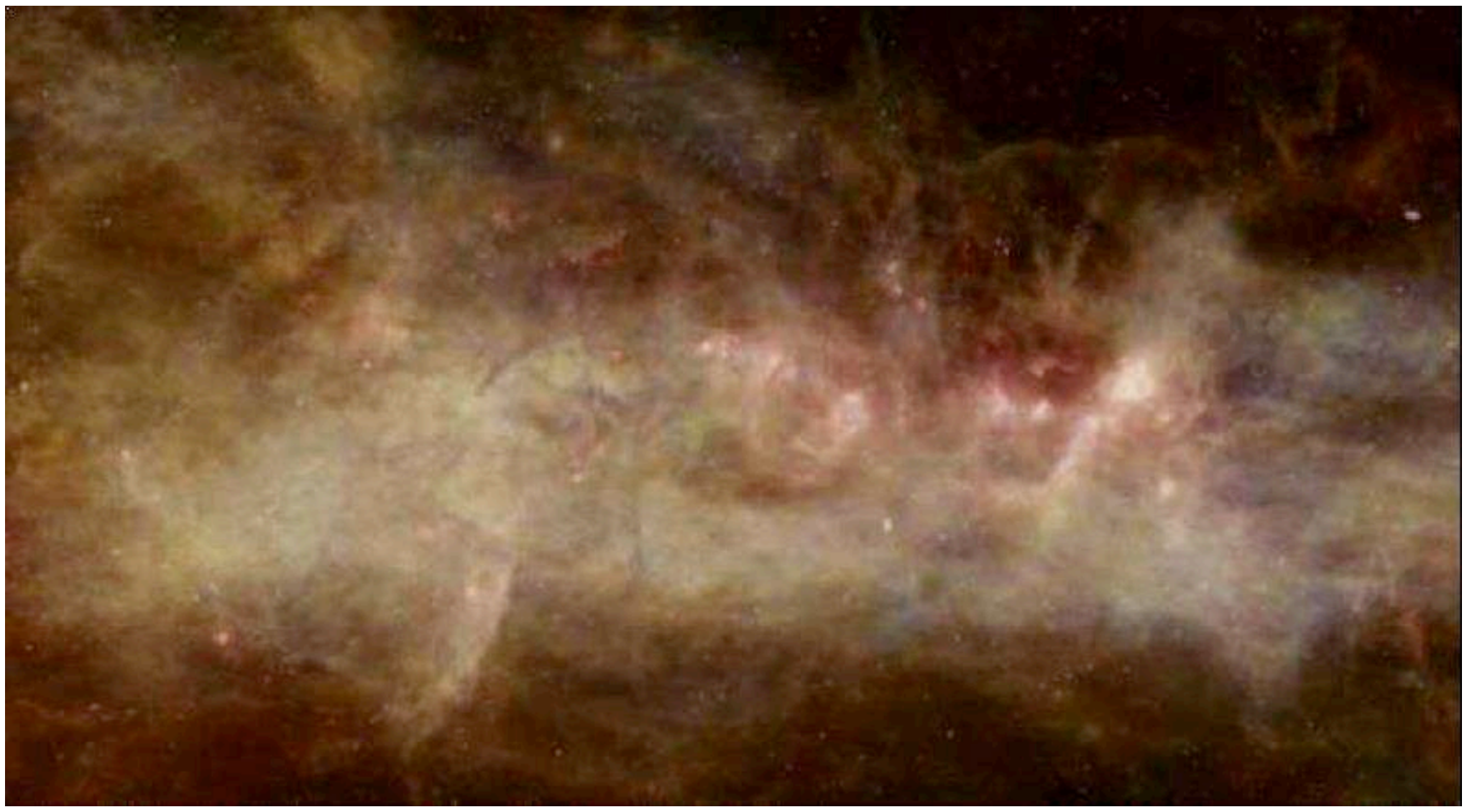

**Figure 69: The midplane of the Milky Way near the constellation of Perseus. 21cm observations of HI from the Canadian Galactic Plane survey (Taylor et al. 2003). Overlaid on these hydrogen data are IRAS images of the dust that intersects our line of sight through the disk. The point-like sources (WRST) are mainly distant galaxies and AGN. (*Figure from J. English, http://www.ras.ucalgary.ca/CGPS/press/aas00/pr/pr_14012000.html*).**

MSE is a unique and powerful tool for studying the interstellar medium (ISM) of the Milky Way. It will play an essential, central role in revolutionary three dimensional ISM mapping achievements that will be boosted by the ESA Gaia parallax distances. In parallel, and in combination with the 3D mapping, it will allow unprecedented specific studies of long-standing and new questions about the nature of the ISM. In all cases, the power of MSE for this science lies in the combination of large aperture, high throughput, high spectral resolution, wide field of view, and high multiplexing:

- *The large aperture and high throughput* increases the number of weak targets that can be individually studied and the distance coverage. In the case of large-scale ISM mapping, MSE's high observing efficiency reduces survey durations to tractable levels, putting it far beyond the reach of any other existing or planned facility/instrument.
- *The high spectral resolution* improves the determination of ISM absorbing columns and/or stellar types used to derive the reddening, and it allows coupling absorptions with emission spectra by means of velocity identification;
- *The wide field of view* facilitates the required coverage over large fractions of the Galaxy;
- *The high multiplexing* increases the number of targets along a given sightline and the spatial resolution of the maps. For detailed study of individual structures, MSE's wide field and dense fibre coverage is well matched to many Galactic features, such as stellar streams, astrospheres, and individual interstellar clouds, targets that would otherwise be accessible in the radio region alone.

#### 4.2.1.1 Towards detailed three dimensional maps of the Galactic ISM

Multi-wavelength observations of the Galactic ISM have never been so detailed nor so abundant as now. High-quality emission maps of the gas and dust have been produced thanks to a wide range of ground- or space-based facilities that operate at almost all wavelengths, including the γ–ray (FERMI), X–ray (ROSAT, Chandra, XMM, Suzaku), UV (GALEX), IR (ISO, Spitzer), sub-mm (Herschel, Planck), mm (JCMT, APEX) and radio (DRAO, LOFAR) regions. Figure 69 shows a composite image of the midplane of the Galaxy near the constellation of Perseus, created from 21 cm data from the Canadian Galactic Plane Survey (Taylor et al. 2003), with IRAS and WRST data overlaid. Typically, datasets usually come in the form of two-dimensional images in specific spectral bands, or in data cubes (i.e., two spatial dimensions plus a spectral, or polarimetric, dimension). Yet, paradoxically, our understanding of the ISM structure remain surprisingly limited, even at the largest spatial scales, primarily because we lack the accurate distances of the features that are responsible for the observed emission and their three-dimensional (3D) distribution.

Building accurate 3D maps of the ISM requires (i) massive amounts of distance-limited absorption data, i.e., stellar surveys at high spectral and spatial resolution covering large fractions of the sky, and (ii) accompanying information on the target distances. MSE and Gaia will provide such a powerful synergy. More precisely, there will be two levels of mapping:

- an initial distance assignment for intervening clouds based on all types of absorption data, dust reddening, gaseous lines or diffuse interstellar bands. The resulting 3D maps will be a general tool of wide use;
- differential mapping, i.e., comparison of the distributions of the various absorbers, e.g., DIBs and dust. This will opens a variety of studies, in particular on the ISM physical properties and the multi-phase structure.

Today our understanding of the origin, interplay and evolution of the various phases of the Galactic ISM is seriously hampered by the lack of realistic 3D density and velocity distributions, since it prevents the construction of quantitative, physically-motivated models. It is hard to believe that the distance to the most prominent features of the X-ray, sub-mm and radio continuum full-sky maps – the so-called North Polar Spur and Loop 1 features – varies from the solar neighborhood to the distance of the galactic center, depending on authors. Consequently, this lack of 3D perspective impacts negatively on a large number of research fields directly linked to the Milky Way ISM: star formation and its link with parental ISM properties, including the role of turbulence, magnetic field, metallicity or dust content; cavities carved by stellar winds and supernovae and their evolutions; mixing of stellar ejecta; large scale evolutionary modeling of the Milky Way structures, bulge, bar, spiral arms, halo properties; tests of accretion versus internal evolution; etc. For all these fields, the 3D structure would be an invaluable ingredient to make full profit of all 2D photometric, spectral and polarimetric datasets.

This situation is also regrettable since a number of major projects making use of prominent astrophysical facilities and/or surveys from ground or space require an increasingly detailed understanding of the Galactic ISM. For instance, interpreting observations of the cosmic microwave background (CMB) and its polarization is strongly dependent on optimal modeling of Galactic dust emission and its spectral characteristics. These, in turn, depend on the spatial distribution and temperature of the dust, and on the intensity of the interstellar radiation field that heats it. To produce realistic models of this radiation field, it is therefore necessary to compute the propagation of photons through the 3D dust and gas distributions. Similarly, understanding the Galactic cosmic rays (GCRs) spectra and their spatial variability requires a detailed knowledge of the 3D distribution of the multiphase ISM. In turn, the analysis and modeling of diffuse γ–ray emission (which is generated by the interaction of cosmic rays with ISM nuclei and by up-scattering of the interstellar radiation field by cosmic electrons) doubly requires a 3D knowledge of the ISM distribution; on the one hand, because it is needed to compute the cosmic ray trajectories accurately, and on the other hand, because the radiation field itself is governed by this distribution. Finally, obscuration by dust poses a significant obstacle to our understanding of stellar populations within the Milky Way. Extinction and reddening are not always able to be estimated to the level of precision that is needed to characterize the properties of distant stars.

Studies of specific astrophysical targets would also benefit from detailed 3D ISM maps. At present the absence of knowledge on their distances, foregrounds, environments and backgrounds limits the identification of their own spectral features, or the studies of the interactions with the ambient medium. In particular, all newly discovered structures around stars (bow-shocks, trails, ionization cavities) would strongly benefit from knowledge of the physical and dynamical properties of the ambient medium. Indeed, searches for particular categories of objects could be more efficiently conducted if one could use 3D maps to identify the most favourable sky regions.

#### 4.2.1.2 The MSE – Gaia perspective

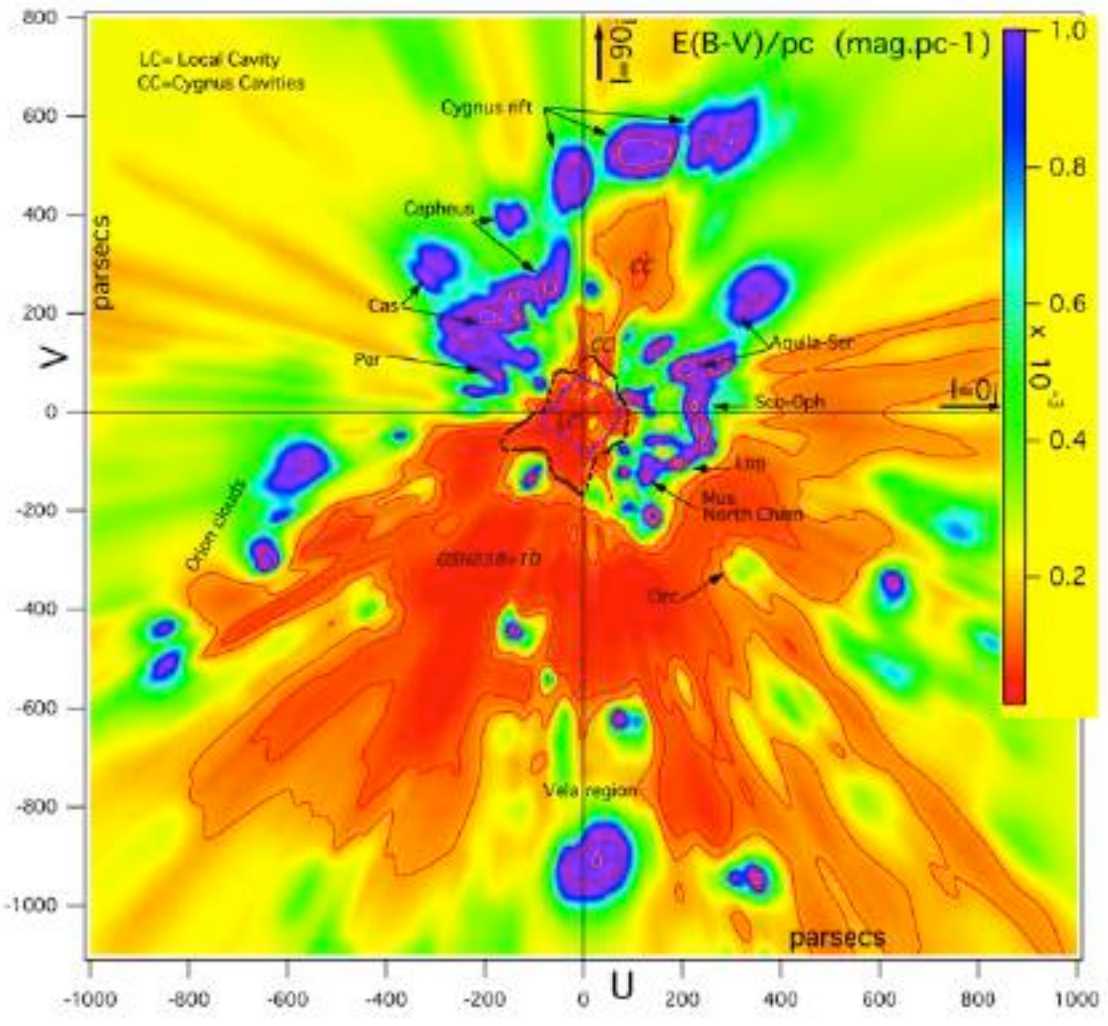

**Figure 70: 3D mapping based on reddening measurements. Planar cut in the 3D opacity distribution inverted from 23 000 (wide + narrow)-band reddening measurements. The Sun is at (0,0). Units are parsecs, red is very low density and violet indicates dense areas. External areas with a homogenous color correspond to the absence of constraining (distance- and longitude-distributed) target stars. The local cavity at the center is surrounded by well-known cloud complexes and other cavities, including a huge region devoid of dust in the third quadrant. The**

**accuracy on cloud distances is on the order of 5 pc to 1kpc, but the angular resolution is of the same order, i.e. very poor. *(Figure from Lallement et al 2014)*.**

3D maps published to date are based on various inversion techniques that have been applied to absorption or extinction data. A full three-dimensional Bayesian method adapted to individual measurements of gas columns (or extinctions) has been developed by Vergely et al. (2001) and recently applied to reddening measurements individually recorded towards 23 000 nearby stars to produce local maps (Lallement et al., 2014, Figure 70). Marshall et al. (2006) has produced the first large scale reddening map by adjusting a modeled Galactic distribution of stars and a 3D dust distribution to 2MASS photometric data. During the last two years the situation has started to evolve very rapidly thanks to the massive spectroscopic and photometric stellar surveys that are coming online. Unprecedented numbers of distance-limited data have been made available that can be used to produce 3D maps of dust or gas (for example, see Sale et al. 2014 for IPHAS-based maps; Zasowski et al. 2015 for SDSS/APOGEE DIB-based maps; Green et al. (2015) for Panstarrs-1 reddening-based maps, Figure 71; the special issue led by Monreal-Ibero & Lallement on 3D maps, Mem. Dell. Soc. Ital., 2015).

It is clear that this situation is about to change dramatically when all current and future mapping efforts start to benefit from the ESA Gaia mission. As discussed in Chapter 1 and Chapter 2, Gaia will measure the positions of all stars down to G~20 mag, and create an extraordinarily precise three-dimensional map of more than a billion stars throughout the Milky Way. This remarkable dataset — once combined with new information on absorption lines, DIBs, reddening and extinction from optical/infrared spectroscopy — has the potential to revolutionize studies of the ISM and its three-dimensional structure. In all present and future maps, accurate distances will replace photometric estimates, and reddening will be more accurately determined simply from removing one unknown parameter in Bayesian reddening computations applied to the photometric data. Most notably, the ESA Gaia mission will in turn benefit immeasurably from detailed Galactic extinction maps that will allow to break degeneracies between reddening and temperature for all faint targets. Clearly, the key synergy for MSE in the study of the Galactic ISM is with Gaia.

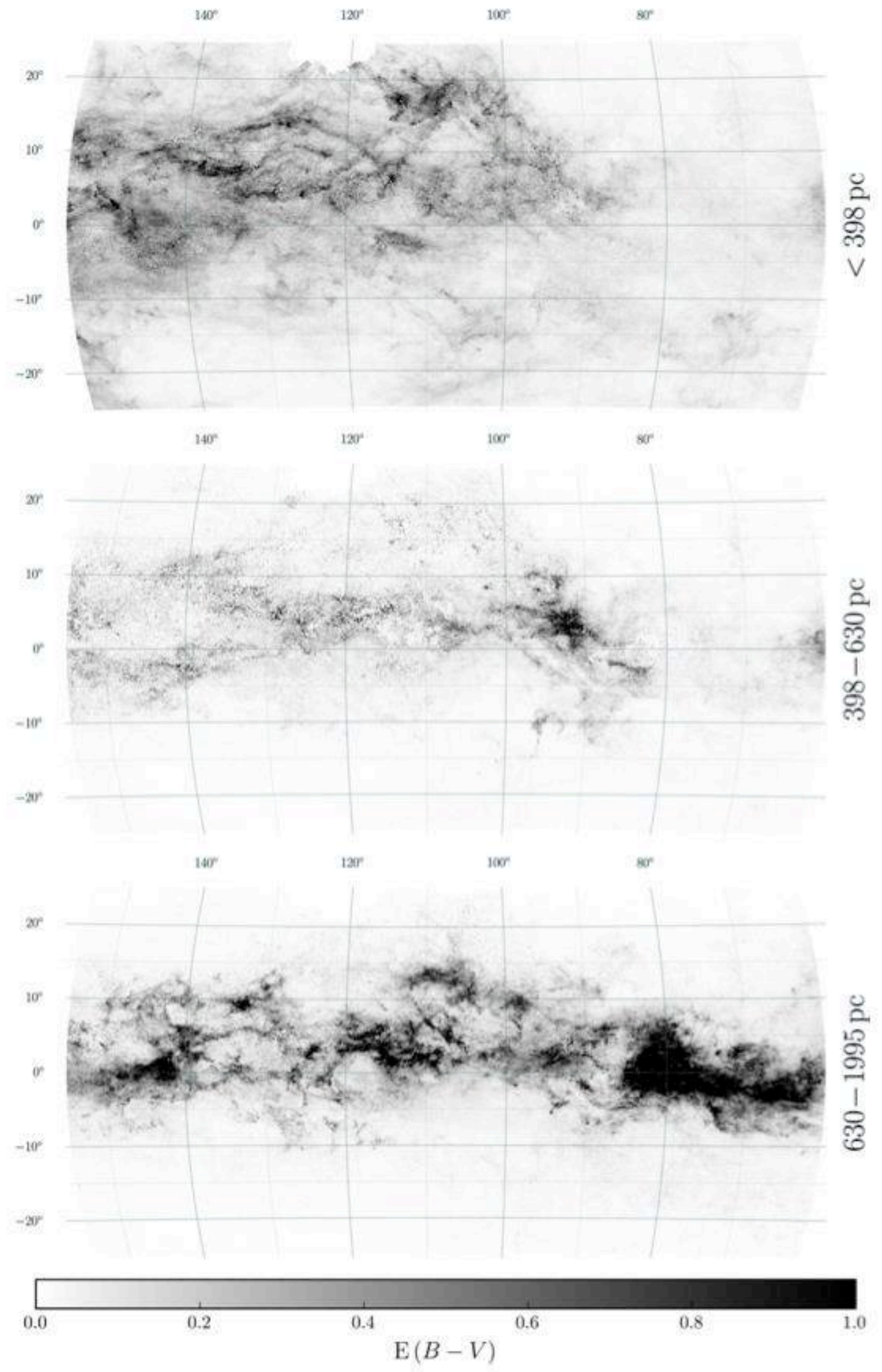

**Figure 71: 3D mapping based on reddening measurements. Dust distribution based on 800 million Panstarrs1/2MASS wide-band photometric data. Shown is reddening towards stars located in three different distance intervals in the Cepheus-Polar region. The angular resolution is excellent (5 arcmin) but radial resolution (distance to clouds) is poor. Both types of maps will strongly benefit from Gaia parallaxes and additional massive spectral data with MSE. *(Figure from Green et al. 2015).***

However, building very accurate maps of larges volumes in the Milky Way and in particular their associated dynamics will be possible only once high resolution massive surveys with high multiplex spectrographs like MSE will enter in action. As a matter of fact, statistical methods based on photometry have limitations in spatial/angular resolution because they require a statistically significant number of targets per unit volume, they suffer from biases associated with the absence or insufficient knowledge of the target metallicity, or of the reddening law applicable to the intervening dust. Most important, they do not provide any kinematic information. In contrast, MSE spectra of individual targets each carry precious information on the line-of-sight velocity structure, and offer the possibility of coupling with the emission spectra (through the velocity identifications). Spectra only provide diagnostics of the physical properties of the clouds by means of line ratios, and, most importantly, correspond to angular and spatial resolution that are only limited by the number of available individual targets. From the technical point of view, new methods of inversion will need to be developed to adapt to MSE spectra, that will be orders of magnitude more numerous than existing high resolution spectral data.

#### 4.2.1.3 MSE and other facilities

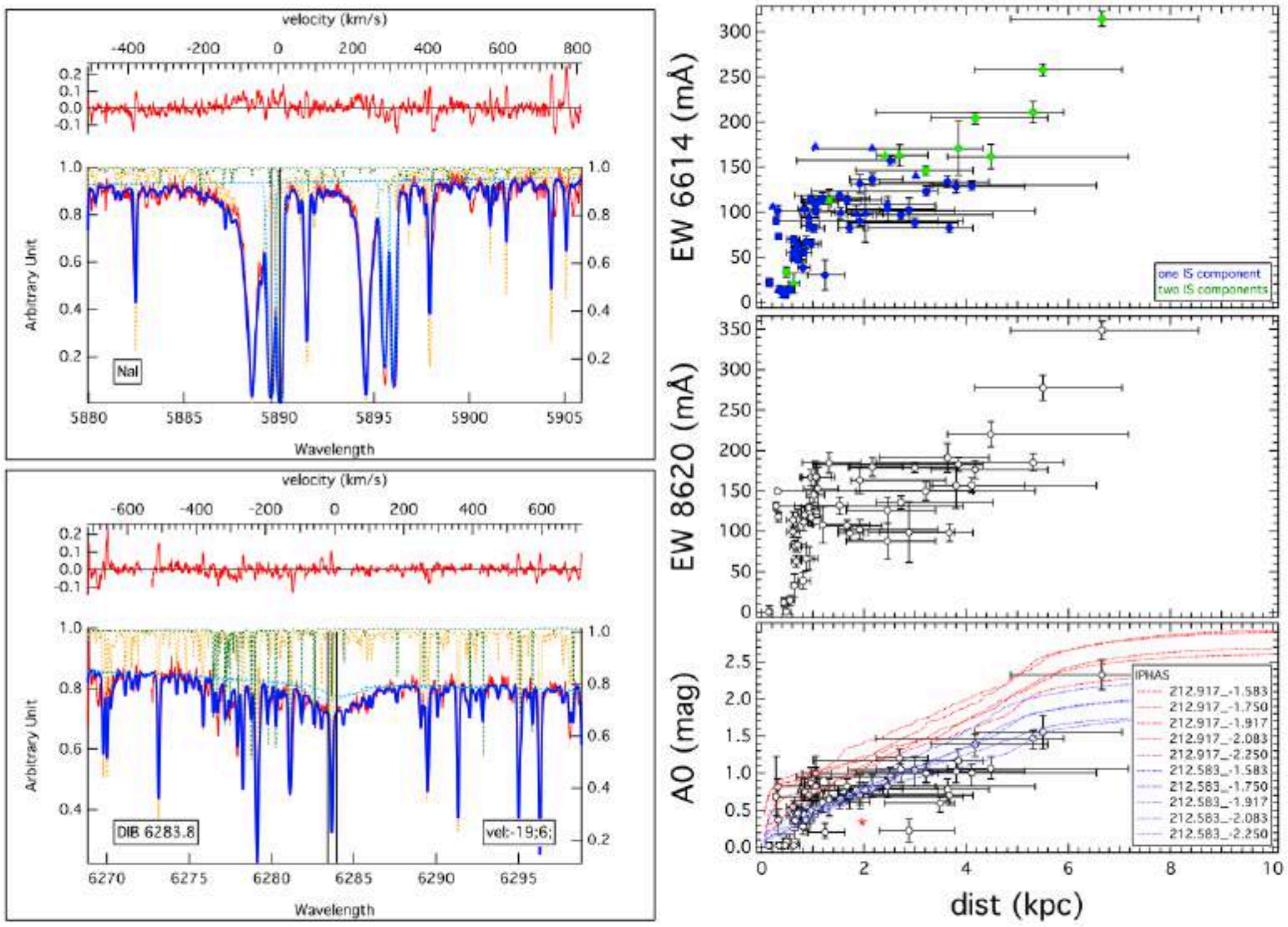

**Figure 72: Mapping potentialities from spectroscopy and photometry: Left: Interstellar NaI doublet (top) and the 6284Å diffuse band (bottom) extracted from a V=14.7 target star at D=2.8 kpc. The R=48 000 VLT/UVES spectrum (red) is modeled with the product of a synthetic stellar spectrum (yellow), an interstellar profile/DIB model with two velocity components (light blue) and a synthetic telluric absorption (green). Right: application of such fitting methods to radial cloud mapping based on the strengths of two DIBs (6284Å, 8620Å; top and middle panels, respectively) and extinction ($A_0$; bottom panel). Distances are derived from the spectroscopic stellar parameters and photometric data. DIBs and extinction vary in a similar way and trace the local Arm and Perseus. Individual spectral data show a stronger variability that may correspond to different physical properties of the clouds or small-scale structure, illustrating the potential of high resolution spectroscopy compared to photometry.** ***(Figures from Puspitarini et al. 2015)***

Building 3D maps of the ISM in the post-Gaia era requires spectroscopy for a vast number of targets, ideally distributed over a range of distances, in as many directions as possible, and at spectral resolutions and SNR levels that are high enough to enable the detection and measurement of key ISM absorption features. Clearly, this is not feasible using single-object spectrographs as the observations would be far too time consuming. It would also be impossible using small-field multi-object spectrographs, since observations need to be carried out over a significant fraction of the sky. In short, a high-resolution mapping of the three-dimensional structure of gas and dust in the Milky Way will require a telescope/spectrograph combination that has high throughput, wide-field, a high level of multiplexing, and high resolutions. MSE will literally produce a paradigm shift for this field.

3D mapping of the ISM based on spectroscopic Gaia target stars is most efficacious when the number of sightlines is maximized. This in turn depends on the number of fibres, the field and survey size, and the telescope/instrument efficiency (i.e., system throughput). High spectral resolution (R ≥ 20000) is a pre-requisite for ISM studies: high resolution disentangles ISM lines or bands from stellar features, allow kinematic analyses, and strongly decreases the lower limit on the quantity of absorbing matter that can be detected since most lines are narrow. Examples of spectral analyses similar to what will be routinely and massively possible with MSE are shown in Figure 72. Lines and diffuse bands extracted from hot or cool star spectra recorded with FLAMES/GIRAFFE or UVES at the ESO-VLT trace the line-of-sight cloud structure at large scale, and profiles agree with those derived from photometric data (in much larger number, as discussed above). Figure 72 also shows how at R ≥ 20 000 it is possible to follow how absorption velocities vary with distance.

The competition to MSE in this field in the immediate future will primarily come from AAT/HERMES, although MSE is much more efficient thanks to its greater wavelength coverage (a factor of ~1.5), multiplexing (a factor of ~8) and throughput (a factor of ~6.5). The Gaia/RVS, Subaru/PFS and Mayall/DESI have spectral resolutions that are too low by factors of ~2 – 5 while VLT/MOONS focuses on the 0.8 – 1.8μm spectral region, thereby missing most of the traditional ISM diagnostics. There is, however, a powerful synergy between MSE and VISTA/4MOST: while VISTA/4MOST is ~5 times slower than MSE in reaching the requisite SNR based on the difference in telescope collecting area, this is roughly compensated by its larger field of view (5 square degrees). The fact that these two facilities are located in different hemispheres is an important consideration when mapping the large-scale structure of the ISM.

A compelling and unique aspect of this science topic is that *any* spectral observation of Galactic and extragalactic targets obtained by MSE contains information on the ISM through the absorption features that are imprinted by intervening matter and through the reddening (or differential absorption). The serendipity of this program therefore enables MSE to obtain a very large and useful dataset for ISM studies. In addition, specific observations and surveys can be envisioned with MSE to make unprecedented progresses in the mapping of the ISM and our understanding of the relevant physics. SRO-03 describes a Galactic Archaeology program for MSE that has, as a basic component, the serendipitous study of MSE using more than a million sightlines spread across the sky. This ISM survey would be extremely efficient since high-quality spectra (i.e., SNR ~ 100) could be obtained in just ~15 min exposures for g = 16 stars of spectral types O, B and A. For reference, a survey of 4950 square degrees covering a swath of 20 degrees centered on the Galactic plane would require ~1000 hrs of MSE, or about two weeks per year over the course of a decade-long survey. Such observations could be undertaken in conditions of very high sky brightness.

In order to understand the multi-phase structure of the Galactic ISM, MSE will access absorption by different tracers to provide the tools to model the physical, dynamical and chemical properties of the molecular, diffuse and ionized Galactic ISM phases. In each MSE individual star spectrum, interstellar gaseous atomic (NaI, CaII, KI, …) and molecular (CH, $CH^+$, CN, C2, ...) absorption lines and diffuse bands (over 400 bands in the optical) can potentially be extracted according to the wavelength range and the target type, and in parallel the reddening can be determined from the spectral/photometric classification and the distance. An interstellar tracer does not need to be strictly proportional to the column density of interstellar hydrogen, or to the total column of dust, to be useful in locating interstellar clouds, since distance assignments are based on the *gradients* of the absorbing columns. It is for this reason that absorption lines arising from any interstellar species, and diffuse bands of all kinds, can potentially be used in a combined analysis (in addition to the reddening). On the other hand, their independent determinations will provide unique information of the interplay between the phases, the role of radiation, energetic particles, magnetic field, etc. Emission data (mainly HI, CO and Hα) can also be used to help phase disentangling. Since absorbing lines contain information on the gas dynamics, both radial velocities and columns (including volume densities from inversion) can be used in models. Angular variations over the sky of the line centers can also be used to constrain the *transverse* motions.

Depending on the range of column densities, some tracers may not be fully appropriate. Strong interstellar lines suffer from saturation close to the disk or towards very opaque clouds, and should thus be replaced by other, weaker, lines for large distances or columns. For nearby targets, information can be extracted from the "classical" and strong lines: NaI (5890 – 5896Å), CaII (3934 – 3968Å), CaI (4227Å) for the gas and from the reddening for the dust. The strongest diffuse bands 5780, 5797, 6284, 6196, 6614Å are also the most appropriate. While the strong lines can still be used above the Galactic plane or in directions where clouds are distributed in velocities, weaker lines are more appropriate for large optical depths (to avoid saturation), for example KI (7699Å), and the weaker diffuse bands (numerous and distributed above 4300Å).

DIBs — whose carriers are still unknown, but are strongly suspected to be macro-molecules in the gas phase — also trace the amount of interstellar matter along a given line of sight. While they have never been considered seriously for mapping purposes, recent work shows that they are ubiquitous, being present even in the spectra of distant cool stars (e.g., Chen et al. 2012), and extractions of lines and DIBs are now able to be performed for any stellar type. This allows us to benefit from all targets and build maps at the highest possible spatial resolution. For reference, we list in Table 6 the major gaseous lines and the interstellar gas phases they are tracing. While the ionized and diffuse atomic ISM fraction is best traced in the optical, the molecular and dense atomic phase is best traced when access to the blue is provided. Molecular clouds are detected with CH (4300Å), $CH^+$ (4230Å) , CN (3870Å) and also metallic lines such as TiI (3630Å), FeI (3860Å), MnI, NiI, AlI (3940Å). Alternatively the molecular phase can be probed with the infrared bands of $C_2$ around 8780Å.

**Table 6: Major ISM diagnostics**

| Feature | Wavelength | Probe |
|---|---|---|
| NaI | 5890 | Strong |
| | 5896 | |
| CaII | 3934 | |
| | 3968 | |
| CaI | 4227 | |
| DIB | 5780 | |
| | 5797 | |
| | 6284 | |
| | 6196 | |
| | 6614 | |
| NaI | 3300 | Faint |
| KI | 7699 | |

| Feature | Wavelength | Probe |
|---|---|---|
| CH | 4300 | Molecular |
| CH+ | 4230 | |
| CN | 3870 | |
| C2 | 8780 | |
| TiII | 3230 | Metallic |
| | 3240 | |
| | 3380 | |
| TiI | 3630 | |
| FeI | 3860 | |
| MnI | 3940? | |
| NiI | 3940? | |
| AlI | 3940? | |

An illustrative example of the MSE potentialities already discussed above is the multi-component absorption study applied to one of the fields of the Gaia-ESO Spectroscopic Survey (GES) and shown in Figure 72. VLT/FLAMES (R = 18 000, 48 000) spectra were adjusted by products of a stellar synthetic spectrum, a DIB or line model and a telluric absorption model, as shown in the left panels. The velocity structure and the absorption strengths are evolving with distance, tracing the spiral arms and their kinematics, as shown in the right panels. MSE survey data, by far superior in angular and magnitude/distance coverage and using Gaia distances, will provide the kinematics of the ISM with unprecedented details along with its 3D distributions.

#### 4.2.1.4 MSE and the ISM: specific studies

In addition to the large scale mapping of the ISM described in the previous section, focused studies in specific regions of the Galaxy with MSE can also provide unique ISM science. We describe below several examples, cognizant that other fields of particular interest do exist or will arise at the time of MSE first light:

*i. Small scale structures seen in polarization, the interface between stars and the ISM*

Detailed information on small-scale (~1”) structures in the ISM is now available from radio polarization data (e.g., Landecker et al. 1999; Ransom et al. 2010; Wolleben et al. 2010; Wolleben 2007). Some of these structures correspond to shells generated by stellar winds or supernovae. Other structures trace the interfaces between evolved stars or planetary nebulae (PNe) and the ambient ISM, including “tails” behind fast-moving PNe. Such interfaces, albeit much larger, are also seen at other wavelengths, most notably in the infrared (e.g., Spitzer “arcs” in Orion) or in the UV (i.e., the spectacular case of Mira’s tail; Wareing et al. 2007). In many (and perhaps most) cases, the origins of the detected polarization signatures are far from clear. The lack of information on the distance and the extent of these objects makes their true nature obscure and undermines efforts to construct quantitative models.

Observations with MSE are ideal for advancing this area of study. Dedicated observations of target stars located in the field and associated absorption measurements would yield more accurate distances to the radio polarization sources, and thus provide accurate physical dimensions. Moreover, the ability of the new observations to detect variations in the velocity structure at very small angular scales would allow for direct comparison to hydrodynamic simulations of the star – ISM interaction, and provide critical information for the construction of rotation measure models. These models, in particular, would give important information on the magnetic-field structure in the interaction and tail regions, and could lead to the construction of better magneto-hydrodynamic models of the star – ISM interaction. Moreover, emission lines might also be detected and their variability studied as a function of the location in the bow-shock and tail structures, when present. Multi-fibre spectra would be particularly adaptable for such observations.

*ii. Supernova Remnants*

Increasingly detailed radio, optical and X-ray observations of supernova remnants have raised a number of important questions. Why are there such large differences in the ratios between X-ray and radio emissions for supernovae of the same type and with comparable radio emissions? Why are there radiative recombination continua (interpreted as freely expanding and recombining gas) in X-ray spectra of mixed-morphology supernova remnants that are known to be interacting with molecular gas (e.g., Miceli 2010)? The nature of the remnant, and the detailed distribution of the ISM surrounding it, are key parameters in understanding these interactions which play a major role in galactic ISM recycling. Observations of the field stars around supernova remnants with MSE will allow the measurement of absorption or emission features that hold clues to the ISM structure, enable correlation studies, and ultimately lead to models that describe the ejecta expansion and shock properties.

*iii. Small scale structure of diffuse molecular clounds and the CH+ problem*

Whether or not diffuse interstellar clouds are “clumpy” remains a hotly debated issue. In particular, any structure in the spatial distribution of the major molecule, $H_2$, is very difficult to identify since the column density of this species can be measured only in the far UV, which requires spectral observations of bright, background stars with O or B spectral types. The degree of $H_2$ clumpiness is critically important for modeling the abundances within these clouds as it affects the penetration of UV photons and thus photo-destruction processes. Because $H_2$ is closely correlated with CH (Federman 1982), observations of the blue lines of this radical can be used as a surrogate for $H_2$. By selecting appropriate clouds at intermediate latitudes where the confusion is minimal and the surface density of background stars is still high, one can map the distribution of CH and infer that of $H_2$. The use of background stars make it possible to probe the spatial structure over a surprisingly broad range of scales, with a dynamic range of about 3000. While the largest separations (∼1.5 degrees or ∼3 pc for a cloud at 100 pc) can be used to delineate the overall geometry of the cloud and its boundaries, the smallest ones (a few arcseconds, or about 0.001 pc) would probe the structure at very small scales.

MSE spectra taken at a resolution of R ≥ 20 000 would be sufficient to reach these objectives because, apart from the main transition at 4300Å, several additional weaker features are available around 3890Å, allowing the measurement of line opacities and, hence, CH column densities even if line profiles are unresolved. Such observations will provide simultaneous, and invaluable, information on the "$CH^+$ problem". The high abundance of $CH^+$ in the ISM is presently not understood: the reaction presents an energy barrier of 4600K and is insufficient to produce the observed amounts of $CH^+$ in the conditions prevailing in diffuse molecular gas. Similarly, the large abundance of $H_2$ in J > 2 rotational states requires energetic processes that have yet to be identified and which can play an important role in the physics and chemistry of diffuse molecular gas. $CH^+$ is thus an important species in that it can be used to investigate the nature of non-thermal processes that may play a key role in the physics of interstellar clouds. Several scenarios involving either shocks (Pineau des Forêts et al. 1986), vortices (Godard et al. 2009) or cloud/intercloud interfaces have been suggested as the additional energy source required to overcome the $CH^+$ formation energy barrier. All these scenarios imply the presence of localized regions heated to higher temperatures but with very different geometries (i.e., shocks and interfaces are essentially 2-D structures) while in the scenario involving turbulence, vortices are distributed over the whole cloud volume. Thus, a detailed study of the spatial distribution of $CH^+$ with a facility like MSE will allow astronomers to identify conclusively the process at work. Furthermore, in this same spectral range, lines from CN are also present, and these can be used to provide additional constraints on cloud models.

*iv. The carriers of the diffuse interstellar bands (DIBs)*

Through detailed mapping and ISM-phase assignment of the hundreds of diffuse bands detectable in the optical, MSE will bring new constraints on the species that are responsible for the DIBs and their conditions of formation. Specifically, a large number of the absorption lines between 4200Å and 9000Å are DIBs. Some DIBs are very tightly correlated with the color excess, some with the atomic gas, and some with the molecular gas. For the most part, DIBs are broader than gaseous lines, but they contain important kinematic information as they are Doppler-shifted according to the radial velocity of their carriers. They can also be quite numerous: optical spectra have been shown to contain as many as 400 DIBs (see, e.g., Hobbs et al. 2009; Friedman et al. 2011). Depending on their relationship to the extinction or the gas columns, their detection can require high SNR and/or highly extincted stars.

At present, no definitive identification exists for any of the carriers, despite decades of study (although large gaseous carbon molecules are strongly favored). Most observations have tended to focus on measuring correlations between specific bands and gas or dust columns. While some families that share some properties have been found, there is a strong variability in the degrees of correlation. Recent studies have shown that the strength of some DIBs is governed by the stellar/interstellar radiation field, reflecting their degree of ionization. The cloud history and physical state (i.e., shocks, ionization, cooling, condensation, shielding against UV photons, etc.) must play a role not only in the cloud chemistry and molecular content, but also in the formation of DIBs.

Building on these studies, MSE would offer a new way to approach the longstanding problem of the identification of DIBs, as well as characterizing the properties of the clouds to which they belong. Determining the detailed spatial distribution of the DIB carriers within a given cloud at high spatial resolution, through spectroscopic measurements of multiple targets, will give information on how those carriers evolve within the cloud. Although challenging, such observations may also reveal DIBs that originate from the same types of molecules since they should display the same spatial structure. No such studies have yet been attempted because they require not just DIB measurements but also absorption data (needed to constrain the physical and chemical structure of the ISM). MSE will therefore be the ideal facility for such a study. Targets should in this case include cool, evolved stars in order to maximize spatial resolution.

### 4.2.2 The Intergalactic Medium

**Science Reference Observation 10 (Appendix J)**

**Linking galaxy evolution with the IGM through tomographic mapping**

*We propose to simultaneously use high-z (z>2.5) Quasi-Stellar Objects (QSOs) and bright galaxy sight-lines to probe the Lyman-α forest and metal content of the IGM at z ~ 2 - 2.5, and target photometrically selected faint galaxies within ~1Mpc of each sight-line to directly identify sources associated with the IGM structures. Moderate resolution (R~5000), deep spectra would be obtained for all sources to target wavelengths from Ly-α to [OIII], including intervening UV absorption lines, at z ~ 2 – 2.5. This will push the capabilities of MSE, requiring good sensitivity and moderate spectral resolution from 3600Å to 1.8μm – with the primary instrumentation requirement of excellent sensitivity at 3600 – 5400Å. Such a study would allow a reconstruction of the gas spatial distribution and of the underlying dark matter distribution, detailed modeling of galactic scale outflows via emission lines, investigations of the complex interplay between metals in galaxies and the IGM, and the first comprehensive, moderate resolution analysis of large samples of galaxies at this epoch.*

The IGM is the reservoir of baryons for galaxy formation, and at high redshift contains most of the baryons in the Universe. Galaxies emit ionizing photons and expel metals and energy through powerful winds which determine the physical state of the gas in the IGM. The relevant physical scale is of order 1 Mpc or less (~2 armins at z ~ 2.5 using standard cosmological parameters); on larger scales, the gas is in the linear regime and probes large scale structures.

Deep observations of high-z quasar and bright galaxy spectra display numerous absorption lines blue-wards of the Lyman-α emission from the quasar (e.g., Rauch 1998). Such features are produced via absorption from line-of-sight intervening gas in a warm photoionised IGM and the resulting "Lyman-α forest" is a primary diagnostic for mapping the distribution and composition of the IGM (Caucci et al. 2008, Lee et al. 2014a,b, Cisewski et al. 2014). Observations of the Ly-α forest have driven numerous investigations of galaxy evolution and cosmology. For example, comparisons of line statistics are a critical test of cosmological simulations: i.e., the IGM's thermal history and opacity is a direct test of reionization of both helium and hydrogen (e.g., Becker & Bolton 2013) while the detection of metal species allows us to probe chemical enrichment and the extent of the circumgalactic medium (CGM; Aracil et al. 2004, Chen et al. 2014).

At high redshift (z > 2), the IGM contains the bulk of the baryons in the Universe. As such, it provides the reservoir of gas available for any galaxy to evolve, with IGM accretion fuelling mass growth via star-formation. By simultaneously probing the IGM structure and composition, and galaxy distributions, nebular gas dynamics and metallicity we can build a complete picture of the interplay between galaxies and their larger scale surroundings. The spatial distribution of the IGM at high-z is also directly related to the underlying dark matter distribution and as such, by fully mapping the IGM, we will allow a complete reconstruction of the matter density field (the cosmic web) at a given epoch.

State-of-the-art work connecting galaxies and absorbers at high redshift has come from the Keck Baryonic Structure Survey (e.g., Steidel et al. 2010) and similar work done at the VLT. To date, the distribution of galaxies around a combined sample of two dozen high-z quasars has been mapped with these two telescopes (see the summary in Rudie et al. 2012; Crighton et al. 2011); These quasar spectra have high-resolution (R ≥ 45000) and the galaxy spectra have resolution R ~ 300 to 1,000. Perhaps surprisingly, it has been found that the flux transmission in the Ly-α forest correlates with Lyman break galaxies up to a scale of 5 Mpc (Adelberger et al. 2003; Crighton et al. 2003). These programs have also revealed the connection between metal absorption lines and the CGM, again to the surprising scale of many hundreds of kpc (Rudie et al. 2012; see Figure 73 for Lyα absorption within ±1000 km $s^{-1}$ of the systemic redshift of 10 galaxies within a projected distance of 100kpc of the line of sight to their background QSOs). Such studies are clearly in their infancy, but are indicators of the powerful combination of wholesale pairings of high-quality QSO spectra with galaxy surveys.

MSE is a powerful facility for the study of the IGM and the galaxies embedded within it. Building on the surveys at Keck and the VLT, the main requirement for increasing the number of absorbers (for association with the foreground galaxies) is the large number of quasars with high-resolution (R ≥ 40000) and high-SNR spectra. In particular, it is important to have wide wavelength coverage so that large numbers of metal species can be associated with a given absorber. In the VLT/UVES and Keck/HIRES archives, high-resolution spectra for ~ 400 quasars are available. Therefore, a campaign with MSE to amass spectra for large numbers of galaxies in these fields would yield a 10 to 20 fold increase in the number of sight lines studied. Since BOSS will observe a large number of quasars brighter than 20 magnitude, it would be possible to identify fields with an excess of quasars for MSE to target. At present, there is no planned or proposed facility that equals the capabilities of MSE in this area.

For larger-scale IGM mapping, a large field of view is required to enable a large number of targets to be simultaneously observed over a large area. This is difficult for non-survey instruments or facilities, and here MSE has a clear advantage. Other large spectroscopic survey instruments either lack the sensitivity (e.g., 4m class, VISTA/4MOST, WHT/WEAVE and Mayall/DESI) or resolution (e.g., R = 2000 at ~ 4000Å for Subaru/PFS). Subaru/PFS, at its very upper limits, could attempt an IGM tomography experiment, however, this does not directly form part of the PFS extragalactic science case (Takada et al., 2014), although IGM tomography could be undertaken as a bi-product of the proposed PFS galaxy evolution survey. Such a survey will not have sufficient resolution to determine the metal content of the IGM, will not identify the faint galaxies associated with the IGM structure at $2 < z < 2.5$ (with a proposed $J < 23.4$ limit) and will not have the sensitivity and wavelength coverage to determine the stellar- or gas-phase metallicity of individual galaxies at this epoch. It is therefore unlikely that this science will be successfully undertaken until the construction of MSE and no planned instrument will be able to complete such a project to the high fidelity level of MSE.

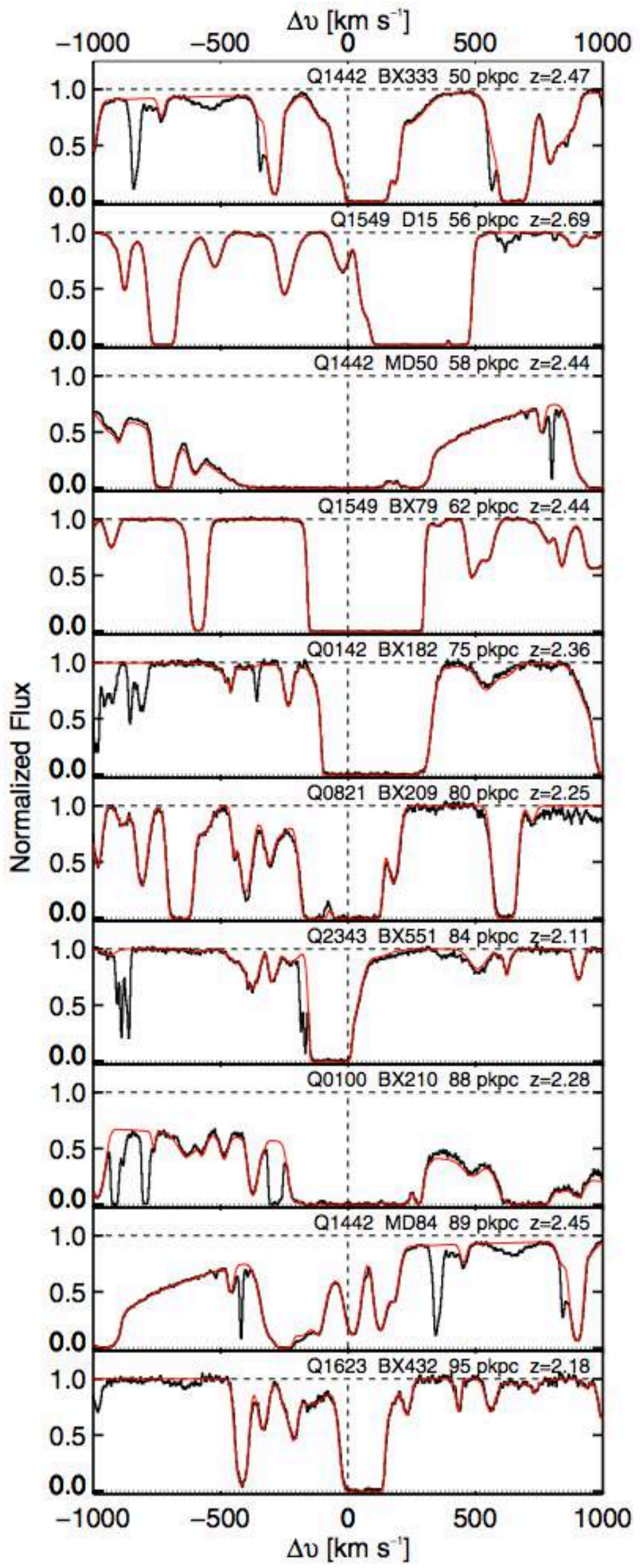

**Figure 73: Lyα absorptions within ±1000 km $s^{-1}$ of the systemic redshift of 10 galaxies located at a projected distance less than 100kpc of the line of sight to background QSOs, from the sample by Rudie et al. (2012). Keck/HIRES data are in black, while the red shows Voigt profile decomposition of the HI absorption near the redshift of the galaxy. The continuum and zero level of the spectrum are shown in dashed and dotted lines, respectively. The systemic redshift of each galaxy is marked by the vertical dashed line at 0 km $s^{-1}$. *(Figure from Rudie et al. 2012).***

The SDSS-III Baryon Oscillation Spectroscopic Survey (BOSS; and its successor the SDSS-IV "Extended BOSS") and Mayall/DESI aim to measure the baryon acoustic oscillation (BAO) scale using the Ly-α forest (amongst other tracers) at z ~ 2 – 3.5 (Delubac et al. 2015). The (e)BOSS survey in particular will collect spectra of 250000 quasars over 10000 $deg^2$, while DESI proposes to survey 14000$deg^2$ and target about five times as many quasars at higher resolution (R ~ 3000 – 4800) with the 4m Mayall telescope (Schlegel et al. 2009). Moderate to high resolution spectra are not required to accurately measure the ~150 comoving Mpc BAO scale (the trade-off between quasar number density and SNR is instead more important, e.g., McQuinn & White 2011), and so BOSS and DESI would not represent a significant competitor to MSE because it chose to undertake a very large Ly-α forest survey at low resolution. A smaller sample of high-resolution spectra will enable correlations in the small-scale IGM structure to be probed by Ly-α absorptions (as induced by, e.g., ionization or temperature fluctuations during He II reionization). This would be an exciting area in which MSE could capitalize.

SRO-11 describes a science program that has multiple aims within the field of IGM studies, with particular focus on connecting the distribution and properties of the IGM to those of the galaxies embedded within it. By simultaneously probing the IGM and galaxies at 2 < z < 2.5, SRO-11 builds a complete picture of both the IGM and galaxy, distribution and composition at this epoch. We will probe the build-up of metals in both environments, and witness the complex interplay between galaxies and their surroundings. We will perform the first detailed high SNR and moderate resolution analysis of a extensive sample of galaxies at z > 2 – increasing sample sizes by orders of magnitude, and fully map the IGM distribution on scales inaccessible to any current or proposed instrument other than MSE.

We now discuss in more detail these key areas of MSE IGM science.

#### 4.2.2.1 The interplay between galaxies and the IGM

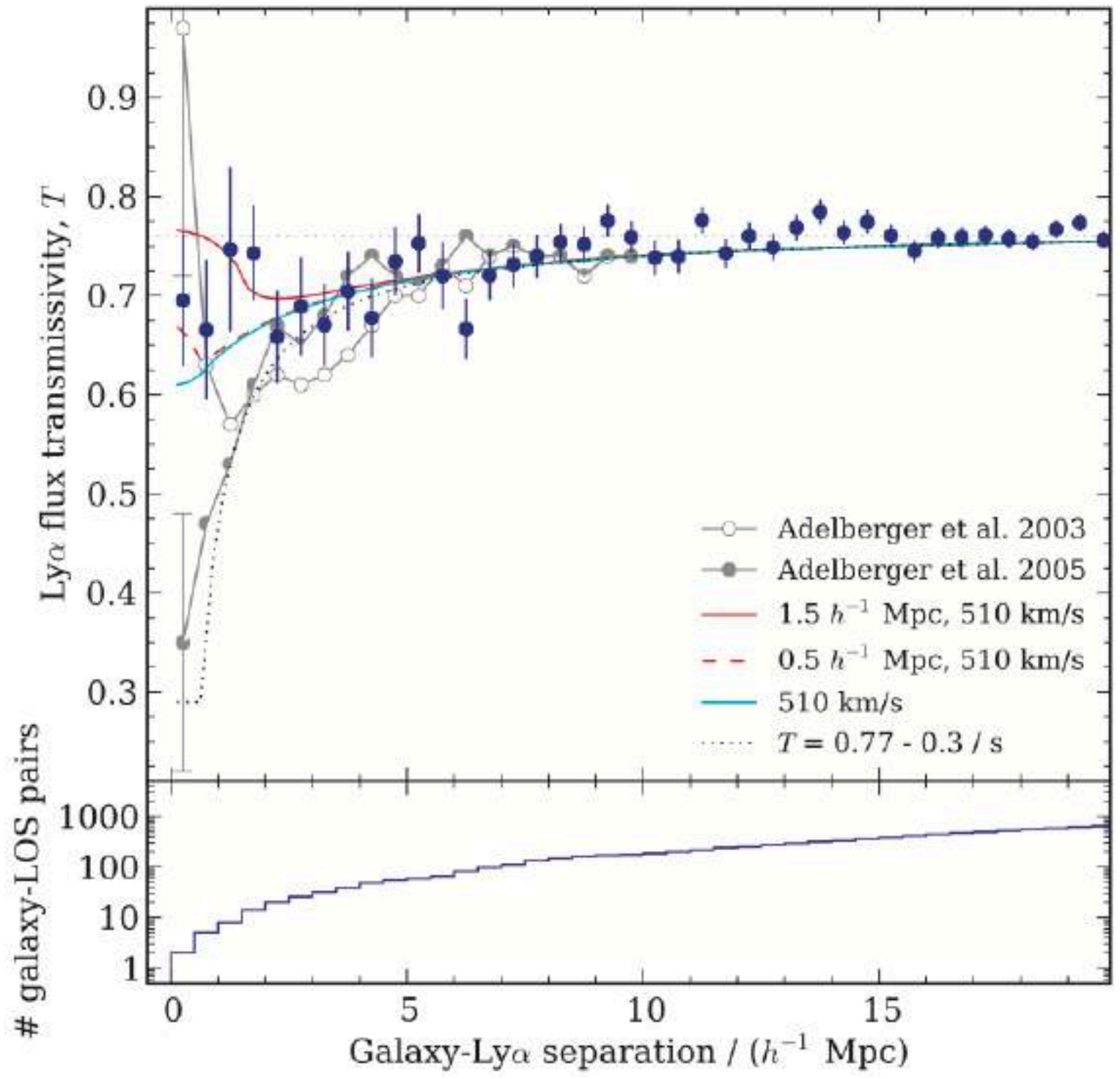

**Figure 74: Mean Lyα forest transmissivity in quasar spectra plotted as a function of distance from the nearest Lyman Break Galaxy from the sample of Crighton et al. (2011). Power-law model correlation functions are also**

**overlaid. A decrease in H I transmission at separations of 2–7 $h^{-1}$ Mpc is seen that is described by a power law with index γ = −1. *(Figure from Crighton et al. 2011).***

The IGM is known to be the major reservoir of baryons at high-redshifts. While mass growth at high-z is likely to be increasingly dominated by mergers (e.g., Conselice 2014), recent results have indicated that in the majority of cases this does little affect star-formation (Robotham et al. 2013, Davies et al. 2015). Therefore, the primary mechanism through which typical galaxies form new stars is through IGM gas accreting onto their halos. Conversely, star-formation in these galaxies emits ionising photons, which heat the IGM and drive superwinds, which expels metals out of the galaxy (e.g., Heckman et al. 1993, Ryan-Weber et al. 2009). In general, however, details of the cycling of gas into, through, and out of galaxies remains poorly understood. An important missing element is an understanding of the gas belonging to the CGM, which is believed to be both the repository of the *inflowing* gas and the receptacle of the feedback of energy and metals generated *within* the galaxy.

Adelberger et al. (2005) used Lyman-Break techniques to select z ∼ 3 galaxies in the fields of quasars. In this way, they were able to study the cross-correlations of Lyα and C IV absorbers seen in quasar spectra with the galaxy population in the field, finding that C IV is detected in most of the galaxies halos up to a few hundred of kiloparsecs. They further found that the cross-correlation function of galaxies and CIV systems appears to be the same as the autocorrelation length of the galaxies, consistent with the idea that CIV systems and galaxies reside in similar parts of the universe. Similarly, they established that Lyα is associated with galaxies up to distances within 5 – 6 $h^{-1}$ Mpc. More surprisingly, their observations hinted at a decrease of this cross-correlation at small distances within 1 $h^{-1}$ Mpc, possibly due to enhanced star formation in these galaxies heating their surrounding IGM.

Crighton et al. (2011) measured the cross-correlation between H I gas causing the Ly-α forest and 1020 Lyman break galaxies at z ∼ 3. In agreement with the Adelberger et al. (2003) power-law relation, they measured the cross- correlation between C IV absorbers and Lyman break galaxies for scales 5 – 15$h^{-1}$ Mpc. This is shown in Figure 74, where an increase in H I absorption compared to the mean absorption level within ∼5$h^{-1}$Mpc of a galaxy can be observed, in agreement with the results of Adelberger et al. (2005). Crighton et al. (2011) argued that the Lyα – Lyman break galaxy cross-correlation can be described by a power law on scales larger than 3$h^{-1}$Mpc. After taking galaxy velocity dispersions into account, the results at <2$h^{-1}$ Mpc are also in good agreement with the results of Adelberger et al. (2005). There is little indication of a region with a transmission spike above the mean IGM value that might indicate the presence of star formation feedback.

More recently, Rudie et al. (2012) studied a sample of 886 star-forming galaxies, located close to background quasar sightlines, in the range 2.0 ≤ z ≤ 2.8. Using high resolution data from Keck, their program aimed to study the gas properties in these systems, such as column density and temperature. From their $N_{HI}$ measurements, they found that absorbers with log $N_{HI}$ > 14.5 $cm^{-2}$ are tightly correlated with the locations of galaxies, while the absorbers with lower column densities are correlated with galaxy positions only on ≥Mpc scales.

Given the differing results in studies of the transmissivity of the Ly-α forest, a new and larger sample of quasar absorbers and galaxy spectra from MSE would represent an important new addition to the field. SRO-11 describes a program to obtain moderate-resolution spectra for galaxies at $2 \leq z \leq 2.5$, lying close to the sightlines of bright quasars to firmly establish the link between foreground galaxies and gas detected in absorption in the quasar spectra. Good coverage is needed for all galaxies within the projected physical separation of 1Mpc. For high spatial completeness, this will require visiting each field several times in order to cover the galaxies around the quasars.

SRO-10 will tackle the key question of understanding how the galaxies embedded in the IGM structure interact with the large scale baryonic distribution. By probing the dynamics of nebular material in galaxies in the vicinity of the IGM sight-lines, SRO-10 will build a picture of how they exchange material with the surrounding environment. Detailed analysis of nebular emission line profiles such as Lyman-α and [OII] (e.g., Verhamme et al. 2008, Weiner et al. 2009) in combination with stellar absorption lines, will allow the identification of galactic scale inflows/outflows and an estimate of the mass exchange between galaxies and their immediate surroundings (e.g., Pettini et al 2001; Bouche et al. 2013). Previously, the large samples of high signal-to-noise and moderate resolution spectra required for this analysis have been constrained to a small number of sources. As such, our understanding of both outflows from super-winds and gas accretion is limited. However, the observing strategy required for IGM tomography (Section 4.2.2.2) lends itself directly to a combined study of the galaxies within the IGM. High signal-to-noise, moderate resolution and good sensitivity over a wide wavelength range are essential for both projects and as such they can be completed simultaneously. With addition of the measurements of galaxy metallicities (Section 4.2.2.3), it will also be possible to directly compare the chemical content of galaxies and the surrounding IGM, probing the pollution of the circumgalactic medium and the buildup in metals outside of galaxies. With comparable resolution spectra between the IGM mapping and galaxy studies, we will be able to directly associate these galaxies with absorption features in the line-of-sight galaxy/QSO spectrum to investigate the properties of individual neutral hydrogen absorbers and the regions they inhabit.

#### 4.2.2.2 Tomographic mapping of the large scale distribution of the intergalactic medium

By targeting a sufficiently large number of bright, background sources it is possible to use the intervening Lyman-α absorption features to fully reconstruct the IGM distribution at a given epoch using Bayesian inversion techniques – so called, IGM tomography (Caucci et al. 2008, Cisewski, et al. 2014, Lee et al., 2014ab, Ozbek et al. 2016). The IGM distribution is directly linked to the dark matter distribution and as such, it is possible to reconstruct the full matter density field on scales of order of the mean separation of lines-of-sight (e.g., Pichon et al 2001, Caucci et al., 2008). Figure 4 shows such a tomographic reconstruction of a small volume of the IGM demonstrating the feasibility and power of this approach, and through SRO-10 this technique will come of age.

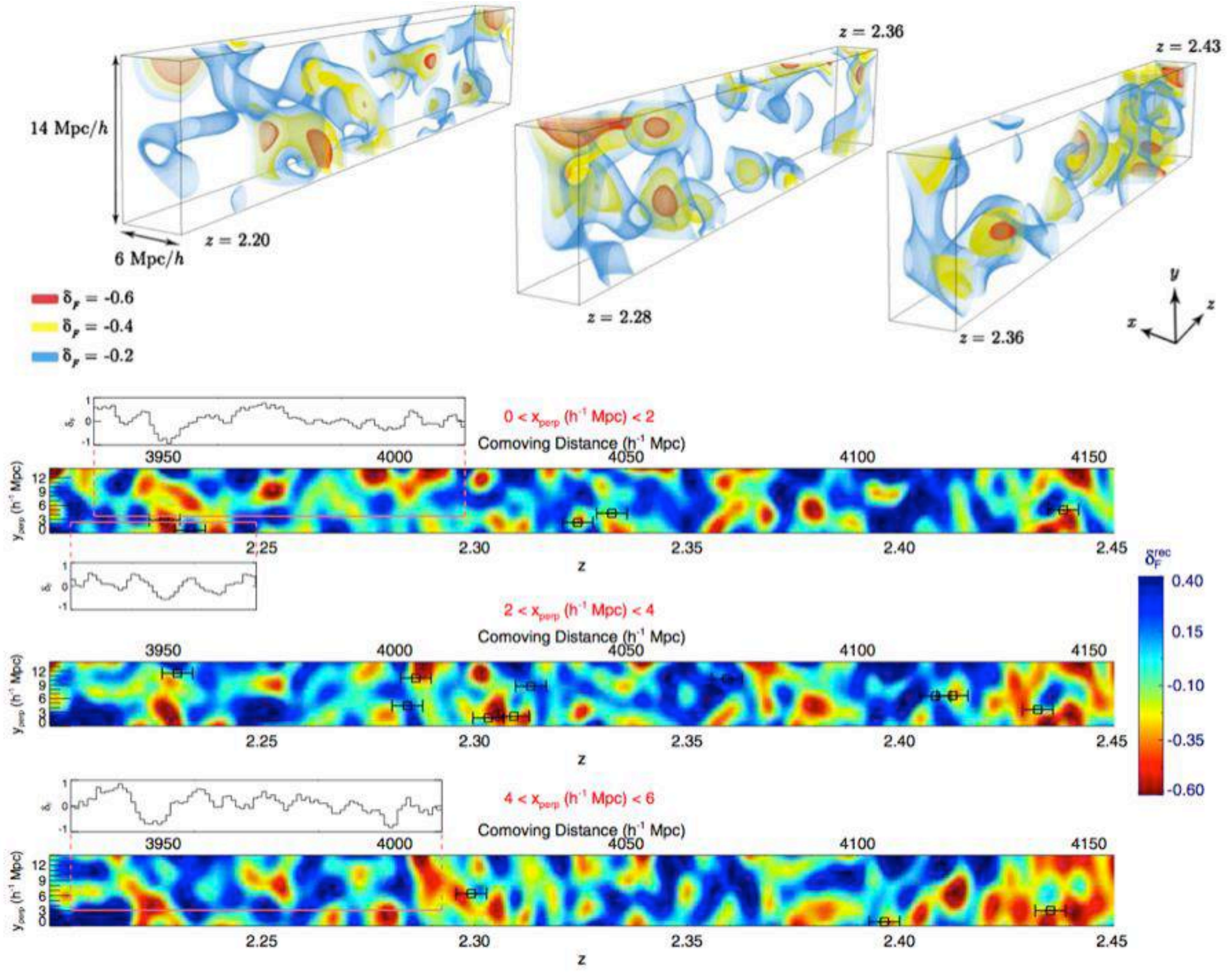

**Figure 75: Tomographic reconstruction of 3D Lyα forest absorption in COSMOS field using LRIS data from Lee et al. (2014a), shown in three redshift segments in 3D (top) and projected over three slices along the R.A. direction (bottom panels). The color scale represents reconstructed Lyα forest transmission such that negative values (red) correspond to overdensities. Square symbols denote positions of coeval galaxies within the map; error bars indicate the $\sigma_v \sim 300$ km $s^{-1}$ uncertainty on their redshifts. Pink solid lines indicate where three of the skewers probe the volume, with inset panels indicating the corresponding 1D absorption spectra that contributed to the tomographic reconstruction *(Figure from Lee et al. 2014a).***

The power of SRO-10 for large scale studies of the IGM will be significant. The Ly-α forest is a powerful probe of structure formation, but there are some important inconsistencies between current, complementary datasets. For example, the amplitude of the power spectrum, $\sigma_8$, inferred from the Ly-α forest flux power spectrum is somewhat larger than the value obtained from the CMB. This may be an indication that there are as yet unidentified physical effects that influence the individual measurements, such as the uncertain thermal state of the IGM (Viel et al. 2009). A large, independent sample of moderate-resolution quasar spectra in the redshift range $2.1 < z < 4.5$ (corresponding to about $0.38 < \lambda_{obs} < 0.67\mu m$ for Ly-α in the observed frame) would probe the flux power spectrum down to the Jeans scale and allow significant progress relative to existing studies. The sample of spectra obtained in SRO-10 will allow an exploration of possible systematic effects and offer valuable insights when compared to low-resolution SDSS data and the much sparser, higher-resolution spectra (see Rollinde et al. 2013).

An additional application of the Ly-α forest observations obtained as part of SRO-10 will be the measurement of its opacity, which is sensitive to the ionization and thermal state of the IGM, and provides an indirect probe of the sources responsible for maintaining the IGM in its highly ionized, post-hydrogen-reionization state. Precise measurements of the redshift evolution of the mean transmission through the Ly-α forest, $F \geq e^{-\tau} >\equiv e^{-\tau_{eff}}$, where $\tau_{eff}$, is the effective optical depth, provide a means to measure the intensity of the UV background (Bolton et al. 2005, Faucher-Giguere et al. 2008). This is an important probe of the ionizing emissivity of sources in the early Universe, and is therefore a key constraint on models of the H I reionization era. In addition, the epoch of singly ionized helium reionization is thought to complete by z ~ 3, roughly coinciding with the peak in quasar numbers (Furlanetto & Oh 2008). An increase in the IGM temperature due to the associated He II photo-heating is observed at z < 4 (Becker et al. 2011, Paris et al. 2011), and will also impact on the observed Ly-α forest opacity.

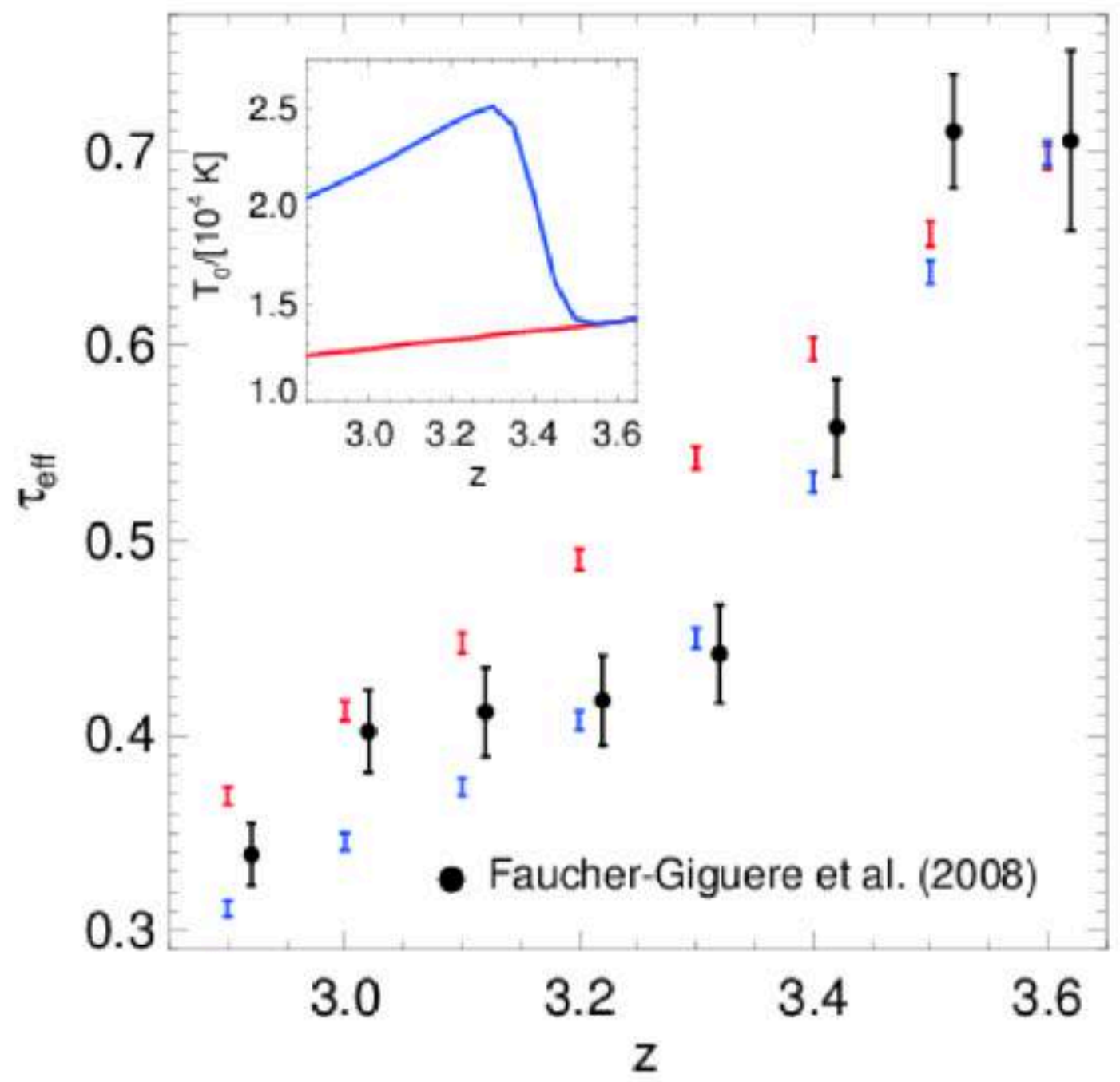

**Figure 76: The Lyα forest effective optical depth plotted against redshift. The filled circles with error bars (1σ statistical only, offset for clarity) display the measurements of Faucher-Giguere et al. (2008) from a sample of 86 ESI/HIRES Keck spectra at 2 ≤ z ≤ 4.2. However, other analyses find no evidence for this feature (e.g., Paris et al. 2011). For comparison, simulated MSE data (at R ~ 5000, SNR = 20) drawn from two different high-resolution hydrodynamical simulations are shown as the red and blue 1σ error bars. The spectra are drawn from models with different IGM thermal histories (see inset) which result in different temperatures at mean density, $T_0$, and hence different $\tau_{eff}$ . Each bin contains a total Lyα forest path length of 4000 proper Mpc (corresponding to ~ 200 independent quasar spectra for a single Δz = 0.1 bin at z = 3.4), and the statistical uncertainties show the standard error on the mean obtained by averaging over three proper Mpc chunks. A large, independently-selected sample of medium- to high-resolution MSE spectra would be well suited for examining the presence of this feature, as well as providing precise measurements of the effective optical depth.**

Measurements of the $\tau_{eff}$ redshift evolution are therefore valuable for improving existing constraints, exploring systematic uncertainties, and investigating the thermal history of the IGM. The latter has an effect on $\tau_{eff}$ through the temperature dependence of the H II recombination rate, such that the number density of neutral hydrogen $n_{HI} \propto T^{-0.7}$. At z ~ 3.2, a "dip" in the otherwise monotonic redshift evolution of $\tau_{eff}$ has previously been measured from spectra obtained with the SDSS survey (Bernardi et al. 2003) and more recently in ESI/HIRES Keck spectra (Faucher-Giguere et al. 2008). It has been suggested that this feature may be due to heating of the IGM during HeII reionization, although it is very difficult to explain this feature theoretically (Bolton et al. 2009). The large, independently-selected sample of moderate- to high-resolution spectra that will be obtained in SRO-10 will enable further examination of this key issue, particularly through targeted follow up at even higher resolution (see Figure 76).

A second, related area is measuring the temperature of the IGM, which is sensitive to the photo-heating of the IGM during reionization. Directly measuring the IGM thermal state requires the relatively narrow thermal widths of Ly-α absorption lines to be resolved:

$$b_{HI} = 12.9kms^{-1}\left(\frac{T}{10^4K}\right)^{0.5}$$

This requires high-resolution (R ~ 40 000) spectroscopy; lower resolution studies at around R ~ 20 000 still resolves the larger scale on which the IGM is pressure (Jeans) smoothed. This "Jeans smoothing" scale is set predominantly by photoheating during reionization. In principle, when combined with measurements of the instantaneous temperature, constraints on the IGM Jeans scale from quasars can thus yield insights into the timing and duration of the reionisation history. The uncertain thermal history is furthermore an important systematic when attempting to measure the instantaneous temperature of the IGM (Becker et al. 2011). The thermal structure of the IGM is also expected to be patchy on large scales due to the inhomogeneous nature of He II reionisation (McQuinn et al. 2009). Studying the line-of-sight variation in the data would thus provide a novel way to test both H I and He II reionization scenarios.

The low number density of sufficiently bright quasars that can be observed at moderate SNR in the high resolution mode of MSE would clearly preclude a very large program such as that described in SRO-10. However, a large number of quasars obtained in a P.I.-led program could still be used very effectively to probe the pressure smoothing scale in the IGM along — and, for sufficiently close quasar pairs, *transverse to* — the line of sight.

#### 4.2.2.3 The build-up of metals at z>2

The campaign to map the distribution of the IGM through Lyα described in SRO-10 and Section 4.2.2.2 also allows us to probe its metallicity. Metal absorption lines from the line of sight IGM are seen in the spectra of high redshift quasars and can be used to map the distribution and evolution of metals in the Universe (Petitjean & Aracil, 2004, Scannapieco et al. 2006). By observing systems with sufficient resolution and sensitivity, we can distinguish these metal absorption lines from those of the Lyman-α forest at the same epoch and map the distribution of metals in the IGM.

With the high signal to noise and moderate resolution observations discussed in Section 4.2.2.2, we will obtain highly robust spectra of a large number of high redshift sources. These spectra will allow the first detailed analysis of both the stellar- and gas- phase metal content of individual galaxies at high-z. Stellar-phase metals of the brighter sources can be obtained through rest-frame UV-continuum stellar absorption lines (using process similar to that discussed in Sommariva et al. 2012), while spectral observations out to 1.8µm will enable the identification of the emission line features required to determine gas-phase metallicities via the $R_{23}$ diagnostic (e.g., Peroux et al. 2013).

Current state of the art observations of galaxies at this epoch consist of either tens of sources observed with low resolution spectrographs (R∼200, e.g., Popesso et al. 2009, Henry et al. 2013), which rely on stacking analysis to identify stellar absorption lines and are limited to statistical analyses of the full population, or small numbers of well studied sources at moderate resolution (R ∼ 1000 – 3000, Belli et al. 2013, Maier et al. 2014). With the proposed IGM mapping observations of SRO-10, we will target ∼140 000 sources at z > 2 with the signal-to-noise required to fully investigate stellar absorption/nebular emission lines in individual galaxies. Target spectra could be further binned in resolution elements to increase signal to noise, while retaining sufficient resolution to identify key features required to determine stellar metallicities. In combination with deep photometric data to derive stellar masses (for example from Subaru/HSC and LSST), we will probe the mass – metallicity relation (e.g., Lara-Lopez et al. 2013a) for a large sample of individual galaxies. Though such a study we would, for the first time, produce a detailed large statistical study of the buildup of metals at z > 2 and witness the formation of the mass – metallicity relation in a robust sample of galaxies.

#### 4.2.2.4 Galaxy-scale absorbers: Damped Lyα systems (DLAs)

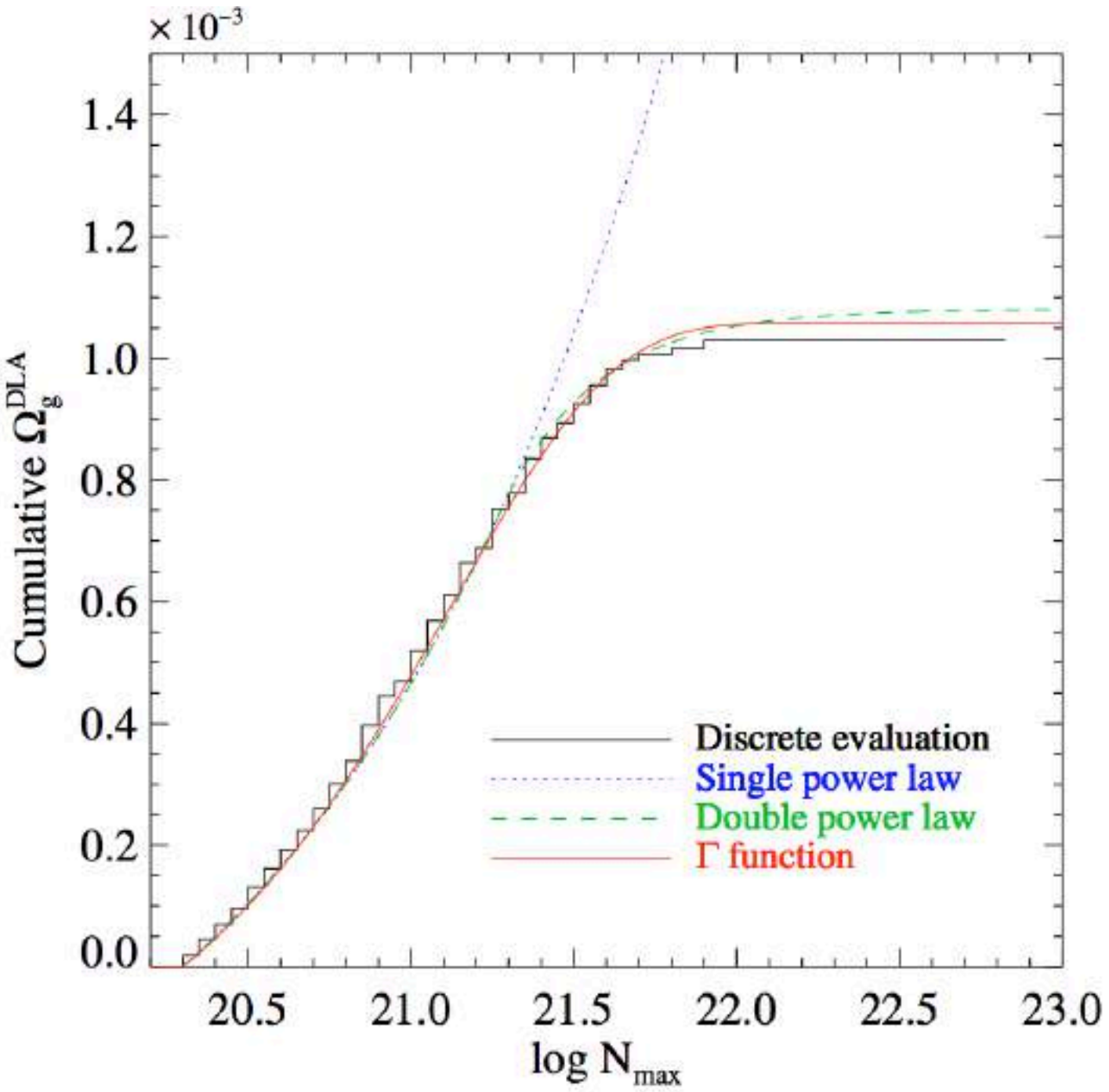

**Figure 77: Cumulative cosmological mass density of neutral gas in DLAs as a function of maximum column density from the study by Noterdaeme et al. 2009 . The apparent flattening of the curve at log *N(H* i) ∼ 21.7 implies convergence.** ***(Figure from Noterdaeme et al. 2009).***

Damped Lyα systems (DLAs) are observable at the highest hydrogen column densities and are expected to trace galaxies. The largest haul of DLAs to date comes from the SDSS (Noterdaeme et al. 2012), permitting an estimate of the cosmic density of neutral gas which is thought to trace the evolution of neutral gas reservoirs for star formation (see Figure 77). However, to move beyond the pure statistics of DLA numbers and HI gas content, high resolution follow-up with spectrographs on 8 – 10m-class VLT/Keck telescopes is required. This is because the metal lines used to determine the metallicities and element ratios are weak, narrow and cannot be easily detected with resolutions of a few thousand. Therefore, although some 1000 DLAs are known, detailed properties such as chemical abundances, kinematics and element ratios are known for just ~10% of these (Zafar et al. 2015). MSE would provide a sample of moderate-resolution spectra that would allow astronomers to increase dramatically the number of metallicity estimates for large samples of DLAs.

Metallicity studies of DLAs have many different applications in astrophysics. The ratio of alpha-to-iron peak elements is known to depend on star formation history, and very metal-poor DLAs offer an alternative to extreme metal-poor stars as diagnostics of early nucleosynthesis. Similarly, the primordial abundance of deuterium constrains Big Bang nucleosynthesis models, while studies of specific elements provide independent constraints on nucleosynthetic origins. Finally, fine structure lines provide independent information on star formation rates, and allow tests for possible time variations in physical constants. For these reasons, the efficient study of ever larger DLA samples is likely to be a priority in the coming decade, and, as we now explain, MSE can play a leading role in such efforts.

Although the last two decades has seen a steady increase in the use of DLAs as probes of chemical abundances at high redshift, the very large number of DLAs that would be available from MSE surveys would allow rare examples of extreme-metallicity DLAs to be identified. At low-metallicities, extremely metal-poor DLAs would provide an excellent complement to the abundance studies based on metal-poor Galactic halo stars. The most metal poor DLAs also offer excellent independent tests of low-metallicity yields (e.g., Akerman et al. 2005). The discovery of a carbon-enhanced metal-poor DLA (Cooke et al. 2011) is currently the only one of its kind, but potentially represents the missing link between Population III enrichment and standard chemical evolution models. At the other end of the scale, extremely metal-enriched DLAs represent the rare cases of galaxies that have self-enriched to close to solar metallicity already by $z \sim 2$ (Kaplan et al. 2010).

For high-resolution observations at the bluest wavelengths, there would be further exciting opportunities. Most notably, at $R \geq 20000$, it is possible to search for molecular gas, which has been detected in only a handful of DLAs to date (Noterdaeme et al. 2008). MSE can detect hundreds of molecule-bearing (mostly $H_2$) galaxies, and there are multiple scientific dividends from such detections. Obviously, this would be the first truly statistical census of $H_2$ at high redshift. Measurement of the various J level populations also provides one of the few ways that the internal temperatures and densities can be measured (Noterdaeme et al. 2010) and the detection of CO gives a measurement of the CMB temperature (Srianand et al. 2008, Noterdaeme et al. 2011). Perhaps of greatest physical importance, follow-up observations of the molecule-bearing DLAs could be used to search for time variation in the fundamental constants (e.g., Murphy et al. 2008). To date, such research has relied on just a few individual absorbers.

The best science results for MSE will require a careful matching of the wavelengths so that the metals and HI are covered for the same redshift range. For example, an excellent line for this work is SiII 1808Å: it is relatively strong, but usually not saturated, and Si is relatively unrefractory (compared with iron, for example). For absorbers in the range $1.8 \leq z \leq 3.5$, the blue low-resolution coverage (for Lyα) would need to be approximately 3400 – 5500Å. The high resolution coverage would need to be ~5000 – 8000Å. This type of study is further motivation for good blue sensitivity (in addition to studies of $H_2$).

In order to calculate the approximate absorption line numbers that will be covered in a survey covering ~10 000 $deg^2$, we adopt a typical quasar colour of $g - i = 0.2$. The bright survey limit is therefore approximately i =19.5. Combined with the quasar luminosity function, this a redshift path is hence $\Delta z \sim 1$ per square degree for a survey with i <19.5. At redshifts of $z \sim 2 - 3$, the number density (i.e, the number per unit redshift path) of DLAs is $n(z) \sim 0.2$. So, for an area of 10000 $deg^2$, we can expect 10 000 x 1 x 0.2 = 2000 DLAs.

With the newly available large number of spectra and images from SDSS in the last few years, it has been possible to undertake a "statistical" approach to some aspects of quasar absorbers studies, including stacking of spectra with low SNR. Examples of this include looking at the lensing effect (Menard & Peroux 2003), dust content (Murphy & Liske 2004; York et al. 2006; Frank & Peroux 2010; Khare et al. 2012; Noterdaeme et al. 2015) in composite spectra, stacking images with absorbers (Zibetti et al. 2005) or statistically correlating galaxies-absorbers (Bouche et al. 2004; Gauthier et al. 2009; Lundgren et al. 2009). With much enlarged samples of this kind, one would be able to divide surveys into data subsets that could be used to study some specific aspects in more details (e.g., influence of metallicity on DLA dust content). Studies such as these have laid the foundation for our understanding of the properties of gas in distant galaxies. However, our current understanding of the ISM at high redshift has clearly reached the point where it is limited by statistics, and this is where MSE can make an enormous impact on this field.

The unique contributions of MSE to this field are two-fold. First, it would be the first facility to provide survey level statistics at high spectral resolution. For instance, a bright-time survey of 10 000 $deg^2$ at the highest resolutions would yield SNR ~ 20 spectra (in 1 hr) at a limit of $g \sim 20.4$. However, in order to convert the measured metal line strengths into useful chemical abundances, the HI content of the absorber must also be measured. The Ly-α line is considerably bluer (at 1216Å) than most of the metal lines, and with equivalent widths (EWs) typically in excess of 10Å can be measured in much lower resolution spectra. The second benefit of the MSE is made through programs like SRO-10, at low/moderate resolution, to measure HI in DLAs over a redshift range of $z \geq 2$. The combination of coverage in both the blue (at lower resolution) and at red wavelengths at higher resolution would yield an extraordinary gain in the statistics of high-redshift DLAs.

## 4.3 Time variable extragalactic astrophysics

### 4.3.1 Reverberation mapping of black holes and the inner parsec of quasars

| **Science Reference Observation 11 (Appendix K)** |
| --- |

**Mapping the Inner Parsec of Quasars with MSE**

*The centre of every massive galaxy in the local Universe hosts a supermassive black hole that likely grew between a redshift of 1 to 3 through active accretion as a luminous quasar. Despite decades of study, the details of the structure and kinematics of the inner parsec of quasars remain elusive. Because of its small angular size, this region is only accessible through time-domain astrophysics. The powerful technique of reverberation mapping takes advantage of the changing emission-line properties of gas near the black hole in response to variations in the luminosity of the black hole's accretion disk to measure the sizes and velocities of the line-emitting regions; with this information, we can map the quasar inner parsec and accurately measure black hole masses. This information is essential for understanding accretion physics and tracing black hole growth over cosmic time; reverberation mapping is the only distance-independent method of measuring black hole masses applicable at cosmological distances. We propose a ground-breaking MSE campaign of ~100 observations of ~5000 quasars over a period of several years (totaling ~600 hours on-sky) to map the inner parsec of these quasars from the innermost broad-line region to the dust-sublimation radius. With high quality spectrophotometry and spectral coverage from 360 nm to 1.8 μm, this unprecedented reverberation-mapping survey will map the structure and kinematics of the inner parsec around a large sample of supermassive black holes actively accreting during the peak quasar era. In addition, a well-calibrated reverberation relation for quasars offers promise for constructing a high-z Hubble diagram to constrain the expansion history of the Universe.*

Active Galactic Nuclei (AGNs) and quasars (by which we mean the most luminous broad-line AGNs) are now thought to lurk at the hearts of all sufficiently massive galaxies. When we observe AGN, we know that the emission comes from metal-enriched gas surrounding a supermassive black hole. MSE will be the world's premier wide-field spectroscopic survey telescope for AGN; a legacy that is ensured by its large collecting area, high throughput and multiplexing, wide spectral coverage, and overall efficiency for survey operations. We have previously discussed AGN and black hole science as it relates to galaxy evolution in Chapter 3, and SRO-08 and SRO-09 in particular details ambitious surveys targeting both galaxies and AGN at high SNR. While their importance to galaxy evolution is significant, AGN, quasars and the supermassive black holes that provide the energy source are fundamentally interesting astrophysical phenomena in their own right, and here we focus on the use of MSE to probe the supermassive black holes and accretion processes that provide the engine to these energetic phenomena.

At the present epoch, supermassive black holes are ubiquitous in the centres of massive galaxies. The black holes grew predominantly around redshifts from 1 to 3, when the universe was approximately a fifth to a half of its current age through active accretion as luminous active galactic nuclei (AGN), also known as quasars. Remarkably, subparsec quasar accretion disks can outshine their host galaxies – thousands of times larger – by two to three orders of magnitude. Despite their small size, black holes are fundamentally linked to their host galaxies as shown through the strong scaling relations between the black hole mass and host galaxy properties. Energy injection during the quasar phase in the form of feedback may regulate these scaling relations; in any case, supermassive black hole growth clearly occurs alongside the build-up of stellar mass in galaxies. Measuring accurate black hole masses and understanding the inner structure of distant quasars is essential to advancing our knowledge of supermassive black hole growth, AGN physics and phenomenology, the mechanisms for launching quasar outflows, and the co-evolution of supermassive black holes and their host galaxies.

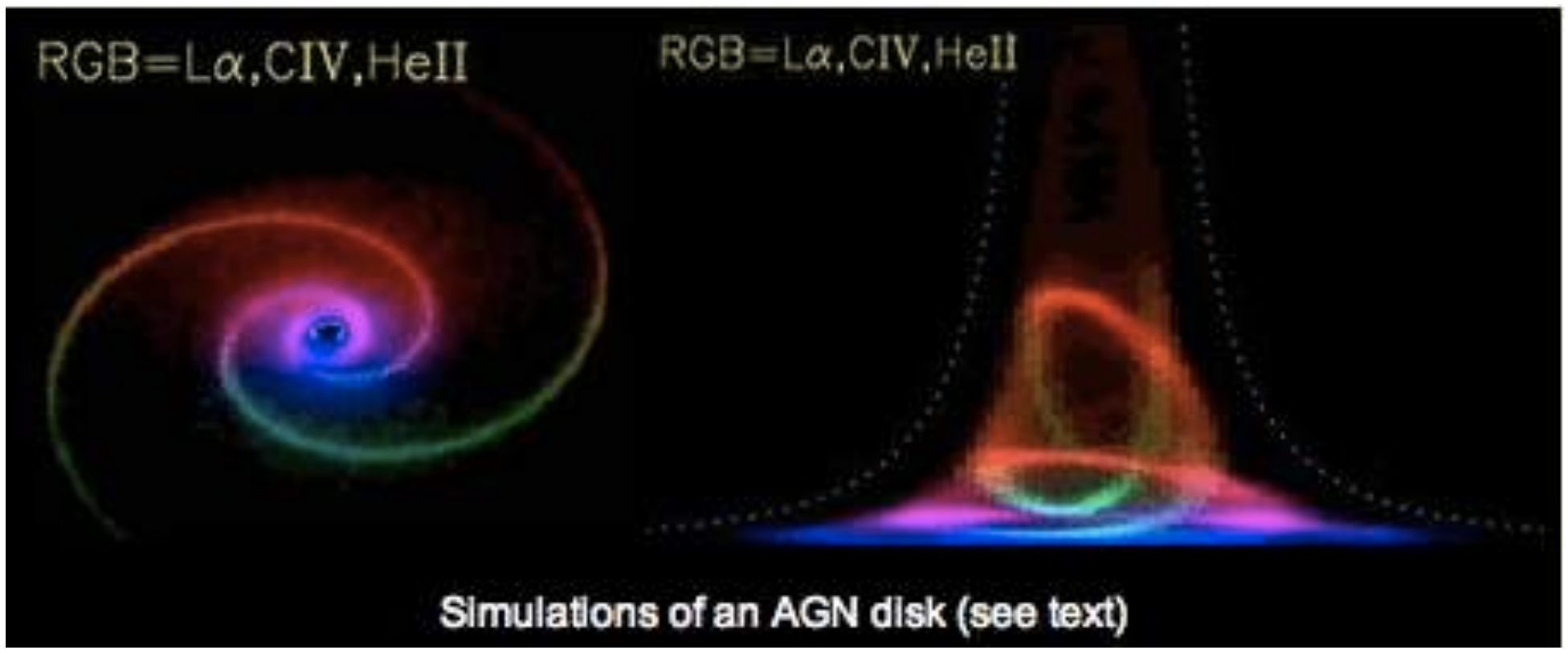

**Figure 78: Left: An image of a quasar where the surrounding broad-line region structure is that of a disk with a spiral density wave. Right: Time delay vs. velocity map for the Lyα (red), C IV (green), and He II (blue) emission lines. When the continuum emission from the accretion disk varies, the gas giving rise to the broad emission lines responds similarly after a characteristic time delay. The spread of velocity components in each emission line is generated by different locations within the broad-line region, and therefore the line as a whole responds with a range of time delays. With sufficient SNR, time sampling, velocity resolution, and flux calibration, the velocity-delay map can be inverted to reconstruct the disk image. (*Image credit:* K. Horne [star-www.st-andrews.ac.uk/astronomy/research/agn.php])**

Though quasars have been studied in radio through X-ray wavelengths for decades, there remain fundamental, open questions about accretion physics. For example, the well-known Shakura & Sunyaev (1973) prescription for describing the light distribution of the UV-optical emitting region of accretion disks underestimates their sizes by factors of several (Blackburne et al. 2011; Jimenez-Vicente et al. 2012; Edelson et al. 2015). The geometry and kinematics of the region generating the broad emission lines – the most prominent features of quasar optical-UV spectra – are still poorly constrained. These distant, cosmic powerhouses have such small angular sizes that they cannot be resolved with existing or near-term technologies; our only access to constraining their structure empirically is through time-domain astrophysics. In particular, we can use time resolution to substitute for angular resolution by measuring the rest-frame time lag of the response of a broad emission line's flux to changes in the continuum illuminating the broad-line region (Blandford & McKee 1982).

Strong, broad, resonance lines are seen from ions with a large range of ionization states, from O VI to Mg II; the gas is photoionized with lower-ionization gas at larger radii out to the dust-sublimation radius. The recombination times in the broad-line region are short, and so the time lag between the continuum and broad-line flux variability corresponds to the light travel time between the central, ionizing UV continuum source and the broad-line region. Measuring this time lag thus provides a characteristic distance between the broad-line region gas and the supermassive black hole, $R_{line}$. The broad emission lines (with widths of thousands of km $s^{-1}$) also reveal the Doppler motions of dense gas close to the central black hole. With a characteristic distance (from the time lag) and velocity (from the line width), the black hole mass is given by $M_{BH} = f(\Delta v)^2 R_{line}/G$ where $G$ is the gravitational constant, $\Delta v$ is a measure of the width of the emission line, and *f* is a factor of order unity which accounts for the geometry and kinematics of the broad-line region (see, e.g., Peterson 2011, and references therein).

Furthermore, for an individual quasar, the time lags and the average continuum luminosities generate an $R_{line}$ value for each line. Comparison of the characteristic time lags from different lines puts powerful constraints on the structure, kinematics, and physical conditions (e.g., gas density and ionization parameter) of the broad-line region gas (Korista & Goad 2004). In the best cases, with appropriate velocity resolution and time sampling, high-fidelity velocity-delay maps (line responsivity as a function of line-of-sight velocity and time delay; see Figure 78) can be obtained with a reverberation-mapping (RM) campaign with a duration $T_{dur} > 6\tau_{line}$, where $\tau_{line}$ is the time delay for a given line. Such data from multiple broad emission lines with varying ionization potentials have the power to image even complex structures in the central regions, for example spiral arms in a disk of accreting material.

Currently, only ~60 local, relatively low-luminosity AGN have RM-based measurements of their black hole masses (Bentz & Katz 2015), and only ~20% of these targets have high-quality data sufficient for producing velocity-delay maps. The largest optical mapping campaign to date is the ongoing (2 yrs to date) SDSS-BOSS program to monitor 849 $i < 21.7$ AGNs with the 2.5-m Apache Point Telescope in a single 7 deg$^2$ field (Shen et al. 2015). Dedicated time-intensive programs that monitor the rest-frame UV through optical of a *handful* of AGN are currently underway with e.g., HST COS (NGC 5548; P.I. Peterson) and the VLT X-Shooter (P.I. Denney). Though these programs will be foundational for taking the next step forward in RM science, MSE promises to go significantly further. Cosmological redshifting means that *optical* spectroscopic RM campaigns of quasars are fundamentally limited by a mismatch between the high quality data at low-*z* and what is accessible for high *z*. The lines sampled in these two regimes – e.g., Hβ and C IV for low and high *z*, respectively – arise from significantly different parts of the broad-line region, and there is large scatter in the black hole masses derived from these two lines in single epoch spectra as a result of their distinct characteristics (Denney 2012). Furthermore, local RM AGN occupy fundamentally different regions of parameter space than typical high-*z* quasars in terms of luminosity, the shape of their ionizing continua, and likely their host galaxy properties. *Accurate reverberation-mapping black hole mass measurements for a large sample of z~2 quasars would enable high-accuracy calibration of black hole masses from single-epoch spectra for the first time.*

SRO-11 describes a dedicated reverberation mapping campaign using MSE that will provide a transformational advance in this field by obtaining ~100 observations of ~5000 quasars over a period of several years (totaling ~600 hours on-sky). These observations will map the inner parsec of these quasars from the innermost broad-line region to the dust-sublimation radius. With high quality spectrophotometry and spectral coverage from 360 nm to 1.8 µm, this unprecedented reverberation-mapping survey will probe the structure and kinematics of the inner parsec around a large sample of supermassive black holes actively accreting during the peak quasar era. The major advantages of MSE over other planned RM programs (e.g., SDSS, OzDES, or 4MOST) are sensitivity (~ 2 – 3 mags deeper in 1 hr) and near-IR capability because of the planned mirror size and instrumentation, and the exquisite Mauna Kea site. The only planned instrument that would be competitive with MSE on these terms is Subaru/PFS; however, the PFS wavelength range only extends to 1.25 mm and there are no current plans to invest the requisite time to a mapping campaign as outlined in this proposal. *The inclusion of near-IR capability (to 1.8 mm) to reverberation-map the rest-frame UV-optical for ~5000 high-z quasars will make MSE a game-changer in this field* (see Figure 79).

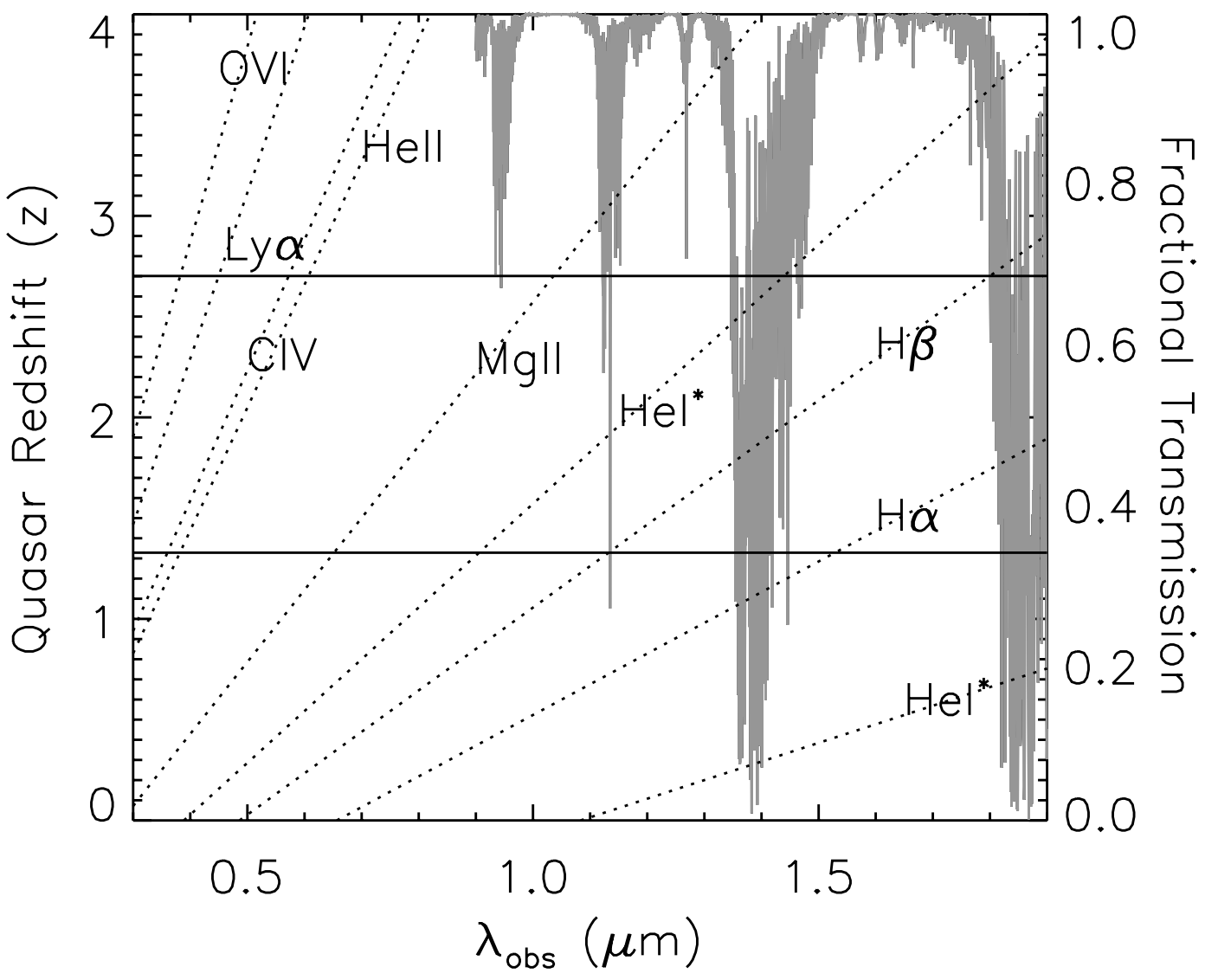

**Figure 79: The observed wavelength of broad emission lines of interest overlaid on a representative atmospheric transmission spectrum for Mauna Kea. Across the peak of the quasar epoch (*z* = 1.3 – 2.7; bounded by horizontal lines) emission lines from O VI to Hα are accessible with wavelength coverage from 360 nm to 1.8 μm; at least one hydrogen line will be at a region of high atmospheric transparency. This range of lines probes size scales from the innermost broad-line region (of order light-days to light-weeks) to the dust sublimation radius (~10 times larger; Mor & Netzer 2012).**

In addition to this core science, SRO-11 will provide a well-calibrated reverberation relation for quasars. This offers the promise for constructing a high-z Hubble diagram to constrain the expansion history of the Universe. Though the demographics of the quasar population have changed remarkably since $z \sim 3$, the structure of luminous quasars shows surprisingly little evolution in fundamentals such as metallicity and spectral energy distribution. They are thus promising objects for constructing a high-*z* Hubble diagram given an appropriate independent estimate of luminosity such as a well-calibrated $R_{line} - L_{cont}$ relation (Bentz et al. 2013). The size of a line-emitting region can be measured from reverberation; this then yields the average quasar luminosity. From the measured flux and the redshift, a Hubble diagram to z ~ 3 can be made from the proposed MSE high-*z* quasar reverberation-mapping campaign, constrain possible time variation in the dark energy equation of state, which would change the distance-redshift relation (Watson et al. 2011, King et al. 2014). Based on SN Ia, the distance-redshift relation is constrained only out to z ~ 1.7. Of course, considerable work remains to be done over the coming decade to ensure that this limit can be reached in a large fraction of randomly selected quasars.

### 4.3.2 Supernovae Ia and massive transient datasets

When fully operational in the 2020s, LSST and SKA1 will give us unprecedented views of the transient universe. LSST will produce orders of magnitude more transients than all of our ground-based spectroscopic assets can hope to follow-up, with estimates of $10^5$ – $10^6$ transient alerts per night.

Transients are a diverse and largely unexplored area of astronomy. Nevertheless, we already have a foretaste of the riches on the horizon from recent wide-field transient studies that work to shallower depths than LSST (e.g., < 16 – 17mags) such as Pan-STARRs, the All-Sky Automated Survey for Supernovae (ASAS-SN), Intermediate Palomar Transient Factory (iPTF), and the Catalina Real-Time Transient Survey (CRTS). Essential to understanding the events discovered by LSST and on-going all-sky transient surveys is spectroscopic follow-up, especially in the cases of rare or previously unknown transients. The importance of this area to the future of the astronomical community was highlighted by the recent NOAO workshop "Spectroscopy in the Era of LSST", held in Tucson in April 2013.

For transient observations, most of the current follow-up programs are performed on object-by-object basis in view of their low number densities and the limited flexibility of multi-object spectrographs. A massive transient follow-up survey, potentially linked with deep spectroscopic survey projects, is only practical on highly multiplexed, rapidly reconfigurable spectrographs with access to large input catalogues. It is here that MSE can dominate the identification of faint transient or variable astrophysical sources of scientific interest, and these observations represent an extreme observing mode for the facility. Rather than directly taking advantage of its massive multiplexing capabilities, transient studies with MSE take advantage of the dedicated spectroscopic nature of the facility to accumulate large datasets over time with only a very small fraction of science fibers ever dedicated to targets. If N targets in every MSE observation (that uses a total of $N_{fibers}$) were observing recently identified transients (say, from LSST or the SKA), the survey efficiency decrease of $(N/N_{fibers})$% would yield the equivalent of N 10-m telescopes doing nothing but transient spectroscopy with their single object spectrographs. This effort would be distinct from target-of-opportunity observing that interrupts ongoing observations; such programs are not well suited to a survey facility like MSE.

A key area of transient science for MSE is in the follow up and characterization of Type Ia supernovae (SNe Ia). These are well established cosmological probes whose luminosity distances provide key constraints on the Dark Energy equation-of-state, w. SNe Ia are geometrical probes that are independent of large-scale structure growth. This, even if future improvements on the Dark Energy Figure-of-Merit (FoM) from SNe Ia are not be as large as from other probes (e.g., weak lensing, BAO, etc.), more data at higher redshifts will be useful for comparing basic cosmological assumptions (particularly if the Universe is not simply described by $\Lambda$CDM). Of course, there is more to cosmology than merely improvements to the Dark Energy FoM, and having larger and more distant SNe Ia samples will, for instance, allow increasingly precise tests of isotropy.

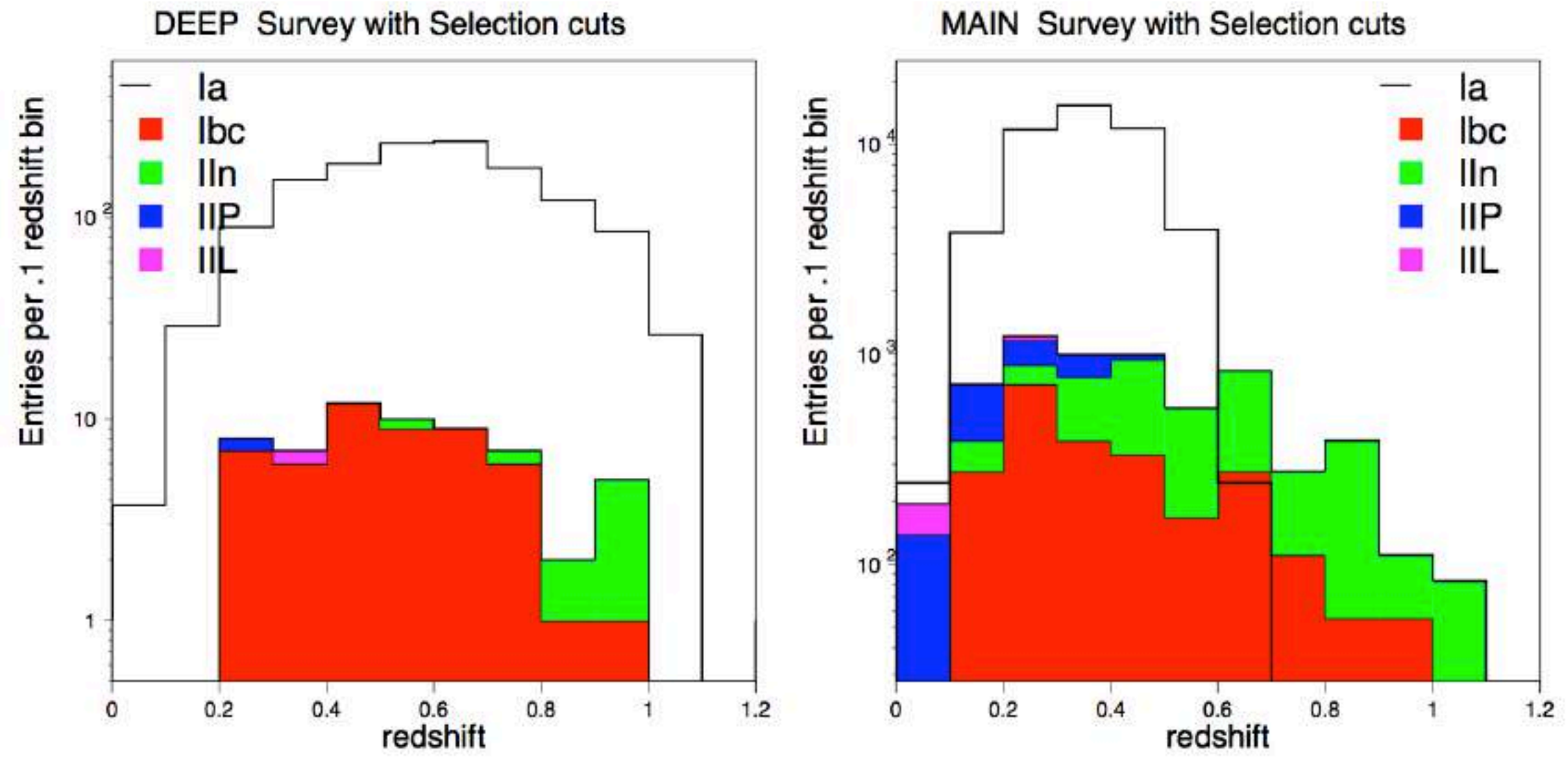

**Figure 80: Predicted rates of Type Ia and non-Ia supernovae as a function of the redshift detected by LSST in the DEEP survey field (left panel) and the main survey (right panel). (*Figure from The LSST Science Book v2).***

Improvements on w to the ∼1% level, using SNe Ia of redshifts z ≤ 0.8, must involve better control of systematic effects: e.g., improved calibrations and better knowledge of the physics driving SN explosions. The analysis of nearby SNe spectra suggests that standardization with spectra alone may be possible. This promises to reduce the intrinsic SnIa dispersions using accurate spectral sub-classifications.

While the number of high-redshift SNe Ia (z > 0.8) is still quite small, samples will improve dramatically with the next generation of imaging surveys (e.g., Euclid, LSST). LSST is aiming to perform high-cadence observations of several "deep drilling fields" in order to obtain well-sampled light curves for ∼ 50 000 SNe Ia per year. Follow-up spectroscopy for these objects will be essential for obtaining their physical properties. Figure 80 shows the predicted number of supernovae of various types in the LSST as a function of redshift in both the deep drilling fields (left panel) and the main survey area (right panel). Given their faintness, a wide-field, large-aperture telescope with a highly-multiplexed spectrograph will be essential. Most of these deep drilling fields will likely be at equatorial latitudes to facilitate this essential spectroscopic follow up. On average, each deep drilling field will contain roughly a thousand SNe Ia per year, so that a single pointing with MSE will contain 150 SNe Ia over the course of a year. Furthermore, Euclid will identify hundreds of SNe (Astier et al. 2014) for which ancillary ground-based follow-up will provide checks on systematics in typing and redshift determination. MSE is poised to become the de facto SNe-followup telescope for LSST and other surveys

Additional, exciting transient science (i.e., beyond SNe Ia cosmology) will certainly benefit from enormous numbers of spectra taken with a wide-field, 10m telescope. Future surveys will deliver thousands of candidates, and efficient spectroscopic follow up will be essential if we are to better understand the physics driving these events. Many of those sources will be run-of-the-mill AGN, typical SNe, or unremarkable examples of known types of stellar variables. But variable-source surveys are already producing exciting science on unusual supernovae (e.g., Chornock et al. 2013), tidal disruptions of stars by supermassive black holes (e.g., Chornock et al.

2014, Holoien et al. 2014), extreme stellar flares (e.g., Schmidt et al. 2014), etc. The transient classes targeted will evolve to keep the science cutting-edge. From the start, typical stellar and quasar variability will be ignored so that the stars and quasars being targeted in various surveys do not dominate the transients being followed up. It is interesting to note that LSST alone will image ~1500 deg$^2$ per night accessible to MSE, in which area there will be ~300 new SNe and >75000 variable stars (Ridgway et al. 2014). Observing programs could be established to monitor single-pointing deep spectroscopic fields during their MSE observing windows. MSE will dominate a new regime of faint transients (probing very early or very late stages, as well as large distances and low luminosities), including fast transients, as it will be able to dedicate more of its time to transient spectroscopy than any other 10-m class telescope.

## 4.4 The dark sector of the Universe

Over the past century, cosmology has evolved from a theoretically motivated science to one driven mainly by observations. With the discovery of the late-time acceleration of the Universe expansion (Riess et al. 1998; Perlmutter et al. 1999) possibly driven by a new form of energy with sufficiently negative pressure, astronomers and physicists have concluded that 96% of the energy density of the universe is in a non baryonic — or "dark" — form. There is a growing realization that understanding these new components of the universe (i.e., the dark matter and dark energy) may require fundamentally new physics. Numerous ideas have been advanced to explain the acceleration and predict its redshift evolution (e.g., see the review of Frieman et al. 2008).

Determining the cosmological world model has thus become one of the hottest questions of contemporary cosmology. Current and expected technologies (in terms of telescopes and instrumentation) offer the possibility to probe a very large fraction of the observable universe (both in imaging and spectroscopy) to a steadily increasing depth. Indeed, the exponentially growing volume of data is now the key factor in supporting the opportunity to test new theories in cosmology. In the 2020s and beyond, the specific constraint of cosmological parameters will be conducted on dedicated facilities/experiments that are optimised to provide measurements to better than 1% via wide-area deep optical and near-IR imaging facilities (e.g., 4MOST, DESI, Euclid, WFIRST, LSST), via combinations of weak lensing (high-resolution imaging), baryonic acoustic oscillations (BAO, mainly using spectroscopic redshifts), and redshift space distortion (RSD) analysis. In these areas, the number of sources rather than redshift accuracy or spectrum continuum signal-to-noise is generally paramount.

The resulting three-dimensional mapping of galaxies provides the information needed to accurately determine the galaxy power spectrum and its time evolution, thereby constraining the various cosmological parameters. Specifically, the isotropic power-spectrum of any large-scale structure tracer — such as galaxies, quasars or the Ly-α forest absorbers — probes universe geometry, the accelerated universal expansion, the matter and energy content of the universe, the law of gravitation on large scales, neutrino masses, and the level of primordial non-Gaussianity. The anisotropic galaxy power-spectrum is caused by an effect called Redshift Space Distortions (RSDs), which also provide a good test for our theory of gravity. In order to reach sub-percent precision, the accurate modeling of astrophysical processes that affect the cosmological measurements is capital.

As described in the rest of this document, the forte of MSE is in the detailed study of astrophysics, and it is this that drives much of the design of the facility. As such, MSE has a significant and important role to play for other dedicated cosmological surveys, in terms of the quantification of the bias of tracers, the measurement of cosmic variance, the calibration of photometric redshifts, and other factors relating to essential calibration. For example, the calibration of photometric redshifts for the Euclid space mission requires deep, ground-based, spectroscopic observations of $10^5$ galaxies to a depth of $RIZ_{AB}$ = 24.5 (Euclid photometric band), and these galaxies must be representative of the sample of galaxies used for weak lensing. This is in addition to follow up spectroscopy required to ensure that the "purity" of the sample (i.e., the fraction of correct redshifts that are measured using on-board slitless spectroscopy) is known to a value better than 0.1%. Strategies to implement this test using the Euclid deep survey require ground-based confirmation; large and complete spectroscopic surveys will be particularly useful for minimizing this important systematic uncertainty (see Figure 81)

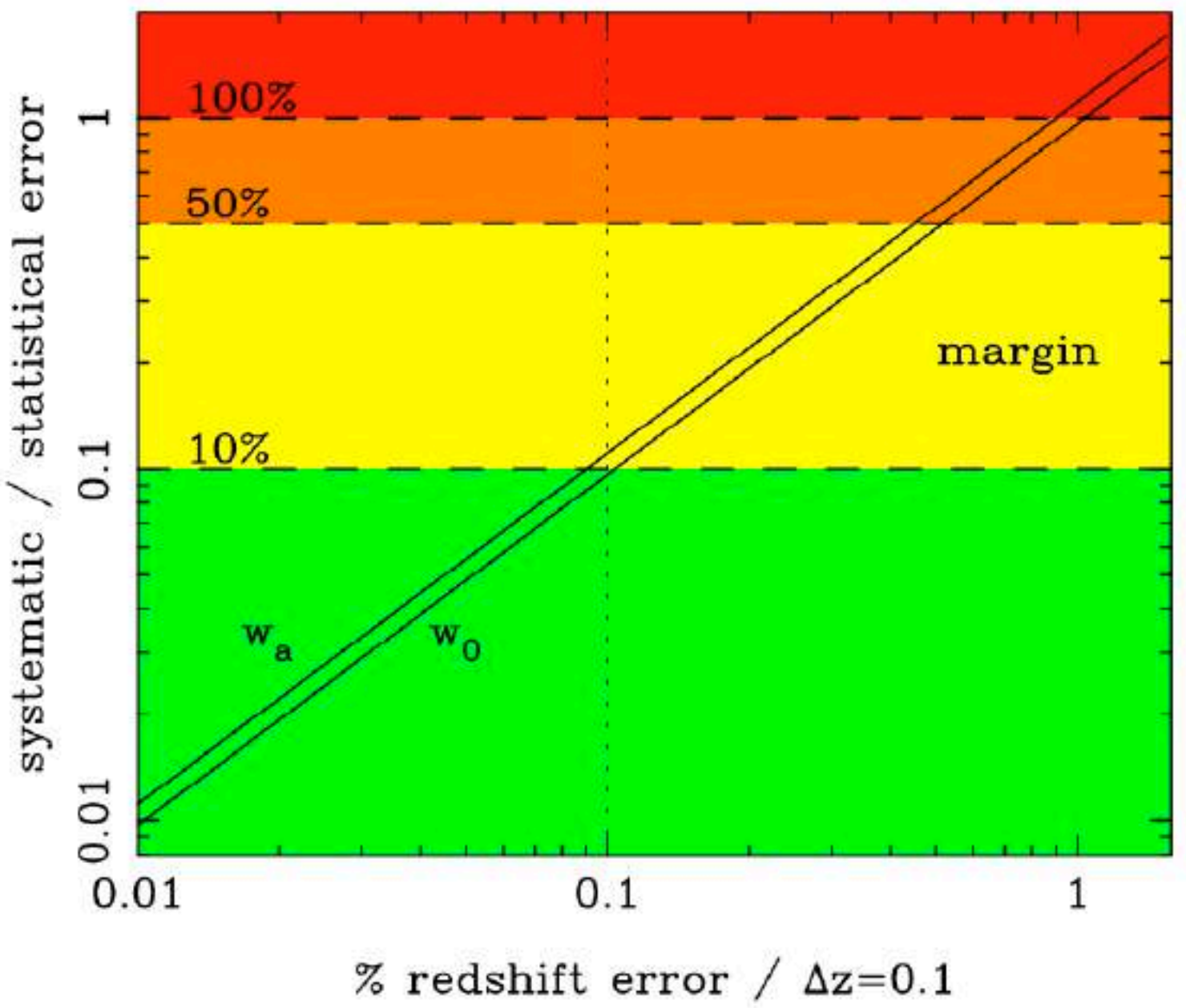

**Figure 81: The predicted systematic error divided by the marginalised statistical error for parameters $w_0$ and $w_a$, from Euclid BAO measurements (including the Planck data), required to normalise the BAO scale (solid lines). These are calculated for different values of the systematic error placed on the mean redshift in a bin of width Δz=0.1. The requirement for Euclid is 0.1% (shows by the vertical dotted line) and can be verified with Euclid slitless spectroscopy using a sample of 140 000 galaxies with purity >99%. Ground based observations of a significant fraction of these same galaxies from a highly complete spectroscopic survey will play an important role. (*Figure from Laureijs et al. 2009*).**

From a stand-alone perspective as well as a calibration one, MSE also has the potential to be a critical tool in the development of our cosmological model. Its wide field of view (to probe large cosmological scales), its high level of multiplexing (numerous tracers targeted simultaneously), its short reconfiguration time (observing efficiency), large aperture and high system throughput (to detect faint sources and increase sample size) are all vital elements to address forefront cosmological questions. But competition is intense, especially with the advent of Mayall/DESI and Subaru/PFS.

In the rest of this section, we explore the potential impact of MSE in the field of cosmology. In keeping with the focus of this document on science that is uniquely enabled and compelling because of the capabilities of MSE, we do not include further discussion of BAO experiments. BAO is a key (if not driving) component of many current programs in cosmology. Such surveys are clearly possible with MSE; indeed, the dedicated operations of MSE makes it the ideal 10m class facility for such a survey. But the expected science gain over what will be available in the mid 2020s is hard to judge in such a competitive field. The capabilities of MSE will nevertheless be available if results from currently planned surveys require future BAO surveys to take place.

As relates to the two elusive components, Dark Matter and Dark Energy, for the former MSE is *the* Astronomical Dark Matter Machine. No facility comes close to obtaining a more comprehensive and complete understanding of the properties of dark matter from sub-dwarf galaxy scales up to the largest cluster scales. For the latter, the capabilities of MSE offer the potential of going beyond the current and planned dark energy experiments through dedicated programs. MSE – when equipped with IFUs – uniquely offers the potential for investigation of the velocity field of stars in distant galaxies, that is essential for advancing RSD analyses and which, when combined with weak-lensing shape measurements, will increase significantly cosmological constraints from weak lensing (Reyes et al. 2010).

### 4.4.1 The dynamics of dark matter halos from sub-luminous to cluster mass scales

#### 4.4.1.1 Identifying the dark sub-halo population through precision velocities

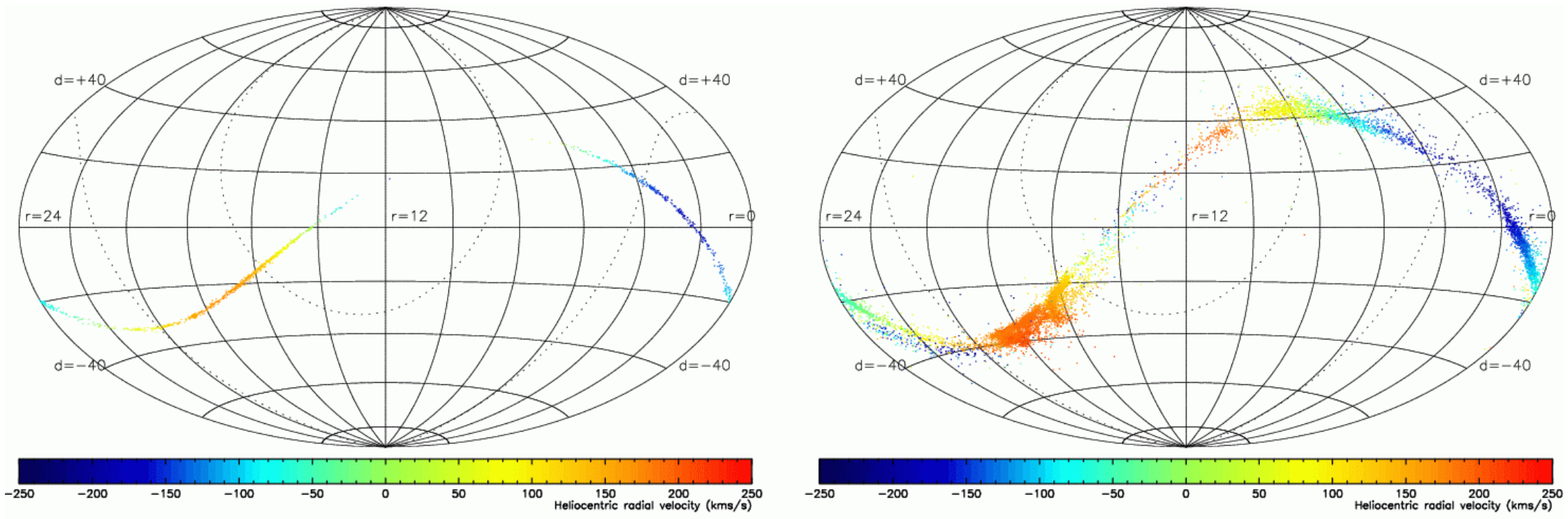

**Figure 82: The effect of dark matter halo sub-structures on low-mass star streams. In a smooth halo (left panel), low-mass stellar streams follow narrow paths on the sky, confined to a narrow range in distance and velocity (shown in colour). In the presence of dark matter substructure (right panel), the potential becomes uneven and the stream path and velocity structure become complex. With Gaia, such studies will now finally be within reach. (*Figure from Ibata et al. 2002*).**

Cold dark matter cosmology predicts that galaxies contain hundreds of dark matter sub-structures (Klypin et al., 1999) with masses similar to dwarf satellite galaxies. Only a small fraction of these dark satellites can be identified with the observed population of satellites, however, which raises the question of where the missing satellites are. A large body of theoretical work has demonstrated that it is possible to hide the vast majority of dark matter satellites by having their baryons expelled during the era of reionization (see, e.g., Kravtsov 2010, and references therein). However, it remains a fundamental prediction of ΛCDM theory that the dark matter clumps exist in large numbers. Possibly the best means to test this prediction is to examine the dynamical influence of such structures directly in nearby galaxies where we possess the richest datasets. Indeed, Ibata et al. (2002) and Johnston et al. (2002) demonstrated that massive dark satellites can strongly perturb fragile structures such as stellar streams. The halo substructures change the host galaxy from a smooth force-field in the absence of CMD lumps (left hand panel of Figure 82) into a "choppy sea" where the stream and its progenitor are tossed hither and thither (right hand panel of Figure 82). The effect of this is that the stream path becomes twisted, which leads to density variations along the stream (Siegal-Gaskins & Valluri 2008, Yoon et al. 2011, Carlberg 2012), kinks in the stream track (Erkal & Belokurov 2015), and accompanying dynamical heating (Ibata et al. 2002, Carlberg 2009). Measurements of this effect allow the mass spectrum of the population of $M \lesssim 10^8 M_\odot$ subhalos to be determined.

The globular cluster stream of highest contrast that is currently known is that of Palomar 5, which is a structure that can be seen directly in SDSS star-count maps of blue point sources. Another good target is the GD-1 stream (Grillmair & Dionatos 2006), a long, thin stream in the northern hemisphere. However, even for these most favourable of objects, only a handful of stars in the streams are bright enough to be detectable by Gaia (their main sequence turnoffs lie at g = 20.2 and g = 18.5, respectively). Thus it is very unlikely that with Gaia alone we will be able to detect many distant halo streams and examine their kinematics in enough detail to detect the presence of dark-matter subhalos. Clearly this is also beyond the capabilities of surveys undertaken on 4m-class telescopes.

However, MSE will allow us to solve this problem. New stellar streams may be found directly in large MSE spectroscopic surveys, or indeed by following up candidate structures found for instance with Euclid or LSST photometry. As with Palomar 5 and GD-1, it is likely that (post-facto) a few giant-branch members can be associated to Gaia stars with proper motion measurements, which would provide an accurate (approximate) orbit for the detected stream.

Simulations in the presence of a large population of expected CDM mini-halos show that there will be localised heating of the globular cluster stream stars by typically a factor of two over and above the intrinsic velocity dispersion of the stream (which in the case of Palomar 5 is about 2 km/s, Kuzma et al. 2015) and changes in the mean velocity along the stream of up to 10 km/s (Carlberg et al. 2015) due to encounters with the larger dark structures. The spatial scale over which the increased dispersion will be present depends on the physical size of the biggest structure that has heated the stream, but are typically larger than about 1 kpc.

In addition, if ΛCDM is correct, there will also be effects from the hundreds of smaller dark matter structures that will make the Galactic potential locally quite irregular. With the mini-halo mass function adopted in Ibata et al. (2002), these cause a heating of the stream of $\sim$5 km/s/Gyr, so that the velocity dispersion will increase approximately uniformly with distance along the tidal tail.

Finding long streams without small-scale velocity structure would provide incontrovertible proof that dark matter is smooth on small (sub-galactic scales), while discovering evidence for heating by dark clumps would be a vindication of the existence of the dark structures and a triumph for dark matter theory. Either result would be a legacy for MSE. SRO-04 describes an observational program, perhaps undertaken as part of the Galactic Archaeology survey of MSE, that will realize this transformational goal.

#### 4.4.1.2 Dwarf galaxies as dark matter laboratories

Given their proximity, large dynamical mass-to-light ratios and early formation times, the Local Group's dwarf galaxies have become attractive targets for investigating the nature of dark matter. Cosmology makes predictions regarding the number and internal structure of the dark matter halos that host dwarf galaxies, while particle physics makes predictions regarding the emission of high-energy photons from dark matter's non-gravitational interactions. Combined with data from X-ray and gamma-ray observatories, MSE's ability to gather large stellar-kinematic samples for faint dwarf galaxies will be crucial for testing all of these predictions and revealing the nature of dark matter in both cosmological and particle physics contexts.

##### 4.4.1.2.1 Cosmological Structure Formation

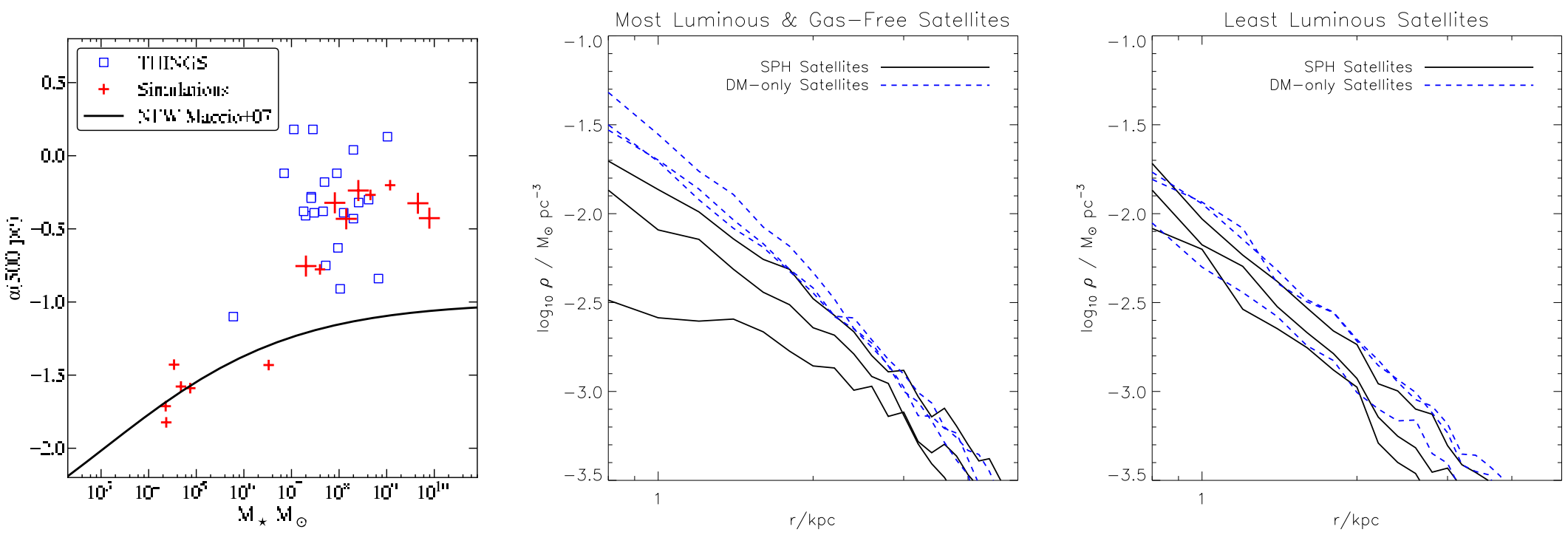

**Figure 83: Dark matter density profiles and effects of baryon-physical processes. Left: Logarithmic slopes (α) of dark matter density profiles, measured at radius r = 500 pc and redshift z = 0 in observed (squares; Oh et al. 2011) and cosmologically-simulated (crosses) field dwarf galaxies, as a function of stellar mass. The simulations use hydrodynamics and star formation prescriptions to study the impact of baryon physics on the structure of cold dark matter (CDM) halos; the solid line indicates the theoretical prediction without baryon physics (i.e., CDM only). (*Figure from Governato et al. 2012*). Middle and right: Dark matter density profiles at z = 0 for simulated Galactic satellites in CDM + baryons (solid black lines) and CDM-only (blue dashed lines) runs. The most luminous satellites $L_V < 10^7 L_{V,\odot}$; similar to Fornax) develop shallower central density profiles, while lower-luminosity satellites $L_V < 10^6 L_{V,\odot}$; similar to Draco) retain centrally cusped profiles. (Figure from Zolotov et al. 2012).**

The current paradigm for cosmological structure formation is built on the hypothesis that dark matter is "cold" and collisionless, enabling hierarchical assembly of dark matter halos beginning on subgalactic scales (White & Rees 1978, Blumenthal et al. 1984, Davis et al. 1985, Navarro, Frenk & White 1997, Springel et al. 2005). Indeed, calculations of the matter power spectrum associated with popular "weakly interacting massive particle" (WIMP; e.g., neutralinos) candidates for the dark matter particle imply that galaxies like the Milky Way should host $\sim 10^{15}$ satellites in the form of individual, self-bound dark matter subhalos, sub-subhalos, ..., and "microhalos" with masses ~ 1 Earth mass (Hofmann et al. 2001, Green et al. 2004, Diemand et al. 2005). Yet the smallest dark matter halos that have been inferred from observations have masses $\sim 10^5 M_\odot$ within their luminous regions (often extrapolated to virial masses $M_{vir} \sim 10^8 M_\odot$), leaving open the possibility of more-complicated dark matter models that would inhibit structure formation on smaller scales (Spergel et al. 2000, Bode et al. 2001, Dalcanton et al. 2001, Gilmore et al. 2007).

Thus for astrophysical purposes, dark matter particle candidates and associated cosmologies can generally be classified according to whether the dark matter particle's properties (e.g., mass and corresponding free-streaming scale, non-gravitational interactions, etc.) play a role in galaxy formation. For the standard cosmology, based on collisionless, cold dark matter (CDM), they do not: non-gravitational interactions are negligible and free-streaming scales are tiny compared to even the smallest galaxies. As a result, the internal structure of CDM halos (i.e., density profile, power spectrum) is similar on all scales (Navarro, Frenk & White 1996, 1997). In contrast, cosmologies based on “warm” dark matter (WDM) invoke larger free-streaming scales that can effectively truncate the matter power spectrum on scales sufficiently large to inhibit the formation of dwarf galaxies. Moreover, non-gravitational self-interactions (e.g., via Yukawa-like potentials, (Loeb et al. 2011) can affect the spatial distribution of dark matter within individual halos (Vogelsberger et al. 2012, Rocha et al. 2013).

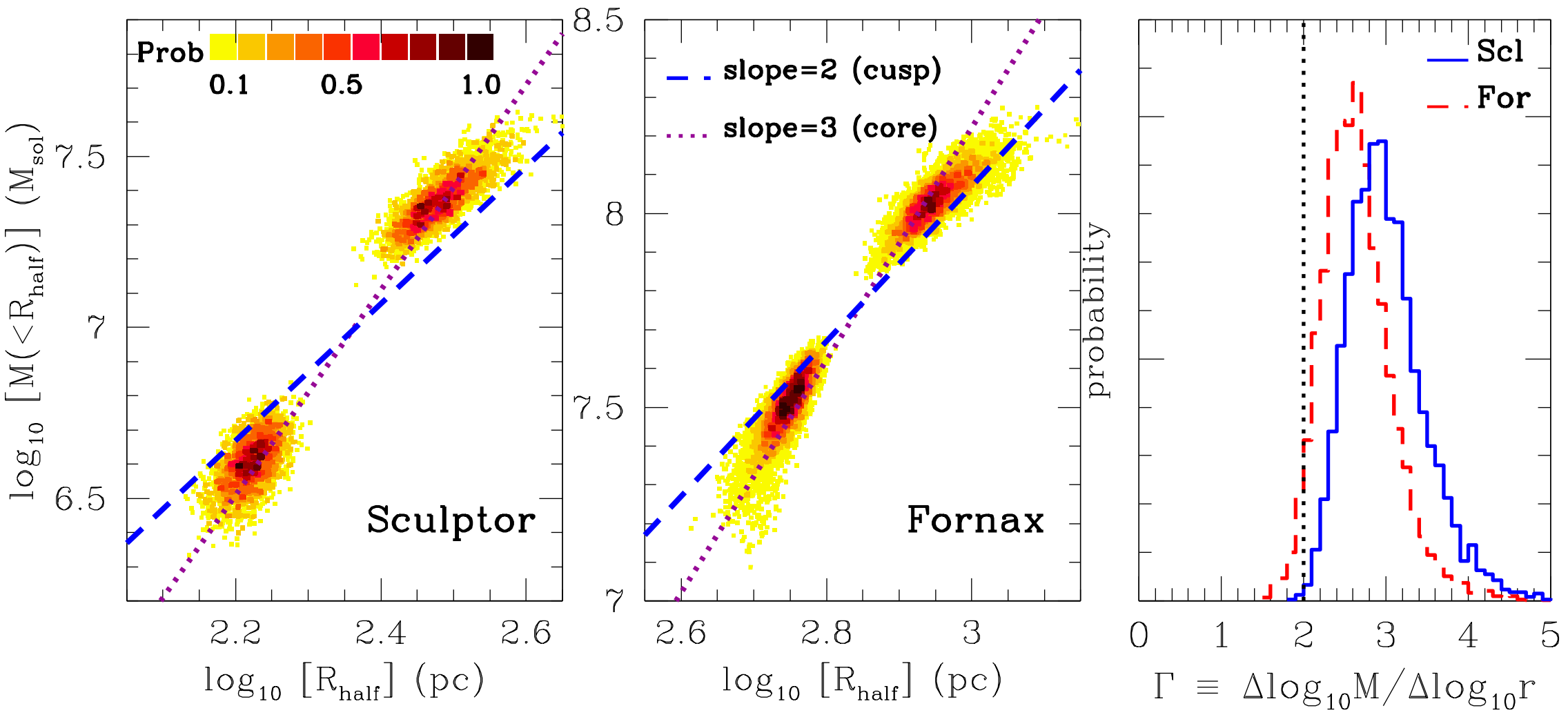

**Figure 84: Left, center: Spectroscopically-derived estimates of half-light radii and dynamical masses enclosed therein, for two chemo-dynamically independent stellar sub-populations in the Fornax $L_V \sim 10^7 L_{V,\odot}$) and Sculptor $L_V \sim 10^6 L_{V,\odot}$) dwarf spheroidals. Overplotted lines indicate central and therefore *maximum* slopes of mass profiles that have a core of constant central density (γ = 3, dotted) and the cusp (γ = 2, long-dashed) characteristic of halos formed in CDM-only simulations. Right: posterior probability distribution for the slope. Results are consistent with cores, which have γ ≤ 3 at all radii, but disfavor cusps γ ≤ 2 at all radii (*Figure from Walker & Penarrubia 2011*).**

Helping to movitave such alternative dark matter models are several apparent discrepancies between CDM-simulated and observed universes, particularly on the smallest galactic scales. Around galaxies with $M_{vir} \sim 10^{12} M_\odot$, cosmological N-body simulations that consider only gravitational interactions among CDM particles generally form ~10 times more subhalos with $M_{vir} \sim 10^8 M_\odot$ than have been detected as luminous dwarf-galactic satellites of either the Milky Way or M31 (Klypin et al. 1999, Moore et al. 1999). Moreover, the most obvious solutions to this “missing satellites” problem – invoking galaxy formation criteria that suppress star formation in all but the most massive subhalos – are limited by the inability of the most massive simulated subhalos to serve as plausible hosts of the most luminous Galactic satellites. The latter have dynamical masses (estimated within their half-light radii) that are systematically smaller than the masses (evaluated at the same radii) of the former (Boylan-Kolchin et al. 2011, 2012). This "too big to fail" problem is likely the symptom of a further discrepancy between the shapes of simulated and observationally-inferred mass-density profiles, ρ(r). CDM halos formed in N-body simulations have central "cusps" characterized by $lim_{\rho\rightarrow\infty}\rho(r) \propto r^{-1}$, whereas rotation curves and stellar kinematics of dwarf galaxies often favor "cores" of uniform density (left-hand panel of Figure 83 and Figure 84; Moore et al. 1994, Flores et al. 1994, Kuzio et al. 2008, de Blok2010, Walker & Penarrubia 2011).

Do these discrepancies falsify the CDM hypothesis, requiring a more-complicated dark matter model? Not necessarily. The relatively recent incorporation of hydrodynamics – along with prescriptions for star formation and energetic feedback therefrom – has demonstrated that "baryon physics" can affect the internal structure and even number of subhalos in a universe otherwise dominated by CDM. Given sufficiently vigorous and bursty star formation, strong winds from massive stars and supernovae drive outflows that remove ISM rapidly compared with local dynamical timescales (Navarro et al. 1996b, Read et al. 2005, Governato et al. 2010, Pontzen et al. 2011, 2014). As a result, the orbits of dark matter particles expand non-adiabatically, potentially transforming central cusps into cores and lowering masses within the half-light radii of dwarf galaxies (Figure 83). This transformation also lowers the binding energies of subhalos, leaving them more vulnerable to tidal disruption. Several N-body+hydro simulations indicate that this scenario might reconcile standard CDM with observations, thereby shifting focus from the nature of dark matter to galaxy formation Brooks et al. 2013, Wetzel et al. 2016).

However, N-body+hydro simulations agree with back-of-the-envelope calculations in predicting that baryon-physical processes become inefficient in the least luminous, most dark-matter-dominated galaxies, such that dwarf galaxies with $L < 10^6 L_\odot$ should retain their primordial CDM cusps (Penarrubia et al. 2012, Garrison-Kimmel et al. 2013; Figure 83). One implication is that questions about the nature of dark matter can finally be separated from considerations of galaxy formation, provided observations can reveal the density profiles of galaxies with $L < 10^6 L_\odot$. The current census of Local Group dwarf galaxies includes ~40 such systems, ~10 of which host more than $10^3$ stars brighter than V ~ 23 (see Chapter 2of this document, also Figure 2. For these galaxies, an MSE survey will deliver stellar-kinematic samples as large as those currently used to distinguish dark matter cores from cusps in dwarf galaxies with $L < 10^6 L_\odot$ (Walker & Penarrubia 2011). Thus MSE will have unprecedented power to constrain the nature of dark matter via observations of dwarf-galactic structure.

#### 4.4.1.2.2 Particle Physics

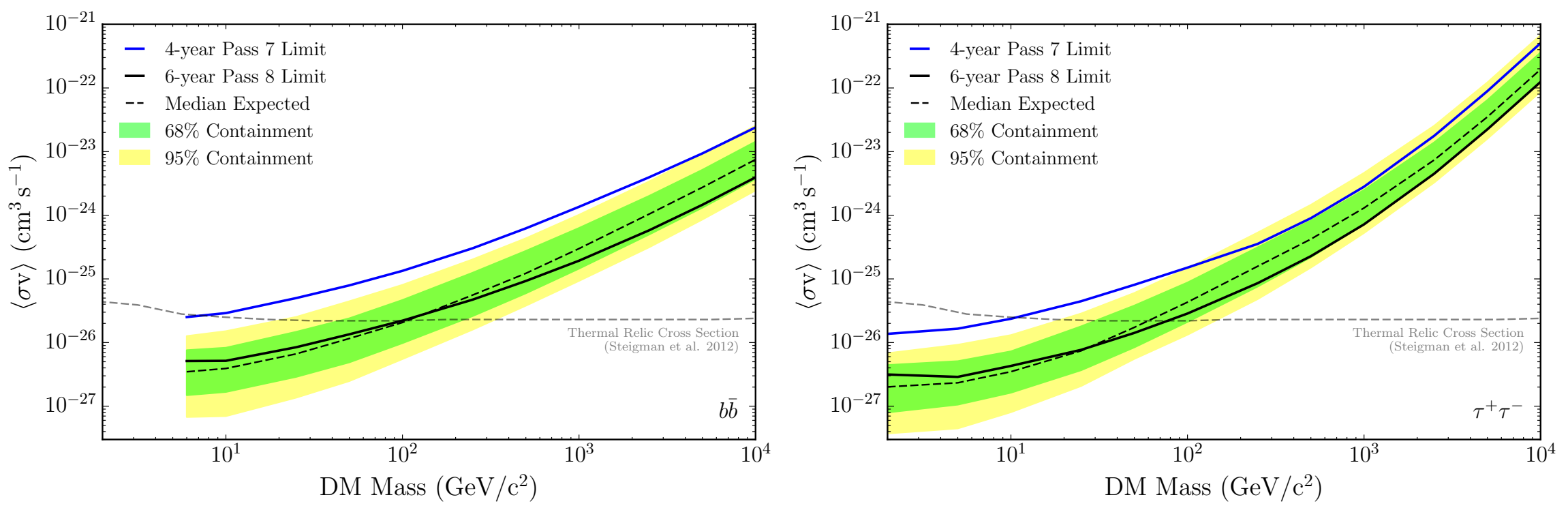

**Figure 85: Upper limits on the cross section for dark matter annihilation into final states of quarks (left) and leptons (right), derived from analysis of gamma-ray data and stellar spectroscopy of 15 Galactic satellites. Dashed lines show the median expected sensitivity, while bands indicate 68% and 95% quantiles. Dashed gray lines indicate the cross section expected for a thermally-produced weakly interacting massive particle (Steigman et al. 2012). (*Figure from Ackerman et al. 2015; see also Geringer-Sameth et al. 2015*).**

Thus far, all empirical evidence for dark matter's existence is derived via interpretation of astronomical observations within the framework of general relativity and/or its Newtonian limit. Confirmation of dark matter's particle nature will require detections of non-gravitational interactions with known (standard model) particles. Possibilities include production at the Large Hadron Collider, “direct” detection of nuclear recoils during rare scattering events in underground mines, and/or “indirect” detection of high-energy photons released as pairs of dark matter particles self-annihilate or individual particles decay. Due to their large dark matter densities and their lack of the astrophysical backgrounds (e.g., supermassive black holes, scattering of cosmic rays off ISM, etc.) that plague searches near the Galactic center, dwarf galaxies represent the cleanest available targets in searches for annihilation and decay signals (Gunn et al. 1978, Lake et al. 1990).

The flux of photons received from annihilation and decay events is proportional to the product of a particle physics factor (annihilation cross-section and decay rate, respectively) and the integral of the dark matter density profile raised to the power n, where n is the number of particles participating in the interaction (n=2 for annihilation and n=1 for decay). Thus, given a measurement of photon flux, or even a non-detection, one can use stellar-kinematic estimates of the dark matter density profile to infer the relevant particle physics properties. For example, Figure 85 shows the most recent constraints on dark matter's annihilation cross section as a function of particle mass (Ackerman et al. 2015; see also Geringer-Sameth et al. 2015). These upper limits are derived by combining non-detections of gamma-rays from the Fermi-LAT with density profiles estimated from stellar-kinematics of fifteen of the Milky Way's dwarf-galactic satellites. For particle masses $M_\chi < 100$ GeV, these limits are beginning to rule out the cross section that would naturally give the cosmologically-required $\Omega_{DM} \sim 0.2$ in the case of thermally-produced WIMPs (Steigman et al. 2012). Stellar-kinematic data have also been used to evaluate the significance of reported decay signals in X-ray observations of individual dwarf spheroidals (Loewenstein et al. 2010, Boyarsky et al. 2010, Jeltema et al. 2015, Ruchayskiy et al. 2015).

Regardless of whether an unambiguous photon signal is ultimately detected, clearly the resulting inferences about particle properties are only as good as estimates of dark matter densities derived from stellar kinematics. This dependence highlights the impact MSE will have on efforts to determine dark matter's particle nature. Among the known dwarf galaxies, the most attractive targets for annihilation/decay searches are also the least luminous ($L_V < 10^3 L_{V,\odot}$), primarily because these happen also to be the nearest. However, while the dynamical masses of these "ultrafaint" systems are also consistent with extremely large dark matter densities (Martin et al. 2007, Simon et al. 2007, 2011), such results are particularly uncertain due to the paltry samples (N << 100) and poorly-constrained systematics like inflation of stellar velocity dispersions by binary stars (McConnachie & Cote 2010). Improving upon existing samples by more than an order of magnitude (see Figure 35 from Section 2.4.1 of this document), MSE will have a major impact in this area, which represents one of the best opportunities to resolve dark matter's particle nature.

#### 4.4.1.3 Unveiling galaxy clusters to z = 1

The mass function of galaxy clusters and its redshift evolution offers a critical test of Dark Energy and/or alternative gravity models because it combines a measurement of the cosmological volume element with the observed growth rate of structure. Standard linear gravitational theory within an expanding universe predicts that overdensities should grow according to

$$\ddot{\delta} + 2H\dot{\delta} - \frac{3}{2}\Omega_M H^2 \delta = 0, \quad \delta = 0.$$

Importantly though, at z < 1, the universe makes a critical transition from a matter-dominated to a Lambda-dominated state. Although gravity drives the development of overdensities, the growth of massive halos is inhibited by the fact that H evolves to a constant value in a Lambda-dominated universe (an effect known as "Hubble drag"). The opposing effects of matter and dark energy forms the basis of cosmological tests based upon galaxy clusters.

A galaxy cluster consists of a dark matter halo (5/6 of the total mass), an intra-cluster medium (ICM) consisting of an atmosphere of hot, diffuse baryons (approximately 1/6 of the total mass) and a population of member galaxies (approximately 1/10th of the ICM mass). Figure 86 shows a visualization of 6 massive clusters where the galaxies, ICM and "dark mass" are all shown. Galaxy clusters may be detected using observational techniques sensitive to each main cluster component: weak-lensing is sensitive to the total cluster mass; the Sunyaev-Zeld'ovich (SZ) decrement and X-ray imaging are sensitive to the cluster ICM; galaxy overdensity studies are sensitive to the member galaxy population.

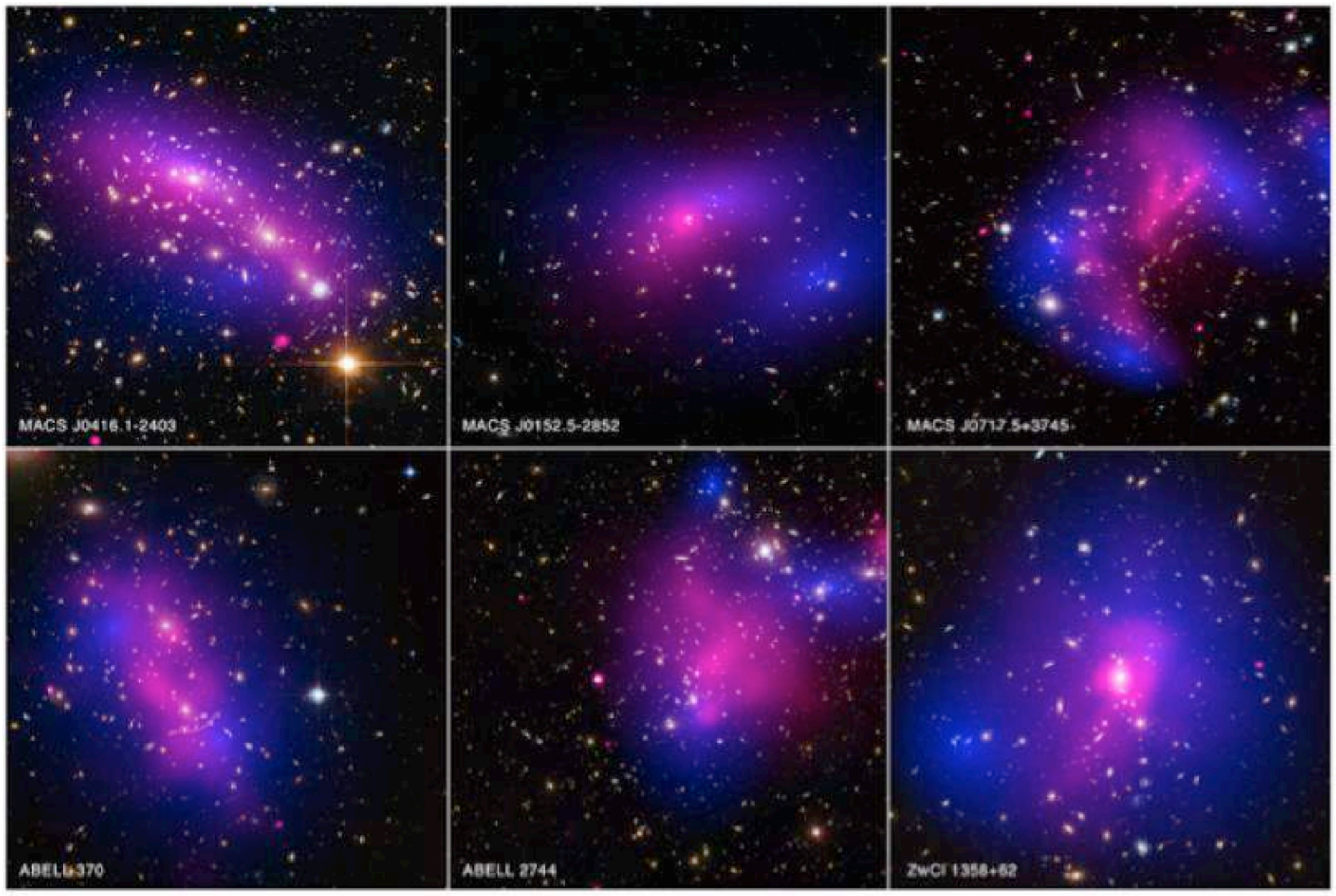

**Figure 86: Composite images of 6 galaxy clusters with X ray images from Chandra (pink) and mass distributions from lensing studies (blue) overlaid on optical HST images.** ***Figure from http://chandra.harvard.edu/photo/2015/dark/. Image credit: X-ray: NASA/CXC/Ecole Polytechnique Federale de Lausanne, Switzerland/D.Harvey & NASA/CXC/Durham Univ/R.Massey; Optical & Lensing Map: NASA, ESA, D. Harvey (Ecole Polytechnique Federale de Lausanne, Switzerland) and R. Massey (Durham University, UK).***

MSE is poised to be an essential tool for cosmological studies based on galaxy clusters through high completeness redshift surveys that will provide the purest sample of clusters with detailed information on their dynamical state for comparison to SZ, X-ray, and galaxy overdensity surveys. Indeed, observations of the cluster abundance to z = 1 and beyond are the current focus of several high-profile projects utilizing these techniques:

- Projects which detect galaxy clusters to z=1 and beyond via their SZ decrement towards the CMB have achieved maturity, detecting samples of several hundred clusters drawn from observations covering up to 2500 square degrees (Bleem et al. 2014). Current upgrades to existing SZ facilities, e.g., SPT-3G and ACT-3G, will permit cluster surveys to be undertaken at greater sensitivity and with faster mapping speeds in the time frame 2016-2020;

- Moderately deep X-ray imaging surveys for galaxy clusters to z = 1 currently cover up to 50 square degrees of the sky and generate samples of several hundred galaxy clusters (e.g., the XXL cluster survey; Pierre et al. 2016, Pacaud et al. 2016). Such studies will be transformed in September 2017 with the launch of the Russo-German space mission Spektr-RG carrying the eROSITA X-ray telescope. This facility will image 27 000 square degrees of the sky and will detect of order 100 000 galaxy clusters to z = 1.

- Blind weak-lensing surveys to detect galaxy clusters are currently in their infancy.

However, weak-lensing studies of individual galaxy clusters currently provide the most accurate method of determining mean masses of samples of galaxy clusters (Applegate et al. 2014). The application of weak lensing to detect cosmic structures will be transformed by the launch of the Euclid space mission. Euclid aims to measure the evolution of the mass correlation function via so-called "cosmic shear" (i.e. weak lensing shear induced by mass distributions typically on scales larger than that of individual clusters). As previously discussed, this mission will image 20 000 square degrees in a single, broad optical-NIR filter with a PSF of ~0.16 arcseconds. In addition, the NISP instrument will generate images in three IR bands at lower PSF to determine galaxy photometric redshifts and perform low resolution slitless spectroscopy at 1 – 2 microns. The Euclid photometric catalogue will also be used to identify galaxy clusters based upon galaxy overdensities in the pseudo-3D space of sky position and galaxy photometric redshift (e.g., Sartoris et al. 2016).

The landscape of galaxy cluster studies over the next two decades will therefore be dominated by large samples of clusters identified using multiple, complementary techniques from observations executed over many thousands of square degrees. The aim of such studies is to obtain precise and accurate knowledge of our cosmological model, particularly the dark energy equation of state, in addition to compiling a detailed picture of how the mass and physical history of galaxy clusters in turn affects the evolutionary history of their member galaxies.

There are two critical issues that affect the extent to which such galaxy cluster samples can be used to test cosmological models:

1) The relationship between the observable used to identify each cluster and its true mass (in this case measured with respect to a common overdensity scale, e.g., $M_{500,c}$).

2) The relationship between a sample of clusters identified using a given observational method (e.g., SZ, X-ray, galaxy overdensity) and the true population of clusters existing in the universe.

Cluster mass — the key parameter whose evolution is predicted by theory and N-body simulations — is a problematic parameter because it cannot be observed directly. Traditionally, cluster masses for large samples of clusters could only be inferred statistically, via various "observable"–mass scaling relations. Popular observables include X-ray temperature/luminosity, SZ decrement, and cluster richness, $N_{gal}$. However, there is currently no mass proxy that is simultaneously accurate (unbiased) and precise. Cluster cosmology benefits hugely from knowing the average cluster mass accurately (for the amplitude of the mass function) and the relative masses of clusters precisely (to get the shape of the mass function). For the absolute calibration, weak lensing currently achieves the highest accuracy. To get relative masses, X-ray gas mass and/or temperature are more useful because they have a smaller intrinsic scatter than WL.

So what can MSE bring to this field of study? Put simply, redshifts. Spectroscopic redshifts are accurate to 0.001 or better. They can be used to fix individual galaxies in space and thus identify clusters, they can be used to determine cluster velocity distributions, and they can be employed to determine the recent star formation history of member galaxies viable the observation of narrow emission and absorption features generated within the photospheres of stars or in the interstellar gas.

Identifying galaxy clusters from large spectroscopic plus photometric data sets generates catalogues which display the arguably the highest purity and lowest contamination of any cluster identification method. The reason for this is that the background of non-cluster galaxies along the line of sight, while large in projected or 2D studies, becomes very low when spread along the redshift axis.

Identifying galaxy clusters with a wide-field extragalactic MSE survey (such as that described in SRO-08) overlapping existing SZ, X-ray or Euclid sky areas offers the possibility of generating a cluster catalogue that can act as a Rosetta Stone for this field, in that it allows the relative bias of SZ, X-ray and/or overdensity cluster identification methods to be understood. For example, do X-ray selected clusters represent a dynamically-relaxed subset of the cluster population above a given mass threshold? Are SZ detected clusters biased toward merging systems? Do weak-lensing identified clusters display an orientation bias which boosts the projected mass of triaxial systems? Each of these physical questions determines how different cluster identification methods view the true cluster population (virialized halos exceeding a given mass threshold; Angulo et al. 2012).

In addition to identifying galaxy groups and clusters to z = 1 and heyond, MSE will also determine the velocity structure of their member galaxies. The observation of thousands of galaxy redshifts in the most massive, nearby clusters offers an opportunity to measure cluster velocity dispersions to a few percent, translating into a $\sim$10% accuracy in mass. This would provide a highly competitive mass calibration among the various cluster mass proxies. The statistical accuracy could be further improved by "stacking" clusters of similar values of observables (e.g., richness, X-ray temperature).

Even in the non-ideal cases where MSE measures only a few 10s to 100s of galaxy velocities per cluster, valuable information on the velocity distribution will still be extracted and will provide a valuable diagnostic of the relaxation state in each cluster. A simple physical definition of the cluster relaxation state is that it describes the time elapsed since the last cluster-scale major merger. Mergers deposit kinetic energy from each progenitor into the velocity distribution of the merged cluster dark matter halo, the ICM and the member galaxies. Each component dissipates this energy via different mechanisms (dynamical friction, gas cooling) and on different timescales. Therefore, using MSE spectroscopic redshifts to obtain quantitative measures of cluster relaxation (for example, using departures from a Gaussian velocity distribution as well as more sophisticated analyses) offers an important diagnostic of the physical source of scatter in cluster scaling relations.

MSE will add two important new dimensions to our understanding of how galaxy clusters trace the underlying mass field: 1) it will provide probably the best (highest purity and lowest contamination) cluster catalogue from which "wavelength bias" engendered by SZ, X-ray or colour-overdensity searches can be understood, 2) MSE will add a quantitative measure of cluster relaxation state to aid our understanding of the source of physical scatter in cluster mass observables. Clearly, this field is an excellent example of the synergies of MSE with next generation multi-wavelength facilities.

### 4.4.2 Peculiar velocities and the growth rate of structures

**Science Reference Observation 12 (Appendix L)**

**Dynamics of the dark and luminous cosmic web during the last three billion years**

*We propose a galaxy redshift and peculiar velocity survey covering up to 24,000 square degrees, i.e. ¾ of the full sky apart from Milky Way. The goal is to measure both the cosmological redshift of about 2 million galaxies and for a subset to obtain their radial peculiar velocities. The survey will measure both early- and late-type galaxies up to redshift z ~ 0.25 using the Fundamental Plane and optical Tully-Fisher techniques. The unprecedented depth of 1 Gpc is significantly deeper than surveys that will be completed in 2025, such as the two Australian southern surveys: the optical multi-fiber TAIPAN and the radio SKA Pathfinder WALLABY survey. This field of research is currently undergoing a revival, and this survey will achieve the same galaxy number density as in the contemporary pioneering proof-of-concept surveys as Cosmicflows-2 and 6dFGSv catalogs, while multiplying the covered universe volume by a factor 150. Only these scales enable strong test of gravitation models.*

*Three major science goals define this program: (i) linear-growth rate of cosmic structures at low redshift, obtained through velocity-velocity comparison and the luminosity fluctuation method, which are free of cosmic variance. This allows one to discriminate between modified theories of gravity at 1% level and yield a significant complement to the high-redshift constraints provided by the coeval spectroscopic surveys achieved e.g., by Subaru/PFS, Mayall/DESI, Euclid, and WFIRST; (ii) galaxy formation will be investigated in regard to unprecedented velocity-cosmic-web's 6D phase-space; (iii) dynamical tests probing the scale of homogeneity, and test for the averaging problem in General relativity and the consequent backreaction conjecture, which potentially provides terms dynamically equivalent to both dark matter and dark energy.*

The peculiar velocities of galaxies are induced by the gravitational pull of the mass content of the nearby universe, and are a direct indicator of the distribution of dark and luminous matter. On large scales, these peculiar motions become coherent bulk flows toward over-dense regions and away from under-dense regions. The velocity bulk flows depend on the amplitude of the matter fluctuations and imprint distinct anisotropic features in the distribution of galaxies in redshift space, generally referred to as redshift space distortions (RSD; Peebles 1980; Kaiser 1987). As depending on the amplitude of the matter fluctuations, peculiar velocities allow us to measure the growth-rate of structures, and provide one of the most powerful probes to distinguish between dark energy and modified gravity models (ESA-ESO WG on Fundamental Cosmology, 2003; Astro2010 Panel Reports "New Worlds, New Horizons in Astronomy and Astrophysics", 2011). Whether the observed large-scale bulk flows are consistent or not with the ΛCDM model is still under debate. Hereafter, we use the phrase "RSD" to refer to peculiar velocity analyses where no ancillary distance information is available, and "peculiar velocities" to refer to analyses where both redshift and distances are known independently.

RSDs are receiving significant attention as a potential cosmological probe. As a tracer of the growth rate of structure, RSDs are in principle capable of distinguishing dark energy from modified gravity models (e.g., Guzzo et al., 2008, Percival & White 2009). Observationally, several projects carried out during the last decade have used RSDs to constrain cosmological parameters: the 2dFGRS (Hawkins et al. 2003), the SDSS LRG (Cabré & Gaztanaga 2009, Samushia et al. 2012), 2SLAQ (Ross et al. 2007) and the VVDS (Guzzo et al. 2008). Most recently measurements have been obtained by WiggleZ (Blake et al. 2012, Contreras et al. 2013) and VIPERS (de la Torre et al. 2013). All major future cosmological spectroscopic surveys now highlight RSDs as one of their probes.

The standard measurement of RSD (by the quadrupole anisotropy in the redshift-space two-point correlation functions of galaxies) is based on galaxy redshift surveys that probe the matter density field traced by a (single) biased population of sources and then it measures the ratio $\beta = f/b$ between the linear growth rate f and the luminosity bias of the tracer, b. Neglecting systematic errors, the sample-variance limit for standard RSD analyses can be approximated by $\sigma \ln \beta \lesssim A b^{-0.7} V^{-0.5} \exp(B/b^2 n)$, in which n is the expected mean number density of galaxies within the survey volume V, and A and B numerical constants (Bianchi et al. 2012). All future spectroscopic surveys will provide tighter cosmological constraints than previously achieved using RSD by measuring a high density of sources at high redshift ($0.8 < z < 2$) and over a much wider field-of-view than current RSD motivated survey. A MSE redshift survey could supersede the Subaru/PFS results (1500 $deg^2$, 4M galaxies) and attain similar precision to Mayall/DESI (14 000 $deg^2$, 25M galaxies) and Euclid (15 000 $deg^2$, 30M galaxies) if collecting more than 10M high-redshift galaxies with luminosity bias $b \geq 1.3$ over 10000 $deg^2$ (see Figure 87, in which two configurations are shown: in the redshift range $0.5 \leq z \leq 2$, and $1 \leq z \leq 2$, both with $n = 0.0003 h^3 Mpc^{-3}$, leading respectively to 15M and 12M spectra). This analysis suggests that RSDs with MSE are a promising area in which to develop key programs, although it is fundamentally necessary for MSE to provide new constraints that other surveys cannot do; clearly Mayall/DESI and Euclid are powerful competitors with RSD.

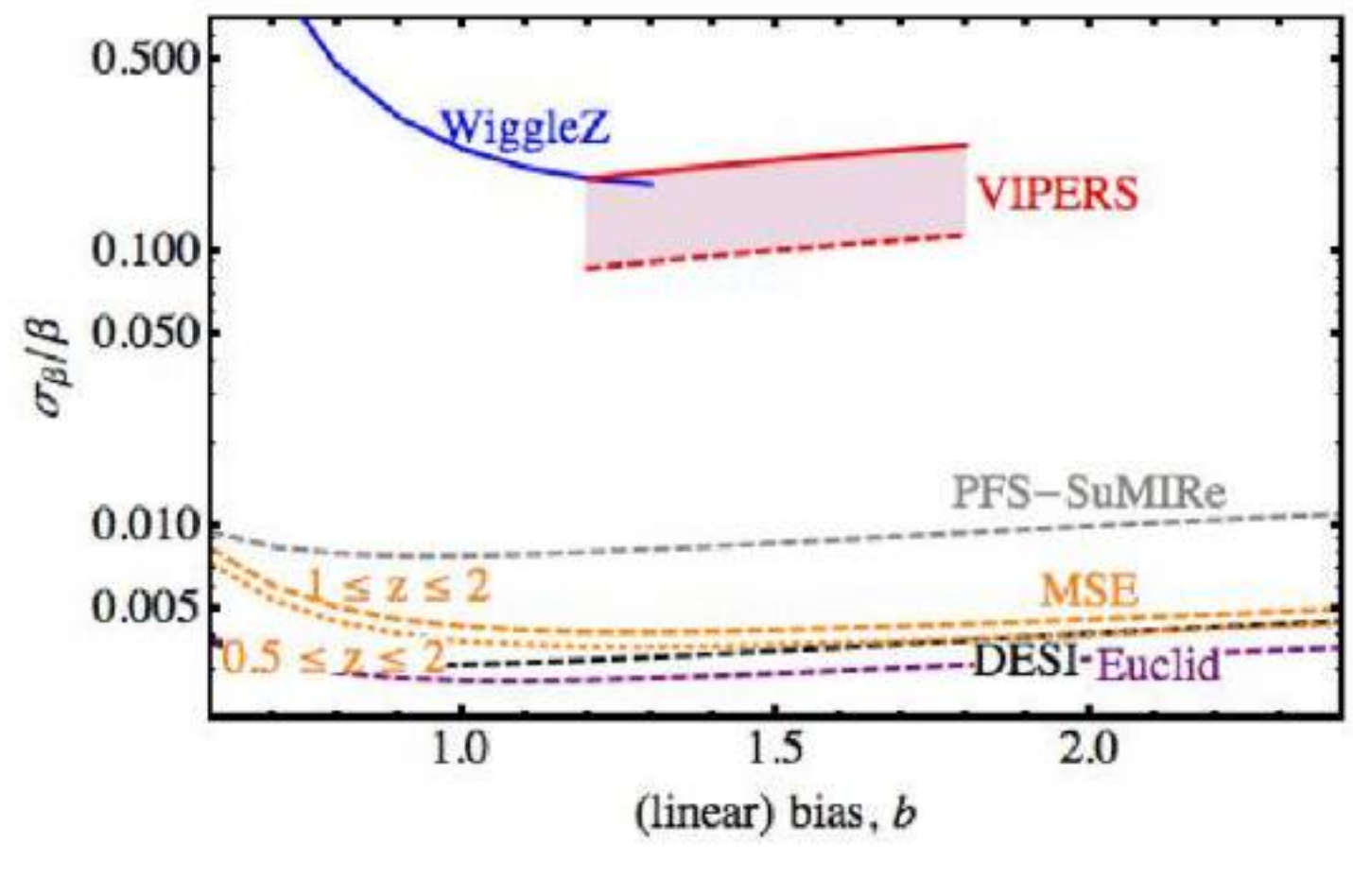

**Figure 87: Relative error on β as a function of the galaxy bias for current and planned (or final stage) surveys (solid and dashed lines, respectively) and for two MSE configurations described in the text.**

However, for all RSD analyses based solely on redshifts, the accurate knowledge of the bias is a central requirement. Moreover, the standard RSD technique is further limited by cosmic variance, especially relevant when dealing with small (local) volumes; multi-tracer techniques (McDonald & Seljak 2009) or higher-order statistics (Marín et al. 2013) are generally used to overcome these problems. Instead, owing to the tight relation between the peculiar motions of galaxies and the underlying mass density fluctuations that depends upon the background cosmology and the growth rate of fluctuations, if distance information is also available independently from redshift, then one can use the full peculiar velocity information to probe structure formation using the so-called "velocity-velocity comparison". Although these kind of measurements are limited to low redshift, this opens up tremendous scientific opportunities, definitely competitive with high-redshift measurements aimed at probing the departures from General Relativity. As described in SRO-12, this program will require a redshift survey of ∼2 million galaxies for which the distances will be independently estimated, using the Tully-Fisher and Fundamental Plane techniques. It is important to stress that while some progress with this type of observation can be made using single fibers, SRO-12 is best suited to the MSE IFU mode, an envisioned early-light capability.

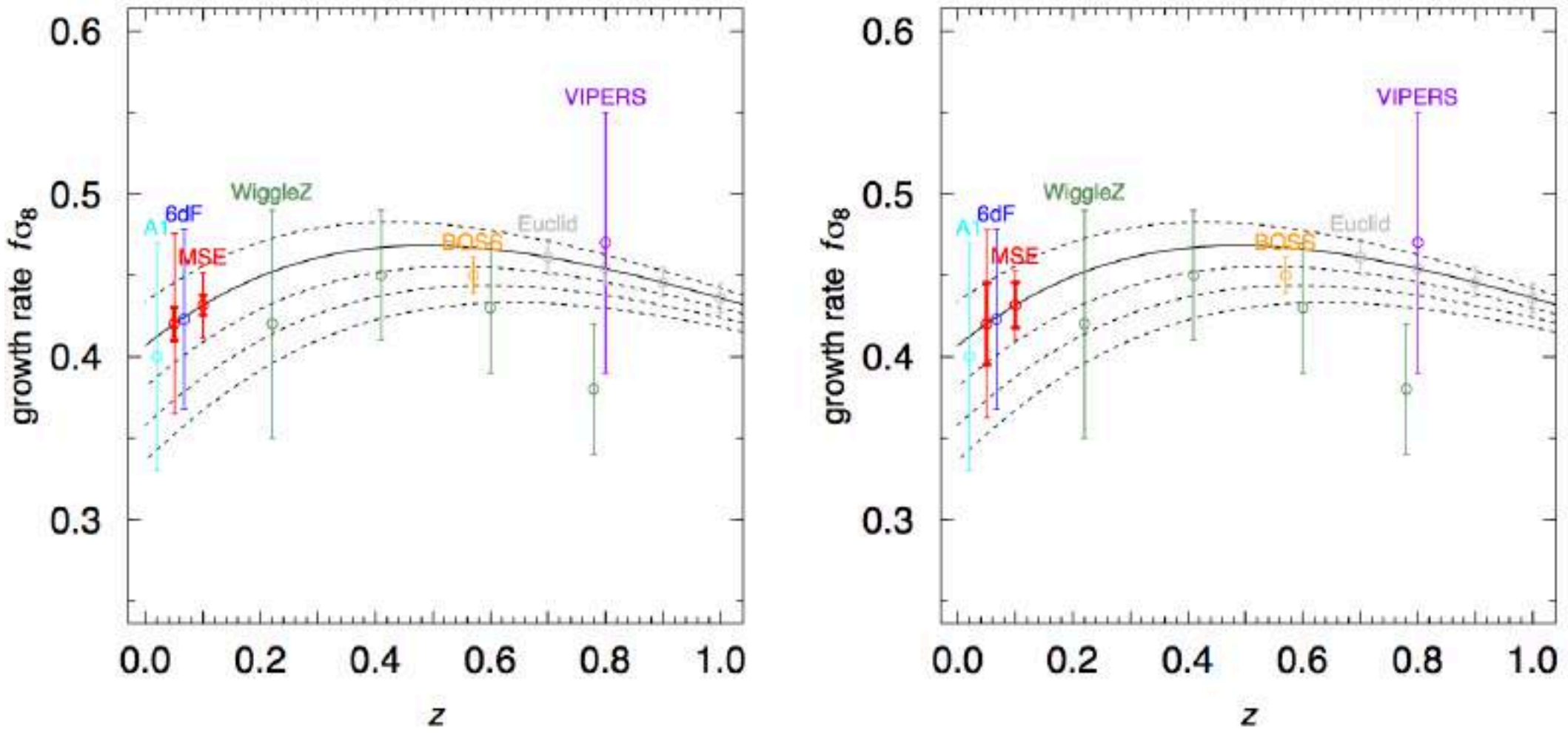

**Figure 88: Linear growth rate at low redshift by velocity-velocity comparison. The predicted MSE Fisher matrix constraints for the survey described in SRO-12 are shown around the fiducial ΛCDM Planck cosmology (solid line) at redshift z < 0.05 and z < 0.1 for 10 000 deg$^2$ and galaxy number density n = 0.03h$^3$Mpc$^{-3}$ (left) and n = 0.003h$^3$Mpc$^{-3}$ (right). Thick and thin (red) error bars account for density-only (RSD) and joint density-velocity MSE constraints, obtained using the same number of galaxies. Dashed (solid) lines correspond to f $\sim\Omega^{\gamma}_{m}$ with γ = 0.5, (0.55), 0.6, 0.65, 0.7 downward.**

A huge advantage of the "velocity-velocity comparison" mentioned above is that it completely eliminates cosmic variance. Indeed, in the case of the hypothetical situation of perfect data, just a comparison of a few points is sufficient to determine the relevant parameters without any assumption about the galaxy bias. This comparison is at the heart of peculiar velocity studies (e.g., Davis et al. 2011), and provides an assessment of gravity and dark energy theories; see Figure 88. As shown by Carrick et al. (2015), the measurement of the normalized linear growth-rate $f\sigma_8$, based on ~ 5000 peculiar velocities with a typical depth of z ~ 0.02, has a precision of 5%; with a much deeper and more comprehensive peculiar velocity and redshift survey from MSE, one might expect to reduce this uncertainty by a factor of ∼ 5, yielding a precision of 1%. In addition to not being affected by cosmic variance, peculiar velocity surveys are also the ideal framework to investigate primordial non-Gaussianities and general-relativistic (gauge) effects specific of large scales (Jeong et al. 2012; Villa et al. 2014), and to test the scale-dependence of the growth rate of structure, a feature prevalent in modified theories of gravity and clustering dark-energy models (Parfrey et al. 2011; see also the review by Clifton et al. 2012).

Following Koda et al. (2014) for a Fisher matrix forecast (see also Johnson et al. 2014), the measurement of $f\sigma_8$ or β from angle-averaged auto- and cross-power spectra of galaxies' density and peculiar velocity can attain arbitrarily high precision. Unlike constraints from RSD analysis by redshift surveys, the relative error of these parameters monotonically decreases with the mean number density of sources, n. Figure 88 shows how a minimal MSE velocity survey covering 10 000 $deg^2$ provides terrific constraints on the growth-rate $f \approx \Omega_m^{\gamma}$, yielding one of the most stringent tests of General Relativity ($\gamma \approx 0.55$) at the level of $\delta\gamma \sim 0.05$. While similar precision will be attained by future spectroscopic redshift surveys at high-redshift such as Subaru/PFS (not shown) and Euclid (which will benefit from a much larger volume but probe the growth-rate in a more challenging regime), only MSE can provide such tight constraints at low redshift. Table 7 reports the Fisher-analysis constraints that can be achieved by SRO-12 with $n = 0.003h^3Mpc^{-3}$ and $n = 0.03h^3Mpc^{-3}$; relative errors can attain the 1% level with a deep and large survey. It illustrates the benefit of the velocity-velocity comparison made possible by the knowledge of peculiar velocities (the small, thick error bars in Figure 88) with respect to standard RSD predictions obtained based on the knowledge of the density only (larger, thin error bars).

**Table 7: Fisher matrix constraints on the rescaled linear growth-rate, β = f/b (where b is the linear bias parameter), and on the amplitude of matter fluctuations, $\sigma_8$, from density-only (RSD) and joint density-velocity (VEL) power spectra. Results for sky coverage of 10000 $deg^2$ (marginalization over the other parameters and using the WMAP-9 reference model). Values for the minimal and optimal setup for SRO-12 are boldfaced.**

| | | $n = 0.003h^3Mpc^{-3}$ | | $n = 0.03h^3Mpc^{-3}$ | |
|---|---|---|---|---|---|
| | | dβ/β | $d\sigma_8/\sigma_8$ | dβ/β | $d\sigma_8/\sigma_8$ |
| z < 0.1 | RSD | 0.151 | 0.137 | 0.145 | 0.132 |
| | VEL | 0.065 | 0.06 | 0.024 | 0.024 |
| z<0.20 | RSD | 0.054 | 0.049 | 0.051 | 0.047 |
| | VEL | **0.036** | **0.032** | 0.015 | 0.014 |
| z<0.25 | RSD | 0.038 | 0.035 | 0.037 | 0.033 |
| | VEL | 0.028 | 0.026 | **0.013** | **0.012** |
| z<0.30 | RSD | 0.029 | 0.026 | 0.028 | 0.025 |
| | VEL | 0.023 | 0.021 | 0.011 | 0.01 |

SRO-12 also allows the application of the *luminosity fluctuation* method. Spatial variations in the luminosity function of galaxies, with luminosities estimated from redshifts instead of distances, are correlated with the peculiar velocity field. Comparing these variations with the peculiar velocities inferred from the redshift surveys could also be a powerful test of our fundamental cosmological paradigm on cosmological scales. The method has been successfully applied to the SDSS Data Release 7 (Feix et al., 2015). An application to MSE with its larger sky coverage and larger number of galaxies should yield results competitive to RSD analysis.

### 4.4.3 Probing homogeneity, isotropy and the back-reaction conjecture

The cosmological standard model relies on the hypothesis that the Universe is homogeneous and isotropic on large scale, without giving any information about this scale. The theory of primordial inflation (which is incorporated into the ΛCDM model) actually predicts a certain level of density fluctuations on all scales, with a power spectrum of scalar fluctuations which is not scale invariant ($n_s \neq 1$) but colored as confirmed by the recent Planck measurements. As a consequence, the transition to the large-scale homogeneity and isotropy is expected to be gradual, attaining the level $\delta\Phi/c^2 \sim 10^{-5}$ at the epoch of last scattering, as inferred from the cosmic microwave background (CMB) radiation. At redshift $z \sim 1100$, the large-scale homogeneity is well supported by the high degree of isotropy of the CMB (Fixsen et al. 1996), under the hypothesis of the cosmological Copernican principle.

At low redshift such symmetries are less evident. As for the isotropy, using hard X-ray data from the HEAO-1 A-2 experiment, Scharf et al. (2000) demonstrated how the dipole anisotropy of the distant X-ray frame ($z \sim 1$) can constrain the amplitude of bulk motions of the universe, probing evidence for a mild anisotropy that indeed could be of Galactic origin. In the local universe, a recent analysis of peculiar velocity data (Hoffman et al. 2015) estimated the bulk flow of the local flow field and showed that it is dominated by the data (as opposed to the assumed prior model) out to $200h^{-1}$Mpc, They also showed that it is consistent in the $X - Y$ supergalactic plane with the CMB dipole velocity down to a very few $kms^{-1}$, as recovered by the Wiener filter velocity field reconstructed from the radial velocity data. With a volume 150 times larger than that currently probed, thus largely encompassing the homogeneity scale, MSE will be unique in being answering the question of whether the cosmological bulk flow exists on 800Mpc scales, as found by some cluster studies. The survey described by SRO-12 can obtain the required data.

Isotropy itself is not sufficient to deduce homogeneity, unless probing it as a function of flux or redshift. Based on simple counting of sources in three-dimensional regions in order to estimate the fractal (Minkowski-Bouligand) dimension and its departure from the Euclidean value $D_2 = 3$, Hogg. et al (2004) assessed the homogeneity scale for the LRG sample from the SDSS at $\sim 70h^{-1}$Mpc in the redshift range $0.2 < z < 0.4$. This value has been confirmed by WiggleZ data (Scrimgeour et al. 2012), proving the transition to the large-scale homogeneity around $\sim 70{-}80h^{-1}$Mpc at redshift $z = 0.2 - 0.8$, with an upper bound around $300h^{-1}$Mpc. This value is consistent with fractal analysis, anomalous diffusion and Shannon entropy methods based numerical-simulation-based studies (e.g., Yadav et al. 2010, Kraljic 2015, Pandey 2013). Very recent measurements based on the BOSS DR12 quasar sample (Laurent et al. 2016) set the homogeneity scale from 250 to $1200h^{-1}$ Mpc at redshift $z = 2.2 – 2.8$.

The most important implication of inhomogeneity is the long-standing "averaging problem" in General Relativity (Ellis 1984), accounting for the role of (small-scale) perturbations on the expansion rate of the spacetime once they enter the nonlinear regime. Whilst in Newtonian cosmology, at the linear order, their effect vanishes on average by construction, in full GR the density fluctuations potentially should have some back-reaction effect on averaged quantities. According to the Buchert formalism (Buchert 2000, Ellis & Buchert 2005), they drive new effective source terms in the Friedmann-like equations accounting for the expansion rate of every spatial domain, D, over which averaged quantities are calculated. A unique cosmological application of peculiar velocity surveys is the direct measurement of the so-called kinematical back-reaction term, $Q_D = \mathrm{Var}[\mathrm{div}\,\vec{v}] - \frac{2}{3}\langle\sigma^2\rangle$ $\mathrm{QD} = \mathrm{Var}[\mathrm{div}\vec{v}]\text{-}\frac{2}{3} < \sigma^2 >$ , the mean and variance being calculated over the domain D (here $\sigma^2$ is the amplitude of the shear velocity tensor). Depending on its sign, this term is dynamically equivalent to dark matter (on small/galactic scales) or to dark energy (on large/cosmological scales). This question is delicate and far from be settled. In the era of precision cosmology, where the immediate goal is validating or disproving the cosmological standard model and the theory of General Relativity, a quantitative measurement of kinematical back-reaction term is mandatory; a non-vanishing value would definitely indicate the necessity to revisit the standard FLRW model. A peculiar velocity survey such as that described by SRO-12 provides the key ingredient to probe this concept; as long as the velocity field in redshift space is irrotational, and so may be derived from a velocity potential (e.g., Nusser & Davis 1994), one can directly infer the kinematical back-reaction term by measuring the Minkowski functionals of the velocity fronts defined by the excursion-set of the velocity potential (Buchert & Carfora 2008). Because of the limited size of past and current velocity surveys, this program has not been addressed on real data yet. Instead, the velocity survey described in SRO-12, owing to its large volume, is expected to yield robust enough results to assess the back-reaction conjecture.

### 4.4.4 Cosmic web dynamics and galaxy formation

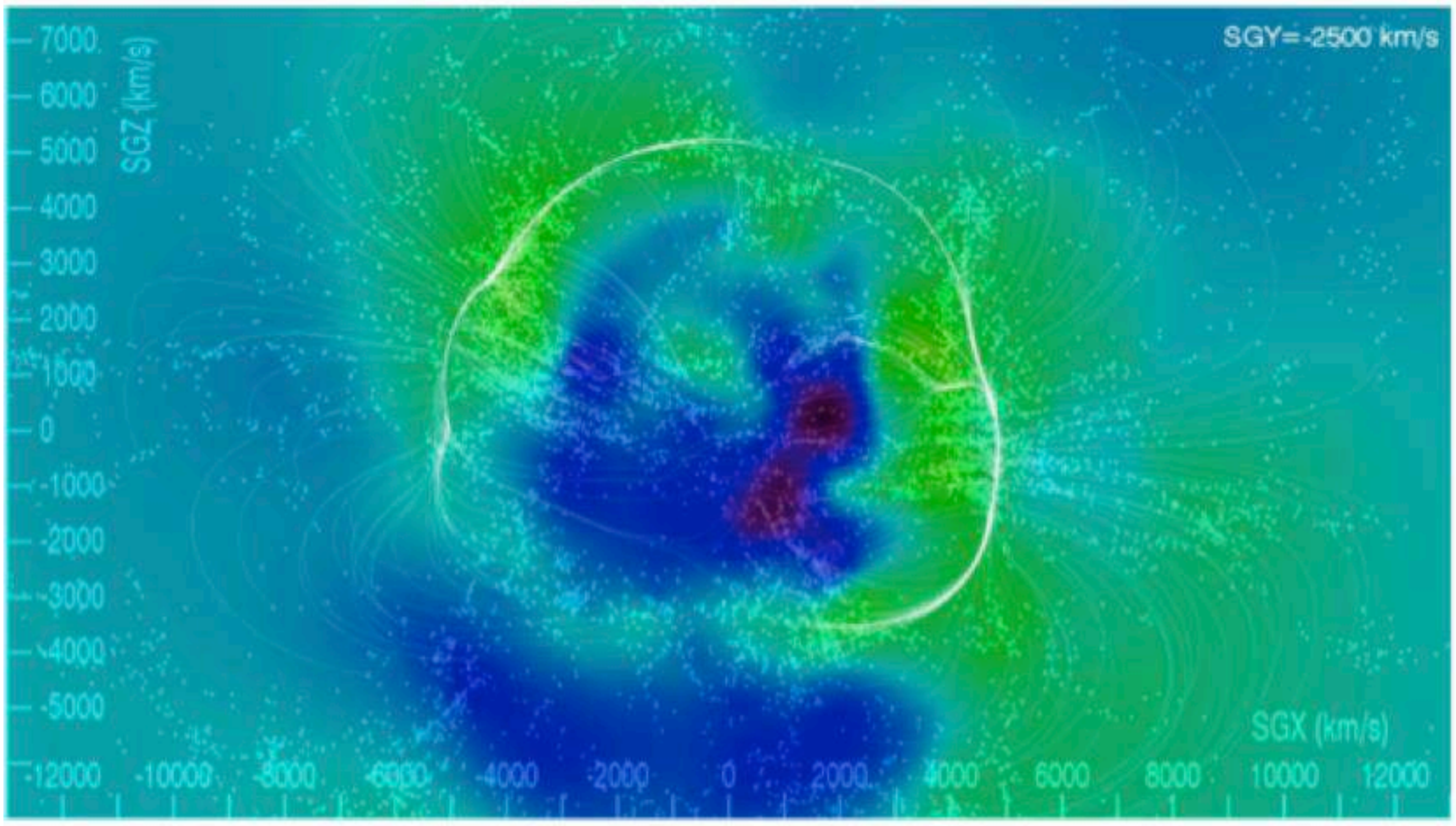

**Figure 89: The matter overdensity field δ computed from the divergence field in Cosmic-flows-2 dataset (Tully et al. 2013) is plotted in color. Galaxies identified by redshift surveys are plotted as white dots. The new technique of defining filaments using cosmic velocity fields gives a new insight to the morphology of the large scale structure. A filament of coherent motion of matter is half the "Arch" seen running from Perseus Pisces to Pavo-Indus structures in redshift surveys and in the cosmic V-web shear tensor. Only the analysis of watersheds allows one to identify the "splitting" zone of this structure. (Figure from Tully et al. 2014).**

It is worth reiterating that the driving science theme for MSE in relation to galaxy formation and evolution is in connecting galaxies to the cosmic web in which they are embedded (Chapter 3).

Peculiar velocity surveys – with non-redshift-based distance information – provide a means to probe the large scale structure's dynamics with MSE through the kinematics of the cosmic-web. SRO-12 details an analysis of extragalactic flows by MSE that will give unprecedented insights into the dark matter role on a large variety of scales, from planes of dwarf galaxies, to groups of galaxies, galaxy clusters, filaments, voids and superclusters (see Figure 89 for an example based on the Cosmicflows-2 dataset; Tully et al. 2014).

With an accurate measurement of peculiar velocities with MSE such as described in SRO-12, one could apply the method devised by Hoffman et al. (2012) to reconstruct the velocity-based cosmic web, or V-web, whose components (sheets, filaments, knots, and voids) are kinematically identified by the local value of the eigenvalues of the velocity shear tensor. Based on N-body simulations (Figure 90), this technique has been proven to resolve cosmic structures down to <0.1$h^{-1}$Mpc, providing much more details than a similar classification algorithm based on the gravitational tidal tensor obtained from the density field (dubbed the "T-web"; Hahn et al. 2007, Forero-Romero et al. 2009), which allows one to achieve only ~1$h^{-1}$Mpc resolution scale. The higher resolution can be explained by the slower evolution of the velocity field away from the linear-regime than the density field, which retains less memory of the initial conditions.

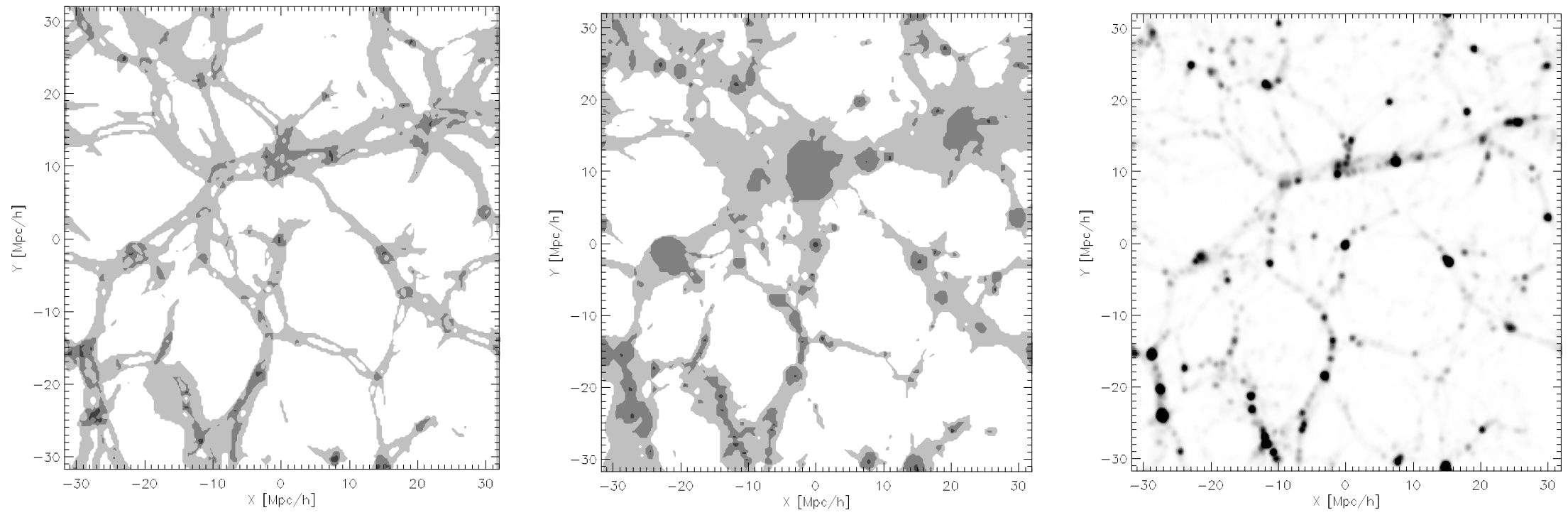

**Figure 90: Cosmic-web components (voids, filaments, sheets, and knots in white, light gray, dark gray, and black, respectively) obtained from the velocity shear tensor (left, the V-web) and gravitational tidal tensor (center, the T-web). The latter, reconstructed from the density field (right), allows ~10 times lower resolution scale. *(Figure adapted from Hoffman et al. 2012).***

Moreover, once the large scale structure (V-web, T-web, density field, …) has been accurately drawn, exploiting the high-resolution spectroscopic capabilities of MSE one can investigate the alignment of the galaxy spin to the cosmic-web filaments, eventually related to the tidal torque theory and its extensions (e.g., Codis et al. 2012; Codis, Pichon & Pogosyan 2015), the segregation of single galaxies and groups with respect to their environment and dynamics, and the implications for galaxy formation (e.g., Hahn et al. 2010). It is a natural extension of the fundamental theme of relating galaxies to the large scale structures described in Chapter 3.

### 4.4.5 Kinematic lensing

Gravitational lensing shears the velocity field of disk galaxies in a non-trivial way, yielding an angle between the maximum- and null-velocity axes that differs from the expected 90° without lensing. Likewise, in presence of lensing, the velocity and morphological position angles do not coincide anymore. Using 7-fibres bundles (IFUs) with total effective diameter 1.8 – 2.7” (or 19-fibres bundles for high-value objects) to map the velocity field of spiral galaxies at redshift z ∼ 0.3 – 1.0, one can directly point-wise estimate the gravitational shear, attaining values as low as typical of cosmological weak-lensing, or larger values as due to galaxy clusters typically located at $z \lesssim 0.5$. This technique, so far proposed in the radio domain and based on high-resolution HI velocity maps, can provide an independent measurement of gravitational lensing of galaxies. Its aim is an improvement in the mass modeling of the large-scale structure and clusters using a smaller number of galaxies than required by traditional (statistical) methods.

Gravitational lensing, one of the best probes to map the spatial distribution of dark matter and explore cosmology, is traditionally measured looking at the deformation and magnification of background images induced by the intervening matter. In weak and intermediate lensing regimes ($\kappa$, $\gamma \sim 0.01 – 0.1$), the required precision is achieved by statistical procedure, averaging the signal over a large number of background sources and looking at the correlation of their shapes. This method is the core of the major weak-lensing projects, e.g., CFHTLenS, RCSLenS, CS82, KiDS, DES, HSC, and will be extraordinarily improved by LSST and ultimately Euclid and WFIRST in the visible and NIR wavelength domain.

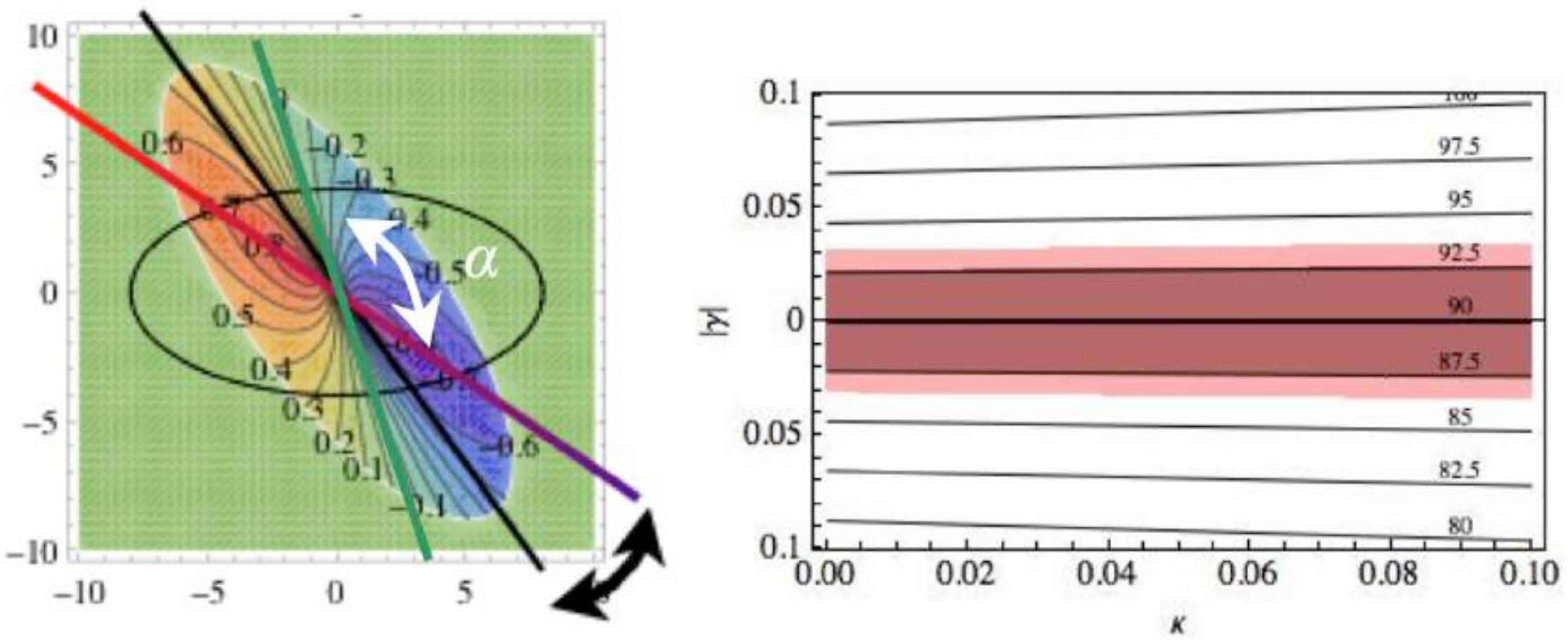

**Figure 91: Kinematic lensing. Left: The effect of gravitational lensing on the velocity field of the background galaxies is to modify from perpendicular the angle between the null velocity axis and maximum velocity axis. Right: Relationship between the lensing shear $\gamma$ and convergence $\kappa$ of the luminosity profile, and the angle $\alpha$ between the kinematic axes; a measurement of the latter with 5 – 7° accuracy (dark-light contours) allows for a point-wise estimate of $\gamma$ of ∼2 – 3%, almost regardless the value of the $\kappa$.**

Alternatively, one may consider the *effect of gravitational lensing on the velocity field* of the background sources, in particular late-type galaxies and fast rotators; gravitational lensing anisotropically stretches the major and minor axes of the luminosity profile of extended sources, and their velocity field. This is illustrated in Figure 91. Here, we consider the background source to be an ideal, late-type galaxy (i.e. a thin circular disk) of angular diameter ∼10’ with a non-trivial inclination with respect to the line-of-sight. It therefore appears as an ellipse in the source plane (black line in Figure 91, left panel) and has a fairly simple rotation curve (here approximating that of NGC 3145). Lensing causes the maximum-velocity axis (a.k.a. the “velocity position angle”, defined by the locations of the maximum and minimum of the velocity map) and the null-velocity axis (colored lines in Figure 91, left panel) to not be perpendicular anymore. Blain (2002) shown that the angle α between the two axes depends on the convergence $\kappa$ and absolute value of the shear $\gamma$ as

$$\alpha = 2\arctan\frac{1-\kappa+\gamma}{1-\kappa-\gamma} \approx \frac{\pi}{2} + 2\gamma - 2\kappa\gamma.$$

This approximation is valid up to the intermediate-lensing regime ($\kappa$, $\gamma \approx 0.1$), typical of weak-lensing by galaxy clusters. To overcome the difficulty in determining the null-velocity axis, one can instead measure the angle between the velocity and photometric position angles, which in the image plane differs from zero in presence of lensing (black arrow in Figure 91, left panel). According to this equation, for a typical accuracy of the velocity position angle, ~5°, and *provided the null-velocity axis is determined with the same accuracy,* one might expect to be able to measure absolute shear as small as $\gamma \sim 0.03 - 0.035$ (Figure 91, right panel), which is the order of magnitude expected for cosmological weak-lensing.

With a perfect velocity measurement based on the full map of the velocity field, such as provided by radio surveys (e.g., SKA), only two lensed galaxies are sufficient to exactly retrieve the intrinsic gravitational shear in weak-lensing limit (Morales 2006). A lower resolution velocity map as probed by high-resolution fibre-bundles (IFUs) in optical bands, eventually improved by a stacking technique of close galaxies, can still bring off the same goal in the *weak- and intermediate-lensing regime*. Gravitational lensing estimated by velocity field will be sensitive to totally different systematics with respect to standard measurements based only on luminosity profile of background sources.

It is worth noting that it is not just in the regime of weak lensing that MSE will have a strong science impact. In the strong lensing regime, MSE will exploit strong science synergies. In SDSS-I, the Sloan Lens ACS Survey (SLACS; Bolton et al. 2006, 2008a) discovered over 100 strong gravitational lens systems through the spectroscopic signature of two redshifts along the same line of sight. Follow-up imaging of spectroscopic lens candidates with the Hubble Space Telescope and high-resolution ground-based facilities has enabled a unique experimental approach to the study of the structure and dynamics of massive galaxies at relatively low redshift (e.g., Koopmans et al. 2006, Bolton et al. 2008b). The BOSS Emission-Line Lens Survey (BELLS: Brownstein et al. 2012, Bolton et al. 2012) is successfully extending this technique to significant cosmological look-back time at redshift z ~ 0.4 – 0.6, thereby probing directly the structural and dynamical *evolution* of LRGs.

A future spectroscopic survey with MSE can allow this technique to be extended to redshifts of z ~ 1 and beyond (for ELG lenses), yielding a direct measurement of massive galaxy structure over an unprecedented cosmic time baseline. ELG lenses would also provide accurate measurements of the total aperture masses of galaxies beyond z = 1, for which masses are currently available only through relatively uncertain stellar-population synthesis techniques. This would provide a powerful new discriminant between competing models for the assembly and evolution of galaxy mass.

## References

Abbas, U., & Sheth, R. K. 2006, MNRAS, 372, 1749
Abbas, U., et al. 2010, MNRAS, 406, 1306
Ackermann, M. & Fermi-LAT Collaboration. 2015, Physical Review Letters, 115, 231301
Adami, C.,et al. 2006, A&A, 459, 679
Adelberger, K. et al., 2003, ApJ, 584, 45
Adelberger, K. L., et al. 2005, ApJ, 629, 636
Akerman, C., et al., 2005, A&A, 440, 499
Alonso, M., et al. 2007, MNRAS, 375, 1017
Andrews, B., Martini, P., 2013, ApJ, 765 140
Angulo, R., et al. 2012, MNRAS, 426, 2046
Ann, H. B., Park, C., & Choi, Y.-Y. 2008, MNRAS, 389, 86
Aoki W. et al. 2013 AJ 145, 13
Aoki W., et al. 2014, Science, 345, 912
Aoki, W., et al. 2009,A&A, 502, 569
Applegate, D., et al. 2014, MNRAS, 439, 48
Aracil, B., et al., 2004, A&A, 419, 811
Asplund M., et al. 2006, ApJ, 644, 229
Astier, P., et al., 2014, A&A, 572, 80
Australian Academy of Science 2015, “Australia in the era of global astronomy”
Bacon, R., Brinchmann, J., Richard, J., et al. 2015, A&A, 575, A75
Bahe, Y. M., McCarthy, I. G., Crain, R. A., & Theuns, T. 2012, MNRAS, 424, 1179
Bahe, Y.M., et al., 2013, MNRAS, 430, 3017
Balcells, M., Benn, C. R., Carter, D., et al. 2010, in Proc. SPIE, Vol. 7735, Ground-based and Airborne Instrumentation for Astronomy III, 77357G
Baldry, I. & Glazebrook 2003 ApJ 593, 258
Baldry, I. K., Driver, S. P., Loveday, J., et al. 2012, MNRAS, 421, 621
Baldry, I. K., et al. 2006, MNRAS, 373, 469
Baldry, I. K., Glazebrook, K., & Driver, S. P. 2008, MNRAS, 388, 945
Balogh, M. L., Christlein, D., Zabludoff, A. I., & Zaritsky, D. 2001, ApJ, 557, 117
Balogh, M. L., et al. 2004, ApJL, 615, L101
Balogh, M. L., et al. 2011, MNRAS, 412, 2303
Balogh, M. L., Navarro, J. F., & Morris, S. L. 2000, ApJ, 540, 113
Balogh, M.L., et al. 2016, MNRAS, 456, 4364
Barklem, P., et al., 2005, A&A, 439, 129
Barnes, J. & Hibbard, J., 2009, AJ, 137, 3071
Barnes, J., 2016, MNRAS, 455, 1957
Barnes, J., Hernquist, L., 1991, ApJ, 370, 65
Bate, N. F., McMonigal, B., Lewis, G. F., Irwin, M. J., Gonzalez-Solares, E., Shanks, T., & Met- calfe, N. 2015, MNRAS, 453, 690
Battaglia, G., et al. 2006, A&A, 459, 423
Battaglia, G., et al. 2008, MNRAS, 383, 183
Battaglia, G., et al. 2011, MNRAS, 411, 1013
Bauer, A.E. et al. 2013, MNRAS 434, 209
Bechtol, K., & The DES Collaboration. 2015, ApJ, 807, 50
Becker G.D, Bolton J.S., 2013, 436, 1023
Becker G.D, Bolton J.S., Haehnelt M.G., & Sargent W.L.W. 2011, 410
Behroozi, Peter S., R.H. Wechsler, & C. Conroy, 2013, ApJ 770, 57
Bekki, K. 2009, MNRAS, 399, 2221
Bekki, K., & Couch, W. J. 2011, MNRAS, 415, 1783
Belli, S., et al., 2013, ApJ, 772, 141
Belokurov, V., et al., 2007, ApJ, 654, 897
Belokurov, V., Walker, M. G., Evans, N. W., Gilmore, G., Irwin, M. J., Mateo, M., Mayer, L., Olszewski, E., Bechtold, J., & Pickering, T. 2009, MNRAS, 397, 1748
Belokurov, V., Zucker, D. B., Evans, N. W., Gilmore, G., Vidrih, S., et al. 2006, ApJ, 642, L137
Bensby T., et al. 2010 A&A, 521, 57
Bensby T., et al. 2011 A&A, 533, 134
Benson, A. J., Bower, R. G., Frenk, C. S., et al. 2003, ApJ, 599, 38
Bentz, M. C., & Katz, S. 2015, PASP, 127, 67
Bentz, M. C., Denney, et al. 2013, ApJ, 767, 149
Berg, D., Skillman, E. & Marble, A. 2011 ApJ 738, 2
Bergeron P., et al. 2005 ApJ 625, 838
Berlind, A. A., & Weinberg, D. H. 2002, ApJ, 575, 587
Bernardi M. et al. 2003, AJ, 125, 32
Bianchi, D., et al., 2012, MNRAS, 427, 2420
Bigiel F., et al. 2010, AJ, 140, 1194
Birnboim, Y., & Dekel, A. 2003, MNRAS, 345, 349
Biviano, A., et al., 2013, A&A, 558, 1
Blackburne, J. A., Pooley, D., Rappaport,S., & Schechter, P. L. 2011, ApJ, 729, 34
Blain, A., 2002, ApJ, 570, 51
Blake, C., et al., 2012, MNRAS, 425, 405
Bland-Hawthorn, J. 2015, in IAU Symposium, Vol. 309, Galaxies in 3D across the Universe, ed. B. L. Ziegler, F. Combes, H. Dannerbauer, & M. Verdugo, 21–28
Blandford, R. D. & McKee, C. F. 1982, ApJ, 255, 419
Blandford, R., et al. 2010, New Worlds, New Horizons in Astronomy and Astrophysics
Bleem, L., et al. 2015, ApJS, 216, 27
Bluck, A., et al., 2014, MNRAS, 441, 599
Blumenthal, G. R., Faber, S. M., Primack, J. R., & Rees, M. J. 1984, Nature, 311, 517
Bode, P., Ostriker, J. P., & Turok, N. 2001, ApJ, 556, 93
Bolton J.S., Haehnelt M.G., Viel M., & Springel V. 2005, MNRAS, 357, 1178
Bolton J.S., Oh S.P., & Furlanetto S.R. 2009, MNRAS, 396, 2405
Bolton, A.S., Burles, S., Koopmans, L.E.V., Treu, T., & Moustakas, L.A. 2006, ApJ,. 638, 703

Bolton, A.S., Burles, S., Koopmans, L.E.V., Treu, T., Gavazzi, R. et al. 2008a, ApJ, 682, 964
Bolton, A.S., et al. 2012, AJ, 144, 144
Bolton, A.S., Treu, T., Koopmans, LV.E., Gavazzi, R., Moustakas, L.A. 2008b, ApJ, 684, 248
Bonaca, A., Geha, M., Kallivayalil, N., 2012, ApJ, 760, 12
Bonifacio, P., et al., 2009, A&A, 502, 1
Bonnivard, V., Combet, C., Daniel, M., Funk, S., Geringer-Sameth, A., Hinton, J. A., Maurin, D., Read, J. I., Sarkar, S., Walker, M. G., & Wilkinson, M. I. 2015, MNRAS, 453, 849
Borucki, W.J., et al., 2010, Science, 327, 977
Bothwell, M., et al. 2013, MNRAS, 433, 1425
Bouche, N., et al. 2004, MNRAS, 354, 25
Bouche, N., et al. 2013, Science, 341, 50
Bournaud, F., Jog, C. J., & Combes, F. 2005, A&A, 437, 69
Bouwens, R., J., et al., 2008, ApJ, 686, 230
Bovy, J., 2016, ApJ, 817, 49
Bower, R. G., et al. 2006, MNRAS, 370, 645
Boyarsky, A., Ruchayskiy, O., Iakubovskyi, D., Walker, M. G., Riemer-Sørensen, S., & Hansen, S. H. 2010, MNRAS, 407, 1188
Boylan-Kolchin, M., Bullock, J. S., & Kaplinghat, M. 2011, MNRAS, 415, L40
—. 2012, MNRAS, 422, 1203
Boylan-Kolchin, M., Bullock, J. S., & Kaplinghat, M. 2012, MNRAS, 422, 1203
Boylan-Kolchin, M., et al., 2015, MNRAS, 453, 1503
Brinchmann, J., et al., 2004, MNRAS, 351, 1151
Brodie & Larsen 2002, AJ, 124, 1410
Bromm, V., & Loeb, A. 2003, Nature, 425, 812
Brooks A.M., et al. 2009, ApJ, 694, 396
Brooks, A. M., Kuhlen, M., Zolotov, A., & Hooper, D. 2013, ApJ, 765, 22
Brott I., et al. 2011, A&A, 530, 116
Brown, W. R., Geller, M. J., Kenyon, S. J., & Diaferio, A. 2010, AJ, 139, 59
Brownstein, J., et al. 2012, ApJ, 744, 41
Buchert, T., Carfora, M., 2008, CQGra, 25, 5001
Buchert, T., et al. 2000, PhRvD, 62, 3525
Buchhave L.A., & Latham, D. 2015, ApJ, 808, 187
Buchhave L.A., et al. 2012, Nature, 486, 375
Buchhave L.A., et al. 2014, Nature, 509, 593
Bundy, K., Bershady, M. A., Law, D. R., et al. 2015, ApJ, 798, 7
Butcher, H., & Oemler, A. 1978, ApJ, 219, 18
Cabre, A., & Gaztanaga, E. 2009, MNRAS 396, 1119
Cappellari et al., 2011b, MNRAS, 416, 1680
Cappellari, M. 2016, ArXiv e-prints, arXiv:1602.04267
Cappellari, M., & Copin, Y. 2003, MNRAS, 342, 345
Cappellari, M., Emsellem, E., Krajnović, D., et al. 2011a, MNRAS, 413, 813
Cappellari, M., McDermid, R. M., Alatalo, K., et al. 2012, Nature, 484, 485
Cappellari, M., Scott, N., Alatalo, K., et al. 2013, MNRAS, 432, 1709
Carlberg, R. G., et al. 2000, ApJL, 532, L1
Carlberg, R., 2009, ApJL 705, L223
Carlberg, R., 2012, ApJ 748, 20
Carlberg, R., et al. 2015, ApJ 808, 15
Carlin, J. L., Grillmair, C. J., Muñoz, R. R., Nidever, D. L., & Majewski, S. R. 2009, ApJ, 702, L9
Carollo D., et al. 2014 ApJ 788, 180
Carrero G., et al. 2015 AJ 149, 12
Carrick, J., et al. 2015, MNRAS, 450, 317
Carter, D., et al. 2008, ApJS, 176, 424
Casteels, K. et al. 2014, MNRAS 445, 1157
Cattaneo, A., Dekel, A., Faber, S. M., & Guiderdoni, B. 2008, MNRAS, 389, 567
Caucci, S., et al., 2008, MNRAS, 386,211
Cayrel et al. 2004, A&A, 416, 1117
Cen, R. 2011, ApJ, 741, 99
Chapman, S., et al. 2006, ApJ, 653, 255
Chen, C., Zhang, J., & Vogeley, M.S. 2009, Mapping the Global Impact of Sloan Digital Sky Survey
Chen, H.-C., et al. 2013, A&A, 550, 62
Chen, H.-C., et al. 2014, MNRAS, 438, 1435
Cheng et al. 2012, ApJ, 746, 149
Chornock, R., et al. 2013, ApJ, 767, 162
Chornock, R., et al. 2014, ApJ, 780, 44
Chou, R., et al., 2011, AJ, 141, 87
Cirasuolo, M., Afonso, J., Bender, R., Bonifacio, P., Evans, C., et al. 2012, arXiv:1208.5780
Cisewski, J., et al., 2014, MNRAS, 440, 2599
Clark, P.C., Glover, S.C.O., Smith, R. J.,et al. 2011, Science, 331, 1040
Clifton, T., et al., 2012, PhR, 513, 1
Codis S., et al. 2012, MNRAS, 427, 3320
Codis S., et al. 2015, MNRAS, 452, 3369
Cohen J.G. et al. 2013 ApJ 778, 56
Coil, A. L., et al. 2006, ApJ, 644, 671
Cole, A. et al. 2014 ApJ, 795, 54
Cole, S., Norberg, P., Baugh, C. M., et al. 2001, MNRAS, 326, 255
Colless, M., Dalton, G., Maddox, S., et al. 2001, MNRAS, 328, 1039
Colless, M., et al. 2001, MNRAS, 328, 1039
Collins, M. L. M., Chapman, S. C., Rich, R. M., Ibata, R. A., Martin, N. F., Irwin, M. J., Bate, N. F., Lewis, G. F., Peñarrubia, J., Arimoto, N., Casey, C. M., Ferguson, A. M. N., Koch, A., McConnachie, A. W., & Tanvir, N. 2013, ApJ, 768, 172
Collins, M., et al., 2011, MNRAS, 413, 1548
Conroy, C., Wechsler, R. H., & Kravtsov, A. V. 2006, ApJ, 647, 201
Conselice, C. J., Bershady, M. A., Dickinson, M., & Papovich, C. 2003, AJ, 126, 1183

Conselice, C. J., Bershady, M. A., Dickinson, M., & Papovich, C. 2003, AJ, 126, 1183
Conselice, C., 2014, ARAA, 52, 291
Contreras, C., et al. 2013, MNRAS, 430, 924
Cooke, R. et al., 2011, MNRAS, 417, 1534
Cooper, M. C., et al. 2006, MNRAS, 370, 198
Cooper, M. C., et al. 2010, MNRAS, 409, 337
Cortese, L., Fogarty, L. M. R., Ho, I.-T., et al. 2014, ApJL, 795, L37
Couch, W. J., Barger, A. J., Smail, I., Ellis, R. S., & Sharples, R. M. 1998, ApJ, 497, 188
Coupon, J., et al. 2012, A&A, 542, 5
Cox, T.J., et al., 2008, MNRAS, 384, 386
Crighton, N. 2011, MNRAS, 414, 28
Crighton, N., et al., 2003, MNRAS, 345, 243
Croom, S. M., Lawrence, J. S., Bland-Hawthorn, J., et al. 2012, MNRAS, 421, 872
Croton, D. J., Springel, V., White, S. D. M., et al. 2006, MNRAS, 365, 11
Croxall, K., et al., 2009, ApJ, 705, 723
Daddi, E., et al. 2007, ApJ, 670, 156
Dalcanton, J. J. & Hogan, C. J. 2001, ApJ, 561, 35
Datson J. et al. 2014 MNRAS, 439, 1028
Davies, L. J. M., Robotham, A. S. G., Driver, S. P., et al. 2015, MNRAS, 452, 616
Davies, L., et al., 2015, MNRAS, 452, 616
Davis, M., Efstathiou, G., Frenk, C. S., & White, S. D. M. 1985, ApJ, 292, 371 de Blok, W. J. G. 2010, Advances in Astronomy, 2010
Davis, T. A., & Bureau, M. 2016, MNRAS, 457, 272
Davis, T. A., Alatalo, K., Sarzi, M., et al. 2011, MNRAS, 417, 882
Davis, T., et al., 2011, ApJ, 741, 67
de Blok, W.J.G., & Walter, F. 2006, AJ, 131, 343
de Burgh-Day, C. O., Taylor, E. N., Webster, R. L., & Hopkins, A. M. 2015a, MNRAS, 451, 2161 —. 2015b, PASA, 32, 40
de la Torre, S., et al. 2013, A&A, 557, 54
De Lucia, G., et al. 2006 , MNRAS, 366, 499
De Rosa, G. et al. 2015, ApJ, 806, 128
De Silva, G., et al. 2007, AJ, 133, 1161
De Silva, G., et al. 2009, A&A, 500, 25
Deason, A. J., Belokurov, V., Evans, N. W., Koposov, S. E., Cooke, R. J., Peñarrubia, J., Laporte, C. F. P., Fellhauer, M., Walker, M. G., & Olszewski, E. W. 2012, MNRAS, 425, 2840
Dekel A., et al. 2009, ApJ, 703, 785
Dekel, A., et al. 2009, Nature, 457, 451
Delubac, T., et al. 2015, A&A, 574, 59
Denney, K. D. 2012, ApJ, 759, 44
Denney, K. D. et al. 2010, ApJ, 721, 715
Diemand, J., Moore, B., & Stadel, J. 2005, Nature, 433, 389
Flores, R. A. & Primack, J. R. 1994, ApJ, 427, L1
Dorman, C., et al., 2012, ApJ, 752, 147
Drake A.J., et al. 2013 ApJ 763, 32
Dressler, A. 1980, ApJ, 236, 351
Dressler, A., Oemler, Jr., A., Couch, W. J., et al. 1997, ApJ, 490, 577
Drinkwater et al. 2003, Nature, 423, 519
Driver, S. P., Hill, D. T., Kelvin, L. S., et al. 2011, MNRAS, 413, 971
Driver, S. P., Hill, D. T., Kelvin, L. S., et al. 2011, MNRAS, 413, 971
Driver, S. P., Wright, A. H., Andrews, S. K., et al. 2016, MNRAS, 455, 3911
Driver, S., & De Propris, R. 2003, A&AS, 285, 175
Driver, S., & Robotham, A., 2010, MNRAS, 407, 2131
Driver, S., et al., 2013, MNRAS, 430, 2622
Driver, S., et al., 2016, MNRAS, 455, 3911
Drlica-Wagner, A., & DES Collaboration. 2015, ApJ, 813, 109
Duc, P.A., et al., 2015, MNRAS, 446, 120
Durrell, P., et al. 2014, ApJ, 794, 103
Edelson, R., et al. 2015, arXiv:1501.05951
Eisenstein D.J., et al. 2006 ApJS, 167, 40
Eke, V. R., Baugh, C. M., Cole, S., Frenk, C. S., & Navarro, J. F. 2006, MNRAS, 370, 1147
Ekstrom S., et al. 2012, A&A, 537, 146
El-Badry et al., 2016, ApJ, 820, 131
Ellingson, E., Lin, H., Yee, H. K. C., & Carlberg, R. G. 2001, ApJ, 547, 609
Ellis, G.F.R, Buchert, T., 2005, PhLA, 347, 38
Ellis, G.F.R., 1984, in General relativity and gravitation, p. 215-288
Ellison, S. L., Patton, D. R., Simard, L., & McConnachie, A. W. 2008, AJ, 135, 1877
Ellison, S., et al., 2011, MNRAS, 418, 2043
Ellison, S., et al., 2013, MNRAS, 435, 3627
Ellison, S., et al., 2015, MNRAS, 448, 221
Ellison, S., Simard, L., Cowan, N., Baldry, I., Patton, D., McConnachie, A. 2009 MNRAS, 396, 1257
Elmegreen, D. et al. 2015, Optimising the US Ground Based Optical and Infrared Astronomy Systems
Emsellem, E., Cappellari, M., Krajnović, D., et al. 2011, MNRAS, 414, 888
Erkal & Belokurov 2015, MNRAS 450, 1136
Fakhouri, O., et al. 2010, MNRAS, 406, 2267
Faucher-Giguere C.-A., Prochaska J.X., Lidz A., Hernquist L., & Zaldarriaga M. 2008, ApJ, 681, 831
Federman, S.R. 1982, ApJ, 257, 125
Feix, M., et al., 2015, PhRvL, 115, 1301
Ferramacho et al. 2014, MNRAS, 442, 2511
Ferrarese, L., et al. 2016, ApJ, arXiv: 1604.06462
Fillingham, S., et al. 2015, MNRAS, 454, 2039
Firnstein M., Przybilla N., 2012 A&A, 543, 80
Fischer, D. & Valenti, J., 2005, ApJ, 622, 1102
Fixsen, D., et al., 1996, ApJ, 473, 576
Fleming et al. 2015, AJ, 149, 143
Fliri, J., & Vals-Gabaud, D. 2012, Ap&SS, 341, 57

Fogarty, L. M. R., Scott, N., Owers, M. S., et al. 2014, MNRAS, 443, 485
—. 2015, MNRAS, 454, 2050
Forero-Romero, J., et al., 2009, MNRAS, 396, 1815
Foucaud, S., et al. 2010, MNRAS, 406, 147
Francois, P., et al. 2007, A&A, 476, 935
Frank, S. & Peroux, C. 2010, MNRAS, 420, 1731
Fraternali, F., Tolstoy, E., Irwin, M. J., & Cole, A. A. 2009, A&A, 499, 121
Frebel A. et al. 2014 ApJ 786, 74
Frebel A., Norris J.E., 2015, ARA&A, 53, 631
Frebel, A., Kirby, E. N., & Simon, J. D. 2010, Nature, 464, 72
Freeman, K., & Bland-Hawthorn, J. 2002, ARA&A, 40, 487
Friedman S,D., York, D.G., McCall, B.J., Dahlstrom, J. Sonnentrucker, P., et al. 2011, ApJ, 727, 33
Friel E.D., et al. 2010 AJ 139, 1942
Friel et al., 2014, A&A, 563, 117
Frieman, J.A., Turner, M.S., & Huterer, D. 2008, ARA&A 46, 385
Frinchaboy P. et al. 2013 ApJ 777, 1
Frinchaboy, P. M., Majewski, S. R., Muñoz, R. R., Law, D. R., Łokas, E. L., Kunkel, W. E., Patterson, R. J., & Johnston, K. V. 2012, ApJ, 756, 74
Fulbright et al. 2007, ApJ, 661, 1152
Fulbright J.P., et al. 2006, ApJ, 636, 821
Furlanetto S.R., & Oh S.P. 2008, ApJ, 681, 1
Furlong, M. et al. 2015 MNRAS 450, 4486
Gao, L., & White, S. D. M. 2007, MNRAS, 377, L5
Garcia Perez A.E., et al. 2013 ApJ 767, 9
Garrison-Kimmel, et al., 2014, MNRAS, 438, 2578
Garrison-Kimmel, S., Rocha, M., Boylan-Kolchin, M., Bullock, J. S., & Lally, J. 2013, MNRAS, 433, 3539
Gauthier, J.-R., et al. 2009, ApJ, 702, 1
Geha, M., Blanton, M., Yan, R. & Tinker, J. 2012 ApJ 757, 85
Geha, M., Guhathakurta, P., Rich, R. M., & Cooper, M. C. 2006, AJ, 131, 332
Geha, M., van der Marel, R. P., Guhathakurta, P., Gilbert, K. M., Kalirai, J., & Kirby, E. N. 2010, ApJ, 711, 361
Genzel, R., Newman, S., Jones, T., et al. 2011, ApJ, 733, 101
Geringer-Sameth, A., Koushiappas, S. M., & Walker, M. 2015, ApJ, 801, 74
Geringer-Sameth, A., Koushiappas, S. M., & Walker, M. G. 2015, Phys. Rev. D, 91, 083535
Ghezzi L., et al. 2010 ApJ, 720, 1290
Gilbank, D. G., et al. 2010, MNRAS, 405, 2419
Gilmore G., et al. 2012, ESO Messenger, 147, 25
Gilmore, G., Wilkinson, M. I., Wyse, R. F. G., Kleyna, J. T., Koch, A., Evans, N. W., & Grebel, E. K. 2007, ApJ, 663, 948
Giovanelli, R., et al. 2005, AJ, 130, 2598
Godard, B., Falgarone, E., & Pineau Des Forets, G. 2009, A&A, 495, 847
Gonzalez et al. 2011 A&A 530, 54
Gonzalez, G., 1997, MNRAS, 284, 403
Governato, F., Brook, C., Mayer, L., Brooks, A., Rhee, G., Wadsley, J., Jonsson, P., Willman, B., Stinson, G., Quinn, T., & Madau, P. 2010, Nature, 463, 203
Governato, F., Zolotov, A., Pontzen, A., Christensen, C., Oh, S. H., Brooks, A. M., Quinn, T., Shen, S., & Wadsley, J. 2012, MNRAS, 2697
Green, A. M., Hofmann, S., & Schwarz, D. J. 2004, MNRAS, 353, L23
Green, G., et al. 2015, ApJ, 810, 25
Greene, J. E., & Ho, L. C. 2004, ApJ, 610, 722
Greif, T.H., Springel, V., White, S.D.M., et al. 2011, ApJ, 737, 75
Grillmair & Dionatos 2006, ApJ 643, L17
Gronnow, A., et al. 2015, MNRAS, 451, 4005
Guerou, A., et al. 2015, ApJ, 804, 70
Gunawardhana et al. 2011, MNRAS 415, 1647
Gunn, J. E., & Gott, III, J. R. 1972, ApJ, 176, 1
Gunn, J. E., Lee, B. W., Lerche, I., Schramm, D. N., & Steigman, G. 1978, ApJ, 223, 1015
Guzzo, L., Pierleoni, M., Meneux, B., Branchini, E., Le F\`evre, O., et al., 2008, Nature, 451, 541
Guzzo, L., Strauss,M. A., Fisher, K. B., Giovanelli, R., & Haynes, M. P. 1997, ApJ, 489, 37
Hahn, O., et al. 2007, MNRAS, 375, 489
Hahn, O., et al. 2010, MNRAS, 405, 274
Haines, C., et al. 2011, MNRAS, 412, 127
Hansen B. et al. 2004 ApJS 155 551
Hansen B. et al. 2007 ApJ 671 380
Harbeck, D., Grebel, E. K., Holtzman, J., Guhathakurta, P., Brandner, W., Geisler, D., Sarajedini, A., Dolphin, A., Hurley-Keller, D., & Mateo, M. 2001, AJ, 122, 3092
Hawkins, E., et al. 2003, MNRAS, 346, 78
Heckman et al., 1993, ASSL, 188,455
Heckman, T. M., Armus, L., & Miley, G. K. 1990, ApJS, 74, 833
Heger A., Woosley S.E., 2010 ApJ 724, 341
Hempel, M. et al., 2014, ESO Messenger, 155, 29
Henry, A., et al., 2013, ApJ, 776, 27
Herwig F., 2005, ARA&A, 43, 435
Higgs, C., R., et al., 2016, MNRAS, 458, 1678
Hill, V., & the DART collaboration, 2012, in Galactic Archaeology: Near Field Cosmology and the formation of the Milky Way, p297
Ho, I.-T., Medling, A. M., Bland-Hawthorn, J., et al. 2016, MNRAS, 457, 1257
Ho, N., Geha, M., Munoz, R. R., Guhathakurta, P., Kalirai, J., Gilbert, K. M., Tollerud, E., Bullock, J., Beaton, R. L., & Majewski, S. R. 2012, ApJ, 758, 124
Hobbs, L.M., York, D.G., Thornburn, J.A., Snow, T.P., Bishof, M., et al 2009, ApJ, 705, 32
Hoffman, Y., et al., 2012, MNRAS, 425, 2049

Hoffman, Y., et al., 2015, MNRAS, 449, 4494
Hofmann, S., Schwarz, D. J., & Stöcker, H. 2001, Phys. Rev. D, 64, 083507
Hogg, D., et al. 2004, ApJ, 601, 29
Hogg, D., et al. 2016, arXiv:160105413
Holoien, T., et al. 2014, MNRAS, 445, 3263
Hopkins, A. M., & Beacom, J. F. 2006, ApJ, 651, 142
Hopkins, A. M., & Beacom, J. F. 2006, ApJ, 651, 142
Hopkins, P. F., Bundy, K., Hernquist, L., Wuyts, S., & Cox, T. J. 2010, MNRAS, 401, 1099
Hopkins, P. F., Croton, D., Bundy, K., et al. 2010, ApJ, 724, 915
Hopkins, P. F., Richards, G. T., & Hernquist, L. 2007, ApJ, 654, 731
Horne, K., et al. 2004, PASP, 116, 465
Howes, L., et al. 2015, Nature, 527, 484
Huff, E. M., Krause, E., Eifler, T., George, M. R., & Schlegel, D. 2013, ArXiv e-prints, arXiv:1311.1489
Hunter I., et al. 2009, A&A, 496, 841
Huxor et al. 2005, MNRAS, 360, 1007
Ibata, R., et al. 2002, MNRAS 332, 915
Ibata, R., et al., 2001, Nature, 412, 49
Ivans I.I., et al. 2003 ApJ 592, 906
Jang & Lee 2014, ApJ, 795, 6
Jeltema, T. & Profumo, S. 2016, MNRAS
Jeong, D., Schmidt, F., Hirata, C., 2012, PhRvD, 85, 3504
Ji et al., 2016, Nature, 531, 610
Ji, I., Peirani, S.,& Yi, S. 2014, A&A, 566, 97
Jimenez-Vicente, J., et al. 2012, ApJ, 751, 106
Jogee, S., Miller, S. H., Penner, K., et al. 2009, ApJ, 697, 1971
Johnson C.I., et al. 2014 AJ 148, 67
Johnson, A., et al., 2014, MNRAS, 444, 3926
Johnston, E. J., Aragón-Salamanca, A., & Merrifield, M. R. 2014, MNRAS, 441, 333
Johnston, K., et al. 1999, ApJ, 512, 109
Johnston, K., et al. 2002, ApJ 570, 656
Just, D. W., Zaritsky, D., Sand, D. J., Desai, V., & Rudnick, G. 2010, ApJ, 711, 192
Kaiser, N., 1987, MNRAS, 227, 1
Kalirai J.S., 2012, Nature, 486, 90
Kalirai, J., et al. 2006, ApJ, 648, 389
Kampczyk, P., et al., 2013, ApJ, 762, 43
Kaplan, K. et al., 2010, PASP, 122, 619
Kaplinghat, M., & Strigari, L. E. 2008, ApJ, 682, L93
Kassin, S. A., Weiner, B. J., Faber, S. M., et al. 2007, ApJL, 660, L35
Kauffmann, G. et al. 2003 MNRAS 346, 1055
Kawata, D., & Mulchaey, J. S. 2008, ApJL, 672, L103
Keller, S., et al., 2014, Nature, 506, 463
Kennicutt, R. 1983 ApJ 272, 54
Keres, D., Katz, N.,Weinberg, D. H., & Dave, R. 2005, MNRAS, 363, 2
Kewley, L. J., Groves, B., Kauffmann, G., & Heckman, T. 2006, MNRAS, 372, 961
Kewley, L. J., Rupke, D., Zahid, H. J., Geller, M. J., & Barton, E. J. 2010, ApJL, 721, L48
Khabiboulline, E., et al. 2014, ApJ, 795, 62
Khare, P. et al., 2012, MNRAS, 419, 1028
Khochfar, S., Emsellem, E., Serra, P., et al. 2011, MNRAS, 417, 845
Kilic M., et al. 2012, MNRAS, 423, 132
Kilic M., et al. 2005 ApJ 633, 1126
King, A. L., et al. 2014, MNRAS, 441, 3454
King, A. L., et al. 2015, arXiv:1504.03031
Kirby E. et al. 2008, ApJ, 685, 43
Kirby E. et al. 2010, ApJS, 191, 352
Kirby, E. N., Boylan-Kolchin, M., Cohen, J. G., Geha, M., Bullock, J. S., & Kaplinghat, M. 2013, ApJ, 770, 16
Kirby, E. N., Cohen, J. G., Simon, J. D., & Guhathakurta, P. 2015, ApJ, 814, L7
Kistler, M. D., Yuksel, H., Beacom, J. F., Hopkins, A. M., & Wyithe, J. S. B. 2009, ApJL, 705, L104
Klypin, A., et al., 2016, MNRAS, 457, 4340
Klypin, A., Kravtsov, A. V., Valenzuela, O., & Prada, F. 1999, ApJ, 522, 82
Klypin, A., Kravtsov, A. V., Valenzuela, O., & Prada, F. 1999, ApJ, 522, 82
Knobel, C., et al. 2009, ApJ, 697, 1842
Koch, A., Kleyna, J. T., Wilkinson, M. I., Grebel, E. K., Gilmore, G. F., Evans, N. W., Wyse, R. F. G., & Harbeck, D. R. 2007, AJ, 134, 566
Koch, A., Wilkinson, M. I., Kleyna, J. T., Irwin, M., Zucker, D. B., Belokurov, V., Gilmore, G. F., Fellhauer, M., & Evans, N. W. 2009, ApJ, 690, 453
Koda, J., et al., 2014, MNRAS, 445, 4267
Koester D., et al. 2014, A&A, 566, 34
Koopmans, L.V.E., Treu, T., Bolton, A.S., Burles, S., & Moustakas, L.A. 2006, ApJ, 649, 599
Koposov, S. E., Belokurov, V., Torrealba, G., & Evans, N. W. 2015, ApJ, 805, 130
Koposov, S. E., Gilmore, G., Walker, M. G., Belokurov, V., Wyn Evans, N., Fellhauer, M., Gieren, W., Geisler, D., Monaco, L., Norris, J. E., Okamoto, S., Peñarrubia, J., Wilkinson, M., Wyse, R. F. G., & Zucker, D. B. 2011, ApJ, 736, 146
Koposov, S., et al., 2008, ApJ, 686, 279
Korista, K. T. & Goad, M. R. 2004, ApJ, 606, 749
Kormendy, J., Ho, L., 2013, ARA&A, 51, 511
Koss, M., et al., 2010, ApJ, 726, 125
Kraljic, D., 2015, MNRAS, 451, 3393
Kravtsov 2010, AdAst 2010, 8
Krumholz, M. R., & Dekel, A. 2012, ApJ, 753, 16
Kubo, J., et al., 2007, ApJ, 671, 466
Kudritzki R.P., et al. 2012 ApJ, 747, 15
Kuzio de Naray, R., McGaugh, S. S., & de Blok, W. J. G. 2008, ApJ, 676, 920
Kuzma et al. 2015, MNRAS 446, 3297

Laevens, B. P. M., Martin, N. F., Bernard, E. J., Schlafly, E. F., Sesar, B., Rix, H.-W., Bell, E. F., Ferguson, A. M. N., Slater, C. T., Sweeney, W. E., Wyse, R. F. G., Huxor, A. P., Burgett, W. S., Chambers, K. C., Draper, P. W., Hodapp, K. A., Kaiser, N., Magnier, E. A., Metcalfe, N., Tonry, J. L., Wainscoat, R. J., & Waters, C. 2015, ApJ, 813, 44
Lake, G. 1990, Nature, 346, 39
Lallement, R., et al., 2014, A&A, 561, 91
Lambas, D., et al., 2003, MNRAS, 346, 1189
Lambas, D., et al., 2012, A&A, 539, 45
Landecker T., Routledge, D., Reynolds, S.P., Smegal, R.J., Borkowski, K.J., et al. 1999, ApJ, 527, 866
Lara-Lopez, M. et al. 2013a, MNRAS 433, 35
Lara-Lopez, M. et al. 2013b, MNRAS 434, 451
Larson, R. B., Tinsley, B. M., & Caldwell, C. N. 1980, ApJ, 237, 692
Larson, R. B., Tinsley, B. M., & Caldwell, C. N. 1980, ApJ, 237, 692
Laureijs, R., et al. 2009, arXiv: 0912.0914
Laurent, P., et al. 2016, arXiv: 1602.09010
Le Fevre, O., et al. 2005, A&A, 439, 877
Le Fevre, O., et al. 2013, A&A, 559, 14
Leauthaud, A., et al. 2012a, ApJ, 744, 159
Leauthaud, A., et al. 2012b, ApJ, 746, 95
Lecureur A., et al. 2007, A&A, 465, 799
Lee et al. 2014a, ApJ, 795, 12
Lee, H., et al. 2006, ApJ, 647, 970
Lee, KG et al., 2014b, ApJ, 788, 49
Lelli, F., et al. 2014, MNRAS, 445, 1694
Letarte, B., et al., 2010, A&A, 523, 17
Lewis, G. F., Ibata, R. A., Chapman, S. C., McConnachie, A., Irwin, M. J., Tolstoy, E., & Tanvir, N. R. 2007, MNRAS, 375, 1364
Lewis, I., Balogh, M., De Propris, R., et al. 2002, MNRAS, 334, 673
Li, C., et al. 2006, MNRAS, 368, 37
Lilly, S. J., et al. 2007, ApJS, 172, 70
Lin, L., et al. 2007, ApJL, 660, L51
Lin, L., et al. 2012, ApJ, 756, 71
Liu et al. 2015, ApJ, 812, 34
Liu Y.J., et al. 2014 ApJ 785, 94
Loeb, A. & Weiner, N. 2011, Physical Review Letters, 106, 171302
Loewenstein, M. & Kusenko, A. 2010, ApJ, 714, 652
Lotz, J. M., Primack, J., & Madau, P. 2004, AJ, 128, 163
Lotz, J., et al. 2010a, MNRAS, 404, 590
Lotz, J., et al. 2010b, MNRAS, 404, 575
Lu, T., Gilbank, D., McGee, S., Balogh, M. & Gallagher, S. 2012 MNRAS 420, 126
Lundgren, B. et al., 2009, ApJ, 698, 819
Madau, P., et al. 1999, ApJ, 514, 648
Madrid, J., & Macchetto, D. 2006, BAAS, vol 38, p. 1286
Madrid, J., & Macchetto, D. 2009, BAAS, vol 49, p.193
Maeder A., Meynet G., 2010, New AR, 54, 32
Magnelli, B., et al. 2011, A&A, 528, A35
Mahajan, S., et al. 2011, MNRAS, 416, 2882
Maier, C., et al., 2014, ApJ, 792, 3
Mandelbaum, R., Seljak, U., Kauffmann, G., Hirata, C. M., & Brinkmann, J. 2006, MNRAS, 368, 715
Marin, F., et al. 2013, MNRAS, 432, 2654
Marshall, D.J., Robin, A.C., Reyl\'e, C., Schultheis, M., & Picaud, S. 2006, A&A, 453, 635
Martig, M., Bournaud, F., Teyssier, R., & Dekel, A. 2009, ApJ, 707, 250
Martin, N. F., Ibata, R. A., Chapman, S. C., Irwin, M., & Lewis, G. F. 2007, MNRAS, 380, 281
Martin, N.F., Ibata, R.A., Chapman, S.C., Irwin, M., & Lewis, G.F. 2007, MNRAS, 380, 281
Martinez-Delgado, D., et al., 2010, AJ, 140, 962
Mateo, M. L. 1998, ARA&A, 36, 435
Mateo, M., Olszewski, E. W., & Walker, M. G. 2008, ApJ, 675, 201
Maulbetsch, C., et al. 2007, ApJ, 654, 53
Mayer et al. 2015, Proceedings of Advancing Astrophysics with the SKA, arXiv:1501.01082
McCarthy, I. G., et al. 2008, MNRAS, 383, 593
McConnachie, A. W. & Côté, P. 2010, ApJ, 722, L209
McConnachie, A. W., Peñarrubia, J., & Navarro, J. F. 2007, MNRAS, 380, L75
McConnachie, A.W., 2012, AJ, 144, 4
McConnachie, A.W., et al. 2005, MNRAS, 356, 979
McConnachie, A.W., et al. 2009, Nature, 461, 66
McDonald, P., & Seljak, U., 2009, JCAP, 10, 7
McElroy, M et al., 2015, MNRAS, 446, 2186
McGee, S. L., Balogh, M. L., Bower, R. G., Font, A. S., & McCarthy, I. G. 2009, MNRAS, 400, 937
McGee, S. L., et al. 2011, MNRAS, 413, 996
McGee, S., Bower, R. & Balogh, M. 2014 MNRAS 442, 105
McGee, S., Goto, R. & Balogh, M. 2014 MNRAS 438, 3188
McQuinn M. et al. 2009, ApJ, 694, 842
McQuinn M., & White M. 2011, MNRAS, 415, 2257
McWilliam A., et al. 2008, AJ, 136, 367
McWilliam, A., Smecker-Hane, T., 2005, ApJ, 622, 29
Melendez, J., et al., 2009, ApJ, 704, 66
Menard, B. & Peroux, C. 2003, A&A, 410, 33
Meneux, B., et al. 2008, A&A, 478, 299
Meneux, B., et al. 2009, A&A, 505, 463
Merluzzi, P., et al., 2010, MNRAS, 402, 753
Miceli M., et al 2010, MmSAI, 75, 282
Michel-Dansac, L., et al. MNRAS, 386, 82
Mihos et al. 2015, ApJL, 812, 34
Minniti, D., Lucas, P.W., Emerson, J.P., Saito, R.K., Hempel, M., et al. 2010, New Astronomy, 15, 433
Mistani, P., et al. 2016, MNRAS,455, 2323
Monreal-Ibero & Lallement, Mem. Dell. Soc. Ital., 2015

Moore, B. 1994, Nature, 370, 629
Moore, B., Ghigna, S., Governato, F., et al. 1999, ApJL, 524, L19
Moore, B., Ghigna, S., Governato, F., Lake, G., Quinn, T., Stadel, J., & Tozzi, P. 1999, ApJ, 524, L19
Mor, R. & Netzer, H. 2012, MNRAS, 420, 526
Morales, M.F. 2006, ApJ 650, L21
Mortazavi, S., et al. 2016, MNRAS, 455, 3058
Moster, B, et al., 2011, MNRAS, 415, 3750
Muñoz et al. 2005, ApJ, 631, L137
Muñoz et al. 2006, ApJ, 649, 201
Munoz et al. 2014, ApJS, 210, 4
Muñoz, R. R., Geha, M., & Willman, B. 2010, AJ, 140, 138
Murphy, M. et al., 2008, MNRAS, 384, 1053
Murphy, M., & Liske, J., 2004, MNRAS, 386, 1192
Naab, T., Oser, L., Emsellem, E., et al. 2014, MNRAS, 444, 3357
Nakamura, F., & Umemura, M. 2001, ApJ, 548, 19
Navarro, Frenk & White, 1997, ApJ, 490, 493
Navarro, Frenk & White. 1996, ApJ, 462, 563
—. 1997, ApJ, 490, 493
Navarro, J. F., Eke, V. R., & Frenk, C. S. 1996, MNRAS, 283, L72
Nikolic, B., et al. 2004, MNRAS, 355, 874
Noeske, K., et al., 2007, ApJ, 660, 47
Nomoto K., et al. 2013 ARAA 51, 457
Norberg, P., et al. 2002a, MNRAS, 332, 827
Norberg, P., et al. 2002b, MNRAS, 332, 827
Noterdaeme, P. et al., 2008, A&A, 481, 327
Noterdaeme, P. et al., 2009, 505, 1087
Noterdaeme, P. et al., 2010, A&A, 523, 80
Noterdaeme, P. et al., 2011, A&A, 526, 7
Noterdaeme, P. et al., 2015, A&A, 577, 24
Nusser, A., Davis, M., 1995, ApJ, 421, 1
Oh, S.-H., de Blok, W. J. G., Brinks, E., Walter, F., & Kennicutt, Jr., R. C. 2011, AJ, 141, 193
Önehag et al. 2014, A&A, 562, 102
Ota, K., 2008, ApJ, 677, 12
Ozbek, M., et al., 2016, MNRAS, 456, 3610
Pacaud, F., et al. 2016, A&A, in press, arXiv: 1512.04264
Pancoast, A., Brewer, B. J., et al. 2014, MNRAS, 445, 3073
Pandey, B., 2013, MNRAS, 430, 3376
Parfrey, K., et al., 2011, PhRvD, 83, 3511
Paris I., et al. 2011, A&A, 530, 50
Park, C., & Choi, Y.-Y. 2009, ApJ, 691, 1828
Park, C., & Hwang, H. S. 2009, ApJ, 699, 1595
Patton, D. R., Ellison, S. L., Simard, L., McConnachie, A. W., & Mendel, J. T. 2011, MNRAS, 412, 591
Patton, D. R., et al. 2002, ApJ, 565, 208
Patton, D.R. & Atfield, J.E., 2008, ApJ, 685, 235
Patton, D.R., et al., 2013, MNRAS, 433, 59
Pawlik, M., et al., 2016, MNRAS, 456, 3032
Peebles, P.J.E., 1980, The Large-Scale Structure of the Universe (Princeton, N.J., Princeton University Press)
Peñarrubia, J., Pontzen, A., Walker, M. G., & Koposov, S. E. 2012, ApJ, 759, L42
Peng et al. 2006, ApJ, 639, 838
Peng, Y.-j., Lilly, S. J., Renzini, A., & Carollo, M. 2012, ApJ, 757, 4
Peng, Y., et al. 2010, ApJ, 721, 193
Peng, Y., Lilly, S. J., Renzini, A., & Carollo, M. 2012, ApJ, 757, 4
Penny L., Gies D.R., 2009, ApJ, 700, 844
Pepe, F.A., Cristiani, S., Rebolo Lopez, R., Santos, N.C., Amorim, A., et al. 2010, SPIE, 7735, 14
Percival, W.J., & White, M., 2009, MNRAS, 393, 297
Perlmutter, S., Aldering, G., Goldhaber, G., Knop, R.A., Nugent, P., et al. 1999, ApJ, 517, 565
Peroux, C., et al., 2013, MNRAS, 436, 2650
Peterson, B. M. 2011, ArXiv1109.4181P
Peth, M., et al., 2016, MNRAS, 458, 963
Petitjean & Aracil, 2004, A&A, 422, 52
Pettini, M., et al., 2001, ApJ, 554, 981
Phleps, S., Peacock, J. A., Meisenheimer, K., & Wolf, C. 2006, A&A, 457, 145
Pichon et al. 2001, MNRAS, 326, 597
Pierre, M., et al. 2016, A&A, in press, arXiv: 1512.04317
Pietrukowicz, P., Udalski, A., Soszy\'nski, I., Nataf, D.M., Wyrzykowski, L., et al. 2012, ApJ, 750, 169
Pineau des Forets G., Flower, D.R., Hartquist, T.W., & Dalgarno, A. 1986, MNRAS, 220, 801
Ploeckinger, S., et al., 2014, MNRAS, 437, 3980
Plummer, H. C. 1911, MNRAS, 71, 460
Poggianti, B., et al. 2006, ApJ, 642, 188
Poggianti, B., et al. 2013, ApJ, 762, 77
Pohlen M., Trujillo I., 2006, A&A, 454, 759
Pontzen, A. & Governato, F. 2011, ArXiv:1106.0499
—. 2014, Nature, 506, 171
Popesso, P., et al., 2009, A&A, 494, 443
Prescott, M. et al. 2011 MNRAS, 417, 1374
Pritchet, C. & van den Bergh, S., 1999, AJ, 118, 883
Pritchet, C., et al. 2010, Unveiling the Cosmos: A vision for Canadian astronomy
Privon, G., et al., 2013, ApJ, 771, 120
Psychogyios, A., et al., 2016, arXiv: 1602.08500
Puspitarini L., et al. 2015, A&A, 573, 35
Querejeta, M., Eliche-Moral, M. C., Tapia, T., et al. 2015, A&A, 579, L2
Rafieferantsoa, M., et al., 2015, MNRAS, 453, 3980
Raichoor, A., & Andreon, S. 2012, A&A, 543, 19
Ramirez I., et al. 2010 A&A, 521, 33
Ransom, R.R., Kothes, R., Wolleben, M., & Landecker, T.L. 2010, ApJ, 724, 946
Rasmussen, J., Mulchaey, J., Bai, L, Ponman, T., Raychaudhury, S., Dariush, A. 2012 ApJ 757, 122
Rauch, M., 1998, ARA&A, 36, 267

Rauer, H., et al. 2014, arXiv: 1310.0696
Read, J. I. & Gilmore, G. 2005, MNRAS, 356, 107
Reines, A., Greene, J. & Geha, M. 2013 ApJ 775, 116
Reyes et al. 2010, Nature, 464, 256
Rich, J. A., Kewley, L. J., & Dopita, M. A. 2011, ApJ, 734, 87
Rich, J. A., Torrey, P., Kewley, L. J., Dopita, M. A., & Rupke, D. S. N. 2012, ApJ, 753, 5
Ridgway et al. 2014, ApJ, 796, 53
Riess, A.G., Filippenko, A.V., Challis, P., Clocchiatti, A., Diercks, a., et al. 1998, AJ, 116, 1009
Rines, K., Diaferio, A. & Natarajan, P. 2007 ApJ 657, 183
Robotham, A. S. G. et al., 2014, MNRAS, 444, 3986
Robotham, A. S. G., Driver, S. P., Davies, L. J. M., et al. 2014, MNRAS, 444, 3986
Robotham, A. S. G., Norberg, P., Driver, S. P., et al. 2011, MNRAS, 416, 2640
Robotham, A., et al, 2013, MNRAS, 431,167
Rocha, M., Peter, A. H. G., Bullock, J. S., Kaplinghat, M., Garrison-Kimmel, S., Oñorbe, J., & Moustakas, L. A. 2013, MNRAS, 430, 81
Rodighiero et al., 2011, ApJ, 739, 40
Rodrigues, T., et al. 2014, MNRAS, 445, 2758
Roederer, I., 2016, AJ, 151, 82
Rollinde, E., et al., 2013, MNRAS, 428, 450
Roskar R., et al. 2008, ApJ, 675, 65
Ross, N.P., da Angela, J., Shanks, T., Wake, D.A., Cannon, R.D., et al. 2007, MNRAS 381, 573
Ruchayskiy, O., Boyarsky, A., Iakubovskyi, D., Bulbul, E., Eckert, D., Franse, J., Malyshev, D., Markevitch, M., & Neronov, A. 2015, ArXiv:1512.07217
Rudie, G. et al., 2012, ApJ, 750, 67
Ryan-Weber, et al. 2009, MNRAS, 395, 1476
Sánchez, S. F., Kennicutt, R. C., Gil de Paz, A., et al. 2012, A&A, 538, A8
Sabater, J., et al., 2013, MNRAS, 430, 638
Sale, S., et al., 2014, MNRAS, 433, 2907
Salim, et al., 2007, ApJS, 173, 267
Samushia, L, et al. 2012, MNRAS, 420, 2102
Sartoris, B., et al. 2016, MNRAS, 459, 1764
Satyapal, S., Secrest, N., McAlpine, W., Ellison, S., Fischer, J. & Rosenberg, J. 2014 ApJ 784, 113
Sbordone et al. 2007, A&A, 465, 815
Scannapieco, C., Wadepuhl, M., Parry, O. H., et al. 2012, MNRAS, 423, 1726
Scannapieco, E., et al. 2006, MNRAS, 365, 615
Scharf, C., et al. 2000, ApJ, 544, 49
Schlegel D.J. et al. 2009, arXiv:0904.0468
Schmidt, K., et al. 2013, MNRAS, 432, 285
Schmidt, S., et al. 2014, ApJ, 781, 24
Schneider R., et al. 2014 MNRAS 442, 1440
Schneider, R., Omukai, K., Bianchi, S., & Valiante, R. 2012, MNRAS, 419, 1566
Schoenrich, R., & Binney, J. 2009, MNRAS, 399, 1145
Schorck, T., et al., 2009, A&A, 507, 817
Scott, C. & Kaviraj, S., 2014, MNRAS, 437, 2137
Scott, N., Fogarty, L. M. R., Owers, M. S., et al. 2015, MNRAS, 451, 2723
Scoville, N., Aussel, H., Brusa, M., et al. 2007, ApJS, 172, 1
Scrimgeour, M., et al. 2012, MNRAS, 425, 116
Scudder, J., et al., 2012, MNRAS, 426, 549
Scudder, J., et al., 2015, MNRAS, 449, 3719
Searle, L., & Zinn, R. 1978, ApJ, 225, 357
Secrest, N., et al., 2015, ApJ, 798, 38
Seljak 2009, Phys. Rev. L., vol. 102, Issue 2,
Seth et al. 2015, Nature, 513, 398
Seymour, N., et al. 2008, MNRAS, 386, 1695
Shafter, A.W., Darnley, M.J., Bode, M.F., & Ciardullo, R. 2012, ApJ, 752, 156
Shakura, N. I. & Sunyaev, R. A. 1973, A&Ap, 24, 337
Shapiro, K. L., Genzel, R., Förster Schreiber, N. M., et al. 2008, ApJ, 682, 231
Sharp, R., & Parkinson, H. 2010, MNRAS, 408, 2495
Shen, Y., et al. 2015, ApJS, 216, 4
Shetrone M.D., et al. 2003 AJ 125, 684
Siegal-Gaskins & Valluri 2008, ApJ 681, 40
Simon, J. D. & Geha, M. 2007, ApJ, 670, 313
Simon, J. D., Geha, M., Minor, Q. E., Martinez, G. D., Kirby, E. N., Bullock, J. S., Kaplinghat, M., Strigari, L. E., Willman, B., Choi, P. I., Tollerud, E. J., & Wolf, J. 2011, ApJ, 733, 46
Simon, J. D., Jacobson, H. R., Frebel, A., Thompson, I. B., Adams, J. J., & Shectman, S. A. 2015, ApJ, 802, 93
Simon, J.D., & Geha, M. 2007, ApJ, 670, 313
Smith, D., 2015, arXiv:1506.05630
Smith, R. 2014 MNRAS 443, 69
Smith, R. J., Lucey, J. R., & Carter, D. 2012a, MNRAS, 426, 2994
Smith, R. J., Lucey, J. R., Price, J., Hudson, M. J., & Phillipps, S. 2012b, MNRAS, 419, 3167
Sobral, D., et al., 2014, MNRAS, 437, 3516
Somerville, R. S., et al. 2004, ApJL, 600, L171
Sommeriva et al., 2012,A&A, 539, 136
Sousa S.G., et al. 2011, A&A, 533, 141
Speagle et al., 2014, ApJS, 214, 15
Spergel, D. N. & Steinhardt, P. J. 2000, Physical Review Letters, 84, 3760
Spitoni E., Matteucci F., 2011, A&A, 531, 72
Springel, V., DiMatteo, T., & Hernquist, L. 2005, ApJL, 620, L79
Springel, V., White, S. D. M., Jenkins, A., Frenk, C. S., Yoshida, N., Gao, L., Navarro, J., Thacker, R., Croton, D., Helly, J., Peacock, J. A., Cole, S., Thomas, P., Couchman, H., Evrard, A., Colberg, J., & Pearce, F. 2005, Nature, 435, 629
Srianand, R. et al. 2008, A&A, 482, 39
Steidel, C. C., et al. 2010, ApJ, 717, 289

Steigman, G., Dasgupta, B., & Beacom, J. F. 2012, Phys. Rev. D, 86, 023506
Stierwalt, S., et al. 2015, ApJ, 805, 2
Stott, J. P., Swinbank, A. M., Johnson, H. L., et al. 2016, MNRAS, 457, 1888
Suda T., Hirschi R., Fujimoto M.Y., 2011, ApJ, 741, 61
Sullivan, M., Le Borgne, D., Pritchet, C. J., et al. 2006, ApJ, 648, 868
Sweet, S., et al., 2014, ApJ, 782, 35
Swinbank, A. M., et al. 2006, MNRAS, 368, 1631
Takada et al. 2014, PASJ, 66, 1
Tal, T., et al., 2009, AJ, 138, 1417
Taylor et al. 2003, AJ, 125, 314
Tempel, E. et al. 2014 A&A 566, 1
Thacker et al. 2016, Unveiling the Cosmos: Canadian astronomy 2016 – 2020
Ting, Y.-S., et al., 2012, MNRAS, 421, 1231
Ting, Y.-S., et al., 2015, ApJ, 807, 104
Tinker, J., Wetzel, A., & Conroy, C. 2011, arXiv: 11075046
Tollerud, E. J., Beaton, R. L., Geha, M. C., Bullock, J. S., Guhathakurta, P., Kalirai, J. S., Majewski, S. R., Kirby, E. N., Gilbert, K. M., Yniguez, B., Patterson, R. J., Ostheimer, J. C., Cooke, J., Dorman, C. E., Choudhury, A., & Cooper, M. C. 2012, ApJ, 752, 45
Tollerud, E. J., Bullock, J. S., Strigari, L. E., & Willman, B. 2008, ApJ, 688, 277
Tolstoy, E., Hill V., Tosi M., 2009, ARA&A
Tolstoy, E., Irwin, M.J., Helmi, A., Battaglia, G., Jablonka, P., et al. 2004, ApJ, 617, L119
Tonnesen, S., Cen, R., 2012, MNRAS, 425, 2313
Torrealba, G., et al., 2016, MNRAS, 459, 2370
Torrey, P., et al. 2012, ApJ, 746, 108
Tremblay P.E., Bergeron P., Gianninas A., 2011, ApJ, 730. 1289
Tremonti, C. et al. 2004 ApJ 613, 898
Treu, T. 2010, ARA&A, 48, 87
Trujillo, I. et al. 2006, ApJ 650, 18
Tully, R.B., et al., 2014, Nature, 513, 71
Tumlinson, J., 2010, ApJ, 708, 1398
Turon, C., Primas, F., Binney, J., Chiappini, C., Drew, J, et al. 2008, Report by the ESA-ESO Working Group on Galactic Populations, Chemistry and Dynamics
van der Wel, A., & van der Marel, R. P. 2008, ApJ, 684, 260
van Dokkum, P. G., & Conroy, C. 2010, Nature, 468, 940
van Uitert, E., et al. 2016, MNRAS, arXiv:1601.06791
Veilleux, S., Cecil, G., & Bland-Hawthorn, J. 2005, ARAA, 43, 769
Venn K.A., et al. 2002, ApJ, 565, 571
Venn K.A., et al. 2004 AJ, 128, 1177
Venn K.A., et al. 2014, ApJ 791, 98
Vergely, J., et al., 2001, A&A, 366, 1016
Verhamme, A., et al., 2008, A&A, 491,89
Viel M., Bolton J.S., & Haehnelt M.G. 2009, MNRAS, 399, L39
Villa, E., et al., 2014, CQGra, 31, 4005
Vogelsberger, M., Zavala, J., & Loeb, A. 2012, MNRAS, 423, 3740
Vulcani, B., et al. 2010, ApJL, 710, L1
Wake, D. A., et al. 2011, ApJ, 728, 46
Walker M., Mateo M., Olszewski E.W., 2009a, AJ, 137, 3100
Walker, M. G. & Peñarrubia, J. 2011, ApJ, 742, 20
Walker, M. G., et al., 2016, ApJ, 819, 53
Walker, M. G., Mateo, M., & Olszewski, E. W. 2008, ApJ, 688, L75
Walker, M. G., Mateo, M., Olszewski, E. W., Bailey, III, J. I., Koposov, S. E., Belokurov, V., & Evans, N. W. 2015a, ApJ, 808, 108
Walker, M. G., Mateo, M., Olszewski, E. W., Gnedin, O. Y., Wang, X., Sen, B., & Woodroofe, M. 2007, ApJ, 667, L53
Walker, M. G., Mateo, M., Olszewski, E. W., Peñarrubia, J., Wyn Evans, N., & Gilmore, G. 2009b, ApJ, 704, 1274
Walker, M. G., Olszewski, E. W., & Mateo, M. 2015b, MNRAS, 448, 2717
Wang, L., Li, C., Kauffmann, G., & De Lucia, G. 2006, MNRAS, 371, 537
Wareing, C.J., Zijlstra, A.A., O'Brien, T.J., & Seibert, M. 2007, ApJ, 670, L125
Weiner, B., et al., 2009, ApJ, 692, 187
Weinmann, S. M., Kauffmann, G., von der Linden, A., & De Lucia, G. 2010, MNRAS, 406, 2249
Weinmann, S. M., van den Bosch, F. C., Yang, X., & Mo, H. J. 2006, MNRAS, 366, 2
Weinmann, S.M., et al. 2012, MNRAS, 426, 2797
Weinmann, S.M., Neistein, E., & Dekel, A. 2011, MNRAS, 1463
Wetzel, A. R., Hopkins, P. F., Kim, J.-h., Faucher-Giguere, C.-A., Keres, D., & Quataert, E. 2016, ArXiv:1602.05957
Wetzel, A., Tinker, J., Conroy, C. & van den Bosch, F. 2013, MNRAS 432, 336
Whitaker et al., 2012, ApJ, 754, 29
White, S. D. M. & Rees, M. J. 1978, MNRAS, 183, 341
White, S. D. M., & Frenk, C. S. 1991, ApJ, 379, 52
Wijesinghe, D. B., et al. 2012, ApJ, 423, 7639
Willman, B., Dalcanton, J.J., Martinez-Delgado, D., West, A.A., Blanton, M.R., et al. 2005, ApJ, 626, L85
Willman, B., Geha, M., Strader, J., Strigari, L. E., Simon, J. D., Kirby, E., Ho, N., & Warres, A. 2011, AJ, 142, 128
Wolf, C., et al. 2009, MNRAS, 393, 1302
Wolleben M., 2007, ApJ, 664, 349
Wolleben M., Fletcher, A., Landecker, T.L., Carretti, E., Dickey, J.M., et al. 2010, ApJ, 724, 48
Wong, K., et al., 2011, ApJ, 728, 119

Woods, D. & Geller, M., 2007, AJ, 134, 527
Woods, D., et al., 2010, AJ, 139, 1857
Xu, C. et al. 2012, ApJ, 760, 72
Yadav, J., et al. 2010, MNRAS, 405, 2009
Yamanoi, H., et al. 2012, AJ, 144, 40
Yang, X., Mo, H. J., & van den Bosch, F. C. 2008, ApJ, 676, 248
Yang, X., Mo, H. J., van den Bosch, F. C., et al. 2007, ApJ, 671, 153
Yong D. et al. 2013a ApJ 762, 26
Yong D. et al. 2013b ApJ 762, 27
Yoon et al. 2011, ApJ 731, 58
York, D. et al., 2006, MNRAS, 367, 945
York, D. G., Adelman, J., Anderson, Jr., J. E., et al. 2000, AJ, 120, 1579
York, D. G., Adelman, J., Anderson, Jr., J. E., et al. 2000, AJ, 120, 1579
Younger J.D., et al., 2007, ApJ, 670, 269
Zafar, T., et al., 2015, A&A, 584, 100
Zahid, H et al., 2013, ApJ, 771, 19
Zahid, H et al., 2014, ApJ, 791, 130
Zasowski, G., et al. 2015, ApJ, 798, 35
Zehavi, I., et al. 2002, ApJ, 571, 172
Zehavi, I., et al. 2011, ApJ, 736, 59
Zhang et al. 2007
Zibetti, S. ApJ, 2005, ApJ, 63L, 105
Zolotov, A., Brooks, A. M., Willman, B., Governato, F., Pontzen, A., Christensen, C., Dekel, A., Quinn, T., Shen, S., & Wadsley, J. 2012, ApJ, 761, 71
Zu, Y., Kochanek, C. S., & Peterson, B. M. 2011, ApJ, 735, 80
Zucker et al. 2006, ApJ, 643, L103